\documentclass{aastex6}
\usepackage{natbib}
\usepackage{graphicx}

\usepackage{graphics}
\usepackage{amssymb}
\usepackage{amsbsy}
\usepackage{amsfonts}
\usepackage{mathtools}
\usepackage{empheq}

\usepackage{amsmath}

\usepackage[makeroom]{cancel}
\usepackage{tikz}

\usepackage{lineno}

\usepackage{hyperref}
\usepackage{bookmark}

\newcommand{\bhat}{\boldsymbol{\hat{b}}}
\newcommand{\bu}{\boldsymbol{u}}
\newcommand{\bV}{\boldsymbol{v}}
\newcommand{\bc}{\boldsymbol{c}}
\newcommand{\bE}{\boldsymbol{E}}
\newcommand{\bb}{\boldsymbol{B}}
\newcommand{\bj}{\boldsymbol{j}}
\newcommand{\bw}{\boldsymbol{w}}

\newcommand{\tbc}{\tilde{\boldsymbol{c}}}
\newcommand{\tc}{\tilde{c}}

\newcommand{\tbh}{\tilde{\boldsymbol{h}}}
\newcommand{\tbbh}{\tilde{\bar{\bar{\boldsymbol{h}}}}}
\newcommand{\hbbh}{\hat{\bar{\bar{\boldsymbol{h}}}}}

\newcommand{\tby}{\tilde{\boldsymbol{y}}}
\newcommand{\ty}{\tilde{y}}

\newcommand{\tbu}{\tilde{\boldsymbol{u}}}

\newcommand{\bA}{\bar{\bar{\boldsymbol{A}}}}
\newcommand{\bp}{\bar{\bar{\boldsymbol{p}}}}
\newcommand{\bq}{\bar{\bar{\boldsymbol{q}}}}
\newcommand{\br}{\bar{\bar{\boldsymbol{r}}}}
\newcommand{\bI}{\bar{\bar{\boldsymbol{I}}}}
\newcommand{\bW}{\bar{\bar{\boldsymbol{W}}}}
\newcommand{\bPi}{\bar{\bar{\boldsymbol{\Pi}}}}
\newcommand{\bQ}{\bar{\bar{\boldsymbol{Q}}}}

\newcommand{\bD}{\bar{\bar{\boldsymbol{D}}}}

\newcommand{\bnu}{\bar{\nu}}
\newcommand{\bY}{\bar{\bar{\boldsymbol{Y}}}}

\newcommand{\vecq}{\vec{\boldsymbol{q}}}
\newcommand{\vecX}{\vec{\boldsymbol{X}}}
\newcommand{\vecQ}{\vec{\boldsymbol{Q}}}

\newcommand{\pr}{\partial}
\newcommand{\nn}{\nonumber}

\newcommand{\bx}{\boldsymbol{x}}
\newcommand{\by}{\boldsymbol{y}}
\newcommand{\bX}{\bar{\bar{\boldsymbol{X}}}}

\newcommand{\kpar}{k_\parallel}

\newcommand{\vpar}{v_\parallel}
\newcommand{\vperp}{v_\perp}

\newcommand{\bee}{\begin{eqnarray}}
\newcommand{\eee}{\end{eqnarray}}

\newcommand{\trace}{\textrm{Tr}}

\newcommand{\sign}{\textrm{sign}}
\newcommand{\erf}{\textrm{erf}}

\usepackage{chngcntr}
\counterwithin{table}{section}
\usepackage{soul}

\begin{document}
\title{Generalized Fluid Models of the Braginskii Type}


\author{P. Hunana\altaffilmark{1,2}, T. Passot\altaffilmark{3}, E. Khomenko\altaffilmark{1,2},
  D. Mart\'inez-G\'omez\altaffilmark{1,2}, M. Collados\altaffilmark{1,2},\\
  A. Tenerani\altaffilmark{4}, G. P. Zank\altaffilmark{5,6}, Y. Maneva\altaffilmark{7}, M. L. Goldstein\altaffilmark{8},   G. M. Webb\altaffilmark{5}.}

\altaffiltext{1}{Instituto de Astrof\'isica de Canarias (IAC), La Laguna, Tenerife, 38205, Spain; peter.hunana@gmail.com}
\altaffiltext{2}{Universidad de La Laguna, La Laguna, Tenerife, 38206, Spain}
\altaffiltext{3}{Universit\'e C\^{o}te d'Azur, Observatoire de la C\^{o}te d'Azur, CNRS, Laboratoire Lagrange, Bd de l'Observatoire,
  CS 34229, 06304 Nice cedex 4, France}
\altaffiltext{4}{Department of Physics, University of Texas, Austin, TX 78712, USA}
\altaffiltext{5}{Center for Space Plasma and Aeronomic Research (CSPAR),
  University of Alabama, Huntsville, AL 35805, USA}
\altaffiltext{6}{Department of Space Science, University of Alabama, Huntsville, AL 35899, USA}
\altaffiltext{7}{Royal Observatory of Belgium, 1180 Brussels, Belgium}
\altaffiltext{8}{Space Science Institute, Boulder, CO 80301, USA}
\altaffiltext{9}{Centre for mathematical Plasma Astrophysics, KU Leuven, 3001 Leuven, Belgium}

\begin{abstract}
Several generalizations of the well-known fluid model of Braginskii (Rev. of Plasma Phys., 1965) are considered. 
We use the Landau collisional operator and the moment method of Grad. We focus on the 21-moment model that is analogous to the Braginskii model, 
and we also consider a 22-moment model. Both models are formulated for general multi-species plasmas with arbitrary masses and temperatures, 
where all the fluid moments are described by their evolution equations. 
The 21-moment model contains two ``heat flux vectors'' (3rd and 5th-order moments) and two ``viscosity-tensors'' (2nd and 4th-order moments). 
The Braginskii model is then obtained as a particular case of a one ion-electron plasma with similar temperatures, 
with de-coupled heat fluxes and viscosity-tensors expressed in a quasi-static approximation. 
We provide all the numerical values of the Braginskii model in a fully analytic form
(together with the 4th and 5th-order moments). For multi-species plasmas, the model makes calculation of
transport coefficients straightforward. Formulation in fluid moments (instead of Hermite moments) is also suitable
for implementation into existing numerical codes. It is emphasized that it is the quasi-static approximation which makes
some Braginskii coefficients divergent in a weakly-collisional regime.
Importantly, we show that the heat fluxes and viscosity-tensors are coupled even in the linear approximation, and that the fully
contracted (scalar) perturbations of the 4th-order moment, which are accounted for in the 22-moment model,
modify the energy exchange rates.
We also provide several Appendices, which can be useful as a guide for deriving the Braginskii model with the moment method of Grad.
\end{abstract}

\maketitle

\vspace{1cm}

\newpage
\tableofcontents
\newpage
\section{Introduction}
The fluid model of \cite{Braginskii1958,Braginskii1965} represents a cornerstone of
plasma transport theory and it is used in many different areas, from solar physics to laboratory plasmas. 
The Braginskii model and its generalizations can be derived through two major classical routes 
1) Chapman-Enskog expansions \citep{ChapmanCowling1939} and 2) the moment method of \cite{Grad1949_1,Grad1949_2,Grad_1958}.
There also exists a more modern route with the projection operator \citep{Krommes2018a,Krommes2018b}.
Both classical routes were originally developed for gases, where the full Boltzmann collisional operator  has to be used.
As was shown by \cite{Landau1936,Landau1937}, for charged particles interacting through Coulomb collisions the Boltzmann
operator can be partially simplified, and this collisional operator is known as the Landau operator.
It is now well-established that for Coulomb collisions both Landau and Boltzmann operators yield the same results,
if in the Boltzmann operator one introduces integration cut-offs that remove the divergences in the same way as the Coulomb logarithm does.   
With the Landau operator, the Boltzmann equation is then typically called the Landau equation. By introducing Rosenbluth potentials
the Landau operator can be re-written into a general Fokker-Planck form, and the name Fokker-Planck equation is often used as well.  
Nevertheless, many authors use the Boltzmann operator during calculations even when Coulomb collisions are considered, because the simplification is not
exceedingly large. Braginskii used the Landau operator. 
Of course, both routes through Chapman-Enskog expansions and the moment method of Grad have sub-variations on how the methods are implemented 
that were developed over the years. For the Chapman-Enskog method where the distribution function is expanded
in Laguerre-Sonine polynomials, see for example \cite{Braginskii1958,Hinton1983,HelanderSigmar2002,Kunz2020}.


Here we use the moment method of Grad, which consists of expanding the distribution function in tensorial Hermite polynomials.
  Concerning only viscosity-tensors and heat fluxes (and neglecting fully contracted scalar perturbations and higher-order tensorial ``anisotropies'' as
  \cite{Balescu1988} calls them), the method of Grad consists of approximating the distribution function as a series 
\begin{equation}
f_a = f_a^{(0)} (1+\chi_a); \qquad \chi_a = \sum_{n=1}^{N}
  \Big[ {h}^{(2n)}_{ij} {H}^{(2n)}_{ij} +{h}^{(2n+1)}_{i} {H}^{(2n+1)}_{i} \Big], \label{eq:One}
\end{equation}
where $f_a^{(0)}$ is Maxwellian,  ``a'' is species index, indices i and j run from 1 to 3,
$H$ are Hermite polynomials and $h$ are Hermite moments. Matrices $h_{ij}^{(2n)}$ are traceless and
can be viewed as viscosity-tensors (stress-tensors),
and vectors ${h}^{(2n+1)}_{i}$ can be viewed as heat fluxes. The series is cut at some chosen $N$, and this distribution function is then used in the Landau
(or Boltzmann) equation, which is integrated to obtain a corresponding fluid model. The usual quasi-static approximation does not have to be applied, and one obtains
evolution equations for all the considered moments. 
For example, prescribing a strict Maxwellian with perturbation
$\chi_a=0$ (or equivalently $N=0$) represents the 5-moment model, with evolution equations for density, fluid velocity and scalar pressure (temperature),  
where stress-tensors and heat fluxes are zero. 
Prescribing $N=1$ represents the 13-moment model, which contains evolution equation for one traceless viscosity
tensor (5 independent components) and an evolution equation for one heat flux vector (3 independent components). This model thus contains the main
ingredients of the model of Braginskii, i.e. the usual viscosity tensor and the usual heat flux vector are present. However, 
prescribing a quasi-static approximation, yields for example the coefficient of parallel electron heat conductivity (for a one-ion electron plasma with ion charge $Z_i=1$)
with value 1.34 instead of the Braginskii value 3.16, meaning the model is not sufficiently precise.   
Prescribing $N=2$ represents 21-moment model, and this model can be viewed as containing evolution equations
for two viscosity tensors and two heat flux vectors. It can be shown that expressing the viscosity tensors and heat fluxes in a 
quasi-static approximation yields a model that is equivalent to \cite{Braginskii1965}. 
In fact, as pointed out for example by \cite{Balescu1988}, the
Hermite polynomials are directly related to the Laguerre-Sonine polynomials; see equation (\ref{eq:Relat1}), and thus
the Chapman-Enskog method and the moment method of Grad have to yield equivalent results at the end. In general, if both heat fluxes and viscosities are accounted for,
an $N$-Laguerre model therefore represents a $(5+8N)$-moment model. 
For a summary of various possible models, see Section \ref{sec:HermX} with Tables \ref{table:Chi1} \& \ref{tab:Chi2}.

Of course, the model of Braginskii can be generalized in many different ways. Naturally, one might focus at the case of
one-ion electron plasma considered by Braginskii, and increase the order of $N$ to study
convergence of transport coefficients with higher-order Laguerre (Hermite) schemes. Several studies of this kind were done in the past
(some numerically imprecise, some considering only unmagnetized plasmas, and some only ion charge $Z_i=1$). 
For example before Braginskii, \cite{Landshoff1949,Landshoff1951} calculated several transport coefficients with models from $N=1$ to $N=4$.
\cite{Kaneko1960} improved the numerical accuracy of Landshoff and also considered $N=5$.
\cite{KanekoTaguchi1978,KanekoYamao1980} performed large calculations with up to a $N=49$.
Perhaps the most comprehensive study to this date was done by \cite{JiHeld2013},
who studied convergence of all the transport coefficients with up to $N=160$. Other useful references can be found in
\cite{Epperlein1986}. The last two studies emphasize that while the transport coefficients parallel to the magnetic field (or for unmagnetized plasma)
converge rapidly for $N\ge 2$, this is not the case for some perpendicular transport coefficients.
For clarity, in the famous work of \cite{SpitzerHarm1953} and the previous work of \cite{Cohen1950} where only unmagnetized plasma is considered 
and viscosity-tensors are neglected,
the perturbation
$\chi_a$ which satisfies the Landau equation was found numerically, and
the model thus technically corresponds to $N=\infty$. Their work is criticized (even though a bit unfairly)
in the monograph of \cite{Balescu1988} (Part 1, p. 266), who calculated all the usual transport coefficients with the moment method of Grad for the
$N=2$ and $N=3$ cases (i.e. the 21-moment model and the 29-moment model). Note that the 3-Laguerre calculations of \cite{Balescu1988} were
shown to be incorrect by \cite{JiHeld2013}, who were able to trace the problem to his analytic collisional matrices (they have also corrected
coefficients in collisional matrices of \cite{Braginskii1958}, which were fortunately not used in his $N=2$ calculation).
That there is a problem with the $N=3$ transport coefficients of \cite{Balescu1988}
can be also seen for example by comparison with \cite{Kaneko1960}. 
Here we focus at the 2-Laguerre approximation used by \cite{Braginskii1965}, i.e. the 21-moment model,
with the goal to extract more physical information from that scheme.

For the 5-moment model and the 13-moment model, the method of Grad was explored in great detail by \cite{Burgers1969} and
\cite{Schunk1975,Schunk1977} (see other references therein). The Boltzmann operator was used and several interaction potentials were considered,
such as collisions between neutral particles (hard sphere interaction), between charges (Coulomb interaction), or an induced dipole interaction
when an ion polarizes a colliding neutral (so called Maxwell molecule interaction).
These models have two important properties that the
 Braginskii model does not have: 1) because the formulation is with evolution equations for stress-tensors
and heat fluxes rather than with quasi-static approximation, these models do not become divergent if a regime of low collisionality is encountered; 2) 
the formulation is as a general multi-fluid description with arbitrary masses $m_a$, $m_b$ and temperatures $T_a$, $T_b$.
Note that the review paper of \cite{Braginskii1965} also contains Section 7 about multi-component plasmas that is
often implicitly cited in the solar literature, but this section should be viewed as heuristic from a perspective that no
heat fluxes or stress-tensors were calculated. In plasma physics, the work of \cite{Braginskii1958,Braginskii1965} is celebrated
for his results for a one ion-electron plasma. 
Here we use the Landau operator and consider only Coulomb collisions. Nevertheless, we will employ the
21-moment model, and we thus improve the precision of the 13-moment model of \cite{Burgers1969}-\cite{Schunk1977} for this interaction potential,
so that the precision matches Braginskii. We will use a restriction that the relative drift velocity between two colliding species
must be small in comparison to their thermal speeds. The same restriction applies for the Braginskii model, for the Burgers-Schunk 13-moment model
(the exception is the Maxwell molecule interaction) and higher-order schemes.
For Coulomb collisions and hard-sphere collisions, only the simplest 5-moment model has been calculated
fully  analytically without this restriction \citep{Burgers1969,Tanenbaum1967,Schunk1977}, yielding the runaway effect. 

Several various multi-fluid descriptions with the level of precision of Braginskii have been considered in the past; 
see for example \cite{Hinton1983}, \cite{Zhdanov2002} (orig. pub. 1982),
\cite{JiHeld2006} (who actually consider general $N$), \cite{Simakov2014,Simakov2016a,Simakov2016b}, or for the case of neoclassical theory (toroidal geometry
applicable to tokamaks) see \cite{Hirshman1977,Hirshman1981}.  
Our model seems to be very close to the model of \cite{Zhdanov2002}, Chapter 8.1, who indeed uses the method of Grad and
calculates the 21-moment model with it. We did not verify full equivalence because of his puzzling notation.  
Even if equivalence is eventually shown for the case of small temperature differences between ions,
we consider a more general case where temperatures of all the species are arbitrary. 
Our clear formulation with fluid moments (instead of Hermite moments) might be also easier to implement into existing numerical
codes. Arbitrary temperatures were also considered by \cite{JiHeld2006}, but we did not verify equivalence with their model either.
We only verified equivalence with their model for the special case of a one ion-electron plasma with  small temperature differences of Braginskii,
by using collisional matrices from \cite{JiHeld2013}. 

Additionally, for all the considered moments we provide the left-hand-sides of our evolution equations in a fully non-linear form,
which is important for direct numerical simulations and  which are not typically given. An important difference then arises even at the linear level, because
 calculations are typically performed with de-coupled viscosity-tensors and heat fluxes, meaning that the two viscosity-tensors interact only with each other, and the
two heat fluxes interact only with each other. We consider coupling between heat fluxes and stress-tensors,
where (even at the linear level in a quasi-static approximation) a heat flux enters a stress-tensor and a stress-tensor enters a heat flux. 
Such couplings are often considered in the collisionless regime; see e.g. 
\cite{Macmahon1965},\cite{Smolyakov1985}, \cite{Ramos2005}, \cite{Goswami2005}, \cite{PSH2012}, \cite{Hunana2019b,Hunana2019a}, 
where the effect is important for example for the perpendicular fast mode, or for the growth rate of the firehose instability (see e.g. Figure 10
in \cite{Hunana2019a}). The coupling might be important also in the highly-collisional regime if sufficiently high frequencies
(or short wavelengths) are considered. The coupling was neglected by \cite{Braginskii1958,Braginskii1965} and \cite{SpitzerHarm1953,Spitzer1962},
and as an example we consider unmagnetized one ion-electron plasma in detail,
and we provide stress-tensors and heat fluxes where this coupling is taken into account.

The coupling between viscosity-tensors and heat fluxes then inevitably leads to the next step, by replacing equation (\ref{eq:One}) with
\begin{equation}
f_a = f_a^{(0)} (1+\chi_a); \qquad \chi_a = \sum_{n=1}^{N}
  \Big[ {h}^{(2n)}_{ij} {H}^{(2n)}_{ij} +{h}^{(2n+1)}_{i} {H}^{(2n+1)}_{i} + h^{(2n)}H^{(2n)} \Big], \label{eq:Two}
\end{equation}
where the scalar hermite moments $h^{(2n)}$ can be viewed as fully contracted (scalar) perturbations of fluid moments.
  The lowest-order moment $h^{(2)}=0$ and all higher-order ones are generally non-zero. Thus, prescribing $N=1$ still yields the 13-moment model, however,
  precribing $N=2$ now yields the 22-moment model. This model is a natural extension of the Braginskii model, because
  it takes into account fully contracted perturbations
  $\widetilde{X}^{(4)}_a =  m_a \int |\bc_a|^{4} (f_a -f_a^{(0)}) d^3v$ of the 4th-order fluid moment. 
 Accounting for the scalar perturbations according to (\ref{eq:Two}), for $N\ge 1$ an $N$-Laguerre model then represents a $(4+9N)$-moment model.
 Another possibility for writing the equation (\ref{eq:Two}) is to separate the matrices
  $\sum_{n=1}^{N} {h}^{(2n)}_{ij} {H}^{(2n)}_{ij}$, and for the vectors and scalars to write the sum from $n=0$, with an imposed requirement that
  $h^{(0)}=0$; $h^{(2)}=0$ \& $h^{(1)}_i=0$ (where the first one is non-trivial).
  This is for example the choice of \cite{Balescu1988}, p. 174, his equations (3.11) \& (3.16).

Finally, the main purpose of this work is to make
the moment method of Grad and the exciting work of Braginskii more understandable, which is reflected in our relatively lengthy Appendix. 

\vspace{0.5cm}

The entire paper is separated into eight Sections and fourteen Appendices. The main paper summarizes the obtained results,
while the Appendices provide the detailed calculations. 

In Section \ref{sec:Section2}, we formulate the entire 21-moment model. We start with a formulation valid for a general collisional operator $C(f_a)$, where both the
left-hand-sides and the collisional right-hand-sides of evolution equations are given in a fully non-linear form.
We then provide collisional contributions for arbitrary masses and temperatures calculated with the Landau operator. Collisional contributions
are calculated in  the usual semi-linear approximation where relative drifts between species are small in comparison to their thermal speeds
(i.e. the runaway effect is not considered), 
and the product of $f_a f_b$ is approximated as $f_a f_b= f_a^{(0)}f_b^{(0)} (1+\chi_a +\chi_b)$, where the ``cross''-contributions $\chi_a\chi_b$ are neglected.
We then provide a simplified model where differences in temperatures between species are small.
For clarity, we also reduce our model to the 13-moment model and we provide  a formulation that is more compact than the one given by
\cite{Burgers1969}-\cite{Schunk1977} (because we only consider Coulomb collisions). We then simplify the evolution equations of our
21-moment model into a semi-linear approximation where viscosity-tensors and heat fluxes are de-coupled, and these are used in Sections 3 \& 4.

In Section \ref{sec:ONEion}, we compare our model to \cite{Braginskii1965} by considering a one ion electron plasma with 
similar temperatures, i.e. where the temperature differences between species are
small with respect to their mean values. 
We provide all the transport coefficients in a fully analytic form, and we verify the entire Table II of \cite{Braginskii1965}
(two of his coefficients are not precise). Parallel electron coefficients (or equivalently for an unmagnetized plasma), can be also found in \cite{Simakov2014}.
We also provide analytic results for the viscosity of the 4th-order fluid moment and the heat flux of the 5th-order fluid moment, which are not
typically given. 

In Section \ref{sec:Electron}, we use the idea of for example \cite{Hinton1983,Zhdanov2002,Simakov2014} that because of the smallness of electron/ion mass ratios,
the \emph{electron} coefficients of Braginskii can be straightforwardly generalized to multiple ion species by introducing an
effective ion charge and effective ion velocity. All the electron analytic coefficients that are given in Section 3 are thus generalized to multi-ion species with
a simple transformation.

In Section \ref{sec:Couplingg1}, we discuss the coupling between viscosity tensors and heat fluxes.  
We provide evolution equations in the semi-linear approximation where this coupling is retained, and we introduce a technique
on how to split the moments into their first and second orders.

In Section \ref{sec:Couplingg}, we consider an example of an unmagnetized one ion-electron plasma and explicitly calculate the coupling of stress-tensors and
heat fluxes. All the results are given in a fully analytic form,  as well as with numerical values for the ion charge $Z_i=1$.

In Section \ref{sec:22main}, we  first formulate the fully non-linear 22-moment model for a general collisional operator. We then provide
  the multi-fluid collisional contributions calculated with the Landau operator in the semi-linear approximation, 
  and we show that the perturbations $\widetilde{X}^{(4)}$ modify the energy exchange rates. We also provide quasi-static solutions for a one ion-electron plasma,
  and we show that the perturbations $\widetilde{X}^{(4)}$ have their own heat conductivities.

In Section \ref{sec:Discussion}, we discuss various topics. 1) We discuss energy conservation. 2) We clarify that from a multi-fluid perspective 
the Braginskii choice of ion collisional time $\tau_i$ should be interpreted as $\tau_i=\tau_{ii}$, and not as $\tau_i=\sqrt{2}\tau_{ii}$.
3) To clarify the higher-order schemes and to double-check our evolution equations,
we calculate the fluid hierarchy for a general $N$, with an unspecified collisional operator.
4) We discuss irreducible and reducible Hermite polynomials and show that both yield the same results.
5) We provide fully non-linear Rosenbluth potentials for the 22-moment model, which might be useful in further studies of the runaway effect with this
scheme. 6) We discuss Hermite closures and their relation to fluid closures, which are required to close the fluid hierarchy.
We also correct our previous erroneous interpretation that Landau fluid closures are necessary to go beyond the 4th-order moment.
 7) We discuss the inclusion of gravity. 
8) We use our multi-fluid formulation to double-check the precision of $m_e/m_i$ expansions. We consider unmagnetized proton-electron plasma, and calculate
the transport coefficients exactly, without using the  smallness of $m_e/m_i$. 
9) We discuss limitations of our approach. 10) We provide conclusions, with examples where our model might be useful.

\vspace{0.5cm}

 Appendix \ref{sec:General} introduces the general concept of tensorial fluid moments and provides an
  evolution equation for an n-th-order fluid moment $\bX^{(n)}_a$ in presence of a general (unspecified) collisional operator, equation (\ref{eq:GenTensor}).
  This evolution equation also remains valid in presence of gravity; see the discussion in Section \ref{sec:gravity}.

 Appendix \ref{sec:Hermite} introduces the tensorial Hermite polynomials of \cite{Grad1949_1,Grad1949_2,Grad_1958}, and discusses in detail the construction 
  of perturbations around the Maxwellian distribution function, i.e. equations (\ref{eq:One}) \& (\ref{eq:Two}),
  which are summarized in Tables \ref{table:Chi1} and \ref{tab:Chi2}. The construction of Hermite closures is addressed as well.

 Appendix \ref{sec:22momentE} derives evolution equations for the 22-moment model (for an unspecified collisional operator), by applying contractions
  at the evolution equations from Appendix \ref{sec:General} and by using decomposition of moments and Hermite closures from Appendix \ref{sec:Hermite}. 

 Appendix \ref{sec:Hierarchy} uses a different technique and instead of applying contractions at equations of Appendix \ref{sec:General},
  simplified fluid hierarchy of a general n-th-order is obtained directly, which only consists of evolution equations for scalars, vectors and matrices. Evaluation
  of these equations for a specific ``n'' recovers the 22-moment equations of Appendix \ref{sec:22momentE}.

 Appendix \ref{sec:BGK} introduces the BGK (relaxation-type) collisional operator of \cite{BGK1954,GK1956}, which greatly
  clarifies the analytic forms of the Braginskii viscosity-tensors and heat fluxes. Viscosities and heat conductivities
  of both models are directly compared in Figures \ref{fig:1}-\ref{fig:2}. The nonlinear solution for the
  viscosity-tensor (with respect to a general direction of magnetic field $\bhat$)
  is addressed in Appendix \ref{sec:BragNonlin}, and Appendix \ref{sec:ambipolar} clarifies
  the ambipolar diffusion between two ion species.

 Appendix \ref{sec:FokkerPlanck} introduces a general (unspecified) Fokker-Planck collisional operator with its
  dynamical friction vector $\boldsymbol{A}_{ab}$ and diffusion tensor $\bD_{ab}$. General relations for the collisional integrals (of n-th-order) are provided,
  which can be used once the $\boldsymbol{A}_{ab}$ \& $\bD_{ab}$ are specified.

 Appendix \ref{sec:5moment} introduces the Landau collisional operator, where the $\boldsymbol{A}_{ab}$ \& $\bD_{ab}$ are expressed in the usual
  form through the Rosenbluth potentials. The 5-moment model (strict Maxwellians) is then considered, and the usual collisional momentum exchange rates $\boldsymbol{R}_{ab}$ and
  energy exchange rates $Q_{ab}$ with the assumption of small drifts between species are derived in detail in Appendices \ref{sec:Rab} \& \ref{sec:Qab}.
  Both contributions are then re-calculated with unrestricted drifts in Appendix \ref{sec:runaway}, where instead of the Rosenbluth potentials, the
  ``center-of-mass'' transformation typically used with the Boltzmann collisional operator has to be used, because the collisional integrals seem to be too complicated
  to calculate directly. This is further discussed in Appendix \ref{sec:difficulties}.

 Appendix \ref{sec:8momentM} considers the 8-moment model, where the simplest heat flux is present,
  and the multi-fluid model of \cite{Burgers1969}-\cite{Schunk1977} is calculated in detail.
  For a direct comparison with Braginskii, a one ion-electron plasma is then considered and quasi-static heat fluxes, together with the
  resulting momentum exchange rates are obtained as well. It is shown that in the limit of strong magnetic field, the perpendicular and
  cross-conductivities $\kappa_\perp$ \& $\kappa_\times$ match the Braginskii model exactly (for both the ion and electron species) and
  only the parallel conductivities $\kappa_\parallel$ are different.

 Appendix \ref{sec:Comparison} compares the parallel heat fluxes and momentum exchange rates of \cite{Braginskii1965}
  with models of \cite{Burgers1969}-\cite{Schunk1977}, \cite{Killie2004}, \cite{Landshoff1949,Landshoff1951} and \cite{SpitzerHarm1953},
  see Tables \ref{eq:TableFr}-\ref{eq:TablePerp}. Useful conversion relations for the results of \cite{Kaneko1960} and \cite{Balescu1988} are
  provided as well. The notation of \cite{SpitzerHarm1953} is clarified in Appendix \ref{sec:Spitzer} and it is shown
  that their model, as well as the model of \cite{Killie2004}, break the Onsager symmetry.

 Appendix \ref{sec:10momentM} calculates in detail the 10-moment multi-fluid model of \cite{Burgers1969}-\cite{Schunk1977},
  where the simplest viscosity-tensor is present. It is shown that in the limit of strong magnetic field, the perpendicular viscosities and
  gyroviscosities $\eta_1,\eta_2,\eta_3,\eta_4$ match the Braginskii model exactly (for both the ion and electron species) and only the
  parallel viscosities $\eta_0$ are different.

 Appendix \ref{sec:HeatFluxB} calculates in detail the momentum exchange rates and
  collisional contributions for the heat fluxes in our 21 \& 22-moment multi-fluid models.
  The calculations are shown on the 11-moment model where only the heat fluxes are present (and viscosities
  and scalar perturbation are absent), because in the semi-linear approximation the calculations can be split.
  Similarly, collisional contributions for viscosity-tensors are calculated in Appendix \ref{sec:BragVisc}, and contributions for the scalar perturbation of the 4th-order moment
  in Appendix \ref{sec:X4coll}.

 Appendix \ref{sec:2species} uses our 21-moment model and calculates heat conductivities
  and viscosities for two examples of an unmagnetized plasma consisting of two ion species (collisions with electrons are neglected).
  The first example (Appendix \ref{sec:2speciesAlpha}) is a plasma consisting of protons and alpha-particles (fully ionized Helium), typical in astrophysical applications.
  The second example (Appendix \ref{sec:DTplasma}) is a deuterium-tritium plasma used in plasma fusion.

\newpage
\section{Multi-fluid generalization of Braginskii (21-moment model)} \label{sec:Section2}
Our model is formulated with heat flux vectors
\begin{eqnarray}
\vecX^{(3)}_a = m_a \int \bc_a|\bc_a|^2 f_a d^3v = 2\vecq_a; \qquad \vecX^{(5)}_a = m_a \int \bc_a |\bc_a|^4 f_a d^3v, \label{eq:Num30}
\end{eqnarray}
and traceless viscosity tensors
\begin{eqnarray}
  \bPi^{(2)}_a &=& m_a \int \big(\bc_a\bc_a-\frac{\bI}{3}|\bc_a|^2\big) f_a d^3v; \qquad
  \bPi^{(4)}_a  = m_a \int \big(\bc_a\bc_a-\frac{\bI}{3}|\bc_a|^2\big)|\bc_a|^2 f_a d^3v, \label{eq:Num31} 
\end{eqnarray}
where the fluctuating velocity $\bc_a=\bV-\bu_a$, and ``a'' is the species index.
We are using free wording because $\vecX^{(5)}_a$ is not really
a heat flux and $\bPi^{(4)}_a$ is not really a viscosity-tensor. 
Also, we use the wording viscosity-tensor and stress-tensor interchangeably in the entire text. 
The species indices are moved freely up and down.
We also define the usual rate-of-strain tensor $\bW_a=(\nabla\bu_a)^S-(2/3)\bI \nabla\cdot\bu_a$, symmetric operator $A_{ij}^S=A_{ij}+A_{ji}$,
 and gravitational acceleration $\boldsymbol{G}$.
All other definitions are addressed in Appendix \ref{sec:General}.
We note that the definition of heat flux in equation (1.21) of \cite{Braginskii1965} contains two well-known misprints
with prime symbols missing on his fluctuating velocities $\bV'$. The heat flux is defined correctly in \cite{Braginskii1958}.

We first present a formulation with a general (unspecified) collisional operator $C(f_a)$. We define (tensorial) collisional contributions 
\begin{eqnarray}
  && \boldsymbol{R}_a = m_a\int \bV C(f_a) d^3v; \qquad Q_a = \frac{m_a}{2}\int|\bc_a|^2 C(f_a)d^3v;\nn\\
  && \bQ^{(2)}_a = m_a \int \bc_a\bc_a C(f_a) d^3v; \qquad \bQ^{(3)}_a = m_a \int \bc_a\bc_a \bc_a C(f_a) d^3v; \nn\\
  && \bQ^{(4)}_a = m_a \int \bc_a\bc_a \bc_a \bc_a C(f_a) d^3v; \qquad \bQ^{(5)}_a = m_a \int \bc_a\bc_a \bc_a \bc_a \bc_a C(f_a) d^3v, \label{eq:Spec}
\end{eqnarray}
where $\boldsymbol{R}_a$ are the usual momentum exchange rates and $Q_a$ are the usual energy exchange rates.  
Then it can be shown that integration of the Boltzmann equation yields the following nonlinear
21-moment model (see details in Appendix \ref{sec:22momentE}), where the basic evolution equations read
\begin{eqnarray}
&& \frac{d_a}{dt} n_a + n_a \nabla\cdot\bu_a =0; \label{eq:Energy20}\\
&& \frac{d_a}{dt}\bu_a +\frac{1}{\rho_a}\nabla\cdot\bp_a -\boldsymbol{G} -\frac{eZ_a}{m_a}\Big( \bE+\frac{1}{c}\bu_a\times\bb\Big)
  = \frac{\boldsymbol{R}_a}{\rho_a}; \label{eq:Energy20x}\\
  && \frac{d_a }{dt}p_a + \frac{5}{3} p_a\nabla\cdot\bu_a
  +\frac{2}{3}\nabla\cdot\vec{\boldsymbol{q}}_a +\frac{2}{3}\bPi^{(2)}_a :(\nabla \bu_a)
  =\frac{2}{3} Q_a, \label{eq:Energy20xx}
\end{eqnarray}
and are accompanied by evolution equations for the stress-tensors and heat flux vectors
\begin{eqnarray}
&&  \frac{d_a\bPi^{(2)}_a}{dt} + \bPi^{(2)}_a \nabla\cdot\bu_a 
  +\Omega_a \big(\bhat\times \bPi^{(2)}_a \big)^S +\big( \bPi^{(2)}_a \cdot\nabla\bu_a\big)^S 
  -\frac{2}{3}\bI(\bPi^{(2)}_a:\nabla\bu_a) \nn\\
  && \qquad +\frac{2}{5}\Big[(\nabla \vecq_a)^S -\frac{2}{3}\bI \nabla\cdot\vecq_a\Big]
 +p_a \bW_a = \bQ^{(2)}_a\,' \equiv \bQ^{(2)}_a -\frac{\bI}{3}\textrm{Tr}\bQ^{(2)}_a; \label{eq:Num1000}
\end{eqnarray}
\begin{eqnarray} 
 && \frac{d_a\vecq_a}{d t} +\frac{7}{5}\vecq_a\nabla\cdot\bu_a  + \frac{7}{5}\vecq_a\cdot\nabla\bu_a +\frac{2}{5}(\nabla\bu_a)\cdot\vecq_a
  +\Omega_a\bhat\times\vecq_a + \frac{5}{2}p_a\nabla\Big(\frac{p_a}{\rho_a}\Big) \nn\\
 &&  +\frac{1}{2}\nabla\cdot \bPi^{(4)}_a 
  -\frac{5}{2}\frac{p_a}{\rho_a}\nabla\cdot\bPi^{(2)}_a
  -\frac{1}{\rho_a}(\nabla\cdot\bp_a)\cdot\bPi^{(2)}_a \nn\\
  && \qquad = \vec{\boldsymbol{Q}}^{(3)}_{a}\,' \equiv \frac{1}{2}\textrm{Tr}\bQ^{(3)}_a-\frac{5}{2}\frac{p_a}{\rho_a}\boldsymbol{R}_a
  -\frac{1}{\rho_a} \boldsymbol{R}_a\cdot\bPi^{(2)}_a; \label{eq:Num1001}
\end{eqnarray}
\begin{eqnarray}
  && \frac{d_a}{dt} \bPi^{(4)}_a +\frac{1}{5}\Big[ (\nabla\vecX^{(5)}_a)^S-\frac{2}{3}\bI(\nabla\cdot\vecX^{(5)}_a)\Big]
  +\frac{9}{7}(\nabla\cdot\bu_a)\bPi^{(4)}_a +\frac{9}{7}(\bPi^{(4)}_a\cdot\nabla\bu_a)^S\nn\\
&&  + \frac{2}{7}\big((\nabla\bu_a)\cdot\bPi^{(4)}_a\big)^S
  -\frac{22}{21}\bI (\bPi^{(4)}_a:\nabla\bu_a)
  +\Omega_a \big( \bhat\times \bPi^{(4)}_a \big)^S
 + 7\frac{p_a^2}{\rho_a} \bW_a \nn\\
&& -\, \frac{14}{5\rho_a} \Big[ \big((\nabla\cdot\bp_a)\vecq_a\big)^S -\frac{2}{3}\bI (\nabla\cdot\bp_a)\cdot\vecq_a\Big] \nn\\
&& = \bQ^{(4)}_a\,' \equiv \trace \bQ^{(4)}_a -\frac{\bI}{3}\trace\trace \bQ^{(4)}_a
 -\frac{14}{5\rho_a}\Big[ (\boldsymbol{R}_a\vecq_a)^S-\frac{2}{3}\bI (\boldsymbol{R}_a\cdot\vecq_a)\Big]; \label{eq:Num1002}
\end{eqnarray}
\begin{eqnarray}
&&  \frac{d_a}{d t}\vecX^{(5)}_a  +\nabla\cdot  \bPi^{(6)}_a 
    +\frac{9}{5}\vecX^{(5)}_a (\nabla\cdot\bu_a) + \frac{9}{5}\vecX^{(5)}_a\cdot\nabla\bu_a
  + \frac{4}{5}(\nabla\bu_a)\cdot\vecX^{(5)}_a +\Omega_a \bhat\times \vecX^{(5)}_a\nn\\
  &&  +70 \frac{p_a^2}{\rho_a}\nabla\Big(\frac{p_a}{\rho_a}\Big) -35 \frac{p_a^2}{\rho_a^2} \nabla\cdot\bPi^{(2)}_a
   -\frac{4}{\rho_a} \big(\nabla\cdot\bp_a\big)\cdot \bPi^{(4)}_a \nn\\
  && =\vecQ^{(5)}_a\,' \equiv \trace\trace \bQ^{(5)}_a -35 \frac{p_a^2}{\rho_a^2}\boldsymbol{R}_a 
  - \frac{4}{\rho_a} \boldsymbol{R}_a\cdot\bPi^{(4)}_a. \label{eq:Num1004}
\end{eqnarray}
The last equation is closed with a fluid closure (derived from a Hermite closure)
\begin{equation}
  \bPi^{(6)}_a  = m_a \int \big(\bc_a\bc_a-\frac{\bI}{3}|\bc_a|^2\big)|\bc_a|^4 f_a d^3v
  = 18 \frac{p_a}{\rho_a} \bPi^{(4)}_a -63 \frac{p_a^2}{\rho_a^2}\bPi^{(2)}_a. \label{ref:Num1010}
\end{equation}  
The system above thus represents a generalized model of \cite{Braginskii1965}, where evolution equations for all the moments are
fully non-linear and valid for a general collisional operator $C(f_a)$. It is a 21-moment model (1 density,
3 velocity, 1 scalar pressure, 3 for each heat flux vector, 5 for each viscosity tensor).

\subsection{Collisional contributions (arbitrary masses and temperatures)} \label{sec:Tarb2}
We use the Landau collisional operator.
All collisional contributions are evaluated in a semi-linear approximation, with an assumption that differences in drift velocities $\bu_b-\bu_a$ are small
with respect to thermal velocities. All the non-linear quantities such as $\vecq_a\cdot(\bu_b-\bu_a)$, including $|\bu_b-\bu_a|^2$ are thus neglected in the multi-fluid
description, which is consistent with models of \cite{Burgers1969} and \cite{Schunk1977}. For energy conservation and a particular
case of a one ion-electron plasma, see Section \ref{sec:energy}.  The wording semi-linear just means that expressions containing pressures and densities
such as $(p_a/\rho_a)\vecq_a$ are retained and not fully linearized with their mean pressure/density values. 
However, for example the last terms of collisional contributions in the equations (\ref{eq:Num1001}), (\ref{eq:Num1002}), (\ref{eq:Num1004}) proportional
to $\boldsymbol{R}_a\vecq_a$, $\boldsymbol{R}_a\cdot\bPi^{(2)}_a$ and $\boldsymbol{R}_a\cdot\bPi^{(4)}_a$ are neglected in the semi-linear approximation.

We introduce the usual reduced mass and reduced temperature
\begin{equation} \label{eq:Num20}
\mu_{ab} = \frac{m_a m_b}{m_a+m_b}; \qquad T_{ab} = \frac{m_a T_b + m_b T_a}{m_a+m_b}, 
\end{equation}
together with collisional frequency (\ref{eq:timeGenM}). The momentum exchange rates are given by
\begin{eqnarray}
  \boldsymbol{R}_{a} 
  &=& \sum_{b\neq a} \nu_{ab} \Big\{ \rho_a (\bu_b-\bu_a) + \frac{\mu_{ab}}{T_{ab}}\Big[V_{ab (1)} \vecq_a 
  - V_{ab (2)} \frac{\rho_a}{\rho_b} \vecq_b  \Big]\nn\\
  && - \frac{3}{56}\Big(\frac{\mu_{ab}}{T_{ab}}\Big)^2
  \Big[ \vecX^{(5)}_a - \frac{\rho_a}{\rho_b} \vecX^{(5)}_b \Big]\Big\}, \label{eq:RabPic}
\end{eqnarray}
with  coefficients that include both masses and temperatures, but which we simply call ``mass-ratio coefficients''
\begin{eqnarray}
  V_{ab (1)} &=& \frac{(21/10)T_am_b+(3/5)T_b m_a}{T_a m_b+T_b m_a};\qquad  V_{ab (2)} = \frac{(3/5)T_am_b+(21/10)T_b m_a}{T_a m_b+T_b m_a}. \label{eq:Energy651}
\end{eqnarray}
These and other mass-ratio coefficients given below come
  from the Landau collisional operator introduced in Appendices \ref{sec:FokkerPlanck} \& \ref{sec:5moment}, where
  one uses perturbed distribution functions of the 21-moment model; see Section \ref{sec:HermX} and Appendix \ref{sec:Hermite},
  with calculations of the collisional integrals in Appendices \ref{sec:HeatFluxB} \& \ref{sec:BragVisc}. 
Energy exchange rates are given by
\begin{equation} \label{eq:Energy23}
Q_{a} = \sum_{b\neq a} 3\rho_a\nu_{ab} \frac{T_b-T_a}{m_a+m_b}, 
\end{equation}
where $|\bu_b-\bu_a|^2$ are neglected as discussed above. 
The heat flux exchange rates are given by
\begin{eqnarray}
   \vec{\boldsymbol{Q}}^{(3)}_{a}\,' 
  &=& -   \Big[ 2\nu_{aa} +\sum_{b\neq a}\nu_{ab} \hat{D}_{ab (1)} \Big]\vecq_a
  +\sum_{b\neq a} \nu_{ab} \hat{D}_{ab (2)} \frac{\rho_a}{\rho_b}\vecq_b \nn\\
  && +   
  \Big[ \frac{3}{70}\nu_{aa}+\sum_{b\neq a}\nu_{ab}\hat{E}_{ab (1)} \Big]\frac{\rho_a}{p_a}\vecX^{(5)}_a
   -\sum_{b\neq a} \nu_{ab}   \hat{E}_{ab (2)}\frac{\rho_b}{p_b}  \frac{\rho_a}{\rho_b}\vecX^{(5)}_b\nn\\
  && -p_a \sum_{b\neq a} \nu_{ab} (\bu_b-\bu_a) \hat{U}_{ab (1)}, \label{eq:FinalQ3}
\end{eqnarray}
with  mass-ratio coefficients
\begin{eqnarray}
  \hat{U}_{ab (1)} &=& \frac{3 m_b (3 T_a m_a+T_a m_b-2 T_b m_a)}{2 (T_a m_b+T_b m_a) (m_a+m_b)};\nn\\
  \hat{D}_{ab (1)} &=& \big\{75 T_a^3 m_a m_b^3 +95 T_a^3 m_b^4 +174 T_a^2 T_b m_a m_b^3 +300 T_a T_b^2 m_a^3 m_b
  +498 T_a T_b^2 m_a^2 m_b^2 +60 T_b^3 m_a^4 \nn\\
  && +104 T_b^3 m_a^3 m_b\big\}\big[20 (T_a m_b +T_b m_a)^3 (m_a+m_b)\big]^{-1};\nn\\
  \hat{D}_{ab (2)} &=& \frac{9 T_a m_b^2 (10 T_a^2 m_a m_b+6 T_a^2 m_b^2 +45 T_a T_b m_a^2+27 T_a T_b m_a m_b -14 T_b^2 m_a^2)}{20 (T_a m_b+T_b m_a)^3 (m_a+m_b)};\nn\\
  \hat{E}_{ab (1)} &=& \frac{3 T_a m_b (19 T_a^2 m_a m_b^2 +23 T_a^2 m_b^3 -2 T_a T_b m_a^2 m_b+36 T_a T_b m_a m_b^2
    +84 T_b^2 m_a^3 +118 T_b^2 m_a^2 m_b)}{560 (T_a m_b +T_b m_a)^3 (m_a+m_b)};\nn\\
  \hat{E}_{ab (2)} &=& \frac{9 T_a T_b m_a m_b^2 (7 T_a m_a+5 T_a m_b-2 T_b m_a) }{112 (T_a m_b+T_b m_a)^3 (m_a+m_b)}. \label{eq:Final_Q3t}
\end{eqnarray}
The 5th-order moment exchange rates are given by
\begin{eqnarray}
\vec{\boldsymbol{Q}}^{(5)}_{a}\,'  
&=&  -\Big[\frac{76}{5} \nu_{aa} +\sum_{b\neq a} \nu_{ab} \hat{F}_{ab (1)} \Big] \frac{p_a}{\rho_a} \vecq_a
+ \sum_{b\neq a} \nu_{ab} \hat{F}_{ab (2)}  \frac{p_a}{\rho_a} \frac{\rho_a}{\rho_b} \vecq_b \nn\\
&& -\Big[\frac{3}{35} \nu_{aa} +\sum_{b\neq a} \nu_{ab} \hat{G}_{ab (1)}\Big]\vecX^{(5)}_a
- \sum_{b\neq a} \nu_{ab} \hat{G}_{ab (2)} \frac{p_a}{p_b}\vecX^{(5)}_b\nn\\
&& - \frac{p_a^2}{\rho_a} \sum_{b\neq a} \nu_{ab} (\bu_b-\bu_a)  \hat{U}_{ab (2)},\label{eq:FinalQ5} 
\end{eqnarray}
with  mass-ratio coefficients
\begin{eqnarray}
  \hat{U}_{ab (2)} &=& \frac{3 m_b (17 T_a^2 m_a m_b +9 T_a^2 m_b^2 +42 T_a T_b m_a^2 +6 T_a T_b m_a m_b -28 T_b^2 m_a^2)}{(T_a m_b+T_b m_a)^2 (m_a+m_b)};\nn\\
  \hat{F}_{ab (1)} &=& \big\{ 855 T_a^5 m_a m_b^4 +759 T_a^5 m_b^5 +2340 T_a^4 T_b m_a^2 m_b^3 +1972 T_a^4 T_b m_a m_b^4 +2640 T_a^3 T_b^2 m_a^3 m_b^2\nn\\
  && +2332 T_a^3 T_b^2 m_a^2 m_b^3 +5880 T_a^2 T_b^3 m_a^4 m_b +3324 T_a^2 T_b^3 m_a^3 m_b^2 -3080 T_a T_b^4 m_a^4 m_b-560 T_b^5 m_a^5\big\}\nn\\
  && \times \big[10 (T_a m_b+T_b m_a)^4 (m_a+m_b) T_a  \big]^{-1};\nn\\
  \hat{F}_{ab (2)} &=& 3 T_a m_b^2 \big\{70 T_a^3 m_a m_b^2 +102 T_a^3 m_b^3 +385 T_a^2 T_b m_a^2 m_b +561 T_a^2 T_b m_a m_b^2 +1890 T_a T_b^2 m_a^3 \nn\\
  && +1446 T_a T_b^2 m_a^2 m_b -588 T_b^3 m_a^3\big\}\big[ 10 (T_a m_b +T_b m_a)^4 (m_a+m_b) \big]^{-1};\nn\\
  \hat{G}_{ab (1)} &=& - \big\{ 565 T_a^4 m_a m_b^4 +533 T_a^4 m_b^5 +1270 T_a^3 T_b m_a^2 m_b^3 +1190 T_a^3 T_b m_a m_b^4  +1020 T_a^2 T_b^2 m_a^3 m_b^2 \nn\\
  && +1152 T_a^2 T_b^2 m_a^2 m_b^3 +3640 T_a T_b^3 m_a^4 m_b +1916 T_a T_b^3 m_a^3 m_b^2 -1400 T_b^4 m_a^5
  -3304 T_b^4 m_a^4 m_b\big\} \nn\\
  && \times \big[ 280 (T_a m_b+T_b m_a)^4 (m_a+m_b)\big]^{-1};\nn\\
  \hat{G}_{ab (2)} &=& -\, \frac{3 T_a T_b m_a m_b^2 (3 T_a^2 m_a m_b -5 T_a^2 m_b^2 -42 T_a T_b m_a^2 -38 T_a T_b m_a m_b
    +12 T_b^2 m_a^2)}{8 (T_a m_b+T_b m_a)^4 (m_a+m_b)}. \label{eq:Final_Q5t}
\end{eqnarray}
Exchange rates for the usual stress-tensor are given by
\begin{eqnarray}
 \bQ_{a}^{(2)}\,' &=&  -\, \frac{21}{10}\nu_{aa} \bPi_a^{(2)} +\frac{9}{70} \nu_{aa}  \frac{\rho_a}{p_a}\bPi_a^{(4)} \nn\\
 && + \sum_{b\neq a} \frac{\rho_a \nu_{ab}}{m_a+m_b} \Big[ - \hat{K}_{ab (1)} \frac{1}{n_a} \bPi_a^{(2)}
   +\hat{K}_{ab (2)} \frac{1}{n_b} \bPi_b^{(2)} \nn\\
   && +L_{ab (1)} \frac{\rho_a}{n_a p_a}\bPi_a^{(4)}  -L_{ab (2)} \frac{\rho_b}{n_b p_b}\bPi_b^{(4)}\Big],
\end{eqnarray}
with  mass-ratio coefficients
\begin{eqnarray}
  \hat{K}_{ab (1)} &=& \frac{10 T_a^2 m_a m_b^2 +15 T_a^2 m_b^3 +35 T_a T_b m_a^2 m_b +42 T_a T_b m_a m_b^2 +10 T_b^2 m_a^3 +12 T_b^2 m_a^2 m_b}{5 (T_a m_b +T_b m_a)^2 m_a};\nn\\
  \hat{K}_{ab (2)} &=& \frac{6 T_a^2 m_a m_b +4 T_a^2 m_b^2 +21 T_a T_b m_a^2 +14 T_a T_b m_a m_b -5 T_b^2 m_a^2}{5 (T_a m_b +T_b m_a)^2};\nn\\
L_{ab (1)} &=& \frac{3 T_a m_b (2 T_a m_a m_b +3 T_a m_b^2 +7 T_b m_a^2 +8 T_b m_a m_b)}{35 (T_a m_b +T_b m_a)^2 m_a};\nn\\
L_{ab (2)} &=& \frac{ 3 m_a T_b (5 T_a m_a +4 T_a m_b -T_b m_a)}{35 (T_a m_b +T_b m_a)^2}.\label{eq:HatK}
\end{eqnarray}
Finally, the 4th-order stress-tensor exchange rates are given by 
\begin{eqnarray}
  \bQ^{(4)}_{a}\,' &=& -\, \frac{53}{20} \nu_{aa} \frac{p_a}{\rho_a} \bPi^{(2)}_a - \frac{79}{140} \nu_{aa} \bPi^{(4)}_a
   + \sum_{b\neq a} \nu_{ab} \Big[ - \hat{M}_{ab (1)} \frac{p_a}{\rho_a} \bPi^{(2)}_a \nn\\
  &&  +\hat{M}_{ab (2)} \frac{p_a^2}{\rho_a p_b} \bPi^{(2)}_b -N_{ab (1)}\bPi^{(4)}_a
    - N_{ab (2)}\frac{p_a^2 \rho_b}{p_b^2\rho_a} \bPi^{(4)}_b \Big],
\end{eqnarray}
with  mass-ratio coefficients
\begin{eqnarray}
  \hat{M}_{ab (1)} &=& \Big\{48 T_a^4 m_a m_b^3 +36 T_a^4 m_b^4 +216 T_a^3 T_b m_a^2 m_b^2 +107 T_a^3 T_b m_a m_b^3 +378 T_a^2 T_b^2 m_a^3 m_b \nn\\
  &&  +36 T_a^2 T_b^2 m_a^2 m_b^2 -315 T_a T_b^3 m_a^3 m_b -70 T_b^4 m_a^4 \Big\}\Big[5 (T_a m_b +T_b m_a)^3 T_a (m_b +m_a)\Big]^{-1}; \nn\\
  \hat{M}_{ab (2)} &=& - \, \Big\{ T_b m_a \big(18 T_a^3 m_a m_b^2 -4 T_a^3 m_b^3 +81 T_a^2 T_b m_a^2 m_b -18 T_a^2 T_b m_a m_b^2 -147 T_a T_b^2 m_a^3 \nn\\
  && -189 T_a T_b^2 m_a^2 m_b +35 T_b^3 m_a^3 \big)\Big\}\Big[5 (T_a m_b +T_b m_a)^3 T_a (m_b +m_a) \Big]^{-1};\nn\\
  N_{ab (1)} &=& -\, \Big\{16 T_a^3 m_a m_b^3 +12 T_a^3 m_b^4 +72 T_a^2 T_b m_a^2 m_b^2 +21 T_a^2 T_b m_a m_b^3 +126 T_a T_b^2 m_a^3 m_b\nn\\
 && \quad -54 T_a T_b^2 m_a^2 m_b^2 -140 T_b^3 m_a^4 -273 T_b^3 m_a^3 m_b\Big\} \Big[35 (T_a m_b +T_b m_a)^3 (m_b +m_a)\Big]^{-1};\nn\\
  N_{ab (2)} &=& -\, \frac{3 T_b^2 m_a^2 (35 T_a^2 m_a m_b +12 T_a^2 m_b^2 -35 T_a T_b m_a^2 -51 T_a T_b m_a m_b +7 T_b^2 m_a^2)}{35 (T_a m_b +T_b m_a)^3 T_a (m_b +m_a)}. \label{eq:Posled20}
\end{eqnarray}
The entire system is now fully specified, and represents a multi-fluid generalization of the model of \cite{Braginskii1965}. 
Coupled with Maxwell's equations, it can be used in multi-fluid numerical simulations. 
 Importantly, when collisional frequencies become small, the right hand sides of evolution equations just
  become small and no coefficients become divergent, which is in contrast to the model of Braginskii,
  where the quasi-static approximation is used for the stress-tensors and heat fluxes.
For a detailed discussion on the limitations of our model in a regime of low-collisionality, see Section \ref{sec:limitations}.
The model of Braginskii is obtained as a particular case
of a one ion-electron plasma with similar temperatures,
in a quasi-static and quasi-linear approximation for the viscosity tensors and heat fluxes, where additionally,
the coupling between viscosity tensors and heat fluxes is neglected.

\newpage
\subsection{Collisional contributions for small temperature differences} \label{sec:Coll-Tequal}
In many instances, it might be satisfactory to consider a situation when the temperature differences between species are small. 
The  mass-ratio coefficients (\ref{eq:Energy651}) then become
\begin{eqnarray}
  V_{ab (1)} &=& \frac{(21/10)m_b+(3/5) m_a}{ m_b+ m_a};\qquad  V_{ab (2)} = \frac{(3/5) m_b+(21/10) m_a}{ m_b+ m_a}, \label{eq:RabV}
\end{eqnarray}
mass-ratio coefficients (\ref{eq:Final_Q3t}) simplify into
\begin{eqnarray}
  \hat{D}_{ab (1)}  &=& \frac{3 m_a^3+(86/5) m_a^2 m_b+(77/10) m_a m_b^2+(19/4) m_b^3 }{(m_a+m_b)^3};\nn\\
  \hat{D}_{ab (2)}  &=& \frac{ (279/20) m_a m_b^2 +(27/10) m_b^3}{(m_a+m_b)^3};\nn\\
  \hat{E}_{ab (1)}  &=& \frac{(9/20) m_a^2 m_b +(6/35) m_a m_b^2+(69/560) m_b^3}{(m_a+m_b)^3};\nn\\
  \hat{E}_{ab (2)}  &=& \frac{(45/112) m_a m_b^2}{(m_a+m_b)^3};\qquad  \hat{U}_{ab (1)}  = \frac{3}{2}  \frac{m_b}{(m_a+m_b)}, \label{eq:FinalQ3c}
\end{eqnarray}
mass-ratio coefficients (\ref{eq:Final_Q5t}) become
\begin{eqnarray}
  \hat{F}_{ab (1)}  &=& \frac{(-56) m_a^4+336 m_a^3 m_b+(1302/5) m_a^2 m_b^2+(1034/5) m_a m_b^3+(759/10) m_b^4}{(m_a+m_b)^4};\nn\\
  \hat{F}_{ab (2)}  &=& \frac{(1953/5) m_a^2 m_b^2 +(1587/10) m_a m_b^3 +(153/5) m_b^4}{(m_a+m_b)^4};\nn\\
  \hat{G}_{ab (1)}  &=& \frac{5 m_a^4-(31/5) m_a^3 m_b-(30/7) m_a^2 m_b^2-(611/140) m_a m_b^3-(533/280) m_b^4}{(m_a+m_b)^4};\nn\\
  \hat{G}_{ab (2)}  &=& \frac{(45/4) m_a^2 m_b^2 +(15/8) m_a m_b^3}{(m_a+m_b)^4};\qquad  \hat{U}_{ab (2)}  = \frac{42 m_a m_b+27 m_b^2}{(m_a+m_b)^2},\label{eq:FinalQ5c}
\end{eqnarray}
 mass-ratio coefficients (\ref{eq:HatK}) become
\begin{eqnarray}
  \hat{K}_{ab (1)} &=& \frac{10 m_a^2 +37 m_a m_b +15 m_b^2}{5 m_a (m_b +m_a)};\qquad
  \hat{K}_{ab (2)} = \frac{4 (4 m_a +m_b)}{5 (m_b +m_a)};\nn\\
  L_{ab (1)} &=& \frac{3 (7 m_a +3 m_b) m_b}{35 m_a (m_b+m_a)}; \qquad
  L_{ab (2)} = \frac{12 m_a}{35 (m_a+m_b)}, \label{eq:Posled23}
\end{eqnarray}
and  mass-ratio coefficients (\ref{eq:Posled20}) simplify into
\begin{eqnarray}
 \hat{M}_{ab (1)} &=& -\, \frac{70 m_a^3 -133 m_a^2 m_b -119 m_a m_b^2 -36 m_b^3}{5 (m_b +m_a)^3 };\qquad
 \hat{M}_{ab (2)} = \frac{4 m_a (28 m_a^2 -m_a m_b +m_b^2)}{5 (m_b +m_a)^3 };\nn\\
  N_{ab (1)} &=& \frac{140 m_a^3 +7 m_a^2 m_b -25 m_a m_b^2 -12 m_b^3}{35 (m_b +m_a)^3};\qquad
  N_{ab (2)} = \frac{12  m_a^2 (7 m_a -3 m_b)}{35 (m_b +m_a)^3}. \label{eq:Posled24}
\end{eqnarray}

\subsection{Reduction to 13-moment model} \label{sec:Schunk13mom}
 As a partial double-check of our calculations,
  neglecting the evolution equations (\ref{eq:Num1002})-(\ref{eq:Num1004}) for $\bPi^{(4)}_a$ \& $\vecX^{(5)}_a$,
  and in the evolution equations
  (\ref{eq:Num1000})-(\ref{eq:Num1001}) for $\bPi^{(2)}_a$ \& $\vecq_a$ prescribing closures (which are derived from Hermite closures) 
\begin{equation} \label{eq:Closure1}
\vecX^{(5)}_a = 28 \frac{p_a}{\rho_a}\vecq_a; \qquad \bPi^{(4)}_a = 7\frac{p_a}{\rho_a} \bPi^{(2)}_a,
\end{equation}
 our 21-moment model simplifies into the 13-moment model, given by collisional contributions
\begin{eqnarray}
  \boldsymbol{R}_{a} 
  &=& \sum_{b\neq a} \nu_{ab} \Big[ \rho_a (\bu_b-\bu_a) + \frac{3}{5}\frac{\mu_{ab}}{T_{ab}} \big( \vecq_a -\frac{\rho_a}{\rho_b}\vecq_b\big) \Big];\nn\\
   \vec{\boldsymbol{Q}}^{(3)}_{a}\,' 
  &=& - \frac{4}{5}\nu_{aa} \vecq_a  +\sum_{b\neq a}\nu_{ab}\Big[ - \hat{D}^*_{ab (1)} \vecq_a
  + \hat{D}^*_{ab (2)} \frac{\rho_a}{\rho_b}\vecq_b  -p_a (\bu_b-\bu_a) \hat{U}_{ab (1)}\Big];\nn\\
 \bQ_{a}^{(2)}\,' &=&  -\, \frac{6}{5}\nu_{aa} \bPi_a^{(2)} 
  + \sum_{b\neq a} \frac{\rho_a \nu_{ab}}{m_a+m_b} \Big[ - \hat{K}^*_{ab (1)} \frac{1}{n_a} \bPi_a^{(2)}
   +\hat{K}^*_{ab (2)} \frac{1}{n_b} \bPi_b^{(2)} \Big], \label{eq:Thierry10}
\end{eqnarray}
 with mass-ratio coefficients
\begin{eqnarray}
\hat{D}^*_{ab (1)} &=& \frac{9 T_a^2 m_a m_b^2 +13 T_a^2 m_b^3 -6 T_a T_b m_a^2 m_b +20 T_a T_b m_a m_b^2 +30 T_b^2 m_a^3 +52 T_b^2 m_a^2 m_b}{10 (m_a +m_b)(T_a m_b +T_b m_a)^2};\nn\\
\hat{D}^*_{ab (2)} &=& \frac{9 T_a m_b^2 (5 T_a m_a +3 T_a m_b -2 T_b m_a)}{10 (m_a +m_b) (T_a m_b +T_b m_a)^2}; \nn\\
\hat{K}^*_{ab (1)} &=& \frac{2 (2 T_a m_a m_b +3 T_a m_b^2 +5 T_b m_a^2 +6 T_b m_a m_b)}{5 m_a (T_a m_b +T_b m_a)};\qquad
\hat{K}^*_{ab (2)} = \frac{2 (3 T_a m_a +2 T_a m_b -T_b m_a)}{5 (T_a m_b +T_b m_a)},
\end{eqnarray}
where $\hat{U}_{ab (1)}$ is unchanged from the 21-moment model. It can be shown that for Coulomb collisions, this model is equivalent to
 equations (44)-(49) of \cite{Schunk1977}, first calculated by \cite{Burgers1969}. For small temperature differences the  mass-ratio coefficients become
\begin{eqnarray}
  \hat{D}^*_{ab (1)} &=& \frac{30 m_a^2 +16 m_a m_b +13 m_b^2 }{10 (m_a +m_b)^2};\qquad \hat{D}^*_{ab (2)} = \frac{27 m_b^2 }{10 (m_a+m_b)^2};\nn\\
  \hat{K}^*_{ab (1)} &=& \frac{2 m_a+ (6/5) m_b}{m_a}; \qquad \hat{K}^*_{ab (2)} = \frac{4}{5}; \qquad  \hat{U}_{ab (1)}  = \frac{3}{2}  \frac{m_b}{(m_a+m_b)}.
\end{eqnarray}
 Our new  21-moment model thus can be viewed as a generalization of the multi-fluid description of
\cite{Burgers1969} \& \cite{Schunk1977}, where the heat fluxes and stress-tensors are described more accurately, and with the same level of precision as in
\cite{Braginskii1965}. Nevertheless, we only use the Landau collisional operator applicable for Coulomb collisions, whereas Burgers-Schunk use
the more general Boltzmann collisional operator and account for several different interaction potentials.

\subsection{Semi-linear approximation (de-coupled stress tensors and heat fluxes)} \label{sec:SemiLinear}
Here we consider the 21-moment model with evolution equations (\ref{eq:Num1000})-(\ref{eq:Num1004}) in the semi-linear approximation,
where additionally viscosity-tensors and heat fluxes are de-coupled.  It will be shown later that 
the contributions introduced by the coupling
are smaller by a factor of $1/\nu_{aa}$. Within the semi-linear approximation we also assume that there are no large-scale gradients of considered fluid moments.
 For example, the de-coupling removes the last terms at the left hand side of equations (\ref{eq:Num1001}), (\ref{eq:Num1002}), (\ref{eq:Num1004})
proportional to $(\nabla p_a)\vecq_a$, $(\nabla p_a)\cdot\bPi^{(2)}_a$ and $(\nabla p_a)\cdot\bPi^{(4)}_a$. We neglect these terms
within the semi-linear approximation also when the coupling is considered (see Sections \ref{sec:Couplingg1} \& \ref{sec:Couplingg}).
In the presence of large-scale gradients
in pressure/temperature these terms might become significant, together with many other terms that are neglected in the semi-linear approximation.
Evolution equations for heat flux vectors simplify into
\begin{eqnarray}
&&  \frac{d_a}{d t}\vecq_a + \Omega_a \bhat\times\vecq_a + \frac{5}{2}p_a \nabla \Big(\frac{p_a}{\rho_a}\Big)= \vec{\boldsymbol{Q}}^{(3)}_{a}\,'; \label{eq:Energy10}\\
&&  \frac{d_a}{d t}\vecX^{(5)}_a +\Omega_a\bhat\times\vecX^{(5)}_a+70\frac{p_a^2}{\rho_a}\nabla\Big(\frac{p_a}{\rho_a}\Big) 
  = \vec{\boldsymbol{Q}}^{(5)}_{a}\,', \label{eq:ExciteEvolve}
\end{eqnarray}
and evolution equations for viscosity-tensors become
\begin{eqnarray}
  && \frac{d_a}{dt} \bPi^{(2)}_a  +\Omega_a \big(\bhat\times \bPi^{(2)}_a \big)^S + p_a \bW_a
  = \bQ_{a}^{(2)}\,' ; \label{eq:Posled21}\\
  && \frac{d_a}{dt} \bPi^{(4)}_a  +\Omega_a \big(\bhat\times \bPi^{(4)}_a \big)^S + 7 \frac{p_a^2}{\rho_a} \bW_a 
  =  \bQ^{(4)}_{a}\,'. \label{eq:Energy22}
\end{eqnarray}
The above system will be used to recover the transport coefficients of \cite{Braginskii1965}. In some instances, it might be actually
advantageous to suppress the non-linearities in numerical simulations, and perform
multi-fluid simulations with system (\ref{eq:Energy10})-(\ref{eq:Energy22}) instead of the system (\ref{eq:Num1000})-(\ref{eq:Num1004}).

\newpage
\section{One ion-electron plasma} \label{sec:ONEion}
\subsection{Ion heat flux \texorpdfstring{$\vecq_a$}{q} of Braginskii (self-collisions)}

Here we consider a one ion-electron plasma of  similar temperatures, which is the choice of \cite{Braginskii1965}.
For the ion heat flux, Braginskii neglects ion-electron collisions. Considering only
self-collisions, evolution equations for ion heat fluxes read
\begin{eqnarray}
  &&  \frac{d_a}{d t}\vecq_a + \Omega_a \bhat\times\vecq_a + \frac{5}{2}p_a \nabla \Big(\frac{p_a}{\rho_a}\Big)= -\frac{4}{5}\nu_{aa} \vecq_a
  +\frac{3}{70}\nu_{aa}\Big( \frac{\rho_a}{p_a}\vecX^{(5)}_a-28\vecq_a\Big); \label{eq:Excite1}\\
&&  \frac{d_a}{d t}\vecX^{(5)}_a +\Omega_a\bhat\times\vecX^{(5)}_a +70\frac{p_a^2}{\rho_a}\nabla\Big(\frac{p_a}{\rho_a}\Big) 
  = -\frac{88}{5}\nu_{aa}\frac{p_a}{\rho_a}\vecq_a-\frac{3}{35}\nu_{aa}\Big(\vecX^{(5)}_a -28\frac{p_a}{\rho_a}\vecq_a\Big).\label{eq:Excite2}
\end{eqnarray}
 Neglecting the evolution equation (\ref{eq:Excite2})  and prescribing closure (\ref{eq:Closure1})
which neglects the second term on the r.h.s. of (\ref{eq:Excite1}),
yields the ion heat flux model of Burgers-Schunk, with the well-known $-4/5$ constant. However, now the equations read
\begin{eqnarray}
  &&  \frac{d_a}{d t}\vecq_a + \Omega_a \bhat\times\vecq_a + \frac{5}{2}p_a \nabla \Big(\frac{p_a}{\rho_a}\Big)= -2\nu_{aa} \vecq_a
  +\frac{3}{70}\nu_{aa} \frac{\rho_a}{p_a}\vecX^{(5)}_a;\nn\\
&&  \frac{d_a}{d t}\vecX^{(5)}_a +\Omega_a\bhat\times\vecX^{(5)}_a +70\frac{p_a^2}{\rho_a}\nabla\Big(\frac{p_a}{\rho_a}\Big) 
  = -\frac{76}{5}\nu_{aa}\frac{p_a}{\rho_a}\vecq_a-\frac{3}{35}\nu_{aa}\vecX^{(5)}_a. \label{eq:Excite3}
\end{eqnarray}
Prescribing the quasi-static approximation (by canceling the $d_a/dt$), yields an analytic solution (see for example a general vector equation (\ref{eq:PPPX})
with solution (\ref{eq:PPP}))
\begin{equation} \label{eq:Braginskii_YES}
\vecq_a = -\kappa_\parallel^a \nabla_\parallel T_a - \kappa_\perp^a \nabla_\perp T_a + \kappa_\times^a \bhat\times\nabla T_a,
\end{equation}
and thermal conductivities
\begin{eqnarray}
\kappa_\parallel^a &=& \frac{125}{32} \frac{p_a}{\nu_{aa} m_a};\nn\\
\kappa_\perp^a &=& \frac{p_a}{\nu_{aa} m_a} \, \frac{2x^2+(648/245)}{x^4+(3313/1225)x^2+(20736/30625)};\nn\\
\kappa_\times^a &=& \frac{p_a}{\nu_{aa} m_a} \, \frac{(5/2)x^3+(2277/490)x}{x^4+(3313/1225)x^2+(20736/30625)}, \label{eq:Thierry51}
\end{eqnarray}
where $x=\Omega_a/\nu_{aa}$. Alternatively, by using numerical values
\begin{eqnarray}
\kappa_\parallel^a &=& 3.906 \frac{p_a}{\nu_{aa} m_a};\nn\\
\kappa_\perp^a &=& \frac{p_a}{\nu_{aa} m_a} \, \frac{2x^2+2.645}{x^4+2.704x^2+0.6771};\nn\\  
\kappa_\times^a &=& \frac{p_a}{\nu_{aa} m_a} \, \frac{(5/2)x^3+4.647x}{x^4+2.704x^2+0.6771},
\end{eqnarray}
which recovers the ion heat flux of \cite{Braginskii1965}, his equation (4.40).
 We use Braginskii notation with vectors $\nabla_\parallel=\bhat\bhat\cdot\nabla$ and
  $\nabla_\perp = \bI_\perp\cdot\nabla = -\bhat\times\bhat\times\nabla$.

\subsection{Ion heat flux \texorpdfstring{$\vecX^{(5)}_a$}{X5} (self-collisions)}
The solution for the vector $\vecX^{(5)}_a$ has a similar form
\begin{equation} 
\vecX^{(5)}_a = \frac{p_a}{\rho_a} \Big[-\kappa_\parallel^{a(5)} \nabla_\parallel T_a - \kappa_\perp^{a(5)} \nabla_\perp T_a + \kappa_\times^{a(5)} \bhat\times\nabla T_a\Big],
\end{equation}
with ``thermal conductivities''
\begin{eqnarray}
  \kappa_\parallel^{a(5)} &=& \underbrace{\frac{2975}{24}}_{123.96}  \frac{p_a}{\nu_{aa} m_a};\nn\\
  \kappa_\perp^{a(5)} &=& \frac{p_a}{\nu_{aa} m_a} \frac{44 x^2+(14688/175)}{x^4+(3313/1225) x^2+ (20736/30625)};\nn\\
  \kappa_\times^{a(5)} &=& \frac{p_a}{\nu_{aa} m_a} \frac{70 x^3+(1086/7)x}{x^4+(3313/1225) x^2+(20736/30625)}. \label{eq:Thierry50}
\end{eqnarray}

\newpage
\subsection{Electron heat flux \texorpdfstring{$\vecq_e$}{q} of Braginskii}
Considering a one-ion electron plasma with  similar temperatures,  and keeping only the dominant term in a $m_e/m_i$ expansion,
the  mass-ratio coefficients (\ref{eq:RabV}), (\ref{eq:FinalQ3c}), (\ref{eq:FinalQ5c}) simplify into
\begin{eqnarray}
&&  V_{ei (1)} = \frac{21}{10}; \quad V_{ei (2)}=\frac{3}{5}; \nn\\
&& \hat{D}_{ei (1)}=\frac{19}{4}; \quad \hat{D}_{ei (2)} = \frac{27}{10};
  \quad \hat{E}_{ei (1)}=\frac{69}{560}; \quad \hat{E}_{ei (2)}=\frac{45}{112}\frac{m_e}{m_i}; \quad \hat{U}_{ei (1)}=\frac{3}{2}\nn\\
  && \hat{F}_{ei (1)} = \frac{759}{10}; \quad \hat{F}_{ei (2)}=\frac{153}{5}; \quad \hat{G}_{ei (1)} = -\,\frac{533}{280};
  \quad \hat{G}_{ei (2)}=\frac{15}{8}\frac{m_e}{m_i};\quad \hat{U}_{ei (2)}=27, \label{eq:Num85}
\end{eqnarray}
collisional exchange rates become
\begin{eqnarray}
  \boldsymbol{R}_e &=& - \rho_e\nu_{ei} \delta\bu + \frac{21}{10}\frac{\rho_e}{p_e} \nu_{ei}\vecq_e
  -\frac{3}{56} \frac{\rho_e^2}{p_e^2}\nu_{ei} \vecX^{(5)}_e;\label{eq:ReExcite}\\
 \vec{\boldsymbol{Q}}^{(3)}_{e}\,' &=&  +\frac{3}{2}p_e\nu_{ei} \delta\bu -\Big[ 2\nu_{ee}+\frac{19}{4}\nu_{ei} \Big]\vecq_e
  +\Big[ \frac{3}{70}\nu_{ee}+\frac{69}{560}\nu_{ei} \Big] \frac{\rho_e}{p_e} \vecX^{(5)}_e;\\
 \vec{\boldsymbol{Q}}^{(5)}_{e}\,' &=& + 27\frac{p_e^2}{\rho_e}\nu_{ei}\delta\bu -\Big[ \frac{76}{5}\nu_{ee}+\frac{759}{10}\nu_{ei}\Big] \frac{p_e}{\rho_e}\vecq_e
  -\Big[ \frac{3}{35}\nu_{ee} -\frac{533}{280}\nu_{ei}\Big] \vecX^{(5)}_e, \label{eq:Num77}
\end{eqnarray}
where $\delta\bu=\bu_e-\bu_i$, and enter the right hand side of the electron momentum equation, and evolution equations for the electron heat flux vectors
\begin{eqnarray}
&&  \frac{d_e}{d t}\vecq_e + \Omega_e \bhat\times\vecq_e + \frac{5}{2}p_e \nabla \Big(\frac{p_e}{\rho_e}\Big)= \vec{\boldsymbol{Q}}^{(3)}_{e}\,';\nn\\
&&  \frac{d_e}{d t}\vecX^{(5)}_e +\Omega_e\bhat\times\vecX^{(5)}_e+70\frac{p_e^2}{\rho_e}\nabla\Big(\frac{p_e}{\rho_e}\Big) 
  = \vec{\boldsymbol{Q}}^{(5)}_{e}\,'. \label{eq:ExciteRit}
\end{eqnarray}
In \cite{Braginskii1965}, the results are expressed through the collisional frequency $\nu_{ei}$, and conversion  with $\nu_{ee}=\nu_{ei}/(Z_i\sqrt{2})$ yields
\begin{eqnarray}
\vec{\boldsymbol{Q}}^{(3)}_{e}\,' &=&  +\frac{3}{2}p_e\nu_{ei} \delta\bu -\Big[ \frac{\sqrt{2}}{Z_i}+\frac{19}{4} \Big]\nu_{ei}\vecq_e
  +\Big[ \frac{3}{70\sqrt{2}Z_i}+\frac{69}{560} \Big]\nu_{ei} \frac{\rho_e}{p_e} \vecX^{(5)}_e;\nn\\
 \vec{\boldsymbol{Q}}^{(5)}_{e}\,' &=& + 27\frac{p_e^2}{\rho_e}\nu_{ei}\delta\bu -\Big[ \frac{76}{5\sqrt{2}Z_i}+\frac{759}{10}\Big]\nu_{ei} \frac{p_e}{\rho_e}\vecq_e
  -\Big[ \frac{3}{35\sqrt{2}Z_i} -\frac{533}{280}\Big]\nu_{ei} \vecX^{(5)}_e. \label{eq:Energy30}
\end{eqnarray}
In a quasi-static approximation, the solution of (\ref{eq:ExciteRit}), (\ref{eq:Energy30}) recovers the famous electron heat flux of \cite{Braginskii1965},
together with vector $\vecX^{(5)}_e$ (which is of course not given by Braginskii). Substituting these results into the momentum exchange rates (\ref{eq:ReExcite}),
recovers the $\boldsymbol{R}_e$ of Braginskii. 

We use the same notation as \cite{Braginskii1965} with $x=\Omega_e/\nu_{ei}$, except (as is the norm in more recent papers) our $\Omega_e$ is 
formulated as a general $\Omega_a$ and is thus negative, whereas in Braginskii $\Omega_e$ is defined as positive.
This yields a simple change of signs in front of the ``cross'' ($\times$) terms with respect to Braginskii. In a quasi-static approximation,
the electron heat flux is split into a thermal and frictional part $\vecq_e=\vecq_e^T+\vecq_e^u$, where
\begin{eqnarray}
\vecq_e^T &=& -\kappa_\parallel^e \nabla_\parallel T_e - \kappa_\perp^e \nabla_\perp T_e + \kappa_\times^e \bhat\times\nabla T_e;\nn\\
\vecq_e^u &=& \beta_0 p_e \delta\bu_\parallel + p_e \delta\bu_\perp \frac{\beta_1'x^2+\beta_0'}{\triangle} 
- p_e \bhat\times\delta\bu \frac{\beta_1''x^3+\beta_0''x}{\triangle}, \label{eq:Thierry63}
\end{eqnarray}
and the heat conductivities are given by
\begin{equation}
\kappa_\parallel^e = \frac{p_e}{m_e \nu_{ei}}\gamma_0; \qquad   
\kappa_\perp^e =  \frac{p_e}{m_e \nu_{ei}} \frac{\gamma_1' x^2+\gamma_0'}{\triangle};
\qquad \kappa_\times^e =  \frac{p_e}{m_e \nu_{ei}}\frac{\gamma_1'' x^3+\gamma_0''x}{\triangle}. \label{eq:BrHF}
\end{equation}
The momentum exchange rates are also split into a thermal and frictional part $\boldsymbol{R}_e=\boldsymbol{R}_e^T+\boldsymbol{R}_e^u$
(thermal force and friction force), according to 
\begin{eqnarray}
  \boldsymbol{R}_e^u &=& -\alpha_0 \rho_e\nu_{ei} \delta\bu_\parallel -\rho_e\nu_{ei}\delta\bu_\perp \Big( 1-\frac{\alpha_1' x^2+\alpha_0'}{\triangle}\Big)
  -\rho_e \nu_{ei}\bhat\times\delta \bu \frac{\alpha_1''x^3 +\alpha_0'' x}{\triangle};\nn\\
  \boldsymbol{R}_e^T &=& -\beta_0 n_e \nabla_\parallel T_e-n_e \nabla_\perp T_e \frac{\beta_1' x^2+\beta_0'}{\triangle}
  +n_e \bhat\times\nabla T_e \frac{\beta_1'' x^3+\beta_0'' x}{\triangle}.
\end{eqnarray}
Instead of a numerical Table II on page 25 of \cite{Braginskii1965}, we provide all the coefficients
in a fully analytic form for a general ion charge $Z_i$, which are given by
\begin{eqnarray}
\alpha_0 &=& \frac{4(16 Z_i^2+61 Z_i\sqrt{2}+72)}{217 Z_i^2+604 Z_i\sqrt{2}+288};  \qquad   
\beta_0 = \frac{30 Z_i (11 Z_i+15\sqrt{2})}{217 Z_i^2+604 Z_i\sqrt{2}+288}; \nn\\
\gamma_0 &=& \frac{25 Z_i(433 Z_i+180\sqrt{2})}{4(217 Z_i^2+604 Z_i\sqrt{2}+288)};
 \label{eq:BragCorr}
\end{eqnarray}

\begin{eqnarray}
 && \triangle = x^4+\delta_1 x^2+\delta_0;\nn\\
 &&   \delta_0 = \Big( \frac{217 Z_i^2+604 Z_i \sqrt{2}+288}{700 Z_i^2}\Big)^2;\nn\\
 &&   \delta_1 = \frac{586601 Z_i^2+ 330152 Z_i \sqrt{2}+106016}{78400 Z_i^2}; \label{eq:Triangle}
\end{eqnarray}
\begin{eqnarray}
   \alpha_1' &=& \frac{9(40337 Z_i + 10996 \sqrt{2})}{78400 Z_i};\nn\\
   \alpha_0' &=& \frac{9(217 Z_i^2+604 Z_i\sqrt{2}+288)(17 Z_i+40\sqrt{2})}{490000 Z_i^3};\nn\\
   \alpha_1'' &=& \frac{477}{280}; \qquad 
   \alpha_0'' = \frac{9(64Z_i^2+151 Z_i\sqrt{2}+253)}{6125 Z_i^2};
\end{eqnarray}
\begin{eqnarray}
  \beta_1' &=& \frac{3 (709 Z_i + 172 \sqrt{2})}{560 Z_i};\nn\\
  \beta_0' &=& \frac{3 (217 Z_i^2+604 Z_i\sqrt{2}+288)(11 Z_i+15\sqrt{2})}{49000 Z_i^3};\nn\\
\beta_1'' &=& \frac{3}{2}; \qquad \beta_0^{''} = \frac{3 (5729 Z_i^2 +6711 Z_i\sqrt{2}+4728)}{19600 Z_i^2};
\end{eqnarray}
\begin{eqnarray}
  \gamma_1' &=& \frac{13 Z_i+4\sqrt{2}}{4 Z_i};\nn\\
  \gamma_0' &=& \frac{(217 Z_i^2+604 Z_i\sqrt{2}+288)(433 Z_i+180\sqrt{2})}{78400 Z_i^3};\nn\\
  \gamma_1'' &=& \frac{5}{2}; \qquad
  \gamma_0'' = \frac{320797 Z_i^2 +202248 Z_i\sqrt{2}+ 72864}{31360 Z_i^2}. \label{eq:BraggCorr1}
\end{eqnarray}

Numerical values for $Z_i=1$ are given in the first column of Table II of \cite{Braginskii1965} and for example  
the parallel coefficients are $\alpha_0=0.5129$; $\beta_0=0.7110$; $\gamma_0=3.1616$, matching his values
exactly. We checked the entire Table II of Braginskii and his table is very precise,
except for two values. For the $\alpha_0$ coefficient, values for $Z_i=2,3$ should be
changed as $0.4408 \to 0.4309$; $0.3965 \to 0.3954$. 
The rest of his table is calculated very accurately, with around handful of irrelevant
last digit rounding changes (such as $3.7703\to 3.7702$ in $\delta_0 (Z_i=1)$, 
$0.2400\to 0.2399$ in $\alpha_0'' (Z_i=3)$; and for $Z_i=4$ charge $0.3752 \to 0.3751$ in $\alpha_0$; $9.055\to9.056$ in $\delta_0$
$0.4478 \to 0.4477$ in $\beta_0'$ etc.). 

Analytic results (\ref{eq:BragCorr}) for parallel coefficients $\alpha_0,\beta_0,\gamma_0$ were also
obtained by \cite{Simakov2014}; see later Section \ref{sec:Electron}.
To triple-check our other results, we re-calculated our approach with analytic collisional matrices of \cite{JiHeld2013}, equations (28a)-(28f),
together with their equations (40)-(44) and other formulas, which yielded the same analytic expressions. 
Unfortunately, the analytic results of \cite{Balescu1988} are written in a such a complicated form,
(see his page 236, with collisional matrices on page 198 and the required conversion equation (5.7.13) on page 270),
that we were able to verify only an analytic match with his parallel coefficients.
The formulation of \cite{Balescu1988} is so different from Braginskii, that Balescu by himself (page 275) only claims a match
of below 1\% for the 21-moment model, not further analyzing possible discrepancies.

\newpage
\subsection{Electron heat flux \texorpdfstring{$\vecX_e^{(5)}$}{X5}}
Similarly to the usual electron heat flux $\vecq_e$,
a quasi-static solution for the heat flux vector $\vecX_e^{(5)}$ has to be split into a thermal and frictional part,
according to
\begin{eqnarray}
  \vecX_e^{(5)T} &=& \frac{p_e}{\rho_e} \Big[-\kappa_\parallel^{e(5)} \nabla_\parallel T_e - \kappa_\perp^{e(5)} \nabla_\perp T_e
    + \kappa_\times^{e(5)} \bhat\times\nabla T_e \Big];\nn\\
\vecX_e^{(5)u} &=& \frac{p_e^2}{\rho_e} \Big[ \beta_0^{(5)}  \delta\bu_\parallel + \frac{\beta_1^{(5)'} x^2+\beta_0^{(5)'}}{\triangle} \delta\bu_\perp
-\frac{\beta_1^{(5)''}x^3+\beta_0^{(5)''}x}{\triangle} \bhat\times\delta\bu\Big], \label{eq:Thierry60}
\end{eqnarray}
with thermal conductivities
\begin{equation}
\kappa_\parallel^{e(5)} = \frac{p_e}{ m_e \nu_{ei}}\gamma_0^{(5)}; \qquad   
\kappa_\perp^{e(5)} =  \frac{p_e}{ m_e \nu_{ei}} \frac{\gamma_1^{(5)'} x^2+\gamma_0^{(5)'}}{\triangle};
\qquad \kappa_\times^{e(5)} =  \frac{p_e}{ m_e \nu_{ei}}\frac{\gamma_1^{(5)''} x^3+\gamma_0^{(5)''} x}{\triangle}. \label{eq:Thierry61}
\end{equation}
The analytic coefficients are given by
\begin{eqnarray}
  \beta_0^{(5)} &=& \frac{840 Z_i (13 \sqrt{2}+12 Z_i )}{217 Z_i^2+ 604 Z_i\sqrt{2} +288};\nn\\
  \beta_1^{(5)'} &=& \frac{3 (5829 Z_i + 1172\sqrt{2})}{280 Z_i};\nn\\
  \beta_0^{(5)'} &=& \frac{3 (217 Z_i^2 +604 Z_i \sqrt{2}+288)(12 Z_i+13 \sqrt{2})}{1750 Z_i^3};\nn\\
  \beta_1^{(5)''} &=& 27; \qquad  \beta_0^{(5)''} = \frac{3 (7611 Z_i^2+8429 Z_i\sqrt{2} +5000)}{700 Z_i^2}, \label{eq:Energy90}
\end{eqnarray}
and
\begin{eqnarray}
  \gamma_0^{(5)} &=& \frac{175 Z_i (204 \sqrt{2}+571 Z_i) }{217 Z_i^2+ 604 Z_i\sqrt{2} +288};\nn\\
  \gamma_1^{(5)'} &=& \frac{113 Z_i +44 \sqrt{2}}{2 Z_i};\nn\\
  \gamma_0^{(5)'} &=& \frac{(217 Z_i^2+604 Z_i\sqrt{2}+288)(571 Z_i+204\sqrt{2})}{2800 Z_i^3};\nn\\
  \gamma_1^{(5)''} &=& 70; \qquad  \gamma_0^{(5)''} = \frac{430783 Z_i^2  +261672 Z_i\sqrt{2} +86880}{ 1120 Z_i^2}, \label{eq:Energy91}
\end{eqnarray}
 with $\triangle$ unchanged and given by (\ref{eq:Triangle}). 
These results were substituted into the momentum exchange rates $\boldsymbol{R}_e$, equation (\ref{eq:ReExcite}),
to obtain the final expression for the friction force and thermal force. Useful relations are
\begin{eqnarray}
  \alpha_0 &=& 1-\frac{21}{10}\beta_0+\frac{3}{56}\beta_0^{(5)};\qquad
  \alpha_1' = \frac{21}{10}\beta_1'-\frac{3}{56}\beta_1^{(5)'};\nn\\
  \alpha_0' &=& \frac{21}{10}\beta_0'-\frac{3}{56}\beta_0^{(5)'};\qquad
  \alpha_1'' = \frac{21}{10}\beta_1'' -\frac{3}{56}\beta_1^{(5)''};\qquad
  \alpha_0'' = \frac{21}{10}\beta_0'' -\frac{3}{56}\beta_0^{(5)''};\nn\\
    \beta_0 &=& \frac{21}{10}\gamma_0-\frac{3}{56}\gamma_0^{(5)};\qquad
  \beta_1' = \frac{21}{10}\gamma_1'-\frac{3}{56}\gamma_1^{(5)'};\nn\\
  \beta_0' &=& \frac{21}{10}\gamma_0'-\frac{3}{56}\gamma_0^{(5)'}; \qquad
  \beta_1'' = \frac{21}{10}\gamma_1''-\frac{3}{56}\gamma_1^{(5)''}; \qquad
  \beta_0'' = \frac{21}{10}\gamma_0''-\frac{3}{56}\gamma_0^{(5)''}.
\end{eqnarray}
For $Z_i=1$, transport coefficients (\ref{eq:Energy90}), (\ref{eq:Energy91}) have numerical values
\begin{eqnarray}
  \beta_0^{(5)} &=& 18.778;\qquad 
  \beta_1^{(5)'} = 80.212;\qquad
  \beta_0^{(5)'} = 70.797;\qquad
  \beta_1^{(5)''} = 27; \qquad  \beta_0^{(5)''} = 105.135;\nn\\
  \gamma_0^{(5)} &=& 110.664;\qquad
  \gamma_1^{(5)'} = 87.613;\qquad
  \gamma_0^{(5)'} = 417.221;\qquad
  \gamma_1^{(5)''} = 70; \qquad  \gamma_0^{(5)''} = 792.610.
\end{eqnarray}

\newpage
\subsection{Ion viscosity \texorpdfstring{$\bPi^{(2)}_a$}{Pi2} of Braginskii (self-collisions)}
Considering self-collisions, evolution equations for the ion viscosity-tensors read
\begin{eqnarray}
  && \frac{d_a}{dt} \bPi^{(2)}_a  +\Omega_a \big(\bhat\times \bPi^{(2)}_a \big)^S + p_a \bW_a
  =  -\, \frac{6}{5}\nu_{aa} \bPi_a^{(2)} +\frac{9}{70} \nu_{aa} \Big(\frac{\rho_a}{p_a} \bPi^{(4)}_a-7  \bPi^{(2)}_a \Big); \label{eq:Energy32}\\
  && \frac{d_a}{dt} \bPi^{(4)}_a  +\Omega_a \big(\bhat\times \bPi^{(4)}_a \big)^S + 7 \frac{p_a^2}{\rho_a} \bW_a 
  =  -\, \frac{33}{5} \nu_{aa} \frac{p_a}{\rho_a} \bPi^{(2)}_a - \frac{79}{140} \nu_{aa} \Big(\bPi^{(4)}_a -7\frac{p_a}{\rho_a} \bPi^{(2)}_a\Big). \label{eq:Energy31}
\end{eqnarray}
Neglecting (\ref{eq:Energy31}) and prescribing closure (\ref{eq:Closure1}) which neglects the second term on the r.h.s. of (\ref{eq:Energy32}),
yields the ion-viscosity model of Burgers-Schunk, with the well-known $-6/5$ constant. However, now the equations read
\begin{eqnarray}
  && \frac{d_a}{dt} \bPi^{(2)}_a  +\Omega_a \big(\bhat\times \bPi^{(2)}_a \big)^S + p_a \bW_a
  =  -\, \frac{21}{10}\nu_{aa} \bPi_a^{(2)} +\frac{9}{70} \nu_{aa} \frac{\rho_a}{p_a} \bPi^{(4)}_a;\nn\\
  && \frac{d_a}{dt} \bPi^{(4)}_a  +\Omega_a \big(\bhat\times \bPi^{(4)}_a \big)^S + 7 \frac{p_a^2}{\rho_a} \bW_a 
  =  -\, \frac{53}{20} \nu_{aa} \frac{p_a}{\rho_a} \bPi^{(2)}_a - \frac{79}{140} \nu_{aa} \bPi^{(4)}_a. \label{eq:Posled12}
\end{eqnarray}
In a quasi-static approximation, solution of (\ref{eq:Posled12}) yields $\bPi^{(2)}_a$ in the following form (see for example
Appendix \ref{sec:BragNonlin})
\begin{eqnarray}
  \bPi^{(2)}_a &=& -\eta_0^a \bW_0 -\eta_1^a\bW_1 -\eta_2^a\bW_2 +\eta_3^a\bW_3+\eta_4^a\bW_4; \label{eq:Energy81}\\
  \bW_0 &=& \frac{3}{2}\big(\bW_a:\bhat\bhat\big) \Big( \bhat\bhat-\frac{\bI}{3}\Big);\nn\\
  \bW_1 &=& \bI_\perp \cdot\bW_a\cdot\bI_\perp +\frac{1}{2}\big( \bW_a:\bhat\bhat\big) \bI_\perp;\nn\\
  \bW_2 &=& \big( \bI_\perp\cdot\bW_a\cdot\bhat\bhat\big)^S;\nn\\
  \bW_3 &=& \frac{1}{2}\big( \bhat\times \bW_a\cdot\bI_\perp\big)^S;\nn\\
  \bW_4 &=& \big(\bhat\times\bW_a\cdot\bhat\bhat\big)^S, \label{eq:Energy80}
\end{eqnarray}
which is equivalent to equations (4.41) \& (4.42) of \cite{Braginskii1965}, after one prescribes in his $\bW_0$ that the matrix $\bW_a$ is traceless.
Alternatively, with respect to $\bhat=(0,0,1)$ (straight magnetic field applied in the
z-direction)
\begin{eqnarray}
  && \Pi_{xx}^{a(2)} = -\frac{\eta_0^a}{2} (W_{xx}^a+W_{yy}^a) -\frac{\eta_1^a}{2}(W_{xx}^a-W_{yy}^a) -\eta_3^a W_{xy}^a  \nn   ;\\
  && \Pi_{xy}^{a(2)} = \frac{\eta_3^a}{2} (W_{xx}^a-W_{yy}^a) -\eta_1^a W_{xy}^a   \nn  ;\\
  && \Pi_{xz}^{a(2)} = -\eta_4^a W_{yz}^a-\eta_2^a W_{xz}^a   \nn ;\\
  && \Pi_{yy}^{a(2)} = -\frac{\eta_0^a}{2} (W_{xx}^a+W_{yy}^a) +\frac{\eta_1^a}{2}(W_{xx}^a-W_{yy}^a) +\eta_3^a W_{xy}^a; \nn \\
  && \Pi_{yz}^{a(2)} = \eta_4^a W_{xz}^a-\eta_2^a W_{yz}^a   \nn  ;\\
  && \Pi_{zz}^{a(2)} = -\eta_0^a W_{zz}^a, \label{eq:Energy82}
\end{eqnarray}
which is equation (2.21) of \cite{Braginskii1965}. The ion viscosities are
\begin{eqnarray}
  \eta_0^a &=&  \frac{1025}{1068} \frac{p_a}{\nu_{aa}};\nn\\
  \eta_2^a &=&  \frac{p_a}{\nu_{aa}}\, \frac{(6/5) x^2+(10947/4900)}{x^4+(79321/19600) x^2 +(71289/30625)} ;\nn\\
  \eta_4^a &=&  \frac{p_a}{\nu_{aa}}\, \frac{x^3+(46561/19600)x}{x^4+(79321/19600)x^2+(71289/30625)},
\end{eqnarray}
where $x=\Omega_a/\nu_{aa}$, and $\eta_1^a(x)=\eta_2^a(2x)$; $\eta_3^a(x)=\eta_4^a(2x)$ 
(The solution is easily obtained for the parallel ``zz'' direction with $\Omega_a=0$, and
for perpendicular directions for example by choosing coupled ``xz'' and ``yz'' directions, and solving 4 equations in 4 unknowns). 
Alternatively, using numerical values
\begin{eqnarray}
\eta_0^a &=&  0.960 \frac{p_a}{\nu_{aa}};\nn\\
\eta_2^a &=& \frac{p_a}{\nu_{aa}}\, \frac{(6/5)x^2+2.234}{x^4+4.047 x^2+2.328};\nn\\
\eta_4^a &=& \frac{p_a}{\nu_{aa}}\, \frac{x^3+2.376 x}{x^4+4.047 x^2+2.328},
\end{eqnarray}
recovering ion viscosities of \cite{Braginskii1965}, his equation (4.44). Numerical values in Braginskii are evaluated precisely, with
the sole exception of one value in the denominator, where his rounded 4.03 should be replaced by 4.05.  

\subsection{Ion viscosity \texorpdfstring{$\bPi^{(4)}_a$}{Pi4} (self-collisions)}
The ion viscosity tensor $\bPi^{(4)}_a$ is given by
\begin{equation} \label{eq:Energy88}
  \bPi^{(4)}_a = \frac{p_a}{\rho_a}\Big[-\eta_0^{a(4)} \bW_0 -\eta_1^{a(4)}\bW_1 -\eta_2^{a(4)}\bW_2 +\eta_3^{a(4)}\bW_3+\eta_4^{a(4)}\bW_4\Big],
\end{equation}
with matrices $\bW_0-\bW_4$ (\ref{eq:Energy80}) unchanged, and viscosities
\begin{eqnarray}
  \eta_0^{a(4)} &=& \frac{8435}{1068}  \frac{p_a}{\nu_{aa}};\nn\\
  \eta_2^{a(4)} &=&  \frac{p_a}{\nu_{aa}}\, \frac{(33/5)x^2+(64347/3500)}{x^4+(79321/19600)x^2+(71289/30625)};\nn\\
  \eta_4^{a(4)} &=&  \frac{p_a}{\nu_{aa}}\, \frac{7x^3+(59989/2800)x}{x^4+(79321/19600)x^2+(71289/30625)},
\end{eqnarray}
where $\eta_1^{a(4)}(x)=\eta_2^{a(4)}(2x)$, $\eta_3^{a(4)}(x)=\eta_4^{a(4)}(2x)$ holds, or with numerical values
\begin{eqnarray}
  \eta_0^{a(4)} &=& 7.898  \frac{p_a}{\nu_{aa}};\nn\\
  \eta_2^{a(4)} &=&  \frac{p_a}{\nu_{aa}}\, \frac{6.600x^2+18.385}{x^4+4.047x^2+2.328};\nn\\
  \eta_4^{a(4)} &=&  \frac{p_a}{\nu_{aa}}\, \frac{7x^3+21.425 x}{x^4+4.047x^2+2.328}.
\end{eqnarray}

\newpage
\subsection{Electron viscosity \texorpdfstring{$\bPi^{(2)}_e$}{Pi2} of Braginskii}
For a one ion-electron plasma with  similar temperatures, the  mass-ratio coefficients (\ref{eq:Posled23}), (\ref{eq:Posled24}) simplify into
\begin{eqnarray}
  \hat{K}_{ei (1)} &=& 3\frac{m_i}{m_e}; \qquad \hat{K}_{ei (2)} =\frac{4}{5}; \qquad L_{ei (1)} = \frac{9}{35} \frac{m_i}{m_e};
  \qquad L_{ei (2)} = \frac{12}{35} \frac{m_e}{m_i};\nn\\
  \hat{M}_{ei (1)} &=& \frac{36}{5}; \qquad \hat{M}_{ei (2)} = \frac{4}{5} \frac{m_e}{m_i}; \qquad
  N_{ei (1)} = -\,\frac{12}{35}; \qquad N_{ei (2)} = -\,\frac{36}{35} \frac{m_e^2}{m_i^2},
\end{eqnarray}
and collisional exchange rates for the viscosity-tensors become
\begin{eqnarray}
  \bQ_{e}^{(2)}\,' &=& -\Big(\frac{21}{10}\nu_{ee} +3\nu_{ei} \Big) \bPi^{(2)}_e
  +\Big( \frac{9}{70}\nu_{ee}+\frac{9}{35}\nu_{ei}\Big) \frac{\rho_e}{p_e} \bPi^{(4)}_e;\nn\\
  \bQ_{e}^{(4)}\,' &=& -\Big( \frac{53}{20}\nu_{ee}+\frac{36}{5}\nu_{ei}\Big) \frac{p_e}{\rho_e}\bPi^{(2)}_e 
  +\Big( -\frac{79}{140}\nu_{ee} +\frac{12}{35}\nu_{ei}\Big) \bPi^{(4)}_e.
\end{eqnarray}
Converting everything to $\nu_{ei}$ with $\nu_{ee}=\nu_{ei}/(Z_i\sqrt{2})$ yields
\begin{eqnarray}
  \bQ_{e}^{(2)}\,' &=& -\Big(\frac{21}{10 Z_i \sqrt{2}} +3 \Big)\nu_{ei} \bPi^{(2)}_e
  +\Big( \frac{9}{70 Z_i \sqrt{2}}+\frac{9}{35}\Big)\nu_{ei} \frac{\rho_e}{p_e} \bPi^{(4)}_e;\nn\\
  \bQ_{e}^{(4)}\,' &=& -\Big( \frac{53}{20 Z_i \sqrt{2}}+\frac{36}{5}\Big)\nu_{ei} \frac{p_e}{\rho_e}\bPi^{(2)}_e 
  +\Big( -\frac{79}{140 Z_i \sqrt{2}} +\frac{12}{35}\Big) \nu_{ei}\bPi^{(4)}_e, \label{eq:Posled32}
\end{eqnarray}
and these contributions enter the r.h.s. of evolution equations
\begin{eqnarray}
  && \frac{d_e}{dt} \bPi^{(2)}_e  +\Omega_e \big(\bhat\times \bPi^{(2)}_e \big)^S + p_e \bW_e
  = \bQ_{e}^{(2)}\,' ; \nn\\
  && \frac{d_e}{dt} \bPi^{(4)}_e  +\Omega_e \big(\bhat\times \bPi^{(4)}_e \big)^S + 7 \frac{p_e^2}{\rho_e} \bW_e 
  =  \bQ^{(4)}_{e}\,'. \label{eq:Posled31}
\end{eqnarray}
In a quasi-static approximation, solution of (\ref{eq:Posled32}), (\ref{eq:Posled31}) yields the electron viscosity tensor $\bPi^{(2)}_e$ in form
 (\ref{eq:Energy81}), (\ref{eq:Energy80}), with electron viscosities
\begin{eqnarray}
  \eta_0^e &=& \frac{p_e}{\nu_{ei}} \,\frac{5 Z_i(408 Z_i +205\sqrt{2})}{6 (192 Z_i^2+301 Z_i\sqrt{2}+178)};\nn\\
  \eta_2^e &=& \frac{p_e}{\nu_{ei}}\Big[\frac{3 \sqrt{2}+6 Z_i}{5Z_i}x^2
    + \frac{3 (192 Z_i^2+301 Z_i \sqrt{2}+178)(408 Z_i+205 \sqrt{2})}{196000 Z_i^3} \Big]/\triangle;\nn\\
  \eta_4^e &=& \frac{p_e}{\nu_{ei}} x \Big[ x^2+ \frac{119520 Z_i^2 + 101784 \sqrt{2}Z_i+46561}{39200 Z_i^2}  \Big]/\triangle;\nn\\
  \triangle &=& x^4 +\frac{212256 Z_i^2 +176376 \sqrt{2} Z_i +79321}{39200 Z_i^2}x^2 
   +\Big( \frac{3(192 Z_i^2+301 Z_i\sqrt{2}+178)}{700 Z^2} \Big)^2, \label{eq:Posled40}
\end{eqnarray}
where $x=\Omega_e/\nu_{ei}$, and relations $\eta_1^{e}(x)=\eta_2^{e}(2x)$, $\eta_3^{e}(x)=\eta_4^{e}(2x)$. For the particular case of $Z_i=1$ these electron viscosities become
\begin{eqnarray}
  \eta_0^e &=& \frac{p_e}{\nu_{ei}} \,\frac{2040 +1025\sqrt{2}}{2220 + 1806\sqrt{2}};\nn\\
  \eta_2^e &=& \frac{p_e}{\nu_{ei}} \Big[\frac{3\sqrt{2}+6}{5} x^2 + \frac{297987}{98000}\sqrt{2}+ \frac{82311}{19600}\Big]/\triangle;\nn\\
  \eta_4^e &=& \frac{p_e}{\nu_{ei}} x \Big[ x^2+ \frac{12723}{4900}\sqrt{2} +\frac{166081}{39200} \Big] / \triangle;\nn\\
  \triangle &=& x^4+ \Big( \frac{22047}{4900}\sqrt{2}+\frac{291577}{39200}\Big)x^2 + \Big( \frac{1431459}{245000}+ \frac{14319}{3500}\sqrt{2}\Big), \label{eq:Energy87}
\end{eqnarray}  
or with numerical values
\begin{eqnarray}
  \eta_0^e &=&  0.73094 \frac{p_e}{\nu_{ei}};\nn\\
  \eta_2^e &=& \frac{p_e}{\nu_{ei}} \big(2.049 x^2+8.500 \big)/\triangle;\nn\\
  \eta_4^e &=& \frac{p_e}{\nu_{ei}} x\big(x^2+7.909 \big)/\triangle;\nn\\
  \triangle &=& x^4+13.801 x^2+11.628,
\end{eqnarray}
recovering the electron viscosity of \cite{Braginskii1965}, his equation (4.45).  
It appears that the Braginskii parallel viscosity value of 0.733 is slightly imprecise and should be 0.731 instead.
The analytic result for parallel viscosity $\eta^e_0$ agrees with \cite{Simakov2014}, and the
value 0.73094 agrees with \cite{JiHeld2013}; see the inset of their Figure 3 (curiously, 
in a more precise 3-Laguerre approximation the coefficient changes to 0.733). Note that for $x\to 0$ viscosity $\eta_2^e\to \eta_0^e$. 
As discussed previously, our $\Omega_e$ is negative and in Braginskii it is positive, yielding an opposite sign in front of $\eta_4^e$.
Braginskii offers electron viscosities only for $Z_i=1$.
The analytic result (\ref{eq:Posled40}) is useful to quickly calculate electron viscosities for any $Z_i$. \cite{JiHeld2013,JiHeld2015} also
provide useful fitting formulas.

\subsection{Electron viscosity \texorpdfstring{$\bPi^{(4)}_e$}{Pi4}}
The solution for electron viscosity tensor $\bPi^{(4)}_e$ has form (\ref{eq:Energy88}) with viscosities
\begin{eqnarray}
  \eta_0^{e(4)} &=& \frac{p_e}{\nu_{ei}} \, \frac{35 Z_i (552 Z_i+241\sqrt{2})}{6(192 Z_i^2+301 Z_i \sqrt{2} +178)};\nn\\
  \eta_2^{e(4)} &=& \frac{p_e}{\nu_{ei}}\Big[ \frac{33 \sqrt{2}+48 Z_i}{10 Z_i} x^2
    +\frac{3 (192 Z_i^2+ 301 Z_i \sqrt{2}+178)(552 Z_i+241\sqrt{2})}{28000 Z_i^3} \Big]/\triangle;\nn\\
  \eta_4^{e(4)} &=& \frac{p_e}{\nu_{ei}} x \Big[ 7 x^2 +\frac{173088 Z_i^2 + 142032 Z_i \sqrt{2} +59989}{5600 Z_i^2}  \Big]/\triangle,
  \label{eq:Energy85}
\end{eqnarray}
where the denominator $\triangle$ is equivalent to (\ref{eq:Posled40}).
For the particular case of $Z_i=1$ these electron viscosities become
\begin{eqnarray}
  \eta_0^{e(4)} &=& \frac{p_e}{\nu_{ei}} \,\frac{35 (241 \sqrt{2}+552)}{6 (301 \sqrt{2}+370)};\nn\\
  \eta_2^{e(4)} &=& \frac{p_e}{\nu_{ei}} \Big[\frac{33 \sqrt{2}+48}{10} x^2+\frac{382983}{14000}\sqrt{2}+\frac{523983}{14000}   \Big]/\triangle;\nn\\
  \eta_4^{e(4)} &=& \frac{p_e}{\nu_{ei}} x \Big[ 7 x^2+ \frac{8877}{350}\sqrt{2}+\frac{233077}{5600} \Big] / \triangle,
\end{eqnarray}
with $\triangle$ equal to (\ref{eq:Energy87}), and with numerical values
\begin{eqnarray}
  \eta_0^{e(4)} &=&  6.546 \frac{p_e}{\nu_{ei}};\nn\\
  \eta_2^{e(4)} &=& \frac{p_e}{\nu_{ei}} \big(9.467 x^2+76.114 \big)/\triangle;\nn\\
  \eta_4^{e(4)} &=& \frac{p_e}{\nu_{ei}} x\big(7 x^2+77.489 \big)/\triangle;\nn\\
  \triangle &=& x^4+13.801 x^2+11.628.
\end{eqnarray}

\newpage
\section{Generalized electron coefficients for multi-species plasmas} \label{sec:Electron}
Here we use the idea of \cite{Simakov2014}, and before that for example by \cite{Zhdanov2002} (orig publ. 1982) and \cite{Hinton1983},  
who pointed out that because of the smallness of mass ratios $m_e/m_i$,
the electron coefficients of \cite{Braginskii1965} can be
straightforwardly generalized for multi-species plasmas. \cite{Simakov2014} considered unmagnetized plasmas and provide analytic parallel coefficients
$\alpha_0$, $\beta_0$, $\gamma_0$ together with the parallel electron viscosity $\eta_0^{e}$. Here we show that the same construction
applies when a magnetic field is present, and that all the electron
coefficients provided in the previous section can be easily generalized in the same way. 
One starts by considering the general multi-species description with collisional contributions given in Section \ref{sec:Coll-Tequal}.
Because of the smallness of $m_e/m_i$,  mass-ratio coefficients for each ion species simplify into (\ref{eq:Num85}). 
One introduces an effective ion charge together with an effective ion velocity
\begin{equation}
Z_{\textrm{eff}} = \frac{\sum_i \nu_{ei}}{\sqrt{2}\nu_{ee}} = \frac{\sum_i n_i Z_i^2}{n_e}; \qquad \langle \bu_i\rangle_{\textrm{eff}} = \frac{\sum_i \nu_{ei}\bu_i }{\sum_i \nu_{ei}},
\end{equation}
and it is straightforward to show that the collisional contributions for a one ion-electron plasma (\ref{eq:ReExcite}), (\ref{eq:Energy30})
are then replaced by
\begin{eqnarray}
  \boldsymbol{R}_e &=& - \rho_e (\sum_i \nu_{ei})(\bu_e -\langle\bu_i\rangle_{\textrm{eff}}) + \frac{21}{10}\frac{\rho_e}{p_e} (\sum_i \nu_{ei})\vecq_e
  -\frac{3}{56} \frac{\rho_e^2}{p_e^2} (\sum_i\nu_{ei}) \vecX^{(5)}_e;\label{eq:Num80}\\
  \vec{\boldsymbol{Q}}^{(3)}_{e}\,' &=&  +\frac{3}{2}p_e (\sum_i \nu_{ei})(\bu_e -\langle\bu_i\rangle_{\textrm{eff}})
  -\Big[ \frac{\sqrt{2}}{Z_{\textrm{eff}}}+\frac{19}{4} \Big]  (\sum_i\nu_{ei}) \vecq_e
  +\Big[ \frac{3}{70\sqrt{2} Z_{\textrm{eff}}}+\frac{69}{560} \Big] (\sum_i\nu_{ei}) \frac{\rho_e}{p_e} \vecX^{(5)}_e; \label{eq:Num82}\\
  \vec{\boldsymbol{Q}}^{(5)}_{e}\,' &=& + 27\frac{p_e^2}{\rho_e}(\sum_i \nu_{ei})(\bu_e -\langle\bu_i\rangle_{\textrm{eff}})
  -\Big[ \frac{76}{5 \sqrt{2}Z_{\textrm{eff}} }+\frac{759}{10} \Big] (\sum_i\nu_{ei})\frac{p_e}{\rho_e}\vecq_e
  -\Big[ \frac{3}{35 \sqrt{2}Z_{\textrm{eff}}} -\frac{533}{280}\Big] (\sum_i\nu_{ei})\vecX^{(5)}_e. \label{eq:Num83}
\end{eqnarray}
Contributions (\ref{eq:Num82}), (\ref{eq:Num83}) enter the right-hand-sides of electron evolution equations (\ref{eq:ExciteRit}).
The system is completely the same as for the one ion-electron plasma, if in (\ref{eq:ReExcite}), (\ref{eq:Energy30}) the following replacement is applied
\begin{equation} \label{eq:miracle}
Z_i \to Z_{\textrm{eff}} ;\qquad \nu_{ei} \to \sum_i \nu_{ei}; \qquad   \delta\bu=\bu_e-\bu_i \to \bu_e - \langle \bu_i\rangle_{\textrm{eff}}.
\end{equation}
If evolution equations can be obtained with the transformation (\ref{eq:miracle}), of course their solution can be obtained with the same transformation as well.
The same transformation applies for the viscous evolution equations (\ref{eq:Posled32}), (\ref{eq:Posled31}) and their solutions. As an example, the generalized 
(thermal) electron heat of \cite{Braginskii1965} for multi-species plasmas reads 
\begin{eqnarray}
  \vecq_e^T &=& -\kappa_\parallel^e \nabla_\parallel T_e - \kappa_\perp^e \nabla_\perp T_e + \kappa_\times^e \bhat\times\nabla T_e;\\
 \kappa_\parallel^e &=& \frac{p_e}{m_e (\sum_i\nu_{ei})}\gamma_0; \qquad   
\kappa_\perp^e =  \frac{p_e}{m_e (\sum_i\nu_{ei})} \frac{\gamma_1' x^2+\gamma_0'}{\triangle};
\qquad \kappa_\times^e =  \frac{p_e}{m_e (\sum_i\nu_{ei})}\frac{\gamma_1'' x^3+\gamma_0''x}{\triangle}; \\
\gamma_0 &=& \frac{25 Z_{\textrm{eff}}(433 Z_{\textrm{eff}}+180\sqrt{2})}{4(217 Z_{\textrm{eff}}^2 +604 Z_{\textrm{eff}}\sqrt{2}+288)}; \qquad
  \gamma_1' = \frac{13 Z_{\textrm{eff}}+4\sqrt{2}}{4 Z_{\textrm{eff}}}; \qquad \gamma_1'' = \frac{5}{2};\nn\\
  \gamma_0' &=& \frac{(217 Z_{\textrm{eff}}^2+604 Z_{\textrm{eff}}\sqrt{2}+288)(433 Z_{\textrm{eff}}+180\sqrt{2})}{78400 Z_{\textrm{eff}}^3};\qquad
  \gamma_0'' = \frac{320797 Z_{\textrm{eff}}^2 +202248 Z_{\textrm{eff}}\sqrt{2}+ 72864}{31360 Z_{\textrm{eff}}^2}; \nn\\
  \triangle &=& x^4+\delta_1 x^2+\delta_0;\qquad \delta_0 = \Big( \frac{217 Z_{\textrm{eff}}^2+604 Z_{\textrm{eff}} \sqrt{2}+288}{700 Z_{\textrm{eff}}^2}\Big)^2;\nn\\
  \delta_1 &=& \frac{586601 Z_{\textrm{eff}}^2+ 330152 Z_{\textrm{eff}} \sqrt{2}+106016}{78400 Z_{\textrm{eff}}^2},
\end{eqnarray}
where $x=\Omega_e/(\sum_i\nu_{ei})$. With recipe (\ref{eq:miracle}) one obtains generalized solutions for the frictional electron heat flux $ \vecq_e^u$,
together with solutions for $\vecX^{(5)}_e$ and viscosity-tensors $\bPi^{(2)}_e$, $\bPi^{(4)}_e$ which are not repeated here.

From the electron momentum equation, the electric field then becomes
\begin{eqnarray}
  \bE  &=& -\frac{1}{c}\bu_e\times\bb-\frac{1}{en_e}\nabla\cdot\bp_e +\frac{m_e}{e}\boldsymbol{G}\nn\\
  && +(\sum_i\nu_{ei})\Big[+\frac{m_e}{e}\big(\langle \bu_i\rangle_{\textrm{eff}} - \bu_e\big)
  +\frac{21}{10}\frac{m_e}{ep_e}\vecq_e
  -\frac{3}{56}\frac{\rho_e^2}{e n_e p_e^2}\vecX^{(5)}_e\Big]
  -\frac{m_e}{e}\frac{d_e \bu_e}{dt}, \label{eq:Num90}
\end{eqnarray}
and expressions for heat fluxes $\vecq_e$ and $\vecX^{(5)}_e$ enter the electric field.

\section{Generalization with coupling of stress-tensors and heat fluxes} \label{sec:Couplingg1}
Here we consider the coupling between viscosity-tensors and heat fluxes.
Using the semi-linear approximation and retaining the coupling,
the 21-moment model (\ref{eq:Num1000})-(\ref{eq:Num1004}) simplifies into
\begin{eqnarray}
  && \frac{d_a}{dt} \bPi^{(2)}_a  +\Omega_a \big(\bhat\times \bPi^{(2)}_a \big)^S + p_a \bW_a + \frac{2}{5}\Big((\nabla \vecq_a)^S -\frac{2}{3}\bI \nabla\cdot\vecq_a\Big)\nn\\
  && \qquad = \bQ_{a}^{(2)}\,' = \bQ^{(2)}_a -\frac{\bI}{3}\textrm{Tr}\bQ^{(2)}_a \label{eq:Nomore110};\\
   &&  \frac{d_a}{d t}\vecq_a + \Omega_a \bhat\times\vecq_a + \frac{5}{2}p_a \nabla \Big(\frac{p_a}{\rho_a}\Big)
  +\frac{1}{2}\nabla\cdot\bPi^{(4)}_a-\frac{5}{2}\frac{p_a}{\rho_a}\nabla\cdot\bPi^{(2)}_a\nn\\
  && \qquad = \vec{\boldsymbol{Q}}^{(3)}_{a}\,' = \frac{1}{2}\textrm{Tr}\bQ^{(3)}_a-\frac{5}{2}\frac{p_a}{\rho_a}\boldsymbol{R}_a;\label{eq:Nomore110X}\\ 
     && \frac{d_a}{dt} \bPi^{(4)}_a +\Omega_a \big( \bhat\times \bPi^{(4)}_a \big)^S  + 7 \frac{p_a^2}{\rho_a} \bW_a
     +\frac{1}{5}\Big[ (\nabla\vecX^{(5)}_a)^S-\frac{2}{3}\bI(\nabla\cdot\vecX^{(5)}_a)\Big] \nn\\
  && \qquad   = \bQ^{(4)}_a\,' = \trace \bQ^{(4)}_a -\frac{\bI}{3}\trace\trace \bQ^{(4)}_a;\label{eq:Nomore103}\\
  &&  \frac{d_a}{d t}\vecX^{(5)}_a +\Omega_a \bhat\times \vecX^{(5)}_a  +70 \frac{p_a^2}{\rho_a}\nabla\Big(\frac{p_a}{\rho_a}\Big)
  +18\frac{p_a}{\rho_a}\nabla\cdot \bPi^{(4)}_a
  -98 \frac{p_a^2}{\rho_a^2}\nabla\cdot\bPi^{(2)}_a  \nn\\ 
  && \qquad   = \vec{\boldsymbol{Q}}^{(5)}_{a}\,' = \trace\trace \bQ^{(5)}_a -35 \frac{p_a^2}{\rho_a^2}\boldsymbol{R}_a. \label{eq:Nomore102}
\end{eqnarray}
 Terms such as $(\nabla p_a)\vecq_a$ were neglected and large-scale gradients are assumed to be small (see Section \ref{sec:SemiLinear}).
The right-hand-sides were given in Sections \ref{sec:Tarb2} \& \ref{sec:Coll-Tequal}, and for one ion-electron plasmas
in Section \ref{sec:ONEion}.  
The system now represents a generalization of \cite{Braginskii1965}, where heat fluxes and stress-tensors are coupled.  
For the highest-level of precision, one should solve dispersion relations directly with the above system, where all the heat fluxes and stress-tensors are
\emph{independent} variables. At the lowest-level of precision, one prescribes the quasi-static approximation and cancels the time-derivatives $d/dt$.
Nevertheless, for sufficiently low frequencies there exists a ``middle-route'' procedure known from the algebra of collisionless models, 
by decomposing each moment into its first and second orders
\begin{eqnarray}
\vecq_a &=& \vecq_a^{(1)}+\vecq_a^{(2)}; \qquad \vecX^{(5)}_a = \vecX^{(5,1)}_a+\vecX^{(5,2)}_a;\nn\\
\bPi^{(2)}_a &=& \bPi^{(2,1)}_a + \bPi^{(2,2)}_a; \qquad \bPi^{(4)}_a = \bPi^{(4,1)}_a + \bPi^{(4,2)}_a,
\end{eqnarray}
and by neglecting the time derivative of the second-order moments. One can consider
\begin{eqnarray}
  && \frac{d_a}{dt} \bPi^{(2,1)}_a  +\Omega_a \big(\bhat\times \bPi^{(2)}_a \big)^S + p_a \bW_a + \frac{2}{5}\Big((\nabla \vecq_a^{(1)})^S -\frac{2}{3}\bI \nabla\cdot\vecq_a^{(1)}\Big)
  = \bQ_{a}^{(2)}\,' ;\label{eq:Nomore206}\\
    && \frac{d_a}{dt} \bPi^{(4,1)}_a +\Omega_a \big( \bhat\times \bPi^{(4)}_a \big)^S  + 7 \frac{p_a^2}{\rho_a} \bW_a
     +\frac{1}{5}\Big[ (\nabla\vecX^{(5,1)}_a)^S-\frac{2}{3}\bI(\nabla\cdot\vecX^{(5,1)}_a)\Big]
 = \bQ^{(4)}_a\,';\label{eq:Nomore207}\\
   &&  \frac{d_a}{d t}\vecq_a^{(1)} + \Omega_a \bhat\times\vecq_a + \frac{5}{2}p_a \nabla \Big(\frac{p_a}{\rho_a}\Big)
  +\frac{1}{2}\nabla\cdot\bPi^{(4,1)}_a-\frac{5}{2}\frac{p_a}{\rho_a}\nabla\cdot\bPi^{(2,1)}_a = \vec{\boldsymbol{Q}}^{(3)}_{a}\,';\label{eq:Nomore204}\\ 
  &&  \frac{d_a}{d t}\vecX^{(5,1)}_a +\Omega_a \bhat\times \vecX^{(5)}_a  +70 \frac{p_a^2}{\rho_a}\nabla\Big(\frac{p_a}{\rho_a}\Big)
   +18\frac{p_a}{\rho_a}\nabla\cdot \bPi^{(4,1)}_a  -98 \frac{p_a^2}{\rho_a^2}\nabla\cdot\bPi^{(2,1)}_a   
   = \vec{\boldsymbol{Q}}^{(5)}_{a}\,', \label{eq:Nomore205}
\end{eqnarray}
 where the collisional contributions on the right hand sides contain full moments $\bPi^{(2)}_a$, $\bPi^{(4)}_a$, $\vecq_a$, $\vecX^{(5)}_a$.
In the collisionless regime, a similar procedure was used for example by \cite{Macmahon1965}, \cite{Smolyakov1985}, \cite{Ramos2005}, \cite{Goswami2005},
\cite{PSH2012}, and 
it is well-known that retaining the time-derivatives $d/dt$ is crucial for recovering the dispersion relation of
perpendicular fast mode with respect to kinetic theory (its wavenumber dependence in the long-wavelength limit).
 It is straightforward to further increase the precision, by for example retaining full $\vecq_a$ \& $\vecX^{(5)}_a$
in the last terms of (\ref{eq:Nomore206}) \& (\ref{eq:Nomore207}),
or by retaining full $\bPi_a^{(2)}$ \& $\bPi_a^{(4)}$ in the last terms of (\ref{eq:Nomore204}) \& (\ref{eq:Nomore205}) (which we do not show).
The procedure and its application is described in detail in \cite{Hunana2019a} (see Sections
5.8 \& 5.9), and the coupling of stress-tensors and heat fluxes is also crucial for the firehose instability  (see Figures 7 \& 10 there; see also 
  figures with simpler models in \cite{HunanaZank2017}).

\newpage
\section{Coupling for unmagnetized one ion-electron plasma.} \label{sec:Couplingg}
We further focus on the particular case of a one ion-electron plasma with  similar temperatures. It is of course possible to algebraically solve
the entire system (\ref{eq:Nomore206})-(\ref{eq:Nomore205}) with a magnetic field present, which will be presented elsewhere. 
Here for clarity and to demonstrate our point we find it sufficient to focus on an unmagnetized plasma.
Equivalently, we thus only consider solutions for parallel moments along the magnetic field, similarly to the heat flux model of
\cite{SpitzerHarm1953}.
 For the heat flux equations (\ref{eq:Nomore206}), (\ref{eq:Nomore207}) it is beneficial to introduce matrices
\begin{equation} \label{eq:Nomore230}
\bY^{(3,1)}_a = (\nabla \vecq_a^{(1)})^S -\frac{2}{3}\bI \nabla\cdot\vecq_a^{(1)}; \qquad \bY^{(5,1)}_a = (\nabla\vecX^{(5,1)}_a)^S-\frac{2}{3}\bI\nabla\cdot\vecX^{(5,1)}_a,
\end{equation}  
which are symmetric and traceless, analogously to matrix $\bW_a$.

\subsection{Ion species (self-collisions)}
For the ion species, the viscosity-tensors have the following form
\begin{eqnarray}
  \bPi^{(2,1)}_a &=& - \frac{1025}{1068} \frac{p_a}{\nu_{aa}} \bW_a;\nn\\
  \bPi^{(2,2)}_a &=& - \frac{1}{\nu_{aa}}\Big[\frac{79}{534}\bY^{(3,1)}_a + \frac{3}{178}\frac{\rho_a}{p_a} \bY^{(5,1)}_a
    + \frac{395}{1068}\frac{\pr \bPi^{(2,1)}_a}{\pr t} + \frac{15}{178} \frac{\rho_a}{p_a}\frac{\pr \bPi^{(4,1)}_a}{\pr t} \Big];\nn\\
  \bPi^{(4,1)}_a &=& - \frac{8435}{1068} \frac{p_a^2}{\rho_a \nu_{aa}}\bW_a;\nn\\
   \bPi^{(4,2)}_a &=& +\frac{1}{\nu_{aa}}\Big[ + \frac{371}{534}\frac{p_a}{\rho_a}\bY^{(3,1)}_a
     -\frac{49}{178}\bY^{(5,1)}_a + \frac{1855}{1068} \frac{p_a}{\rho_a} \frac{\pr \bPi^{(2,1)}_a}{\pr t}
     - \frac{245}{178}\frac{\pr \bPi^{(4,1)}_a}{\pr t}   \Big], \label{eq:Energy200}
\end{eqnarray}
and heat fluxes become
\begin{eqnarray}
  \vecq_a^{(1)} &=& - \frac{125}{32} \frac{p_a}{m_a \nu_{aa}} \nabla T_a;\nn\\
  \vecq_a^{(2)} &=& +\frac{1}{\nu_{aa}}\Big[ +\frac{515}{96}\frac{p_a}{\rho_a} \nabla\cdot\bPi^{(2,1)}_a - \frac{95}{96}\nabla\cdot \bPi^{(4,1)}_a
    - \frac{5}{48}\frac{\pr \vecq^{(1)}_a}{\pr t}  - \frac{5}{96} \frac{\rho_a}{p_a}\frac{\pr \vecX^{(5,1)}_a}{\pr t} \Big];\nn\\
  \vecX^{(5,1)}_a &=& - \frac{2975}{24} \frac{p_a^2}{\rho_a m_a \nu_{aa}} \nabla T_a;\nn\\
  \vecX^{(5,2)}_a &=& +  \frac{1}{\nu_{aa}}\Big[  +\frac{p^2_a}{\rho^2_a} \frac{13825}{72}\nabla\cdot\bPi^{(2,1)}_a
    - \frac{2485}{72}\frac{p_a}{\rho_a} \nabla\cdot\bPi^{(4,1)}_a +\frac{665}{36} \frac{p_a}{\rho_a}\frac{\pr \vecq^{(1)}_a}{\pr t}
    - \frac{175}{72}\frac{\pr\vecX^{(5,1)}_a}{\pr t}     \Big].
\end{eqnarray}
The model is fully specified and closed, and can be used in the given form. Nevertheless, it is possible to further apply the semi-linear approximation,
in which case  the viscosity corrections simplify into
\begin{eqnarray}
  \bPi^{(2,2)}_a &=& +\underbrace{\frac{45575}{17088}}_{2.6671} \frac{p_a}{m_a\nu_{aa}^2}\Big[2\nabla \nabla T_a -\frac{2}{3}\bI\nabla^2 T_a\Big] 
   + \underbrace{\frac{1164025}{1140624}}_{1.0205} \frac{p_a}{\nu_{aa}^2}\frac{\pr \bW_a}{\pr t};\nn\\
  \bPi^{(4,2)}_a &=& +\underbrace{\frac{536725}{17088}}_{31.4095} \frac{p_a^2}{\rho_a m_a \nu_{aa}^2}\Big[2\nabla \nabla T_a -\frac{2}{3}\bI\nabla^2 T_a\Big] 
  + \underbrace{\frac{10498075}{1140624}}_{9.2038} \frac{p_a^2}{\rho_a \nu_{aa}^2}\frac{\pr \bW_a}{\pr t}, \label{eq:energy300}
\end{eqnarray}
 and the heat flux corrections become 
\begin{eqnarray}
  \vecq_a^{(2)} &=&  + \underbrace{\frac{45575}{17088}}_{2.6671} \frac{p_a^2}{\rho_a \nu_{aa}^2}\nabla\cdot\bW_a
  +\underbrace{\frac{31625}{4608}}_{6.8631}\frac{p_a}{m_a \nu_{aa}^2}\frac{\pr \nabla T_a}{\pr t} ;\nn\\
  \vecX^{(5,2)}_a &=&  +\underbrace{\frac{1131725}{12816}}_{88.3056} \frac{p_a^3}{\rho_a^2 \nu_{aa}^2}\nabla\cdot\bW_a
  +\underbrace{\frac{791875}{3456}}_{229.1305} \frac{p_a^2}{\rho_a m_a \nu_{aa}^2}\frac{\pr \nabla T_a}{\pr t}. \label{eq:energy301}
\end{eqnarray}

\newpage
\subsection{Electron species}
For the electron species, it is useful to introduce denominator
\begin{equation}
D_1 = 192 Z_i^2+301 \sqrt{2}Z_i+178,
\end{equation}  
and solutions for the stress-tensors are
\begin{eqnarray}
  \bPi^{(2,1)}_e &=& -
  \frac{ 5 Z_i(205 \sqrt{2}+408 Z_i)}{6 D_1 } \frac{p_e}{\nu_{ei}} \bW_e;\nn\\
  \bPi^{(2,2)}_e &=& -\frac{1}{D_1 \nu_{ei}}\Big[ \frac{Z_i}{3}(79 \sqrt{2}-96 Z_i) \bY^{(3,1)}_e +3Z_i(\sqrt{2}+4 Z_i) \frac{\rho_e}{p_e} \bY^{(5,1)}_e\nn\\
    && \qquad + \frac{5}{6} Z_i (79 \sqrt{2}-96 Z_i) \frac{\pr \bPi^{(2,1)}_e}{\pr t}
    +15 Z_i (\sqrt{2} +4 Z_i) \frac{\rho_e}{p_e} \frac{\pr \bPi^{(4,1)}_e}{\pr t} \Big] ;\nn\\
  \bPi^{(4,1)}_e &=& - \frac{35Z_i (241 \sqrt{2}+552Z_i)}{6 D_1} \frac{p_e^2}{\rho_e \nu_{ei}}\bW_e;\nn\\
  \bPi^{(4,2)}_e &=& +\frac{1}{D_1 \nu_{ei}}\Big[   \frac{7}{3} Z_i(53 \sqrt{2}+288 Z_i) \frac{p_e}{\rho_e}\bY^{(3,1)}_e
    - 7 Z_i (7 \sqrt{2}+20 Z_i) \bY^{(5,1)}_e \nn\\
    &&\qquad  + \frac{35}{6} Z_i (53 \sqrt{2}+288 Z_i) \frac{p_e}{\rho_e}\frac{\pr \bPi^{(2,1)}_e}{\pr t} 
    -35 Z_i (7 \sqrt{2}+20 Z_i) \frac{\pr \bPi^{(4,1)}_e}{\pr t}  \Big],
\end{eqnarray}
 with matrices $\bY_e$ defined by (\ref{eq:Nomore230}). For the heat fluxes it is useful to define denominator
\begin{equation}
D_2 = 217 Z_i^2 + 604 Z_i\sqrt{2}+288,
\end{equation}
together with $\delta\bu=\bu_e-\bu_i$, and the results read
\begin{eqnarray}
  \vecq_e^{(1)} &=& - \frac{25 Z_i (180 \sqrt{2}+433 Z_i)}{4D_2} \frac{p_e}{m_e\nu_{ei}}\nabla T_e +\frac{30 Z_i (15\sqrt{2}+11 Z_i)}{D_2} p_e \delta\bu ;\nn\\
  \vecq_e^{(2)} &=& +\frac{1}{D_2 \nu_{ei}}\Big[ \frac{5}{4} Z_i (1236 \sqrt{2}+4097 Z_i) \frac{p_e}{\rho_e} \nabla\cdot\bPi^{(2,1)}_e
  - \frac{5}{4} Z_i (228 \sqrt{2}+709 Z_i) \nabla\cdot\bPi^{(4,1)}_e\nn\\
 && \qquad  - \frac{5}{2} Z_i(12 \sqrt{2}-533 Z_i)  \frac{\pr \vecq^{(1)}_e}{\pr t}
  - \frac{15}{4} Z_i (4\sqrt{2}+23 Z_i)  \frac{\rho_e}{p_e} \frac{\pr \vecX^{(5,1)}_e}{\pr t}\Big];\nn\\
  \vecX^{(5,1)}_e &=& - \frac{175 Z_i (204 \sqrt{2}+571 Z_i)}{D_2} \frac{p_e^2}{\rho_e m_e \nu_{ei}} \nabla T_e
  + \frac{840 Z_i (13 \sqrt{2}+12 Z_i)}{D_2} \frac{p^2_e}{\rho_e} \delta\bu;\nn\\
  \vecX^{(5,2)}_e &=& \frac{1}{D_2\nu_{ei}} \Big[ +175 Z_i (316 \sqrt{2}+1103 Z_i)  \frac{p_e^2}{\rho_e^2}\nabla\cdot\bPi^{(2,1)}_e
    - 35 Z_i (284 \sqrt{2}+951 Z_i)  \frac{p_e}{\rho_e} \nabla\cdot\bPi^{(4,1)}_e \nn\\
 &&\qquad    +70 Z_i (76 \sqrt{2}+759 Z_i) \frac{p_e}{\rho_e} \frac{\pr \vecq^{(1)}_e}{\pr t}
    -175 Z_i (4 \sqrt{2}+19 Z_i)\frac{\pr \vecX^{(5,1)}_e}{\pr t} \Big].
\end{eqnarray}
The system is now fully specified and can be used in this form. For the particular case of $Z_i=1$ numerical values become
\begin{eqnarray}
  \bPi^{(2,1)}_e &=& - 0.7309 \frac{p_e}{\nu_{ei}}\bW_e; \nn\\
  \bPi^{(2,2)}_e &=& -\frac{1}{\nu_{ei}}\Big[ 0.006587 \bY^{(3,1)}_e + 0.02041 \frac{\rho_e}{p_e} \bY^{(5,1)}_e + 0.01647\frac{\pr \bPi^{(2,1)}_e}{\pr t}
    +0.1021\frac{\rho_e}{p_e} \frac{\pr \bPi^{(4,1)}_e}{\pr t}\Big];\nn\\
  \bPi^{(4,1)}_e &=& - 6.5455 \frac{p_e^2}{\rho_e \nu_{ei}}\bW_e; \nn\\
  \bPi^{(4,2)}_e &=& +\frac{1}{\nu_{ei}}\Big[ 1.0644\frac{p_e}{\rho_e}\bY^{(3,1)}_e - 0.2630\bY^{(5,1)}_e +2.6609 \frac{p_e}{\rho_e}\frac{\pr \bPi^{(2,1)}_e}{\pr t}
    -1.3152 \frac{\pr \bPi^{(4,1)}_e}{\pr t}\Big];  
\end{eqnarray} 
\begin{eqnarray}
  \vecq_e^{(1)} &=& - 3.1616 \frac{p_e}{m_e\nu_{ei}}\nabla T_e +0.7110 p_e \delta\bu ;\nn\\
  \vecq_e^{(2)} &=& + \frac{1}{\nu_{ei}} \Big[ 5.3754\frac{p_e}{\rho_e} \nabla\cdot\bPi^{(2,1)}_e
  - 0.9486 \nabla\cdot\bPi^{(4,1)}_e  + 0.9492  \frac{\pr \vecq^{(1)}_e}{\pr t}
  - 0.07906 \frac{\rho_e}{p_e} \frac{\pr \vecX^{(5,1)}_e}{\pr t}\Big];\nn\\
  \vecX^{(5,1)}_e &=& - 110.664 \frac{p_e^2}{\rho_e m_e \nu_{ei}} \nabla T_e
  + 18.7783 \frac{p^2_e}{\rho_e} \delta\bu;\nn\\
  \vecX^{(5,2)}_e &=& \frac{1}{\nu_{ei}} \Big[ 199.554  \frac{p_e^2}{\rho_e^2}\nabla\cdot\bPi^{(2,1)}_e
    - 34.831  \frac{p_e}{\rho_e} \nabla\cdot\bPi^{(4,1)}_e 
  + 44.625 \frac{p_e}{\rho_e} \frac{\pr \vecq^{(1)}_e}{\pr t}
    -3.1747 \frac{\pr \vecX^{(5,1)}_e}{\pr t} \Big].
\end{eqnarray}
By further applying the quasi-linear approximation,  corrections to the electron viscosities become
\begin{eqnarray}
  \bPi^{(2,2)}_e &=&  
   + \frac{25 Z_i^2(119520 Z_i^2+101784 Z_i\sqrt{2}+46561)}{18 D_1^2} \frac{p_e}{\nu_{ei}^2}\frac{\pr \bW_e}{\pr t}\nn\\
  && - \frac{10 Z_i^2(11040 Z_i^2+15557 Z_i\sqrt{2}+ 8922)}{D_2 D_1} \frac{p_e}{\nu_{ei}} \Big[\big(\nabla\delta\bu \big)^S -\frac{2}{3}\bI\nabla\cdot\delta\bu \Big]\nn\\
  && + \frac{25 Z_i^2(534000 Z_i^2 +366451 Z_i \sqrt{2} + 131256)}{12 D_2 D_1} \frac{p_e}{\nu_{ei}^2 m_e} \Big[2\nabla \nabla T_e -\frac{2}{3}\bI\nabla^2 T_e\Big]; \\
 \bPi^{(4,2)}_e &=&  
   + \frac{175 Z_i^2(173088 Z_i^2 +142032 Z_i\sqrt{2}+59989)}{18 D_1^2}\frac{p_e^2}{\nu_{ei}^2 \rho_e}\frac{\pr \bW_e}{\pr t}\nn\\
  && - \frac{70 Z_i^2 (16992 Z_i^2 +23993 Z_i\sqrt{2}+13698)}{D_2 D_1} \frac{p_e^2}{\nu_{ei} \rho_e} \Big[\big(\nabla\delta\bu \big)^S -\frac{2}{3}\bI\nabla\cdot\delta\bu \Big] \nn\\
  && + \frac{175 Z_i^2(834576 Z_i^2 +603679 Z_i\sqrt{2} +220824)}{12 D_2 D_1} \frac{p_e^2}{\nu_{ei}^2 m_e \rho_e} \Big[2\nabla \nabla T_e -\frac{2}{3}\bI\nabla^2 T_e\Big];  \label{eq:energy306}
\end{eqnarray}
 together with corrections for the heat fluxes
\begin{eqnarray}
  \vecq_e^{(2)} &=& \frac{25 Z_i^2 (534000 Z_i^2+366451 Z_i\sqrt{2} +131256)}{12 D_1 D_2}
  \frac{p_e^2}{\rho_e \nu_{ei}^2} \nabla\cdot\bW_e \nn\\
  && - \frac{75 Z_i^2(5729 Z_i^2 +6711 Z_i \sqrt{2}+ 4728)}{D_2^2}\frac{p_e}{\nu_{ei}}\frac{\pr\delta\bu}{\pr t}\nn\\
 && + \frac{125 Z_i^2 (320797 Z_i^2 +202248 Z_i\sqrt{2} + 72864)}{8 D_2^2} \frac{p_e}{\nu_{ei}^2 m_e} \frac{\pr \nabla T_e}{\pr t}    ;\nn\\
  \vecX^{(5,2)}_e &=& \frac{175 Z_i^2 (712272 Z_i^2 +463249 Z_i\sqrt{2} +155208)}{3 D_1 D_2}  \frac{p_e^3}{\rho_e^2 \nu_{ei}^2}\nabla\cdot\bW_e\nn\\
  && - \frac{2100 Z_i^2 (7611 Z_i^2 +8429 Z_i\sqrt{2}+5000)}{D_2^2}  \frac{p_e^2}{\nu_{ei}\rho_e} \frac{\pr\delta\bu}{\pr t} \nn\\
  && + \frac{875 Z_i^2(430783 Z_i^2+261672 Z_i\sqrt{2}+86880)}{2D_2^2} \frac{p_e^2}{\nu_{ei}^2 \rho_e m_e}\frac{\pr \nabla T_e}{\pr t}.
\end{eqnarray}
 For the ion charge $Z_i=1$, the numerical values read
\begin{eqnarray}
  \bPi^{(2,2)}_e &=& + 0.6801 \frac{p_e}{\nu_{ei}^2}\frac{\pr \bW_e}{\pr t}
  - 0.3880 \frac{p_e}{\nu_{ei}} \Big[\big(\nabla\delta\bu \big)^S -\frac{2}{3}\bI\nabla\cdot\delta\bu \Big] \nn\\
  && + 2.2799 \frac{p_e}{\nu_{ei}^2 m_e} \Big[2\nabla \nabla T_e -\frac{2}{3}\bI\nabla^2 T_e\Big]; \label{eq:energy305}\\
  \bPi^{(4,2)}_e &=&  
   + 6.6638 \frac{p_e^2}{\nu_{ei}^2 \rho_e}\frac{\pr \bW_e}{\pr t}
   - 4.1827 \frac{p_e^2}{\nu_{ei} \rho_e} \Big[\big(\nabla\delta\bu \big)^S -\frac{2}{3}\bI\nabla\cdot\delta\bu \Big] \nn\\
  && + 25.7440 \frac{p_e^2}{\nu_{ei}^2 m_e \rho_e} \Big[2\nabla \nabla T_e -\frac{2}{3}\bI\nabla^2 T_e\Big];  \label{eq:energy306xx}
\end{eqnarray}
 together with
\begin{eqnarray}
  \vecq_e^{(2)} &=& 2.2799
  \frac{p_e^2}{\rho_e \nu_{ei}^2} \nabla\cdot\bW_e -0.8098 \frac{p_e}{\nu_{ei}}\frac{\pr\delta\bu}{\pr t}
  +5.7487 \frac{p_e}{\nu_{ei}^2 m_e} \frac{\pr \nabla T_e}{\pr t};\nn\\
  \vecX^{(5,2)}_e &=& 82.1278 \frac{p_e^3}{\rho_e^2 \nu_{ei}^2}\nabla\cdot\bW_e -27.8859 \frac{p_e^2}{\nu_{ei}\rho_e} \frac{\pr\delta\bu}{\pr t}
  +210.2318 \frac{p_e^2}{\nu_{ei}^2 \rho_e m_e}\frac{\pr \nabla T_e}{\pr t}. \label{eq:energy307}
\end{eqnarray}
The rate-of-strain tensor $\bW_e$ obviously enters the electron heat fluxes, even in a quasi-static approximation. 

\subsection{Momentum exchange rates}
Collisional momentum exchange rates $\boldsymbol{R}_e=-\boldsymbol{R}_i$ given by (\ref{eq:ReExcite})
can also be split into the first and second order
$\boldsymbol{R}_e = \boldsymbol{R}_e^{(1)}+ \boldsymbol{R}_e^{(2)}$, according to 
\begin{eqnarray}
  \boldsymbol{R}_e^{(1)} &=& - \rho_e\nu_{ei} \delta\bu  + \frac{21}{10}\frac{\rho_e}{p_e} \nu_{ei}\vecq_e^{(1)}
  -\frac{3}{56} \frac{\rho_e^2}{p_e^2}\nu_{ei} \vecX^{(5,1)}_e ;\nn\\
  \boldsymbol{R}_e^{(2)} &=& + \frac{21}{10}\frac{\rho_e}{p_e} \nu_{ei}\vecq_e^{(2)}
  -\frac{3}{56} \frac{\rho_e^2}{p_e^2}\nu_{ei} \vecX^{(5,2)}_e. \label{eq:Thierry111}
\end{eqnarray}
Then by using results given in the previous section 
\begin{eqnarray}
  \boldsymbol{R}_e^{(1)} &=& 
  - \nu_{ei} \rho_e \frac{(D_2-153Z_i^2-360 Z_i\sqrt{2})}{D_2} \delta\bu
  -\frac{30 Z_i(15\sqrt{2}+11 Z_i)}{D_2} n_e \nabla T_e;\nn\\
 \boldsymbol{R}_e^{(2)} &=& + \frac{6 Z_i (47 \sqrt{2}+69 Z_i)}{D_2}  \nabla\cdot \bPi^{(2,1)}_e
  - \frac{6 Z_i (11 \sqrt{2}+13 Z_i)}{D_2} \frac{\rho_e}{p_e}  \nabla\cdot \bPi^{(4,1)}_e\nn\\
&&  - \frac{12 Z_i (29 \sqrt{2}+4 Z_i)}{D_2} \frac{\rho_e}{p_e} \frac{\pr \vecq^{(1)}_e}{\pr t}
  + \frac{3 Z_i (2 \sqrt{2}-Z_i)}{D_2} \frac{\rho_e^2}{p_e^2} \frac{\pr \vecX^{(5,1)}_e}{\pr t},
\end{eqnarray}
or for a particular case of $Z_i=1$
\begin{eqnarray}
\boldsymbol{R}_e^{(2)} &=& + 0.5980  \nabla\cdot \bPi^{(2,1)}_e
  - 0.1261 \frac{\rho_e}{p_e}  \nabla\cdot \bPi^{(4,1)}_e\nn\\
&&  - 0.3974 \frac{\rho_e}{p_e} \frac{\pr \vecq^{(1)}_e}{\pr t}
  + 0.004036 \frac{\rho_e^2}{p_e^2} \frac{\pr \vecX^{(5,1)}_e}{\pr t}.
\end{eqnarray}
Finally, at a semi-linear level
\begin{eqnarray}
  \boldsymbol{R}_e^{(2)} &=& \frac{10 Z_i^2 (11040 Z_i^2+15557 Z_i\sqrt{2}+8922)}{D_1 D_2} \frac{p_e}{\nu_{ei}} \nabla\cdot\bW_e\nn\\
  && - \frac{720 Z_i^2 (64 Z_i^2+151 Z_i\sqrt{2}+253)}{D_2^2} \rho_e \frac{\pr (\delta\bu)}{\pr t}  \nn\\
  && + \frac{75 Z_i^2 (5729 Z_i^2 +6711 Z_i\sqrt{2}+4728)}{D_2^2} \frac{n_e}{ \nu_{ei}} \frac{\pr \nabla T_e}{\pr t},
\end{eqnarray}
and for $Z_i=1$ the full momentum exchange rates become
\begin{eqnarray}
  \boldsymbol{R}_e &=& 
  - 0.5129\nu_{ei} \rho_e \delta\bu   -0.7110 n_e \nabla T_e + 0.3880 \frac{p_e}{\nu_{ei}} \nabla\cdot\bW_e
  -0.2068 \rho_e \frac{\pr (\delta\bu)}{\pr t} + 0.8098 \frac{n_e}{\nu_{ei}} \frac{\pr \nabla T_e}{\pr t}, \label{eq:energy999}
\end{eqnarray}
where $\delta\bu=\bu_e-\bu_i$.
Only the first two terms of (\ref{eq:energy999}) were considered by \cite{Braginskii1965} and \cite{SpitzerHarm1953} (the latter having slightly 
different proportionality constants; see Appendix \ref{sec:Comparison}). A further generalization by keeping the full
  $\bPi_a^{(2)}$ \& $\bPi_a^{(4)}$ viscosity tensors in the last terms of (\ref{eq:Nomore204}) \& (\ref{eq:Nomore205}) brings another 3 terms to $\boldsymbol{R}_e$
  (not shown). Naturally, in a highly-collisional regime ($\nu_{ei}\gg \omega$) all additional terms are small in comparison to
the first two terms of (\ref{eq:energy999}).
Nevertheless, at higher frequencies
(shorter length-scales) these additional contributions might become significant. Interestingly, 
the rate-of-strain tensor $\bW_e$ enters the momentum exchange rates
 (even at the linear level), with contribution $\nabla\cdot\bW_e=\nabla^2\bu_e+(1/3)\nabla(\nabla\cdot\bu_e)$. 
 Note that some terms are  proportional to $1/\nu_{ei}$ and become unbounded (divergent) in a regime of low-collisionality,
which is a consequence of the expansion procedure (i.e. a quasi-static approximation).
 Evolution equations (\ref{eq:Nomore110})-(\ref{eq:Nomore102}) are of course well-defined in the regime of low collisionality.

\newpage
\section{Multi-fluid 22-moment model} \label{sec:22main}
Here we consider a natural generalization of the 21-moment model, by accounting for 
a fully contracted perturbation of the 4th-order fluid moment $X_{ijkl}^{a(4)}=m_a \int c_i^a c_j^a c_k^a c_l^a f_a d^3v$. The fully contracted
(scalar) moment is decomposed into its Maxwellian core and a perturbation $\widetilde{X}^{(4)}_a$ (denoted with tilde), according to
\begin{equation} \label{eq:Thierry215}
  X^{(4)}_a =  m_a \int |\bc_a|^{4} f_a d^3v = 15 \frac{p^2_a}{\rho_a}+\widetilde{X}^{(4)}_a,
\end{equation}
meaning a definition $\widetilde{X}^{(4)}_a =  m_a \int |\bc_a|^{4} (f_a -f_a^{(0)}) d^3v$, where $f_a^{(0)}$ is Maxwellian. The scalar
perturbation $\widetilde{X}^{(4)}_a$ enters the decomposition of the 4th-order moment
\begin{eqnarray}
  X_{ijkl}^{a(4)} &=& \frac{1}{15}\Big(15\frac{p_a^2}{\rho_a} +\widetilde{X}^{(4)}_a \Big)
  \big(\delta_{ij}\delta_{kl}+\delta_{ik}\delta_{jl}+\delta_{il}\delta_{jk}\big)\nn\\
  && +\frac{1}{7} \Big[ \Pi_{ij}^{a(4)} \delta_{kl}+\Pi_{ik}^{a(4)}\delta_{jl}+\Pi_{il}^{a(4)}\delta_{jk}
  +\Pi_{jk}^{a(4)}\delta_{il}+\Pi_{jl}^{a(4)}\delta_{ik}+\Pi_{kl}^{a(4)}\delta_{ij} \Big] +\sigma_{ijkl}^{a(4)}\,', \label{eq:Thierry80}
\end{eqnarray}
where we neglect the traceless tensor $\sigma_{ijkl}^{a(4)}\,'$, 
and the entire model now represents the 22-moment model.
 The fully non-linear model is given by evolution equations (\ref{eq:Energy20})-(\ref{eq:Num1000}) which are unchanged,
together with
\begin{eqnarray} 
 && \frac{d_a\vecq_a}{d t} +\frac{7}{5}\vecq_a\nabla\cdot\bu_a  + \frac{7}{5}\vecq_a\cdot\nabla\bu_a +\frac{2}{5}(\nabla\bu_a)\cdot\vecq_a
  +\Omega_a\bhat\times\vecq_a + \frac{5}{2}p_a\nabla\Big(\frac{p_a}{\rho_a}\Big) \nn\\
 && +\frac{1}{6}\nabla \widetilde{X}^{(4)}_a +\frac{1}{2}\nabla\cdot \bPi^{(4)}_a 
  -\frac{5}{2}\frac{p_a}{\rho_a}\nabla\cdot\bPi^{(2)}_a
  -\frac{1}{\rho_a}(\nabla\cdot\bp_a)\cdot\bPi^{(2)}_a \nn\\
  && \qquad = \vec{\boldsymbol{Q}}^{(3)}_{a}\,' \equiv \frac{1}{2}\textrm{Tr}\bQ^{(3)}_a-\frac{5}{2}\frac{p_a}{\rho_a}\boldsymbol{R}_a
  -\frac{1}{\rho_a} \boldsymbol{R}_a\cdot\bPi^{(2)}_a; \label{eq:Thierry90}
\end{eqnarray}
\begin{eqnarray}
  && \frac{d_a}{dt} \bPi^{(4)}_a +\frac{1}{5}\Big[ (\nabla\vecX^{(5)}_a)^S-\frac{2}{3}\bI(\nabla\cdot\vecX^{(5)}_a)\Big]
  +\frac{9}{7}(\nabla\cdot\bu_a)\bPi^{(4)}_a +\frac{9}{7}(\bPi^{(4)}_a\cdot\nabla\bu_a)^S\nn\\
&&  + \frac{2}{7}\big((\nabla\bu_a)\cdot\bPi^{(4)}_a\big)^S
  -\frac{22}{21}\bI (\bPi^{(4)}_a:\nabla\bu_a)
 -\, \frac{14}{5\rho_a} \Big[ \big((\nabla\cdot\bp_a)\vecq_a\big)^S -\frac{2}{3}\bI (\nabla\cdot\bp_a)\cdot\vecq_a\Big] \nn\\
 && +\Omega_a \big( \bhat\times \bPi^{(4)}_a \big)^S
 + \frac{7}{15}\big(15\frac{p_a^2}{\rho_a}+\widetilde{X}^{(4)}_a \big)\bW_a \nn\\
&& = \bQ^{(4)}_a\,' \equiv \trace \bQ^{(4)}_a -\frac{\bI}{3}\trace\trace \bQ^{(4)}_a
 -\frac{14}{5\rho_a}\Big[ (\boldsymbol{R}_a\vecq_a)^S-\frac{2}{3}\bI (\boldsymbol{R}_a\cdot\vecq_a)\Big]; \label{eq:Thierry91}
\end{eqnarray}
\begin{eqnarray}
  && \frac{d_a}{dt} \widetilde{X}^{(4)}_a +\nabla\cdot\vecX^{(5)}_a  -20 \frac{p_a}{\rho_a}\nabla\cdot\vecq_a
  +\frac{7}{3}\widetilde{X}^{(4)}_a(\nabla\cdot\bu_a) +4\big(\bPi^{(4)}_a -5\frac{p_a}{\rho_a}\bPi^{(2)}_a\big):\nabla\bu_a \nn\\
 && -\frac{8}{\rho_a} (\nabla\cdot\bp_a)\cdot \vecq_a 
    = \widetilde{Q}^{(4)}_a\,' \equiv \trace\trace \bQ^{(4)}_a -20 \frac{p_a}{\rho_a}Q_a-\frac{8}{\rho_a} \boldsymbol{R}_a \cdot\vecq_a; \label{eq:Thierry92}
\end{eqnarray}
\begin{eqnarray}
&&  \frac{d_a}{d t}\vecX^{(5)}_a +\frac{1}{3}\nabla\widetilde{X}^{(6)}_a +\nabla\cdot \bPi^{(6)}_a\nn\\
  &&  +\frac{9}{5}\vecX^{(5)}_a (\nabla\cdot\bu_a) + \frac{9}{5}\vecX^{(5)}_a\cdot\nabla\bu_a
  + \frac{4}{5}(\nabla\bu_a)\cdot\vecX^{(5)}_a +\Omega_a \bhat\times \vecX^{(5)}_a\nn\\
&&  +70 \frac{p_a^2}{\rho_a}\nabla\Big(\frac{p_a}{\rho_a}\Big) -35 \frac{p_a^2}{\rho_a^2} \nabla\cdot\bPi^{(2)}_a -\frac{7}{3\rho_a}\big(\nabla\cdot\bp^a\big)\widetilde{X}^{(4)}_a
  -\frac{4}{\rho_a} \big(\nabla\cdot\bp^a\big)\cdot \bPi^{(4)}_a \nn\\
  && =\vecQ^{(5)}_a\,' \equiv \vecQ^{(5)}_a -35 \frac{p_a^2}{\rho_a^2}\boldsymbol{R}_a -\frac{7}{3\rho_a} \boldsymbol{R}_a \widetilde{X}^{(4)}_a
  - \frac{4}{\rho_a} \boldsymbol{R}_a\cdot\bPi^{(4)}_a. \label{eq:Thierry93}
\end{eqnarray}
 The last equation (\ref{eq:Thierry93}) is closed with closure (\ref{ref:Num1010}) for the stress-tensor $\bPi^{(6)}_a$,
together with a closure  for the scalar perturbation (derived from a Hermite closure)
\begin{equation} \label{eq:Thierry70}
\widetilde{X}^{(6)}_a =  m_a \int |\bc_a|^{6} (f_a -f_a^{(0)}) d^3v = 21 \frac{p_a}{\rho_a}\widetilde{X}^{(4)}_a.  
\end{equation}

 In the semi-linear approximations the 22-moment model reads 
\begin{eqnarray}
  && \frac{d_a}{dt} \bPi^{(2)}_a  +\Omega_a \big(\bhat\times \bPi^{(2)}_a \big)^S + p_a \bW_a
  + \frac{2}{5}\Big((\nabla \vecq_a)^S -\frac{2}{3}\bI \nabla\cdot\vecq_a\Big)\nn\\
  && \qquad = \bQ_{a}^{(2)}\,' \label{eq:Nomore110XXa};\\
   &&  \frac{d_a}{d t}\vecq_a + \Omega_a \bhat\times\vecq_a + \frac{5}{2}p_a \nabla \Big(\frac{p_a}{\rho_a}\Big)
  +\frac{1}{2}\nabla\cdot\bPi^{(4)}_a-\frac{5}{2}\frac{p_a}{\rho_a}\nabla\cdot\bPi^{(2)}_a\nn\\
  && \qquad +\frac{1}{6}\nabla \widetilde{X}^{(4)}_a
  = \vec{\boldsymbol{Q}}^{(3)}_{a}\,';\label{eq:Nomore110XX}\\ 
     && \frac{d_a}{dt} \bPi^{(4)}_a +\Omega_a \big( \bhat\times \bPi^{(4)}_a \big)^S  + 7 \frac{p_a^2}{\rho_a} \bW_a
     +\frac{1}{5}\Big[ (\nabla\vecX^{(5)}_a)^S-\frac{2}{3}\bI(\nabla\cdot\vecX^{(5)}_a)\Big] \nn\\
     && \qquad   = \bQ^{(4)}_a\,' ;\label{eq:Nomore103X}\\
  && \frac{d_a}{dt} \widetilde{X}^{(4)}_a +\nabla\cdot\vecX^{(5)}_a  -20 \frac{p_a}{\rho_a}\nabla\cdot\vecq_a
     = \widetilde{Q}^{(4)}_a\,' = \trace\trace \bQ^{(4)}_a -20 \frac{p_a}{\rho_a}Q_a; \label{eq:PPosled12X} \\   
  &&  \frac{d_a}{d t}\vecX^{(5)}_a +\Omega_a \bhat\times \vecX^{(5)}_a  +70 \frac{p_a^2}{\rho_a}\nabla\Big(\frac{p_a}{\rho_a}\Big)
  +18\frac{p_a}{\rho_a}\nabla\cdot \bPi^{(4)}_a
  -98 \frac{p_a^2}{\rho_a^2}\nabla\cdot\bPi^{(2)}_a  \nn\\ 
  && \qquad + 7\frac{p_a}{\rho_a}\nabla \widetilde{X}^{(4)}_a
  = \vec{\boldsymbol{Q}}^{(5)}_{a}\,'. \label{eq:Nomore102X}
\end{eqnarray}
 As discussed in Section \ref{sec:SemiLinear}, in the semi-linear approximation we are neglecting terms such as $(\nabla p_a)\vecq_a$
  which might become significant in the presence of large-scale gradients, together with other terms that are neglected.
In comparison to the 21-moment model given by (\ref{eq:Nomore110})-(\ref{eq:Nomore102}), evolution equations
(\ref{eq:Nomore110XXa}) \& (\ref{eq:Nomore103X}) for stress-tensors $\bPi^{(2)}_a$ \& $\bPi^{(4)}_a$ remain unchanged.
Importantly, collisional contributions $\boldsymbol{R}_a$;
$\bQ_{a}^{(2)}\,'$; $\vec{\boldsymbol{Q}}^{(3)}_{a}\,'$; $\bQ_{a}^{(4)}\,'$; $\vec{\boldsymbol{Q}}^{(5)}_{a}\,'$  
given in Section \ref{sec:Tarb2} remain unchanged as well. The only differences are:
1) scalar perturbations $\widetilde{X}^{(4)}_a$ now enter the left hand sides of evolution equations (\ref{eq:Nomore110XX}) \& (\ref{eq:Nomore102X}) for
heat fluxes $\vecq_a$ \& $\vecX^{(5)}_a$; 2) a new evolution equation  (\ref{eq:PPosled12X}) for scalar $\widetilde{X}^{(4)}_a$ is present, with
collisional contributions $\widetilde{Q}^{(4)}_a\,'$ that need to be specified; 
3) the energy exchange rates $Q_a$ entering the scalar pressure equation (\ref{eq:Energy20xx}) are modified, and given below.

\newpage
\subsection{Collisional contributions (arbitrary temperatures)} \label{sec:SummaryQ4} 
The energy exchange rates entering equation (\ref{eq:Energy20xx}) are now given by
\begin{eqnarray}
 Q_{a} = \sum_{b\neq a} Q_{ab} =\sum_{b\neq a} \frac{\rho_a \nu_{ab}}{(m_a+m_b)} \Big\{ 3(T_b-T_a)+ \hat{P}_{ab (1)}\frac{\rho_a}{n_a p_a} \widetilde{X}^{(4)}_a
    - \hat{P}_{ab (2)} \frac{\rho_b}{n_b p_b} \widetilde{X}^{(4)}_b     \Big\}, \label{eq:Thierry38}
\end{eqnarray}
with  mass-ratio coefficients
\begin{eqnarray}
 \hat{P}_{ab (1)} = \frac{3 T_a m_b (5 T_b m_b +4 T_b m_a -T_a m_b  )}{40 (T_a m_b +T_b m_a)^2};\qquad
\hat{P}_{ab (2)} = \frac{3T_b m_a (5 T_a m_a +4 T_a m_b -T_b m_a  )}{40 (T_a m_b +T_b m_a)^2}. \label{eq:Thierry388}
\end{eqnarray}
Interestingly, scalar perturbations $\widetilde{X}^{(4)}_a$ thus enter the energy exchange rates. For self-collisions all the contributions
naturally disappear.  As discussed also later in Section \ref{sec:energy}, for multi-fluid models the conservation of energy
  $Q_{ab}+Q_{ba}=(\bu_b-\bu_a)\cdot\boldsymbol{R}_{ab}$ is satisfied only approximately, because in the semi-linear approximation the differences in
  drifts $\bu_b-\bu_a$ are assumed to be small, meaning $Q_{ab}+Q_{ba}=0$ holds. 
  To satisfy the energy conservation exactly, the collisional integrals would have to be calculated non-linearly with unrestriced drifts (i.e. with the runaway effect).
  Nevertheless, for a plasma consisting of only two species
  (such as a one ion-electron plasma) the conservation of energy can be imposed by hand, by calculating $Q_{ab}$
  according to (\ref{eq:Thierry38}), (\ref{eq:Thierry388}) and prescribing $Q_{ba}=-Q_{ab}+(\bu_b-\bu_a)\cdot\boldsymbol{R}_{ab}$.  

Collisional exchange rates entering evolution equation (\ref{eq:PPosled12X}) are given by
\begin{eqnarray}
  \widetilde{Q}^{(4)}_{a}\,' &=& -\frac{4}{5} \nu_{aa} \widetilde{X}^{(4)}_a 
  +\sum_{b\neq a} \nu_{ab} \Big\{ -\frac{p_a^2}{\rho_a} \frac{(T_b-T_a)}{T_a} \hat{S}_{ab (0)}
-\widetilde{X}^{(4)}_a \hat{S}_{ab (1)} 
 +\frac{p_a^2 \rho_b}{p_b^2 \rho_a}  \widetilde{X}^{(4)}_b \hat{S}_{ab (2)}\Big\}, \label{eq:Thierry39}
\end{eqnarray}
with  mass-ratio coefficients
\begin{eqnarray}
  \hat{S}_{ab (0)} &=& \frac{36  T_a m_a m_b}{(T_a m_b +T_b m_a)(m_b+m_a)};\nn\\
  \hat{S}_{ab (1)} &=& -\, \Big\{m_a \big(17 T_a^3 m_b^3 -36 T_a^2 T_b m_a m_b^2 -69 T_a^2 T_b m_b^3 +12 T_a T_b^2 m_a^2 m_b
  -48 T_a T_b^2 m_a m_b^2 \nn\\
  && \quad -40 T_b^3 m_a^3 -84 T_b^3 m_a^2 m_b\big)\Big\}\Big[10 (T_a m_b +T_b m_a)^3 (m_b+m_a)\Big]^{-1};\nn\\
  \hat{S}_{ab (2)} &=& \frac{3 T_b^2 m_a^2 m_b (7 T_a m_a +4 T_a m_b -3 T_b m_a)}{2 (T_a m_b +T_b m_a)^3 (m_b+m_a)}, \label{eq:Thierry37}
\end{eqnarray}
where the self-collisional contributions are represented by the first term of (\ref{eq:Thierry39}).
\subsubsection{Small temperature differences}
For small temperature differences the  mass-ratio coefficients become 
\begin{eqnarray}
  \hat{P}_{ab (1)} &=& \frac{3m_b}{10 (m_b+m_a)}; \qquad \hat{P}_{ab (2)} = \frac{3m_a}{10(m_b+m_a)};\nn\\
  \hat{S}_{ab (0)} &=& \frac{36 m_a m_b}{(m_b+m_a)^2};\qquad
  \hat{S}_{ab (1)} = \frac{2 m_a (10 m_a^2 +8 m_a m_b +13 m_b^2)}{5 (m_b +m_a)^3};\qquad 
  \hat{S}_{ab (2)} = \frac{6 m_a^2 m_b}{(m_b+m_a)^3}, \label{eq:Thierry55}
\end{eqnarray}
and for example for self-collisions $\hat{S}_{aa (1)} = 31/20$ and $\hat{S}_{aa (2)} = 3/4$.
We further consider a one ion-electron plasma.

\newpage
\subsection{Ion species (self-collisions)}
In a quasi-static approximation the solution of equation (\ref{eq:PPosled12X}) becomes  
\begin{equation}
\widetilde{X}^{(4)}_a = -\,\frac{5}{4\nu_{aa}} \Big[\nabla\cdot\vecX^{(5)}_a  -20 \frac{p_a}{\rho_a}\nabla\cdot\vecq_a\Big].
\end{equation}
The quasi-static solution is thus completely determined by the heat fluxes $\vecq_a$ \& $\vecX^{(5)}_a$ and for a magnetized plasma it has
the following form
\begin{eqnarray}
  \widetilde{X}^{(4)}_a &=& -\,\frac{5}{4\nu_{aa}}
  \Big\{\nabla\cdot\Big[\frac{p_a}{\rho_a} \Big(-\kappa_\parallel^{a(5)} \nabla_\parallel T_a - \kappa_\perp^{a(5)} \nabla_\perp T_a
      + \kappa_\times^{a(5)} \bhat\times\nabla T_a\Big)\Big] \nn\\
  && \quad   -20 \frac{p_a}{\rho_a}\nabla\cdot\Big( -\kappa_\parallel^a \nabla_\parallel T_a
  - \kappa_\perp^a \nabla_\perp T_a + \kappa_\times^a \bhat\times\nabla T_a\Big)\Big\}, \label{eq:Thierry52}
\end{eqnarray}
where the thermal conductivities are given by (\ref{eq:Thierry51}), (\ref{eq:Thierry50}).

It feels natural to define thermal conductivities (of the moment $\widetilde{X}^{(4)}_a$)
\begin{equation}
  \kappa_\parallel^{a(4)} = \frac{5}{4}\big(\kappa_\parallel^{a(5)}-20\kappa_\parallel^a\big); \qquad
  \kappa_\perp^{a(4)} = \frac{5}{4}\big(\kappa_\perp^{a(5)}-20\kappa_\perp^a\big); \qquad 
  \kappa_\times^{a(4)} = \frac{5}{4}\big(\kappa_\times^{a(5)}-20\kappa_\times^a\big),
\end{equation}
and result (\ref{eq:Thierry52}) then transforms into
\begin{eqnarray}
  \widetilde{X}^{(4)}_a &=& -\,\frac{p_a}{\nu_{aa}\rho_a}
  \nabla\cdot\Big[  - \kappa_\parallel^{a(4)} \nabla_\parallel T_a - \kappa_\perp^{a(4)} \nabla_\perp T_a
    + \kappa_\times^{a(4)} \bhat\times\nabla T_a\Big]\nn\\
  && -\frac{5}{4\nu_{aa}}\Big(-\kappa_\parallel^{a(5)} \nabla_\parallel T_a - \kappa_\perp^{a(5)} \nabla_\perp T_a
      + \kappa_\times^{a(5)} \bhat\times\nabla T_a\Big)\cdot\nabla\Big(\frac{p_a}{\rho_a}\Big), \label{eq:Thierry53}
\end{eqnarray}
with thermal conductivities 
\begin{eqnarray}
  {\kappa}_\parallel^{a(4)} &=& \frac{1375}{24} \frac{p_a}{\nu_{aa}m_a}; \nn\\
  {\kappa}_\perp^{a(4)} &=& \frac{p_a}{\nu_{aa}m_a}\frac{5 x^2+ (9504/245)}{x^4+(3313/1225) x^2+ (20736/30625)}; \nn\\ 
  {\kappa}_\times^{a(4)} &=& \frac{p_a}{\nu_{aa}m_a}\frac{25x^3+(3810/49)x}{x^4+(3313/1225) x^2+ (20736/30625)}. \label{eq:Thierry681}
\end{eqnarray}
The second term of (\ref{eq:Thierry53}) is strictly non-linear and may be neglected for simplicity. The solution
for $\widetilde{X}^{(4)}_a$ thus can be written as a divergence of a heat flux vector defined by the expression in the
square brackets of (\ref{eq:Thierry53}). We have used Braginskii notation with vectors
$\nabla_\parallel=\bhat\bhat\cdot\nabla$ and $\nabla_\perp = \bI_\perp\cdot\nabla = -\bhat\times\bhat\times\nabla$. 

The result (\ref{eq:Thierry53}) can be further simplified in the semi-linear approximation, where one may use
$\nabla\cdot(\bhat\times\nabla T_a)=0$ and so
\begin{equation}
  \widetilde{X}^{(4)}_a = +\frac{p_a}{\nu_{aa}\rho_a} \Big[  \kappa_\parallel^{a(4)} \nabla_\parallel^2 T_a + \kappa_\perp^{a(4)} \nabla_\perp^2 T_a \Big],
\end{equation}
with scalars $\nabla_\parallel^2=\bhat\bhat:\nabla\nabla$ and $\nabla_\perp^2=\nabla^2-\nabla_\parallel^2$, and for
zero magnetic field 
\begin{equation}
\widetilde{X}^{(4)}_a = + \underbrace{\frac{1375}{24}}_{57.292} \frac{p_a^2}{\nu_{aa}^2\rho_a m_a} \nabla^2 T_a.
\end{equation}
Note that the result is proportional to $1/\nu_{aa}^2$ and thus small in a highly-collisional regime.

\newpage
\subsection{Electron species (one ion-electron plasma)}
 Here we consider a one ion-electron plasma with small temperature differences. Similar to Braginskii, an exact energy conservation can be imposed by
 hand, according to
\begin{eqnarray}
Q_{ie} = \frac{\rho_i\nu_{ie}}{m_i}\Big[ 3(T_e-T_i) +\frac{3}{10}m_e \Big( \frac{\widetilde{X}^{(4)}_i}{p_i}-\frac{\widetilde{X}^{(4)}_e}{p_e}\Big)\Big]; 
\qquad Q_{ei}=-Q_{ie}-(\bu_e-\bu_i)\cdot\boldsymbol{R}_{ei}. \label{eq:Thierry278e}
\end{eqnarray}
 The electron coefficients (\ref{eq:Thierry55}) become
$\hat{S}_{ei(1)}=(26/5)(m_e/m_i)$ and $\hat{S}_{ei(2)}= 6 (m_e/m_i)^2$, and collisional contributions 
(\ref{eq:Thierry39}) have a simple form
\begin{equation} \label{eq:Thierry278}
  \widetilde{Q}^{(4)}_{e}\,' = -\frac{4}{5} \nu_{ee} \widetilde{X}^{(4)}_e,
\end{equation}
determined solely by the electron-electron collisions. A quasi-static solution of equation (\ref{eq:PPosled12X}) then becomes  
\begin{equation}
\widetilde{X}^{(4)}_e = -\,\frac{5\sqrt{2}Z_i}{4\nu_{ei}} \Big[\nabla\cdot\vecX^{(5)}_e  -20 \frac{p_e}{\rho_e}\nabla\cdot\vecq_e\Big],
\end{equation}
where we have used $\nu_{ee}=\nu_{ei}/(Z_i\sqrt{2})$. The electron heat fluxes are given by (\ref{eq:Thierry63}) and (\ref{eq:Thierry60}) and are of course determined by
both electron-electron and electron-ion collisions. The full solution thus consists of thermal and frictional parts
$\widetilde{X}^{(4)}_e = \widetilde{X}^{(4)T}_e +\widetilde{X}^{(4)u}_e$, where
\begin{eqnarray}
  \widetilde{X}^{(4)T}_e &=& -\,\frac{5\sqrt{2} Z_i}{4\nu_{ei}}
  \Big\{\nabla\cdot\Big[\frac{p_e}{\rho_e} \Big(-\kappa_\parallel^{e(5)} \nabla_\parallel T_e - \kappa_\perp^{e(5)} \nabla_\perp T_e
      + \kappa_\times^{e(5)} \bhat\times\nabla T_e\Big)\Big] \nn\\
  && \quad   -20 \frac{p_e}{\rho_e}\nabla\cdot\Big( -\kappa_\parallel^e \nabla_\parallel T_e
  - \kappa_\perp^e \nabla_\perp T_e + \kappa_\times^e \bhat\times\nabla T_e\Big)\Big\}; \label{eq:Thierry64}\\
  \widetilde{X}^{(4)u}_e &=& -\,\frac{5Z_i\sqrt{2}}{4\nu_{ei}}
  \Big\{\nabla\cdot\Big[\frac{p_e^2}{\rho_e} \Big( \beta_0^{(5)}  \delta\bu_\parallel + \frac{\beta_1^{(5)'} x^2+\beta_0^{(5)'}}{\triangle} \delta\bu_\perp
-\frac{\beta_1^{(5)''}x^3+\beta_0^{(5)''}x}{\triangle} \bhat\times\delta\bu  \Big)\Big] \nn\\
  && \quad   -20 \frac{p_e}{\rho_e}\nabla\cdot\Big( \beta_0 p_e \delta\bu_\parallel + p_e \delta\bu_\perp \frac{\beta_1'x^2+\beta_0'}{\triangle} 
- p_e \bhat\times\delta\bu \frac{\beta_1''x^3+\beta_0''x}{\triangle}\Big)\Big\}, \label{eq:Thierry65}
\end{eqnarray}
with $\delta\bu=\bu_e-\bu_i$. It is again natural to define electron thermal conductivities
(of the moment $\widetilde{X}^{(4)}_e$)
\begin{equation}
  \kappa_\parallel^{e(4)} = \frac{5\sqrt{2}Z_i}{4}\big(\kappa_\parallel^{e(5)}-20\kappa_\parallel^e\big); \qquad
  \kappa_\perp^{e(4)} = \frac{5 \sqrt{2}Z_i}{4}\big(\kappa_\perp^{e(5)}-20\kappa_\perp^e\big); \qquad 
  \kappa_\times^{e(4)} = \frac{5\sqrt{2}Z_i}{4}\big(\kappa_\times^{e(5)}-20\kappa_\times^e\big),
\end{equation}
together with transport coefficients
\begin{eqnarray}
&&  \beta_0^{(4)} = \frac{5\sqrt{2}Z_i}{4}\big(\beta_0^{(5)}-20\beta_0\big); \qquad \beta_1^{(4)'} = \frac{5\sqrt{2}Z_i}{4}\big(\beta_1^{(5)'}-20\beta_1'\big);
  \qquad \beta_0^{(4)'} = \frac{5\sqrt{2}Z_i}{4}\big(\beta_0^{(5)'}-20\beta_0'\big);\nn\\
  && \beta_1^{(4)''} = \frac{5\sqrt{2}Z_i}{4}\big(\beta_1^{(5)''}-20\beta_1''\big);\qquad \beta_0^{(4)''} = \frac{5\sqrt{2}Z_i}{4}\big(\beta_0^{(5)''}-20\beta_0''\big);\\
&&  \gamma_0^{(4)} = \frac{5\sqrt{2}Z_i}{4}\big(\gamma_0^{(5)}-20\gamma_0\big); \qquad \gamma_1^{(4)'} = \frac{5\sqrt{2}Z_i}{4}\big(\gamma_1^{(5)'}-20\gamma_1'\big);
  \qquad \gamma_0^{(4)'} = \frac{5\sqrt{2}Z_i}{4}\big(\gamma_0^{(5)'}-20\gamma_0'\big);\nn\\
  && \gamma_1^{(4)''} = \frac{5\sqrt{2}Z_i}{4}\big(\gamma_1^{(5)''}-20\gamma_1''\big);\qquad \gamma_0^{(4)''} = \frac{5\sqrt{2}Z_i}{4}\big(\gamma_0^{(5)''}-20\gamma_0''\big).  
\end{eqnarray}
The thermal and frictional parts then become
\begin{eqnarray}
  \widetilde{X}^{(4)T}_e &=& -\,\frac{p_e}{\nu_{ei}\rho_e}
  \nabla\cdot \Big(-\kappa_\parallel^{e(4)} \nabla_\parallel T_e - \kappa_\perp^{e(4)} \nabla_\perp T_e
  + \kappa_\times^{e(4)} \bhat\times\nabla T_e\Big)\nn\\
  && -\,\frac{5\sqrt{2}Z_i}{4\nu_{ei}}  \Big(-\kappa_\parallel^{e(5)} \nabla_\parallel T_e - \kappa_\perp^{e(5)} \nabla_\perp T_e
      + \kappa_\times^{e(5)} \bhat\times\nabla T_e\Big)\cdot\nabla\Big(\frac{p_e}{\rho_e}\Big); \label{eq:Thierry66}\\
  \widetilde{X}^{(4)u}_e &=& -\,\frac{p_e}{\nu_{ei}\rho_e}
  \nabla\cdot\Big( \beta_0^{(4)} p_e \delta\bu_\parallel + \frac{\beta_1^{(4)'} x^2+\beta_0^{(4)'}}{\triangle} p_e\delta\bu_\perp
  -\frac{\beta_1^{(4)''}x^3+\beta_0^{(4)''}x}{\triangle} p_e\bhat\times\delta\bu  \Big)\nn\\
  && -\,\frac{5\sqrt{2}Z_i}{4\nu_{ei}}
  \nabla\cdot \Big( \beta_0^{(5)}  p_e\delta\bu_\parallel + \frac{\beta_1^{(5)'} x^2+\beta_0^{(5)'}}{\triangle} p_e \delta\bu_\perp
-\frac{\beta_1^{(5)''}x^3+\beta_0^{(5)''}x}{\triangle} p_e \bhat\times\delta\bu  \Big)\cdot\nabla\Big(\frac{p_e}{\rho_e}\Big), \label{eq:Thierry67}
\end{eqnarray}
where the second terms of (\ref{eq:Thierry66}) \& (\ref{eq:Thierry67}) are purely non-linear and may be neglected for simplicity. 
The thermal conductivities are
\begin{equation}
\kappa_\parallel^{e(4)} = \frac{p_e}{ m_e \nu_{ei}}\gamma_0^{(4)}; \qquad   
\kappa_\perp^{e(4)} =  \frac{p_e}{ m_e \nu_{ei}} \frac{\gamma_1^{(4)'} x^2+\gamma_0^{(4)'}}{\triangle};
\qquad \kappa_\times^{e(4)} =  \frac{p_e}{ m_e \nu_{ei}}\frac{\gamma_1^{(4)''} x^3+\gamma_0^{(4)''} x}{\triangle}, \label{eq:Thierry68}
\end{equation}
and the transport coefficients become
\begin{eqnarray}
  &&  \beta_0^{(4)} = \frac{150 Z_i^2 \sqrt{2}(16\sqrt{2}+29Z_i)}{217 Z_i^2 +604 Z_i \sqrt{2}+288};\qquad
   \beta_1^{(4)'} = -\, \frac{3 \sqrt{2}(548 \sqrt{2}+1261 Z_i)}{224};\nn\\
  && \beta_0^{(4)'} = \frac{3 \sqrt{2}(217 Z_i^2 +604 Z_i \sqrt{2}+288) (16 \sqrt{2}+29 Z_i)}{9800 Z_i^2};\nn\\
  && \beta_1^{(4)''} = -\, \frac{15 Z_i \sqrt{2}}{4};\qquad
   \beta_0^{(4)''} = \frac{3 \sqrt{2}(3079Z_i^2+3181 Z_i\sqrt{2}+1420)}{490 Z_i};
\end{eqnarray}
\begin{eqnarray}
  &&  \gamma_0^{(4)} = \frac{250 Z_i^2 \sqrt{2}(66 \sqrt{2}+229 Z_i)}{217 Z_i^2 +604 Z_i \sqrt{2}+288}; \qquad
   \gamma_1^{(4)'} = \frac{5 \sqrt{2}(4\sqrt{2}-17Z_i)}{8}; \nn\\
  && \gamma_0^{(4)'} =\frac{ \sqrt{2}(217 Z_i^2 +604 Z_i \sqrt{2}+288) (66 \sqrt{2}+229 Z_i)}{1960 Z_i^2};\nn\\
  && \gamma_1^{(4)''} = 25 Z_i \sqrt{2};\qquad 
    \gamma_0^{(4)''} = \frac{\sqrt{2} (176437 Z_i^2 +102558 Z_i \sqrt{2}+30480)}{784 Z_i};  
\end{eqnarray}
\begin{eqnarray}
 && \triangle = x^4+\delta_1 x^2+\delta_0;\nn\\
 &&   \delta_0 = \Big( \frac{217 Z_i^2+604 Z_i \sqrt{2}+288}{700 Z_i^2}\Big)^2;\qquad
    \delta_1 = \frac{586601 Z_i^2+ 330152 Z_i \sqrt{2}+106016}{78400 Z_i^2},
\end{eqnarray}
and with numerical values for $Z_i=1$
\begin{eqnarray}
    \beta_0^{(4)} &=& 8.0576;\qquad
   \beta_1^{(4)'} = -38.5624; \qquad
   \beta_0^{(4)'} = 30.3787;\qquad
   \beta_1^{(4)''} = -5.3033;\qquad
   \beta_0^{(4)''} = 77.9054; \nn\\
   \gamma_0^{(4)} &=& 83.8471;\qquad
   \gamma_1^{(4)'} = -10.0260; \qquad
   \gamma_0^{(4)'} = 316.1179;\qquad
   \gamma_1^{(4)''} = 35.3553;\qquad
   \gamma_0^{(4)''} = 634.8735;\nn\\
    \delta_0 &=& 3.7702; \qquad
   \delta_1 = 14.7898. 
\end{eqnarray}
At the semi-linear level the solution becomes
\begin{eqnarray}
  \widetilde{X}^{(4)T}_e &=& +\frac{p_e}{\nu_{ei}\rho_e} \Big[  \kappa_\parallel^{e(4)} \nabla_\parallel^2 T_e + \kappa_\perp^{e(4)} \nabla_\perp^2 T_e \Big];\\
  \widetilde{X}^{(4)u}_e &=& -\,\frac{p_e^2}{\nu_{ei}\rho_e}
  \Big[ \beta_0^{(4)} \nabla\cdot\delta\bu_\parallel + \frac{\beta_1^{(4)'} x^2+\beta_0^{(4)'}}{\triangle} \nabla\cdot\delta\bu_\perp
    -\frac{\beta_1^{(4)''}x^3+\beta_0^{(4)''}x}{\triangle} \nabla\cdot(\bhat\times\delta\bu)\Big],
\end{eqnarray}
and for zero magnetic field
\begin{eqnarray}
  \widetilde{X}^{(4)}_e &=&  \gamma_0^{(4)} \frac{p_e^2}{\nu_{ei}^2 \rho_e m_e} \nabla^2 T_e 
   - \beta_0^{(4)} \frac{p_e^2}{\nu_{ei}\rho_e} \nabla\cdot\delta\bu.
\end{eqnarray}

\newpage
\section{Discussion \& Conclusions} \label{sec:Discussion}
Here we discuss various topics that we find of importance. 

\subsection{Energy conservation} \label{sec:energy}
Collisional integrals were calculated in a
semi-linear approximation, where all quantities such as $\vecq_a\cdot(\bu_b-\bu_a)$ or $|\bu_b-\bu_a|^2$ are neglected and considered small.
This approach is typically used for calculations with Landau or Boltzmann collisional operators, and
is for example used in the models of \cite{Burgers1969} and \cite{Schunk1977}.   
Importantly, an exact energy conservation $Q_{ab}+Q_{ba}=(\bu_b-\bu_a)\cdot\boldsymbol{R}_{ab}$ cannot be achieved, because
the collisional integrals would have to be calculated non-linearly.
An exact conservation of energy can be achieved only
in two particular cases, the first being a one ion-electron plasma (or a two-species plasma) where the conservation of energy can be imposed by hand,
according to
\begin{equation} \label{eq:Thierry279}
Q_{ie}=3 n_e \nu_{ei}(T_e-T_i)\frac{m_e}{m_i}; \qquad Q_{ei}=-Q_{ie}-(\bu_e-\bu_i)\cdot\boldsymbol{R}_{ei},
\end{equation}  
which is the choice of \cite{Braginskii1965}; see his equation (2.18).
Such a construction cannot be done in general for multi-species plasmas, and conservation of energy is thus satisfied only approximately.

The second particular case is by neglecting all heat fluxes and stress-tensors, and considering only
a  5-moment model with perturbation $\chi_a=0$.
In this specific example of collisions between strict Maxwellians, multi-fluid calculations can be done for unrestricted 
drifts (see \cite{Burgers1969}, \cite{Schunk1977}, and our Appendix \ref{sec:runaway}),  yielding 
\begin{eqnarray}
  \boldsymbol{R}_{ab} &=& \rho_a\nu_{ab}(\bu_b-\bu_a) \Phi_{ab}; \label{eq:Energy50}\\
  Q_{ab} &=& \rho_a\nu_{ab}\Big[ 3\frac{T_b-T_a}{m_a+m_b} \Psi_{ab} + \frac{m_b}{m_a+m_b}|\bu_b-\bu_a|^2 \Phi_{ab}\Big], \label{eq:Energy51}
\end{eqnarray}
where one defines functions
\begin{eqnarray}
 \Psi_{ab}&=& e^{-\epsilon^2};\qquad \Phi_{ab} = \Big( \frac{3}{4}\sqrt{\pi}\frac{\textrm{erf}(\epsilon)}{\epsilon^3}-\frac{3}{2}\frac{e^{-\epsilon^2}}{\epsilon^2}\Big);
  \qquad \epsilon = \frac{|\bu_b-\bu_a|}{\sqrt{v_{\textrm{th} a}^2+v_{\textrm{th} b}^2}},\label{eq:Energy51m}
\end{eqnarray}
thermal velocities $v_{\textrm{th} a}^2=2T_a/m_a$, and collisional frequencies (\ref{eq:timeGenM}).
Because $\rho_a \nu_{ab}=\rho_b \nu_{ba}$ holds, both momentum and energy are conserved. Collisional exchange rates (\ref{eq:Energy50}), (\ref{eq:Energy51})
represent the ``runaway'' effect, and the function $\Phi_{ab}$ is directly related to the Chandrasekhar function; for further
details see Appendix \ref{sec:runaway} and Figure \ref{fig:Phi}.

For a particular case when differences in drift velocities $|\bu_b-\bu_a|$ become much smaller than thermal velocities so that $\epsilon \ll 1$, functions
$\Phi_{ab}\to 1$ and $\Psi_{ab}\to 1$ and $\boldsymbol{R}_{ab} = \rho_a\nu_{ab}(\bu_b-\bu_a)$.
To correctly account for small $|\bu_b-\bu_a|^2$ contributions in $Q_{ab}$ while keeping the differences in temperatures unrestricted
is achieved by $\Psi_{ab}=1-\epsilon^2$, yielding the following equivalent forms
\begin{eqnarray} 
  Q_{ab} &=& \rho_a\nu_{ab} \Big[3 \frac{T_b-T_a}{m_a+m_b} \Big(1-\frac{|\bu_b-\bu_a|^2}{\frac{2T_a}{m_a}+\frac{2T_b}{m_b}}\Big)
  +\frac{m_b}{m_b+m_a} |\bu_b-\bu_a|^2\Big];\\
  Q_{ab} &=& \rho_a\nu_{ab} \Big[ 3 \frac{T_b-T_a}{m_a+m_b}+ \frac{3}{2}\Big( \frac{T_a m_b}{T_a m_b+ T_b m_a}
  -\frac{1}{3}\frac{m_b}{m_b+m_a}\Big)|\bu_b-\bu_a|^2\Big];\label{eq:Energy52}
\end{eqnarray}
 see also (\ref{eq:weirdX}). 
Energy is still conserved. When additionally the differences in temperatures are small as well
(with respect to their mean temperature), the frictional part simplifies into
\begin{equation} \label{eq:Energy61}
  Q_{ab} = \rho_a\nu_{ab} \Big[ 3\frac{T_b-T_a}{m_a+m_b}+ \frac{m_b}{m_a+m_b}|\bu_b-\bu_a|^2\Big].
\end{equation}
One can of course neglect the runaway effect from the beginning, and account for small $|\bu_b-\bu_a|^2$ contributions
either through the center-of-mass velocity transformation, as is for example done in the Appendix of \cite{Braginskii1965},
or by using the Rosenbluth potentials; see Appendix \ref{sec:Rab}, \ref{sec:Qab}.

Note that considering the 22-moment model, the fully contracted scalar perturbations $\widetilde{X}^{(4)}$
  modify the energy conservation, according to 
\begin{eqnarray}
&& Q_{ab} = \frac{\rho_a \nu_{ab}}{(m_a+m_b)} \Big[ 3(T_b-T_a)+ \hat{P}_{ab (1)}\frac{\rho_a}{n_a p_a} \widetilde{X}^{(4)}_a
  - \hat{P}_{ab (2)} \frac{\rho_b}{n_b p_b} \widetilde{X}^{(4)}_b     \Big];\nn\\
&& \hat{P}_{ab (1)} = \frac{3 T_a m_b (5 T_b m_b +4 T_b m_a -T_a m_b  )}{40 (T_a m_b +T_b m_a)^2};\qquad
\hat{P}_{ab (2)} = \frac{3T_b m_a (5 T_a m_a +4 T_a m_b -T_b m_a  )}{40 (T_a m_b +T_b m_a)^2},\label{eq:Thierry35}
\end{eqnarray}
 and for only two species one can again impose an exact energy conservation by hand; see e.g. (\ref{eq:Thierry278e}).   

\subsection{Collisional frequencies for ion-electron plasma} \label{section:ColFreq}
The Landau collisional operator yields the following collisional frequencies (see for example \cite{Hinton1983} or our Appendix \ref{sec:Rab})
\begin{equation} \label{eq:timeGenM}
\nu_{ab}=\tau^{-1}_{ab} = \frac{16}{3}\sqrt{\pi}\frac{n_b e^4 Z_a^2 Z_b^2 \ln\Lambda}{m_a^2(v_{\textrm{th}a}^2+v_{\textrm{th}b}^2)^{3/2}} \Big( 1+\frac{m_a}{m_b}\Big),
\end{equation}
where $v_{\textrm{th}a}^2=2T_a/m_a$, and $\rho_a \nu_{ab}= \rho_b \nu_{ba}$ holds.
Equivalently, in the form of \cite{Burgers1969} \& \cite{Schunk1977}
\begin{equation} \label{eq:timeSchunk}
\nu_{ab}= \frac{16}{3}\sqrt{\pi} \Big(\frac{\mu_{ab}}{2T_{ab}}\Big)^{3/2} \frac{m_b n_b}{m_a+m_b} \frac{e^4 Z_a^2 Z_b^2 \ln\Lambda}{\mu_{ab}^2},
\end{equation}
where the reduced mass $\mu_{ab}$ and reduced temperature $T_{ab}$ are defined in (\ref{eq:Num20}). 
For a particular case of self-collisions
\begin{equation}
\nu_{aa}= \frac{4}{3}\sqrt{\pi}\frac{n_a e^4 Z_a^4 \ln\Lambda}{T_a^{3/2} \sqrt{m_a}}. 
\end{equation}
For a particular case of $T_a=T_b=T$,
\begin{equation} \label{eq:Brag_time}
\nu_{ab} = \frac{4}{3}\sqrt{2\pi}\frac{n_b e^4 Z_a^2 Z_b^2 \ln\Lambda}{T^{3/2}} \frac{\sqrt{\mu_{ab}}}{m_a}, 
\end{equation}
which identifies with equation (7.6) of \cite{Braginskii1965} (after one uses $\nu_{ab}=n_b \mu_{ab}\alpha_{ab}'/m_a$).
For a particular case of a one ion-electron plasma, collisional frequencies simplify into
\begin{eqnarray}
  &&  \nu_{ii}= \frac{4}{3}\sqrt{\pi}\frac{n_i e^4 Z_i^4\ln\Lambda}{T^{3/2}_i \sqrt{m_i}}; \qquad
\nu_{ie}= \frac{4}{3}\sqrt{2\pi}\frac{n_e e^4 Z_i^2\ln\Lambda}{T^{3/2}_e \sqrt{m_i}} \sqrt{\frac{m_e}{m_i}}; \nn \\
&&  \nu_{ee}= \frac{4}{3}\sqrt{\pi}\frac{n_e e^4 \ln\Lambda}{T^{3/2}_e \sqrt{m_e}};\qquad
\nu_{ei}= \frac{4}{3}\sqrt{2\pi}\frac{n_i e^4 Z_i^2\ln\Lambda}{T^{3/2}_e \sqrt{m_e}}, \label{eq:Posible1}
\end{eqnarray}
where one assumes $T_i/m_i\ll T_e/m_e$, so the ions cannot be extremely hot.
Obviously, $\nu_{ii}\gg \nu_{ie}$ (by a factor of $\sqrt{m_i/m_e}$ for equal temperatures and $Z_i=1$),
however $\nu_{ee}\sim \nu_{ei}$, with exact relation $\nu_{ei}=Z_i\sqrt{2}\nu_{ee}$ after one uses $n_e=Z_in_i$.
The relation $\rho_i \nu_{ie}=\rho_e \nu_{ei}$ holds exactly in (\ref{eq:Posible1}). Note the important difference
that while $\nu_{ei}$ contains a factor of $\sqrt{2}$, the $\nu_{ii}$ does not.
Thus, comparing \cite{Braginskii1965} expressions (2.5i) and (2.5e) with definitions (\ref{eq:Posible1}), Braginskii clearly uses
\begin{equation} \label{eq:BragNat}
\tau_{i}=\tau_{ii};\qquad \tau_{e}=\tau_{ei},
\end{equation}
 which also agrees with his definition (7.6), equivalent to our (\ref{eq:Brag_time}).

However, very often when considering ion-electron plasma, a different definition of $\nu_{ab}$ is used without the reduced mass,
in the following form
\begin{eqnarray}
m_a \ll m_b:\qquad  \nu_{ab}=\tau_{ab}^{-1} &=& \frac{16}{3}\sqrt{\pi} \frac{n_be^4Z_a^2Z_b^2\ln\Lambda}{m_a^2 v_{\textrm{th} a}^3}
  = \frac{4}{3}\sqrt{2\pi} \frac{n_be^4Z_a^2Z_b^2\ln\Lambda}{T_a^{3/2}\sqrt{m_a}}, \label{eq:nu_abW}
\end{eqnarray}
which for example agrees with the Appendix of \cite{HelanderSigmar2002} (page 277, after using cgs units $\epsilon_0\to 1/(4\pi)$).
We have added the $m_a\ll m_b$ designation even though it is not present
in \cite{HelanderSigmar2002}, because obviously it is the only way how to obtain (\ref{eq:nu_abW}) from the general
(\ref{eq:timeGenM}).
Importantly, $\rho_a \nu_{ab}\neq \rho_b \nu_{ba}$, and if one would use (\ref{eq:nu_abW}) to calculate $\nu_{ie}$,
the result would be erroneous. Instead, the $\nu_{ie}$ must be
calculated from $\nu_{ei}$ so that the momentum is conserved. Technically, (\ref{eq:nu_abW}) should not be used for self-collisions either.
Nevertheless, using (\ref{eq:nu_abW}) yields the following collisional frequencies
\begin{eqnarray}
  &&  \nu_{ii}= \frac{4}{3}\sqrt{2\pi}\frac{n_i e^4 Z_i^4 \ln\Lambda}{T^{3/2}_i \sqrt{m_i}}; \qquad
   \nu_{ie} = \frac{m_e n_e}{m_i n_i} \nu_{ei} = \frac{4}{3}\sqrt{2\pi}\frac{n_e e^4 Z_i^2 \ln\Lambda}{T^{3/2}_e\sqrt{m_i}} \sqrt{\frac{m_e}{m_i}}; \nn\\
&&  \nu_{ee}= \frac{4}{3}\sqrt{2\pi}\frac{n_e e^4 \ln\Lambda}{T^{3/2}_e \sqrt{m_e}};\qquad
\nu_{ei}= \frac{4}{3}\sqrt{2\pi}\frac{n_i e^4 Z_i^2\ln\Lambda}{T^{3/2}_e\sqrt{m_e}}.\label{eq:FUN}
\end{eqnarray} 
Now $\nu_{ii}$ contains a factor of $\sqrt{2}$, leading to an interpretation that Braginskii uses
\begin{equation}
\tau_i=\sqrt{2}\tau_{ii};\qquad \tau_e=\tau_{ei}. 
\end{equation}
Also, now for $Z_i=1$ relation $\nu_{ee}=\nu_{ei}$ holds. These definitions of collisional frequencies are
used in a majority of the modern plasma literature, where one argues that it seems unnatural to introduce asymmetry between
$\nu_{ii}$ and $\nu_{ei}$ (see e.g. Part 1 of \cite{Balescu1988}, p.192, p.274). 
Obviously, for multi-species
plasmas collisional frequencies (\ref{eq:timeGenM}) have to be used, and we thus find it much more natural to use
the original \cite{Braginskii1965} definitions (\ref{eq:Posible1}), (\ref{eq:BragNat}) also for an ion-electron plasma.  
Of course, for the Landau operator both approaches yield the same results because the collisional integrals are properly
calculated. However, a difference arises for the phenomenological operators such as the BGK or the Dougherty (Lenard-Bernstein) operators,
where for example one needs to add $\nu_{ee}+\nu_{ei}$. Calculating this addition according to (\ref{eq:FUN}) would be incorrect, and
one has to use (\ref{eq:Posible1}) instead. Comparison of Braginskii viscosities and heat conductivities with the BGK operator can
be found in Appendix \ref{sec:BGKcomparision}.

\subsection{Fluid hierarchy} \label{sec:Fhierarchy}
Even though we do not calculate the collisional integrals for general n-th order moments, we find it useful to discuss the fluid hierarchy and 
formulate it for a general collisional operator $C(f_a)$. One defines heat flux vectors, stress-tensors and fully
contracted moments according to
\begin{eqnarray}
  \vecX^{(2n+1)}_a &=& m_a \int \bc_a|\bc_a|^{2n} f_a d^3v; \nn\\
   \bPi^{(2n)}_a &=& m_a \int \big(\bc_a\bc_a-\frac{\bI}{3}|\bc_a|^{2}\big) |\bc_a|^{2n-2} f_a d^3v;\nn\\
  X^{(2n)}_a &=&  m_a \int |\bc_a|^{2n} f_a d^3v = (2n+1)!! \frac{p^n_a}{\rho^{n-1}_a}+\widetilde{X}^{(2n)}_a, \label{eq:Num50}
\end{eqnarray}
together with collisional contributions
\begin{eqnarray}
   \vec{\boldsymbol{\mathcal{Q}}}^{(2n+1)}_a &=& m_a \int |\bc_a|^{2n} \bc_a C(f_a) d^3v;\nn\\
   \bar{\bar{\boldsymbol{\mathcal{Q}}}}^{(2n)}_a &=& m_a \int |\bc_a|^{2n-2} \bc_a \bc_a C(f_a) d^3v;\nn\\
    Q^{(2n)}_a &=& m_a \int |\bc_a|^{2n} C(f_a) d^3v;\qquad Q_a = \frac{m_a}{2} \int |\bc_a|^{2} C(f_a) d^3v, \label{eq:Thierry25}
\end{eqnarray}
where to prevent incompatibility with the previous notation, for vectors and matrices
we use $\boldsymbol{\mathcal{Q}}$ (mathcal Q) instead of $\boldsymbol{Q}$. The new notation fixes the problem that
for example $\vecQ^{(3)}_a\,'$ was used for the right hand side of evolution equation of the heat flux $\vecq_a$, and not for $\vecX^{(3)}_a$.
It also clarifies that in the vector notation the matrix $\bar{\bar{\boldsymbol{\mathcal{Q}}}}^{(2n)}_a=\trace\trace\ldots \trace \bQ_a^{(2n)}$.
Note that $Q^{(2)}_a=2Q_a$; $X^{(2)}_a=3p_a$; $\widetilde{X}^{(2)}_a=0$; $\vecX^{(3)}_a=2\vecq_a$ and $\vecX^{(1)}_a=0$. 

Fully non-linear evolution equations are given in Appendix \ref{sec:Hierarchy}; see (\ref{eq:Num8})-(\ref{eq:Num10}).
In the semi-linear approximation, these simplify into evolution equations for vectors valid for $n\ge 1$
\begin{eqnarray}
  &&  \frac{d_a}{d t}\vecX^{(2n+1)}_a 
  +\frac{1}{3}\nabla \widetilde{X}^{(2n+2)}_a +\nabla\cdot\bPi^{(2n+2)}_a  -\frac{(2n+3)!!}{3} \frac{p_a^n}{\rho_a^n} \nabla\cdot\bPi^{(2)}_a
  +\Omega_a \bhat\times \vecX^{(2n+1)}_a\nn\\
  && +(2n+3)!! \frac{(n)}{3} \frac{p_a^{n}}{\rho_a^{n-1}}\nabla\Big(\frac{p_a}{\rho_a} \Big)
  = \vec{\boldsymbol{\mathcal{Q}}}^{(2n+1)}_a\,' = \vec{\boldsymbol{\mathcal{Q}}}^{(2n+1)}_a  -\frac{(2n+3)!!}{3} \frac{p_a^n}{\rho_a^n}\boldsymbol{R}_a, \label{eq:Num11}
\end{eqnarray}
stress-tensors valid for $n\ge 1$
\begin{eqnarray}
  && \frac{d_a}{dt} \bPi^{(2n)}_a +\frac{1}{5}\Big[ \big(\nabla\vecX^{(2n+1)}_a\big)^S -\frac{2}{3}\bI \nabla\cdot\vecX^{(2n+1)}_a \Big]
  +\Omega_a \big( \bhat\times \bPi^{(2n)}_a\big)^S \nn\\
  && + \frac{(2n+3)!!}{15} \frac{p^n_a}{\rho_a^{n-1}} \bW_a
   = \bar{\bar{\boldsymbol{\mathcal{Q}}}}^{(2n)}_a\,' = \bar{\bar{\boldsymbol{\mathcal{Q}}}}^{(2n)}_a -\frac{\bI}{3}Q^{(2n)}_a, \label{eq:Num12}
\end{eqnarray}
and scalar perturbations valid for $n\ge 2$
\begin{eqnarray}
&& \frac{d_a}{dt} \widetilde{X}^{(2n)}_a +\nabla\cdot\vecX^{(2n+1)}_a 
   -(2n+1)!! \frac{(2n)}{3}\Big(\frac{p_a}{\rho_a} \Big)^{n-1} \nabla\cdot\vecq_a\nn\\
  && = \widetilde{Q}^{(2n)}_a\,' = Q^{(2n)}_a - (2n+1)!! \frac{(2n)}{3}\Big(\frac{p_a}{\rho_a} \Big)^{n-1} Q_a, \label{eq:Num13}
\end{eqnarray}
where (n) without species index should not be confused with the number density. 
Equation (\ref{eq:Num13}) is also valid for $n=1$, but it is identically zero.
In comparison to previous notation  
$\vec{\boldsymbol{\mathcal{Q}}}^{(3)}_a\,' = 2 \vecQ^{(3)}_a\,'$; $\vec{\boldsymbol{\mathcal{Q}}}^{(5)}_a\,' = \vecQ^{(5)}_a\,'$;
$\bar{\bar{\boldsymbol{\mathcal{Q}}}}^{(2)}_a\,'=\bQ^{(2)}_a\,'$; and $\bar{\bar{\boldsymbol{\mathcal{Q}}}}^{(4)}_a\,'=\bQ^{(4)}_a\,'$. 

\subsection{Reducible \& irreducible Hermite polynomials} \label{sec:HermX}
The \emph{irreducible} Hermite polynomials $H(\tbc)$ (notation without tilde) are usually defined through Laguerre-Sonine polynomials
$L(\tbc)$ (see for example equation (G1.4.4), page 326 of \cite{Balescu1988})
\begin{eqnarray}
  H^{(2n)} (\tc) &=& L_n^{(1/2)} (\frac{\tc^2}{2});\qquad 
  H^{(2n+1)}_i (\tbc) = \sqrt{\frac{3}{2}} \tc_i L_n^{(3/2)}(\frac{\tc^2}{2});\nn\\
  H^{(2n)}_{ij} (\tbc) &=&  \sqrt{\frac{15}{8}}(\tc_i\tc_j-\frac{\tc^2}{3}\delta_{ij})L_{n-1}^{(5/2)}(\frac{\tc^2}{2}), \label{eq:Relat1}
\end{eqnarray}
where we use tilde for the normalized fluctuating velocity $\tbc = \sqrt{m_a/T_a}\bc_a$ with species indices dropped.
In our calculations we find it more natural to use the \emph{reducible} Hermite polynomials $\tilde{H}(\tbc)$ (notation with tilde) of Grad
defined according to
\begin{equation}
\tilde{H}^{(m)}_{r_1 r_2\ldots r_m}(\tbc) = (-1)^m e^{\frac{\tc^2}{2}} \frac{\pr}{\pr \tc_{r_1}}  \frac{\pr}{\pr \tc_{r_2}}\cdots  \frac{\pr}{\pr \tc_{r_m}} e^{-\frac{\tc^2}{2}}.   
\end{equation}
By applying a sufficient number of contractions then yields definitions of fully contracted scalars, vectors and matrices  
\begin{equation}
  \tilde{H}^{(2n)} = \tilde{H}^{(2n)}_{r_1 r_1 \ldots r_n r_n};\qquad \tilde{H}^{(2n+1)}_i = \tilde{H}^{(2n+1)}_{i r_1 r_1 \ldots r_n r_n};\qquad
  \tilde{H}^{(2n)}_{ij} = \tilde{H}^{(2n)}_{i j r_1 r_1 \ldots r_{n-1} r_{n-1}},
\end{equation}  
together with conveniently defined traceless matrices (notation with hat)
\begin{equation}
\hat{H}_{ij}^{(2n)} = \tilde{H}_{ij}^{(2n)}-\frac{1}{3}\delta_{ij}\tilde{H}^{(2n)}.
\end{equation}
The relation between irreducible and reducible Hermite polynomials then can be shown to be  
\begin{eqnarray}
H^{(2n)}  &=& \Big(\frac{1}{ 2^n n! (2n+1)!!}\Big)^{1/2 }\tilde{H}^{(2n)} ;\qquad   
H^{(2n+1)}_i  = \Big(\frac{3}{2^n n! (2n+3)!!}\Big)^{1/2} \tilde{H}_i^{(2n+1)} ;\nn\\
H^{(2n)}_{ij}  &=& \Big(\frac{15}{2^n (n-1)! (2n+3)!!} \Big)^{1/2} \hat{H}_{ij}^{(2n)}, \label{eq:Nomore3000}
\end{eqnarray}
and both approaches use essentially the same polynomials, the only difference is the location of normalization factors. 
The reducible Hermite polynomials are used to define Hermite moments
\begin{eqnarray}
  \tilde{h}^{(2n)} = \frac{1}{n_a}\int f_a \tilde{H}^{(2n)} d^3c; \qquad
  \tilde{h}^{(2n+1)}_i = \frac{1}{n_a}\int f_a \tilde{H}^{(2n+1)}_i d^3c; \qquad
  \hat{h}^{(2n)}_{ij} = \frac{1}{n_a}\int f_a \hat{H}^{(2n)}_{ij} d^3c,
\end{eqnarray}
and analogously for the irreducible ones. Note that the scalar $\tilde{h}^{(2)}=0$, 
and we thus often use $\hat{h}^{(2)}_{ij}=\tilde{h}^{(2)}_{ij}=(1/n_a)\int f_a \tilde{H}^{(2)}_{ij} d^3c $. Finally, by using orthogonality relations one obtains 
perturbation $\chi_a$ of the distribution function $f_a=f_a^{(0)}(1+\chi_a)$ around Maxwellian $f_a^{(0)}$, in the following form
\begin{eqnarray}
  \chi_a &=& \sum_{n=1}^{N}
  \Big[ \frac{15}{2^n (n-1)! (2n+3)!!}\hat{h}^{(2n)}_{ij} \hat{H}^{(2n)}_{ij}
    +\frac{1}{2^n n! (2n+1)!!} \tilde{h}^{(2n)} \tilde{H}^{(2n)} \nn\\
    && + \frac{3}{2^n n! (2n+3)!!} \tilde{h}^{(2n+1)}_{i} \tilde{H}^{(2n+1)}_{i} \Big]; \label{eq:Nomore777}\\
  \chi_a &=& \sum_{n=1}^{N}
  \Big[ {h}^{(2n)}_{ij} {H}^{(2n)}_{ij} +{h}^{(2n)} {H}^{(2n)} +{h}^{(2n+1)}_{i} {H}^{(2n+1)}_{i} \Big],
\end{eqnarray}
and the two approaches are equivalent. Alternatively, because $\hat{h}^{(2n)}_{ij}$ are traceless, 
it is possible to use $\hat{h}^{(2n)}_{ij} \hat{H}^{(2n)}_{ij}=\hat{h}^{(2n)}_{ij} \tilde{H}^{(2n)}_{ij}$. Note that $2^n n! (2n+1)!!=(2n+1)!$.
The 13-moment model of Burgers-Schunk is obtained by $N=1$. Prescribing $N=2$ yields the 22-moment model  
\begin{equation}
\chi_a =  \frac{1}{2}\hat{h}_{ij}^{(2)} \hat{H}_{ij}^{(2)} + \frac{1}{10} \tilde{h}_{i}^{(3)} \tilde{H}_{i}^{(3)}
  +\frac{1}{28} \hat{h}_{ij}^{(4)} \hat{H}_{ij}^{(4)} + \frac{1}{120}\tilde{h}^{(4)} \tilde{H}^{(4)}  +\frac{1}{280} \tilde{h}_{i}^{(5)} \tilde{H}_{i}^{(5)},
\end{equation}
with  Hermite polynomials
\begin{eqnarray}
  \tilde{H}^{(3)}_i &=& \tc_i ( \tc^2-5); \qquad \tilde{H}^{(5)}_i = \tc_i ( \tc^4-14\tc^2+35);\nn\\
\hat{H}^{(2)}_{ij} &=& \big( \tc_i\tc_j-\frac{\delta_{ij}}{3}\tc^2\big);\qquad 
\hat{H}^{(4)}_{ij} = \big( \tc_i\tc_j-\frac{\delta_{ij}}{3}\tc^2\big) (\tc^2-7); \qquad
\tilde{H}^{(4)} = \tc^4-10\tc^2+15,
\end{eqnarray}
and neglecting $\tilde{h}^{(4)}=0$ (meaning $\widetilde{X}^{(4)}=0$) yields the 21-moment model.

\newpage
Transformation from Hermite to fluid moments is done according to
\begin{eqnarray}
&& \tbh^{(3)}_a = \frac{2}{p_a}\sqrt{\frac{m_a}{T_a}}\vecq_a;\qquad   
 \tbh^{(5)}_a = \frac{1}{p_a}\sqrt{\frac{m_a}{T_a}}\Big( \frac{m_a}{T_a} \vecX^{(5)}_a-28\vecq_a \Big); \nn\\
 &&  \hbbh^{(2)}_a =  \tbbh^{(2)}_a = \frac{1}{p_a}\bPi^{(2)}_a;\qquad  \hbbh^{(4)}_a = \frac{\rho_a}{p_a^2} \bPi^{(4)}_a - \frac{7}{p_a}\bPi^{(2)}_a;
 \qquad \tilde{h}^{(4)}_a = \frac{\rho_a}{p_a^2} \widetilde{X}^{(4)}_a. \label{eq:Energy753}
\end{eqnarray}
Various models are summarized
  in Tables \ref{table:Chi1} \& \ref{tab:Chi2}. In Table \ref{table:Chi1} the perturbation $\chi_a$ is given in reducible Hermite moments,
and in Table \ref{tab:Chi2} the perturbation is given in fluid moments.

\begin{table}[ht!]
\centering
\begin{tabular}{| l | l |}
\hline
Model name         &    Corresponding perturbation of $f_a=f_a^{(0)}(1+\chi_a)$ in Hermite moments\\
\hline    
5-moment;    &    $\chi_a = 0$;\\
8-moment;    &   $\chi_a = \frac{1}{10} \tilde{h}_i^{(3)} \tilde{H}_{i}^{(3)};$ \\
10-moment;   &   $\chi_a = \frac{1}{2}\tilde{h}_{ij}^{(2)}\tilde{H}_{ij}^{(2)};$ \\
13-moment;   &   $\chi_a = \frac{1}{2}\tilde{h}_{ij}^{(2)}\tilde{H}_{ij}^{(2)}  + \frac{1}{10} \tilde{h}_i^{(3)}\tilde{H}_i^{(3)};$ \\
20-moment;  &   $ \chi_a = \frac{1}{2}\tilde{h}_{ij}^{(2)} \tilde{H}_{ij}^{(2)} + \frac{1}{6} \tilde{h}_{ijk}^{(3)} \tilde{H}_{ijk}^{(3)}; $\\
21-moment;   &   $ \chi_a = \frac{1}{2}\tilde{h}_{ij}^{(2)}\tilde{H}_{ij}^{(2)}  + \frac{1}{10} \tilde{h}_i^{(3)}\tilde{H}_i^{(3)}
  +\frac{1}{28}\hat{h}_{ij}^{(4)}\tilde{H}_{ij}^{(4)} + \frac{1}{280} \tilde{h}_i^{(5)}\tilde{H}_i^{(5)}$; \\
22-moment;   &   $\chi_a= \frac{1}{2}\tilde{h}_{ij}^{(2)}\tilde{H}_{ij}^{(2)}  + \frac{1}{10} \tilde{h}_i^{(3)}\tilde{H}_i^{(3)}
  +\frac{1}{28}\hat{h}_{ij}^{(4)}\tilde{H}_{ij}^{(4)} +\frac{1}{120} \tilde{h}^{(4)}\tilde{H}^{(4)}+ \frac{1}{280} \tilde{h}_i^{(5)}\tilde{H}_i^{(5)};$\\
  \hline
9-moment;     &  $\chi_a = \frac{1}{10} \tilde{h}_i^{(3)} \tilde{H}_{i}^{(3)}+\frac{1}{120} \tilde{h}^{(4)}\tilde{H}^{(4)};$ \\ 
11-moment;    &  $\chi_a = \frac{1}{10} \tilde{h}_i^{(3)} \tilde{H}_{i}^{(3)}+\frac{1}{280} \tilde{h}_i^{(5)}\tilde{H}_i^{(5)};$\\
12-moment;    &  $\chi_a = \frac{1}{10} \tilde{h}_i^{(3)} \tilde{H}_{i}^{(3)}+\frac{1}{120} \tilde{h}^{(4)}\tilde{H}^{(4)}+\frac{1}{280} \tilde{h}_i^{(5)}\tilde{H}_i^{(5)};$\\
15-moment;    &  $ \chi_a = \frac{1}{2}\tilde{h}_{ij}^{(2)}\tilde{H}_{ij}^{(2)}  + \frac{1}{28}\hat{h}_{ij}^{(4)}\tilde{H}_{ij}^{(4)};$\\
16-moment;    &  $ \chi_a = \frac{1}{2}\tilde{h}_{ij}^{(2)}\tilde{H}_{ij}^{(2)}  + \frac{1}{28}\hat{h}_{ij}^{(4)}\tilde{H}_{ij}^{(4)}
+ \frac{1}{120} \tilde{h}^{(4)}\tilde{H}^{(4)};$\\
\hline
\end{tabular}
\caption{Summary of various models with the perturbation $\chi_a$ given in reducible Hermite moments. Species indices ``a'' are dropped. The upper half of the
  table contains ``major'' models, and the lower half contains other possibilities.  Note that the 16-moment model should not be confused with
the anisotropic (bi-Maxwellian based) 16-moment model described in Section \ref{sec:limitations}.}
\label{table:Chi1}
\end{table}
\begin{table}[ht!]
\centering
\begin{tabular}{| l | l |}
\hline
Model name         &    Corresponding perturbation of $f_a=f_a^{(0)}(1+\chi_a)$ in fluid moments\\
\hline    
5-moment;    &   $\chi_a = 0;$\\
8-moment;    &   $\chi_a = -\,\frac{m_a}{p_a T_a} \big(\vecq_a\cdot\bc_a\big) \Big( 1-\frac{m_a}{5 T_a}c_a^2\Big);$ \\
10-moment;   &   $\chi_a = \frac{m_a}{2p_a T_a}\big(\bPi_a^{(2)}:\bc_a\bc_a\big);$\\
13-moment;   &   $\chi_a = \frac{m_a}{2p_a T_a}\big(\bPi_a^{(2)}:\bc_a\bc_a\big)  -\frac{m_a}{p_a T_a}\big(\vecq_a\cdot\bc_a \big) \Big( 1-\frac{m_a}{5 T_a}c_a^2\Big);$ \\
20-moment;   &   $\chi_a = \frac{m_a}{2p_a T_a}\big(\bPi_a^{(2)}:\bc_a\bc_a\big)
     +\frac{m_a^2}{6p_a T_a^2}\big(\bc_a\cdot\bq_a:\bc_a\bc_a\big)  -\frac{m_a}{p_a T_a} \big(\vecq_a\cdot\bc_a\big);$\\
21-moment;   &   $ \chi_a = \frac{1}{2p_a} \big(\bPi_a^{(2)}:\tbc_a\tbc_a\big) +\frac{1}{28}\Big[
    \frac{\rho_a}{p_a^2} \big(\bPi^{(4)}_a:\tbc_a\tbc_a \big)  -\frac{7}{p_a} \big(\bPi_a^{(2)}:\tbc_a\tbc_a\big)\Big] (\tc_a^2-7)$\\
             &   $\qquad +\frac{1}{5 p_a}\sqrt{\frac{m_a}{T_a}} (\vecq_a\cdot\tbc_a)(\tc_a^2-5) + \frac{1}{280 p_a}\sqrt{\frac{m_a}{T_a}}
\Big[\frac{\rho_a}{p_a} (\vecX^{(5)}_a\cdot\tbc_a) -28 (\vecq_a\cdot\tbc_a) \Big] (\tc_a^4-14\tc_a^2+35);$ \\
22-moment;   &  $ \chi_a = \frac{1}{2p_a} \big(\bPi_a^{(2)}:\tbc_a\tbc_a\big) +\frac{1}{28}\Big[
    \frac{\rho_a}{p_a^2} \big(\bPi^{(4)}_a:\tbc_a\tbc_a \big)  -\frac{7}{p_a} \big(\bPi_a^{(2)}:\tbc_a\tbc_a\big)\Big] (\tc_a^2-7)$\\
             &   $\qquad +\frac{1}{5 p_a}\sqrt{\frac{m_a}{T_a}} (\vecq_a\cdot\tbc_a)(\tc_a^2-5) + \frac{1}{280 p_a}\sqrt{\frac{m_a}{T_a}}
\Big[\frac{\rho_a}{p_a} (\vecX^{(5)}_a\cdot\tbc_a) -28 (\vecq_a\cdot\tbc_a) \Big] (\tc_a^4-14\tc_a^2+35)$ \\
             & $\qquad +\frac{1}{120}\frac{\rho_a}{p_a^2} \widetilde{X}^{(4)}_a (\tc_a^4-10\tc^2+15);$\\
\hline
\end{tabular}
\caption{Summary of various models with the perturbation $\chi_a$ given in fluid moments. Results for the 21 \& 22-moment models
  are written with normalized $\tbc_a = \sqrt{m_a/T_a}\bc_a$.}
\label{tab:Chi2}
\end{table}

\newpage
\subsection{Rosenbluth potentials (22-moment model)}
Here we summarize the Rosenbluth potentials, defined according to 
\begin{eqnarray}
  H_b (\bV) &=& \int \frac{f_b(\bV')}{|\bV'-\bV|}d^3v';\qquad  G_b (\bV) = \int |\bV'-\bV| f_b(\bV') d^3v', 
\end{eqnarray}
where the first potential should not be confused with the irreducible Hermite polynomials. 
For the 22-moment model the fully non-linear results read 
\begin{eqnarray}
  H_b (\bV) &=& n_b \sqrt{\frac{m_b}{T_b}} \Big\{ \frac{1}{\ty}\erf\Big(\frac{\ty}{\sqrt{2}}\Big)
    -\sqrt{\frac{2}{\pi}} \frac{e^{-\ty^2/2}}{10} \Big( \tby\cdot\tbh^{(3)}_b+(\ty^2-5)\frac{\tby\cdot\tbh^{(5)}_b}{28}\Big) \nn\\
 && +\frac{1}{2} (\hbbh^{(2)}_b :\tby\tby) \Big[ \frac{3}{\ty^5}\erf\Big(\frac{\ty}{\sqrt{2}}\Big)
    -\sqrt{\frac{2}{\pi}}\big(\frac{1}{\ty^2}+\frac{3}{\ty^4}\big)e^{-\ty^2/2}\Big] \nn\\
    && \qquad -\frac{1}{28} (\hbbh^{(4)}_b :\tby\tby)  \sqrt{\frac{2}{\pi}} e^{-\ty^2/2}
   +\frac{1}{120} \tilde{h}^{(4)}_b (3-\ty^2)\sqrt{\frac{2}{\pi}} e^{-\ty^2/2} \Big\};\\
  G_b (\bV) &=& n_b \sqrt{\frac{T_b}{m_b}} \Big\{ \sqrt{\frac{2}{\pi}}e^{-\ty^2/2} +\big(\ty+\frac{1}{\ty}\big)\erf\big(\frac{\ty}{\sqrt{2}}\big) \nn\\
  && +\Big(\frac{\erf(\ty/\sqrt{2})}{5\ty^3} - \sqrt{\frac{2}{\pi}} \frac{e^{-\ty^2/2}}{5\ty^2} \Big)\tby\cdot\tbh^{(3)}_b
    - \sqrt{\frac{2}{\pi}} \frac{e^{-\ty^2/2}}{140} \tby\cdot\tbh^{(5)}_b\nn\\     
  && \qquad -\frac{1}{2}(\hbbh^{(2)}_b :\tby\tby) \Big[ \frac{3}{\ty^4} \sqrt{\frac{2}{\pi}} e^{-\ty^2/2}
    +\big(\frac{1}{\ty^3}-\frac{3}{\ty^5}\big)\erf\Big(\frac{\ty}{\sqrt{2}}\Big) \Big] \nn\\
  && \qquad -\frac{1}{14} (\hbbh^{(4)}_b :\tby\tby)\Big[ \sqrt{\frac{2}{\pi}}\big( \frac{1}{\ty^2}+\frac{3}{\ty^4}\big) e^{-\ty^2/2}
      -\frac{3}{\ty^5}\erf\Big(\frac{\ty}{\sqrt{2}}\Big) \Big] 
    -\frac{1}{60} \tilde{h}^{(4)}_b \sqrt{\frac{2}{\pi}} e^{-\ty^2/2}  \Big\},
\end{eqnarray}
where we use the variable
\begin{equation}
\tby = \sqrt{\frac{m_b}{T_b}}(\bV-\bu_b).
\end{equation}
These Rosenbluth potentials are used to calculate the dynamical friction vector $\boldsymbol{A}_{ab}$ and the diffusion
tensor $\bD_{ab}$, which then form the Landau collisional operator, according to
\begin{eqnarray}
&&  \boldsymbol{A}_{ab}(\bV) = 2\frac{c_{ab}}{m_a^2}\big(1+\frac{m_a}{m_b}\big)\frac{\pr H_b(\bV)}{\pr\bV}; \qquad \bD_{ab} (\bV)
  =2\frac{c_{ab}}{m_a^2}\frac{\pr^2 G_b(\bV)}{\pr\bV\pr\bV}; \qquad c_{ab} = 2\pi e^4 Z_a^2 Z_b^2\ln\Lambda;\nn\\
&& C_{ab}(f_a,f_b) = -\,\frac{\pr}{\pr \bV} \cdot \Big[ \boldsymbol{A}_{ab}f_a-\frac{1}{2} \frac{\pr}{\pr\bV}\cdot(\bD_{ab} f_a)\Big]. 
\end{eqnarray}
The dynamical friction vectors and diffusion tensors can be found in the Appendix;  see equations (\ref{eq:Energy750})-(\ref{eq:Energy751}); 
(\ref{eq:DynVec})-(\ref{eq:DiffTensor}) and (\ref{eq:Thierry120})-(\ref{eq:Thierry121}). For clarity, we split the calculations
 into heat fluxes (Appendix \ref{sec:HeatFluxB}), viscosities (Appendix \ref{sec:BragVisc}) and scalar perturbations (Appendix \ref{sec:X4coll}).
These results are fully non-linear and could be potentially useful to construct more sophisticated models that could capture
collisional effects beyond the semi-linear approximation, or perhaps to explore the runaway effect numerically.    
All the equations can be transformed from Hermite moments to  fluid moments by (\ref{eq:Energy753}).

\newpage
\subsection{Hermite closures} \label{sec:HC}
The general hierarchy of evolution equations (\ref{eq:Num11})-(\ref{eq:Num13}) needs to be closed with appropriate closures at the last retained fluid moment. 
A correct form of a fluid closure is obtained in the Hermite space, by cutting the perturbation $\chi_a$ given by (\ref{eq:Nomore777}) at
an appropriate $N$.  For example, the 22-moment model is obtained with
 Hermite closures $\tilde{h}^{(6)}_a=0$ and $\tilde{h}^{a(6)}_{ij}=0$, which translate into fluid closures (\ref{eq:Thierry70}) \& (\ref{ref:Num1010}).

It is useful to summarize closures for higher-order moments, with details given in Appendix \ref{sec:Hermite}.
It can be shown that for vectors and scalars, fluid closures derived from Hermite closures read
\begin{eqnarray}
\vecX^{(2n+1)}_a &=& \sum_{m=1}^{n-1} (-1)^{m+n+1} \Big(\frac{p_a}{\rho_a}\Big)^{n-m} \frac{n!}{m!(n-m)!} \frac{(2n+3)!!}{(2m+3)!!} \vecX^{(2m+1)}_a;\nn\\  
\widetilde{X}^{(2n)}_a &=& \sum_{m=2}^{n-1} (-1)^{m+n+1} \Big(\frac{p_a}{\rho_a}\Big)^{n-m} \frac{n!}{m!(n-m)!} \frac{(2n+1)!!}{(2m+1)!!} \widetilde{X}^{(2m)}_a,
\label{eq:ClosureN}
\end{eqnarray}
together with closures for stress-tensors
\begin{eqnarray}
  \bPi_{a}^{(2n)} &=& \sum_{m=0}^{n-2} (-1)^{m+n} \Big(\frac{p_a}{\rho_a}\Big)^{n-m-1} \frac{(n-1)!}{m!(n-m-1)!}\frac{(2n+3)!!}{(2m+5)!!} \bPi_{a}^{(2m+2)},
 \label{eq:ClosureN2} 
\end{eqnarray}
where the result is zero if the upper summation index is less than the lower summation index, yielding 
closures $\vecX^{(3)}_a = 0$; $\widetilde{X}^{(4)}_a = 0$ and $\bPi_{a}^{(2)}=0$. The 
closures are summarized bellow in Tables \ref{table:HC1} and \ref{table:HC5}.

\begin{table}[ht!]
\centering
\begin{tabular}{| l | l |}
\hline
Hermite closures         &    Fluid closures \\
\hline    
$\tilde{h}_i^{(3)}=0$;    &    $X^{(3)}_i = 0$;\\
$\tilde{h}^{(4)}=0$;      &    $\widetilde{X}^{(4)} = 0$;\\
$\tilde{h}_i^{(5)}=0$;    &    $X^{(5)}_i = 14 \frac{p}{\rho} X^{(3)}_i$;\\
$\tilde{h}^{(6)}=0$;      &   $\widetilde{X}^{(6)} = 21 \frac{p}{\rho}\widetilde{X}^{(4)}$;\\
$\tilde{h}_i^{(7)}=0$;    &   $X^{(7)}_i = 27 \frac{p}{\rho}X^{(5)}_i-189 \frac{p^2}{\rho^2} X_i^{(3)}$;\\
$\tilde{h}^{(8)}=0$;      &   $\widetilde{X}^{(8)} = 36\frac{p}{\rho} \widetilde{X}^{(6)} -378 \frac{p^2}{\rho^2} \widetilde{X}^{(4)}$;\\  
$\tilde{h}_i^{(9)}=0$;    &   $X^{(9)}_i = 44\frac{p}{\rho}X^{(7)}_i -594\frac{p^2}{\rho^2}X^{(5)}_i+2772 \frac{p^3}{\rho^3} X^{(3)}_i$.\\
\hline
\end{tabular}
\caption{Summary of (MHD) Hermite closures, together with corresponding fluid closures. Species indices ``a'' are dropped. The usual heat flux $q_i=X_i^{(3)}/2$.
  Note that beyond the 4th-order moment both classes start to differ. It can be shown that erroneously prescribing closures at the last retained moment such as
  $X^{(5)}_i=0$ or $\widetilde{X}^{(6)}=0$ leads to unphysical instabilities (unless one prescribes $X^{(3)}_i$ or $\widetilde{X}^{(4)}=0$ as well),
  which is later demonstrated in Appendix \ref{sec:HC-MHD}, Table \ref{table:Herm10}.
  A general form for closures corresponding to $\tilde{h}^{(2n+1)}_i=0$ and $\tilde{h}^{(2n)}=0$ is given by (\ref{eq:ClosureN}). An analogous
  table can be constructed for CGL parallel closures; see Appendix \ref{sec:HC-CGL}, Table \ref{table:Herm2}.}
\label{table:HC1}
\end{table}

\begin{table}[ht!]
\centering
\begin{tabular}{| l | l |}
\hline
Hermite closures         &    Fluid closures \\
\hline
$\hat{h}^{(2)}_{ij}=0$;      &   $\Pi_{ij}^{(2)}=0$;\\
$\hat{h}^{(4)}_{ij}=0$;      &   $\Pi_{ij}^{(4)}=7\frac{p}{\rho}\Pi_{ij}^{(2)}$;\\
$\hat{h}^{(6)}_{ij}=0$;      &   $\Pi_{ij}^{(6)}=18\frac{p}{\rho}\Pi_{ij}^{(4)}-63\frac{p^2}{\rho^2}\Pi_{ij}^{(2)}$;\\
$\hat{h}^{(8)}_{ij}=0$;      &   $\Pi_{ij}^{(8)}=33\frac{p}{\rho}\Pi_{ij}^{(6)}-297\frac{p^2}{\rho^2}\Pi_{ij}^{(4)}+693\frac{p^3}{\rho^3}\Pi_{ij}^{(2)}$;\\  
\hline
\end{tabular}
\caption{Similar to Table \ref{table:HC1}, but for Hermite closures $\hat{h}^{(2n)}_{ij}=0$. A general form for closures corresponding
  to $\hat{h}^{(2n)}_{ij}=0$ is given by (\ref{eq:ClosureN2}). }
\label{table:HC5}
\end{table}

Here we need to address one incorrect interpretation  
that we used in some of our previous papers. In the last paragraph of 
\cite{HunanaPRL2018}, and also in \cite{Hunana2019b,Hunana2019a} it is claimed that
Landau fluid closures are necessary to go beyond the 4th-order moment in the fluid hierarchy. This interpretation was
obtained in the CGL framework for parallel moments by considering closures at the last retained moment $\widetilde{X}^{(2n)}_a=0$ and
$\vecX^{(2n+1)}_{\parallel a}=0$.
It was shown (see detailed proof in Section 12.2 in \cite{Hunana2019a}) that beyond the 4th-order moment,
all fluid models become unstable if these closures are used. The proof is constructed correctly.
What is incorrect is the interpretation, that the proof implies that 
Landau fluid closures are required to overcome this issue. The much simpler Hermite closures overcome this difficulty
as well. 

In another words, beyond the 4th-order moment it is not possible to cut the fluid hierarchy by simply neglecting
the next order moment with closures such as $\vecX^{(5)}_a=0$ or $\widetilde{X}^{(6)}_a=0$, and such closures should be viewed as erroneous.
For the CGL model the closures have different coefficients than for the MHD model because the moments are defined differently 
(a brief summary is given in Appendix \ref{sec:HC-CGL}, Table \ref{table:Herm2}). The CGL closures will be
addressed in detail  in a separate publication.

Importantly, the problem also disappears when one decouples the fluid hierarchy. For example, higher-order
Laguerre (Hermite) schemes that are typically used to obtain more precise transport coefficients for $\vecq_a$ and $\bPi^{(2)}_a$,
neglect all the scalar perturbations $\widetilde{X}^{(4)}_a=\cdots =\widetilde{X}^{(2n)}_a=0$,
together with neglecting coupling between heat fluxes and stress-tensors. 
In our formulation this yields a system
\begin{eqnarray}
  &&  \frac{d_a}{d t}\vecX^{(2n+1)}_a  +\Omega_a \bhat\times \vecX^{(2n+1)}_a 
   +(2n+3)!! \frac{(n)}{3} \frac{p_a^{n}}{\rho_a^{n-1}}\nabla\Big(\frac{p_a}{\rho_a} \Big) \nn\\
  && \qquad = \vec{\boldsymbol{\mathcal{Q}}}^{(2n+1)}_a  -\frac{(2n+3)!!}{3} \frac{p_a^n}{\rho_a^n}\boldsymbol{R}_a; \label{eq:Num15}\\
  && \frac{d_a}{dt} \bPi^{(2n)}_a  +\Omega_a \big( \bhat\times \bPi^{(2n)}_a\big)^S 
   + \frac{(2n+3)!!}{15} \frac{p^n_a}{\rho_a^{n-1}} \bW_a \nn\\
  && \qquad = \bar{\bar{\boldsymbol{\mathcal{Q}}}}^{(2n)}_a -\frac{\bI}{3}Q^{(2n)}_a. \label{eq:Num16}
\end{eqnarray}
Closures (\ref{eq:ClosureN}), (\ref{eq:ClosureN2}) are not required, because the equations are de-coupled. We did not calculate
collisional contributions for higher-order moments, but in the semi-linear approximation equations
(\ref{eq:Num15})-(\ref{eq:Num16}) remain de-coupled and represent two independent hierarchies.   
An essential feature of the Landau (or the Boltzmann) collisional operator is that the operator couples all the heat fluxes together,
and it also couples all the stress-tensors together. Thus by going higher and higher in the fluid
hierarchy does not create new contributions in a quasi-static approximation,
but yields increasingly precise transport coefficients for $\vecq_a$ and $\bPi^{(2)}_a$.
Also, because the momentum exchange rates $\boldsymbol{R}_{a}$ contain contributions from all the heat fluxes
$\vecX^{(3)}_a\ldots \vecX^{(2n+1)}_a$, they become increasingly precise as well. System (\ref{eq:Num15})-(\ref{eq:Num16}) nicely
clarifies how higher-order schemes can be viewed.  Reinstating the coupling between heat fluxes and viscosity-tensors introduces additional
contributions but does not change the transport coefficients of the de-coupled system. A brief comparison of
various models is presented in Appendix \ref{sec:Comparison}. 

\subsection{Inclusion of gravity} \label{sec:gravity}
We have not explicitly considered the force of gravity during our calculations in the Appendix, nevertheless, its inclusion is trivial. 
With the gravitational acceleration $\boldsymbol{G}$ included, the Boltzmann equation reads
\begin{equation}
  \frac{\pr f_a}{\pr t}+ \bV\cdot\nabla f_a + \Big[\boldsymbol{G}+\frac{eZ_a}{m_a}(\bE+\frac{1}{c}\bV\times\bb)\Big]\cdot\nabla_v f_a
  =C(f_a).
\end{equation}
We use big $\boldsymbol{G}$ instead of small $\boldsymbol{g}$ to clearly distinguish it from the heat flux $\boldsymbol{q}$.
Gravity does not enter the collisional operator, and collisional integrals with the right hand side are not effected.
Gravity enters the left hand side, and when the Boltzmann equation is integrated gravity of course enters the fluid hierarchy of moments.
With the two exceptions of the density equation and the pressure tensor equation, gravity enters evolution equations for
all other moments, analogously as the electric field does. An explicit collisionless equation for a general n-th order moment with the
electric field present is for example
equation (12.13) of \cite{Hunana2019a}. Because no Maxwell's equations are used in deriving the fluid hierarchy,
the presence of gravity can be accounted for by simply replacing
\begin{equation}
\frac{eZ_a}{m_a}\bE \to \boldsymbol{G}+\frac{eZ_a}{m_a}\bE.
\end{equation}  
Furthermore, such a hierarchy is not very useful because the evolution equation for an n-th order moment is coupled with
``n'' momentum equations. Subtracting these momentum equations yields final equation (12.16) in \cite{Hunana2019a}, where the electric field is
not present, meaning that gravity is not present either. In other words, the collisionless equation (12.16) of Hunana et al., as well
as our new collisional equation (\ref{eq:GenTensor}) remain valid in the presence of gravitational force.  
The inclusion of gravity to the entire model is thus achieved trivially by adding $-\boldsymbol{G}$ into the left hand side of
the momentum equation (\ref{eq:Energy20x}) (which we have done), and no additional calculations are required. 
In the main text, the only other equation which contains gravity is the electric field equation (\ref{eq:Num90}).

\subsection{Precision of \texorpdfstring{$m_e/m_i$}{me/mi} expansions (unmagnetized proton-electron plasma)} \label{sec:PEexact}
The multi-fluid formulation is also an excellent tool to double-check the precision of $m_e/m_i$ expansions.   
It is possible to again consider a one ion-electron plasma, but this time calculate the transport coefficients precisely,
without any expansions in the smallness of $m_e/m_i$. As an example we consider
an unmagnetized proton-electron plasma ($Z_p=1$, $m_p/m_e=1836.15267$) with  similar  temperatures $T_e=T_p=T_{ep}$.
Charge neutrality implies $n_e=n_p$ and so $p_e=p_p$.
We however maintain $\nabla T_e \neq \nabla T_p$, because the gradients can be different. We first calculate heat fluxes.
For clarity, we are solving 4 coupled evolution equations, which are explicitly given
in Appendix \ref{sec:2species}; see equations (\ref{eq:Num91})-(\ref{eq:Num92}).

Precise calculation should not use simplified collisional times (\ref{eq:Posible1}) where expansions in $m_e/m_i$ have been made,
but exact collisional times (\ref{eq:timeGenM}) with
numerical values $\nu_{ee} = 0.707299 \nu_{ep}$ and $\nu_{pp} = 0.0165063\nu_{ep}$ (we take $\ln\Lambda$ to be constant).
The quasi-static approximation then yields heat fluxes
\begin{eqnarray}
  \vecq_e &=&  \Big[-3.159370 \nabla T_e +  8.301\times 10^{-6}\nabla T_p\Big] \frac{p_e}{m_e\nu_{ep}}  
  +0.711046 p_e \delta\bu;\nn\\
  \vecX^{(5)}_e &=& \Big[-110.5793 \nabla T_e
   +1.376 \times 10^{-3} \nabla T_p \Big] \frac{p_e^2}{\rho_e m_e \nu_{ep}} 
  +18.78249 \frac{p_e^2}{\rho_e}\delta\bu; \nn\\
  \vecq_p &=& \Big[ -3.302411 \nabla T_p +0.2516\times 10^{-3}  \nabla T_e \Big] \frac{p_p}{m_p \nu_{pp}}  
  +0.206535 \times 10^{-4} p_p \delta\bu; \nn \\
  \vecX^{(5)}_p &=&  \Big[ -103.3984 \nabla T_p +  0.7863 \times 10^{-2} \nabla T_e\Big] \frac{p_p^2}{\rho_p m_p \nu_{pp}}
  +0.646475 \times 10^{-3} \frac{p_p^2}{\rho_p} \delta\bu, 
\end{eqnarray}
where $\delta\bu=(\bu_e-\bu_p)$. For the electron heat flux $\vecq_e$, note the difference of the thermal conductivity 3.1594 from the Braginskii value 3.1616.
The difference is caused by calculating the  mass-ratio coefficients (\ref{eq:FinalQ3c}), (\ref{eq:FinalQ5c}) exactly without $m_e/m_p$ expansions, together
with slightly different ratios of frequencies 
(a less-precise calculation with neglecting proton-proton collisions by $\nu_{pp}=0$ and using simplified $\nu_{ee}=\nu_{ep}/\sqrt{2}$ yields 3.1600).

For the proton heat flux $\vecq_p$, the relatively large difference between the thermal conductivity 3.302 and Braginskii self-collisional value $125/32=3.906$ is caused
by the proton-electron collisions. Similarly for the $\vecX^{(5)}_p$, where the self-collisional value is $2975/24=123.96$.
Calculating the coupled system exactly has a nice advantage that one can
calculate the momentum exchange rates in two different ways
\begin{eqnarray}
  \boldsymbol{R}_{e} 
  &=&  \nu_{ep} \Big\{ -\rho_e \delta\bu + \frac{\mu_{ep}}{T_{ep}}\Big[V_{ep (1)} \vecq_e 
  - V_{ep (2)} \frac{\rho_e}{\rho_p} \vecq_p  \Big]  - \frac{3}{56}\Big(\frac{\mu_{ep}}{T_{ep}}\Big)^2
  \Big[ \vecX^{(5)}_e - \frac{\rho_e}{\rho_p} \vecX^{(5)}_p \Big]\Big\};\nn\\ 
  \boldsymbol{R}_{p} 
  &=& \nu_{pe} \Big\{ +\rho_p \delta\bu + \frac{\mu_{ep}}{T_{ep}}\Big[V_{pe (1)} \vecq_p 
  - V_{pe (2)} \frac{\rho_p}{\rho_e} \vecq_e  \Big] - \frac{3}{56}\Big(\frac{\mu_{ep}}{T_{ep}}\Big)^2
  \Big[ \vecX^{(5)}_p - \frac{\rho_p}{\rho_e} \vecX^{(5)}_e \Big]\Big\},
\end{eqnarray}
and both options yield the same result
\begin{equation}
\boldsymbol{R}_e = -\boldsymbol{R}_p =  -0.711046 n_e \nabla T_e -0.2065 \times 10^{-4} n_e \nabla T_p  -0.513306 \rho_e \nu_{ep} \delta\bu.
\end{equation}

Viscosities  of proton-electron plasma are
  (for clarity, we are solving 4 equations in 4 unknowns, explicitly given by (\ref{eq:Posled21x})-(\ref{eq:Energy22xxxx}))
\begin{eqnarray}
  \bPi^{(2)}_e &=& \Big[-0.730622  \bW_e  -0.2800\times 10^{-2}\bW_p\Big]\frac{p_e}{\nu_{ei}};\nn\\
  \bPi^{(4)}_e &=& \Big[-6.542519 \bW_e   +3.1509 \times 10^{-2} \bW_p \Big] \frac{p_e^2}{\rho_e \nu_{ei}};\nn\\
  \bPi^{(2)}_p &=& \Big[-0.892105 \bW_p   -0.4621 \times 10^{-4} \bW_e \Big] \frac{p_p}{\nu_{pp}};\nn\\
  \bPi^{(4)}_p &=& \Big[-7.250870 \bW_p   -0.3759 \times 10^{-3} \bW_e \Big] \frac{p_p^2}{\rho_p \nu_{pp}}, \label{eq:Num777x}
\end{eqnarray}
and for proton species the relatively large differences from self-collisional values $1025/1068=0.960$ and $8435/1068=7.898$ are
again caused by proton-electron collisions. 
 In Appendix \ref{sec:2species}, we consider another examples of coupling between two species, and we calculate heat fluxes and
viscosities for protons \& alpha particles (fully ionized Helium), and for the deuterium-tritium plasma used in plasma fusion. 

\newpage
\subsection{Limitations of our approach} \label{sec:limitations}
 It is important to clarify the limitations of our model. In the highly-collisional regime, our limitations
are the same as for the model of \cite{Braginskii1965}. For example, we describe only Coulomb collisions and we do not take into account ionization \& recombination
and radiative transfer.
 Additionally, our approach shows that coupling of stress-tensors and heat fluxes should be ideally investigated with the 22-moment model.
  Even though this model is fully formulated in Section \ref{sec:22main}, including its 
  collisional contributions calculated  with the Landau operator, we did not use this model to further explore the resulting coupling.
  
\subsubsection{Weakly collisional regime: expansions around bi-Maxwellians}  
  The situation becomes more complicated in the weakly-collisional regime 
  where there might not be enough collisions to keep the distribution function sufficiently close to the equilibrium Maxwellian $f_a^{(0)}$.
  The distribution function might evolve
  to such an extent that the core assumptions in the entire derivation break down, i.e. equation (\ref{eq:One}) looses its validity.
  A better approach is then to consider expansions similar to equation (\ref{eq:One}), but  around a bi-Maxwellian $f_a^{(0)}$ (see e.g. \cite{Oraevskii1968,ChoduraPohl1971,DemarsSchunk1979,BarakatSchunk1982}, and references therein),
which can handle much larger departures from the highly-collisional Maxwellian distribution.
In order to point out the differences and difficulties associated with this approach, it is of interest to briefly describe
how expansions around an anisotropic bi-Maxwellian would look like.
The simplest anisotropic model is known as the CGL, after the pioneering work of 
Chew, Goldberger and Low \citep{Chew1956}. The difference with our current approach starts with
the decomposition of the pressure tensor $p_{ij}^a$ defined in (\ref{eq:defTensor}), and the decomposition reads
\begin{eqnarray}
\textrm{isotropic:}  \qquad   \bp_a &=& p_a \bI+\bPi_a^{(2)}; \label{eq:Pi1}\\
\textrm{anisotropic:} \qquad  \bp_a &=& p_{\parallel a}\bhat\bhat + p_{\perp a}(\bI-\bhat\bhat)+\bPi_a^{(2)\textrm{CGL}} \label{eq:Pi2}\\
 &=& p_a \bI +(p_{\parallel a}-p_{\perp a}) \Big( \bhat\bhat-\frac{\bI}{3}\Big) +\bPi_a^{(2)\textrm{CGL}},\nn
\end{eqnarray}
with scalar pressures
\begin{eqnarray}
  p_{\parallel a} &=&\bp_a:\bhat\bhat = m_a \int c_{\parallel a}^2 f_a d^3v; \qquad
  p_{\perp a} = \bp_a :\bI_\perp/2 = \frac{m_a}{2}\int |\bc_{\perp a}|^2 f_a d^3v. \label{eq:p_proj}
\end{eqnarray}
Directly from the above definitions, the stress-tensors satisfy
\begin{equation}
\textrm{Tr}\bPi^{(2)}_a = \textrm{Tr}\bPi_a^{(2)\textrm{CGL}}=0; \qquad \bPi^{(2)}_a:\bhat\bhat\neq 0; \qquad \bPi_a^{(2)\textrm{CGL}}:\bhat\bhat=0, \label{eq:Pi-properties}
\end{equation}   
 and while  $\bPi_a^{(2)}$ has 5 independent components,  $\bPi_a^{(2)\textrm{CGL}}$ has only 4. 

The decomposition of the heat flux tensor $q_{ijk}^a$ defined by equation (\ref{eq:defTensor}) is slightly more complicated. 
In an arbitrary-collisional regime one needs to define two heat flux vectors 
\begin{equation}
  \boldsymbol{S}^\parallel_a = \bq_a:\bhat\bhat = m_a \int c_{\parallel a}^2 \bc_a f_a d^3v; \qquad
  \boldsymbol{S}^\perp_a = \bq_a:\bI_\perp/2 = \frac{m_a}{2}\int |\bc_{\perp a}|^2\bc_a f_a d^3v. \label{eq:DefS}
\end{equation}
These heat flux vectors are further split by projecting them along the $\bhat$, which defines the \emph{gyrotropic} (scalar) heat fluxes $q_{\parallel a}$ \& $q_{\perp a}$,
and the perpendicular projection defines the \emph{non-gyrotropic heat flux vectors} $\boldsymbol{S}^\parallel_{\perp a}$ \& $\boldsymbol{S}^\perp_{\perp a}$,
according to
\begin{eqnarray}
  q_{\parallel a} &=& \bhat\cdot\boldsymbol{S}^\parallel_a = m_a \int c_{\parallel a}^2 c_{\parallel a} f_a d^3v;\qquad
  q_{\perp a} = \bhat\cdot\boldsymbol{S}^\perp_a = \frac{m_a}{2}\int |\bc_{\perp a}|^2 c_{\parallel a} f_a d^3v; \nn\\
  \boldsymbol{S}^\parallel_{\perp a} &=& \bI_\perp\cdot\boldsymbol{S}^\parallel_a = m_a \int c_{\parallel a}^2 \bc_{\perp a} f_a d^3v;
  \qquad \boldsymbol{S}^\perp_{\perp a}=\bI_\perp\cdot\boldsymbol{S}^\perp_a = \frac{m_a}{2}\int |\bc_{\perp a}|^2\bc_{\perp a} f_a d^3v.
\end{eqnarray}
The following relations then hold $\boldsymbol{S}^\parallel_a = q_{\parallel a}\bhat+\boldsymbol{S}^\parallel_{\perp a}$; \&  
$\boldsymbol{S}^\perp_a = q_{\perp a}\bhat+\boldsymbol{S}^\perp_{\perp a}$; together with 
$\bhat\cdot\boldsymbol{S}^{\parallel}_{\perp a}=0$; \& $\bhat\cdot\boldsymbol{S}^{\perp}_{\perp a}=0$.
The two different decompositions of the entire heat flux tensor then read
\begin{eqnarray}
\textrm{isotropic:} \qquad \bq_a &=& \frac{2}{5}\Big[ \vecq_a \bI\Big]^S + \bar{\bar{\boldsymbol{\sigma}}}_a'; \label{eq:Hiso} \\ 
\textrm{anisotropic:}\qquad  \bq_a &=& q_{\parallel a} \bhat\bhat\bhat + q_{\perp a} \Big[ \bhat\bI_\perp\Big]^S
  +\Big[ \boldsymbol{S}^\parallel_{\perp a} \bhat\bhat  \Big]^S + \frac{1}{2}\Big[ \boldsymbol{S}^\perp_{\perp a} \bI_\perp\Big]^S
  +\bar{\bar{\boldsymbol{\sigma}}}_a; \label{eq:HF_decomp}\nn\\
  &=& q_{\perp a}\big[ \bhat\bI\big]^S + (q_{\parallel a}-3q_{\perp a})\bhat\bhat\bhat
  +\frac{1}{2}\Big[\boldsymbol{S}^\perp_{\perp a} \bI\Big]^S
  +\Big[ \big(\boldsymbol{S}^\parallel_{\perp a} -\frac{\boldsymbol{S}^\perp_{\perp a}}{2}\big)\bhat\bhat\Big]^S+\bar{\bar{\boldsymbol{\sigma}}}_a, \label{eq:best}
\end{eqnarray}
where both $\bar{\bar{\boldsymbol{\sigma}}}_a'$ and $\bar{\bar{\boldsymbol{\sigma}}}_a$ are traceless.
Neglecting these traceless contributions, the isotropic approach accounts for 3 (out of 10) scalar components of $\bq_a$
 and represents a 13-moment model
(1 density, 3 velocity, 1 scalar pressure, 5 stress-tensor components, 3 heat flux $\vecq_a$ components).
The anisotropic approach accounts for 6 scalar components of $\bq_a$ and represents a 16-moment model,
 described by sixteen scalar evolution equations
(1 density, 3 velocity, 2 scalar pressures, 4 stress-tensor components,
  3 for each heat flux vector $\boldsymbol{S}^\parallel_a$ \& $\boldsymbol{S}^\perp_a$.) 
Unfortunately, such a complicated decomposition of the heat flux is necessary 
in an arbitrary-collisional regime, and we only used decomposition (\ref{eq:Hiso}).
For clarity, direct relation with the usual heat flux vector $\vecq_a$  reads
\begin{equation} 
  \vec{\boldsymbol{q}}_a = \frac{1}{2}\boldsymbol{S}^{\parallel}_a+\boldsymbol{S}^\perp_a = 
  \Big( \frac{1}{2}q_{\parallel a} +q_{\perp a} \Big)\bhat +\frac{1}{2} \boldsymbol{S}^{\parallel}_{\perp a}+\boldsymbol{S}^{\perp}_{\perp a}.
\end{equation}
Note that both $q_{\parallel a}$ \& $q_{\perp a}$ denote components along the $\bhat$. The highly-collisional limit is achieved by $q_{\parallel a}=3q_{\perp a}$ and
$\boldsymbol{S}^\parallel_{\perp a}=\boldsymbol{S}^\perp_{\perp a}/2$, in which case $\vecq_a=(5/2)q_{\perp a}\bhat+(5/4)\boldsymbol{S}^\perp_{\perp a}$ or
equivalently $\vecq_a=(5/6)q_{\parallel a}\bhat+(5/2)\boldsymbol{S}^\parallel_{\perp a}$. We used the same notation as for example collisionless papers by
\cite{PassotSulem2007,SulemPassot2015,Hunana2019b,Hunana2019a}.

These anisotropic decompositions must be retained in an arbitrary-collisional regime.
However, calculations with the Landau (Boltzmann) collisional operators then become very complicated. 
Notably, \cite{ChoduraPohl1971,DemarsSchunk1979,BarakatSchunk1982} used the anisotropic 16-moment model as described above
and calculated the collisional contributions for several interaction potentials.
Judging from the papers above,
maintaining the precision of our current model (where the 4th \& 5th order moments are considered) and extending it to an anisotropic (bi-Maxwellian) regime
seems to be so complicated, that it might not be worth the effort. 
 Curiously, in a simplified spherically symmetric radial geometry \cite{Cuperman1980,Cuperman1981,CupermanDryer1985}
  considered what seems like a mixture of anisotropic and isotropic moments, with anisotropic temperatures, isotropic heat flux vector,
  and the parallel (anisotropic) perturbation of the 4th-order moment (which we call $\widetilde{r}_{\parallel\parallel a}$).

\subsubsection{Landau fluid closures for the collisionless case}
\begin{figure*}[!htpb]
  $$\includegraphics[width=0.49\linewidth]{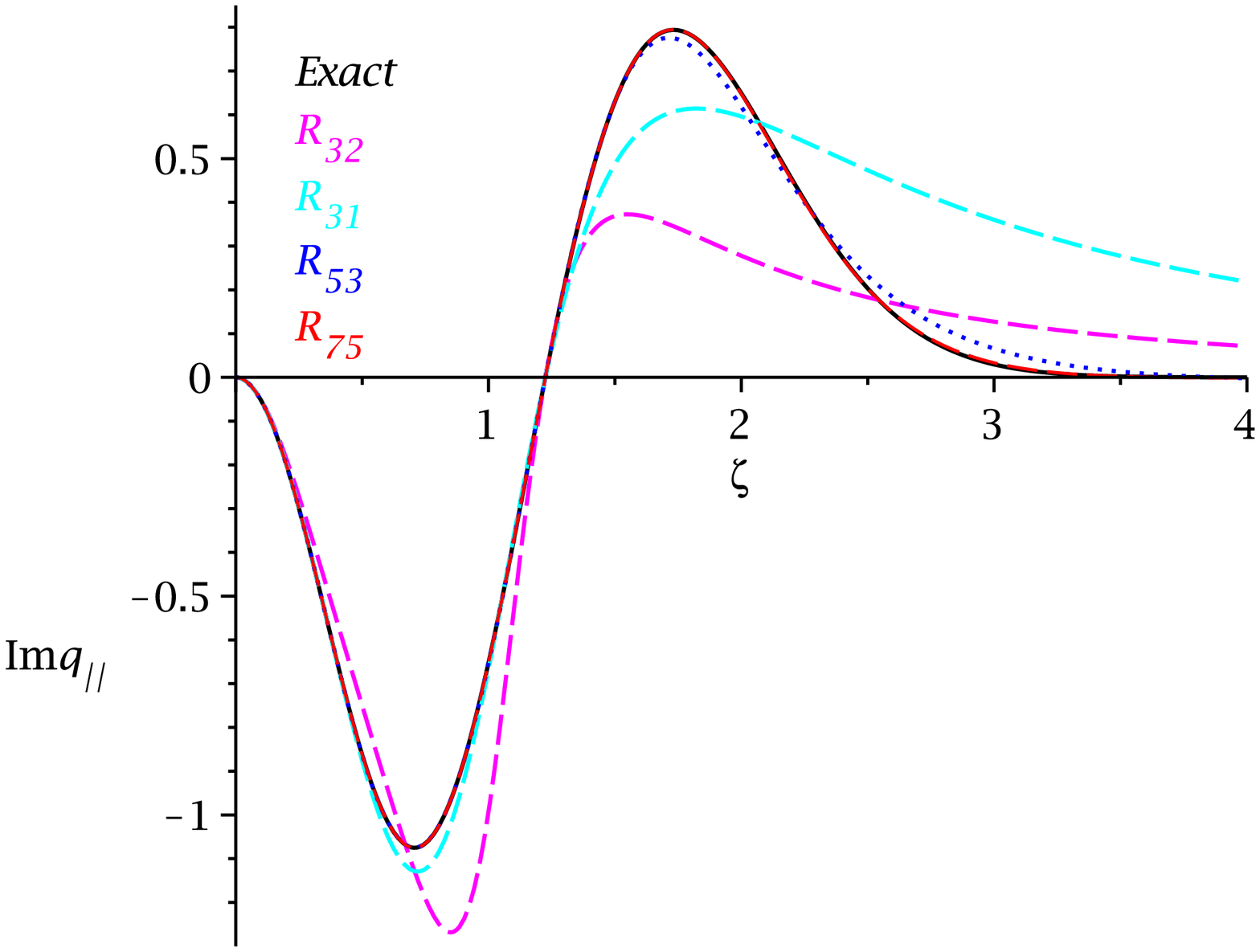}\hspace{0.02\textwidth}\includegraphics[width=0.49\linewidth]{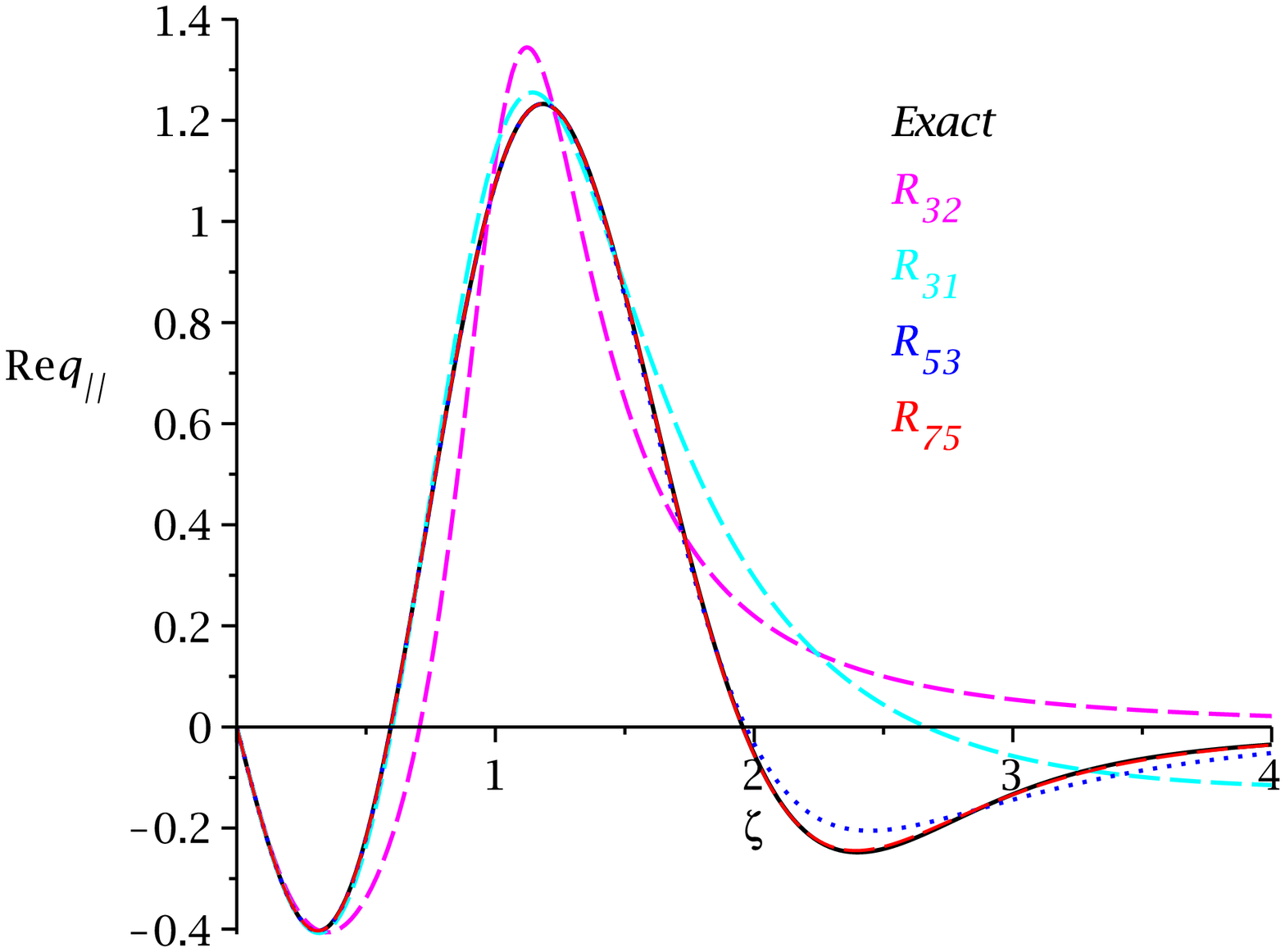}$$
  \caption{Comparison of normalized collisionless heat fluxes $\hat{q}_{\parallel a} = \zeta_a+(2\zeta_a^3-3\zeta_a) R(\zeta_a)$ in a weakly-damped regime
    with real valued $\zeta_a$. Left panel: imaginary part of $\hat{q}_{\parallel a}$. Right panel: real part of $\hat{q}_{\parallel a}$. 
    Colors are described in the text. Our Braginskii-type models do not contain these
  collisionless heat fluxes.}
  \label{fig:HeatFlux}
\end{figure*}

In contrast to the free-streaming formula of \cite{Hollweg1974,Hollweg1976}, 
in plasma physics the collisionless heat flux is typically associated with the phenomenon of Landau damping.
For example, collisionless linear kinetic theory expanded around bi-Maxwellian with  mean \emph{zero drifts} in gyrotropic limit yields
in Fourier space  perturbation of the distribution function $f_a=f_a^{(0)}(1+\chi_a)$ in the following form
\begin{equation}
  \chi_a =  
  \frac{B_\parallel^{(1)}}{B_0} \frac{m_a}{2 T_{\perp a}^{(0)}}\bigg[ v_\perp^2  + \frac{T_{\perp a}^{(0)}}{T_{\parallel a}^{(0)}}\frac{\kpar \vpar \vperp^2}{(\omega-\kpar\vpar)} \bigg]
  +\Phi\frac{eZ_a}{T_{\parallel a}^{(0)}} \frac{\kpar\vpar}{(\omega-\kpar\vpar)}, \label{eq:Thierry456}
\end{equation}
 with the electrostatic potential $\Phi=iE_\parallel^{(1)}/k_\parallel$. Integrating (\ref{eq:Thierry456}) then yields a parallel collisionless heat flux
\begin{equation}
q_{\parallel a}^{(1)} = - v_{\textrm{th}\parallel a} n_a^{(0)} T_{\parallel a}^{(0)} \sign(\kpar)\Big( \zeta_a+2\zeta_a^3 R(\zeta_a) -3\zeta_a R(\zeta_a)\Big)
\Big[ \frac{B_\parallel^{(1)}}{B_0}\frac{T_{\perp a}^{(0)}}{T_{\parallel a}^{(0)}} + \Phi\frac{e Z_a}{T_{\parallel a}^{(0)}} \Big], \label{eq:HFlux}
\end{equation}
with variable $\zeta_a=\omega/(|k_\parallel|v_{\textrm{th}\parallel a})$;  parallel thermal speed $v_{\textrm{th}\parallel a}=\sqrt{2T_{\parallel a}/m_a}$,
 plasma response function $R(\zeta_a)=1+\zeta_aZ(\zeta_a)$ and
plasma dispersion function $Z(\zeta_a) =  i\sqrt{\pi} \exp(-\zeta_a^2) [ 1+\textrm{erf}(i\zeta_a) ]$. Such a kinetic answer can be expressed in fluid variables
by searching for Landau fluid closures,  for example by replacing the $R(\zeta_a)$ with its 3-pole Pad\'e approximants 
\begin{eqnarray}
  R_{3,2}(\zeta_a) &=& \frac{1-i\frac{\sqrt{\pi}}{2}\zeta_a}{1-i\frac{3\sqrt{\pi}}{2}\zeta_a-2\zeta_a^2+i\sqrt{\pi} \zeta_a^3};\qquad  
  R_{3,1}(\zeta_a) = \frac{1-i\frac{(4-\pi)}{\sqrt{\pi}}\zeta_a}{1-i\frac{4}{\sqrt{\pi}}\zeta_a-2\zeta_a^2+2i\frac{(4-\pi)}{\sqrt{\pi}}\zeta_a^3}.
\end{eqnarray}
The use of Pad\'e approximants allows one to express (\ref{eq:HFlux}) through lower-order moments and eliminate the explicit dependence on $\zeta_a$,
yielding collisionless heat fluxes in Fourier space
\begin{eqnarray}
  R_{3,2}(\zeta_a): \qquad q_{\parallel a}^{(1)} &=& - i \frac{2}{\sqrt{\pi}} n_{a}^{(0)} v_{\textrm{th}\parallel a} \sign(\kpar) T_{\parallel a}^{(1)}; \label{eq:Thierry450}\\
  R_{3,1}(\zeta_a): \qquad q_{\parallel a}^{(1)} &=& \frac{3\pi-8}{4-\pi} \,p^{(0)}_{\parallel a} u^{(1)}_{\parallel a}
  -i\frac{\sqrt{\pi}}{4-\pi}n^{(0)}_a v_{\textrm{th}\parallel a}\sign(\kpar) T^{(1)}_{\parallel a},\label{eq:Thierry451}
\end{eqnarray}  
 where $T^{(1)}_{\parallel a}$ is perturbed temperature, and $u^{(1)}_{\parallel a}$ is perturbed fluid velocity (mean value $u^{(0)}_{\parallel a}=0$ is assumed).
The heat flux closure (\ref{eq:Thierry450}) was obtained by \cite{HammettPerkins1990} \& \cite{Snyder1997} and closure (\ref{eq:Thierry451}) is equation (2) in
\cite{HunanaPRL2018} (or equation (3.211) in \cite{Hunana2019b}). 
In real space these collisionless heat fluxes become
\begin{eqnarray}
 R_{3,2}(\zeta_a): \qquad  q_{\parallel a} (z) &=& -\frac{2}{\pi^{3/2}} n_a^{(0)} v_{\textrm{th} \parallel a} \textrm{V.P.}
  \int_{0}^\infty \frac{T^{(1)}_{\parallel a}(z+z')-T^{(1)}_{\parallel a}(z-z')}{z'}dz'; \label{eq:Thierry450x}\\
R_{3,1}(\zeta_a): \qquad  q_{\parallel a}(z) &=& \frac{3\pi-8}{4-\pi}p^{(0)}_{\parallel a} u^{(1)}_{\parallel a}
  -\frac{n^{(0)}_a v_{\textrm{th} \parallel a}}{\sqrt{\pi}(4-\pi)}\textrm{V.P.}
  \int_{0}^\infty \frac{T^{(1)}_{\parallel a}(z+z')-T^{(1)}_{\parallel a}(z-z')}{z'}dz',\label{eq:Thierry451x}
\end{eqnarray}
where the non-locality presents itself as an integral over the entire magnetic field line, where temperatures everywhere along that field line
matter to determine the heat flux at a specific spatial point.  Note that the thermal part of (\ref{eq:Thierry451x}) is almost two times larger then
 (\ref{eq:Thierry450x}). The Cauchy principal value can be replaced by $\lim_{\epsilon\to +0}\int_\epsilon^\infty$. 
This approach thus indeed allows one to have expressions for  collisionless heat fluxes in a quasi-static approximation. However, as is well-known
these expressions are not very precise with respect to kinetic theory.
 For example, the precision can be easily compared by plotting normalized heat fluxes $\hat{q}_{\parallel a} = \zeta_a+2\zeta_a^3 R(\zeta_a) -3\zeta_a R(\zeta_a)$,
  which is shown in Figure \ref{fig:HeatFlux}. Weakly-damped regime with real valued $\zeta_a$ is considered.
  The left panel shows the imaginary part of $\hat{q}_{\parallel a}$ and the right panel
  shows the real part of $\hat{q}_{\parallel a}$.  
 Exact kinetic heat flux is solid black line, heat flux $R_{3,2}$ is dashed magenta line and heat flux $R_{3,1}$ is dashed cyan line.
 For comparison, higher-order fluid models with approximants $R_{5,3}$ (dotted blue line) and $R_{7,5}$ (dashed red line) 
 are shown as well (see equation (A11) \& (A38) in \cite{Hunana2019b}). 
 The $R_{5,3}$ model represents a dynamic closure at the 4th-order moment and the $R_{7,5}$ model
 represents a dynamic closure at the 6th-order moment, given by equation (5) \& (8) of Hunana et al. 2018. The heat fluxes in these higher-order models are thus described
 by their usual evolution equations, nevertheless, their precision can be compared with the same technique. 
 Which quasi-static heat flux is a better choice depends on the value of $\zeta_a$, because 
 the $R_{3,1}$ has a higher power-series precision (for small $\zeta_a$) and the $R_{3,2}$ has a higher asymptotic-series precision (for large $\zeta_a$).
 Regime $\zeta_a\ll 1$ can be viewed as isothermal and regime $\zeta_a\gg 1$ can be viewed as adiabatic.
 In the left panel of Figure \ref{fig:HeatFlux} the $R_{3,1}$ is more precise up to roughly $\zeta_a=2.3$ and in the right panel up to $\zeta_a=1.6$. For larger $\zeta_a$
 values than shown, the $R_{3,1}$ heat flux converges much slower to the correct zero values than the $R_{3,2}$, especially for the real part.

 The major obstacle in precision for the quasi-static heat fluxes of Landau fluid models actually comes from the perpendicular heat flux $q_{\perp a}$ (which is along the $\bhat$), because only a closure of \cite{Snyder1997} with a crude Pad\'e approximant $R_1(\zeta_a) = 1/(1-i\sqrt{\pi}\zeta_a)$ is available.
As a consequence, for large $\zeta_a$ values the quasi-static heat flux $q_{\perp a}$ fails to disappear and instead converges to a constant value. To recover the adiabatic behavior for $q_{\perp a}$, one has to abandon the idea of quasi-static $q_{\perp a}$
and consider its evolution equation, with a closure at the 4th-order moment.
There is a vast amount of literature about Landau fluids with various approaches; see e. g.
\cite{HammettPerkins1990,Hammett1992,Snyder1997,Snyder2001,Goswami2005,PassotSulem2007,PSH2012,
  SulemPassot2015,JosephDimits2016,HunanaPRL2018,JiJoseph2018,Chen2019,Wang2019} and references therein, where some authors also include
collisional effects. For a simple introductory guide to collisionless Landau fluids see \cite{Hunana2019b}.
  As a side note, Landau fluid closures are not constructed with any specific mode in mind
(as incorrectly criticized for example by \cite{Scudder2021}).
The closures are constructed universally for all the modes, so that
numerical simulations can be performed; see e.g. \cite{PerronePassot2018}.
Interestingly, as discussed by \cite{Meyrand2019}, from a non-linear perspective the effect of Landau damping might be canceled
out by the effect of plasma echo. From a linear perspective, the presence of drifts also modifies the Landau damping, because the variable $\zeta_a$ which enters the
plasma response function $R(\zeta_a)$ then contains the drift velocity $u_{\parallel a}$. For sufficiently large drifts the sound mode
can be generated by the current-driven ion-acoustic instability; see e.g. \cite{GurnettBhattacharjee2005} p. 368, or \cite{Fitzpatrick} p. 258;
and for a 3-component plasma which allows the net current to be zero by the ion-ion
  (or the electron-ion \& electron-electron) acoustic instability, see \cite{Gary1993} p. 44-55.

\subsubsection{Ion-sound wave damping in homogeneous media: comparison of various models}\label{sec:ion-sound}

\begin{figure*}[!htpb]
	$$\includegraphics[width=0.55\linewidth]{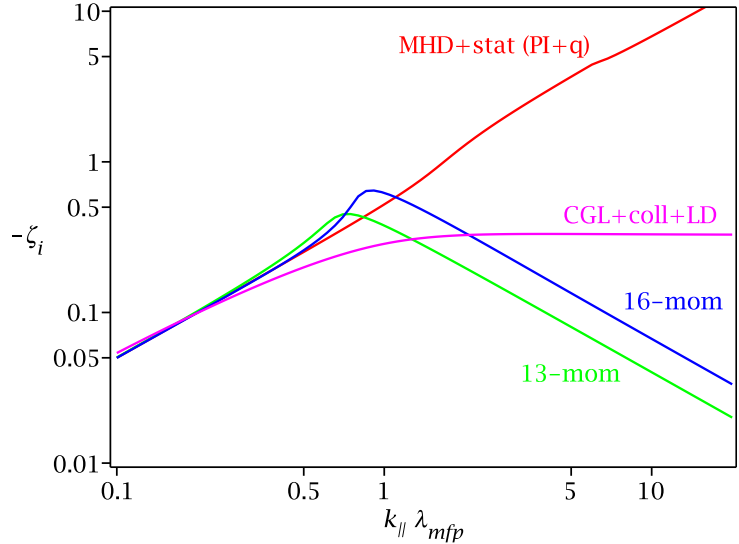}$$
  \caption{Normalized damping rate $\zeta_i=\omega_i/(|k_\parallel| v_{\textrm{th}\parallel})$ for a parallel propagating ion-sound wave as
    a function of $k_\parallel \lambda_{\textrm{mfp}}$, where $\lambda_{\textrm{mfp}}$ is a mean-free-path. Red line: Braginskii-type (isotropic) 13-moment model
    with quasi-static stress-tensor and heat flux; Green line: Braginskii-type 13-moment model with evolution equations for the stress-tensor and heat flux;
    Blue line: bi-Maxwellian 16-moment model with evolution equations for parallel and perpendicular pressures and (gyrotropic) heat fluxes;
    Magenta line: Landau fluid model with quasi-static heat fluxes of \cite{Snyder1997}. }
	\label{fig:Sound2}
\end{figure*}

 To further clarify our limitations, it is useful to explore the linear properties of  waves propagating along the ambient
  magnetic field (assumed to be straight and aligned with the $z$-coordinate) in a homogeneous medium, in regimes that range from
  the highly-collisional to the weakly-collisional ones.
  Let us in particular consider the damping of a monochromatic ion-sound wave of parallel wavenumber $k_\|$ in a proton-electron plasma where
  the electrons are assumed to be cold. The latter assumption is not physically appropriate
  because kinetic theory is not well-defined for cold electrons (see e.g. discussion in \cite{Hunana2019b} p. 73)
  but it allows one to simplify the presentation with a goal to describe the general behavior
  and not to provide precise values of the damping rates.  Four different models are compared in Figure \ref{fig:Sound2}, all using the heuristic
  BGK collisional operator,
  which leads to much simpler calculations for  models with a distribution function expanded around a bi-Maxwellian.
 The x-axis shows $k_\| \lambda_{\textrm{mfp}}$ where $\lambda_{\textrm{mfp}}= v_{{\textrm{th}} \|}/\nu$ is the ion mean-free-path
    and $\nu$ is the collisional frequency, so that $k_\| \lambda_{\textrm{mfp}}\ll 1$ represents a highly-collisional regime and $k_\| \lambda_{\textrm{mfp}}\gg 1$
    represents a weakly-collisional regime.
  The y-axis shows a damping rate as an imaginary part of $\zeta=\omega/(|k_\parallel| v_{\textrm{th}\parallel})$.
    The usual isotropic 13-moment model (green line) and the anisotropic 16-moment model (blue line) with all the moments described by their time-dependent (dynamical) evolution equations
    were discussed after equation (\ref{eq:best}). For the parallel sound mode at the linear level considered here, the 13-moment model is reduced to evolution equations
    for $\rho,u_z,p,\Pi_{zz},q_z$ (we consider the case where $\Pi_{zz}$ and $q_z$ are coupled) and
    the 16-moment model reduces to evolution equations for $\rho,u_z,p_\parallel,p_\perp,q_\parallel,q_\perp$ (we consider mean equal pressures $p_\parallel^{(0)}=p_\perp^{(0)}$).
    Figure \ref{fig:Sound2} shows that these two models behave in a similar way:
    both reach a maximum damping rate around $k_\| \lambda_{\textrm{mfp}}\sim 0.5-1$ 
    and converge towards zero damping rate in the collisionless regime (with only a small shift in $k_\| \lambda_{\textrm{mfp}}$ between them).
    In contrast, the red line, corresponding to the 13-moment model with the $\Pi_{zz}$ \& $q_z$ taken in the quasi-static approximation,
    shows that  the damping rate does not reach a
    maximum and instead continues to increase in a weakly-collisional regime,
    and around $k_\| \lambda_{\textrm{mfp}}\sim 6.3$ the sound mode stops existing (it becomes non-propagating with zero real frequency).
    This is consequence of the quasi-static approximation for the stress-tensor $\Pi_{zz}\sim 1/\nu$,
    which in the collisionless regime becomes unbounded (the parallel heat flux
  $q_z\sim 1/\nu$ becomes unbounded as well, but this simply reflects an isothermal behavior with no damping present).
  While a vanishing damping is preferred against a quantity that blows up in a weakly-collisional regime,
  all three models are technically incorrect, because Landau damping provides a significant contribution for the damping rate
 as the plasma becomes weakly collisional. 
  To illustrate  the importance of Landau damping, we display by a magenta line the damping rate obtained with a Landau fluid
    model which contains evolution equations for $\rho,u_z,p_\parallel,p_\perp$, but where the quasi-static $q_\parallel$ \& $q_\perp$ are given by the collisionaly modified
    3+1 closures of \cite{Snyder1997}, i.e. their  equations (48)-(49),
    which for isotropic mean temperatures $T_{\parallel a}^{(0)}=T_{\perp a}^{(0)}$ considered here, are equivalent to
  (our thermal speed contains a factor of 2, which is not the case in that paper)
\begin{eqnarray}
 q_{\parallel a}^{(1)} &=& -\,\frac{\frac{4}{3\pi-8}n_a^{(0)} v_{\textrm{th}\parallel a}^2 }{\bar{\nu}_a +\frac{2\sqrt{\pi}}{3\pi-8}v_{\textrm{th}\parallel a} |\kpar|}
  i\kpar T_{\parallel a}^{(1)};\qquad
  q_{\perp a}^{(1)} = - \,\frac{\frac{1}{2}n_{a}^{(0)} v_{\textrm{th}\parallel a}^2}{\bar{\nu}_a+\frac{\sqrt{\pi}}{2}v_{\textrm{th}\parallel a}|\kpar|} i\kpar
    T_{\perp a}^{(1)}, \label{eq:Snyder1}
\end{eqnarray}
where in general $\bar{\nu}_a=\sum_b \nu_{ab}$. Technically, closures (\ref{eq:Snyder1}) are only applicable to a weakly-collisional regime because
$q_{\parallel a}\neq 3 q_{\perp a}$ in the highly-collisional limit. 
  In spite of this and the additional difficulty associated with the cold electron limit considered here, an interesting point is that the behavior
of the damping rate is very close to the prediction of the three other models in the highly-collisional regime, while the damping rate
converges to a constant value
  in the collisionless case. This is in fact analogous to the case of the damping of a pure sound wave in rarefied media,
  which was considered by \cite{Stubbe1994} and \cite{Stubbe1999}.
  In the former paper, the result of an experiment by \cite{MeyerSessler1957}
  (measuring the damping length of a sound wave of a given frequency $\omega$, emitted at one end of the domain filled with a rarefied neutral gas)
  are compared with various theoretical models. The results are very similar to those presented here, and show in particular that the damping is
  dominated by a nonlocal effect analogous to Landau damping when $2\nu/\omega $ decreases below unity (see Fig. 6 and 7 of \cite{Stubbe1994}).
  This simple result on the damping of an ion-sound wave shows that, in a homogeneous medium, a Braginskii-type model provides reasonable predictions
  as long as the typical wavelength is larger than the mean free path, or equivalently, when its frequency stays below the collision frequency.
  More sophisticated models are needed in the weakly collisional case, that should retain new contributions originating from a Landau-fluid closure.

\subsubsection{Large gradients and large drifts}
It is now of interest to consider inhomogeneous situations, where other applicability conditions apply for the Braginskii-type models.
In high energy-density laser-produced plasmas,
there are often situations relevant for inertial confinement fusion experiments, where the typical electron mean-free-path
becomes of the order of the typical scale of electron temperature gradients, or even larger. In this case the usual Braginskii formulas,
used for example for the Nernst effect (see e.g. \cite{Lancia2014}) become invalid and have  to be replaced by nonlocal expressions.
In this context an explicit nonlocal formula was proposed by \cite{Luciani1983} for the electron thermal heat flux due to steep temperature gradients,
offering an improvement (in the one-dimensional case) for the Spitzer-Braginskii heat flux,
where one required proportionality constant is obtained by fitting from Fokker-Planck simulations.
Further extension to three dimensions was proposed by \cite{Schurtz2000}, but it is to be noted that this approach is not appropriate
in the very weakly collisional case, as for example in the Solar corona when the density has significantly decreased.

Additional complications arise in a regime of weak collisionality.
In space physics, the collisionless heat flux
is typically associated with the free-streaming formula of \cite{Hollweg1974,Hollweg1976}
\begin{equation} \label{eq:Hollweg}
	\vecq_e^{\textrm{Hollweg}} = \frac{3}{2}p_e \bu_{sw} \alpha,
\end{equation} 
where one multiplies the thermal energy of one electron $(3/2)T_e$ (we take $k_b=1$ in the entire paper) with the number density $n_e$ and the solar wind
speed $\bu_{sw}$. The free ``bugger factor'' $\alpha$ as \cite{Hollweg1974} calls it,
is dependent on a given form of an electron distribution function where the tail had departed and run away. Note that the parallel \emph{frictional} heat fluxes
(i.e. due to \emph{small} differences in drifts $\delta\bu$) of \cite{SpitzerHarm1953} and \cite{Braginskii1965} are also independent of collisional frequencies,
even though derived from collisions, and up to the numerical values have the same form as (\ref{eq:Hollweg}).
As a side note, in the numerical model of \cite{SpitzerHarm1953} the frictional heat flux
is technically incorrect because it does not satisfy the Onsager symmetry; see our Tables \ref{eq:TablePar2} \& \ref{eq:TableTr}, which was already criticized
by \cite{Balescu1988} (p. 268).
 Of course, in our usual fluid formalism a tail of a distribution function can not suddenly depart. Even though our model contains
	evolution equations for the perturbation of the 4th-order moment (i.e. a ``reduced kurtosis'' which describes if a distribution is tail-heavy or tail-light)
	and also for the 5th-order moment (sometimes called a hyper-skewness), our distribution functions still have to remain sufficiently close to Maxwellian.   
	For the isotropic 5-moment model (i.e. strict Maxwellians),
	the runaway effect is just represented through collisional contributions $\boldsymbol{R}_{ab}$ and $Q_{ab}$
	which decrease to zero for large drifts
	(see equations (\ref{eq:Energy50})-(\ref{eq:Energy51m}) derived in Appendix \ref{sec:runaway};
	see also \cite{Dreicer1959,Tanenbaum1967,Burgers1969,Schunk1977,Balescu1988}). We note that for sufficiently large drifts between species various
	instabilities can develop with a subsequent development of turbulence, which should restrict the runaway effect long before relativistic effects. 
	Importantly, it is unclear how the heat flux collisional contributions $\vecQ^{(3)}_{ab}\,'$ (and higher) would look like for unrestricted drifts,
	because the collisional integrals seem exceedingly complicated. Even if calculated, only the drifts between species would be 
	allowed to be unrestricted, the distribution of each species will have to be restricted to remain close to Maxwellian.  
For the simplest CGL plasmas (i.e. considering colliding strict bi-Maxwellians with no stress-tensors or heat fluxes), the corresponding
collisional integrals where numerically evaluated for selected cases by \cite{BarakatSchunk1981}.
For a further particular case of unrestricted drifts only \emph{along} the magnetic field and of Coulomb collisions, \cite{Hellinger2009} obtained exact analytic
forms for the collisional integrals (for $p_\parallel$ \& $p_\perp$),
which are however expressed through a double hypergeometric functions. Judging from the two papers above, a proper extension of our model 
to an anisotropic regime with unrestricted drifts seems overly complicated.
 Other approach for the heat flux modeling was presented by \cite{Canullo1996}.

\subsubsection{Comments on the positivity of the perturbed distribution function}
 An additional complication arises in a low-collisionality regime in the presence of sufficiently strong large-scale gradients.
 Perturbations of the distribution function considered in equation (\ref{eq:One})  might become so large, that the corresponding model might become invalid.
 The distribution function around which to expand is indeed not well-defined in this case.
Strictly speaking, in a weakly-collisional
(or a collisionless) regime, one should abandon the construction of fluid models derived from the Boltzmann equation, and perform kinetic simulations by directly evolving the Boltzmann equation.
 Perhaps the best example is a radially expanding flow, such as the solar corona with emerging solar wind, where the spherical expansion
creates strong large-scale gradients and simultaneously drives the system towards a collisionless regime.
It seems that in this extreme case it might be indeed possible (but not with certainty) that the underlying distribution function can even become negative, $f_a<0$,
which is of course unphysical. We expect that our 21 \& 22-moment models might fail in this particular situation, 
even if evolution equations are retained, but as discussed below, we were unfortunately not able to reach clear conclusion and
further research is needed to clearly establish the areas of validity.

The $f_a<0$ was criticized for example by \cite{Scudder2021,Cranmer2021} (and references therein), on an example of 8-moment models in
a quasi-static approximation. It is in fact questionable if the $f_a<0$ can be shown in a quasi-static approximation.
It is necessary to distinguish between two different cases, depending if large-scale gradients are present or absent during the
transition into the low-collisionality regime.
In the homogeneous case, the situation is clear because one needs to describe the presence of waves with frequencies $\omega$,
and neglecting the time-derivative $d/dt$ in the evolution equations automatically imposes requirement $\omega \ll \nu$, i.e.
the collisional frequencies $\nu$ must remain sufficiently large. In this case, it is erroneous to simply take the quasi-static heat flux $\vecq_a\sim 1/\nu$,
evaluate it for some arbitrarily small $\nu$ and claim that $f_a<0$. Instead, it is necessary to retain the evolution equations with $d\vecq_a/dt$; see e.g. 
(\ref{eq:Excite3}), (\ref{eq:ExciteRit}) or the coupled system (\ref{eq:Nomore110XXa})-(\ref{eq:Nomore102X}),
which precludes one from reaching direct interpretation that $f_a<0$ (unless one calculates the eigenvector and shows otherwise).
The negativity of the distribution function
may not take place, and as a consequence, the procedure seems inadequate for disproving the moment method of Grad in a homogeneous low-collisionality regime.
The situation is much less clear when large-scale gradients are present, as in the example of the solar wind expansion.
In that case, it is possible to argue that keeping the evolution equations 
and solving an initial value problem might help only temporarily, because the system eventually has to converge to some stationary solution,
which might show that $f_a<0$. Such a possibility seems to be implied by the simple 1D radially expanding quasi-static models (see e.g. \cite{Cranmer2021},
and references therein). The quasi-static approximation can however be questioned in this case as well, but from a different perspective. 
Introducing a heat flux or a stress-tensor is analogous to introducing a new degree of freedom into a system, and if this new degree of freedom is not restricted in any way, 
it might of course yield an unphysical system. In plasma physics, degrees of freedom are usually restricted by associated instabilities that
develop, which can not be revealed in a quasi-static approximation (even if an instability is non-propagating).
Useful examples are the anisotropic CGL and 16-moment models described above.
Using a quasi-static approximation, one might erroneously conclude that the temperature anisotropy can grow without bounds in these models, whereas considering
evolution equations reveals the firehose and mirror instabilities, which can restrict the anisotropy. Similar situation might be applicable here,
where sufficiently large drifts (and possibly large heat fluxes and stress-tensors) might cause various instabilities and also development of
turbulence, but further clarifications are needed if our fluid models contain some of these instabilities,
especially considering that our collisional contributions are valid only when differences in drifts between species are
much smaller than their thermal velocities. In this regard, it is not clear if it is appropriate to neglect the Alfv\'enic fluctuations
in the radially expanding models. Finally, it is also not clear if it is physically meaningful to show $f_a<0$ by skipping the stress-tensor in the expansions of Grad (which is a 2nd-order moment before the 3rd-order heat flux moment), because its contributions to a total $f_a$
might be significant. For a sufficient proof that the $f_a$ can become negative, it might be necessary to consider at least the 13-moment model,
where both stress-tensors and heat fluxes are retained.

\newpage
\subsection{Conclusions}
We have discussed various generalizations of the 21-moment model of \cite{Braginskii1958,Braginskii1965}:
1) We have presented a multi-fluid formulation 
for arbitrary masses $m_a$ \& $m_b$ and arbitrary temperatures $T_a$ \& $T_b$. 2) All the
fluid moments are described by their evolution equations,  whose left hand sides are given 
in a fully non-linear form. 3) Formulation with evolution equations has an important consequence that the model  
does not become divergent (unbounded) if a regime of low-collisionality is encountered.
4) For a one ion-electron plasma we have provided all the  Braginskii transport coefficients in a fully analytic form
 for a general ion charge $Z_i$ (and arbitrary strength of magnetic field). 
5) We have also provided fully analytic higher-order transport coefficients (for $\bPi^{(4)}$ and $\vecX^{(5)}$), which are not typically given.
6) All the electron coefficients were further
generalized to multi-ion plasmas. 7) We have considered coupling between viscosity-tensors and heat fluxes, where a heat flux enters a
viscosity-tensor and a viscosity-tensor enters a heat flux. As a consequence, we have introduced new higher-order physical
effects even for the simplest case of unmagnetized one ion-electron plasma of \cite{SpitzerHarm1953}.
For example, the electron rate of strain tensor $\bW_e$ enters the electron heat fluxes even linearly, and thus, it subsequently linearly enters the
momentum  exchange rates; see equation (\ref{eq:energy999}).  
8) We have formulated the 22-moment model which is a natural extension of the 21-moment model, where one takes into account fully contracted
scalar perturbations
$\widetilde{X}^{(4)}_a$ entering the decomposition of the 4-th order moment $X_{ijkl}^{a(4)}$; see equation (\ref{eq:Thierry80}).
Collisional contributions for this model with arbitrary masses and temperatures are given in Section \ref{sec:SummaryQ4}
and supplement those given in Section \ref{sec:Tarb2} for the
21-moment model. Interestingly, scalar perturbations $\widetilde{X}^{(4)}_a$ modify the energy exchange rates,
see equation (\ref{eq:Thierry38}) or (\ref{eq:Thierry35}). In the quasi-static approximation, scalar perturbations $\widetilde{X}^{(4)}_a$
can be written as divergence of heat flux vectors with their own heat conductivities;
see for example solutions for a one ion-electron plasma with the ion heat conductivities (\ref{eq:Thierry681}) and
the electron heat conductivities (\ref{eq:Thierry68}). These corrections remain small in the highly-collisional regime, but might
become significant at small wavelengths and/or at large frequencies.

Our model can be useful for direct numerical simulations, as well as for quick calculation of transport coefficients
in a quasi-static approximation. We provide three examples for coupling between two species. Thermal conductivities
and viscosities for unmagnetized proton-electron plasma (without $m_e/m_p$ expansions) were presented in Section \ref{sec:PEexact},
and two examples for proton-alpha particles and deuterium-tritium were moved to Appendix \ref{sec:2species}. 
Our model can also be useful from an observational perspective. 
For example, the parallel thermal heat flux $\vecq_e$ of \cite{Braginskii1965} and \cite{SpitzerHarm1953} (they differ only by 3.16 vs 3.20 factors
rounded as 3.2) is sometimes analyzed in observational studies;  see e.g. \cite{Salem2003,Bale2013,Halekas2021} and \cite{Verscharen2019} (page 61).
It is also measured in  (exospheric) kinetic numerical simulations \citep{Landi2014}.
Our model suggests that it would be beneficial to analyze both parallel heat fluxes, which for $Z_i=1$ read 
\begin{equation}
  \vecq_{e } = \frac{\vecX^{(3)}_{e}}{2} = -3.2 \frac{p_e}{m_e \nu_{ei}} \nabla T_e;\qquad 
  \vecX^{(5)}_{e } = - 110.7 \frac{p_e^2}{\rho_e m_e \nu_{ei}} \nabla T_e, \label{eq:Observ}
\end{equation}
and which can be analyzed with the same techniques. 
For long parallel mean-free-paths (low collisionality regime), both heat fluxes naturally have to become non-local and independent of the mean-free-path.
Our limitations are described in Section \ref{sec:limitations}, and ``flattening/saturation'' of heat fluxes due to the runaway effect and
Landau damping is not captured in our model.  Our model is aimed at the highly-collisional regime and in the
low-collisionality regime our heat fluxes are just described by their evolution equations, where
the collisional right hand sides are small. Nevertheless, it would be interesting to see if
 in observational studies or kinetic simulations the $\vecX^{(5)}_e$ could be
described by a free-streaming formula similar to the one of \cite{Hollweg1974,Hollweg1976}, in a form $\vecX^{(5)}_e=(3/2) (p_e^2/\rho_e) \bu_{sw}\alpha_5$, where the
``bugger factor'' $\alpha_5$ has to be determined from a given form of a distribution function, or if such a concept does not apply for $\vecX^{(5)}_e$.
 As a side note, concerning collisionless heat fluxes for plasmas where spherical expansion and large drifts are not present and Landau damping dominates,
our model actually implies
that a correct interpretation should not be that Landau damping diminishes/saturates the heat flux in a low-collisionality regime.  The
correct interpretation is that Landau damping creates the collisionless heat flux. Collisionless Landau fluid closures for quasi-static
parallel scalar $X^{(5)}_\parallel$ can be found in \cite{Hunana2019b} (p. 84). 
 In addition to (\ref{eq:Observ}), it might be also useful to analyze the scalar perturbation, which for $Z_i=1$ reads
\begin{equation} \label{eq:Thierry280}
\widetilde{X}^{(4)}_e = + 83.8  \frac{p_e^2}{\nu_{ei}^2 \rho_e m_e} \nabla^2 T_e.
\end{equation}

Our multi-fluid model might be also useful for modeling of enrichment of minor ion abundancies 
in stellar atmospheres, because of the very precise thermal force (thermal diffusion).
Let us summarize the thermal force description in three major models: the model of 
\cite{Burgers1969}-\cite{Schunk1977}, the model of \cite{Killie2004}, and our model. Of course, all three models are formulated as a general multi-fluid,
but for a simplicity of the discussion let us simplify and compare only thermal forces given by 
\begin{eqnarray}
 \textrm{Burgers-Schunk:} \qquad \boldsymbol{R}_{e}^T &=& +\frac{3}{5}\frac{\rho_{e}}{p_{e}} \nu_{ei} \vecq_e;  \label{eq:TFcorr1}\\
 \textrm{Killie et al.:} \qquad \boldsymbol{R}_{e}^T &=& +\frac{6}{35}\frac{\rho_{e}}{p_{e}} \nu_{ei} \vecq_e;\label{eq:TFcorr2}\\
  \textrm{present paper:}\qquad \boldsymbol{R}_e^T &=&  + \frac{21}{10}\frac{\rho_e}{p_e} \nu_{ei}\vecq_e
  -\frac{3}{56} \frac{\rho_e^2}{p_e^2}\nu_{ei} \vecX^{(5)}_e. \label{eq:TFcorr}
\end{eqnarray}
Note that the viscosity-tensors are not required to describe the thermal force,
and focusing only at the heat fluxes, instead of the 13-moment  model of Burgers-Schunk, one can consider only the 8-moment model. 
Similarly, instead of our 21 \& 22-moment models, one can consider only the 11-moment model.
In general, the parallel thermal heat flux is given by $\vecq_e=-\gamma_0p_e/(m_e\nu_{ei})\nabla T_e$ and
the resulting parallel thermal force by $\boldsymbol{R}_{e}^T=-\beta_0 n_e\nabla T_e$,
with coefficients $\gamma_0$ and $\beta_0$. From the work of \cite{SpitzerHarm1953}, for $Z_i=1$ the correct coefficient of thermal conductivity is $\gamma_0=3.203$ and
the correct coefficient of thermal force is $\beta_0=0.703$. 
The model of Burgers-Schunk (\ref{eq:TFcorr1}) has thermal conductivity $\gamma_0=1.34$,
and with that value it describes
the thermal force  actually quite accurately,  yielding $\beta_0=0.804$
(for other $Z_i$ values see comparison in Table \ref{eq:TableTr} in Appendix \ref{sec:Comparison}).
However, a problem arises if one uses the correct value of thermal conductivity $\gamma_0=3.2$ in the expression (\ref{eq:TFcorr1}), which
overestimates the thermal force. 
\cite{Killie2004} developed a different 8-moment model, where the expansion is done differently than in equation (\ref{eq:One}),
with the goal to improve the heat flux and the thermal
force of Burgers-Schunk. The model is described in Appendix \ref{sec:Killie}. For $Z_i=1$, its heat flux value is $\gamma_0=3.92$, which greatly improves the
model of Burgers-Schunk and for that value it also improves the thermal force,  yielding $\beta_0=0.672$. 
Additionally, now one can use the correct $\gamma_0=3.2$ value in expression (\ref{eq:TFcorr2}) and the thermal force will be roughly
correct (and 7/2 times smaller than Burgers-Schunk). However, as we point out in Appendix \ref{sec:Comparison}
(see Table \ref{eq:TablePar2}), the model of \cite{Killie2004} breaks the Onsager symmetry between
the frictional heat flux and the thermal force. The numerical model of \cite{SpitzerHarm1953} also does not satisfy
the Onsager symmetry and its frictional heat flux is technically incorrect, even though in this case the discrepancies are small.
 Our model satisfies the Onsager symmetry, it has thermal conductivity $\gamma_0=3.1616$ and thermal force $\beta_0=0.711$
 (the same as Braginskii). In summary, our multi-fluid model has a very precise thermal force (\ref{eq:TFcorr}) with precision equal to \cite{Braginskii1965},
and we thus offer an improvement to the multi-fluid models of \cite{Burgers1969}-\cite{Schunk1977} and \cite{Killie2004}.

\section{Acknowledgments}
This work was supported by the European Research Council in the frame
of the Consolidating Grant ERC-2017-CoG771310-PI2FA ``Partial Ionisation: Two-Fluid Approach'', led by Elena Khomenko.
Anna Tenerani acknowledges support of the NASA Heliophysics Supporting Research Grant \#80NSSC18K1211.
We acknowledge support of the NSF EPSCoR RII-Track-1 Cooperative Agreement No. OIA-1655280 ``Connecting the Plasma Universe to Plasma
Technology in Alabama'', led by Gary P. Zank. Gary M. Webb was funded in part by NASA grant 80NSSC19K0075.

\newpage
\appendix

\section{General evolution equations} \label{sec:General}
We consider Boltzmann equation (in CGS units)
\begin{equation} \label{eq:Vlasov}
  \frac{\pr f_a}{\pr t}+ \bV\cdot\nabla f_a +\frac{eZ_a}{m_a}(\bE+\frac{1}{c}\bV\times\bb)\cdot\nabla_v f_a
  =C(f_a),
\end{equation}
where ``a'' is a species index and $C(f_a)=\sum_b C_{ab}(f_a,f_b)$ is the Landau collisional operator, so equation (\ref{eq:Vlasov}) is called the Landau equation. 
One defines the usual number density $n_a = \int f_a d^3 v$;
density $\rho_a=m_a n_a$, fluid velocity $\bu_a=(1/n_a) \int \bV f_a d^3 v$ and fluctuating velocity $\bc_a=\bV-\bu_a$,
and further defines the pressure tensor $\bp_a$, heat flux tensor $\bq_a$, 4th-order moment $\br_a$, and 5th-order \& 6th-order moments $\bX^{(5)}_a$,
$\bX^{(6)}_a$ according to
\begin{eqnarray}
  \bp_a &=& m_a \int \bc_a\bc_a f_a d^3 v;\qquad
  \bq_a = m_a \int \bc_a\bc_a\bc_a f_a d^3 v;\qquad
  \br_a = m_a \int \bc_a\bc_a\bc_a\bc_a f_a d^3 v; \label{eq:defTensor}\\
  \bX_a^{(5)} &=& m_a \int \bc_a\bc_a\bc_a\bc_a \bc_a f_a d^3 v; \qquad  \bX_a^{(6)} = m_a \int \bc_a\bc_a\bc_a\bc_a \bc_a \bc_a f_a d^3 v.
\end{eqnarray}
Writing of the tensor product $\otimes$ is suppressed everywhere and
$\bc_a\bc_a=\bc_a\otimes\bc_a$. For complicated fluid models the species index $'a'$ often blurs the clarity of the tensor algebra, and thus 
in the vector notation (\ref{eq:defTensor}) we emphasize tensors of second-rank and above with the double overbar symbol.
Sometimes we move the index $'a'$ freely up and down (which here does not represent any mathematical operation),
and in the index notation the index $'a'$ is often dropped completely, so for example $p_{ij}^a = m_a \int c_i^a c_j^a f_a d^3 v$
and $p_{ij} = m \int c_ic_j f d^3 v$ are equivalent. The Einstein summation convention does not apply for the species index $'a'$,
and summations over other particle species are written down explicitly. The divergence is defined through the first index $(\nabla\cdot\bp_a)_j=\pr_i p^a_{ij}$.

Here we do not consider ionization and recombination processes and the Landau collisional operator conserves
the number of particles $\int C(f_a) d^3v=0$ for each species. One defines a unit vector in the direction of magnetic field
$\bhat=\bb/|\bb|$, cyclotron frequency $\Omega_a=eZ_a|\bb|/(m_a c)$ and convective derivative $d_a/dt=\pr/\pr t+\bu_a\cdot\nabla$.
It is also useful to define a symmetric operator $'S'$,
which acts on a matrix as $A_{ij}^S=A_{ij}+A_{ji}$ and on a tensor of 3rd-rank as $A_{ijk}^S=A_{ijk}+A_{jki}+A_{kij}$, i.e. it cycles around
all indices. We often use operator trace $\textrm{Tr}$ and unit matrix $\bI$, where $\textrm{Tr}\bA=\bI:\bA$, and operator
':' represents double contraction. We also use $\bI_\perp=\bI-\bhat\bhat$.

To derive the model of \cite{Braginskii1965} with the moment method of Grad, it is necessary to consider the evolution equation for the 5th-order moment
$\bX^{(5)}_a$ and perform a closure at $\bX^{(6)}_a$. Integrating (\ref{eq:Vlasov}) over velocity space yields the the following hierarchy of evolution equations
\begin{eqnarray}
&& \frac{\pr n_a}{\pr t} + \nabla\cdot(n_a\bu_a) =0; \label{eq:density1} \\
&& \frac{\pr \bu_a}{\pr t} + \bu_a\cdot\nabla\bu_a +\frac{1}{\rho_a}\nabla\cdot\bp_a -\frac{eZ_a}{m_a}\Big( \bE+\frac{1}{c}\bu_a\times\bb\Big)
  = \frac{\boldsymbol{R}_a}{\rho_a}; \label{eq:momentum1}\\
&&  \frac{\pr\bp_a}{\pr t}  +\nabla\cdot \big( \bq_a+\bu_a\bp_a \big) +\Big[ \bp_a\cdot\nabla\bu_a
    +\Omega_a\bhat\times\bp_a \Big]^S =\bQ^{(2)}_a; \label{eq:PR_tensor}\\
&&  \frac{\pr \bq_a}{\pr t}  +\nabla\cdot \big( \br_a+\bu_a\bq_a \big) +\Big[ \bq_a\cdot\nabla\bu_a
    +\Omega_a\bhat\times\bq_a-\frac{1}{\rho_a} (\nabla\cdot\bp_a) \bp_a \Big]^S =\bQ^{(3)}_a-\frac{1}{\rho_a}\Big[ \boldsymbol{R}_a \bp_a\Big]^S; \label{eq:QR_tensor}\\
 && \frac{\pr}{\pr t}\br_{a}+\nabla\cdot\big( \bX^{(5)}_a +\bu_a \br_a\big)
  +\Big[ \br_a \cdot\nabla\bu_a +\Omega_a \bhat\times \br_a-\frac{1}{\rho_a}\big(\nabla\cdot\bp_a\big) \bq_a\Big]^S
   = \bQ^{(4)}_a-\frac{1}{\rho_a}\big[ \boldsymbol{R}_a \bq_a \big]^S;\label{eq:Thierry-R}\\
 && \frac{\pr}{\pr t}\bX^{(5)}_a+\nabla\cdot\big( \bX^{(6)}_a +\bu_a \bX^{(5)}_a\big)
  +\Big[ \bX^{(5)}_a\cdot\nabla\bu_a +\Omega_a \bhat\times \bX^{(5)}_a-\frac{1}{\rho_a}\big(\nabla\cdot\bp_a\big) \br_a\Big]^S 
   = \bQ^{(5)}_a-\frac{1}{\rho_a}\big[ \boldsymbol{R}_a \br_a \big]^S, \label{eq:X5_tensor}
\end{eqnarray}
where the collisional contributions on the right hand sides are given by (\ref{eq:Spec}). 
It is also possible to define a general n-th-order moment $\bX^{(n)}_a$ and collisional contributions $\bQ^{(n)}_a$
\begin{equation} \label{eq:bXn}
X^{(n)}_{r_1 r_2 \ldots r_n} = m \int c_{r_1} c_{r_2}\ldots c_{r_n} f d^3 v; \qquad Q^{(n)}_{r_1 r_2\ldots r_n} = m \int c_{r_1} c_{r_2}\ldots c_{r_n} C(f) d^3 v,
\end{equation}  
together with symmetric operator 'S' that cycles around all of its indices
\begin{equation}
\big[ X^{(n)}\big]^S_{r_1 r_2 r_3 \ldots r_n} = X^{(n)}_{r_1 r_2 r_3 \ldots r_n} + X^{(n)}_{r_2 r_3 \ldots r_n r_1} + X^{(n)}_{r_3\ldots r_n r_1 r_2}+ \cdots\cdots + X^{(n)}_{r_n r_1 r_2 r_3\ldots r_{n-1}},
\end{equation}
(so that it contains ``n'' terms) and derive the following evolution equation for $\bX^{(n)}_a$
\begin{eqnarray}
 && \frac{\pr}{\pr t} \bX^{(n)}_a +\nabla\cdot \big( \bX^{(n+1)}_a+\bu_a\bX^{(n)}_a \big) +\Big[ \bX^{(n)}_a\cdot\nabla\bu_a
    +\Omega_a\bhat\times\bX^{(n)}_a-\frac{1}{\rho_a} (\nabla\cdot\bp_a) \bX^{(n-1)}_a \Big]^S \nn\\
 && =\bQ^{(n)}_a-\frac{1}{\rho_a}\Big[ \boldsymbol{R}_a \bX^{(n-1)}_a\Big]^S, \label{eq:GenTensor}
\end{eqnarray}
valid for $n\ge 2$. The left hand side of (\ref{eq:GenTensor}) is equal to the collisionless equation (12.16) of \cite{Hunana2019a}.
Evolution equations (\ref{eq:PR_tensor})-(\ref{eq:X5_tensor}) then can be obtained easily by
evaluation of (\ref{eq:GenTensor}). Note that definition (\ref{eq:bXn}) yields
$\bX^{(2)}=\bp$, $\bX^{(3)}=\bq$, $\bX^{(4)}=\br$, however, $\bX^{(1)}=0$.

As was pointed out already by \cite{Grad1949_1,Grad1949_2}, who developed the moment approach considering rarified gases, because fluid moments are symmetric in all of their indices,
a general n-th order moment $\bX^{(n)}$ contains $\binom{n+2}{n}=(n+1)(n+2)/2$ distinct
(scalar) components. So the density has 1, velocity has 3, pressure tensor has 6, heat flux tensor has 10,
$\bX^{(4)}$ has 15, and $\bX^{(5)}$ has 21 scalar components. The system (\ref{eq:density1})-(\ref{eq:X5_tensor}) thus represents 56-moment model. 

\newpage
\section{Tensorial Hermite Decomposition} \label{sec:Hermite}
\setcounter{equation}{0}
In the famous work of \cite{Grad1949_1,Grad1949_2,Grad_1958},
the so-called \emph{tensorial} Hermite decomposition is used, which is a generalization of the 1D version. 
The 1D Hermite polynomials of order ``m'' are defined as
\begin{equation}
{H}^{(m)}(x) = (-1)^m e^{\frac{x^2}{2}} \frac{d^m}{dx^m} e^{-\frac{x^2}{2}},
\end{equation}
and evaluated step by step as ${H}^{(0)}=1$; ${H}^{(1)}=x$; ${H}^{(2)}=x^2-1$; ${H}^{(3)}=x^3-3x$;
${H}^{(4)}=x^4-6x^2+3$; ${H}^{(5)}=x^5-10x^3+15x$. So polynomials of even-order contain only even-powers of $x$ and polynomials of odd-order
contain only odd-powers of $x$. These polynomials are orthogonal to each other by
\begin{equation} \label{eq:Hortho}
\frac{1}{\sqrt{2\pi}}\int_{-\infty}^\infty {H}^{(n)}(x) {H}^{(m)}(x) e^{-\frac{x^2}{2}} dx = n! \delta_{nm}.
\end{equation}  
Note that the ``weight'' $\exp(-x^2/2)$ was used by Grad instead of quantum-mechanical $\exp(-x^2)$. 
Of course, it is important to use the correct weight with both classes of Hermite polynomials.
Curiously, if the weight is accidentally mismatched (i.e. by using $\exp(-x^2)$ in our (\ref{eq:Hortho}) or $\exp(-x^2/2)$ in the quantum version),
in addition to naturally wrong numerical constants, the even-even and odd-odd couples of polynomials are not orthogonal any more! 
Generalization to tensors for isotropic Maxwellian distribution reads
\begin{equation} \label{eq:HerM}
\tilde{H}^{(m)}_{r_1,r_2\ldots r_m}(\tbc) = (-1)^m e^{\frac{\tc^2}{2}} \frac{\pr}{\pr \tc_{r_1}}  \frac{\pr}{\pr \tc_{r_2}}\cdots  \frac{\pr}{\pr \tc_{r_m}} e^{-\frac{\tc^2}{2}}.   
\end{equation}
We use the same notation as \cite{Balescu1988}, where \emph{reducible} Hermite polynomials are denoted with tilde, and
\emph{irreducible} polynomials have no tilde. We added tilde on normalized $\tbc$ to make transitioning to usual fluid moments straightforward.  
Then explicit evaluation step by step gives
\begin{eqnarray}
&&  \tilde{H}^{(0)}(\tbc) = 1;\nn\\
&&  \tilde{H}^{(1)}_i(\tbc) = \tc_i;\nn\\
&&  \tilde{H}^{(2)}_{ij}(\tbc) = \tc_i \tc_j -\delta_{ij};\nn\\
&&  \tilde{H}^{(3)}_{ijk}(\tbc) = \tc_i \tc_j \tc_k - \big( \delta_{ij} \tc_k +\delta_{jk} \tc_i +\delta_{ik} \tc_j \big);\nn\\
  &&  \tilde{H}^{(4)}_{ijkl}(\tbc) = \tc_i \tc_j \tc_k \tc_l - \big( \delta_{ij}\tc_k\tc_l +\delta_{jk}\tc_l\tc_i +\delta_{kl}\tc_i\tc_j
  +\delta_{li}\tc_j\tc_k+\delta_{ik}\tc_j\tc_l +\delta_{jl}\tc_i\tc_k\big)\nn\\
  && \qquad \qquad \quad+ \delta_{ij}\delta_{kl} + \delta_{ik}\delta_{jl} +\delta_{il}\delta_{jk},\label{eq:H5pol2}
\end{eqnarray}
and quickly starts to grow
\begin{eqnarray}
\tilde{H}^{(5)}_{ijklm}(\tbc) &=& \tc_i\tc_j\tc_k\tc_l\tc_m -\big( \delta_{ij}\tc_k\tc_l\tc_m +\delta_{jk}\tc_l\tc_i\tc_m
  +\delta_{kl}\tc_i\tc_j\tc_m +\delta_{li}\tc_j\tc_k\tc_m +\delta_{ik}\tc_j\tc_l\tc_m \nn\\
  && +\delta_{jl}\tc_i\tc_k\tc_m +\delta_{im}\tc_j\tc_k\tc_l+\delta_{jm}\tc_i\tc_k\tc_l+\delta_{km}\tc_i\tc_j\tc_l
  +\delta_{lm}\tc_i\tc_j\tc_k \big)\nn\\
  && +\delta_{ij}\delta_{kl}\tc_m +\delta_{ik}\delta_{jl}\tc_m +\delta_{il}\delta_{jk}\tc_m 
  +\delta_{ij}\delta_{km}\tc_l  +\delta_{ij}\delta_{lm}\tc_k \nn\\
  && +\delta_{jk}\delta_{lm}\tc_i +\delta_{jk}\delta_{im}\tc_l  +\delta_{kl}\delta_{im}\tc_j 
  +\delta_{kl}\delta_{jm}\tc_i  +\delta_{li}\delta_{jm}\tc_k \nn\\
  && +\delta_{li}\delta_{km}\tc_j 
   +\delta_{ik}\delta_{jm}\tc_l+\delta_{ik}\delta_{lm}\tc_j +\delta_{jl}\delta_{im}\tc_k +\delta_{jl}\delta_{km}\tc_i. \label{eq:H5pol}
\end{eqnarray}  
The choice of Grad with $\exp(-x^2/2)$ has a great benefit, because no numerical constants are present
in the entire hierarchy of Hermite polynomials, which is not the case for the weight $\exp(-x^2)$.
Here, numerical factors appear only after one applies contractions (trace) at the above  expressions. Similarly to the 1D case, 
polynomials of even-order contain only terms with even number of velocities $\tc$, and polynomials of odd-order only terms
with odd number of $\tc$. 

For Maxwellian distribution, the normalized velocity is
\begin{equation}
\tbc = \sqrt{\frac{m_a}{T_a}} (\bV-\bu_a) = \sqrt{\frac{m_a}{T_a}}\bc,
\end{equation}
where for simplicity we suppress to write species index ``a'' for velocity $\bc$ in the expressions that follow, and for many other variables as well
(the Hermite decomposition is done indendently for each species, and species variable ``a'' just makes expressions more blurry).
It is possible to work both in normalized and
physical units. The entire distribution function can be written as
\begin{equation} \label{eq:PerturbB}
  f_a = f^{(0)}_a(1+\chi_a) = n_a\Big(\frac{m_a}{T_a} \Big)^{3/2} \phi^{(0)} (1+\chi_a); \qquad \phi^{(0)} = \frac{e^{-\frac{\tc^2}{2}}}{(2\pi)^{3/2}},
\end{equation}
where $\chi_a$ represents the wanted perturbation. One can go quickly between physical and normalized units by
\begin{equation} \label{eq:PerturbB1}
\int f_a (\bc) d^3c = n_a \int \phi^{(0)} (1+\chi_a(\tbc)) d^3\tc.
\end{equation}  
The tensorial polynomials are again orthogonal to each other, where by using ``weight''
$\phi^{(0)}$
\begin{eqnarray}
  \int \phi^{(0)} \tilde{H}^{(0)} \tilde{H}^{(0)} d^3\tc &=& 1;\nn\\
  \int \phi^{(0)} \tilde{H}^{(1)}_{i} \tilde{H}^{(1)}_{j} d^3\tc &=& \delta_{ij};\nn\\
  \int \phi^{(0)} \tilde{H}^{(2)}_{i j} \tilde{H}^{(2)}_{k l} d^3\tc &=& \delta_{ik}\delta_{jl} + \delta_{il}\delta_{jk};\nn\\ 
   \int \phi^{(0)} \tilde{H}^{(3)}_{r_1 r_2 r_3} \tilde{H}^{(3)}_{s_1 s_2 s_3} d^3\tc &=& \delta_{r_1 s_1}\delta_{r_2 s_2}\delta_{r_3 s_3}
    + \delta_{r_1 s_1}\delta_{r_2 s_3}\delta_{r_3 s_2} + \delta_{r_1 s_2}\delta_{r_2 s_1}\delta_{r_3 s_3}\nn\\
&& +\delta_{r_1 s_2}\delta_{r_2 s_3}\delta_{r_3 s_1}
  + \delta_{r_1 s_3}\delta_{r_2 s_1}\delta_{r_3 s_2} +\delta_{r_1 s_3}\delta_{r_2 s_2}\delta_{r_3 s_1}, \label{eq:examp}
\end{eqnarray}
and expressions quickly become long
\begin{eqnarray}
  \int \phi^{(0)} \tilde{H}^{(4)}_{r_1 r_2 r_3 r_4} \tilde{H}^{(4)}_{s_1 s_2 s_3 s_4} d^3\tc &=& +\delta_{r_1 s_1}\delta_{r_2 s_2}\delta_{r_3 s_3}\delta_{r_4 s_4}
  + \delta_{r_1 s_1}\delta_{r_2 s_2}\delta_{r_3 s_4}\delta_{r_4 s_3} + \delta_{r_1 s_1}\delta_{r_2 s_3}\delta_{r_3 s_2}\delta_{r_4 s_4}\nn\\
&&  + \delta_{r_1 s_1}\delta_{r_2 s_3}\delta_{r_3 s_4}\delta_{r_4 s_2} + \delta_{r_1 s_1}\delta_{r_2 s_4}\delta_{r_3 s_2}\delta_{r_4 s_3}
  + \delta_{r_1 s_1}\delta_{r_2 s_4}\delta_{r_3 s_3}\delta_{r_4 s_2}\nn\\
  &&   + \delta_{r_1 s_2}\delta_{r_2 s_1}\delta_{r_3 s_3}\delta_{r_4 s_4} + \delta_{r_1 s_2}\delta_{r_2 s_1}\delta_{r_3 s_4}\delta_{r_4 s_3}
  + \delta_{r_1 s_2}\delta_{r_2 s_3}\delta_{r_3 s_1}\delta_{r_4 s_4}\nn\\
  &&  + \delta_{r_1 s_2}\delta_{r_2 s_3}\delta_{r_3 s_4}\delta_{r_4 s_1} + \delta_{r_1 s_2}\delta_{r_2 s_4}\delta_{r_3 s_1}\delta_{r_4 s_3}
  + \delta_{r_1 s_2}\delta_{r_2 s_4}\delta_{r_3 s_3}\delta_{r_4 s_1}\nn\\
  &&  + \delta_{r_1 s_3}\delta_{r_2 s_1}\delta_{r_3 s_2}\delta_{r_4 s_4} + \delta_{r_1 s_3}\delta_{r_2 s_1}\delta_{r_3 s_4}\delta_{r_4 s_2}
  + \delta_{r_1 s_3}\delta_{r_2 s_2}\delta_{r_3 s_1}\delta_{r_4 s_4}\nn\\
  &&  + \delta_{r_1 s_3}\delta_{r_2 s_2}\delta_{r_3 s_4}\delta_{r_4 s_1} + \delta_{r_1 s_3}\delta_{r_2 s_4}\delta_{r_3 s_1}\delta_{r_4 s_2}
  + \delta_{r_1 s_3}\delta_{r_2 s_4}\delta_{r_3 s_2}\delta_{r_4 s_1} \nn\\
  &&  + \delta_{r_1 s_4}\delta_{r_2 s_1}\delta_{r_3 s_2}\delta_{r_4 s_3} + \delta_{r_1 s_4}\delta_{r_2 s_1}\delta_{r_3 s_3}\delta_{r_4 s_2}
  + \delta_{r_1 s_4}\delta_{r_2 s_2}\delta_{r_3 s_1}\delta_{r_4 s_3}\nn\\
  &&  + \delta_{r_1 s_4}\delta_{r_2 s_2}\delta_{r_3 s_3}\delta_{r_4 s_1} + \delta_{r_1 s_4}\delta_{r_2 s_3}\delta_{r_3 s_1}\delta_{r_4 s_2}
   + \delta_{r_1 s_4}\delta_{r_2 s_3}\delta_{r_3 s_2}\delta_{r_4 s_1}. \label{eq:exampX}
\end{eqnarray}  
The general orthogonality can be written by introducing multi-indices $\boldsymbol{r}=r_1\ldots r_n$ and $\boldsymbol{s}=s_1\dots s_m$
\begin{equation} \label{eq:ortho}
\int \phi^{(0)} \tilde{H}^{(n)}_{\boldsymbol{r}} \tilde{H}^{(m)}_{\boldsymbol{s}} d^3\tc = \delta_{mn} \boldsymbol{\delta}^{(n)}_{\boldsymbol{r}\boldsymbol{s}}, 
\end{equation}
where the new symbol $\boldsymbol{\delta}^{(n)}_{\boldsymbol{r}\boldsymbol{s}}$ is equal to one if indices $r_1\ldots r_n$ are a permutation
of $s_1\ldots s_n$ and otherwise it is zero. In other words, for $n=m$ the right hand side contains $n!$ terms, where each of these terms
has a form $\delta_{r_1 s_1}\delta_{r_2 s_2}\ldots \delta_{r_n s_n}$ and to calculate the other terms keep $r$-indices fixed, and do all the
possible permutations with $s$-indices (or vice versa).
A particular case of (\ref{eq:ortho}) reads
\begin{equation} \label{eq:ortho2}
m\neq 0:\qquad \int \phi^{(0)} \tilde{H}^{(m)}_{\boldsymbol{s}} d^3\tc =0, 
\end{equation}
i.e. integral over a single Hermite polynomial with weight $\phi^{(0)}$ is zero.

The goal of the Hermite expansion is to find perturbation of the
distribution function $\chi_a$ in (\ref{eq:PerturbB}). For the most general decomposition,
one can chose to express the perturbation $\chi_a$ as a sum of Hermite polynomials
\begin{eqnarray}
  \chi_a &=& \sum_{m=1}^\infty A_{r_1 r_2 \ldots r_m}^{(m)} \tilde{H}_{r_1 r_2 \ldots r_m}^{(m)} \nn\\
  &=& A_{r_1}^{(1)}\tilde{H}_{r_1}^{(1)} + A_{r_1 r_2}^{(2)}\tilde{H}_{r_1 r_2}^{(2)} + A_{r_1 r_2 r_3}^{(3)}\tilde{H}_{r_1 r_2 r_3}^{(3)}
  + A_{r_1 r_2 r_3 r_4}^{(4)}\tilde{H}_{r_1 r_2 r_3 r_4}^{(4)} +\cdots,  \label{eq:HermiteA1}
\end{eqnarray}
where coefficients $A_{r_1 r_2 \ldots r_m}^{(m)}$ need to be found. Note that full contractions over all indices are present and the result is a scalar. 
Multiplying (\ref{eq:HermiteA1}) by weight $\phi^{(0)}$ and polynomial $\tilde{H}_{s_1 s_2 \ldots s_n}^{(n)}$ and integrating over $d^3\tc$ by using
orthogonality (\ref{eq:ortho}) then yields
\begin{eqnarray}
\int \chi_a \phi^{(0)} \tilde{H}_{\boldsymbol{s}}^{(n)} d^3\tc = A_{\boldsymbol{r}}^{(n)} \delta^{(n)}_{\boldsymbol{r}\boldsymbol{s}} = n! A_{\boldsymbol{s}}^{(n)}, 
\end{eqnarray}
where the last equality holds because coefficient $A_{\boldsymbol{s}}^{(n)}$ is a fluid variable and symmetric in all of its indices. Coefficients
$A_{\boldsymbol{s}}^{(n)}$ are thus found according to
\begin{eqnarray}
  A_{\boldsymbol{s}}^{(n)} = \frac{1}{n!} \int \chi_a \phi^{(0)} \tilde{H}_{\boldsymbol{s}}^{(n)}(\tbc) d^3\tc =
  \frac{1}{n!} \int (1+\chi_a) \phi^{(0)} \tilde{H}_{\boldsymbol{s}}^{(n)}(\tbc) d^3\tc
  = \frac{1}{n!} \Big[\underbrace{\frac{1}{n_a} \int f_a \tilde{H}_{\boldsymbol{s}}^{(n)}(\tbc) d^3 c}_{\tilde{h}^{(n)}_{\boldsymbol{s}}}\Big], \label{eq:HermiteA2}
\end{eqnarray}
where we have used orthogonality relation (\ref{eq:ortho2}) and changed the integration variable to $d^3c$ with (\ref{eq:PerturbB1}).
Quantities in the brackets of (\ref{eq:HermiteA2}) are called \emph{Hermite moments} $\tilde{h}^{(n)}_{\boldsymbol{s}}$.
The entire Hermite expansion then can be summarized
into two easy steps:\\
\noindent 
1) Calculate Hermite moments
\begin{equation}
  \tilde{h}_{r_1 r_2\ldots r_m}^{a(m)} = \frac{1}{n_a} \int f_a \tilde{H}_{r_1 r_2 \ldots r_m}^{a(m)}(\tbc) d^3c, \label{eq:Hermite3}
\end{equation}
2) the final perturbation is
\begin{equation} \label{eq:Hermite1}
\chi_a = \sum_{m=1}^\infty \frac{1}{m!} \tilde{h}_{r_1 r_2 \ldots r_m}^{a(m)} \tilde{H}_{r_1 r_2 \ldots r_m}^{a(m)}(\tbc). 
\end{equation} 
It is useful to omit writing the species indices ``a'' on both $\tilde{h}$ and $\tilde{H}$, as well as on fluid moments,
we will keep the species index only for $n_a,m_a,T_a,p_a$. The final perturbations will be written in a full form.

By using definitions of general fluid moments, one straightforwardly calculates Hermite moments
\begin{eqnarray}
  \tilde{h}^{(1)}_i &=& \frac{1}{n_a} \int f_a \tilde{H}^{(1)}_i d^3c = 0;\nn\\
  \tilde{h}_{ij}^{(2)} &=& \frac{1}{n_a} \int f_a \tilde{H}^{(2)}_{ij} d^3c = \frac{1}{p_a}\Pi_{ij}^{(2)};\nn\\
  \tilde{h}_{ijk}^{(3)} &=& \frac{1}{n_a} \int f_a \tilde{H}_{ijk}^{(3)} d^3c = \frac{1}{p_a}\sqrt{\frac{m_a}{T_a}} q_{ijk}, \label{eq:HerMom}
\end{eqnarray}
together with
\begin{eqnarray}
  \tilde{h}_{ijkl}^{(4)} &=&  \frac{1}{n_a} \int f_a \tilde{H}_{ijkl}^{(4)}(\bc) d^3c 
 = \frac{\rho_a}{p_a^2} r_{ijkl}  +\delta_{ij}\delta_{kl}+\delta_{ik}\delta_{jl}+\delta_{il}\delta_{jk}\nn\\
  && \quad -\frac{1}{p_a}\big( \delta_{ij} p_{kl}+\delta_{jk}p_{li} +\delta_{kl} p_{ij} +\delta_{li} p_{jk} +\delta_{ik} p_{jl}+\delta_{jl}p_{ik}\big)\nn\\
  &=& \frac{\rho_a}{p_a^2} r_{ijkl}  -\big( \delta_{ij}\delta_{kl}+\delta_{ik}\delta_{jl}+\delta_{il}\delta_{jk}\big)\nn\\
 && \quad -\frac{1}{p_a}\big( \delta_{ij} \Pi_{kl}^{(2)}+\delta_{jk} \Pi_{li}^{(2)} +\delta_{kl} \Pi_{ij}^{(2)}
 +\delta_{li} \Pi_{jk}^{(2)} +\delta_{ik} \Pi_{jl}^{(2)}+\delta_{jl}\Pi_{ik}^{(2)}\big),
\end{eqnarray}
and
\begin{eqnarray}
  \tilde{h}_{ijklm}^{(5)} &=& \frac{1}{n_a} \int f_a \tilde{H}_{ijklm}^{(5)}(\bc) d^3c = \frac{\rho_a^{3/2}}{p_a^{5/2}} X^{(5)}_{ijklm}
  -\frac{\rho_a^{1/2}}{p_a^{3/2}}\Big( \delta_{ij} q_{k l m} +\delta_{jk} q_{l i m}
  +\delta_{kl} q_{i j m} +\delta_{li} q_{j k m} \nn\\
  && +\delta_{ik} q_{j l m} 
   +\delta_{jl} q_{i k m} +\delta_{im} q_{j k l}+\delta_{jm} q_{i k l}+\delta_{km} q_{i j l}
   +\delta_{lm} q_{i j k} \Big).  \label{eq:HerMomL}
\end{eqnarray}

\newpage
\subsection{Usual perturbations of Grad}
\subsubsection{20-moment model}
By using the definition of the perturbation (\ref{eq:Hermite1}) and cutting the hierarchy at
\begin{equation}
\chi_a = \tilde{h}_{i}^{(1)}\tilde{H}_i^{(1)} + \frac{1}{2}\tilde{h}_{ij}^{(2)} \tilde{H}_{ij}^{(2)} + \frac{1}{6} \tilde{h}_{ijk}^{(3)} \tilde{H}_{ijk}^{(3)},
\end{equation}  
yields the 20-moment perturbation of Grad 
\begin{equation} \label{eq:20mom_perturb}
  \textrm{20-moment}:\qquad \chi_a = \frac{m_a}{2p_a T_a}\big(\bPi_a^{(2)}:\bc_a\bc_a\big) +\frac{m_a^2}{6p_a T_a^2}\big(\bc_a\cdot\bq_a:\bc_a\bc_a\big)
  -\frac{m_a}{p_a T_a} \big(\vecq_a\cdot\bc_a\big),
\end{equation}
where one defines vector $\vecq_a=(1/2)\trace\bq_a$.

\subsubsection{13-moment model}
To quickly obtain the simplified 13-moment model of Grad, one can use 
$\bq=(2/5)(\vecq\bI)^S+\boldsymbol{\sigma}'$ with $\boldsymbol{\sigma}'$ neglected (validity of this equation is shown below) and calculating
$\bc\cdot\bq:\bc\bc=(6/5)(\vecq\cdot\bc)c^2$ yields the 13-moment model 
\begin{equation} \label{eq:MomPi}
\textrm{13-moment}:\qquad \chi_a = \frac{m_a}{2p_a T_a}\big(\bPi_a^{(2)}:\bc_a\bc_a\big)  -\frac{m_a}{p_a T_a}\big(\vecq_a\cdot\bc_a \big) \Big( 1-\frac{m_a}{5 T_a}c_a^2\Big).
\end{equation}
To re-derive the heat flux contribution in the 13-moment model from scratch can be done by using a contracted Hermite polynomial 
\begin{equation} \label{eq:cica}
\tilde{H}^{(3)}_i \equiv \delta_{jk} \tilde{H}^{(3)}_{ijk} = \tc_i (\tc^2-5).  
\end{equation}
However, one needs to be careful about the normalization constant because applying
contractions $\delta_{r_1 r_2}$ and $\delta_{s_1 s_2}$ on (\ref{eq:examp}) yields
\begin{eqnarray}
   \int \phi^{(0)} \tilde{H}^{(3)}_{i} \tilde{H}^{(3)}_{j} d^3\tc = 10 \delta_{i j},
\end{eqnarray}
which can be also verified by direct calculation. (Note that it is important to apply contractions
on (\ref{eq:examp}) as stated
above, and not accidentally as $\delta_{r_1 s_1}\delta_{r_2 s_2}$ which would yield an erroneous coefficient 20, the contraction 
must satisfy definition (\ref{eq:cica})). Then one calculates Hermite moment
\begin{equation}
  \tilde{h}_i^{(3)} = \frac{1}{n_a} \int f_a \tilde{H}_{i}^{(3)} d^3c = \frac{2}{p_a} \sqrt{\frac{m_a}{T_a}} \vec{q}_i,
\end{equation}
(which is equal to $\tilde{h}_{ikk}^{(3)}$) and the heat flux perturbation becomes 
\begin{equation} \label{eq:8mom_perturb}
\textrm{8-moment}:\qquad \chi_a = \frac{1}{10} \tilde{h}_i^{(3)} \tilde{H}_{i}^{(3)} = -\,\frac{m_a}{p_a T_a} \big(\vecq_a\cdot\bc_a\big) \Big( 1-\frac{m_a}{5 T_a}c_a^2\Big),
\end{equation}
recovering (\ref{eq:MomPi}).

\subsubsection{Double-checking the fluid moments}
By using the 8-moment perturbation (\ref{eq:8mom_perturb}) (or the perturbation of the 13-moment model (\ref{eq:MomPi})),
it is possible to calculate the heat flux moment,
for example by switching to normalized units
and using integral (\ref{eq:Quse}) valid for any vector $\vecq$, yielding 
\begin{eqnarray}
\textrm{13-moment}:\qquad  q_{ijk} &=& m_a \int c_i c_j c_k f_a^{(0)}(1+\chi_a) d^3 c =
  -\int \tc_i \tc_j \tc_k (\vecq\cdot\tbc)\big(1-\frac{\tc^2}{5} \big) \phi^{(0)} d^3\tc \nn\\
  &=& \frac{2}{5} \big[\bI\vecq\big]^S_{ijk}.
\end{eqnarray}
In contrast, by using the 20-moment perturbation (\ref{eq:20mom_perturb}) and integral (\ref{eq:Qtriple}) yields identity $\bq=\bq$, as it should be. Thus,
the full heat flux tensor can be decomposed as
\begin{equation} \label{eq:PicRit}
\bq = \frac{2}{5} \big[\bI\vecq\big]^S + \boldsymbol{\sigma}',
\end{equation}  
where $\boldsymbol{\sigma}'$ represents the highest-order irreducible part of the heat flux tensor,
and by applying trace at (\ref{eq:PicRit}) it can be verified that $\boldsymbol{\sigma}'$ is traceless.    
Calculation of the 4th-order moment $\br$ yields (with either 10, 13 or 20-moment model)
\begin{eqnarray}
  r_{ijkl} &=& m_a \int c_i c_j c_k c_l f_a^{(0)}(1+\chi_a) d^3 c\nn\\
  &=& \frac{p_a^2}{\rho_a}\int \tc_i\tc_j \tc_k\tc_l \phi^{(0)}d^3\tc    + \frac{p_a}{2\rho_a} \int \tc_i\tc_j \tc_k\tc_l \tbc\tbc:\bPi^{(2)} \phi^{(0)}d^3\tc\nn\\
  &=&  \frac{p_a^2}{\rho_a}\Big[ \delta_{ij}\delta_{kl} +\delta_{ik}\delta_{jl} +\delta_{jk}\delta_{il} \Big]
  +\frac{p_a}{\rho_a}\Big[ \delta_{ij}\Pi_{kl}^{(2)}+\delta_{ik}\Pi_{jl}^{(2)} +\delta_{il}\Pi_{jk}^{(2)}+\delta_{jk}\Pi_{il}^{(2)}
    +\delta_{jl}\Pi_{ik}^{(2)} +\delta_{kl}\Pi_{ij}^{(2)}\Big], \label{eq:X4mom}
\end{eqnarray}
where one can use integrals (\ref{eq:Intt1}), (\ref{eq:Intt2}). Applying trace at (\ref{eq:X4mom}) yields
\begin{eqnarray}
  \trace\br = 5 \frac{p_a^2}{\rho_a}\bI+7\frac{p_a}{\rho_a} \bPi^{(2)};\qquad \trace\trace\br = 15 \frac{p_a^2}{\rho_a}.
\end{eqnarray}

If one does not want to use our provided integrals from Appendix \ref{sec:IntPic} (or wants to verify them),
all the needed integrals can be calculated by using the powerfull orthogonality theorem. As an example 
\begin{eqnarray}
  \int \tc_i\tc_j \tc_k \tilde{H}^{(3)}_{lmn} \phi^{(0)} d^3\tc =
  \int \tilde{H}_{ijk}^{(3)} \tilde{H}_{lmn}^{(3)}\phi^{(0)} d^3\tc
  + \int  \big( \delta_{ij} \tc_k +\delta_{jk} \tc_i +\delta_{ik} \tc_j \big) \tilde{H}_{lmn}^{(3)}\phi^{(0)} d^3\tc,
\end{eqnarray}
where the first term is calculated with orthogonality (\ref{eq:examp}), and the second term is zero
 (because all the resulting terms can be re-written as $\tilde{H}_i^{(1)} \tilde{H}_{lmn}^{(3)}$ which yields zero after integration;
see also integral (\ref{eq:ThierryI})).  

In some calculations, one actually does not need to work with the complicated right hand side of (\ref{eq:examp}), because
once the integral is calculated, the result is going to be applied on $\tilde{h}_{s_1 s_2 s_3}^{(3)}$, which is a fluid variable and symmetric in all of
its indices. Let us demonstrate it by using the 20-moment heat flux perturbation
\begin{eqnarray} \label{eq:pert1}
\chi_a =  \frac{1}{6} \tilde{h}_{s_1 s_2 s_3}^{(3)} \tilde{H}_{s_1 s_2 s_3}^{(3)}(\tbc),
\end{eqnarray}
and calculate the heat flux moment again, this time with the Hermite variables
\begin{eqnarray}
  q_{r_1 r_2 r_3} &=& m_a \int c_{r_1} c_{r_2} c_{r_3} f_a^{(0)}(1+\chi_a) d^3 c\nn\\
  &=& \frac{p_a}{6}  \sqrt{\frac{T_a}{m_a}} \tilde{h}_{s_1 s_2 s_3}^{(3)} \int \tc_{r_1} \tc_{r_2} \tc_{r_3} \tilde{H}^{(3)}_{s_1 s_2 s_3} \phi^{(0)} d^3\tc \nn\\
  &=&  \frac{p_a}{6}  \sqrt{\frac{T_a}{m_a}} \tilde{h}_{s_1 s_2 s_3}^{(3)} \int \tilde{H}^{(3)}_{r_1 r_2 r_3} \tilde{H}^{(3)}_{s_1 s_2 s_3} \phi^{(0)} d^3\tc \nn\\
  &=&  \frac{p_a}{6}  \sqrt{\frac{T_a}{m_a}} \tilde{h}_{s_1 s_2 s_3}^{(3)} \boldsymbol{\delta}^{(3)}_{(r_1 r_2 r_3)(s_1 s_2 s_3)}\nn\\
  &=& p_a\sqrt{\frac{T_a}{m_a}} \tilde{h}_{r_1 r_2 r_3}^{(3)}.
\end{eqnarray}  
In the derivation, we did not use the complicated right hand side of (\ref{eq:examp}), we only used
$\tilde{h}_{\boldsymbol{r}}^{(n)}\boldsymbol{\delta}^{(n)}_{\boldsymbol{rs}}=n!\tilde{h}_{\boldsymbol{s}}^{(n)}$, and the factor of $3!$ cancelled out as well. 

Similarly, using the same perturbation (\ref{eq:pert1}) one can derive the 5th-order fluid moment $\bX^{(5)}$,
by using the Hermite polynomial $\tilde{H}^{(5)}_{ijklm}$, equation (\ref{eq:H5pol}), according to
\begin{eqnarray}
  X^{(5)}_{r_1 r_2 r_3 r_4 r_5} &=& m_a \int c_{r_1} c_{r_2} c_{r_3} c_{r_4} c_{r_5} f_a^{(0)}(1+\chi_a) d^3 c\nn\\
  &=& \frac{p_a}{6}  \Big(\frac{T_a}{m_a}\Big)^{3/2}
  \tilde{h}_{s_1 s_2 s_3}^{(3)} \int \tc_{r_1} \tc_{r_2} \tc_{r_3}\tc_{r_4}\tc_{r_5} \tilde{H}^{(3)}_{s_1 s_2 s_3} \phi^{(0)} d^3\tc \nn\\
  &=& \frac{p_a}{6}  \Big(\frac{T_a}{m_a}\Big)^{3/2} \tilde{h}_{s_1 s_2 s_3}^{(3)} \int \Big[
    \delta_{r_1 r_2}\tc_{r_3}\tc_{r_4}\tc_{r_5} +\delta_{r_2 r_3}\tc_{r_4} \tc_{r_1} \tc_{r_5}
    +\delta_{r_3 r_4 }\tc_{r_1} \tc_{r_2} \tc_{r_5} \nn\\
  && \qquad +\delta_{r_4 r_1} \tc_{r_2} \tc_{r_3} \tc_{r_5} +\delta_{r_1 r_3}\tc_{r_2} \tc_{r_4} \tc_{r_5} 
    +\delta_{r_2 r_4} \tc_{r_1} \tc_{r_3} \tc_{r_5}  +\delta_{r_1 r_5} \tc_{r_2} \tc_{r_3} \tc_{r_4} \nn\\
  &&\qquad  +\delta_{r_2 r_5} \tc_{r_1} \tc_{r_3} \tc_{r_4} +\delta_{r_3 r_5} \tc_{r_1} \tc_{r_2} \tc_{r_4}
    +\delta_{r_4 r_5} \tc_{r_1} \tc_{r_2} \tc_{r_3} \Big]\tilde{H}^{(3)}_{s_1 s_2 s_3} \phi^{(0)} d^3\tc \nn\\
  &=& p_a \Big(\frac{T_a}{m_a}\Big)^{3/2} \Big[
    \delta_{r_1 r_2} \tilde{h}_{r_3 r_4 r_5}^{(3)} +\delta_{r_2 r_3} \tilde{h}_{r_4 r_1 r_5}^{(3)} 
    +\delta_{r_3 r_4 } \tilde{h}_{r_1 r_2 r_5}^{(3)} \nn\\
  && \qquad +\delta_{r_4 r_1} \tilde{h}_{r_2 r_3 r_5}^{(3)} +\delta_{r_1 r_3} \tilde{h}_{r_2 r_4 r_5}^{(3)} 
    +\delta_{r_2 r_4} \tilde{h}_{r_1 r_3 r_5}^{(3)}  +\delta_{r_1 r_5} \tilde{h}_{r_2 r_3 r_4}^{(3)} \nn\\
  &&\qquad  +\delta_{r_2 r_5} \tilde{h}_{r_1 r_3 r_4}^{(3)} +\delta_{r_3 r_5} \tilde{h}_{r_1 r_2 r_4}^{(3)}
    +\delta_{r_4 r_5} \tilde{h}_{r_1 r_2 r_3}^{(3)} \Big].
\end{eqnarray}
Or rewriten with the heat fluxes according to (\ref{eq:HerMom}) and using usual indices
\begin{eqnarray}
X^{(5)}_{ijklm} &=& \frac{p_a}{\rho_a}\Big[ \delta_{ij} q_{k l m} +\delta_{jk} q_{l i m}
  +\delta_{kl} q_{i j m} +\delta_{li} q_{j k m}  +\delta_{ik} q_{j l m} \nn\\ 
&&   +\delta_{jl} q_{i k m} +\delta_{im} q_{j k l}+\delta_{jm} q_{i k l}+\delta_{km} q_{i j l}
  +\delta_{lm} q_{i j k} \Big], \label{eq:X5mom}
\end{eqnarray}
and by using heat flux decomposition (\ref{eq:PicRit}) with $\boldsymbol{\sigma}'$ neglected
\begin{eqnarray}
  X^{(5)}_{ijklm} &=& \frac{4}{5}\frac{p_a}{\rho_a} \Big[ q_i \big( \delta_{jk}\delta_{lm}+\delta_{kl}\delta_{jm}+\delta_{jl}\delta_{km}\big)
     + q_j \big( \delta_{kl}\delta_{im} +\delta_{il}\delta_{km}+\delta_{ik}\delta_{lm}\big)\nn\\
    && + q_k \big( \delta_{ij}\delta_{lm}+\delta_{il}\delta_{jm}+\delta_{jl}\delta_{im}\big)
     + q_l \big( \delta_{ij}\delta_{km}+\delta_{jk}\delta_{im}+\delta_{ik}\delta_{jm}\big)\nn\\
    && + q_m \big( \delta_{ij}\delta_{kl}+\delta_{jk}\delta_{il}+\delta_{ik}\delta_{jl}\big)\Big]. \label{eq:Nomore23}
\end{eqnarray}
Applying contractions at (\ref{eq:X5mom}) yields
\begin{eqnarray}
  \big[\trace \bX^{(5)} \big]_{ijk} &=& \frac{p_a}{\rho_a}\Big[ 2\big(\bI\vecq\big)^S +9\bq\Big]_{ijk}
  = \frac{28}{5}\frac{p_a}{\rho_a} \big( \bI\vecq\big)^S_{ijk} + 9\frac{p_a}{\rho_a} \sigma_{ijk}\,';\nn\\
  \vecX^{(5)} &=& \trace \trace \bX^{(5)}  =  28\frac{p_a}{\rho_a}\vecq. 
\end{eqnarray}


\newpage
\subsection{Higher-order perturbations (full \texorpdfstring{$\bX^{(4)}$}{X4} and \texorpdfstring{$\bX^{(5)}$}{X5} moments)}
By using the technique described above, it is possible to use the following higher-order perturbation
\begin{equation}
  \chi_a = \frac{1}{2!} \tilde{h}_{s_1 s_2}^{(2)} \tilde{H}_{s_1 s_2}^{(2)}+ \frac{1}{3!} \tilde{h}_{s_1 s_2 s_3}^{(3)} \tilde{H}_{s_1 s_2 s_3}^{(3)}
  + \frac{1}{4!} \tilde{h}_{s_1 s_2 s_3 s_4}^{(4)} \tilde{H}_{s_1 s_2 s_3 s_4}^{(4)}+\frac{1}{5!} \tilde{h}_{s_1 s_2 s_3 s_4 s_5}^{(5)} \tilde{H}_{s_1 s_2 s_3 s_4 s_5}^{(5)},
  \label{eq:scaryP}
\end{equation}
and directly calculate fluid moments (we use $\bX^{(4)}$ instead of $\br$ from now on)
\begin{eqnarray}
  \big[\bX^{(4)}\big]_{r_1 r_2 r_3 r_4} &=& \frac{p_a^2}{\rho_a} \Big[ \tilde{h}_{r_1 r_2 r_3 r_4}^{(4)}
    +\delta_{r_1 r_2}\tilde{h}^{(2)}_{r_3 r_4}+\delta_{r_2 r_3}\tilde{h}^{(2)}_{r_1 r_4}
    +\delta_{r_3 r_4}\tilde{h}^{(2)}_{r_1 r_2} \nn\\
    && \qquad + \delta_{r_1 r_4}\tilde{h}^{(2)}_{r_2 r_3}+\delta_{r_1 r_3}\tilde{h}^{(2)}_{r_2 r_4} + \delta_{r_2 r_4}\tilde{h}^{(2)}_{r_1 r_3}\nn\\
    && \qquad +\delta_{r_1 r_2}\delta_{r_3 r_4}+\delta_{r_1 r_3}\delta_{r_2 r_4}+\delta_{r_2 r_3}\delta_{r_1 r_4}\Big], \label{eq:X4full}
\end{eqnarray}    
and
\begin{eqnarray}
  \big[\bX^{(5)}\big]_{r_1 r_2 r_3 r_4 r_5} &=& p_a \Big(\frac{T_a}{m_a} \Big)^{3/2} \Big[\tilde{h}_{r_1 r_2 r_3 r_4 r_5}^{(5)}
     +\delta_{r_1 r_2} \tilde{h}_{r_3 r_4 r_5}^{(3)} +\delta_{r_2 r_3} \tilde{h}_{r_4 r_1 r_5}^{(3)} 
    +\delta_{r_3 r_4 } \tilde{h}_{r_1 r_2 r_5}^{(3)} \nn\\
  && \qquad +\delta_{r_4 r_1} \tilde{h}_{r_2 r_3 r_5}^{(3)} +\delta_{r_1 r_3} \tilde{h}_{r_2 r_4 r_5}^{(3)} 
    +\delta_{r_2 r_4} \tilde{h}_{r_1 r_3 r_5}^{(3)}  +\delta_{r_1 r_5} \tilde{h}_{r_2 r_3 r_4}^{(3)} \nn\\
  &&\qquad  +\delta_{r_2 r_5} \tilde{h}_{r_1 r_3 r_4}^{(3)} +\delta_{r_3 r_5} \tilde{h}_{r_1 r_2 r_4}^{(3)}
    +\delta_{r_4 r_5} \tilde{h}_{r_1 r_2 r_3}^{(3)} \Big]. \label{eq:X5full}
\end{eqnarray}
Both results contain new contributions, represented by the $\tilde{h}_{r_1 r_2 r_3 r_4}^{(4)}$ and $\tilde{h}_{r_1 r_2 r_3 r_4 r_5}^{(5)}$.

It is useful to introduce notation where by applying contraction at a tensor, the contracted indices will be suppressed,
so for example $\tilde{h}_i^{(3)}\equiv \tilde{h}_{ikk}^{(3)}$, or $X^{(4)}_{ij} \equiv X^{(4)}_{ijkk}$ and $X^{(4)} \equiv X^{(4)}_{iikk}$.
We define all the contractions without any additional factors, with the sole exception of the heat flux vector $\vecq$ where
the additional factor of $1/2$ is present, to match its usual definition. 
To emphasize this
difference, in the index notation we thus keep an arrow on the components of the heat flux vector $\vec{q}_i$, to clearly distinguish it from the
contracted tensor $q_{ijk}$.

By applying contractions at (\ref{eq:X4full}), (\ref{eq:X5full}) then yields 
\begin{eqnarray}
X^{(4)}_{ij}  &=& \frac{p_a^2}{\rho_a} \big[ \tilde{h}^{(4)}_{ij} +7 \tilde{h}^{(2)}_{ij} +5\delta_{ij}\big];\nn\\
X^{(4)}      &=& \frac{p_a^2}{\rho_a} \big[ \tilde{h}^{(4)}+15\big];\nn\\  
X^{(5)}_{ijk} &=& \frac{p_a^2}{\rho_a} \sqrt{\frac{T_a}{m_a}}\Big[ \tilde{h}^{(5)}_{ijk}
    +\delta_{ij} \tilde{h}^{(3)}_{k}  +\delta_{jk} \tilde{h}^{(3)}_{i}  +\delta_{ik} \tilde{h}^{(3)}_{j} +9\tilde{h}^{(3)}_{ijk}\Big];\nn\\
X^{(5)}_{i}  &=& \frac{p_a^2}{\rho_a} \sqrt{\frac{T_a}{m_a}}\Big[ \tilde{h}^{(5)}_{i} +14 \tilde{h}^{(3)}_{i}\Big],
\end{eqnarray}
and applying contractions at the Hermite moments (\ref{eq:HerMom})-(\ref{eq:HerMomL}) yields
\begin{eqnarray}
&&  \tilde{h}_{ij}^{(2)} = \frac{1}{p_a}\Pi_{ij}^{(2)}; \qquad  \tilde{h}_i^{(3)} = \frac{2}{p_a} \sqrt{\frac{m_a}{T_a}} \vec{q}_i;\nn\\
&&  \tilde{h}_{ij}^{(4)} = \frac{\rho_a}{p_a^2} X_{ij}^{(4)} -5\delta_{ij}-\frac{7}{p_a}\Pi_{ij}^{(2)};\nn\\
&&  \tilde{h}^{(4)} =  \frac{\rho_a}{p_a^2} X^{(4)} -15 ;\nn\\   
&& \tilde{h}_{ijk}^{(5)} = \frac{1}{p_a}\sqrt{\frac{m_a}{T_a}} \Big[ \frac{\rho_a}{p_a}X_{ijk}^{(5)}
    -\big( 2\delta_{ij}\vec{q}_k+2\delta_{jk}\vec{q}_i+2\delta_{ik}\vec{q}_j+9 q_{ijk}\big)\Big];\nn\\
&&  \tilde{h}_i^{(5)} = \frac{1}{p_a}\sqrt{\frac{m_a}{T_a}} \Big( \frac{\rho_a}{p_a} X^{(5)}_{i} - 28  \vec{q}_i \Big).  
\end{eqnarray}

\subsubsection{Viscosity \texorpdfstring{$\Pi^{(4)}_{ij}$}{Pi4} of the 4th-order moment \texorpdfstring{$X_{ij}^{(4)}$}{X4}} \label{sec:Pi4}
The usual visocity tensor is defined as a traceless matrix
\begin{equation}
\Pi_{ij}^{(2)} = m_a \int \big( c_i c_j -\frac{1}{3}\delta_{ij} c^2 \big) f_a d^3c.
\end{equation}  
Similarly, it is beneficial to introduce a traceless viscosity tensor of the 4th-order fluid moment
\begin{equation}
  \Pi_{ij}^{(4)} = m_a \int \big( c_i c_j -\frac{1}{3}\delta_{ij} c^2 \big) c^2 f_a d^3c. 
\end{equation}
In another words, the moment $X_{ij}^{(4)}$ is decomposed as
\begin{equation}  \label{eq:DefinePi4}
  X_{ij}^{(4)} = \frac{\delta_{ij}}{3} X^{(4)} +\Pi_{ij}^{(4)},
\end{equation}
where the fully contracted $X^{(4)} = m_a \int c^4 f_a d^3c$. Scalar $X^{(4)}$ is further decomposed to its ``core'' Maxwellian part, and additional 
perturbation $\widetilde{X}^{(4)}$ (with wide tilde) according to
\begin{equation}
X^{(4)} = 15 \frac{p_a^2}{\rho_a} +\widetilde{X}^{(4)},
\end{equation}
and the corresponding Hermite moments thus become
\begin{eqnarray}
  \tilde{h}_{ij}^{(4)} &=& \frac{\rho_a}{p_a^2} \frac{\delta_{ij}}{3} \widetilde{X}^{(4)} +  \frac{\rho_a}{p_a^2}\Pi_{ij}^{(4)} - \frac{7}{p_a}\Pi_{ij}^{(2)};\nn\\
  \tilde{h}^{(4)} &=& \frac{\rho_a}{p_a^2} \widetilde{X}^{(4)}.
\end{eqnarray}
It is important to emphasize that depending on the choice of perturbation $\chi_a$, in general $\widetilde{X}^{(4)}$ is non-zero. 
However, this perturbation is not required to derive the model of \cite{Braginskii1965}, and for example 
\cite{Balescu1988}  prescribes irreducible $h^{(4)}=0$. In the next section we will consider simplified perturbations and derive the
above results in a more direct manner, nevertheless, the more general case (\ref{eq:scaryP}) is a very useful guide that it is 
possible to consider perturbations with non-zero $h^{(4)}$. 

Finally, because the reducible matrix $\tilde{h}_{ij}^{(4)}$ is not traceless in general (unless one prescribes Hermite closure
$\tilde{h}^{(4)}=0$ which makes it traceless by definition), it is useful to introduce traceless
\begin{equation}
\hat{h}_{ij}^{(4)} = \tilde{h}_{ij}^{(4)} - \frac{\delta_{ij}}{3}\tilde{h}^{(4)} = \frac{\rho_a}{p_a^2}\Pi_{ij}^{(4)} - \frac{7}{p_a}\Pi_{ij}^{(2)},  
\end{equation} 
where we used hat instead of tilde. 

\subsubsection{Simplified perturbations (21-moment model)}
Instead of working with very complicated perturbations (\ref{eq:scaryP}), it was shown by \cite{Balescu1988} that to obtain
the model of \cite{Braginskii1965}, it is suffucient to work with simplified
\begin{equation} \label{eq:Irr21mom}
  \chi_a = h_{ij}^{(2)} H_{ij}^{(2)} + h_{i}^{(3)} H_{i}^{(3)}  +h_{ij}^{(4)} H_{ij}^{(4)} + h_{i}^{(5)} H_{i}^{(5)}.
\end{equation}
Perturbation (\ref{eq:Irr21mom}) is written with \emph{irreducible} Hermite polynomials (notation without tilde), discussed in the next section.
This perturbation represents 21-moment model, and recovers both the stress-tensor and the heat flux of Braginskii. 
However, the connection between irreducible and reducible Hermite polynomials can be very blurry at first,
and we continue with \emph{reducible} Hermite polynomials. 

Applying contractions at the hierarchy of reducible polynomials (\ref{eq:H5pol}) yields 
\begin{eqnarray}
  \tilde{H}^{(3)}_i &=& \tc_i ( \tc^2-5); \qquad \tilde{H}^{(5)}_i = \tc_i ( \tc^4-14\tc^2+35);\nn\\
   \tilde{H}^{(2)}_{ij} &=& \tc_i \tc_j -\delta_{ij};\qquad
 \tilde{H}^{(4)}_{ij} = \tc_i \tc_j(\tc^2-7) -\delta_{ij}(\tc^2-5). 
\end{eqnarray}  
By using these polynomials, the Hermite moments then calculate 
\begin{eqnarray}
  &&  \tilde{h}_i^{(3)} = \frac{2}{p_a} \sqrt{\frac{m_a}{T_a}} \vec{q}_i;\qquad
  \tilde{h}_i^{(5)} = \frac{1}{p_a}\sqrt{\frac{m_a}{T_a}} \Big( \frac{\rho_a}{p_a} X^{(5)}_{i} - 28  \vec{q}_i \Big);\nn\\
&&  \tilde{h}_{ij}^{(2)} = \frac{1}{p_a}\Pi_{ij}^{(2)};\qquad
  \hat{h}_{ij}^{(4)} = \frac{\rho_a}{p_a^2}\Pi_{ij}^{(4)} - \frac{7}{p_a}\Pi_{ij}^{(2)}, \nn
\end{eqnarray}
of course recovering previous results. The reducible Hermite polynomials satisfy following orthogonality relations
\begin{eqnarray}
  && \int  \tilde{H}^{(3)}_{i} \tilde{H}^{(3)}_{j} \phi^{(0)} d^3\tc = 10 \delta_{i j};\qquad 
   \int \tilde{H}^{(5)}_i \tilde{H}^{(5)}_j \phi^{(0)} d^3\tc = 280\delta_{ij}; \nn\\
&& \int \tilde{H}^{(2)}_{i j} \tilde{H}^{(2)}_{k l} \phi^{(0)} d^3\tc = \delta_{ik}\delta_{jl} + \delta_{il}\delta_{jk};\nn\\
&& \int \tilde{H}^{(4)}_{ij} \tilde{H}^{(4)}_{kl} \phi^{(0)} d^3\tc = 14 \big(\delta_{ik}\delta_{jl}+\delta_{il}\delta_{jk}\big)+4\delta_{ij}\delta_{kl},  
\end{eqnarray}
and because Hermite moments $\tilde{h}^{(2)}_{kl}$, $\hat{h}^{(4)}_{kl}$ are symmetric and traceless
\begin{eqnarray}
\tilde{h}^{(2)}_{kl} \int \tilde{H}^{(2)}_{i j} \tilde{H}^{(2)}_{k l} \phi^{(0)} d^3\tc = 2 \tilde{h}^{(2)}_{ij};\qquad
 \hat{h}^{(4)}_{kl} \int \tilde{H}^{(4)}_{ij} \tilde{H}^{(4)}_{kl} \phi^{(0)} d^3\tc = 28 \hat{h}^{(4)}_{ij}.  
\end{eqnarray}
Thus, a perturbation which can be directly derived from the hierarchy of reducible Hermite polynomials
(with no reference to irreducible Hermite polynomials or Laguerre-Sonine polynomials) reads
\begin{equation}
  \chi_a = \frac{1}{2}\tilde{h}_{ij}^{(2)} \tilde{H}_{ij}^{(2)} + \frac{1}{10} \tilde{h}_{i}^{(3)} \tilde{H}_{i}^{(3)}
  +\frac{1}{28} \hat{h}_{ij}^{(4)} \tilde{H}_{ij}^{(4)} + \frac{1}{280} \tilde{h}_{i}^{(5)} \tilde{H}_{i}^{(5)}, \label{eq:RedXi}
\end{equation}
where each term is calculated as
\begin{eqnarray}
  \frac{1}{10} \tilde{h}_i^{(3)} \tilde{H}_i^{(3)} &=& \frac{1}{5 p_a}\sqrt{\frac{m_a}{T_a}} (\vecq_a\cdot\tbc_a)(\tc_a^2-5);\nn\\
  \frac{1}{280} \tilde{h}_i^{(5)} \tilde{H}_i^{(5)} &=& \frac{1}{280 p_a}\sqrt{\frac{m_a}{T_a}}
  \Big[\frac{\rho_a}{p_a} (\vecX^{(5)}_a \cdot\tbc_a) -28 (\vecq_a\cdot\tbc_a) \Big] (\tc_a^4-14\tc_a^2+35);\nn\\ 
  \frac{1}{2}\tilde{h}_{ij}^{(2)} \tilde{H}_{ij}^{(2)} &=& \frac{1}{2p_a} \big(\bPi_a^{(2)}:\tbc_a\tbc_a\big);\nn\\
  \frac{1}{28}\hat{h}_{ij}^{(4)} \tilde{H}_{ij}^{(4)} &=& \frac{1}{28}\Big[
    \frac{\rho_a}{p_a^2} \big(\bPi^{(4)}_a:\tbc_a\tbc_a \big)  -\frac{7}{p_a} \big(\bPi_a^{(2)}:\tbc_a\tbc_a\big)\Big] (\tc_a^2-7), \label{eq:Fic1}
\end{eqnarray}
with normalized velocity $\tbc_a=\sqrt{m_a/T_a}\bc_a$. Bellow we show that perturbation (\ref{eq:RedXi}), (\ref{eq:Fic1})
is equivalent to the perturbation of Balescu (\ref{eq:Irr21mom})
obtained with irreducible polynomials.
The final perturbation of the 21-moment model which recovers \cite{Braginskii1965} thus reads
\begin{eqnarray}
  \chi_a &=& \frac{1}{2p_a} \big(\bPi_a^{(2)}:\tbc_a\tbc_a\big) +\frac{1}{28}\Big[
    \frac{\rho_a}{p_a^2} \big(\bPi^{(4)}_a:\tbc_a\tbc_a \big)  -\frac{7}{p_a} \big(\bPi_a^{(2)}:\tbc_a\tbc_a\big)\Big] (\tc_a^2-7)\nn\\
&&  +\frac{1}{5 p_a}\sqrt{\frac{m_a}{T_a}} (\vecq_a\cdot\tbc_a)(\tc_a^2-5) + \frac{1}{280 p_a}\sqrt{\frac{m_a}{T_a}}
  \Big[\frac{\rho_a}{p_a} (\vecX^{(5)}_a \cdot\tbc_a) -28 (\vecq_a\cdot\tbc_a) \Big] (\tc_a^4-14\tc_a^2+35).
\end{eqnarray}

Finally, because $\hat{h}_{ij}^{(4)}$ is traceless, its double contraction with $\tilde{H}_{ij}^{(4)}$ makes the part of this
polynomial proportional to $\delta_{ij}$ redundant in the final perturbation. It is possible to define another traceless polynomial
(with hat instead of tilde)
\begin{equation}
\hat{H}_{ij}^{(4)} = \tilde{H}_{ij}^{(4)} -\frac{\delta_{ij}}{3}\tilde{H}^{(4)} =  \big(\tc_i \tc_j -\frac{\delta_{ij}}{3}\tc^2 \big)(\tc^2-7), \label{eq:Nomore63}
\end{equation}  
and replace the following term in the perturbation (\ref{eq:RedXi})
\begin{equation}
\hat{h}_{ij}^{(4)} \tilde{H}_{ij}^{(4)} = \hat{h}_{ij}^{(4)} \hat{H}_{ij}^{(4)},
\end{equation}
where the part of (\ref{eq:Nomore63}) proportional to $\delta_{ij}$ is still supressed in the final perturbation.
However, the traceless definition (\ref{eq:Nomore63}) makes it possible to now directly define the traceless Hermite moment $\hat{h}_{ij}^{(4)}$
as an integral over $\hat{H}_{ij}^{(4)}$
\begin{equation}
\hat{h}_{ij}^{(4)} = \frac{1}{n_a}\int f_a \hat{H}_{ij}^{(4)} d^3 c. 
\end{equation}  
This is the main motivation behind irreducible Hermite polynomials, as is further clarified below. 

\newpage
\subsection{Irreducible Hermite polynomials} \label{sec:Irreducible}
In the work of \cite{Balescu1988}, the \emph{irreducible} Hermite polynomials are defined through Laguerre-Sonine polynomials, according to
(see equation (G1.4.4), page 326 of Balescu)
\begin{eqnarray}
  H^{(2n)} (\tc) &=& L_n^{1/2} (\frac{\tc^2}{2});\nn\\
  H^{(2n+1)}_i (\tbc) &=& \sqrt{\frac{3}{2}} \tc_i L_n^{3/2}(\frac{\tc^2}{2});\nn\\
  H^{(2n)}_{ij} (\tbc) &=&  \sqrt{\frac{15}{8}}(\tc_i\tc_j-\frac{\tc^2}{3}\delta_{ij})L_{n-1}^{5/2}(\frac{\tc^2}{2}).
\end{eqnarray}
To recover the model \cite{Braginskii1965}, one only needs (see Table 4.1, page 327 of Balescu)
\begin{eqnarray} 
&&  H^{(3)}_i = \frac{1}{\sqrt{10}}\tc_i (\tc^2-5);\qquad 
  H^{(5)}_i = \frac{1}{2\sqrt{70}}\tc_i (\tc^4-14\tc^2+35);\nn\\
&&  H^{(2)}_{ij} = \frac{1}{\sqrt{2}}\big(\tc_i \tc_j -\frac{1}{3}\tc^2\delta_{ij}\big);\qquad  
  H^{(4)}_{ij} = \frac{1}{2\sqrt{7}}(\tc_i\tc_j-\frac{1}{3}\tc^2\delta_{ij})(\tc^2-7), \label{Odd_Ir}
\end{eqnarray}
yielding Hermite moments 
\begin{eqnarray}
  h^{(3)}_i &=& \sqrt{\frac{2}{5}} \frac{1}{p_a} \sqrt{\frac{m_a}{T_a}} \vec{q}_i;\qquad
  h^{(5)}_i =  \frac{1}{2\sqrt{70} p_a} \sqrt{\frac{m_a}{T_a}} \Big[ \frac{\rho_a}{p_a} X^{(5)}_i -28 \vec{q}_i \Big];\nn\\
  h^{(2)}_{ij} &=& \frac{1}{\sqrt{2} p_a} \Pi_{ij}^{(2)};\qquad
  h^{(4)}_{ij} = \frac{1}{2\sqrt{7} p_a}\Big[
    \frac{\rho_a}{p_a} \Pi^{(4)}_{ij} - 7 \Pi_{ij}^{(2)}\Big].
\end{eqnarray}
Furthermore, orthogonal relations are
\begin{equation}
  \int \phi^{(0)} H_i^{(2n+1)} H_j^{(2n+1)} d^3c =\delta_{ij}; \qquad
   h_{kl}^{(2n)}\int \phi^{(0)} H_{ij}^{(2n)} H_{kl}^{(2n)} d^3c = h_{ij}^{(2n)},
\end{equation}  
yielding perturbation (\ref{eq:Irr21mom}), which then recovers perturbation (\ref{eq:RedXi}), (\ref{eq:Fic1}) obtained with
reducible polynomials. Both approaches are therefore equivalent, which is further addressed in Section \ref{sec:NthGen}. 
\subsubsection{Higher-order tensorial ``anisotropies''}
It is useful to clarify what contributions are obtained by using irreducible Hermite polynomials
\begin{eqnarray}
&&  {H}^{(3)}_{ijk}(\tbc) = \tc_i \tc_j \tc_k - \frac{1}{5}\tc^2\big( \delta_{ij} \tc_k +\delta_{jk} \tc_i +\delta_{ik} \tc_j \big);\nn\\
&&  {H}^{(4)}_{ijkl}(\tbc) = \tc_i \tc_j \tc_k \tc_l - \frac{1}{7}\tc^2 \big( \delta_{ij}\tc_k\tc_l +\delta_{jk}\tc_l\tc_i +\delta_{kl}\tc_i\tc_j
  +\delta_{li}\tc_j\tc_k+\delta_{ik}\tc_j\tc_l +\delta_{jl}\tc_i\tc_k\big)\nn\\
&& \qquad \qquad +\frac{1}{35}\tc^4\big( \delta_{ij}\delta_{kl} + \delta_{ik}\delta_{jl} +\delta_{il}\delta_{jk}\big), \label{eq:AnisoBalescu}
\end{eqnarray}
which \cite{Balescu1988} calls ``anisotropies'' (even though they are valid as a perturbation for isotropic Maxwellian).  
Importantly, applying trace on (\ref{eq:AnisoBalescu}) yields zero. 
The corresponding Hermite moments calculate
\begin{eqnarray}
  h^{(3)}_{ijk} &=& \frac{1}{p_a} \sqrt{\frac{m_a}{T_a}} \Big[ \bq-\frac{2}{5}\big(\bI\vecq\big)^S\Big]_{ijk}
  =\frac{1}{p_a} \sqrt{\frac{m_a}{T_a}} \sigma'_{ijk};\qquad   h^{(4)}_{ijkl} = \frac{\rho_a}{p_a^2} \sigma^{(4)'}_{ijkl},
\end{eqnarray}
and directly yield the highest-order irreducible parts.

\newpage
\subsection{Decomposition of \texorpdfstring{$X^{(4)}_{ijkl}$}{X(4)ijkl}} \label{sec:NomoreX}
We continue with the \emph{reducible} Hermite polynomials.
To decompose the full 4th-order fluid moment $X^{(4)}_{ijkl}$, it is necessary to consider the following perturbation  (i.e. the 16-moment model) 
\begin{equation}
  \chi_a = \frac{1}{2}\tilde{h}^{(2)}_{ij}\tilde{H}^{(2)}_{ij}  +\frac{1}{28}\hat{h}^{(4)}_{ij}\tilde{H}^{(4)}_{ij} + \frac{1}{120}\tilde{h}^{(4)}\tilde{H}^{(4)},
  \label{eq:Nomore21}
\end{equation}  
and by using this perturbation to calculate $X^{(4)}_{ijkl}$. In comparision to the previous perturbation of the 21-moment model, the last term
with Hermite polynomial $\tilde{H}^{(4)}$ is new. It is derived with orthogonality relation $\int \phi^{(0)} H^{(4)} H^{(4)} d^3\tc =120$.  
We will need the following integrals. Applying contraction
$\delta_{r_3 r_4}$ at the orthogonality relation (\ref{eq:exampX}) yields
\begin{eqnarray}
 && \int  \tilde{H}^{(4)}_{r_1 r_2} \tc_{s_1} \tc_{s_2}\tc_{s_3}\tc_{s_4} \phi^{(0)} d^3\tc =  \int  \tilde{H}^{(4)}_{r_1 r_2} \tilde{H}^{(4)}_{s_1 s_2 s_3 s_4} \phi^{(0)} d^3\tc =\nn\\
  && \qquad +2\delta_{r_1 s_1}\delta_{r_2 s_2}\delta_{s_3 s_4}
   + 2\delta_{r_1 s_1}\delta_{r_2 s_3}\delta_{s_2 s_4}  + 2\delta_{r_1 s_1}\delta_{r_2 s_4}\delta_{s_2 s_3}\nn\\
  &&\qquad   + 2\delta_{r_1 s_2}\delta_{r_2 s_1}\delta_{s_3 s_4}   + 2\delta_{r_1 s_2}\delta_{r_2 s_3}\delta_{s_1 s_4}  + 2\delta_{r_1 s_2}\delta_{r_2 s_4}\delta_{s_1 s_3}
  \nn\\
 &&\qquad  + 2\delta_{r_1 s_3}\delta_{r_2 s_1}\delta_{s_2 s_4}  + 2\delta_{r_1 s_3}\delta_{r_2 s_2}\delta_{s_1 s_4}  + 2\delta_{r_1 s_3}\delta_{r_2 s_4}\delta_{s_1 s_2}
    \nn\\
  && \qquad + 2\delta_{r_1 s_4}\delta_{r_2 s_1}\delta_{s_2 s_3}  + 2\delta_{r_1 s_4}\delta_{r_2 s_2}\delta_{s_1 s_3}
   + 2\delta_{r_1 s_4}\delta_{r_2 s_3}\delta_{s_1 s_2}, \label{eq:Nomore3}
\end{eqnarray}
and further applying traceless $\hat{h}^{(4)}_{r_1 r_2}$ at (\ref{eq:Nomore3}) leads to
\begin{eqnarray}
 && \hat{h}^{(4)}_{r_1 r_2}\int  \tilde{H}^{(4)}_{r_1 r_2} \tc_{s_1} \tc_{s_2}\tc_{s_3}\tc_{s_4} \phi^{(0)} d^3\tc \nn\\
  && =  4 \Big[ \hat{h}_{s_1 s_2}^{(4)}\delta_{s_3 s_4}
   + \hat{h}_{s_1 s_3}^{(4)}\delta_{s_2 s_4}  + \hat{h}_{s_1 s_4}^{(4)}\delta_{s_2 s_3}
 +  \hat{h}_{s_2 s_3}^{(4)}\delta_{s_1 s_4}+  \hat{h}_{s_2 s_4}^{(4)}\delta_{s_1 s_3} + \hat{h}_{s_3 s_4}^{(4)}\delta_{s_1 s_2}\Big]. \label{eq:Nomore1}
\end{eqnarray}
Applying contraction $\delta_{r_1 r_2}$ at (\ref{eq:Nomore3}) and multiplying by $\tilde{h}^{(4)}$ yields
\begin{eqnarray}
 \tilde{h}^{(4)} \int  \tilde{H}^{(4)} \tc_{s_1} \tc_{s_2}\tc_{s_3}\tc_{s_4} \phi^{(0)} d^3\tc
  = 8 \tilde{h}^{(4)} \Big[ \delta_{s_1 s_2} \delta_{s_3 s_4} + \delta_{s_1 s_3}\delta_{s_2 s_4} +\delta_{s_1 s_4}\delta_{s_2 s_3}\Big]. \label{eq:Nomore11}
\end{eqnarray}
Similarly, 
\begin{eqnarray}
  && \tilde{h}^{(2)}_{r_1 r_2} \int  \tilde{H}^{(2)}_{r_1 r_2} \tc_{s_1} \tc_{s_2}\tc_{s_3}\tc_{s_4} \phi^{(0)} d^3\tc \nn\\
  && = 2 \Big[ \tilde{h}_{s_1 s_2}^{(2)}\delta_{s_3 s_4}
   + \tilde{h}_{s_1 s_3}^{(2)}\delta_{s_2 s_4}  + \tilde{h}_{s_1 s_4}^{(2)}\delta_{s_2 s_3}
+  \tilde{h}_{s_2 s_3}^{(2)}\delta_{s_1 s_4}+  \tilde{h}_{s_2 s_4}^{(2)}\delta_{s_1 s_3} + \tilde{h}_{s_3 s_4}^{(2)}\delta_{s_1 s_2}\Big]. \label{eq:Nomore2}
\end{eqnarray}  
Results (\ref{eq:Nomore1}), (\ref{eq:Nomore11}), (\ref{eq:Nomore2}) allow us to calcuate the $X^{(4)}_{ijkl}$ moment, which becomes
\begin{eqnarray}
  X^{(4)}_{ijkl} &=& m_a \int f_a^{(0)} \Big[ 1+\frac{1}{2}\tilde{h}^{(2)}_{r_1 r_2}\tilde{H}^{(2)}_{r_1 r_2}
    +\frac{1}{28}\hat{h}^{(4)}_{r_1 r_2}\tilde{H}^{(4)}_{r_1 r_2} + \frac{1}{120}\tilde{h}^{(4)}\tilde{H}^{(4)}  \Big] c_i c_j c_k c_l d^3c \nn\\ 
  &=& + \frac{1}{15}\frac{p_a^2}{\rho_a}\Big(15+\tilde{h}^{(4)}\Big)\Big( \delta_{ij}\delta_{kl} +\delta_{ik}\delta_{jl} +\delta_{jk}\delta_{il} \Big)\nn\\
  &&  + \frac{1}{7} \frac{p_a^2}{\rho_a} \Big[ \big( \hat{h}_{ij}^{(4)} +7 \tilde{h}_{ij}^{(2)} \big) \delta_{kl}
    +\big( \hat{h}_{ik}^{(4)} +7 \tilde{h}_{ik}^{(2)} \big) \delta_{jl}
    +\big( \hat{h}_{il}^{(4)} +7 \tilde{h}_{il}^{(2)} \big) \delta_{jk}\nn\\
    && \quad +\big( \hat{h}_{jk}^{(4)} +7 \tilde{h}_{jk}^{(2)} \big) \delta_{il}
    + \big( \hat{h}_{jl}^{(4)} +7 \tilde{h}_{jl}^{(2)} \big) \delta_{ik}
    +\big( \hat{h}_{kl}^{(4)} +7 \tilde{h}_{kl}^{(2)} \big) \delta_{ij}\Big]. \label{eq:Nomore20}
\end{eqnarray}
Form (\ref{eq:Nomore20}) nicely shows how various parts of perturbation (\ref{eq:Nomore21}) contribute to the decomposition, including the
new $\tilde{h}^{(4)}$. Prescribing Hermite closures $\hat{h}_{ij}^{(4)}=0$, $\tilde{h}^{(4)}=0$ recovers decomposition (\ref{eq:X4mom}) used
in the Burgers-Schunk model. Finally, rewritten with fluid moments
\begin{equation}
  \Pi^{(4)}_{ij} = \frac{p_a^2}{\rho_a} \big( \hat{h}_{ij}^{(4)} +7 \tilde{h}_{ij}^{(2)} \big); \qquad \widetilde{X}^{(4)}=\frac{p_a^2}{\rho_a}\tilde{h}^{(4)};
  \qquad X^{(4)} =15\frac{p_a^2}{\rho_a}+\widetilde{X}^{(4)},
\end{equation}
and representing all other terms that were not obtained from (\ref{eq:Nomore21}) by traceless $\sigma^{(4)'}_{ijkl}$
(which represents the highest-order irreducible part of $X_{ijkl}^{(4)}$), the decomposition becomes   
\begin{eqnarray}
  X_{ijkl}^{(4)} &=& \frac{1}{15} X^{(4)} \big(\delta_{ij}\delta_{kl}+\delta_{ik}\delta_{jl}+\delta_{il}\delta_{jk}\big)\nn\\
  && +\frac{1}{7} \Big[ \Pi_{ij}^{(4)} \delta_{kl}+\Pi_{ik}^{(4)}\delta_{jl}+\Pi_{il}^{(4)}\delta_{jk}
  +\Pi_{jk}^{(4)}\delta_{il}+\Pi_{jl}^{(4)}\delta_{ik}+\Pi_{kl}^{(4)}\delta_{ij} \Big] + \sigma_{ijkl}^{(4)'}, \label{eq:Nomore5}
\end{eqnarray}
or equivalently
\begin{eqnarray}
  X_{ijkl}^{(4)} &=& -\,\frac{1}{35} X^{(4)} \big(\delta_{ij}\delta_{kl}+\delta_{ik}\delta_{jl}+\delta_{il}\delta_{jk}\big)\nn\\
  && +\frac{1}{7} \Big[ X^{(4)}_{ij}\delta_{kl}+X^{(4)}_{ik}\delta_{jl}+X^{(4)}_{il}\delta_{jk}+X^{(4)}_{jk}\delta_{il}
  +X^{(4)}_{jl}\delta_{ik}+X^{(4)}_{kl}\delta_{ij} \Big] + \sigma_{ijkl}^{(4)'}.
\end{eqnarray}
Decomposition (\ref{eq:Nomore5}) is equivalent to equation (30.22) of \cite{Grad_1958}. Essentially, any tensorial moment can be decomposed by
subtracting all the possible contractions of that moment. Note that simply prescribing closure $\Pi_{ij}^{(4)}=0$ in (\ref{eq:Nomore5})
would be erroneous, unless one also prescribes $\Pi_{ij}^{(2)}=0$ as well. Correct simplification of (\ref{eq:Nomore5}) is
obtained by prescribing Hermite closure $\hat{h}_{ij}^{(4)}=0$, meaning by prescribing fluid closure $\Pi_{ij}^{(4)}=7(p_a/\rho_a)\Pi_{ij}^{(2)}$.
Additionally, one can also prescribe Hermite closure $\tilde{h}^{(4)}=0$, which is equivalent to fluid closure $\widetilde{X}^{(4)}=0$.

\subsection{Decomposition of \texorpdfstring{$X^{(5)}_{ijklm}$}{X(5)ijklm}} \label{sec:NomoreX2}
We only use simplified perturbation
\begin{equation} \label{eq:Nomore24}
\chi_a =  \frac{1}{10}\tilde{h}_i^{(3)} \tilde{H}_i^{(3)} + \frac{1}{280}\tilde{h}_i^{(5)} \tilde{H}_i^{(5)}.
\end{equation}
By using this perturbation, it is possible to calculate the 5th-order fluid moment
\begin{eqnarray}
  X^{(5)}_{s_1 s_2 s_3 s_4 s_5} &=& \frac{1}{35} \frac{p_a^{5/2}}{\rho_a^{3/2}} \Big[
   \big(\tilde{h}_{s_1}^{(5)} + 14\tilde{h}_{s_1}^{(3)}\big) \big( \delta_{s_2 s_3}\delta_{s_4 s_5}+\delta_{s_2 s_4}\delta_{s_3 s_5} +\delta_{s_2 s_5}\delta_{s_3 s_4}\big) \nn\\
  && +\big(\tilde{h}_{s_2}^{(5)}+ 14\tilde{h}_{s_2}^{(3)}\big) \big( \delta_{s_1 s_3}\delta_{s_4 s_5}+\delta_{s_1 s_4}\delta_{s_3 s_5} +\delta_{s_1 s_5}\delta_{s_3 s_4}\big) \nn\\
  && +\big(\tilde{h}_{s_3}^{(5)}+ 14\tilde{h}_{s_3}^{(3)}\big) \big( \delta_{s_1 s_2}\delta_{s_4 s_5}+\delta_{s_1 s_4}\delta_{s_2 s_5} +\delta_{s_1 s_5}\delta_{s_2 s_4}\big) \nn\\
  && +\big(\tilde{h}_{s_4}^{(5)}+ 14\tilde{h}_{s_4}^{(3)}\big) \big( \delta_{s_1 s_2}\delta_{s_3 s_5}+\delta_{s_1 s_3}\delta_{s_2 s_5} +\delta_{s_1 s_5}\delta_{s_2 s_3}\big) \nn\\
   && +\big(\tilde{h}_{s_5}^{(5)}+ 14\tilde{h}_{s_5}^{(3)}\big) \big( \delta_{s_1 s_2}\delta_{s_3 s_4}+\delta_{s_1 s_3}\delta_{s_2 s_4} +\delta_{s_1 s_4}\delta_{s_2 s_3}\big)\Big].
\end{eqnarray}
Because we considered simplified perturbation (\ref{eq:Nomore24}), we do not consider full decomposition with $\sigma^{(5)}\,'$. 
Prescribing Hermite closure $\tilde{h}^{(5)}_i=0$ yields previously obtained decomposition (\ref{eq:Nomore23}). Finally, by switching from Hermite to fluid moments
\begin{eqnarray}
&&  \tilde{h}_i^{(3)} = 2 \frac{\rho_a^{1/2}}{p_a^{3/2}} \vec{q}_i;\qquad
  \tilde{h}_i^{(5)} = \frac{\rho_a^{1/2}}{p_a^{3/2}} \Big( \frac{\rho_a}{p_a} X^{(5)}_{i} - 28  \vec{q}_i \Big);
\qquad \tilde{h}_i^{(5)} + 14 \tilde{h}_i^{(3)} = \frac{\rho_a^{3/2}}{p_a^{5/2}} X^{(5)}_{i},   
\end{eqnarray}
the decomposition becomes
\begin{eqnarray}
  X^{(5)}_{s_1 s_2 s_3 s_4 s_5} &=& \frac{1}{35} \Big[ 
  X_{s_1}^{(5)} \big( \delta_{s_2 s_3}\delta_{s_4 s_5}+\delta_{s_2 s_4}\delta_{s_3 s_5} +\delta_{s_2 s_5}\delta_{s_3 s_4}\big) \nn\\
  && +X_{s_2}^{(5)} \big( \delta_{s_1 s_3}\delta_{s_4 s_5}+\delta_{s_1 s_4}\delta_{s_3 s_5} +\delta_{s_1 s_5}\delta_{s_3 s_4}\big) \nn\\
  && +X_{s_3}^{(5)} \big( \delta_{s_1 s_2}\delta_{s_4 s_5}+\delta_{s_1 s_4}\delta_{s_2 s_5} +\delta_{s_1 s_5}\delta_{s_2 s_4}\big) \nn\\
  && +X_{s_4}^{(5)} \big( \delta_{s_1 s_2}\delta_{s_3 s_5}+\delta_{s_1 s_3}\delta_{s_2 s_5} +\delta_{s_1 s_5}\delta_{s_2 s_3}\big) \nn\\
  && +X_{s_5}^{(5)} \big( \delta_{s_1 s_2}\delta_{s_3 s_4}+\delta_{s_1 s_3}\delta_{s_2 s_4} +\delta_{s_1 s_4}\delta_{s_2 s_3}\big)\Big].
\end{eqnarray}
As a double-check, applying contraction $\delta_{s_4 s_5}$ at the last expression yields
\begin{eqnarray}
 X^{(5)}_{s_1 s_2 s_3} = \frac{1}{5} \Big[  X_{s_1}^{(5)} \delta_{s_2 s_3} +  X_{s_2}^{(5)} \delta_{s_1 s_3} +  X_{s_3}^{(5)} \delta_{s_1 s_2}\Big],
\end{eqnarray}
and applying another contraction yields an identity. Note that it is not possible to perform closure
$\vecX^{(5)}=0$, such a closure would be erroneous (unless $\vecq=0$ is prescribed as well). Instead, one needs to perform closure at the
Hermite moment $\tilde{h}_i^{(5)}=0$, or in another words, the correct closure is $\vecX^{(5)}=28 (p_a/\rho_a)\vecq$.

\newpage
\subsection{Table of useful integrals} \label{sec:IntPic}
The Hermite polynomials allows one to built the hierarchy of following integrals.
One introduces weight
\begin{equation}
\phi^{(0)} = \frac{e^{-\tc^2/2}}{(2\pi)^{3/2}},
\end{equation}
and for any odd ``m'' the following integral holds
\begin{equation} \label{eq:rule1}
m=\textrm{odd}: \qquad \int \tc_{r_1}\tc_{r_2}\tc_{r_3}\ldots \tc_{r_m}\phi^{(0)}d^3\tc =0.
\end{equation}
The validity of  (\ref{eq:rule1}) can be shown by using ``common-sense'' symmetries and Gaussian integration,
or by rewriting the integral with pairs of Hermite polynomials, one of even-order and one of odd-order,
$\tilde{H}^{(r_m+1)/2}\tilde{H}^{(r_m-1)/2}$ (where the result of integration is zero), and a hierarchy of lower-order
integrals that will also be odd-even pairs, yielding zero. 

A particular case of the orthogonality theorem is, that for any $m\neq 0$, an integral over any single Hermite polynomial with
weight $\phi^{(0)}$ is zero
\begin{equation} \label{eq:rule2}
m\neq 0:\qquad \int \tilde{H}^{(m)}_{r_1 r_2 r_3 \ldots r_m} \phi^{(0)} d^3\tc =0.  
\end{equation}  
The two rules (\ref{eq:rule1}), (\ref{eq:rule2}) allow one to calculate integrals for any even ``m'' number of
velocities $\tbc$, such as $\tbc\tbc\tbc\tbc\tbc\tbc$, which would be otherwise
very difficult to do just by using ``common-sense'' symmetries and Gaussian integration. Actually, for ``m'' being even,
quicker than using (\ref{eq:rule2}) is to rewrite the integrals into $\tilde{H}^{(m/2)}\tilde{H}^{(m/2)}$,
and use ortogonality relations (\ref{eq:examp}).  A very useful integral also reads
\begin{equation} \label{eq:ThierryI}
m < n:\qquad \int  \tc_{r_1}\ldots \tc_{r_m} H^{(n)}_{s_1\ldots s_n} \phi^{(0)}d^3\tc =0,
\end{equation}
which validity is easily shown by rewriting the $\tc_{r_1}\ldots \tc_{r_m}$ with $H^{(m)}_{r_1\ldots r_m}$
  (where the result of integration is zero) and a hierarchy of lower-order Hermite polynomials where the result of
integration is also zero.

It is possible to build the following table when ``m'' is even
\begin{eqnarray}
  \int  \phi^{(0)} d^3\tc &=& 1;\nn\\
  \int \tc_i\tc_j  \phi^{(0)} d^3\tc &=&\delta_{ij};\nn\\
  \int \tc_i\tc_j \tc_k \tc_l  \phi^{(0)} d^3\tc
  &=& \delta_{ij}\delta_{kl} +\delta_{ik}\delta_{jl} +\delta_{jk}\delta_{il}; \label{eq:Intt1}
\end{eqnarray}
\begin{eqnarray}
 && \int  \tc_{r_1}\tc_{r_2}\tc_{r_3}\tc_{s_1}\tc_{s_2}\tc_{s_3} \phi^{(0)} d^3 \tc =
   \delta_{r_1 s_1}\big( \delta_{r_2 s_2}\delta_{r_3 s_3} +\delta_{r_2 s_3}\delta_{r_3 s_2}\big) \nn\\
&&\qquad  +\delta_{r_1 s_2}\big( \delta_{r_2 s_1}\delta_{r_3 s_3}+\delta_{r_2 s_3}\delta_{r_3 s_1}\big) 
   +\delta_{r_1 s_3} \big( \delta_{r_2 s_1}\delta_{r_3 s_2} +\delta_{r_2 s_2}\delta_{r_3 s_1}\big)\nn\\
   && \qquad +\delta_{r_1 r_2} \big( \delta_{s_1 s_2}\delta_{r_3 s_3} +\delta_{s_2 s_3}\delta_{r_3 s_1} + \delta_{s_3 s_1} \delta_{r_3 s_2}\big) \nn\\
   &&\qquad  +\delta_{r_1 r_3} \big( \delta_{s_1 s_2}\delta_{r_2 s_3}+\delta_{s_2 s_3}\delta_{r_2 s_1}+\delta_{s_3 s_1}\delta_{r_2 s_2}\big)\nn\\  
  &&\qquad  +\delta_{r_2 r_3}\big( \delta_{s_1 s_2}\delta_{r_1 s_3} + \delta_{s_2 s_3}\delta_{r_1 s_1} +\delta_{s_3 s_1}\delta_{r_1 s_2}\big).
\end{eqnarray}
These integrals can be used to obtain other useful integrals, for example valid for any (3-dimensional) vector $\vecq$
\begin{eqnarray}
&& \int \tc_i\tc_j \tc_k (\tbc\cdot\vecq) \phi^{(0)} d^3\tc
  = \delta_{ij}\vec{q}_{k}+\delta_{jk}\vec{q}_{i} +\delta_{ki}\vec{q}_{j} =\big[\bI\vecq\big]^S_{ijk};\\
 && \int  \tc_{r_1}\tc_{r_2}\tc_{r_3}\tc_{s_1}\tc_{s_2}(\tbc\cdot\vecq) \phi^{(0)} d^3 \tc =
   \delta_{r_1 s_1}\big( \delta_{r_2 s_2} \vec{q}_{r_3} +\vec{q}_{r_2}\delta_{r_3 s_2}\big) \nn\\
&&\qquad  +\delta_{r_1 s_2}\big( \delta_{r_2 s_1} \vec{q}_{r_3}+\vec{q}_{r_2}\delta_{r_3 s_1}\big) 
   +\vec{q}_{r_1} \big( \delta_{r_2 s_1}\delta_{r_3 s_2} +\delta_{r_2 s_2}\delta_{r_3 s_1}\big)\nn\\
   && \qquad +\delta_{r_1 r_2} \big( \delta_{s_1 s_2}\vec{q}_{r_3} +\vec{q}_{s_2}\delta_{r_3 s_1} + \vec{q}_{s_1} \delta_{r_3 s_2}\big) \nn\\
   &&\qquad  +\delta_{r_1 r_3} \big( \delta_{s_1 s_2}\vec{q}_{r_2}+\vec{q}_{s_2}\delta_{r_2 s_1}+\vec{q}_{s_1}\delta_{r_2 s_2}\big)\nn\\  
  &&\qquad  +\delta_{r_2 r_3}\big( \delta_{s_1 s_2}\vec{q}_{r_1} + \vec{q}_{s_2}\delta_{r_1 s_1} +\vec{q}_{s_1}\delta_{r_1 s_2}\big).
\end{eqnarray}
and by further contractions
\begin{eqnarray}
&&  \int  \tc_{i}\tc_{j}\tc_{k} \tc^2 (\tbc\cdot\vecq) \phi^{(0)} d^3 \tc
  = 7 \big( \delta_{ij}\vec{q}_k + \delta_{jk} \vec{q}_i +\delta_{ki}\vec{q}_j\big) = 7 \big[\bI\vecq\big]^S_{ijk};\\  
&& \int  \tc_{i}\tc_{j}\tc_{k} \Big(1-\frac{\tc^2}{5}\Big) (\tbc\cdot\vecq) \phi^{(0)} d^3 \tc
  = -\frac{2}{5} \big[\bI\vecq\big]^S_{ijk}. \label{eq:Quse}
\end{eqnarray}
As a quick double-check of the above results, by performing further contractions
\begin{eqnarray}
&&  \int \tc_i \tc^2 (\tbc\cdot\vecq) \phi^{(0)} d^3\tc
  = 5\vec{q}_i;\qquad
    \int  \tc_{i} \tc^4 (\tbc\cdot\vecq) \phi^{(0)} d^3 \tc
  = 35 \vec{q}_i,
\end{eqnarray}
which is easy to verify directly.  

Similarly, for a triple contraction with any fully symmetric 3rd-rank tensor $\bq$
\begin{eqnarray}
&&   \int \tc_i\tc_j\tc_k(\tbc\tbc: \bq\cdot\tbc) \phi^{(0)} d^3 \tc
   = 6\big( q_{ijk} + \vec{q}_i \delta_{jk} +\vec{q}_j \delta_{ik} +\vec{q}_k\delta_{ij}\big); \label{eq:Qtriple}\\
&&    \int \tc_i \tc^2(\tbc\tbc: \bq\cdot\tbc) \phi^{(0)} d^3 \tc = 42 \vec{q}_i,
\end{eqnarray}
where one defines vector $\vecq=(1/2)\trace\bq$. Finally, for any ($3\times 3$) matrix $\bA$
\begin{eqnarray}
&&  \int  \tc_{i}\tc_{j} (\tbc\tbc: \bA) \phi^{(0)} d^3 \tc
  =  A_{ij}+A_{ji} +(\trace\bA)\delta_{ij};\\
&&  \int  \tc_{r_1}\tc_{r_2}\tc_{r_3}\tc_{s_1} (\tbc\tbc: \bA) \phi^{(0)} d^3 \tc
  = \delta_{r_1 s_1}\big( A_{r_2 r_3}+A_{r_3 r_2}\big)\nn\\
&& \qquad +\delta_{r_2 s_1}\big( A_{r_1 r_3}+A_{r_3 r_1}\big) +\delta_{r_3 s_1}\big( A_{r_1 r_2}+A_{r_2 r_1}\big)\nn\\
&& \qquad +\delta_{r_1 r_2} \big(A_{r_3 s_1}+ A_{s_1 r_3} + (\trace\bA) \delta_{r_3 s_1}\big) \nn\\
&& \qquad +\delta_{r_1 r_3} \big( A_{r_2 s_1} +A_{s_1 r_2} +(\trace\bA)\delta_{r_2 s_1}\big) \nn\\
&& \qquad +\delta_{r_2 r_3} \big( A_{r_1 s_1} +A_{s_1 r_1} +(\trace\bA) \delta_{r_1 s_1}\big),
\end{eqnarray}  
and by further contractions
\begin{eqnarray}
&&  \int  \tc^2 (\tbc\tbc: \bA) \phi^{(0)} d^3 \tc =  5\trace\bA;\\  
&&   \int  \tc_{i}\tc_{j} \tc^2 (\tbc\tbc: \bA) \phi^{(0)} d^3 \tc
  = 7\big( A_{ij}+A_{ji} +(\trace\bA)\delta_{ij}\big);\\
&& \int  \tc^4 (\tbc\tbc: \bA) \phi^{(0)} d^3 \tc =  35\trace\bA,
\end{eqnarray}
and so for symmetric traceless matrix $\bPi$
\begin{eqnarray}
&&  \int  \tc_{i}\tc_{j} (\tbc\tbc: \bPi) \phi^{(0)} d^3 \tc
  = 2 \Pi_{ij};\\ 
&&   \int  \tc_{i}\tc_{j} \tc^2 (\tbc\tbc: \bPi) \phi^{(0)} d^3 \tc
  = 14 \Pi_{ij};\\
 &&  \int \tc_i\tc_j\tc_k\tc_l (\tbc\tbc: \bPi) \phi^{(0)} d^3 \tc\nn\\
 && \qquad = 2\big( \delta_{ij}\Pi_{kl}+\delta_{ik}\Pi_{jl} +\delta_{il}\Pi_{jk}+\delta_{jk}\Pi_{il}+\delta_{jl}\Pi_{ik} +\delta_{kl}\Pi_{ij}\big).\label{eq:Intt2}
\end{eqnarray}  
A curious reader might find the following integrals useful
\begin{eqnarray}
&& \int \tc_i \tc_j \tc_k \tc_l \tc^2 \phi^{(0)} d^3\tc = 7 \Big[ \delta_{ij}\delta_{kl} + \delta_{ik}\delta_{jl} +\delta_{il}\delta_{jk}\Big];\nn\\
  && \int \tc_i \tc_j \tc_k \tc_l (\tc^2-7) \phi^{(0)} d^3\tc = 0;\nn\\
  && \int \tc_i \tc_j (\tc^2-5)\phi^{(0)} d^3\tc =0;\nn\\
  && \int (\tc^2-3)\phi^{(0)} d^3\tc =0. 
\end{eqnarray}

\newpage
\subsection{General n-th order perturbation} \label{sec:NthGen}
The hierarchy of simplified \emph{reducible} Hermite polymomials (with tilde) can be calculated directly from (\ref{eq:HerM}) as
\begin{eqnarray}
\tilde{H}^{(1)}_i &=& \tc_i;\nn\\  
\tilde{H}^{(2)}_{ij} &=& \tc_i \tc_j -\delta_{ij};\nn\\
\tilde{H}^{(3)}_i   &=& \tc_i ( \tc^2-5);  \nn\\
\tilde{H}^{(4)}_{ij} &=& \tc_i \tc_j(\tc^2-7) -\delta_{ij}(\tc^2-5);\nn\\
\tilde{H}^{(5)}_i   &=& \tc_i ( \tc^4-14\tc^2+35);\nn\\
\tilde{H}^{(6)}_{ij} &=& \tc_i\tc_j (\tc^4-18\tc^2+63) -\delta_{ij}(\tc^4-14\tc^2+35);\nn\\
\tilde{H}^{(7)}_{i} &=& \tc_i (\tc^6-27\tc^4+189\tc^2-315);\nn\\ 
\tilde{H}^{(8)}_{ij} &=& \tc_i\tc_j (\tc^6-33\tc^4+297\tc^2-693) -\delta_{ij}(\tc^6-27\tc^4+189\tc^2-315);\nn\\
\tilde{H}^{(9)}_{i} &=& \tc_i(\tc^8-44\tc^6+594\tc^4-2772\tc^2+3465), \label{eq:Nomore60}
\end{eqnarray}
and fully contracted ones for the even orders are
\begin{eqnarray}
\tilde{H}^{(2)} &=& \tc^2-3;\nn\\
\tilde{H}^{(4)} &=& \tc^4-10\tc^2+15;\nn\\ 
\tilde{H}^{(6)} &=& \tc^6-21\tc^4+105\tc^2-105;\nn\\  
\tilde{H}^{(8)} &=& \tc^8-36\tc^6+378\tc^4-1260\tc^2+945.
\end{eqnarray}
The even-order polynomials $\tilde{H}^{(2n)}_{ij}$ can be rewritten into 
\begin{eqnarray}
\tilde{H}^{(2)}_{ij} &=& \big( \tc_i\tc_j-\frac{\delta_{ij}}{3}\tc^2\big) +\frac{\delta_{ij}}{3} \tilde{H}^{(2)};\nn\\
\tilde{H}^{(4)}_{ij} &=& \big( \tc_i\tc_j-\frac{\delta_{ij}}{3}\tc^2\big) (\tc^2-7) +\frac{\delta_{ij}}{3} \tilde{H}^{(4)};\nn\\
\tilde{H}^{(6)}_{ij} &=& \big( \tc_i\tc_j-\frac{\delta_{ij}}{3}\tc^2\big) (\tc^4-18\tc^2+63) +\frac{\delta_{ij}}{3} \tilde{H}^{(6)};\nn\\
\tilde{H}^{(8)}_{ij} &=& \big( \tc_i\tc_j-\frac{\delta_{ij}}{3}\tc^2\big) (\tc^6-33\tc^4+297\tc^2-693) +\frac{\delta_{ij}}{3} \tilde{H}^{(8)}. \label{eq:BalGen1}
\end{eqnarray}
The orthogonality relations can be calculated as
\begin{eqnarray}
&& \int \phi^{(0)} \tilde{H}^{(1)}_i\tilde{H}^{(1)}_j d^3\tc = \delta_{ij};\qquad  
 \int \phi^{(0)} \tilde{H}^{(2)}\tilde{H}^{(2)} d^3\tc = 6;\nn\\
&& \int \phi^{(0)} \tilde{H}^{(3)}_i\tilde{H}^{(3)}_j d^3\tc = 10 \delta_{ij};\qquad
 \int \phi^{(0)} \tilde{H}^{(4)}\tilde{H}^{(4)} d^3\tc = 120;\nn\\  
&& \int \phi^{(0)} \tilde{H}^{(5)}_i\tilde{H}^{(5)}_j d^3\tc = 280 \delta_{ij};\qquad 
 \int \phi^{(0)} \tilde{H}^{(6)}\tilde{H}^{(6)} d^3\tc = 5040;\nn\\
&& \int \phi^{(0)} \tilde{H}^{(7)}_i\tilde{H}^{(7)}_j d^3\tc = 15120 \delta_{ij};\qquad
 \int \phi^{(0)} \tilde{H}^{(8)}\tilde{H}^{(8)} d^3\tc = 362880;\nn\\
&& \int \phi^{(0)} \tilde{H}^{(9)}_i\tilde{H}^{(9)}_j d^3\tc = 1330560 \delta_{ij},\nn\\  
\end{eqnarray}
together with
\begin{eqnarray}
&& \hat{h}_{kl}^{(2)} \int \phi^{(0)} \tilde{H}^{(2)}_{ij}\tilde{H}^{(2)}_{kl} d^3\tc = 2\tilde{h}_{ij}^{(2)};\qquad
 \hat{h}_{kl}^{(4)} \int \phi^{(0)} \tilde{H}^{(4)}_{ij}\tilde{H}^{(4)}_{kl} d^3\tc = 28\hat{h}_{ij}^{(4)};\nn\\
&& \hat{h}_{kl}^{(6)} \int \phi^{(0)} \tilde{H}^{(6)}_{ij}\tilde{H}^{(6)}_{kl} d^3\tc = 1008\hat{h}_{ij}^{(6)};\qquad
 \hat{h}_{kl}^{(8)} \int \phi^{(0)} \tilde{H}^{(8)}_{ij}\tilde{H}^{(8)}_{kl} d^3\tc = 66528\hat{h}_{ij}^{(8)},
\end{eqnarray}
where we used traceless Hermite moments (with hat)
\begin{equation} \label{eq:Nomore70}
\hat{h}_{ij}^{(2n)} = \tilde{h}_{ij}^{(2n)}-\frac{1}{3}\delta_{ij}\tilde{h}^{(2n)}, 
\end{equation}
with $\tilde{h}^{(2)}=0$ (so that $\hat{h}^{(2)}_{ij}=\tilde{h}^{(2)}_{ij}$). Perturbation of the distribution function then becomes
\begin{eqnarray}
  \chi_a &=& \frac{1}{2}\tilde{h}_{ij}^{(2)}\tilde{H}_{ij}^{(2)}  + \frac{1}{10} \tilde{h}_i^{(3)}\tilde{H}_i^{(3)}
  +\frac{1}{28}\hat{h}_{ij}^{(4)}\tilde{H}_{ij}^{(4)} +\frac{1}{120} \tilde{h}^{(4)}\tilde{H}^{(4)}+ \frac{1}{280} \tilde{h}_i^{(5)}\tilde{H}_i^{(5)}\nn\\
  && +\frac{1}{1008}\hat{h}_{ij}^{(6)}\tilde{H}_{ij}^{(6)} +\frac{1}{5040} \tilde{h}^{(6)}\tilde{H}^{(6)}
  +\frac{1}{15120} \tilde{h}_i^{(7)}\tilde{H}_i^{(7)}\nn\\
  && +\frac{1}{66528}\hat{h}_{ij}^{(8)}\tilde{H}_{ij}^{(8)} +\frac{1}{362880} \tilde{h}^{(8)}\tilde{H}^{(8)}
  +\frac{1}{1330560} \tilde{h}_i^{(9)}\tilde{H}_i^{(9)}+\cdots. \label{eq:Nomore67}
\end{eqnarray}
Corresponding perturbation with the irreducible polynomials reads
\begin{eqnarray}
  \chi_a &=& {h}_{ij}^{(2)}{H}_{ij}^{(2)}  + {h}_i^{(3)}{H}_i^{(3)}
  +{h}_{ij}^{(4)}{H}_{ij}^{(4)} +{h}^{(4)}{H}^{(4)}+ {h}_i^{(5)}{H}_i^{(5)}\nn\\
  && {h}_{ij}^{(6)}{H}_{ij}^{(6)} +{h}^{(6)}{H}^{(6)}
  +{h}_i^{(7)}{H}_i^{(7)} +{h}_{ij}^{(8)}{H}_{ij}^{(8)} +{h}^{(8)}{H}^{(8)}
  +{h}_i^{(9)}{H}_i^{(9)}+\cdots,
\end{eqnarray}
i.e. no normalization constants are explicitly present. Now one then can clearly see the motivation behind
the definition of irreducible polynomials of \cite{Balescu1988}, where
direct relation between irreducible (no tilde) and reducible (tilde) Hermite polynomials can be shown to be
\begin{eqnarray}
H^{(2n)}  &=& \Big(\frac{1}{ 2^n n! (2n+1)!!}\Big)^{1/2 }\tilde{H}^{(2n)} ;\nn\\  
H^{(2n+1)}_i  &=& \Big(\frac{3}{2^n n! (2n+3)!!}\Big)^{1/2} \tilde{H}_i^{(2n+1)} ;\nn\\
H^{(2n)}_{ij}  &=& \Big(\frac{15}{2^n (n-1)! (2n+3)!!} \Big)^{1/2} \Big(\tilde{H}_{ij}^{(2n)}-\frac{1}{3}\delta_{ij}\tilde{H}^{(2n)} \Big). \label{eq:Nomore30}
\end{eqnarray}
Up to the normalization constants (which can be viewed as coming from the orthogonality relations), scalar and vector polynomials are equivalent to each other.
The only difference is for matrices $H^{(2n)}_{ij}$, where the irreducible polynomials are defined as traceless. Multiplying (\ref{eq:Nomore30}) by
$f_a/n_a$ and integrating over $d^3 c$ yields analogous relations for the Hermite moments
\begin{eqnarray}
h^{(2n)}  &=& \Big(\frac{1}{ 2^n n! (2n+1)!!}\Big)^{1/2 }\tilde{h}^{(2n)} ;\nn\\  
h^{(2n+1)}_i  &=& \Big(\frac{3}{2^n n! (2n+3)!!}\Big)^{1/2} \tilde{h}_i^{(2n+1)} ;\nn\\
h^{(2n)}_{ij}  &=& \Big(\frac{15}{2^n (n-1)! (2n+3)!!} \Big)^{1/2}
\Big(\underbrace{\tilde{h}_{ij}^{(2n)}-\frac{1}{3}\delta_{ij}\tilde{h}^{(2n)}}_{\hat{h}_{ij}^{(2n)}} \Big).  \label{eq:Nomore31}
\end{eqnarray}
Importantly, because $\hat{h}_{ij}^{(2n)}$ is traceless, multiplying  (\ref{eq:Nomore30}) and (\ref{eq:Nomore31}) yields
\begin{equation}
h^{(2n)}_{ij} H^{(2n)}_{ij} = \frac{15}{2^n (n-1)! (2n+3)!!} \,\hat{h}_{ij}^{(2n)} \tilde{H}_{ij}^{(2n)}. \label{eq:Nomore69}
\end{equation}
The two approaches with reducible and irreducible polynomials thus yield the same result, with the only difference being the location of
normalization constants. Furthermore, it feels natural to define traceless polynomials (with hat instead of tilde)
\begin{equation}
\hat{H}_{ij}^{(2n)}= \tilde{H}_{ij}^{(2n)}-\frac{1}{3}\delta_{ij}\tilde{H}^{(2n)},
\end{equation}
and on the r.h.s. of (\ref{eq:Nomore69}) replace
\begin{equation}
\hat{h}^{(2n)}_{ij} \tilde{H}_{ij}^{(2n)} = \hat{h}^{(2n)}_{ij} \hat{H}_{ij}^{(2n)}, 
\end{equation}  
which holds because $\hat{h}_{ij}^{(2n)}$ is traceless. The main advantage of introducing polynomials $\hat{H}_{ij}^{(2n)}$ is,
that instead of calculating $\hat{h}^{(2n)}_{ij}$ from its definition (\ref{eq:Nomore70}), one can directly define
\begin{equation}
\hat{h}^{(2n)}_{ij} = \frac{1}{n_a} \int f_a \hat{H}_{ij}^{(2n)} d^3c. 
\end{equation}  
Then the two approaches are indeed equivalent because the same polynomials are used,
with the location of normalization constants being an ad-hoc choice.

\newpage
From the Appendix of \cite{Balescu1988}, one can guess and then verify the following generalizations for the reducible polynomials 
\begin{eqnarray}
  \tilde{H}^{(2n)} &=& \sum_{m=0}^n (-1)^{m+n} \frac{n!}{m!(n-m)!}\frac{(2n+1)!!}{(2m+1)!!}\tc^{2m}; \label{eq:PPic1}\\
  \tilde{H}^{(2n+1)}_i &=& \tc_i \sum_{m=0}^n (-1)^{m+n} \frac{n!}{m!(n-m)!}\frac{(2n+3)!!}{(2m+3)!!}\tc^{2m};\\
  \tilde{H}^{(2n)}_{ij} &=& \tc_i \tc_j \Big(\sum_{m=0}^{n-1} (-1)^{m+n-1} \frac{(n-1)!}{m!(n-m-1)!}\frac{(2n+3)!!}{(2m+5)!!}\tc^{2m}\Big)
  -\delta_{ij}\frac{\tilde{H}_k^{(2n-1)}}{\tc_k}\nn\\
  && = \big(\tc_i \tc_j -\frac{\delta_{ij}}{3}\tc^2\big)\Big(\sum_{m=0}^{n-1} (-1)^{m+n-1} \frac{(n-1)!}{m!(n-m-1)!}\frac{(2n+3)!!}{(2m+5)!!}\tc^{2m}\Big)                             +\frac{\delta_{ij}}{3} \tilde{H}^{(2n)}; \label{eq:PPic2}\\
  \hat{H}^{(2n)}_{ij} &=&
  \big(\tc_i \tc_j -\frac{\delta_{ij}}{3}\tc^2\big)\Big(\sum_{m=0}^{n-1} (-1)^{m+n-1} \frac{(n-1)!}{m!(n-m-1)!}\frac{(2n+3)!!}{(2m+5)!!}\tc^{2m}\Big).
\end{eqnarray}
Applying trace at (\ref{eq:PPic2}) yields (\ref{eq:PPic1}). Similarly, the orthogonal relations are
\begin{eqnarray}
&& \int \phi^{(0)} \tilde{H}^{(2n)} \tilde{H}^{(2m)} d^3\tc = 2^n n! (2n+1)!! \,\delta_{nm};\label{eq:OrtH1}\\  
&& \int \phi^{(0)} \tilde{H}^{(2n+1)}_i \tilde{H}^{(2m+1)}_j d^3\tc = \frac{2^n n! (2n+3)!!}{3} \,\delta_{ij} \delta_{nm};\\
  && \int \phi^{(0)} \tilde{H}^{(2n)}_{ij} \tilde{H}^{(2m)}_{kl} d^3\tc =  \frac{2^{n-1} (n-1)! (2n+1)!!}{15}
  \Big[ (2n+3)\big(\delta_{ik}\delta_{jl}+\delta_{il}\delta_{jk}\big) +2(n-1)\delta_{ij}\delta_{kl} \Big]\delta_{nm};  \label{eq:OrtH2}\\
  && \int \phi^{(0)} \hat{H}^{(2n)}_{ij} \hat{H}^{(2m)}_{kl} d^3\tc =  \frac{2^{n-1} (n-1)! (2n+3)!!}{15}
  \Big[ \delta_{ik}\delta_{jl}+\delta_{il}\delta_{jk}-\frac{2}{3}\delta_{ij}\delta_{kl} \Big]\delta_{nm},\label{eq:OrtH3}
\end{eqnarray}
and applying $\delta_{ij}\delta_{kl}$ on (\ref{eq:OrtH2}) recovers (\ref{eq:OrtH1}). Note that the orders of Hermite moments
``m'' and ``n'' are 1-dimensional and $\delta_{nn}=1$. In contrast, for the indices $\delta_{ii}=3$ applies.
 Also note that $n!=n!!(n-1)!!$ and $2^n n! = (2n)!!$, implying $2^n n! (2n+1)!!=(2n+1)!$. 
Applying traceless
$\hat{h}_{kl}^{(2n)}$ on (\ref{eq:OrtH2}) or (\ref{eq:OrtH3}) yields orthogonal relation
\begin{eqnarray}
  \hat{h}_{kl}^{(2n)}\int \phi^{(0)} \hat{H}^{(2n)}_{ij} \hat{H}^{(2n)}_{kl} d^3\tc
  &=& \hat{h}_{kl}^{(2n)}\int \phi^{(0)} \tilde{H}^{(2n)}_{ij} \tilde{H}^{(2n)}_{kl} d^3\tc \nn\\
  &=&  \frac{2^{n} (n-1)! (2n+3)!!}{15}\hat{h}^{(2n)}_{ij}.
\end{eqnarray}
Finally, the general perturbation then can be written as
\begin{eqnarray}
  \chi_a &=& \sum_{n=1}^{\infty}
  \Big[ \frac{15}{2^n (n-1)! (2n+3)!!}\hat{h}^{(2n)}_{ij} \hat{H}^{(2n)}_{ij}
    +\frac{1}{2^n n! (2n+1)!!} \tilde{h}^{(2n)} \tilde{H}^{(2n)} \nn\\
    && + \frac{ 3}{2^n n! (2n+3)!!} \tilde{h}^{(2n+1)}_{i} \tilde{H}^{(2n+1)}_{i} \Big], \label{eq:Nomore40}
\end{eqnarray}
where for the first term $\tilde{h}^{(2)}=0$ (and so $\hat{h}^{(2)}_{ij}=\tilde{h}^{(2)}_{ij}$). Alternativelly,
$\hat{h}^{(2n)}_{ij} \hat{H}^{(2n)}_{ij}=\hat{h}^{(2n)}_{ij} \tilde{H}^{(2n)}_{ij}$. Perturbation
(\ref{eq:Nomore40}) is equivalent to perturbation with irreducible polynomials
\begin{eqnarray}
  \chi_a &=& \sum_{n=1}^{\infty}
  \Big[ {h}^{(2n)}_{ij} {H}^{(2n)}_{ij} +{h}^{(2n)} {H}^{(2n)} +{h}^{(2n+1)}_{i} {H}^{(2n+1)}_{i} \Big], \label{eq:Nomore41}
\end{eqnarray}
where again $h^{(2)}=0$.

\newpage
\subsection{Hierarchy of MHD Hermite closures} \label{sec:HC-MHD}
Let us use the 3rd-order moment $X^{(3)}_i=2q_i$ instead of the heat flux, so that no additional factors are present (also note that $X^{(2)}=3p$).
The even-order moments are decomposed according to
\begin{equation}
  X_{ij}^{(2n)} = \frac{\delta_{ij}}{3} X^{(2n)} + \Pi_{ij}^{(2n)},
\end{equation}
where the scalar part $X^{(2n)}$ is further decomposed into its Maxwellian ``core'' and perturbation $\widetilde{X}^{(2n)}$ (with wide tilde) as
\begin{equation}
  X^{(2n)} = (2n+1)!! \frac{p^n}{\rho^{n-1}}+\widetilde{X}^{(2n)},\label{eq:Nomore65}
\end{equation} 
so for example
\begin{equation}
  X^{(4)} = 15 \frac{p^2}{\rho}+\widetilde{X}^{(4)}; \qquad X^{(6)} = 105 \frac{p^3}{\rho^2}+\widetilde{X}^{(6)};
  \qquad X^{(8)} = 945 \frac{p^4}{\rho^3}+\widetilde{X}^{(8)}. \label{eq:Nomore61}
\end{equation}
Then by using Hermite polynomials (\ref{eq:Nomore60})-(\ref{eq:BalGen1}), one
calculates hierarchy of Hermite moments
\begin{eqnarray}
  \tilde{h}^{(3)}_i &=& \frac{\rho^{1/2}}{p^{3/2}} X^{(3)}_i;\qquad \tilde{h}^{(4)} = \frac{\rho}{p} \widetilde{X}^{(4)};\nn\\
 \tilde{h}^{(5)}_i &=& \frac{\rho^{1/2}}{p^{3/2}}\Big[ \frac{\rho}{p}X^{(5)}_i - 14 X^{(3)}_i \Big];\qquad 
 \tilde{h}^{(6)} = \frac{\rho}{p^2} \Big[\frac{\rho}{p} \widetilde{X}^{(6)}-21 \widetilde{X}^{(4)}\Big];\nn\\
\tilde{h}^{(7)}_i &=& \frac{\rho^{1/2}}{p^{3/2}}\Big[ \frac{\rho^2}{p^2}X^{(7)}_i -27\frac{\rho}{p}X^{(5)}_i+189 X_i^{(3)}\Big];\nn\\
 \tilde{h}^{(8)} &=& \frac{\rho}{p^2}\Big[\frac{\rho^2}{p^2} \widetilde{X}^{(8)}-36\frac{\rho}{p} \widetilde{X}^{(6)} +378 \widetilde{X}^{(4)}\Big];\nn\\
\tilde{h}^{(9)}_i &=& \frac{\rho^{1/2}}{p^{3/2}} \Big[ \frac{\rho^3}{p^3}X^{(9)}_i -44\frac{\rho^2}{p^2}X^{(7)}_i+594\frac{\rho}{p}X^{(5)}_i
    -2772 X^{(3)}_i\Big], \label{eq:Thierry-1}
\end{eqnarray}
together with
\begin{eqnarray}
  \hat{h}_{ij}^{(4)} &=& \frac{1}{p}\Big[ \frac{\rho}{p}\Pi_{ij}^{(4)}-7\Pi_{ij}^{(2)}\Big];\nn\\
  \hat{h}_{ij}^{(6)} &=& \frac{1}{p}\Big[ \frac{\rho^2}{p^2}\Pi_{ij}^{(6)}-18\frac{\rho}{p}\Pi_{ij}^{(4)} +63\Pi_{ij}^{(2)}\Big];\nn\\
  \hat{h}_{ij}^{(8)} &=& \frac{1}{p}\Big[ \frac{\rho^3}{p^3}\Pi_{ij}^{(8)}-33\frac{\rho^2}{p^2}\Pi_{ij}^{(6)} +297 \frac{\rho}{p}\Pi_{ij}^{(4)}
   -693 \Pi_{ij}^{(2)} \Big]. \label{eq:Thierry-2}
\end{eqnarray}
Prescribing the last retained Hermite moment to be zero, then yields corresponding fluid closures which are summarized 
in Section \ref{sec:HC}, Tables \ref{table:HC1} \& \ref{table:HC5}.


\subsubsection*{Propagation along the B-field (ion-acoustic mode)}
For a propagation parallel to the mean magnetic field which is applied in the z-direction, linearized equations without collisions read
\begin{eqnarray}
&& \frac{\pr \rho}{\pr t} +\rho_0 \pr_z u_z =0;\qquad \quad \frac{\pr u_z}{\pr t}+\frac{1}{\rho_0}\pr_z p=0;\nn\\
&& \frac{\pr p}{\pr t}+\frac{5}{3}p_0 \pr_z u_z+\frac{2}{3}\pr_z q_z =0;\nn\\
&& \frac{\pr q_z}{\pr t}+\frac{1}{6}\pr_z X^{(4)} -\frac{5}{2}\frac{p_0}{\rho_0}\pr_z p =0;\nn\\
&& \frac{\pr X^{(4)}}{\pr t}+\pr_z X^{(5)}_z +\frac{7}{3}X^{(4)}_0 \pr_z u_z =0; \nn\\
&& \frac{\pr X^{(5)}_z}{\pr t}+\frac{1}{3}\pr_z X^{(6)} -\frac{7}{3}\frac{X^{(4)}_0}{\rho_0}\pr_z p=0, \label{eq:PPPosled} 
\end{eqnarray}
where all the variables are scalars.  We are neglecting collisions and viscosities, to make direct comparison with the CGL
model in the next section. The even-order moments are decomposed into a Maxwellian ``core'' and tilde
perturbations with (\ref{eq:Nomore61}), so their mean values are $X^{(4)}_0=15 p_0^2/\rho_0$ and $X^{(6)}_0=105 p_0^3/\rho_0^2$.
These moments are thus linearized according to
\begin{equation}
  X^{(4)} \overset{\textrm{lin.}}{=} X^{(4)}_0\Big(2\frac{p}{p_0}-\frac{\rho}{\rho_0}\Big)+\widetilde{X}^{(4)};
  \qquad X^{(6)} \overset{\textrm{lin.}}{=} X^{(6)}_0\Big(3\frac{p}{p_0}-2\frac{\rho}{\rho_0}\Big)+\widetilde{X}^{(6)}, 
\end{equation}
and the last three equations of (\ref{eq:PPPosled}) then become
\begin{eqnarray}
&& \frac{\pr q_z}{\pr t}+\frac{1}{6}\pr_z \widetilde{X}^{(4)} +\frac{5}{2}\frac{p_0}{\rho_0}\big(\pr_z p -\frac{p_0}{\rho_0}\pr_z\rho\big)=0;\nn\\
&& \frac{\pr \widetilde{X}^{(4)}}{\pr t}+\pr_z X^{(5)}_z -20 \frac{p_0}{\rho_0}\pr_z q_z =0; \nn\\
&& \frac{\pr X^{(5)}_z}{\pr t}+\frac{1}{3}\pr_z \widetilde{X}^{(6)} +70 \frac{p_0^2}{\rho_0^2}\big(\pr_z p -\frac{p_0}{\rho_0}\pr_z\rho\big)=0. \label{eq:Nomore62}  
\end{eqnarray}
Prescribing closure at the last retained moment, yields dispersion relations in the variable
$\zeta=\omega/(|k_\parallel| v_{th})$ that are summarized in Table \ref{table:Herm10}.

\begin{table}[ht!]
\centering
\begin{tabular}{| l | l | l |}
\hline
Closure         &   Dispersion relation  &  Solution $\pm\zeta =$                                                      \\
\hline 
$\tilde{h}^{(3)}_z=0$;   & $\zeta^2-5/6 =0$;                                               &   $0.913$;                          \\  
$\tilde{h}^{(4)}=0$;   & $\zeta^4- (5/3)\zeta^2+(5/12)=0$;                    &   $0.553; \quad 1.166$;                         \\   
$\tilde{h}^{(5)}_z=0$;   & $\zeta^4-(7/3)\zeta^2+(35/36)=0$;                   &   $0.737; \quad 1.338$;                         \\  
$\tilde{h}^{(6)}=0$;   & $\zeta^6-(7/2)\zeta^4+ (35/12)\zeta^2-(35/72)=0$;  &   $0.471; \quad 0.966; \quad 1.531$;          \\  
\hline
${X}^{(5)}_z=0$;   & $\zeta^4-(35/36)=0$;                               &   $0.99; \quad 0.99\,i$;                         \\
$\widetilde{X}^{(6)}=0$;   & $\zeta^6-(35/12)\zeta^2+(35/36)=0$;       &   $0.59; \quad 1.23;\quad 1.36\,i$;                         \\
\hline
\end{tabular}
\caption{Summary of Hermite closures and corresponding dispersion relations for parallel propagating ion-acoustic mode (electrons are cold),
  where $\zeta=\omega/(|k_\parallel|v_{\textrm{th}})$. With Hermite closures (upper half of the table), no spurious instabilities are present.
  Unphysical instabilities appear if one prescribes
  erroneous fluid closures at the last retained moment $X^{(5)}_z=0$ or $\widetilde{X}^{(6)}=0$ (lower half of the table). However, if one prescribes
  at the same time $\widetilde{X}^{(6)}=0$ and $\widetilde{X}^{(4)}=0$, the system is again well-defined with dispersion relation equivalent
to closure $\tilde{h}^{(4)}=0$.}
\label{table:Herm10}
\end{table}

The example clearly demonstrates that Landau fluid closures are actually \emph{not} required to
go beyond the 4th-order moment, which contradicts a claim in the last paragraph of \cite{HunanaPRL2018}, and also
in various parts of \cite{Hunana2019b,Hunana2019a} (see e.g. Section 12.2 in Part 1). Obviously, closures $X^{(5)}_z=0$ or $\widetilde{X}^{(6)}=0$ are not allowed
by the fluid hierarchy (unless $q_z=0$ or $\widetilde{X}^{(4)}=0$ as well). Instead, for moments of order $n\ge 5$,
one needs to construct ``classical'' closures at the Hermite moments. Nevertheless, all the Landau fluid closures reported
in the above papers are constructed correctly. 

Out of curiosity, prescribing closures with a free parameter ``$a$'' as $X^{(5)}_z=28 a (p_0/\rho_0) q_z$ or
$\widetilde{X}^{(6)}=21 a (p_0/\rho_0) \widetilde{X}^{(4)}$ yields the following dispersion relations
\begin{eqnarray}
&&\zeta^4-\frac{7a}{3}\zeta^2+\frac{35a}{18}-\frac{35}{36}=0;\label{eq:Prd1}\\
&&\zeta^6-\frac{7 a}{2}\zeta^4+\Big( \frac{35 a}{6}-\frac{35}{12}\Big)\zeta^2- \frac{35 a}{24}+\frac{35}{36}=0.\label{eq:Prd2}
\end{eqnarray}
The $X^{(5)}_z$ closure with dispersion relation (\ref{eq:Prd1}) yields an instability for $a<1/2$, and the $\widetilde{X}^{(6)}$ closure with
(\ref{eq:Prd2}) yields an instability for $a<2/3$. There is therefore a lot of closures which do not create these unphysical
instabilities. 

Finally, the situation is saved by completely de-coupling the odd and even moments, for example prescribing
$\widetilde{X}^{(6)}=\widetilde{X}^{(4)}=0$, so that equations (\ref{eq:Nomore62}) are replaced by
\begin{eqnarray}
&& \frac{\pr q_z}{\pr t} +\frac{5}{2}\frac{p_0}{\rho_0}\big(\pr_z p -\frac{p_0}{\rho_0}\pr_z\rho\big)=0;\nn\\
&& \frac{\pr X^{(5)}_z}{\pr t} +70 \frac{p_0^2}{\rho_0^2}\big(\pr_z p -\frac{p_0}{\rho_0}\pr_z\rho\big)=0.   
\end{eqnarray} 
Dispersion relation of this model is equivalent to closure $\tilde{h}^{(4)}=0$.

\newpage
\subsection{Hierarchy of CGL (parallel) Hermite closures} \label{sec:HC-CGL}
The hierarchy of 1D Hermite polynomials calculates (with weight $\exp(-\tc^2/2)$)
\begin{eqnarray}
{H}^{(1)} &=& \tc;\nn\\  
{H}^{(2)} &=& \tc^2-1 ;\nn\\
{H}^{(3)} &=& \tc (\tc^2-3);\nn\\
{H}^{(4)} &=& \tc^4-6 \tc^2+3;\nn\\
{H}^{(5)} &=& \tc (\tc^4-10 \tc^2+15);\nn\\
{H}^{(6)} &=& \tc^6-15 \tc^4+45 \tc^2-15;\nn\\
{H}^{(7)} &=& \tc (\tc^6-21 \tc^4+105 \tc^2-105);\nn\\ 
{H}^{(8)} &=& \tc^8-28 \tc^6+210 \tc^4-420 \tc^2+105;\nn\\
{H}^{(9)} &=& \tc (\tc^8-36 \tc^6+378 \tc^4-1260 \tc^2+945), \label{eq:Nomore42}
\end{eqnarray}
further yielding the following hierarchy of Hermite moments
\begin{eqnarray}
  h^{(1)} &=& 0;\qquad  h^{(2)} = 0;\nn\\
  h^{(3)} &=& \frac{\rho^{1/2}}{p^{3/2}}X^{(3)};\qquad  h^{(4)} = \frac{\rho}{p^{2}}\widetilde{X}^{(4)};\nn\\
  h^{(5)} &=& \frac{\rho^{1/2}}{p^{3/2}}\Big( \frac{\rho}{p}X^{(5)} -10 X^{(3)}\Big);\qquad
  h^{(6)} = \frac{\rho}{p^{2}}\Big(\frac{\rho}{p} \widetilde{X}^{(6)}-15\widetilde{X}^{(4)}\Big);\nn\\
  h^{(7)} &=& \frac{\rho^{1/2}}{p^{3/2}}\Big( \frac{\rho^2}{p^2}X^{(7)}-21\frac{\rho}{p}X^{(5)}+105 X^{(3)}\Big);\nn\\
  h^{(8)} &=& \frac{\rho}{p^{2}}\Big(\frac{\rho^2}{p^2} \widetilde{X}^{(8)} -28\frac{\rho}{p}\widetilde{X}^{(6)}+210 \widetilde{X}^{(4)}\Big);\nn\\
  h^{(9)} &=& \frac{\rho^{1/2}}{p^{3/2}}\Big( \frac{\rho^3}{p^3}X^{(9)} -36 \frac{\rho^2}{p^2}X^{(7)} +378 \frac{\rho}{p}X^{(5)} -1260 X^{(3)}\Big),
\end{eqnarray}  
where the even moments were separated into
\begin{eqnarray}
  X^{(4)} &=& 3 \frac{p^2}{\rho}+\widetilde{X}^{(4)}; \qquad X^{(6)} = 15 \frac{p^3}{\rho^2}+\widetilde{X}^{(6)};\qquad
  X^{(8)} = 105 \frac{p^4}{\rho^3}+\widetilde{X}^{(8)};\nn\\
  X^{(2n)} &=& (2n-1)!! \frac{p^n}{\rho^{n-1}}+\widetilde{X}^{(2n)}.  \label{eq:PPPosled10}
\end{eqnarray}
This yields the hierarchy of Hermite closures summarized in Table \ref{table:Herm2}.
Note the difference of (\ref{eq:PPPosled10}) with the isotropic (MHD) decomposition (\ref{eq:Nomore65})
 (in the 3D CGL geometry one typically uses notation $X^{(4)}=r_{\parallel\parallel}$). 

\begin{table}[ht!]
\centering
\begin{tabular}{| l | l |}
\hline
Hermite closures         &    Fluid closures \\
\hline    
${h}^{(3)}=0$;    &    $X^{(3)} = 0$;\\
${h}^{(4)}=0$;    &    $\widetilde{X}^{(4)} = 0$;\\
${h}^{(5)}=0$;    &    $X^{(5)} = 10 \frac{p}{\rho} X^{(3)}$;\\
${h}^{(6)}=0$;    &    $\widetilde{X}^{(6)} = 15 \frac{p}{\rho}\widetilde{X}^{(4)}$;\\
${h}^{(7)}=0$;    &    $X^{(7)} = 21 \frac{p}{\rho}X^{(5)}-105 \frac{p^2}{\rho^2} X^{(3)}$;\\
${h}^{(8)}=0$;    &    $\widetilde{X}^{(8)} = 28\frac{p}{\rho} \widetilde{X}^{(6)} -210 \frac{p^2}{\rho^2} \widetilde{X}^{(4)}$;\\  
${h}^{(9)}=0$;    &    $X^{(9)} = 36\frac{p}{\rho}X^{(7)} -378\frac{p^2}{\rho^2}X^{(5)}+1260 \frac{p^3}{\rho^3} X^{(3)}$.\\
\hline
\end{tabular}
\caption{Summary of Hermite closures for parallel CGL moments, together with corresponding fluid closures.
   The usual parallel heat flux $q_\parallel=X^{(3)}$. Note that beyond the 4th-order moment both classes start to differ.  A general form 
  corresponding to $h^{(2n+1)}=0$ and $h^{(2n)}=0$ is given by (\ref{eq:ClosureNN}).}
\label{table:Herm2}
\end{table}

Hermite polynomials (\ref{eq:Nomore42}) can be written in a general form 
\begin{eqnarray}
  H^{(2n+1)} &=& \sum_{m=0}^{n} (-1)^{n-m} \frac{(2n+1)!}{2^{n-m}(2m+1)! (n-m)!} \tc^{2m+1};\nn\\
  H^{(2n)}   &=& \sum_{m=0}^{n} (-1)^{n-m} \frac{(2n)!}{2^{n-m}(2m)! (n-m)!} \tc^{2m}.
\end{eqnarray}
Then it can be shown that prescribing Hermite closure $h^{(2n+1)}=0$ or $h^{(2n)}=0$, is equivalent to prescribing fluid closure
\begin{eqnarray}
  X^{(2n+1)} &=& \sum_{m=1}^{n-1} (-1)^{n-m+1} \frac{(2n+1)!}{2^{n-m}(2m+1)! (n-m)!} \Big(\frac{p}{\rho}\Big)^{n-m} X^{(2m+1)};\nn\\
 \widetilde{X}^{(2n)} &=& \sum_{m=2}^{n-1} (-1)^{n-m+1} \frac{(2n)!}{2^{n-m}(2m)! (n-m)!} \Big(\frac{p}{\rho}\Big)^{n-m} \widetilde{X}^{(2m)}.
  \label{eq:ClosureNN}
\end{eqnarray}
By using equations (12.49)-(12.54) from \cite{Hunana2019a}, we calculated the corresponding dispersion relations, which are
summarized in Table \ref{table:Herm3}.

\begin{table}[ht!]
\centering
\begin{tabular}{| l | l | l |}
\hline
Closure         &   Dispersion relation  &  Solution $\pm\zeta =$                                                                   \\
\hline 
${h}^{(3)}=0$;   & $\zeta^2-3/2 =0$;                                               &   $1.225$;                                       \\  
${h}^{(4)}=0$;   & $\zeta^4-3\zeta^2+3/4 =0$;                                      &   $0.525; \quad 1.651$;                           \\   
${h}^{(5)}=0$;   & $\zeta^4-5\zeta^2+15/4=0$;                                      &   $0.959; \quad 2.020$;                           \\  
${h}^{(6)}=0$;   & $\zeta^6-(15/2)\zeta^4+(45/4)\zeta^2-15/8 =0$;                  &   $0.436; \quad 1.336; \quad 2.351$;              \\  
${h}^{(7)}=0$;   & $\zeta^6-(21/2)\zeta^4+(105/4)\zeta^2-105/8 =0$;                &   $0.816; \quad 1.674; \quad 2.652$;              \\  
${h}^{(8)}=0$;   & $\zeta^8-14 \zeta^6+(105/2) \zeta^4-(105/2) \zeta^2+105/16=0$;  &   $0.381; \quad 1.157; \quad 1.982; \quad 2.931$;  \\
${h}^{(9)}=0$;   & $\zeta^8-18 \zeta^6+(189/2) \zeta^4-(315/2)\zeta^2+945/16=0$;   &   $0.724; \quad 1.469; \quad 2.267; \quad 3.191$.  \\
\hline
\end{tabular}
\caption{Summary of Hermite closures and corresponding dispersion relations for parallel propagating ion-acoustic mode (electrons are cold),
  where $\zeta=\omega/(|k_\parallel|v_{\textrm{th}})$. No spurious instabilities are present. Spurious instabilities occur if one
  prescribes at the last retained moment closures $\widetilde{X}^{(2n)}=0$ or $X^{(2n+1)}=0$.}
\label{table:Herm3}
\end{table}
Curiously, from \cite{Hunana2019b} the not ``well-behaved'' Pad\'e approximants of plasma dispersion function $R(\zeta)$ that contain no Landau damping
read
\begin{eqnarray}
R_{4,5}(\zeta) &=& \frac{1-(2/3)\zeta^2}{1-4\zeta^2+(4/3)\zeta^4};\nn\\  
R_{6,9}(\zeta) &=&  \frac{1-(8/5)\zeta^2+(4/15)\zeta^4}{1-6\zeta^2+4\zeta^4-(8/15)\zeta^6};\nn\\
R_{8,13}(\zeta) &=& \frac{1-(94/35)\zeta^2+(20/21)\zeta^4-(8/105)\zeta^6}{1-8\zeta^2+8\zeta^4-(32/15)\zeta^6+(16/105)\zeta^8}.\label{eq:Nomore50}
\end{eqnarray}
Comparing (\ref{eq:Nomore50}) with Table \ref{table:Herm3}, one comes to a non-obvious observation that
denominators of the above approximants are equal to dispersion relations obtained
with Hermite closures $h^{(4)}=0$,  $h^{(6)}=0$ and $h^{(8)}=0$. This observation is analogous with Landau fluid closures
when electrons are cold; see equation (3.358) of \cite{Hunana2019b}. 
Thus, it is expected that for proton-electron plasma with finite temperatures
(and with electron inertia retained) these three dispersion relations will be equivalent to
\begin{eqnarray}
\frac{T_{\parallel e}^{(0)}}{T_{\parallel p}^{(0)}} R_{n,n'}(\zeta_p) + R_{n,n'}(\zeta_e) =0,
\end{eqnarray}
which we did not verify.

\newpage
\section{Evolution equations for 22-moment model} \label{sec:22momentE}
\setcounter{equation}{0}
Here we use evolution equations (\ref{eq:density1})-(\ref{eq:X5_tensor}), and by applying contractions at these equations we obtain
the 22-moment model in detail.  
The pressure tensor is decomposed as $p_{ij}^a=p_a\delta_{ij} +\Pi^{a(2)}_{ij}$, where the scalar pressure $p_a=p_{ii}^a/3$.
Instead of considering full moments $X^{(3)}_{ijk}$, $X^{(4)}_{ijkl}$, $X^{(5)}_{ijklm}$, $X^{(6)}_{ijklmn}$, one only considers contracted vectors and matrices
\begin{equation}
X^{a(3)}_{i} = X^{a(3)}_{ijj}; \qquad X^{a(4)}_{ij} = X^{a(4)}_{ijkk}; \qquad X^{a(5)}_{i} = X^{a(5)}_{ijjkk}; \qquad X^{a(6)}_{ij} = X^{a(6)}_{ijkkll}.
\end{equation}
The even-order moments are decomposed by separating the traceless viscosity-tensors $\Pi^{(2n)}_{ij}$ 
\begin{equation} \label{eq:Energy44}
X^{a(4)}_{ij}=\frac{\delta_{ij}}{3} X^{a(4)}+\Pi^{a(4)}_{ij} ;\qquad X^{a(6)}_{ij}=\frac{\delta_{ij}}{3} X^{a(6)}+\Pi^{a(6)}_{ij},
\end{equation}
where the fully contracted (scalars) $X^{a(4)}=X^{a(4)}_{iijj}$, $X^{a(6)}=X^{a(6)}_{iijjkk}$.
The scalars are further decomposed into their ``Maxwellian core'' and
a perturbation around this core (which is denoted by wide tilde)
\begin{equation} \label{eq:Energy45}
X^{(4)}_a=\trace\trace\br_a=15\frac{p_a^2}{\rho_a}+\widetilde{X}^{(4)}_a; \qquad X^{(6)}_a=\trace\trace\trace \bX^{(6)}_a =105\frac{p_a^3}{\rho_a^2}+\widetilde{X}^{(6)}_a.
\end{equation}
As in \cite{Braginskii1965}, we use notation with the Boltzmann constant $k_B=1$,
and the temperature is defined as $T_a=p_a/n_a$. Note that $m_a/T_a=\rho_a/p_a$.


\subsection{Decomposition of moments}
The heat flux tensor $q_{ijk}$ and moments $X^{(4)}_{ijkl}$, $X^{(5)}_{ijklm}$ are decomposed according to (see Appendix \ref{sec:Hermite}) 
\begin{eqnarray}
  q_{ijk}^a &=& \frac{2}{5} \Big[ \bI \vecq^a \big]_{ijk}^S;\label{eq:Energy41}\\
  X_{ijkl}^{a(4)} &=& \frac{1}{15}\Big(15\frac{p_a^2}{\rho_a} +\widetilde{X}^{a(4)}\Big) \big(\delta_{ij}\delta_{kl}+\delta_{ik}\delta_{jl}+\delta_{il}\delta_{jk}\big)\nn\\
  && +\frac{1}{7} \Big[ \Pi_{ij}^{a(4)} \delta_{kl}+\Pi_{ik}^{a(4)}\delta_{jl}+\Pi_{il}^{a(4)}\delta_{jk}
  +\Pi_{jk}^{a(4)}\delta_{il}+\Pi_{jl}^{a(4)}\delta_{ik}+\Pi_{kl}^{a(4)}\delta_{ij} \Big];\label{eq:Energy42}\\
  X^{a(5)}_{s_1 s_2 s_3 s_4 s_5} &=& \frac{1}{35} \Big[ 
  X_{s_1}^{a(5)} \big( \delta_{s_2 s_3}\delta_{s_4 s_5}+\delta_{s_2 s_4}\delta_{s_3 s_5} +\delta_{s_2 s_5}\delta_{s_3 s_4}\big) \nn\\
  && +X_{s_2}^{a(5)} \big( \delta_{s_1 s_3}\delta_{s_4 s_5}+\delta_{s_1 s_4}\delta_{s_3 s_5} +\delta_{s_1 s_5}\delta_{s_3 s_4}\big) \nn\\
  && +X_{s_3}^{a(5)} \big( \delta_{s_1 s_2}\delta_{s_4 s_5}+\delta_{s_1 s_4}\delta_{s_2 s_5} +\delta_{s_1 s_5}\delta_{s_2 s_4}\big) \nn\\
  && +X_{s_4}^{a(5)} \big( \delta_{s_1 s_2}\delta_{s_3 s_5}+\delta_{s_1 s_3}\delta_{s_2 s_5} +\delta_{s_1 s_5}\delta_{s_2 s_3}\big) \nn\\
  && +X_{s_5}^{a(5)} \big( \delta_{s_1 s_2}\delta_{s_3 s_4}+\delta_{s_1 s_3}\delta_{s_2 s_4} +\delta_{s_1 s_4}\delta_{s_2 s_3}\big)\Big], \label{eq:Energy40}
\end{eqnarray}
where the highest-order irreducible parts  of moments (\ref{eq:Energy41})-(\ref{eq:Energy40}) denoted as $\sigma^{(3)'}_{ijk}$, $\sigma^{(4)'}_{ijkl}$,
$\sigma^{(5)'}_{ijklm}$ are neglected (which provides the reduction from 56-moment model to 22-moment model). 

\subsection{Evolution equation for scalar pressure \texorpdfstring{$p_a$}{p}}
By using decomposition $\bp_a=p_a\bI+\bPi^{(2)}_a$, evolution equation for scalar pressure $p_a$ is obtained by applying 
$(1/3)\textrm{Tr}$ on the pressure tensor equation (\ref{eq:PR_tensor}), yielding
\begin{eqnarray}
 \frac{\pr p_a}{\pr t} +\bu_a\cdot\nabla p_a + \frac{5}{3} p_a\nabla\cdot\bu_a
  +\frac{2}{3}\nabla\cdot\vec{\boldsymbol{q}}_a +\frac{2}{3}\bPi^{(2)}_a :(\nabla \bu_a)
  = \frac{1}{3}\textrm{Tr}\bQ^{(2)}_a=\frac{2}{3} Q_a. \label{eq:pressure}
\end{eqnarray}
Alternativelly, by using temperature $T_a=p_a/n_a$ yields the following equation
\begin{equation} \label{eq:BragT}
  \frac{3}{2}n_a\frac{d_a T_a}{d t} + p_a\nabla\cdot\bu_a +\nabla\cdot\vecq_a +\bPi^{(2)}_a :(\nabla \bu_a) =\frac{1}{2}\textrm{Tr}\bQ^{(2)}_a
  = Q_a, 
\end{equation}
which identifies with equation (2.3) of \cite{Braginskii1965}. The collisional energy exchange rates
\begin{equation}
Q_a = \frac{m_a}{2}\int|\bc_a|^2 C(f_a)d^3v.
\end{equation}

\subsection{Evolution equation for viscosity tensor \texorpdfstring{$\bPi^{(2)}_a$}{Pi2}}
Evolution equation for the  usual viscosity-tensor is obtained  
by subtracting $\bI$ times (\ref{eq:pressure}) from (\ref{eq:PR_tensor}), yielding
\begin{eqnarray}
&&  \frac{d_a\bPi^{(2)}_a}{dt} + \bPi^{(2)}_a \nabla\cdot\bu_a 
  +\Omega_a \big(\bhat\times \bPi^{(2)}_a \big)^S +\big( \bPi^{(2)}_a \cdot\nabla\bu_a\big)^S 
  -\frac{2}{3}\bI(\bPi^{(2)}_a:\nabla\bu_a)+\nabla\cdot\bq_a
  -\frac{2}{3}\bI\nabla\cdot\vec{\boldsymbol{q}}_a\nn\\
&&\qquad  +p_a \Big[(\nabla \bu_a)^S -\frac{2}{3}\bI \nabla\cdot\bu_a\Big] = \bQ^{(2)}_a -\frac{\bI}{3}\textrm{Tr}\bQ^{(2)}_a. \label{eq:Pi_evol}
\end{eqnarray}
It is possible to define the well-known rate-of-strain tensor
\begin{equation} \label{eq:Stress}
  \bW_a = (\nabla \bu_a)^S -\frac{2}{3}\bI \nabla\cdot\bu_a.
\end{equation}
Equation (\ref{eq:Pi_evol}) is exact. 
By using heat flux decomposition (\ref{eq:Energy41}) yields $\nabla\cdot\bq_a=(2/5)((\nabla\vecq_a)^S+\bI\nabla\cdot\vecq_a)$, and so
equation (\ref{eq:Pi_evol}) becomes
\begin{eqnarray}
&&  \frac{d_a\bPi^{(2)}_a}{dt} + \bPi^{(2)}_a \nabla\cdot\bu_a 
  +\Omega_a \big(\bhat\times \bPi^{(2)}_a \big)^S +\big( \bPi^{(2)}_a \cdot\nabla\bu_a\big)^S 
  -\frac{2}{3}\bI(\bPi^{(2)}_a:\nabla\bu_a)+\frac{2}{5}\Big[(\nabla \vecq_a)^S -\frac{2}{3}\bI \nabla\cdot\vecq_a\Big]\nn\\
&&\qquad  +p_a \bW_a = \bQ^{(2)}_a\,' \equiv \bQ^{(2)}_a -\frac{\bI}{3}\textrm{Tr}\bQ^{(2)}_a, \label{eq:Energy43}
\end{eqnarray}
which for example identifies with equations (39)-(40) of \cite{Schunk1977}.
It is possible to define
\begin{eqnarray}
  \bW_a^q  &=& \frac{2}{5}\Big[(\nabla \vecq_a)^S -\frac{2}{3}\bI \nabla\cdot\vecq_a\Big],\label{eq:Stress2}
\end{eqnarray}
where we used a heat flux superscript '$q$' to differentiate it from (\ref{eq:Stress}). 
As a double check, applying trace on (\ref{eq:Energy43}) yields that both sides are zero.

\subsection{Evolution equation for heat flux vector \texorpdfstring{$\vecq_a$}{q}}
Evolution equation for $\vecq_a$ is obtained by applying $(1/2)\textrm{Tr}$ on (\ref{eq:QR_tensor}),
yielding 
\begin{eqnarray} 
 && \frac{d_a\vecq_a}{d t} +\vecq_a\nabla\cdot\bu_a  + \vecq_a\cdot\nabla\bu_a +\bq_a:\nabla\bu_a
  +\Omega_a\bhat\times\vecq_a + \frac{1}{2}\textrm{Tr}\nabla\cdot \br_a-\frac{1}{\rho_a}\Big[ \frac{3}{2}p_a \nabla\cdot\bp_a
    + (\nabla\cdot\bp_a)\cdot\bp_a \Big] \nn\\
  &&\qquad =\frac{1}{2}\textrm{Tr}\Big[\bQ^{(3)}_a-\frac{p_a}{\rho_a}(\boldsymbol{R}_a\bI)^S\Big]
  -\frac{1}{\rho_a} \boldsymbol{R}_a\cdot\bPi^{(2)}_a, \label{eq:QR_tensor3}
\end{eqnarray}
where $\textrm{Tr}( \boldsymbol{R}_a \bI)^S=5\boldsymbol{R}_a$. This equation is exact.
By using heat flux decomposition (\ref{eq:Energy41}) yields
\begin{equation}
\bq_a:\nabla\bu_a=(2/5)\big[\vecq_a\cdot\nabla\bu_a+(\nabla\bu_a)\cdot\vecq_a+\vecq_a\nabla\cdot\bu_a\big],
\end{equation}
and applying trace at decomposition (\ref{eq:Energy42}) yields
\begin{equation}
\trace\br_a = 5 \frac{p_a^2}{\rho_a}\bI +  \frac{\bI}{3}\widetilde{X}^{(4)}_a + \bPi^{(4)}_a,
\end{equation}
which is of course equivalent to decomposition (\ref{eq:Energy44}), (\ref{eq:Energy45}).
Note that a closure $\textrm{Tr}\br_a = 5\frac{p^2_a}{\rho_a} \bI$ can be viewed as an isotropic analogy of the anisotropic bi-Maxwellian ``normal'' closure
$r_{\parallel\parallel a} = \frac{3p_{\parallel a}^2}{\rho}$, $r_{\parallel\perp a} = \frac{p_{\parallel a} p_{\perp a}}{\rho_a}$, 
$r_{\perp\perp a} = \frac{2 p_{\perp a}^2}{\rho_a}$ with $p_{\parallel a}=p_{\perp a}=p_a$,
because the following general identity holds for any gyrotropic distribution function
$\textrm{Tr}\br^{\textrm{g}}_a=r_{\parallel\parallel a}\bhat\bhat+r_{\parallel\perp a}(\bI+\bhat\bhat)+2r_{\perp\perp a}(\bI-\bhat\bhat)$.
Then one calculates
\begin{equation}
  \frac{1}{2}\textrm{Tr}\nabla\cdot \br_a = \frac{5}{2}\nabla\Big(\frac{p_a^2}{\rho_a}\Big) +\frac{1}{6}\nabla \widetilde{X}^{(4)}_a
  +\frac{1}{2}\nabla\cdot \bPi^{(4)}_a,
\end{equation}
together with
\begin{eqnarray}
  \frac{1}{2}\textrm{Tr}\nabla\cdot \br_a-\frac{1}{\rho_a}\Big[ \frac{3}{2}p_a \nabla\cdot\bp_a
    + (\nabla\cdot\bp_a)\cdot\bp_a \Big] &=& \frac{5}{2}p_a\nabla\Big(\frac{p_a}{\rho_a}\Big)
  +\frac{1}{6}\nabla \widetilde{X}^{(4)}_a +\frac{1}{2}\nabla\cdot \bPi^{(4)}_a \nn\\
  && -\frac{5}{2}\frac{p_a}{\rho_a}\nabla\cdot\bPi^{(2)}_a
  -\frac{1}{\rho_a}(\nabla\cdot\bp_a)\cdot\bPi^{(2)}_a,
\end{eqnarray}
and evolution equation (\ref{eq:QR_tensor3}) becomes
\begin{eqnarray} 
 && \frac{d_a\vecq_a}{d t} +\frac{7}{5}\vecq_a\nabla\cdot\bu_a  + \frac{7}{5}\vecq_a\cdot\nabla\bu_a +\frac{2}{5}(\nabla\bu_a)\cdot\vecq_a
  +\Omega_a\bhat\times\vecq_a + \frac{5}{2}p_a\nabla\Big(\frac{p_a}{\rho_a}\Big) \nn\\
 && +\frac{1}{6}\nabla \widetilde{X}^{(4)}_a +\frac{1}{2}\nabla\cdot \bPi^{(4)}_a 
  -\frac{5}{2}\frac{p_a}{\rho_a}\nabla\cdot\bPi^{(2)}_a
  -\frac{1}{\rho_a}(\nabla\cdot\bp_a)\cdot\bPi^{(2)}_a \nn\\
  && \qquad = \vec{\boldsymbol{Q}}^{(3)}_{a}\,' \equiv \frac{1}{2}\textrm{Tr}\bQ^{(3)}_a-\frac{5}{2}\frac{p_a}{\rho_a}\boldsymbol{R}_a
  -\frac{1}{\rho_a} \boldsymbol{R}_a\cdot\bPi^{(2)}_a. \label{eq:Energy46}
\end{eqnarray}
As a double-check, reducing the 22-moment model into 13-moment model with closures $\widetilde{X}^{(4)}_a=0$ and $\bPi^{(4)}_a=7(p_a/\rho_a)\bPi^{(2)}_a$,
so that
\begin{equation}
\frac{1}{2}\nabla\cdot \bPi^{(4)}_a 
-\frac{5}{2}\frac{p_a}{\rho_a}\nabla\cdot\bPi^{(2)}_a \overset{\textrm{13-m}}{\to} \frac{p_a}{\rho_a}\nabla\cdot\bPi^{(2)}_a
+\frac{7}{2}\bPi^{(2)}_a\cdot\nabla\Big(\frac{p_a}{\rho_a}\Big),
\end{equation}
then evolution equation (\ref{eq:Energy46}) recovers equations (39)-(40) of \cite{Schunk1977}.

\subsection{Evolution equation for viscosity-tensor \texorpdfstring{$\bPi^{(4)}_a$}{Pi4}}
Nonlinear evolution equation for the 4th-order moment $r_{ijkl}^a = X^{a(4)}_{ijkl}$ is given by (\ref{eq:Thierry-R}).
First, we need to obtain evolution equation for matrix
$(\trace\br^a)_{ij}=X^{a(4)}_{ij}$, which is further decomposed into (\ref{eq:Energy44}) \& (\ref{eq:Energy45}).
Applying trace at (\ref{eq:Thierry-R}) yields
\begin{eqnarray}
  && \frac{d_a}{dt} \trace \br_a +\nabla\cdot\big( \trace\bX^{(5)}_a\big) +(\nabla\cdot\bu_a)\trace\br_a +2\br_a:\nabla\bu_a \nn\\
  && +\Big[ (\trace\br_a)\cdot\nabla\bu_a +\Omega_a\bhat\times(\trace\br_a) -\frac{2}{\rho_a}(\nabla\cdot\bp_a)\vecq_a \Big]^S
  -\frac{2}{\rho_a}(\nabla\cdot\bp_a)\cdot\bq_a \nn\\
  && = \trace \bQ^{(4)}_a -\frac{2}{\rho_a}\Big[ \big(\boldsymbol{R}_a \vecq_a\big)^S +\boldsymbol{R}_a\cdot \bq_a \Big]. \label{eq:PPosled1}
\end{eqnarray}
As a quick double-check, equation (\ref{eq:PPosled1}) appears equivalent to equation (3.4.35), page 154 of \cite{Balescu1988} (after accounting for different
normalization constants of $1/2$ and adding a missing ``s'' index to his 4th-order moment $S_{rsnm}$). 
Applying another trace at  (\ref{eq:PPosled1}) yields
\begin{eqnarray}
  && \frac{d_a}{dt} X^{(4)}_a +\nabla\cdot\vecX^{(5)}_a + (\nabla\cdot\bu_a) X^{(4)}_a +4(\trace\br_a):\nabla\bu_a
  -\frac{8}{\rho_a} (\nabla\cdot\bp_a)\cdot \vecq_a \nn\\
  && = \trace\trace \bQ^{(4)}_a -\frac{8}{\rho_a} \boldsymbol{R}_a \cdot\vecq_a. \label{eq:PPosled2}
\end{eqnarray}
To obtain evolution equation for matrix $\Pi^{a(4)}_{ij}$, we need to subtract $(\bI/3)$ times (\ref{eq:PPosled2}) from (\ref{eq:PPosled1}).
For example, we need to calculate
\begin{eqnarray}
  X_{ijk}^{a(5)} &=& \frac{1}{5} \Big( X_i^{a(5)}\delta_{jk}+ X_j^{a(5)}\delta_{ik} + X_k^{a(5)}\delta_{ij} \Big);\nn\\
  \pr_k  X_{kij}^{a(5)} &=&  \frac{1}{5} \Big( \pr_j X_i^{a(5)} +\pr_i X_j^{a(5)} +\delta_{ij} \pr_k X_k^{a(5)} \Big);\nn\\
  (\pr_k  X_{kij}^{a(5)}) -\frac{\delta_{ij}}{3} \pr_k X_k^{a(5)} &=& \frac{1}{5} \Big( \pr_j X_i^{a(5)} +\pr_i X_j^{a(5)} -\frac{2}{3}\delta_{ij} \pr_k X_k^{a(5)} \Big),
\end{eqnarray}  
together with
\begin{eqnarray}
  \br_a:\nabla\bu_a &=& \frac{1}{15}X^{(4)}_a\Big((\nabla\bu_a)^S+ \bI(\nabla\cdot\bu_a)\Big) \nn\\
  && +\frac{1}{7} \Big[ \bPi^{(4)}_a (\nabla\cdot\bu_a) +\bI (\bPi^{(4)}_a:\nabla\bu_a) + \big( \bPi^{(4)}_a\cdot\nabla\bu_a\big)^S
    + \big( (\nabla\bu_a)\cdot \bPi^{(4)}_a \big)^S \Big];\nn\\
  \trace \br_a:\nabla\bu_a &=& \frac{1}{3}X^{(4)}_a(\nabla\cdot\bu_a) +\bPi^{(4)}_a:\nabla\bu_a;\nn\\
  2 \br_a:\nabla\bu_a -\frac{\bI}{3}4 \trace \br_a:\nabla\bu_a &=& \frac{2}{15}X^{(4)}_a\Big((\nabla\bu_a)^S -\frac{7}{3} \bI(\nabla\cdot\bu_a)\Big)
  -\frac{22}{21} \bI (\bPi^{(4)}_a:\nabla\bu_a)\nn\\
  && +\frac{2}{7} \Big[ \bPi^{(4)}_a (\nabla\cdot\bu_a) + \big( \bPi^{(4)}_a\cdot\nabla\bu_a\big)^S
    + \big( (\nabla\bu_a)\cdot \bPi^{(4)}_a \big)^S \Big],
\end{eqnarray}
and useful identities are
\begin{eqnarray}
  \big[ (\trace\br_a)\cdot\nabla\bu_a \big]^S &=& \frac{1}{3}X^{(4)}_a (\nabla\bu_a)^S +\big[\bPi^{(4)}_a\cdot\nabla\bu\big]^S;\nn\\
  \big[ \bhat\times(\trace\br_a)   \big]^S &=& \big[ \bhat\times \bPi^{(4)}_a  \big]^S.
\end{eqnarray}
The heat flux  contributions calculate
\begin{equation}
  (\nabla\cdot\bp_a)\cdot\bq_a = \frac{2}{5}\Big[ \big((\nabla\cdot\bp_a)\vecq_a\big)^S +\bI (\nabla\cdot\bp_a)\cdot\vecq_a\Big],\nn
\end{equation}
so the heat fluxes are added as
\begin{eqnarray}
&&   -\,2 \big((\nabla\cdot\bp_a)\vecq_a\big)^S -2(\nabla\cdot\bp_a)\cdot\bq_a
  +\bI\frac{8}{3} (\nabla\cdot\bp_a)\cdot\vecq_a \nn\\
&&  = -\, \frac{14}{5} \Big[ \big((\nabla\cdot\bp_a)\vecq_a\big)^S -\frac{2}{3}\bI (\nabla\cdot\bp_a)\cdot\vecq_a\Big].
\end{eqnarray}
The fully nonlinear evolution equation for matrix $\bPi^{(4)}_a$ thus reads 
\begin{eqnarray}
  && \frac{d_a}{dt} \bPi^{(4)}_a +\frac{1}{5}\Big[ (\nabla\vecX^{(5)}_a)^S-\frac{2}{3}\bI(\nabla\cdot\vecX^{(5)}_a)\Big]
  +\frac{9}{7}(\nabla\cdot\bu_a)\bPi^{(4)}_a +\frac{9}{7}(\bPi^{(4)}_a\cdot\nabla\bu_a)^S\nn\\
&&  + \frac{2}{7}\big((\nabla\bu_a)\cdot\bPi^{(4)}_a\big)^S
  -\frac{22}{21}\bI (\bPi^{(4)}_a:\nabla\bu_a)
 -\, \frac{14}{5\rho_a} \Big[ \big((\nabla\cdot\bp_a)\vecq_a\big)^S -\frac{2}{3}\bI (\nabla\cdot\bp_a)\cdot\vecq_a\Big] \nn\\
 && +\Omega_a \big( \bhat\times \bPi^{(4)}_a \big)^S
 + \frac{7}{15}\big(15\frac{p_a^2}{\rho_a}+\widetilde{X}^{(4)}_a \big)\Big[ (\nabla\bu_a)^S-\frac{2}{3}\bI(\nabla\cdot\bu_a)\Big] \nn\\
&& = \bQ^{(4)}_a\,' \equiv \trace \bQ^{(4)}_a -\frac{\bI}{3}\trace\trace \bQ^{(4)}_a
 -\frac{14}{5\rho_a}\Big[ (\boldsymbol{R}_a\vecq_a)^S-\frac{2}{3}\bI (\boldsymbol{R}_a\cdot\vecq_a)\Big]. \label{eq:r1stTraceF}
\end{eqnarray}
At the semi-linear level (while keeping the $d/dt$) evolution equation (\ref{eq:r1stTraceF}) simplifies into
\begin{eqnarray}
  && \frac{d_a}{dt} \bPi^{(4)}_a +\frac{1}{5}\Big[ (\nabla\vecX^{(5)}_a)^S-\frac{2}{3}\bI(\nabla\cdot\vecX^{(5)}_a)\Big]
  +\Omega_a \big( \bhat\times \bPi^{(4)}_a \big)^S \nn\\
&&  + 7 \frac{p_a^2}{\rho_a} \Big[ (\nabla\bu_a)^S-\frac{2}{3}\bI(\nabla\cdot\bu_a)\Big] 
 = \bQ^{(4)}_a\,' = \trace \bQ^{(4)}_a -\frac{\bI}{3}\trace\trace \bQ^{(4)}_a.\label{eq:r1stTraceF2}
\end{eqnarray}
Finally, neglecting the coupling between heat fluxes and viscosities (which is the choice of Braginskii),   
the simplest evolution equation reads
\begin{eqnarray}
  && \frac{d_a}{dt} \bPi^{(4)}_a  +\Omega_a \big(\bhat\times \bPi^{(4)}_a \big)^S + 7 \frac{p_a^2}{\rho_a} \bW_a 
  = \bQ^{(4)}_a\,', \label{eq:r1stTraceFF}
\end{eqnarray}
where $\bW_a = (\nabla\bu_a)^S-(2/3)\bI(\nabla\cdot\bu_a)$ is the usual rate-of-strain tensor.

\subsection{Evolution equation for perturbation \texorpdfstring{$\widetilde{X}^{(4)}_a$}{\textasciitilde X4}}
Fully non-linear evolution equation (\ref{eq:PPosled2}) for $X^{(4)}_a$ reads
\begin{eqnarray}
  && \frac{d_a}{dt} X^{(4)}_a +\nabla\cdot\vecX^{(5)}_a  
  +\frac{7}{3}X^{(4)}_a(\nabla\cdot\bu_a) +4\bPi^{(4)}_a:\nabla\bu_a 
  -\frac{8}{\rho_a} (\nabla\cdot\bp_a)\cdot \vecq_a \nn\\
  && = \trace\trace \bQ^{(4)}_a -\frac{8}{\rho_a} \boldsymbol{R}_a \cdot\vecq_a. \label{eq:PPosled10}
\end{eqnarray}
Then by using $X^{(4)}_a=15 (p_a^2/\rho_a)+\widetilde{X}^{(4)}_a$ with
\begin{equation}
  \frac{d_a}{dt} \Big(\frac{p_a^2}{\rho_a}\Big) = \frac{p_a}{\rho_a}\Big[
    -\frac{7}{3}p_a\nabla\cdot\bu_a -\frac{4}{3}\nabla\cdot\vecq_a-\frac{4}{3}\bPi^{(2)}_a:\nabla\bu_a+\frac{4}{3}Q_a\Big],
\end{equation}
one obtains fully non-linear evolution equation for $\widetilde{X}^{(4)}_a$
\begin{eqnarray}
  && \frac{d_a}{dt} \widetilde{X}^{(4)}_a +\nabla\cdot\vecX^{(5)}_a  -20 \frac{p_a}{\rho_a}\nabla\cdot\vecq_a
  +\frac{7}{3}\widetilde{X}^{(4)}_a(\nabla\cdot\bu_a) +4\big(\bPi^{(4)}_a -5\frac{p_a}{\rho_a}\bPi^{(2)}_a\big):\nabla\bu_a \nn\\
 && -\frac{8}{\rho_a} (\nabla\cdot\bp_a)\cdot \vecq_a 
    = \widetilde{Q}^{(4)}_a\,' \equiv \trace\trace \bQ^{(4)}_a -20 \frac{p_a}{\rho_a}Q_a-\frac{8}{\rho_a} \boldsymbol{R}_a \cdot\vecq_a, \label{eq:PPosled11}
\end{eqnarray}
and at the semi-linear level
\begin{eqnarray}
  && \frac{d_a}{dt} \widetilde{X}^{(4)}_a +\nabla\cdot\vecX^{(5)}_a  -20 \frac{p_a}{\rho_a}\nabla\cdot\vecq_a
     = \widetilde{Q}^{(4)}_a\,' = \trace\trace \bQ^{(4)}_a -20 \frac{p_a}{\rho_a}Q_a. \label{eq:PPosled12}
\end{eqnarray}
Collisional contributions can be found in Section \ref{sec:SummaryQ4}; see equation (\ref{eq:Thierry39}).


\subsection{Evolution equation for heat flux vector \texorpdfstring{$\vecX^{(5)}_a$}{X5}}
Applying trace twice at (\ref{eq:X5_tensor}) yields
\begin{eqnarray}
&&  \frac{\pr}{\pr t}\trace\trace \bX^{a(5)} +\nabla\cdot\big( \trace\trace \bX^{a(6)}\big)
  +\nabla\cdot\big(\bu^a \trace\trace \bX^{a(5)}\big) +\big(\trace\trace\bX^{a(5)}\cdot\nabla\big)\bu^a\nn\\
  &&  +4\big(\trace \bX^{a(5)}\big):\nabla\bu^a +\Omega_a \bhat\times \big(\trace\trace\bX^{a(5)}\big)
  -\frac{1}{\rho_a}\Big[ \big(\nabla\cdot\bp^a\big)\trace\trace \bX^{a(4)}
    +4\big(\nabla\cdot\bp^a\big)\cdot \trace\bX^{a(4)}\Big] \nn\\
  && = \trace\trace\bQ^{a(5)}-\frac{1}{\rho_a}\Big[ \boldsymbol{R}^a \trace\trace\bX^{a(4)}
    +4 \boldsymbol{R}^a\cdot \trace \bX^{a(4)} \Big].
\end{eqnarray}
By using definition of vectors $\vecX^{(5)}=\trace\trace \bX^{(5)}$, $\vecQ^{(5)}=\trace\trace\bQ^{(5)}$ and
\begin{eqnarray}
  X^{(5)}_{ijk} &=& \frac{1}{5}\Big[ X^{(5)}_i\delta_{jk}+ X^{(5)}_j\delta_{ik} + X^{(5)}_k\delta_{ij}\Big];\nn\\
  X^{(5)}_{ijk}\pr_j u_k &=& \frac{1}{5}\Big[ X^{(5)}_i \nabla\cdot\bu_a + X^{(5)}_j \pr_j u_i^a + X^{(5)}_k \pr_i u_k^a\Big],
\end{eqnarray}  
together with decompositions (\ref{eq:Energy44}) \& (\ref{eq:Energy45}),   
the fully non-linear evolution equation becomes 
\begin{eqnarray}
&&  \frac{d_a}{d t}\vecX^{(5)}_a +\frac{1}{3}\nabla\widetilde{X}^{(6)}_a +\nabla\cdot \bPi^{(6)}_a\nn\\
  &&  +\frac{9}{5}\vecX^{(5)}_a (\nabla\cdot\bu_a) + \frac{9}{5}\vecX^{(5)}_a\cdot\nabla\bu_a
  + \frac{4}{5}(\nabla\bu_a)\cdot\vecX^{(5)}_a +\Omega_a \bhat\times \vecX^{(5)}_a\nn\\
&&  +70 \frac{p_a^2}{\rho_a}\nabla\Big(\frac{p_a}{\rho_a}\Big) -35 \frac{p_a^2}{\rho_a^2} \nabla\cdot\bPi^{(2)}_a -\frac{7}{3\rho_a}\big(\nabla\cdot\bp^a\big)\widetilde{X}^{(4)}_a
  -\frac{4}{\rho_a} \big(\nabla\cdot\bp^a\big)\cdot \bPi^{(4)}_a \nn\\
  && =\vecQ^{(5)}_a\,' \equiv \vecQ^{(5)}_a -35 \frac{p_a^2}{\rho_a^2}\boldsymbol{R}_a -\frac{7}{3\rho_a} \boldsymbol{R}_a \widetilde{X}^{(4)}_a
  - \frac{4}{\rho_a} \boldsymbol{R}_a\cdot\bPi^{(4)}_a. \label{eq:Nomore100}
\end{eqnarray}
Because we do not go higher in the hierarchy, the model is closed with closures (see equations (\ref{eq:Thierry-1}) \& (\ref{eq:Thierry-2})
or Section \ref{sec:HC} with Tables \ref{table:HC1} \& \ref{table:HC5}) 
\begin{equation}
\widetilde{X}^{(6)}_a = 21 \frac{p_a}{\rho_a} \widetilde{X}^{(4)}_a; \qquad \bPi^{(6)}_a = 18 \frac{p_a}{\rho_a} \bPi^{(4)}_a -63 \frac{p_a^2}{\rho_a^2}\bPi^{(2)}_a.  
\end{equation}
At a  semi-linear level equation (\ref{eq:Nomore100}) becomes
\begin{eqnarray}
  &&  \frac{d_a}{d t}\vecX^{(5)}_a +7\frac{p_a}{\rho_a}\nabla\widetilde{X}^{(4)}_a +18\frac{p_a}{\rho_a}\nabla\cdot \bPi^{(4)}_a
  -98 \frac{p_a^2}{\rho_a^2}\nabla\cdot\bPi^{(2)}_a   \nn\\
&&  +\Omega_a \bhat\times \vecX^{(5)}_a  +70 \frac{p_a^2}{\rho_a}\nabla\Big(\frac{p_a}{\rho_a}\Big) 
   =\vecQ^{(5)}_a\,' = \vecQ^{(5)}_a -35 \frac{p_a^2}{\rho_a^2}\boldsymbol{R}_a. \label{eq:Nomore101}
\end{eqnarray}

\newpage
\section{Simplified general fluid hierarchy} \label{sec:Hierarchy}
\setcounter{equation}{0}
Previously, we introduced a full fluid hierarchy in Section \ref{sec:General}, which contains n-dimensional moments $X^{(n)}_{ijk\ldots n}$.
By applying contractions at these moments  in Appendix \ref{sec:22momentE}, we have derived evolution equations for the 22-moment model.
Instead of doing that, it is of course possible to obtain evolution equations for the contracted moments directly from the
Boltzmann equation. This simplified hierarchy is formulated with heat fluxes (vectors) and stress-tensors (matrices)
\begin{eqnarray}
\vecX^{(2n+1)}_a = m_a \int \bc_a|\bc_a|^{2n} f_a d^3v; \qquad 
\bPi^{(2n)}_a = m_a \int \big(\bc_a\bc_a-\frac{\bI}{3}|\bc_a|^{2}\big) |\bc_a|^{2n-2} f_a d^3v,
\end{eqnarray}
together with fully contracted scalars which are decomposed into a Maxwellian core and perturbation (notation with tilde)
\begin{equation}
  X^{(2n)}_a =  m_a \int |\bc_a|^{2n} f_a d^3v = (2n+1)!! \frac{p^n_a}{\rho^{n-1}_a}+\widetilde{X}^{(2n)}_a, \label{eq:Num41}
\end{equation}
meaning a definition $\widetilde{X}^{(2n)}_a =  m_a \int |\bc_a|^{2n} (f_a -f_a^{(0)}) d^3v$, where $f_a^{(0)}$ is Maxwellian.
In another words, one considers matrices
\begin{equation}
 X^{a(2n)}_{ij} =  m_a \int |\bc_a|^{2n-2}c^a_i c^a_j f_a d^3v = \frac{\delta_{ij}}{3} X^{a(2n)} +\Pi^{a(2n)}_{ij}, \label{eq:Num40}
\end{equation}
which are decomposed into fully contracted scalars and stress-tensors. 
Note that $\vecX^{(1)}_a=0$ and $\widetilde{X}^{(2)}_a=0$.

Unfortunatelly, the traditional definition of the heat flux vector $\vecq_a=(1/2)\trace \bq_a$ which contains a factor of $1/2$,
goes against the general ideology that no additional factors are introduced by contractions. Also, we have previously reserved
vector $\vecQ^{(3)}_a\,'$ for the right hand side of the heat flux $\vecq_a$ evolution equation, and not for $\vecX^{(3)}_a$.
Obviously, our previous notation is not ideal for generalization to an n-th  order moments. To circumvent all the problems with
  the previous definitions, we define new collisional contributions for heat fluxes and stress-tensors
with $\mathcal{Q}$ (mathcal of Q), as vectors and matrices     
\begin{eqnarray}
   \mathcal{Q}^{a(2n+1)}_i &=& m_a \int |\bc_a|^{2n} c^a_i C(f_a) d^3v ;\nn\\
   \mathcal{Q}^{a(2n)}_{ij} &=& m_a \int |\bc_a|^{2n-2} c^a_i c^a_j C(f_a) d^3v ; \label{eq:Thierry20}
\end{eqnarray}
together with fully contracted
\begin{eqnarray}
    Q^{(2n)}_a &=& m_a \int |\bc_a|^{2n} C(f_a) d^3v;\qquad Q_a = \frac{m_a}{2} \int |\bc_a|^{2} C(f_a) d^3v. \label{eq:Thierry21}
\end{eqnarray}
The energy exchange rates $Q_a$ contain the traditional factor of $1/2$, and $Q^{(2)}_a=2Q_a$. The momentum
exchange rates $\boldsymbol{R}_a=m_a \int \bV C(f_a) d^3v$. In the vector notation matrix
$\bar{\bar{\boldsymbol{\mathcal{Q}}}}^{a(2n)}=\trace\trace\ldots \trace \bQ^{a(2n)}$.

Then, direct integration of the Boltzmann equation and subtraction of momentum equations yields evolution equations for scalars
\begin{eqnarray}
&&  \frac{\pr}{\pr t} X^{a(2n)} +\pr_k( u^a_k X^{a(2n)}) + \pr_k X^{a(2n+1)}_k 
  +(2n) X^{a(2n)}_{ik} \pr_k u^a_i \nn\\
&& \qquad -\frac{(2n)}{\rho_a} (\nabla\cdot\bp^a)_k X^{a(2n-1)}_k = Q^{a(2n)} -\frac{(2n)}{\rho_a} R^a_k X^{a(2n-1)}_k,\label{eq:Num1}
\end{eqnarray}
where $(n)$ without species index should not be confused with the number density,  
evolution equations for vectors
\begin{eqnarray}
&&  \frac{\pr}{\pr t}X^{a(2n+1)}_i +\pr_k(u_k^a X^{a(2n+1)}_i)+\pr_k X^{a(2n+2)}_{ki}  +X^{a(2n+1)}_k \pr_k u_i^a
  +(2n) X^{a(2n+1)}_{ijk} \pr_k u_j^a \nn\\
  && -\frac{(2n)}{\rho_a}(\nabla\cdot\bp^a)_k X^{a(2n)}_{ki} -\frac{1}{\rho_a}(\nabla\cdot\bp^a)_i X^{a(2n)}
  +\Omega_a (\bhat\times \vecX^{a(2n+1)})_i \nn\\
  && = \mathcal{Q}^{a(2n+1)}_i -\frac{1}{\rho_a} R^a_i X^{a(2n)}-\frac{(2n)}{\rho_a} R^a_k X^{a(2n)}_{ki}, \label{eq:Num2}
\end{eqnarray}
and matrices
\begin{eqnarray}
&&  \frac{\pr}{\pr t} X^{a(2n)}_{ij} + \pr_k (u_k^a X^{a(2n)}_{ij}) +\pr_k X^{a(2n+1)}_{kij} + (2n-2)X^{a(2n)}_{ijkl}(\pr_k u_l^a)\nn\\
  && + \Big[ X^{a(2n)}_{ik} \pr_k u_j^a +\Omega_a (\bhat\times \bX^{a(2n)})_{ij} - \frac{1}{\rho_a} (\nabla\cdot\bp^a)_i X^{a(2n-1)}_j \Big]^S   
   -\frac{(2n-2)}{\rho_a}  (\nabla\cdot \bp^a)_k X^{a(2n-1)}_{kij}\nn\\
  && = \mathcal{Q}^{a(2n)}_{ij} -\frac{1}{\rho_a}\Big[ R_i^a X^{a(2n-1)}_j \Big]^S -\frac{ (2n-2)}{\rho_a} R_k^a X^{a(2n-1)}_{kij}, \label{eq:Num3}
\end{eqnarray}
which are valid for $n\ge 1$. For example evaluating (\ref{eq:Num1}) for $n=1$ yield evolution equation for scalar pressure $p_a$.
Applying trace at (\ref{eq:Num3}) recovers (\ref{eq:Num1}).

Matrices $X_{ij}^{a(2n)}$ are then decomposed according to (\ref{eq:Num40}) where stress-tensors $\Pi^{a(2n)}_{ij}$ are traceless, 
and higher-order tensors are decomposed according to (where tensors $\sigma$ are neglected, which is the core of the hierarchy simplification)
\begin{eqnarray}
  X^{a(2n+1)}_{ijk} &=& \frac{1}{5}\Big[ X^{a(2n+1)}_i\delta_{jk}+ X^{a(2n+1)}_j\delta_{ik} + X^{a(2n+1)}_k\delta_{ij}\Big];\label{eq:Num5}\\
  X_{ijkl}^{a(2n)} &=& \frac{1}{15} X^{a(2n)} \big(\delta_{ij}\delta_{kl}+\delta_{ik}\delta_{jl}+\delta_{il}\delta_{jk}\big)\nn\\
  && +\frac{1}{7} \Big[ \Pi_{ij}^{a(2n)} \delta_{kl}+\Pi_{ik}^{a(2n)}\delta_{jl}+\Pi_{il}^{a(2n)}\delta_{jk}
  +\Pi_{jk}^{a(2n)}\delta_{il}+\Pi_{jl}^{a(2n)}\delta_{ik}+\Pi_{kl}^{a(2n)}\delta_{ij} \Big]. \label{eq:Num6}
\end{eqnarray}
Applying trace at (\ref{eq:Num5}) yields identity, and applying trace at (\ref{eq:Num6}) yields
decomposition (\ref{eq:Num40}). Evolution equations for fully contracted moments (scalars) then become
\begin{eqnarray}
&&  \frac{d_a}{d t} X^{(2n)}_a + \nabla\cdot\vecX^{(2n+1)}_a +\frac{(2n+3)}{3} X^{(2n)}_a\nabla\cdot\bu_a
  +(2n) \bPi^{(2n)}_a:\nabla\bu_a \nn\\
&& \qquad -\frac{(2n)}{\rho_a} (\nabla\cdot\bp_a)\cdot\vecX^{(2n-1)}_a = Q^{(2n)}_a -\frac{(2n)}{\rho_a}\boldsymbol{R}_a\cdot\vecX^{(2n-1)}_a, 
\end{eqnarray}
for heat fluxes (vectors)
\begin{eqnarray}
  &&  \frac{d_a}{d t}\vecX^{(2n+1)}_a + \frac{(2n+5)}{5}\Big[\vecX^{(2n+1)}_a \nabla\cdot\bu_a +\vecX^{(2n+1)}_a\cdot \nabla \bu_a\Big]
  +\frac{(2n)}{5}(\nabla\bu_a)\cdot \vecX^{(2n+1)}_a \nn\\
  && +\frac{1}{3}\nabla X^{(2n+2)}_a +\nabla\cdot\bPi^{(2n+2)}_a
  -\frac{(2n+3)}{3\rho_a}(\nabla\cdot\bp_a) X^{(2n)}_a -\frac{(2n)}{\rho_a}(\nabla\cdot\bp_a)\cdot\bPi^{(2n)}_a \nn\\
  && +\Omega_a \bhat\times \vecX^{(2n+1)}_a 
   = \vec{\boldsymbol{\mathcal{Q}}}^{(2n+1)}_a -\frac{(2n+3)}{3\rho_a} \boldsymbol{R}_a X^{(2n)}_a-\frac{(2n)}{\rho_a}\boldsymbol{R}_a\cdot\bPi^{(2n)}_a,
\end{eqnarray}
and for stress-tensors (matrices)
\begin{eqnarray}
  && \frac{d_a}{dt} \bPi^{(2n)}_a +\frac{1}{5}\Big[ \big(\nabla\vecX^{(2n+1)}_a\big)^S -\frac{2}{3}\bI \nabla\cdot\vecX^{(2n+1)}_a \Big]
  +\frac{(2n+5)}{7} \bPi^{(2n)}_a (\nabla\cdot\bu_a)\nn\\
  && +\Big[ \frac{(2n+5)}{7} \big(\bPi^{(2n)}_a \cdot\nabla\bu_a\big)^S +\frac{(2n-2)}{7} \big((\nabla\bu_a)\cdot \bPi^{(2n)}_a  \big)^S
    -\frac{2(4n+3)}{21}\bI (\bPi^{(2n)}_a:\nabla\bu_a) \Big]\nn\\
   && -\frac{(2n+3)}{5\rho_a} \Big[ \big( (\nabla\cdot\bp_a)\vecX^{(2n-1)}_a\big)^S -\frac{2}{3}\bI (\nabla\cdot\bp_a)\cdot\vecX^{(2n-1)}_a \Big]\nn\\
  && +\Omega_a \big( \bhat\times \bPi^{(2n)}_a\big)^S + \frac{(2n+3)}{15} X^{(2n)}_a \bW_a\nn\\
  && = \bar{\bar{\boldsymbol{\mathcal{Q}}}}^{(2n)}_a\, ' \equiv
  \bar{\bar{\boldsymbol{\mathcal{Q}}}}^{(2n)}_a -\frac{\bI}{3}Q^{(2n)}_a -\frac{(2n+3)}{5\rho_a} \Big[ \big( \boldsymbol{R}_a \vecX^{(2n-1)}_a\big)^S
    -\frac{2}{3}\bI \boldsymbol{R}_a \cdot \vecX^{(2n-1)}_a \Big]. \label{eq:Num8}
\end{eqnarray}
By applying trace at equation (\ref{eq:Num8}) it can be verified that it is traceless. 

The fully contracted scalar variables are then decomposed into a Maxwellian core and perturbation (with tilde) according to (\ref{eq:Num41}),
yielding evolution equation for scalars
\begin{eqnarray}
&& \frac{d_a}{dt} \widetilde{X}^{(2n)}_a +\nabla\cdot\vecX^{(2n+1)}_a +\frac{(2n+3)}{3}\widetilde{X}^{(2n)}_a \nabla\cdot\bu_a +(2n) \bPi^{(2n)}_a:\nabla\bu_a \nn\\
  && -(2n+1)!! \frac{(2n)}{3}\Big(\frac{p_a}{\rho_a} \Big)^{n-1} \Big[ \nabla\cdot\vecq_a +\bPi^{(2)}_a:\nabla\bu_a\Big]
  -\frac{(2n)}{\rho_a} (\nabla\cdot\bp_a)\cdot\vecX^{(2n-1)}_a \nn\\
  && = \widetilde{Q}^{(2n)}_a\,' \equiv
  Q^{(2n)}_a - (2n+1)!! \frac{(2n)}{3}\Big(\frac{p_a}{\rho_a} \Big)^{n-1} Q_a -\frac{(2n)}{\rho_a}\boldsymbol{R}_a\cdot\vecX^{(2n-1)}_a, \label{eq:Num9}
\end{eqnarray}
and heat fluxes
\begin{eqnarray}
  &&  \frac{d_a}{d t}\vecX^{(2n+1)}_a + \frac{(2n+5)}{5}\Big[\vecX^{(2n+1)}_a \nabla\cdot\bu_a +\vecX^{(2n+1)}_a\cdot \nabla \bu_a\Big]
  +\frac{(2n)}{5}(\nabla\bu_a)\cdot \vecX^{(2n+1)}_a \nn\\
  && +\frac{1}{3}\nabla \widetilde{X}^{(2n+2)}_a +\nabla\cdot\bPi^{(2n+2)}_a
  -\frac{(2n+3)}{3\rho_a}(\nabla\cdot\bp_a) \widetilde{X}^{(2n)}_a -\frac{(2n)}{\rho_a}(\nabla\cdot\bp_a)\cdot\bPi^{(2n)}_a \nn\\
  && +(2n+3)!! \frac{(n)}{3} \frac{p_a^{n}}{\rho_a^{n-1}}\nabla\Big(\frac{p_a}{\rho_a} \Big)
  -\frac{(2n+3)!!}{3} \frac{p_a^n}{\rho_a^n} \nabla\cdot\bPi^{(2)}_a 
   +\Omega_a \bhat\times \vecX^{(2n+1)}_a \nn\\ 
&& =\vec{\boldsymbol{\mathcal{Q}}}^{(2n+1)}_a\,' \equiv \vec{\boldsymbol{\mathcal{Q}}}^{(2n+1)}_a -\frac{(2n+3)}{3\rho_a} \boldsymbol{R}_a \widetilde{X}^{(2n)}_a
  -\frac{(2n+3)!!}{3} \frac{p_a^n}{\rho_a^n}\boldsymbol{R}_a
  -\frac{(2n)}{\rho_a}\boldsymbol{R}_a\cdot\bPi^{(2n)}_a. \label{eq:Num10}
\end{eqnarray}
Evolution equation for stress-tensors (\ref{eq:Num8}) contains only one trivial term with $X^{(2n)}_a$, where
\begin{equation*}
\frac{(2n+3)}{15} X^{(2n)}_a \bW_a = \frac{(2n+3)!!}{15} \frac{p^n_a}{\rho_a^{n-1}} \bW_a + \frac{(2n+3)}{15} \widetilde{X}^{(2n)}_a \bW_a,
\end{equation*}
and we do not re-write the full equation. Equations (\ref{eq:Num8})-(\ref{eq:Num10}) are valid for $n\ge 1$, where
for $n=1$ (\ref{eq:Num9}) reduces to zero, so this equation is meaningfull only for $n \ge 2$. In the semi-linear approximation,
the hierarchy simplifies into (\ref{eq:Num11})-(\ref{eq:Num13}).

\newpage
\section{BGK collisional operator} \label{sec:BGK}
\setcounter{equation}{0}
Before calculations with the Landau collisional operator, it is beneficial to first get familiar with the
heuristic relaxation-type operator known as BGK, after Bhatnagar-Gross-Krook \citep{BGK1954,GK1956}, written in the following form
\begin{equation} \label{eq:BGK_multi}
C(f_a) = \sum_b C_{ab}(f_a) = -\sum_b \nu_{ab} (f_a-f^{(0)}_{ab}).
\end{equation}
The Maxwellian $f^{(0)}_{ab}$ has two indices and is defined as
\begin{equation} \label{eq:MaxBGK}
f_{ab}^{(0)} = n_a \Big( \frac{m_a}{2\pi T_{a}} \Big)^{3/2} \exp \Big( -\frac{m_a |\bV-\bu_{b}|^2}{2T_{a}}\Big).
\end{equation}
Note that only velocity $\bu_b$ has index ``b'' and that temperature, mass and density has index ``a''. To account for different temperatures
is possible by considering generalized BGK operators of \cite{Haack2017}. 
The simple BGK operator yields momentum and energy exchange rates 
\begin{equation}
\boldsymbol{R}_{ab} = \rho_a\nu_{ab}  (\bu_b-\bu_a); \qquad  Q_{ab} = \frac{1}{2}\rho_a \nu_{ab}|\bu_b-\bu_a|^2,
\end{equation}
where both the momentum and energy are conserved (note that for heuristic operators it is advisable to directly calculate
both $\boldsymbol{R}_{ab}$ and $\boldsymbol{R}_{ba}$ together with $Q_{ab}$ and $Q_{ba}$ to verify that they are well defined).
This BGK operator also satisfies the Boltzmann H-theorem, which for multi-species plasmas has a general form
\begin{equation} \label{eq:Htheorem2}
\int C_{ab}(f_a)\ln f_a d^3v + \int C_{ba}(f_b)\ln f_b d^3v \le 0,
\end{equation}
where the equality is true only if $f_a$ and $f_b$ are Maxwellians. 
For the BGK operator, each part of the H-theorem (\ref{eq:Htheorem2}) is satisfied independently.  
It can be shown that $\int (f_a-f_{ab}^{(0)}) \ln f_{ab}^{(0)} d^3v = 0$, and subtracting this integral from the first term
of (\ref{eq:Htheorem2}) yields
\begin{eqnarray}
  \int C_{ab}(f_a) \ln f_a d^3v &=& \nu_{ab} \int (f_{ab}^{(0)}-f_a) \ln f_a d^3v - \underbrace{\nu_{ab}\int (f_{ab}^{(0)}-f_a) \ln f_{ab}^{(0)} d^3v}_{0} \nn\\
  &=& \nu_{ab}\int (f_{ab}^{(0)}-f_a) \ln \Big( \frac{f_a}{f_{ab}^{(0)}}\Big) d^3v \le 0,
\end{eqnarray}
where in the last step one uses that for any real numbers $a>0$ and $b>0$ the following identity holds
$ (a-b)\ln (b/a) \le 0$ (the identity is easily verified, because for $a > b$ the first term is positive and the logarithm is negative, and
for $a < b$ the first term is negative and the logarithm is positive; the identity is equal to zero only if $a = b$).

The BGK collisional contributions calculate
\begin{eqnarray}
  \bQ^{(2)}_{ab} = m_a\int \bc_a\bc_a C_{ab}(f_a)d^3v &=& -\nu_{ab} \bPi_a^{(2)} +\nu_{ab}\rho_a \delta\bu \delta\bu; \label{eq:Q2ab}\\
  \bQ^{(3)}_{ab} = m_a\int \bc_a\bc_a \bc_a C_{ab}(f_a)d^3v &=& -\nu_{ab}\bq_a +\nu_{ab}p_a \big[\delta\bu\bI\big]^S
  +\nu_{ab} \rho_a \delta\bu \delta\bu \delta\bu, 
\end{eqnarray}
where $\delta\bu=\bu_b-\bu_a$.

\subsection{Viscosity-tensor \texorpdfstring{$\bPi^{(2)}_a$}{Pi2}}
Collisional contributions that enter the r.h.s. of evolution equation (\ref{eq:Energy43}) are
\begin{equation}
  \bQ^{(2)}_a\,' \equiv  \bQ^{(2)}_a -\frac{\bI}{3}\textrm{Tr}\bQ^{(2)}_a = -\bnu_a \bPi^{(2)}_a - \bW_a^{\textrm{frict}},
\end{equation}
where we defined
\begin{eqnarray}
  \bnu_a &=& \sum_b\nu_{ab}; \\
  \bW_a^{\textrm{frict}} &=& -\rho_a \sum_b\nu_{ab}\big( \delta\bu\delta\bu-\frac{\bI}{3}|\delta\bu|^2\big), \label{eq:Stress3}
\end{eqnarray}
and where superscript 'frict' means frictional contributions due to $\delta\bu$. The frictional contributions are only non-linear,
but we keep them to show that it is possible to take them into account. Using quasi-static approximation, 
evolution equation (\ref{eq:Energy43}) can be simplified into 
\begin{equation} \label{eq:PtensorA}
  (\bhat\times\bPi^{(2)}_a)^S +\frac{\bnu_a}{\Omega_a}\bPi^{(2)}_a = -\frac{1}{\Omega_a} \big(p_a \bW_a+\bW_a^q+ \bW_a^{\textrm{frict}}\big),
\end{equation}
where matrices $\bW_a$ and $\bW_a^q$ are given by (\ref{eq:Stress}), (\ref{eq:Stress2}).
Equation (\ref{eq:PtensorA}) can be directly solved. Nevertheless, the stress-tensor of Braginskii does not contain
heat flux contributions, or frictional contributions. To understand the solution of Braginskii more clearly, let us first solve the above equation
only with the matrix $\bW_a$.

The simplest quasi-static $\bPi^{(2)}_a$ is thus obtained by solving
\begin{equation} \label{eq:PtensorF5X}
  (\bhat\times\bPi^{(2)}_a)^S +\frac{\bnu_a}{\Omega_a}\bPi^{(2)}_a = -\frac{p_a}{\Omega_a} \bW_a.
\end{equation}
For any traceless and symmetric matrix $\bW_a$, solution of (\ref{eq:PtensorF5X}) reads  (see details in Section \ref{sec:BragNonlin}) 
\begin{eqnarray}
  \bPi^{(2)}_a &=& -\eta_0^a \bW_0 -\eta_1^a\bW_1 -\eta_2^a\bW_2 +\eta_3^a\bW_3+\eta_4^a\bW_4;\nn\\
  \bW_0 &=& \frac{3}{2}\big(\bW_a:\bhat\bhat\big) \Big( \bhat\bhat-\frac{\bI}{3}\Big);\nn\\
  \bW_1 &=& \bI_\perp \cdot\bW_a\cdot\bI_\perp +\frac{1}{2}\big( \bW_a:\bhat\bhat\big) \bI_\perp;\nn\\
  \bW_2 &=& \big( \bI_\perp\cdot\bW_a\cdot\bhat\bhat\big)^S;\nn\\
  \bW_3 &=& \frac{1}{2}\big( \bhat\times \bW_a\cdot\bI_\perp\big)^S;\nn\\
  \bW_4 &=& \big(\bhat\times\bW_a\cdot\bhat\bhat\big)^S, \label{eq:Stress_genFX}
\end{eqnarray}
with BGK viscosity coefficients
\begin{eqnarray} \label{eq:eta_BGK_pX}
  \eta_0^a &=& \frac{p_a}{\bnu_a}; \quad \eta_1^a = \frac{p_a\bnu_a}{4\Omega_a^2+\bnu_a^2};
  \quad \eta_2^a = \frac{p_a\bnu_a}{\Omega_a^2+\bnu_a^2};
\quad \eta_3^a = \frac{2p_a\Omega_a}{4\Omega_a^2+\bnu_a^2}; \quad \eta_4^a = \frac{p_a\Omega_a}{\Omega_a^2+\bnu_a^2}.
\end{eqnarray}
 Coefficient $\eta_0$ is called the parallel viscosity, $\eta_1,\eta_2$ perpendicular viscosities, and
$\eta_3,\eta_4$ gyroviscosities. Importantly, the BGK solution (\ref{eq:Stress_genFX}) is identical
to the form of \cite{Braginskii1965} viscosity-tensor,  his equations (4.41)-(4.42), only his viscosities are different.
A comparison is presented in the next section. All four matrices $\bW_0,\ldots\bW_4$ are traceless and $\bW_0+\bW_1+\bW_2=\bW_a$. 

When magnetic field is zero, so $\Omega_a=0$ and $\eta_0^a=\eta_1^a=\eta_2^a$,
the stress tensor (\ref{eq:Stress_genFX}) simplifies into $\bPi^{(2)}_a =-\eta_0^a \bW_a$ 
and contributes to the momentum equations in a familiar form
\begin{equation}
  \bb=0:\qquad \nabla\cdot\bPi^{(2)}_a = -\nabla\cdot(\eta_0^a\bW_a)=-\eta_0^a\Big( \nabla^2\bu_a +\frac{1}{3}\nabla (\nabla\cdot\bu_a)\Big)-(\nabla\eta_0^a)\cdot\bW_a,
\end{equation}
analogously to the viscosity of Navier-Stokes equations (the last term can be neglected if $\eta_0^a$ is spatially independent).
  
If the mean magnetic field is sufficiently strong so that its curvature can be neglected, (\ref{eq:Stress_genFX})
can be evaluated with respect to $\bhat_0=(0,0,1)$, yielding
\begin{eqnarray}
  && \Pi_{xx}^{(2)a} = -\frac{\eta_0^a}{2} (W_{xx}^a+W_{yy}^a) -\frac{\eta_1^a}{2}(W_{xx}^a-W_{yy}^a) -\eta_3^a W_{xy}^a  \nn   ;\\
  && \Pi_{xy}^{(2)a} = \frac{\eta_3^a}{2} (W_{xx}^a-W_{yy}^a) -\eta_1^a W_{xy}^a   \nn  ;\\
  && \Pi_{xz}^{(2)a} = -\eta_4^a W_{yz}^a-\eta_2^a W_{xz}^a   \nn ;\\
  && \Pi_{yy}^{(2)a} = -\frac{\eta_0^a}{2} (W_{xx}^a+W_{yy}^a) +\frac{\eta_1^a}{2}(W_{xx}^a-W_{yy}^a) +\eta_3^a W_{xy}^a; \nn \\
  && \Pi_{yz}^{(2)a} = \eta_4^a W_{xz}^a-\eta_2^a W_{yz}^a   \nn  ;\\
  && \Pi_{zz}^{(2)a} = -\eta_0^a W_{zz}^a, \label{eq:Pi_gen_BR}
\end{eqnarray}
which is equation (2.21) of \cite{Braginskii1965}. 
As a double check, adding $\Pi_{xx}^{(2)a}+\Pi_{yy}^{(2)a}+\Pi_{zz}^{(2)a}=-\eta_0^a(W_{xx}^a+W_{yy}^a+W_{zz}^a)=0$, so the stress tensor is
indeed traceless (even though all the diagonal components are non-zero). For strong magnetic field $\Omega_a\gg \bnu_a$ viscosities
(\ref{eq:eta_BGK_pX}) simplify into
\begin{equation}
\eta_0^a = \frac{p_a}{\bnu_a}; \quad \eta_1^a = \frac{1}{4}\frac{p_a\bnu_a}{\Omega_a^2}; \quad \eta_2^a = 4\eta_1^a;
\quad \eta_3^a = \frac{p_a}{2\Omega_a}; \quad \eta_4^a = 2\eta_3^a.
\end{equation}

Considering only self-collisions, the BGK viscosity coefficients (\ref{eq:eta_BGK_pX}) were first recovered
by \cite{Kaufman1960}, even though he does not write them explicitly, and one needs
to get them from rearranging his equations (12)-(15) into form (\ref{eq:Pi_gen_BR}). The same results for $\eta_0-\eta_3$ can also be found for example
in \cite{HelanderSigmar2002} (p. 86); see also \cite{ZankBook2014} (p. 164), however, $\eta_4$ coefficient is erroneously related to $\eta_3=2\eta_4$,
which is a valid relation only in the limit when $x=\Omega_a/\bnu_a$ is small (i.e. a weak magnetic field).
Correct relations are $\eta_3^a(x)=\eta_4^a(2x)$ and $\eta_1^a(x)=\eta_2^a(2x)$, valid for both the BGK and Braginskii solutions. 

Now one can consider more general (\ref{eq:PtensorA}), with heat flux contributions $\bW_a^q$ and 
frictional contributions $\bW_a^{\textrm{frict}}$. Solution of (\ref{eq:PtensorA}) is analogous to (\ref{eq:Stress_genFX})
because all matrices on the r.h.s. are traceless and symmetric. However, it is useful to
rewrite the solution into a different form by defining new matrix
\begin{equation}
\widetilde{\bW}_a = (\nabla\bu_a)^S+\frac{2}{5 p_a}(\nabla\vecq_a)^S,
\end{equation}
and the stress-tensor then reads
\begin{eqnarray}
  \bPi^{(2)}_a &=& -\eta_0^a \bW_0 -\eta_1^a\bW_1 -\eta_2^a\bW_2 +\eta_3^a\bW_3+\eta_4^a\bW_4;\nn\\
  \bW_0 &=& \Big[\frac{3}{2}\big(\widetilde{\bW}_a:\bhat\bhat\big) -\nabla\cdot\bu_a-\frac{2}{5 p_a}\nabla\cdot\vecq_a\Big]\Big( \bhat\bhat-\frac{\bI}{3}\Big)\nn\\
        && -\frac{3}{2}\frac{\rho_a}{p_a}\Big[\sum_b\nu_{ab}\big( \delta u_\parallel^2 -\frac{1}{3}|\delta\bu|^2\big)\Big]\Big(\bhat\bhat-\frac{\bI}{3}\Big); \nn\\
  \bW_1 &=& \bI_\perp \cdot\widetilde{\bW_a}\cdot\bI_\perp
  +\Big[\frac{1}{2}\big( \widetilde{\bW}_a:\bhat\bhat\big) - \nabla\cdot\bu_a-\frac{2}{5 p_a}\nabla\cdot\vecq_a\Big]\bI_\perp\nn\\
       && -\frac{\rho_a}{p_a} \sum_b \nu_{ab}\big(\delta\bu_\perp\delta\bu_\perp-\frac{\bI_\perp}{2}|\delta\bu_\perp|^2\big);\nn\\
  \bW_2 &=& \big( \bI_\perp\cdot\widetilde{\bW}_a\cdot\bhat\bhat\big)^S
         -\frac{\rho_a}{p_a} \sum_b \nu_{ab}\big[\delta u_\parallel \bhat\delta\bu_\perp\big]^S;\nn\\
  \bW_3 &=& \frac{1}{2}\big( \bhat\times \widetilde{\bW}_a\cdot\bI_\perp\big)^S
         -\frac{\rho_a}{2p_a} \sum_b \nu_{ab} \big[(\bhat\times\delta\bu)\delta\bu_\perp\big]^S;\nn\\
  \bW_4 &=& \big(\bhat\times\widetilde{\bW}_a\cdot\bhat\bhat\big)^S
         -\frac{\rho_a}{p_a}\sum_b\nu_{ab}\big[(\bhat\times \delta\bu)\delta u_\parallel\bhat \big]^S, \label{eq:Stress_genFXa}
\end{eqnarray}
with viscosities (\ref{eq:eta_BGK_pX}). Prescribing $\vecq_a=0$ and $\delta\bu=0$ of course recovers (\ref{eq:Stress_genFX}).

\subsection{Heat flux vector \texorpdfstring{$\vecq_a$}{q}}
We consider the 13-moment model where evolution equation (\ref{eq:Energy46}) becomes 
\begin{eqnarray} 
 && \frac{d_a\vecq_a}{d t} +\frac{7}{5}\vecq_a\nabla\cdot\bu_a  + \frac{7}{5}\vecq_a\cdot\nabla\bu_a +\frac{2}{5}(\nabla\bu_a)\cdot\vecq_a
  +\Omega_a\bhat\times\vecq_a + \frac{5}{2}p_a\nabla\Big(\frac{p_a}{\rho_a}\Big) \nn\\
 &&  +\frac{p_a}{\rho_a}\nabla\cdot\bPi^{(2)}_a
+\frac{7}{2}\bPi^{(2)}_a\cdot\nabla\Big(\frac{p_a}{\rho_a}\Big)
  -\frac{1}{\rho_a}(\nabla\cdot\bp_a)\cdot\bPi^{(2)}_a, \nn\\
  && \qquad = \vec{\boldsymbol{Q}}^{(3)}_{a}\,' \equiv \frac{1}{2}\textrm{Tr}\bQ^{(3)}_a-\frac{5}{2}\frac{p_a}{\rho_a}\boldsymbol{R}_a
  -\frac{1}{\rho_a} \boldsymbol{R}_a\cdot\bPi^{(2)}_a, \label{eq:Energy466}
\end{eqnarray}
and the BGK collisional contributions calculate
\begin{equation} \label{eq:Thierry266}
\frac{1}{2}\textrm{Tr} \bQ^{(3)}_{ab} -\frac{5}{2}\frac{p_a}{\rho_a}\boldsymbol{R}_{ab} = -\nu_{ab}\vecq_a
+\frac{\nu_{ab}}{2} \rho_a \delta\bu |\delta\bu|^2.
\end{equation}
In a quasi-static approximation (\ref{eq:Energy466}) can be simplified into
\begin{eqnarray}
  \bhat\times\vecq_a +\frac{\bnu_a}{\Omega_a}\vecq_a  &=& -\,\frac{1}{\Omega_a}\Big[
    \frac{5}{2}p_a\nabla\Big( \frac{p_a}{\rho_a}\Big) +\frac{p_a}{\rho_a}\nabla\cdot\bPi^{(2)}_a 
   +\frac{7}{2}\bPi^{(2)}_a\cdot\nabla\Big( \frac{p_a}{\rho_a}\Big)  -\frac{1}{\rho_a}(\nabla\cdot\bp_a)\cdot\bPi^{(2)}_a\nn\\
   && +\frac{1}{\rho_a} \boldsymbol{R}_a\cdot\bPi^{(2)}_a
   -\sum_b\frac{\nu_{ab}}{2} \rho_a \delta\bu |\delta\bu|^2\Big]. \label{eq:QstatP}
\end{eqnarray}
A general vector equation (where $\vec{\boldsymbol{a}}$ is an unspecified vector, unrelated to the species index)
\begin{equation} \label{eq:PPPX}
  \bhat\times\vecq_a +\frac{\bnu_a}{\Omega_a}\vecq_a  =  - \frac{\vec{\boldsymbol{a}}}{\Omega_a},
\end{equation}
has the following exact solution
 (split the equation to parallel and perpendicular parts
  $\vecq_a=\vecq_{\parallel a}+\vecq_{\perp a}$ \& $\vec{\boldsymbol{a}}=\vec{\boldsymbol{a}}_\parallel+\vec{\boldsymbol{a}}_\perp$
  with $\bhat\times \vecq_{\parallel a}=0$; apply $\bhat\times$ on the perpendicular part, use $\bhat\times (\bhat\times\vecq_{\perp_a})=-\vecq_{\perp a}$,
  and solve the two coupled perpendicular equations by eliminating the $\bhat\times\vecq_{\perp a}$)
\begin{equation}
  \vecq_a = -\frac{1}{\bnu_a} (\vec{\boldsymbol{a}}\cdot\bhat) \bhat +\frac{\Omega_a}{\Omega_a^2+\bnu_a^2}\bhat\times\vec{\boldsymbol{a}}
  -\frac{\bnu_a}{\Omega_a^2+\bnu_a^2} \vec{\boldsymbol{a}}_{\perp}. \label{eq:PPP}
\end{equation}
 Note that $\bhat\times\vec{\boldsymbol{a}}=\bhat\times\vec{\boldsymbol{a}}_\perp$. 
Result (\ref{eq:PPP}) represents solution of equation (\ref{eq:QstatP}). For zero magnetic field $\vecq_a = -\vec{\boldsymbol{a}} /\bnu_a$.
The BGK frictional contributions due to $\delta\bu$ are only non-linear, in contrast,
the electron heat flux of Braginskii contains frictional $\delta\bu$ contributions that are linear. At the semi-linear level,
(\ref{eq:QstatP}) simplifies into
\begin{eqnarray}
  \bhat\times\vecq_a +\frac{\bnu_a}{\Omega_a}\vecq_a  &=& -\,\frac{1}{\Omega_a}\Big[
    \frac{5}{2}p_a\nabla\Big( \frac{p_a}{\rho_a}\Big) +\frac{p_a}{\rho_a}\nabla\cdot\bPi^{(2)}_a\Big],
\end{eqnarray}    
with a solution again given by (\ref{eq:PPP}). The BGK operator can therefore account for linear (!)
contributions of the stress-tensor $\bPi_a^{(2)}$ that enters the heat flux $\vecq_a$, similarly to the previous
result (\ref{eq:Stress_genFXa}) where the heat $\vecq_a$ flux entered the stress-tensor $\bPi_a^{(2)}$. 
Such a coupling is typically neglected with the Landau collisional operator. 

The simplest BGK heat flux is a solution of equation
\begin{equation}
\bhat\times\vecq_a +\frac{\bnu_a}{\Omega_a}\vecq_a  =  - \frac{5}{2}\frac{p_a}{\Omega_a m_a}  \nabla T_a, \label{eq:QR_tensor6X}
\end{equation}
and the solution reads
\begin{equation} \label{eq:Braginskii_qX}
\vecq_a = -\kappa_\parallel^a \nabla_\parallel T_a - \kappa_\perp^a \nabla_\perp T_a + \kappa_\times^a \bhat\times\nabla T_a,
\end{equation}
with thermal conductivities
\begin{equation}
 \kappa_\parallel^a = \frac{5}{2}\frac{p_a}{\bnu_a m_a}; \qquad
  \kappa_\perp^a = \frac{5}{2}\frac{p_a}{m_a}\frac{\bnu_a}{(\Omega_a^2+\bnu_a^2)}; \qquad
  \kappa_\times^a = \frac{5}{2}\frac{p_a}{m_a}\frac{\Omega_a}{(\Omega_a^2+\bnu_a^2)}.\label{eq:HF_BGK2X}\\
\end{equation}
We use the Braginskii notation with vector $\nabla_\parallel=\bhat\bhat\cdot\nabla$.
If magnetic field is zero, so that $\Omega_a=0$ and $\kappa_\parallel^a=\kappa_\perp^a$, 
the solution simplifies into $\vecq_a = -\kappa_\parallel^a \nabla T_a$.

\subsection{BGK vs Braginskii comparison} \label{sec:BGKcomparision}
\begin{figure*}[!htpb]
  \includegraphics[width=0.42\linewidth]{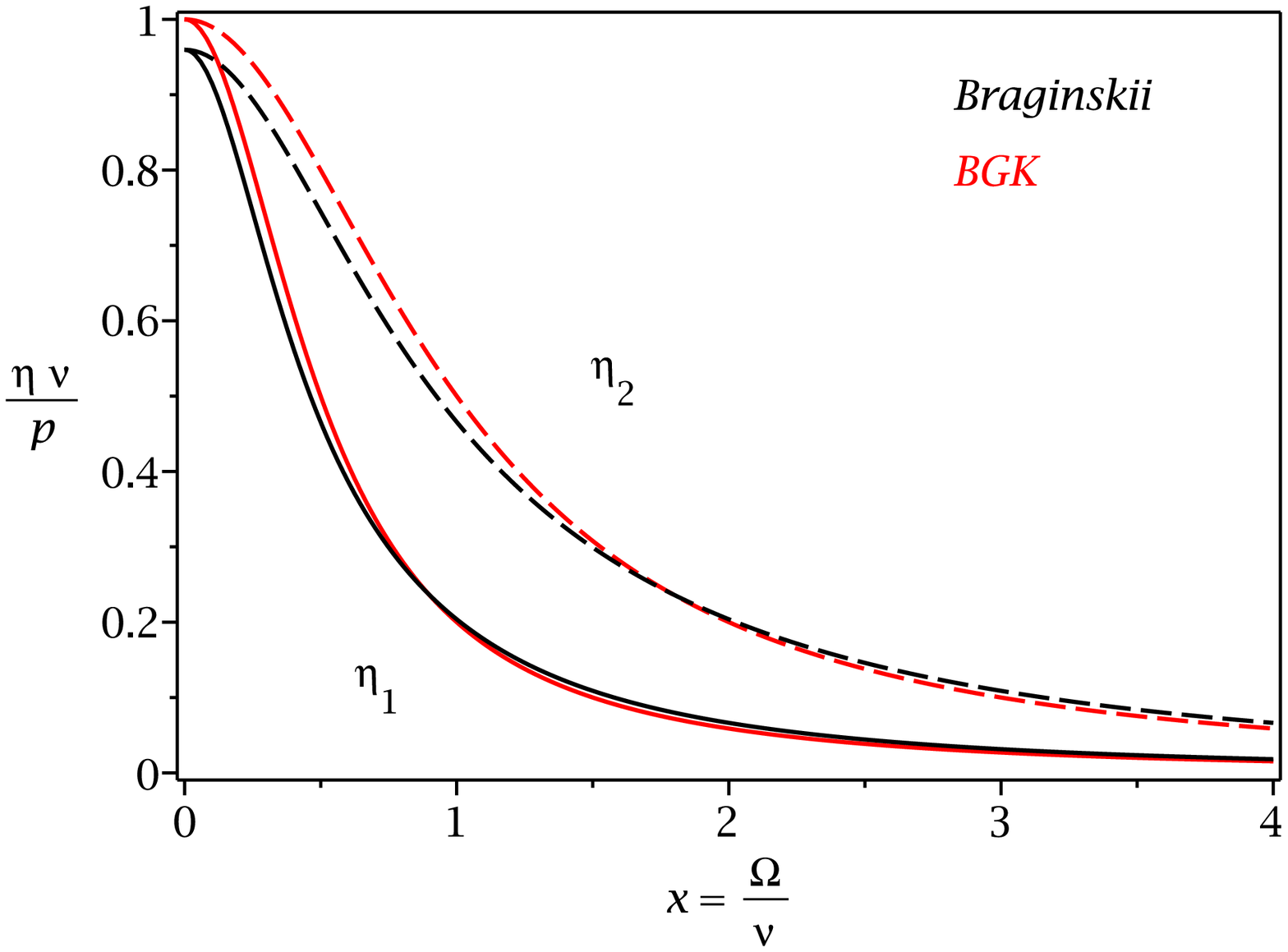}\hspace{0.1\textwidth}\includegraphics[width=0.42\linewidth]{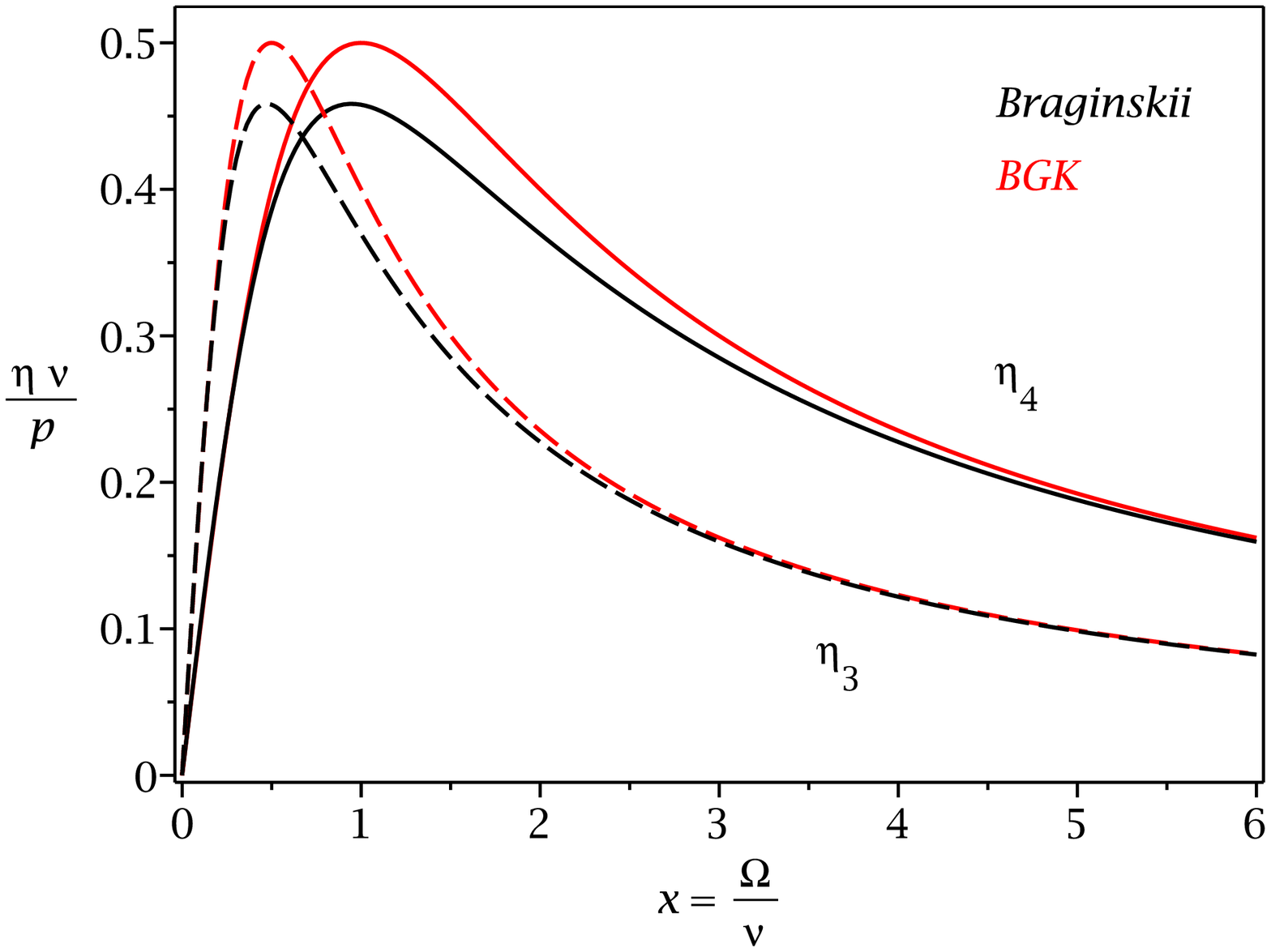}  
  \caption{Ion viscosities of the BGK model (red) and of the Braginskii model (black) normalized as $\hat{\eta}^i=\eta^i \nu_{ii}/p_i$
    vs. ratio $x=\Omega_i/\nu_{ii}$.
    Left panel: perpendicular viscosities $\eta_1^i,\eta_2^i$. Right panel: gyroviscosities $\eta_3^i,\eta_4^i$.
    For large values of $x$, the BGK asymptotic profiles
    for $\hat{\eta}^i_3=1/(2x)$ and $\hat{\eta}_4^i=1/x$ become independent
    of collisional frequencies and match the asymptotic profiles of Braginskii exactly.
    BGK asymptotic profiles for $\hat{\eta}^i_1=1/(4x^2)$ and $\hat{\eta}^i_2=1/x^2$ have correct functional
    dependence, but differ from the Braginskii asymptotes by a proportionality constant.
  The BGK operator reproduces the ion viscosity of Braginskii with surprisingly good accuracy.} \label{fig:1}
\end{figure*}

\begin{figure*}[!htpb]
  \includegraphics[width=0.42\linewidth]{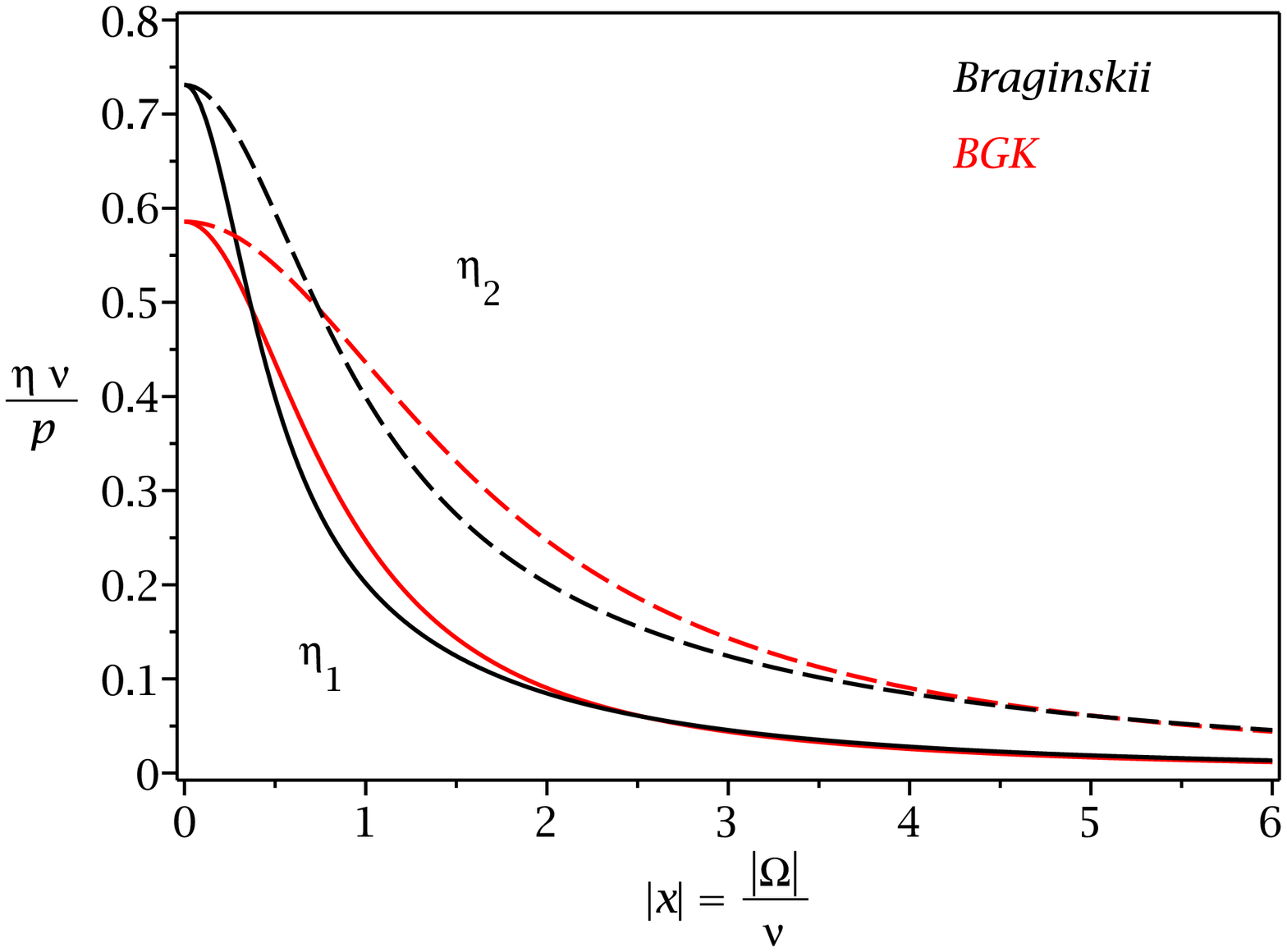}\hspace{0.1\textwidth}\includegraphics[width=0.42\linewidth]{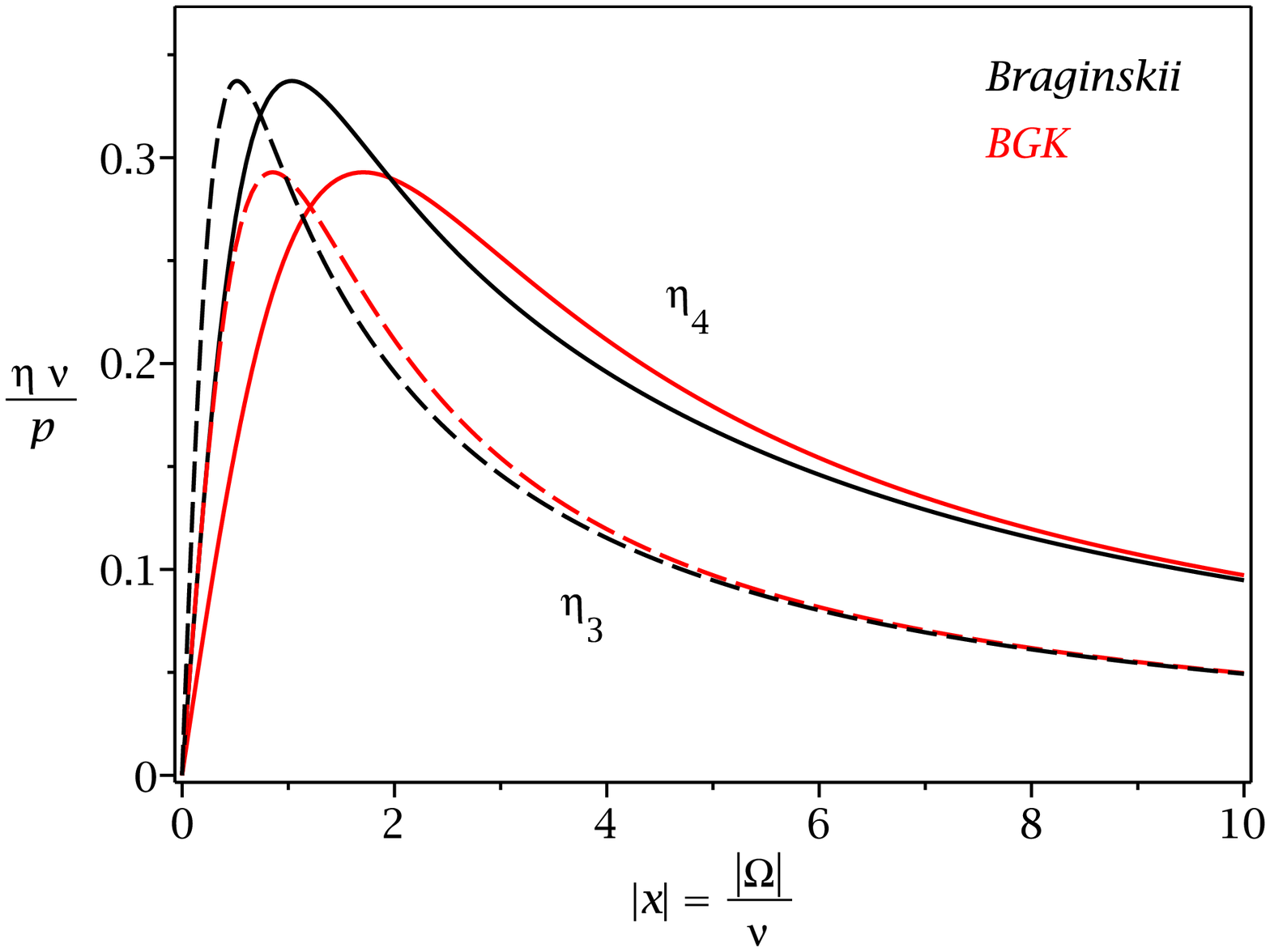}  
  \caption{Electron viscosities normalized as $\hat{\eta}^e=\eta^e \nu_{ei}/p_e$ vs. ratio $|x|=|\Omega_e|/\nu_{ei}$. Results are less
  precise than for ions in Figure \ref{fig:1}, especially for small values of $x$, nevertheless, the same conclusions are obtained.} \label{fig:11}
\end{figure*}

\begin{figure*}[!htpb]
  \includegraphics[width=0.42\linewidth]{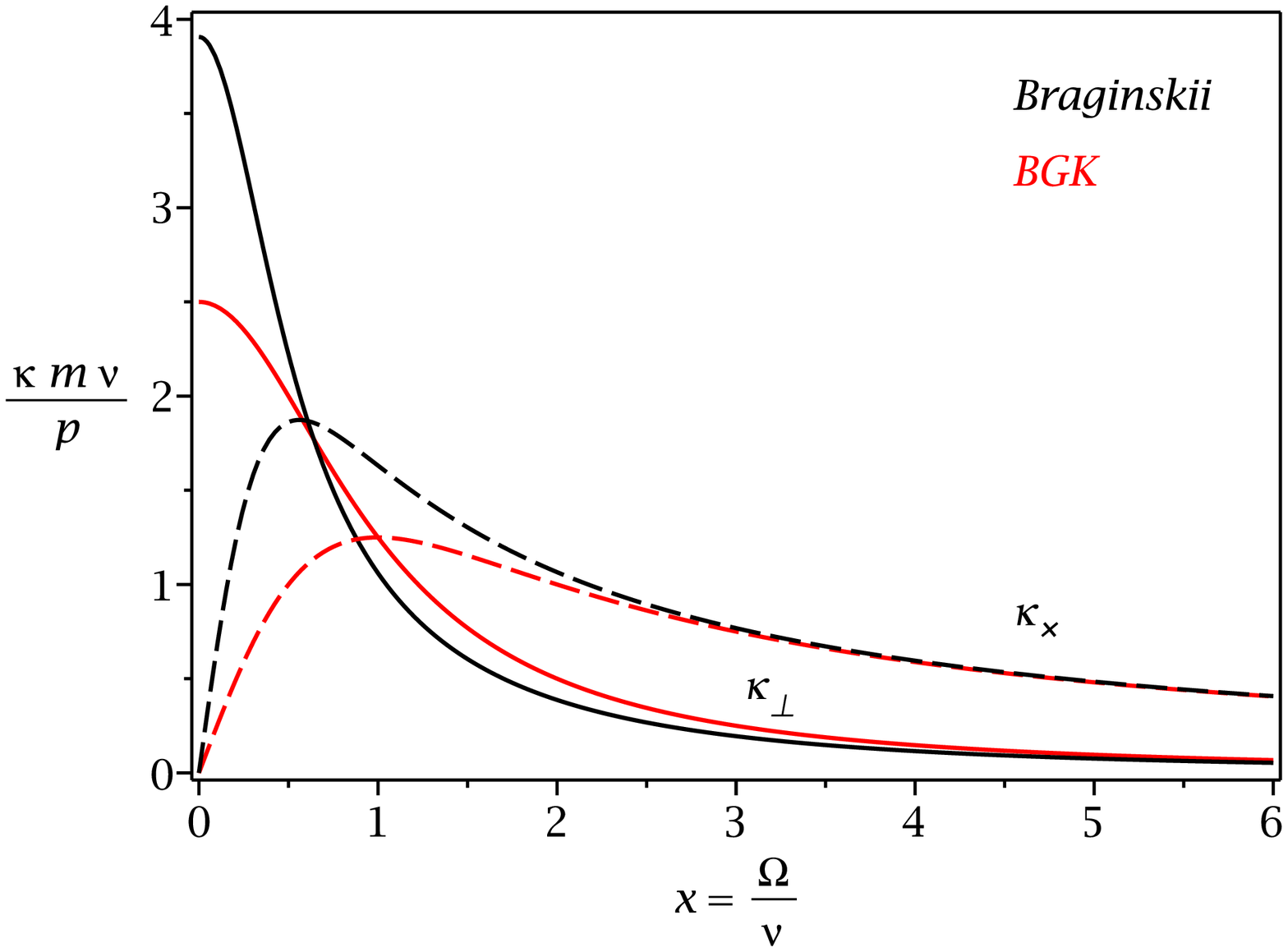}\hspace{0.1\textwidth}\includegraphics[width=0.42\linewidth]{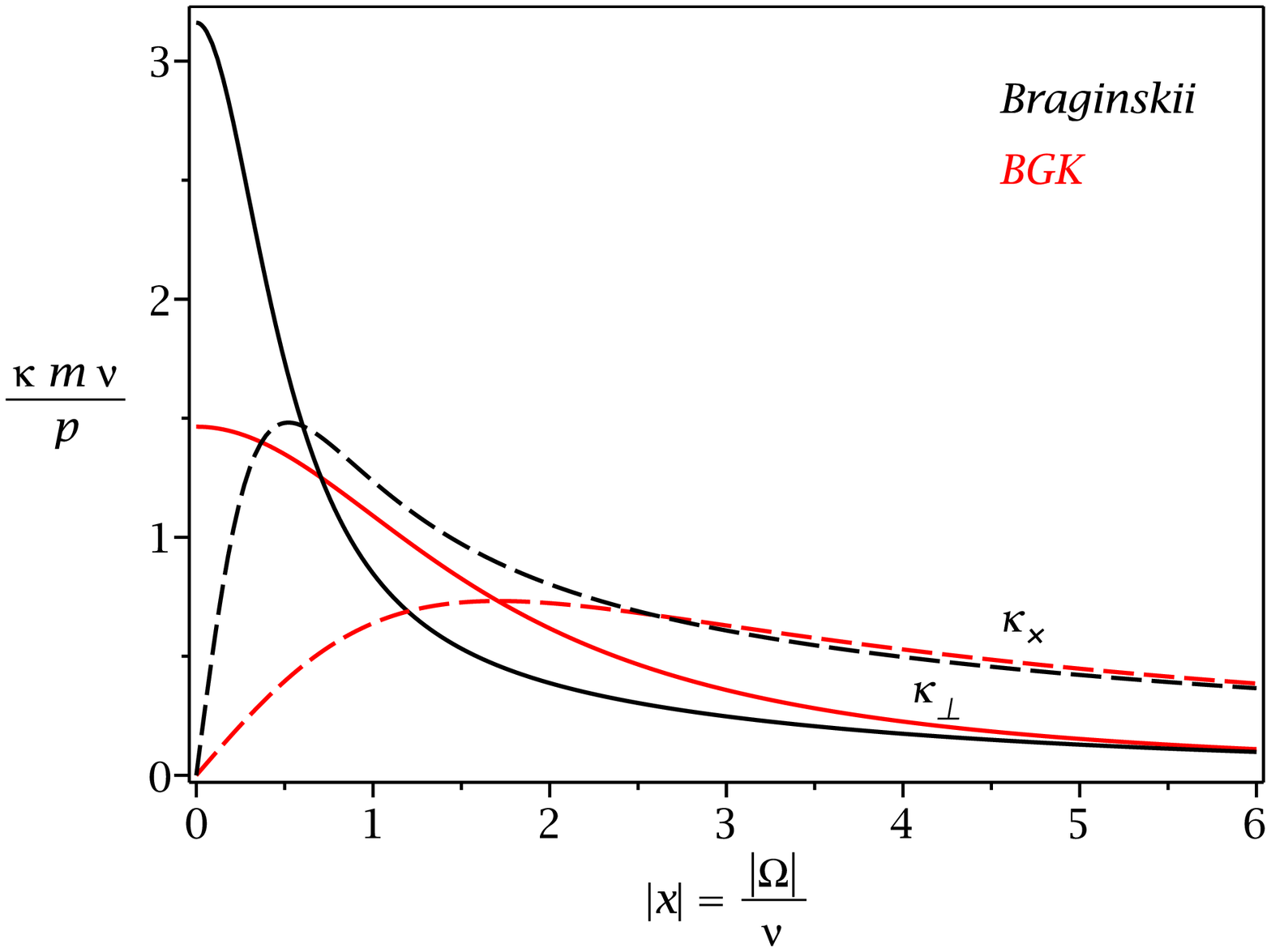}  
  \caption{Heat conductivities $\kappa^a_\perp$ and $\kappa^a_\times$.  
    Left panel: ion species, normalized as $\kappa^i m_i \nu_{ii}/p_i$. Right panel: electron species, normalized as $\kappa^e m_e \nu_{ei}/p_e$.
    For large values of $x$, the BGK asymptotic profiles $\kappa^a_\times$ (dashed lines) match the Braginskii results exactly,
    whereas for $\kappa_\perp^a$ (solid lines) the results differ by a proportionality constant.
    } \label{fig:2}
\end{figure*}

Here we compare the BGK viscosities and heat conductivities with those of \cite{Braginskii1965} for a one ion-electron
plasma with ion charge $Z_i=1$. The BGK viscosities (\ref{eq:eta_BGK_pX}) contain $\bnu_a=\sum_b \nu_{ab}$,
and in general should be added according to
\begin{eqnarray}
&& \bnu_i = \nu_{ii}+\nu_{ie} = \nu_{ii} \Big( 1+\frac{\sqrt{2}}{Z_i}  \sqrt{\frac{m_e}{m_i}} \Big(\frac{T_i}{T_e}\Big)^{3/2}\Big); \nn\\
&&  \bnu_e = \nu_{ee}+\nu_{ei} = \nu_{ei} \Big( 1+\frac{1}{Z_i\sqrt{2}}\Big). \label{eq:timesX}
\end{eqnarray}
However, for the ion species Braginskii neglects ion-electron collisions and thus  
$\bnu_i= \nu_{ii}$ and $\bnu_e = 1.707 \nu_{ei}$; see Section \ref{section:ColFreq}. Using Braginskii
notation with one-index $\nu_{i}=\nu_{ii}$ and $\nu_e=\nu_{ei}$ then implies  
\begin{equation} \label{eq:First}
\bnu_a = \alpha_a \nu_a; \quad \textrm{where}\quad \alpha_i=1; \quad \alpha_e=1.707,
\end{equation}
and introducing quantity $x=\Omega_a/\nu_a$ the BGK viscosities (\ref{eq:eta_BGK_pX}) become
\begin{eqnarray}
  \eta_0^a &=& \frac{p_a}{\alpha_a\nu_a}; \quad \eta_1^a = \frac{p_a}{\nu_a}\frac{\alpha_a}{4x^2+\alpha_a^2}; \quad
  \eta_2^a = \frac{p_a}{\nu_a}\frac{\alpha_a}{x^2+\alpha_a^2};
\quad \eta_3^a = \frac{p_a}{\nu_a}\frac{2x}{4x^2+\alpha_a^2}; \quad \eta_4^a = \frac{p_a}{\nu_a}\frac{x}{x^2+\alpha_a^2}.\label{eq:Kauf2}
\end{eqnarray}
Note that $\eta_1^a(x)=\eta_2^a(2x)$ and $\eta_3^a(x)=\eta_4^a(2x)$. Similarly, the BGK heat conductivities (\ref{eq:HF_BGK2X}) become
\begin{eqnarray}
  \kappa_\parallel^a &=& \frac{5}{2\alpha_a}\frac{p_a}{\nu_a m_a}; \qquad
  \kappa_\perp^a = \frac{5}{2}\frac{p_a}{\nu_a m_a}\frac{\alpha_a}{(x^2+\alpha_a^2)}; \qquad
  \kappa_\times^a = \frac{5}{2}\frac{p_a}{\nu_a m_a}\frac{x}{(x^2+\alpha_a^2)}.\label{eq:HF_BGK2XX}
\end{eqnarray}
Viscosities and heat conductivities for Braginskii are given in the main text. 
Ion viscosities are compared in Figure \ref{fig:1}, electron viscosities in Figure \ref{fig:11} and
heat conductivities in Figure \ref{fig:2}. A small value of $x$ represents weak magnetic field and large value of $x$ represents
strong magnetic field. 

\newpage
\subsection{Nonlinear stress-tensor decomposition} \label{sec:BragNonlin}
Here we want to consider BGK equation for the stress-tensor (\ref{eq:PtensorF5X}) 
\begin{equation} \label{eq:Num499}
  (\bhat\times\bPi)^S +\frac{\nu}{\Omega}\bPi = -\frac{p}{\Omega} \bW,
\end{equation}
and clarify solution (\ref{eq:Stress_genFX}). Species indices are dropped and both $\bPi$ and $\bW$ are symmetric and traceless. 
First we need to learn how to decompose any general matrix. It is useful for a moment to consider 
undefined matrices $\bW$ and $\bPi$ which are not necessarily symmetric nor traceless.
\subsubsection{Decomposition of a matrix}
We will work both in the reference frame of magnetic field lines ($\bhat_0=(0,0,1)$) which nicely guides and clarifies the calculations, and
also in a laboratory reference frame with general $\bhat$. In the reference frame of magnetic field lines one uses matrices  
\begin{eqnarray}
\bhat\bhat = \left( \begin{array}{ccc}
 0 & 0 & 0 \\
 0 & 0 & 0 \\
 0 & 0 & 1
\end{array} \right); \qquad
\bI_\perp = \bI-\bhat\bhat =\left( \begin{array}{ccc}
 1 & 0 & 0 \\
 0 & 1 & 0 \\
 0 & 0 & 0
\end{array} \right);\qquad
\bI^\times = \left( \begin{array}{ccc}
 0 & -1 & 0 \\
 +1 & 0 & 0 \\
 0 & 0 & 0
\end{array} \right),
\end{eqnarray}
where the last matrix is defined as $\bhat\times\bW=(\bI^\times) \cdot \bW$. Then one takes a general matrix $\bW$, and
starts multiplying it with matrices $\bhat\bhat$ and $\bI_\perp$ from the left and right, yielding
a general decomposition
\begin{eqnarray}
  \bW &=& \bW_0\,'+\bW_1\,'+\bW_2;\label{eq:Num510}\\
  \bW_0\,' &=& \bhat\bhat\cdot\bW\cdot\bhat\bhat = (\bW:\bhat\bhat)\bhat\bhat;\nn\\ 
  \bW_1\,' &=& \bI_\perp\cdot\bW\cdot\bI_\perp;\nn\\
  \bW_2 &=& \bI_\perp \cdot\bW\cdot\bhat\bhat+\bhat\bhat\cdot\bW\cdot\bI_\perp = (\bI_\perp \cdot\bW\cdot\bhat\bhat)^S.\nn
\end{eqnarray}
In the reference frame of magnetic field lines  
\begin{eqnarray}
\bW_0\,' = \left( \begin{array}{ccc}
 0 & 0 & 0 \\
 0 & 0 & 0 \\
 0 & 0 & W_{zz}
\end{array} \right); \qquad
\bW_1\,' = \left( \begin{array}{ccc}
 W_{xx} & W_{xy} & 0 \\
 W_{yx} & W_{yy} & 0 \\
 0 & 0 & 0
\end{array} \right);\qquad 
\bW_2 = \left( \begin{array}{ccc}
 0 & 0 & W_{xz} \\
 0 & 0 & W_{yz} \\
 W_{zx} & W_{zy} & 0
\end{array} \right),
\end{eqnarray}
and adding these matrices together obviously yields the full matrix $\bW$. However, the decomposition (\ref{eq:Num510}) also works in the
laboratory reference frame with general $\bhat$, as can be verified by adding the general matrices together. 
It is possible to consider an alternative decomposition, according to
\begin{eqnarray}
  \bW &=& \bW_0+\bW_1+\bW_2; \label{eq:Num511}\\
  \bW_0 &=& (\bW:\bhat\bhat)\bhat\bhat +\frac{1}{2}(\bW:\bI_\perp) \bI_\perp;\nn\\ 
  \bW_1 &=& \bI_\perp\cdot\bW\cdot\bI_\perp -\frac{1}{2}(\bW:\bI_\perp) \bI_\perp;\nn\\
  \bW_2 &=& (\bI_\perp \cdot\bW\cdot\bhat\bhat)^S,\nn
\end{eqnarray}
where in the reference frame of magnetic field lines
\begin{eqnarray}
\bW_0  = \left( \begin{array}{ccc}
 \frac{1}{2}(W_{xx}+W_{yy}) & 0 & 0 \\
 0 & \frac{1}{2}(W_{xx}+W_{yy}) & 0 \\
 0 & 0 & W_{zz}
\end{array} \right); \qquad
\bW_1  = \left( \begin{array}{ccc}
 \frac{1}{2}(W_{xx}-W_{yy})  & W_{xy} & 0 \\
 W_{yx} &  -\frac{1}{2}(W_{xx}-W_{yy}) & 0 \\
 0 & 0 & 0
\end{array} \right).
\end{eqnarray}
Decomposition (\ref{eq:Num511}) again works for general $\bhat$, and
in comparison to the previous decomposition $\bW_0\,'+\bW_1\,'=\bW_0+\bW_1$. The advantage is that if $\bW$ is traceless then all 3 matrices are
traceless. It is useful to re-arrange
(\ref{eq:Num511}) by separating the trace of $\bW$ with 
$(\bW:\bI_\perp)\bI_\perp = (\bW:\bI)\bI_\perp-(\bW:\bhat\bhat)\bI_\perp$, yielding decomposition
\begin{eqnarray}
  \bW &=& \bW_0+\bW_1+\bW_2;\\
  \bW_0 &=& \frac{3}{2}(\bW:\bhat\bhat)\Big(\bhat\bhat-\frac{\bI}{3}\Big) +\frac{1}{2}(\bW:\bI) \bI_\perp;\nn\\ 
  \bW_1 &=& \bI_\perp\cdot\bW\cdot\bI_\perp +\frac{1}{2}(\bW:\bhat\bhat)\bI_\perp-\frac{1}{2}(\bW:\bI) \bI_\perp;\nn\\
  \bW_2 &=& (\bI_\perp \cdot\bW\cdot\bhat\bhat)^S.\nn
\end{eqnarray}
The same decomposition is used for the stress-tensor $\bPi$
\begin{eqnarray}
  \bPi &=& \bPi_0+\bPi_1+\bPi_2;\label{eq:Num501}\\
  \bPi_0 &=& \frac{3}{2}(\bPi:\bhat\bhat)\Big(\bhat\bhat-\frac{\bI}{3}\Big) +\frac{1}{2}(\bPi:\bI) \bI_\perp;\nn\\ 
  \bPi_1 &=& \bI_\perp\cdot\bPi\cdot\bI_\perp +\frac{1}{2}(\bPi:\bhat\bhat)\bI_\perp-\frac{1}{2}(\bPi:\bI) \bI_\perp;\nn\\
  \bPi_2 &=& (\bI_\perp \cdot\bPi\cdot\bhat\bhat)^S. \nn
\end{eqnarray}
Let us solve for $\bPi_0$. By applying $:\bhat\bhat$ and $:\bI$ at equation (\ref{eq:Num499}) and using identities
\begin{equation}
(\bhat\times\bPi)^S :\bhat\bhat =0; \qquad (\bhat\times\bPi)^S :\bI =0,
\end{equation}
yields 
\begin{equation}
\bPi:\bhat\bhat = -\frac{p}{\nu}\bW:\bhat\bhat; \qquad \bPi:\bI = -\frac{p}{\nu}\bW:\bI,
\end{equation}
and plugging these results into (\ref{eq:Num501}) yields the final solution for the parallel stress-tensor
\begin{equation}
  \bPi_0 = -\frac{p}{\nu} \Big[\frac{3}{2}(\bW:\bhat\bhat)\Big(\bhat\bhat-\frac{\bI}{3}\Big) +\frac{1}{2}(\bW:\bI) \bI_\perp\Big]
  = -\frac{p}{\nu} \bW_0.
\end{equation}
The solution is valid for any general matrix $\bW$ (not necessarily symmetric or traceless).
If this result is compared with the expression (4.42) of \cite{Braginskii1965} given bellow by (\ref{eq:W0_Brag}), one notices
\begin{eqnarray}
  \bW_0^{\textrm{BR}} = (\ref{eq:W0_Brag}) = \frac{3}{2} (\bW:\bhat\bhat)\Big(\bhat\bhat-\frac{\bI}{3}\Big) +\frac{1}{2}(\bW:\bI)\Big(\frac{\bI}{3}-\bhat\bhat\Big)
  \neq \bW_0,
\end{eqnarray}
and his result is valid only if $\bW$ is traceless (which it is).
The reason why Braginskii left his result in form (\ref{eq:W0_Brag}) and did not simplify it with $\bW:\bI=0$
is likely an alternative form (\ref{eq:W0_Brag2}).

\subsubsection{Symmetric and traceless matrices}
We further consider only symmetric and traceless matrices $\bW$ and $\bPi$, so all previous expressions are simplified with
$\bW:\bI=0$, $\bPi:\bI=0$ and the BGK parallel stress-tensor $\bPi_0=-(p/\nu) \bW_0$. For clarity, it is useful to write several possible forms for
\begin{eqnarray}
  \bW_0 &=& \frac{3}{2}(\bW:\bhat\bhat)\Big(\bhat\bhat-\frac{\bI}{3}\Big); \label{eq:W00}\\
        &=& \frac{3}{2}\Big[ \bW:\Big(\bhat\bhat-\frac{\bI}{3}\Big)\Big] \Big(\bhat\bhat-\frac{\bI}{3}\Big); \label{eq:W0_Brag}\\
  &=& \frac{3}{2}\Big[ (\nabla\bu)^S:\Big(\bhat\bhat-\frac{\bI}{3}\Big)\Big] \Big(\bhat\bhat-\frac{\bI}{3}\Big); \label{eq:W0_Brag2}\\
  &=& 3\Big[ (\nabla\bu):\Big(\bhat\bhat-\frac{\bI}{3}\Big)\Big] \Big(\bhat\bhat-\frac{\bI}{3}\Big); \label{eq:W0_Fitz}\\
  &=& 3\Big[ \bhat\cdot(\nabla\bu)\cdot\bhat -\frac{1}{3}\nabla\cdot\bu\Big] \Big(\bhat\bhat-\frac{\bI}{3}\Big).
\end{eqnarray}
Braginskii uses (\ref{eq:W0_Brag}), for example \cite{Fitzpatrick} uses (\ref{eq:W0_Fitz}), and we use (\ref{eq:W00}). In the reference frame of 
magnetic field lines
\begin{equation}
\bW_0 = \frac{3}{2}W_{zz}  \left( \begin{array}{ccc}
    -1/3, & 0, & 0 \\
    0, & -1/3, & 0 \\
    0, & 0, & +2/3
  \end{array} \right); \qquad
\bPi_0 
  = \frac{p}{\nu} W_{zz}
  \left( \begin{array}{ccc}
    1/2, & 0, & 0 \\
    0, & 1/2, & 0 \\
    0, & 0, & -1
  \end{array} \right). \label{eq:Lin_PIBR0}
\end{equation}

To solve equation (\ref{eq:Num499}) it is
beneficial to introduce  two other matrices $\bW_3$, $\bW_4$ by decomposing 
\begin{eqnarray}
  (\bhat\times\bW)^S &=& 2\bW_3 + \bW_4;\label{eq:Num513}\\
  2\bW_3 &=& (\bhat\times\bW\cdot\bI_\perp)^S;\nn\\
  \bW_4 &=& (\bhat\times\bW\cdot\bhat\bhat)^S,\nn
\end{eqnarray}
where in the reference frame of magnetic field lines 
\begin{eqnarray}
2\bW_3  = \left( \begin{array}{ccc}
 -2W_{xy} & W_{xx}-W_{yy} & 0 \\
 W_{xx}-W_{yy} & 2W_{xy} & 0 \\
 0 & 0 & 0
\end{array} \right); \qquad
\bW_4  = \left( \begin{array}{ccc}
 0  & 0 & -W_{yz} \\
 0 &  0 & W_{xz} \\
 -W_{yz} & W_{xz} & 0
\end{array} \right).
\end{eqnarray}
Decomposition (\ref{eq:Num513}) is again valid for general $\bhat$ which is easily verified by using $\bI_\perp+\bhat\bhat=\bI$,
and the stress-tensor is decomposed in the same way  
\begin{eqnarray}
  (\bhat\times\bPi)^S &=& 2\bPi_3 + \bPi_4;\label{eq:Num514}\\
  2\bPi_3 &=& (\bhat\times\bPi\cdot\bI_\perp)^S;\nn\\
  \bPi_4 &=& (\bhat\times\bPi\cdot\bhat\bhat)^S.\nn
\end{eqnarray}
Finally, by applying $\bhat\times$ at matrices $\bW_0\ldots \bW_4$ yields the following identities
\begin{eqnarray}
&&  (\bhat\times \bW_0)^S =0; \qquad \bhat\times\bW_1 = \bW_3; \qquad (\bhat\times\bW_2)^S = \bW_4; \nn\\
&&  \bhat\times \bW_3 = -\bW_1; \qquad (\bhat\times\bW_4)^S = -\bW_2,\label{eq:Num5041}
\end{eqnarray}
which are easy to verify in a general reference frame with $\bhat$. The same identities hold for the stress-tensor 
\begin{eqnarray}
&&  (\bhat\times \bPi_0)^S =0; \qquad \bhat\times\bPi_1 = \bPi_3; \qquad (\bhat\times\bPi_2)^S = \bPi_4; \nn\\
&&  \bhat\times \bPi_3 = -\bPi_1; \qquad (\bhat\times\bPi_4)^S = -\bPi_2. \label{eq:Num504}
\end{eqnarray}

\subsubsection*{Final solution}
Now we are ready to solve equation (\ref{eq:Num499}), which is rewritten as
\begin{equation} \label{eq:Num500}
  2\bPi_3+\bPi_4 +\frac{\nu}{\Omega}\big(\bPi_0+\bPi_1+\bPi_2\big) = -\frac{p}{\Omega} \big(\bW_0+\bW_1+\bW_2\big).
\end{equation} 
One solution $\bPi_0=-(p/\nu)\bW_0$ has already been obtained, and can be eliminated from (\ref{eq:Num500}).
For the rest of the equation, the most straightforward approach is to be guided by the reference frame of magnetic field lines,
which shows that the system (\ref{eq:Num500}) can be directly split into two independent equations
\begin{eqnarray}
&&  2\bPi_3 +\frac{\nu}{\Omega}\bPi_1 = -\frac{p}{\Omega}\bW_1 \label{eq:Num502};\\
&&  \bPi_4 + \frac{\nu}{\Omega}\bPi_2 =-\frac{p}{\Omega}\bW_2 \label{eq:Num503}.
\end{eqnarray}
In the general reference frame, the split can be achieved for example by applying $\bI_\perp\cdot$ from left \& right at (\ref{eq:Num500}),
which by using identities $\bI_\perp\cdot\bPi_4\cdot\bI_\perp=0$; $\bI_\perp\cdot\bPi_2\cdot\bI_\perp=0$ and $\bI_\perp\cdot\bW_2\cdot\bI_\perp=0$
yields (\ref{eq:Num502}) and subsequently (\ref{eq:Num503}). The split significantly simplifies the ``inversion procedure''.

Furthermore, by applying $\bhat\times$ at (\ref{eq:Num502}), applying $\bhat\times$ together with symmetric operator at (\ref{eq:Num503}),
and using identities (\ref{eq:Num5041})-(\ref{eq:Num504}) 
then gives
\begin{eqnarray}
-2\bPi_1 +\frac{\nu}{\Omega}\bPi_3 = -\frac{p}{\Omega}\bW_3; \label{eq:Num505}\\
 -\bPi_2 + \frac{\nu}{\Omega}\bPi_4 =-\frac{p}{\Omega}\bW_4 \label{eq:Num506}.
\end{eqnarray}
Equations (\ref{eq:Num502}), (\ref{eq:Num505}) are coupled and can be treated as 2 equations in 2 unknowns, and similarly
equations (\ref{eq:Num503}), (\ref{eq:Num506}), finally yielding solutions
\begin{eqnarray}
  \bPi_1 &=& -\frac{p\nu}{4\Omega^2+\nu^2} \bW_1 + \frac{2p\Omega}{4\Omega^2+\nu^2}\bW_3;\\
  \bPi_2 &=& -\frac{p\nu}{\Omega^2+\nu^2}\bW_2 +\frac{p\Omega}{\Omega^2+\nu^2}\bW_4.
\end{eqnarray}
The entire solution for the stress-tensor $\bPi=\bPi_0+\bPi_1+\bPi_2$ thus reads 
\begin{eqnarray}
  \bPi= -\frac{p}{\nu}\bW_0 -\frac{p\nu}{4\Omega^2+\nu^2}\bW_1 -\frac{p\nu}{\Omega^2+\nu^2}\bW_2
  +\frac{2p\Omega}{4\Omega^2+\nu^2}\bW_3+\frac{p\Omega}{\Omega^2+\nu^2}\bW_4.
\end{eqnarray}

\subsection{BGK operator and electric field}
The BGK operator is also an excellent tool to clarify various processes in fully ionized or partially ionized plasmas.
Here we want to clarify the Ohmic (magnetic) diffusion together with the ambipolar diffusion, both caused by the
momentum exchange rates
\begin{equation} \label{eq:BGK_R}
\boldsymbol{R}_{a} = \sum_{b\neq a} \rho_a \nu_{ab} (\bu_b-\bu_a).
\end{equation}
From the BGK perspective, one does not need to worry about complicated Landau and Boltzmann operators, 
and simply ``adopt'' correct collisional frequencies; see for example Appendix C of \cite{Schunk1977}.
Momentum exhange rates (\ref{eq:BGK_R}) are actually
the correct answer if relative drift velocities are small and one considers the 5-moment model (i.e. if 
the heat flux is neglected).

We restrict our focus on spatial scales much longer than the Debye length.
The displacement current is neglected, the Gauss's law $\nabla\cdot\bE=4\pi e\sum_aZ_an_a$ is replaced by
the charge neutrality and no condition is placed on $\nabla\cdot\bE$. The Maxwell's equations then read
\begin{eqnarray}
&&  \sum_a Z_a n_a =0; \qquad \boldsymbol{j}=\sum_a e Z_a n_a\bu_a =\frac{c}{4\pi}\nabla\times\bb;\label{eq:Max1}\\
&&  \frac{\pr \bb}{\pr t}=-c\nabla\times\bE;\qquad \nabla\cdot\bb=0. \label{eq:Max}
\end{eqnarray}
By focusing on spatial and temporal scales of the ion and neutral species,
we do not need to resolve the electron motion. In the electron momentum equation the electron inertia
represented by $d_e\bu_e/dt$ is neglected (which does not mean that $m_e=0$, relations $\rho_a\nu_{ab}=\rho_b\nu_{ba}$ still hold),
and the electric field is expressed as
\begin{equation} \label{eq:Ebasic}
\bE  = -\frac{1}{c}\bu_e\times\bb-\frac{1}{en_e}\nabla\cdot\bp_e+\frac{\boldsymbol{R}_e}{en_e}.
\end{equation}
Momentum equations for ions then become
\begin{equation}
  \rho_i\frac{d_i\bu_i}{dt}+\nabla\cdot\bp_i+\frac{Z_in_i}{n_e}\nabla\cdot\bp_e -\frac{eZ_in_i}{c}(\bu_i-\bu_e)\times\bb
  =\boldsymbol{R}_i+\frac{Z_in_i}{n_e}\boldsymbol{R}_e. \label{eq:momI}
\end{equation}  
Also, by using (\ref{eq:Max1}), the electron density $n_e$ and electron velocity $\bu_e$ is expressed as
\begin{equation}
  n_e = \sum_i Z_i n_i; \qquad \bu_e = \frac{1}{n_e}\sum_i Z_in_i\bu_i -\frac{\boldsymbol{j}}{en_e};
  \qquad \boldsymbol{j}=\frac{c}{4\pi}\nabla\times\bb, \label{eq:reduce} 
\end{equation}
where the summations are over ion species. The electron density equation
$\pr n_e/\pr t+\nabla\cdot(n_e\bu_e)=0$ becomes redundant, because
multiplying all the density equations for charges (including electrons) by $Z_a$ and summing them together yields a requirement
$\nabla\cdot(\sum_aZ_an_a\bu_a)=0$, which is satisfied by
$\nabla\cdot\boldsymbol{j}=0$ in (\ref{eq:reduce}) automatically. Expressions (\ref{eq:reduce}) and (\ref{eq:Ebasic}) then can
be substituted to all other equations (which is easy to do numerically), and the occurence of $\bE,\bu_e,n_e$ in the
entire model is thus elliminated.   

For a particular case of $\boldsymbol{R}_e$ given by (\ref{eq:BGK_R}), the electric field (\ref{eq:Ebasic}) then becomes
\begin{eqnarray}
  \bE &=& -\frac{1}{cn_e}\big(\sum_i Z_i n_i\bu_i\big)\times\bb +\frac{\boldsymbol{j}\times\bb}{cen_e} -\frac{1}{en_e}\nabla\cdot\bp_e
  +\frac{m_e}{e^2 n_e} \boldsymbol{j}\sum_{a\neq e}\nu_{ea}\nn\\
  && +\frac{m_e}{e} \Big[ \big(\sum_{a\neq e}\nu_{ea}\bu_a\big) -\frac{1}{n_e}\big(\sum_{a\neq e}\nu_{ea}\big)\big(\sum_i Z_i n_i\bu_i\big) \Big]. \label{eq:EfieldX}
\end{eqnarray}
Summations over 'a' include both ions and neutrals.
Terms on the r.h.s. can be called the convective term, the Hall term, the electron pressure term, the Ohmic term, and a mixed
collisional term due to ion and neutral velocities.
When (\ref{eq:EfieldX}) is used in the induction equation, the Ohmic term ($\sim\boldsymbol{j}$) becomes directly diffusive through
identity $\nabla\times(\eta_B\nabla\times\bb)=-\eta_B\nabla^2\bb +\nabla(\eta_B)\times(\nabla\times\bb)$,
where one defines a coefficient of magnetic diffusion $\eta_B=(\sum_{a\neq e}\nu_{ea}) m_e c^2/(4\pi e^2 n_e)$. In contrast, no other term
in (\ref{eq:EfieldX}) is directly diffusive in this sense. Nevertheless, the so-called ambipolar diffusion due to differences
in velocities $\bu_a$ between different species is still present implicitly, which can be
shown by solving dispersion relations. The explicit presence of ambipolar diffusion caused by $\sim-(\boldsymbol{j}\times\bb)\times\bb=\boldsymbol{j}_\perp|\bb|^2$ 
is revealed by a construction of a single fluid model, formulated with respect to the center-of-mass velocity of all the species.
In general, ambipolar diffusion between two species with indices $(a,b)$ exists if
\begin{equation}
\frac{Z_a}{m_a} \neq \frac{Z_b}{m_b},
\end{equation}
which is demonstrated in Section \ref{sec:ambipolar}.

In partially ionized solar plasmas one often focuses on a two-fluid model formulated with center-of-mass velocities
for the ion species $\langle\bu_i\rangle=(\sum_i\rho_i\bu_i)/\sum_i\rho_i$ and for the neutral species
$\langle\bu_n\rangle=(\sum_n\rho_n\bu_n)/\sum_n\rho_n$. Velocities for each species are thus decomposed into
$\bu_i=\langle\bu_i\rangle+\bw_i$, $\bu_n=\langle\bu_n\rangle+\bw_n$ where $\bw_i,\bw_n$ represent drifts, and
because $\langle\bu_i\rangle$, $\langle\bu_n\rangle$ can be pulled out in front of the summations 
the electric field (\ref{eq:EfieldX}) transforms into
\begin{eqnarray}
  \bE &=& -\frac{1}{c}\langle\bu_i\rangle\times\bb
  -\frac{1}{cn_e}\big(\sum_i Z_i n_i\bw_i\big)\times\bb +\frac{\boldsymbol{j}\times\bb}{cen_e} -\frac{1}{en_e}\nabla\cdot\bp_e
  +\frac{m_e}{e^2 n_e} \boldsymbol{j}\sum_{a\neq e}\nu_{ea}\nn\\
  && +\big(\langle\bu_n\rangle-\langle\bu_i\rangle\big)\frac{m_e}{e}\sum_n\nu_{en}\nn\\
&&  +\frac{m_e}{e}\Big[ \big(\sum_{a\neq e}\nu_{ea}\bw_a\big) -\frac{1}{n_e}\big(\sum_{a\neq e}\nu_{ea}\big)\big(\sum_i Z_i n_i\bw_i\big) \Big]. \label{eq:lastT}
\end{eqnarray}
Electric field (\ref{eq:lastT}) still represents multi-fluid electric field, where one considers
separate evolution equations for all the drifts $\bw_a$. To obtain a two-fluid electric field these drifts have to be
somehow elliminated, which is of course not straightforward to justify.
In partially ionized solar plasmas the usual justification is that 1) one takes into account only the first
ionization degree, with all the ions having $Z_i=1$; 2) by precribing 
that on average $\sum_i n_i\bw_i=0$ (which for example eliminates ambipolar diffusion between different ions) together
with $\sum_n n_n\bw_n=0$;
3) that all the species have roughly the same temperature which by using collisional freqencies
$\nu_{ei}=n_i f(T)/\sqrt{m_e}$ yields $\sum_i \nu_{ei}\bw_i=0$; 4) that all the neutrals have roughly same cross-sections
(radii $r_n$) which by using $\nu_{en}=n_nf(T)r_n^2/\sqrt{m_e}$ yields $\sum_n \nu_{en}\bw_n=0$. 
The two-fluid electric field thus reads
\begin{eqnarray}
  \bE &=& -\frac{1}{c}\langle\bu_i\rangle\times\bb  +\frac{\boldsymbol{j}\times\bb}{cen_e} -\frac{1}{en_e}\nabla\cdot\bp_e
  +\frac{m_e}{e^2 n_e} \boldsymbol{j}\sum_{a\neq e}\nu_{ea}\nn\\
  && +\big(\langle\bu_n\rangle-\langle\bu_i\rangle\big)\frac{m_e}{e}\sum_n\nu_{en}. \label{eq:lastFF}
\end{eqnarray}
The center-of-mass velocity for ions $\langle\bu_i\rangle$ can be freely replaced by the center-of-mass velocity for all the charges
$\langle\bu_c\rangle$ (which includes electrons).
Then electric field (\ref{eq:lastFF}) is almost identical to equation (115) of \cite{Khomenko2014}, except that  
the $\sum_n\nu_{en}$ in the last term of (\ref{eq:lastFF}) is replaced by $(\sum_n\nu_{en})-(\sum_i\sum_n\nu_{in})$ in that paper.
The difference arises from an alternative approach in that paper, where the electron inertia is not
neglected from the beginning, but instead the electric field is derived by first summing momentum equations for all the species together, and
prescribing quasi-static current $\boldsymbol{j}$. Then, subsequent expansion in mass-ratios retains contributions from $\boldsymbol{R}_i$. Nevertheless,
the missing contributions are small $\nu_{in}\ll \nu_{en}$, explaining the small difference between these two approaches.  

For a particular case of only one ion species and one neutral species,  
so that $n_e=Z_i n_i$ and $\bu_e=\bu_i-\boldsymbol{j}/(en_e)$,  
the electric field (\ref{eq:EfieldX}) simplifies into
\begin{eqnarray}
  \bE &=& -\frac{1}{c}\bu_i\times\bb +\frac{\boldsymbol{j}\times\bb}{cen_e} -\frac{1}{en_e}\nabla\cdot\bp_e
  +\frac{m_e}{e^2 n_e} \boldsymbol{j} (\nu_{ei}+\nu_{en})\nn\\
  && \quad+\frac{m_e}{e} \nu_{en}(\bu_n-\bu_i);\\
 \frac{\pr\bb}{\pr t} &=& \nabla\times(\bu_i\times\bb)-\nabla\times\Big(\frac{\boldsymbol{j}}{en_e}\times\bb\Big)
  +\frac{c}{e}\nabla\times\Big(\frac{1}{n_e}\nabla\cdot\bp_e\Big) \nn\\
&& \quad -\nabla\times \big(\eta_B\nabla\times\bb\big) -\nabla\times\Big[\frac{cm_e}{e}\nu_{en}(\bu_n-\bu_i)\Big],\label{eq:core2}
\end{eqnarray}
with the coefficient of magnetic diffusion $\eta_B=(\nu_{ei}+\nu_{en}) m_e c^2/(4\pi e^2 n_e)$.

\subsection{Ambipolar diffusion of two ion species} \label{sec:ambipolar}
Here we consider a two-fluid model consisting of two different ion species with species indices $(i,j)$, so the charge neutrality reads $n_e=Z_in_i+Z_jn_j$.
A particular case consisting of one ion and one neutral species can be obtained by prescribing $Z_j=0$ and index $j=n$ (or $Z_i=0$ and $i=n$).  
The momentum equations are
\begin{eqnarray}
&&  \rho_i\frac{d_i\bu_i}{dt}+\nabla\cdot\bp_i+\frac{Z_in_i}{n_e}\nabla\cdot\bp_e -\frac{eZ_in_i}{c}\frac{Z_j n_j}{n_e}(\bu_i-\bu_j)\times\bb
  -\frac{Z_i n_i}{cn_e}\boldsymbol{j}\times\bb
  =\boldsymbol{R}_i+\frac{Z_in_i}{n_e}\boldsymbol{R}_e;\label{eq:2fluid-1}\\
&&  \rho_j\frac{d_j\bu_j}{dt}+\nabla\cdot\bp_j+\frac{Z_jn_j}{n_e}\nabla\cdot\bp_e +\frac{eZ_in_i}{c}\frac{Z_j n_j}{n_e}(\bu_i-\bu_j)\times\bb
  -\frac{Z_j n_j}{cn_e}\boldsymbol{j}\times\bb
  =\boldsymbol{R}_j+\frac{Z_jn_j}{n_e}\boldsymbol{R}_e,
\end{eqnarray}
with the collisional right hand sides
\begin{eqnarray}
  \boldsymbol{R}_i+\frac{Z_in_i}{n_e}\boldsymbol{R}_e &=& -(\bu_i-\bu_j)\Big[\rho_i\nu_{ij} +\rho_e\nu_{ei}\Big(\frac{Z_j n_j}{n_e}\Big)^2
    +\rho_e \nu_{ej} \Big(\frac{Z_in_i}{n_e}\Big)^2 \Big] \nn\\
  &&  - \boldsymbol{j}\frac{m_e}{en_e}(\nu_{ei}Z_j n_j-\nu_{ej} Z_i n_i);\nn\\
  \boldsymbol{R}_j+\frac{Z_jn_j}{n_e}\boldsymbol{R}_e &=& (\bu_i-\bu_j)\Big[\rho_i\nu_{ij} 
    +\rho_e \nu_{ei} \Big(\frac{Z_jn_j}{n_e}\Big)^2 +\rho_e\nu_{ej}\Big(\frac{Z_i n_i}{n_e}\Big)^2 \Big] \nn\\
  &&  + \boldsymbol{j}\frac{m_e}{en_e}(\nu_{ei} Z_j n_j-\nu_{ej}Z_i n_i),
\end{eqnarray}
and the electric field (which determines the induction equation) reads
\begin{eqnarray}
  \bE &=& -\frac{1}{cn_e}\big(Z_i n_i\bu_i+Z_jn_j\bu_j\big)\times \bb +\frac{\boldsymbol{j}\times\bb}{cen_e}-\frac{1}{en_e}\nabla\cdot\bp_e
  +\frac{m_e}{e^2 n_e}\boldsymbol{j}(\nu_{ei}+\nu_{ej})\nn\\
  &&+\frac{m_e}{en_e}(\bu_i-\bu_j)\big( Z_jn_j\nu_{ei}-Z_in_i\nu_{ej}\big). \label{eq:2fluid-2}
\end{eqnarray}   
The ambipolar diffusion term $-\boldsymbol{j}\times\bb\times\bb=\boldsymbol{j}_\perp |\bb|^2$ is not directly present in the
electric field, and the only term which directly causes magnetic diffusion in the induction equation is the Ohmic term $(\sim\boldsymbol{j})$. 
Nevertheless, the ambipolar diffusion is still present implicitly, which can be shown by solving dispersion relations,
or by constructing a single-fluid model. 

By using the same notation as \cite{Zaqarashvili2011} and introducing center-of-mass velocity $\boldsymbol{V}=(\rho_i\bu_i+\rho_j\bu_j)/\rho$ where
the total density $\rho=\rho_i+\rho_j$, and difference in velocities $\bw=\bu_i-\bu_j$, so that $\bu_i=\boldsymbol{V}+(\rho_j/\rho)\bw$,
$\bu_j=\boldsymbol{V}-(\rho_i/\rho)\bw$, yields momentum equations
\begin{eqnarray}
  && \rho\frac{\pr\boldsymbol{V}}{\pr t} + \rho\boldsymbol{V}\cdot\nabla\boldsymbol{V}
  +\nabla\cdot(\bp_i+\bp_j+\bp_e)
  -\frac{1}{c}\boldsymbol{j}\times\bb +\nabla\cdot\Big(\frac{\rho_i\rho_j}{\rho}\bw\bw\Big)=0; \label{eq:2fluid-3}\\
  && \frac{\pr\bw}{\pr t}+\bw\cdot\nabla\boldsymbol{V}+\boldsymbol{V}\cdot\nabla\bw +\frac{\rho_j}{\rho}\bw\cdot\nabla\bw
  -\bw\cdot\nabla\big(\frac{\rho_i}{\rho}\bw\big)-\frac{eZ_in_iZ_jn_j}{cn_e}\frac{\rho}{\rho_i\rho_j} \bw\times\bb\nn\\
  && \quad  +\frac{1}{\rho_i}\nabla\cdot\bp_i-\frac{1}{\rho_j}\nabla\cdot\bp_j
  +\frac{1}{n_e}\Big(\frac{Z_in_i}{\rho_i}-\frac{Z_jn_j}{\rho_j}\Big)\Big(\nabla\cdot\bp_e -\frac{1}{c}\boldsymbol{j}\times\bb\Big)\nn\\
  && \quad = -\bw \frac{\rho}{\rho_i\rho_j}\Big[\rho_i\nu_{ij} +\rho_e\nu_{ei}\Big(\frac{Z_j n_j}{n_e}\Big)^2
    +\rho_e \nu_{ej} \Big(\frac{Z_in_i}{n_e}\Big)^2 \Big] 
   - \boldsymbol{j}\frac{\rho}{\rho_i\rho_j}\frac{m_e}{en_e}(\nu_{ei} Z_j n_j-\nu_{ej}Z_i n_i), \label{eq:2fluid-4}
\end{eqnarray}
with electric field
\begin{eqnarray}
  \bE &=& -\frac{1}{c}\boldsymbol{V}\times\bb -\frac{1}{cn_e\rho}(Z_in_i\rho_j-Z_jn_j\rho_i)\bw\times\bb
  +\frac{\boldsymbol{j}\times\bb}{cen_e}-\frac{1}{en_e}\nabla\cdot\bp_e
  +\frac{m_e}{e^2 n_e}\boldsymbol{j}(\nu_{ei}+\nu_{ej})\nn\\
  &&+\frac{m_e}{en_e}\bw\big( Z_jn_j\nu_{ei}-Z_in_i\nu_{ej}\big).\label{eq:2fluid-5}
\end{eqnarray}
System (\ref{eq:2fluid-3})-(\ref{eq:2fluid-5}) is of course equivalent to (\ref{eq:2fluid-1})-(\ref{eq:2fluid-2}). 
However, in a particular case when the collisions are very frequent, the r.h.s. of (\ref{eq:2fluid-4}) becomes
very large, and by neglecting all the ``inertial'' terms in the first line of (\ref{eq:2fluid-4}) with $\bw$, allows one to obtain
an explicit expression for the velocity difference
\begin{eqnarray}
\bw &=& 
\frac{1}{D}\Big[ -\boldsymbol{j}\frac{m_e}{en_e}(\nu_{ei} Z_j n_j-\nu_{ej}Z_i n_i) -\frac{\rho_j}{\rho}\nabla\cdot\bp_i+ \frac{\rho_i}{\rho}\nabla\cdot\bp_j
- \frac{1}{\rho n_e}\big(Z_in_i\rho_j-Z_jn_j\rho_i\big)\Big(\nabla\cdot\bp_e -\frac{1}{c}\boldsymbol{j}\times\bb\Big)
\Big],\label{eq:Zaq}
\end{eqnarray}
where we defined denominator
\begin{equation} \label{eq:denominator}
D= \Big[\rho_i\nu_{ij} +\rho_e\nu_{ei}\Big(\frac{Z_j n_j}{n_e}\Big)^2
  +\rho_e \nu_{ej} \Big(\frac{Z_in_i}{n_e}\Big)^2 \Big].
\end{equation}
For frequent collisions only the first term in (\ref{eq:Zaq}) $\sim \boldsymbol{j}$ is finite, and all other terms are small. Nevertheless, the sought-after term
is the last term in (\ref{eq:Zaq}) $\sim \boldsymbol{j}\times\bb$, because when (\ref{eq:Zaq}) is used in (\ref{eq:2fluid-5}) it creates the ambipolar term
$\sim -\boldsymbol{j}\times\bb\times\bb$. The single-fluid electric field reads 
\begin{eqnarray}
  \bE &=& -\frac{1}{c}\boldsymbol{V}\times\bb
  +\boldsymbol{j}\frac{m_e}{e^2n_e}\Big[ \nu_{ei}+\nu_{ej}-\frac{m_e}{n_eD}(\nu_{ei} Z_j n_j-\nu_{ej}Z_i n_i)^2\Big]\nn\\
 && + \frac{\boldsymbol{j}\times\bb}{cen_e}\Big[ 1+\frac{2m_e}{n_e\rho D}(Z_in_i\rho_j-Z_jn_j\rho_i)(\nu_{ei} Z_j n_j-\nu_{ej}Z_i n_i)\Big]\nn\\ 
  && -\,\frac{\boldsymbol{j}\times\bb\times\bb}{c^2 n_e^2\rho^2 D} (Z_in_i\rho_j-Z_jn_j\rho_i)^2 \nn\\
  && -\frac{\nabla\cdot\bp_e}{en_e}\Big[1+\frac{m_e}{\rho n_e D}\big( Z_jn_j\nu_{ei}-Z_in_i\nu_{ej}\big)\big(Z_in_i\rho_j-Z_jn_j\rho_i\big) \Big]\nn\\
  && +\frac{m_e}{en_e\rho D}\big( Z_jn_j\nu_{ei}-Z_in_i\nu_{ej}\big)\Big[ -\rho_j\nabla\cdot\bp_i+ \rho_i\nabla\cdot\bp_j\Big]\nn\\
  && -\frac{1}{cn_e\rho^2 D}(Z_in_i\rho_j-Z_jn_j\rho_i)\Big[ -\rho_j\nabla\cdot\bp_i+ \rho_i\nabla\cdot\bp_j
    - \frac{1}{n_e}\big(Z_in_i\rho_j-Z_jn_j\rho_i\big)\nabla\cdot\bp_e \Big]\times\bb. \label{eq:Efunny}
\end{eqnarray}
Importantly, the sign in front of the ambipolar term is negative, and because
$-\boldsymbol{j}\times\bb\times\bb=+\boldsymbol{j}_\perp |\bb|^2$, the term indeed creates diffusion in the induction equation.
It is possible to define a coefficient of ambipolar diffusion
\begin{equation} \label{eq:Ambi}
\eta_A = \frac{|\bb|^2}{4\pi\rho} A = V_A^2 A; \quad \textrm{where} \quad A=\frac{\big( Z_i n_i\rho_j-Z_j n_j\rho_i \big)^2}{n_e^2 \rho D},
\end{equation}
and $V_A$ is the Alfv\'en speed. As a double check, prescribing zero charge for one of the species, electric field (\ref{eq:Efunny}) identifies with equation (A.10) of
\cite{Zaqarashvili2011} (for example, our denominator simplifies to $D=\alpha_{in}+\alpha_{en}=\alpha_n$). Also, 
$\eta_A=|\bb|^2\rho_n^2/(4\pi\rho^2(\rho_i\nu_{in}+\rho_e\nu_{en}))$ identifies with the usual coefficient of ambipolar diffusion; see for
example equation (20) in \cite{Khomenko2012} (after switching to cgs units with $\mu_0\to 4\pi$). 
The ambipolar diffusion exists if
\begin{equation} \label{eq:pretty}
\frac{Z_i}{m_i}\neq \frac{Z_j}{m_j}.
\end{equation}
It is important to emphasize that the reduction to a single-fluid model was obtained by assuming that collisions
are sufficiently frequent, and the ambipolar diffusion (as well as other terms) now contain a denominator $D$, which can be simplified into $D=\rho_i\nu_{ij}$.
So when collisional frequencies $\nu_{ij}$ become small, leads to an artificial ``explosion'' of the ambipolar diffusion.
This is nicely demonstrated in the figures of \cite{Zaqarashvili2011} plotted with respect to a wavenumber  
$\bar{k}\sim k/\nu$, where it is shown that for a single-fluid description, the ambipolar diffusion in a collisionless regime (when $\bar{k}$ becomes large)
yields cut-off frequencies for waves. The mechanism is completely analogous to the ``explosion'' of the Braginskii stress-tensor or the heat
flux vector in a collisionless regime. In contrast, as they show in their two-fluid figures, 
no ``explosion'' of the ambipolar diffusion is present. The effect is further discussed in \cite{Zaqarashvili2012}.

\subsubsection{Damping of Alfv\'en waves}
For example, considering Alfv\'en waves at long-wavelengths and focusing only on the ambipolar diffusion
(with the Hall-term, Ohmic terms and pressure terms neglected), the induction equation reads
\begin{equation}
\frac{\pr\bb}{\pr t} = \nabla\times(\boldsymbol{V}\times\bb) + \nabla\times\Big[\eta_A (\nabla\times\bb)_\perp\Big], 
\end{equation}
with the coefficient of ambipolar diffusion (\ref{eq:Ambi}).
This yields the following dispersion relation for Alfv\'en waves
\begin{equation} \label{eq:Zaq1}
\omega^2+i\omega V_a^2\kpar^2 A-V_A^2\kpar^2=0,
\end{equation}
with solutions
\begin{equation} \label{eq:Zaq2}
\omega = \pm V_A |\kpar|\sqrt{1-\Big( \frac{V_A\kpar A}{2}\Big)^2} -i\frac{V_A^2\kpar^2 A}{2}.
\end{equation}  
Obviously, the Alfv\'en waves are damped and for wavenumbers $\kpar\ge 2/(V_A A)$ 
the real part of the frequency even becomes zero, so the wave stops existing (i.e. cut-off wavenumber). For the particular case
of one specie being neutral, the quantity $A=\rho_n^2/(\rho \alpha_n)$, which can be approximated as
$A=\rho_n^2/(\rho \alpha_{in})=\zeta_n^2/(\zeta_i\nu_{in})$. Then  
expressions (\ref{eq:Zaq1}), (\ref{eq:Zaq2}) identify with equations (44)-(47) of \cite{Zaqarashvili2011}, however, 
one needs to use their definition $\nu_{in}=\alpha_{in}/\rho$ instead of the more logical (and correct) $\nu_{in}=\alpha_{in}/\rho_i$.

\newpage
\section{General Fokker-Planck collisional operator} \label{sec:FokkerPlanck}
\setcounter{equation}{0}
For Coulomb collisions, the Boltzmann collisional operator can be approximated by a general Fokker-Planck type of collisional operator,
\begin{equation} \label{eq:FP_operator}
C_{ab}(f_a,f_b) = -\nabla_v\cdot \Big[ \boldsymbol{A}_{ab}f_a-\frac{1}{2}\nabla_v\cdot(\bD_{ab} f_a)\Big],
\end{equation}
where higher-order derivatives in velocity space are neglected, and where $\boldsymbol{A}$ is called a dynamical friction vector
and $\bD$ is called a diffusion tensor. In space physics and astrophysics, various
approximations for $\boldsymbol{A}$ and $\bD$ are used, and if a collisional operator has form (\ref{eq:FP_operator}),
then equation (\ref{eq:Vlasov}) is summarily called the Fokker-Planck equation. 
Summation over all the species (including self-collisions) then defines the full operator $C(f_a)=\sum_b C_{ab}(f_a,f_b)$ which can be also
  written as $C(f_a) = -\nabla_v\cdot [ \boldsymbol{A}_{a}f_a-(1/2)\nabla_v\cdot(\bD_{a} f_a)]$, where one 
defines $\boldsymbol{A}_a = \sum_b \boldsymbol{A}_{ab}$ \& $\bD_a = \sum_b \bD_{ab}$.
The Fokker-Planck operators work extremely well for any collisional
 process where collisions with a small scattering angle dominate, and where a lot of subsequent collisions gradually yield (in a sense of
a random walk) a significant deviation from a particle original velocity direction. This is exactly the case for scattering by the electrostatic
Coulomb force, where the Rutherford scattering cross-section is proportional to $1/\sin^4(\chi/2)$, and heavily dominated by events with a
small scattering angle $\chi$.

For any tensor $\bX$, a general Fokker-Planck operator can be integrated according to
\begin{equation}
\int \bX C_{ab}(f_a,f_b)d^3v = \int f_a \boldsymbol{A}_{ab}\cdot\frac{\pr\bX}{\pr\bV}d^3v + \frac{1}{2}\int f_a \bD_{ab} : \frac{\pr}{\pr\bV}\frac{\pr\bX}{\pr\bV}d^3v,
\end{equation}
and for clarity explicitly in the index notation
\begin{equation}
\int \bX C_{ab}(f_a,f_b)d^3v = \int f_a A^{ab}_i \frac{\pr\bX}{\pr v_i}d^3v + \frac{1}{2}\int f_a D^{ab}_{ij} \frac{\pr}{\pr v_i}\frac{\pr\bX}{\pr v_j}d^3v.
\end{equation}
Useful identities are
\begin{eqnarray}
  \frac{\pr|\bV|}{\pr v_i} = \frac{v_i}{|\bV|};  \quad \frac{\pr|\bc|}{\pr v_i} = \frac{c_i}{|\bc|};
  \quad \frac{\pr|\bV|^2}{\pr v_i} =2 v_i;\quad \frac{\pr|\bc|^2}{\pr v_i} =2 c_i,
\end{eqnarray}
and the  tensorial collisional contributions defined in (\ref{eq:Spec}) can be calculated according to  
\begin{eqnarray}
  \boldsymbol{R}_{ab} &=& m_a \int f_a \boldsymbol{A}_{ab} d^3v; \label{eq:FP1}\\
  Q_{ab} &=& m_a \int f_a \boldsymbol{A}_{ab}\cdot\bc_a d^3v + \frac{m_a}{2}\int f_a \textrm{Tr} \bD_{ab} d^3v; \label{eq:FP2}\\
  \bQ_{ab}^{(2)} &=& m_a \int f_a \big[ \boldsymbol{A}_{ab}\bc_a\big]^S d^3v +\frac{m_a}{2}\int f_a \big[\bD_{ab}\big]^S d^3v;\\
  \bQ_{ab}^{(3)} &=& m_a \int f_a \big[ \boldsymbol{A}_{ab}\bc_a\bc_a\big]^S d^3v +\frac{m_a}{2}\int f_a \big[\bD_{ab}^S \bc_a \big]^S d^3v. \label{eq:FP}
\end{eqnarray}
If the diffusion tensor is symmetric then $\bD^S_{ab}=2\bD_{ab}$
(For clarity, the symmetric operator does not act on species indices and in general $\bD_{ab}\neq \bD_{ba}$ similarly to $\nu_{ab}\neq\nu_{ba}$,
the symmetric operator acts as $(\bD^{ab}_{ij})^S=\bD^{ab}_{ij}+\bD^{ab}_{ji}$).
The 4th \& 5th-order collisional contributions are
\begin{eqnarray}
  \big(\bQ_{ab}^{(4)}\big)_{ijkl} &=& m_a \int f_a \big[ \boldsymbol{A}_{ab}\bc_a\bc_a\bc_a\big]^S_{ijkl} d^3v
  +\frac{m_a}{2}\int f_a \Big[ \big[\bD_{ab}^S \bc_a \bc_a \big]^S_{ijkl} + (\bD^{ab})^S_{ik}c_j^a c_l^a + (\bD^{ab})^S_{jl} c_i^a c_k^a\Big] d^3v;\label{eq:Thierry30}\\
  \big(\bQ_{ab}^{(5)}\big)_{ijklm} &=& m_a \int f_a \big[ \boldsymbol{A}_{ab}\bc_a\bc_a\bc_a\bc_a\big]^S_{ijklm} d^3v
  +\frac{m_a}{2}\int f_a \Big[ \big[\bD_{ab}^S \bc_a \bc_a \bc_a\big]^S_{ijklm} + (\bD^{ab})^S_{ik}c_j^a c_l^a c_m^a \nn\\
    && + (\bD^{ab})^S_{il} c_j^a c_k^a c_m^a + (\bD^{ab})^S_{jl} c_i^a c_k^a c_m^a + (\bD^{ab})^S_{jm} c_i^a c_k^a c_l^a
    + (\bD^{ab})^S_{km} c_i^a c_j^a c_l^a   \Big] d^3v.\label{eq:Thierry31}
\end{eqnarray}  
The first integral in (\ref{eq:Thierry30}) proportional to $\boldsymbol{A}_{ab}$ contains 4 terms, and the second integral in
  (\ref{eq:Thierry30}) proportional to $\bD^{ab}$ contains 12 terms. Similarly, the first integral in (\ref{eq:Thierry31}) contains 5 terms
  and the second integral in (\ref{eq:Thierry31}) contains 20 terms. The second integrals in (\ref{eq:Thierry30})-(\ref{eq:Thierry31}) can be written simply by
picking two indices for $\bD^{ab}$ and giving the rest of indices to $\bc_a\bc_a$ \& $\bc_a\bc_a\bc_a$. 
Generalization to an n-th order collisional contributions defined in (\ref{eq:bXn})
is done naturally by introducing a set of indices $R=\{r_1\ldots r_n\}$  together with an ordered set $(s_1,s_2)$, and writing
\begin{eqnarray}
 \big(\bQ_{ab}^{(n)}\big)_{r_1 r_2 \ldots r_n} &=& m_a \int f_a \big[ A^{ab}_{r_1} c^a_{r_2}\ldots c^a_{r_n}\big]^S d^3v 
 +\frac{m_a}{2}\int f_a \Big[ D^{ab}_{s_1 s_2} c_{s_3}\ldots c_{s_n} \Big]d^3v; \nn\\
 && \textrm{where} \quad (s_1,s_2)\in R=\{r_1\ldots r_n\}; \quad \textrm{and} \quad s_3\ldots s_n\in R\setminus \{s_1,s_2\},
\end{eqnarray}
so that the first integral contains (n) terms, and the second integral contains $2\binom{n}{2}=n(n-1)$ terms.  Alternatively,
one can replace the ordered set $(s_1,s_2)$ by a non-ordered set $\{s_1,s_2\}$ and include the symmetric operator on $\bD^{ab}$.

It is useful to write collisional contributions for contracted vectors, matrices and 
scalars, by assuming symmetric $\bD^{ab}$. We use definitions from Section {\ref{sec:Fhierarchy}}, see equation (\ref{eq:Thierry25}),
which were also used in Appendix \ref{sec:Hierarchy}; see equation (\ref{eq:Thierry20}) \& (\ref{eq:Thierry21}). This yields collisional contributions
for vectors
\begin{eqnarray}
  \vec{\boldsymbol{\mathcal{Q}}}^{(2n+1)}_{ab} &=& m_a \int \Big[ (2n)(\boldsymbol{A}^{ab}\cdot\bc_a)\bc_a|\bc_a|^{2n-2}
    + \boldsymbol{A}^{ab} |\bc_a|^{2n} \Big] f_a d^3v\nn\\
  && + m_a \int \Big[ (2n)(n-1)(\bD^{ab}:\bc_a\bc_a)\bc_a |\bc_a|^{2n-4} + (n)(\trace\bD^{ab})\bc_a |\bc_a|^{2n-2} \nn\\
  && \qquad  +(2n)(\bc_a\cdot\bD^{ab}) |\bc_a|^{2n-2}\Big] f_a d^3v;
\end{eqnarray}
matrices
\begin{eqnarray}
  \bar{\bar{\boldsymbol{\mathcal{Q}}}}^{(2n)}_{ab} &=& m_a \int \Big[ (\boldsymbol{A}^{ab} \bc_a)^S |\bc_a|^{2n-2}
    +(2n-2) (\boldsymbol{A}^{ab}\cdot \bc_a)\bc_a\bc_a |\bc_a|^{2n-4}\Big] f_a d^3v \nn\\
  && + m_a \int \Big[  \bD^{ab}|\bc_a|^{2n-2} +(2n-2) (\bD^{ab}\cdot\bc_a\bc_a)^S |\bc_a|^{2n-4}\nn\\
    &&  +(n-1)(\trace \bD^{ab})\bc_a\bc_a |\bc_a|^{2n-4}
    + (n-1)(2n-4) (\bD^{ab}:\bc_a\bc_a)\bc_a\bc_a|\bc_a|^{2n-6}  \Big] f_a d^3v; \label{eq:Thierry22}
\end{eqnarray}
and scalars
\begin{eqnarray}
  Q^{(2n)}_{ab} &=& m_a \int \Big[ (2n)(\boldsymbol{A}^{ab}\cdot \bc_a) |\bc_a|^{2n-2} + (n)(\trace\bD^{ab})|\bc_a|^{2n-2}\nn\\
    && +(2n) (n-1)(\bD^{ab}:\bc_a\bc_a)|\bc_a|^{2n-4}\Big] f_a d^3v, \label{eq:Thierry23}
\end{eqnarray}
all valid for $n\ge 1$. Applying trace at (\ref{eq:Thierry22}) recovers (\ref{eq:Thierry23}).

\newpage
\section{Landau collisional operator (5-moment model)} \label{sec:5moment}
\setcounter{equation}{0}
For Coulomb collisions, a very accurate collisional operator was obtained by \cite{Landau1936,Landau1937} in the 
following form (see for example equation (1.2) in \cite{Braginskii1958})
\begin{equation}
  C_{ab}(f_a,f_b) = -\, \frac{2\pi e^4 Z_a^2 Z_b^2\ln\Lambda}{m_a}\frac{\pr}{\pr\bV}\cdot \int  \bar{\bar{\boldsymbol{V}}}\cdot
  \Big[ \frac{f_a(\bV)}{m_b}\frac{\pr f_b(\bV')}{\pr \bV'}-\frac{f_b(\bV')}{m_a}\frac{\pr f_a(\bV)}{\pr\bV}\Big]d^3v'; \label{eq:Landau}
\end{equation}
\begin{equation}
\bar{\bar{\boldsymbol{V}}} = \frac{\bI}{|\bV-\bV'|}-\frac{(\bV-\bV')(\bV-\bV')}{|\bV-\bV'|^3}.
\end{equation}
With this collisional operator, equation (\ref{eq:Vlasov}) is known as the Landau equation. 
The Landau collisional operator is sometimes called
the Landau collisional integral because (\ref{eq:Landau}) contains integral over $d^3v'$ (i.e. it is an integro-differential operator).
The operator can be rewritten
into the general Fokker-Planck form (\ref{eq:FP_operator}) by introducing Rosenbluth potentials
\begin{equation} 
  H_b(\bV)=\int \frac{f_b(\bV')}{|\bV-\bV'|}d^3v'; \quad \textrm{and}\quad G_b(\bV)=\int f_b(\bV')|\bV-\bV'|d^3v',
\end{equation}
yielding (see for example equations (7)-(8) of \cite{Hinton1983})
\begin{equation}
  \boldsymbol{A}_{ab} = 2\frac{c_{ab}}{m_a^2}\big(1+\frac{m_a}{m_b}\big)\frac{\pr H_b(\bV)}{\pr\bV}; \qquad \bD_{ab} 
  =2\frac{c_{ab}}{m_a^2}\frac{\pr^2 G_b(\bV)}{\pr\bV\pr\bV}; \qquad c_{ab} = 2\pi e^4 Z_a^2 Z_b^2\ln\Lambda. \label{eq:AD}
\end{equation}
Useful identities are
\begin{equation}
\frac{\pr}{\pr\bV}\cdot\bar{\bar{\boldsymbol{V}}} = -2\frac{\bV-\bV'}{|\bV-\bV'|^3}=-\,\frac{\pr}{\pr\bV'}\cdot\bar{\bar{\boldsymbol{V}}};
  \qquad \frac{\pr^2}{\pr\bV\pr\bV}|\bV-\bV'|=\bar{\bar{\boldsymbol{V}}},
\end{equation}
and it is easy to verify that (\ref{eq:FP_operator}), (\ref{eq:AD}) recovers the Landau operator (\ref{eq:Landau})
(after one uses the Gauss-Ostrogradsky divergence theorem in velocity $d^3v'$, which makes the associated integral to vanish).
By using Laplacian $\nabla^2_v=\nabla_v\cdot\nabla_v$, the following identity implies
\begin{equation} \label{eq:check1}
\nabla^2_v \frac{1}{|\bV-\bV'|} = -4\pi \delta(\bV-\bV'); \quad => \quad \nabla^2_v H_b(\bV) = -4\pi f_b(\bV).
\end{equation}
The Rosenbluth potential $H_b(\bV)$ is thus completely analogous to the electrostatic potential $\Phi(\bx)$
(with a Poisson equation $\nabla^2 \Phi(\bx) =-4\pi \rho_c(\bx)$, where $\rho_c(\bx)$ is the charge spatial distribution),
here just used in velocity space. Also, because of identity $\nabla^2_v |\bV-\bV'|=2/|\bV-\bV'|$, the Rosenbluth potentials are related by
\begin{equation} \label{eq:Ros3}
  H_b = \frac{1}{2}\nabla^2_v G_b; \quad =>
  \quad \boldsymbol{A}_{ab}=\frac{1}{2}\big(1+\frac{m_a}{m_b}\big)\frac{\pr}{\pr\bV}\cdot\bD_{ab}.
\end{equation}
  
However, the structure of Rosenbluth potentials implies that the Landau operator is quite complicated. Already in the simplest example when
prescribing Maxwellian $f_b=n_b/(\pi^{3/2}v_{\textrm{th} b}^3)\exp(-y^2)$ with the (vector) variable $\by=(\bV-\bu_b)/v_{\textrm{th} b}$ and scalar $y=|\by|$, yields
Rosenbluth potentials 
\begin{eqnarray}
  H_b (\bV) &=& \frac{n_b}{v_{\textrm{th} b}}\frac{\textrm{erf}(y)}{y};\label{eq:Ros}\\
  G_b (\bV) &=& n_b v_{\textrm{th} b} \Big[ \frac{1}{\sqrt{\pi}}e^{-y^2}+\big(\frac{1}{2y}+y\big)\textrm{erf}(y)\Big], \label{eq:Ros2}
\end{eqnarray}
where the error function $\textrm{erf}(y)=(2/\sqrt{\pi})\int_0^y e^{-z^2}dz$ is present. 
These Rosenbluth potentials make collisional contributions (\ref{eq:FP1}), (\ref{eq:FP2}) difficult to calculate.

For clarity on how the $H_b$ is obtained, it is useful to introduce (vector) variable $\bx=(\bV'-\bV)/v_{\textrm{th} b}$, and scalar $x=|\bx|$
and change the integration into $d^3v'=v_{\textrm{th} b}^3 d^3x $, so that 
\begin{equation}
  H_b (\bV) = \frac{n_b}{\pi^{3/2} v_{\textrm{th} b}^3} \int_{-\infty}^\infty \frac{e^{-\frac{|\bV'-\bu_b|^2}{v_{\textrm{th} b}^2}}}{|\bV'-\bV|}d^3v'
  =  \frac{n_b}{\pi^{3/2} v_{\textrm{th} b}} \int_{-\infty}^\infty \frac{e^{-|\bx+\by|^2}}{x} d^3x.
\end{equation}  
In the last integral the variable $\by$ is a constant (because $\bV$ and $\bu_b$ are constants).
One introduces spherical co-ordinates in the $\bx$-space  with orthogonal unit vectors $\hat{\boldsymbol{e}}_1, \hat{\boldsymbol{e}}_2, \hat{\boldsymbol{e}}_3$,
where the direction of vector $\by$ forms axis $\hat{\boldsymbol{e}}_3=\by/y$, so that the vector
\begin{equation}\label{eq:ypic}
\bx = x \sin\theta \cos\phi \hat{\boldsymbol{e}}_1 + x\sin\theta\sin\phi \hat{\boldsymbol{e}}_2 + x\cos\theta \hat{\boldsymbol{e}}_3. 
\end{equation}
In this reference frame $\by=(0,0,y)$ and so $|\bx+\by|^2=x^2+y^2+2xy\cos\theta$. Then one  
can calculate the integral in spherical co-ordinates $d^3x = x^2 \sin\theta dx d\theta d\phi$, yielding
\begin{eqnarray}
&&  \int_{-\infty}^\infty \frac{e^{-|\bx+\by|^2}}{x} d^3x = 2\pi \int_0^\infty \int_0^\pi x e^{-(x^2+y^2)} \sin\theta e^{-2xy\cos\theta} d\theta dx \nn\\
  &&  =  2\pi \int_0^\infty x e^{-(x^2+y^2)} \frac{1}{2xy} \Big( e^{+2xy} -e^{-2xy} \Big) dx
  = \frac{\pi}{y} \int_0^\infty \Big( e^{-(x-y)^2} - e^{-(x+y)^2} \Big) dx \nn\\
  && = \frac{\pi}{y} \Big( \int_{-y}^\infty e^{-z^2} dz -\int_{y}^\infty e^{-z^2} dz \Big)
  = \frac{\pi}{y}\int_{-y}^y e^{-z^2} dz = \frac{2\pi}{y}\int_0^y e^{-z^2} dz =\frac{\pi^{3/2}}{y}\textrm{erf}(y),
\end{eqnarray}
recovering (\ref{eq:Ros}). The result can be verified by calculating (\ref{eq:check1}). 
Similarly, the potential $G_b$ can be obtained by calculating
\begin{eqnarray}
&&  \int_{-\infty}^\infty x e^{-|\bx+\by|^2} d^3x = \frac{\pi}{y} \int_0^\infty x^2 \Big( e^{-(x-y)^2} - e^{-(x+y)^2} \Big) dx =
  \frac{\pi}{y} \Big( \int_{-y}^\infty (z+y)^2 e^{-z^2} dz -\int_{y}^\infty (z-y)^2 e^{-z^2} dz \Big)\nn\\
  && =\frac{\pi}{y}\Big(2\int_0^y z^2 e^{-z^2} dz +4y\int_y^\infty ze^{-z^2} dz +2y^2\int_0^y e^{-z^2} dz\Big)
  = \pi^{3/2} \big( y+\frac{1}{2y}\big)\textrm{erf}(y) +\pi e^{-y^2},
\end{eqnarray}  
recovering (\ref{eq:Ros2}), and which can be verified to satisfy (\ref{eq:Ros3}).

Note that because $\erf(0)=0$, the error function can be actually defined as an indefinite integral
\begin{equation}
  \frac{2}{\sqrt{\pi}}\int e^{-x^2}dx=\erf(x); \qquad \frac{2}{\sqrt{\pi}}\int e^{-\frac{(x+a)^2}{b^2}}dx=\frac{\erf(x+a)}{b^2}.\nn
\end{equation}
Useful relations are $\erf(-x)=-\erf(x)$ and $\erf(\infty)=1$. Then the calculations above can be done more elegantly, for example
\begin{equation}
\int_0^\infty e^{-(x-y)^2} dx = \frac{\sqrt{\pi}}{2}\erf(x-y)\Big|_{x=0}^{x=\infty} = \frac{\sqrt{\pi}}{2}\big(1+\erf(y)\big).\nn  
\end{equation}  
\subsection{Momentum exchange rates \texorpdfstring{$\boldsymbol{R}_{ab}$}{Rab}} \label{sec:Rab}
To obtain the momentum exchange rates $\boldsymbol{R}_{ab}$, one needs to calculate
\begin{eqnarray}
  \boldsymbol{R}_{ab} &=& m_a \int f_a \boldsymbol{A}_{ab} d^3v
  = 2\frac{c_{ab}}{m_a}\big(1+\frac{m_a}{m_b}\big)\int f_a \frac{\pr H_b}{\pr\bV} d^3v;\nn\\
  &=& - 2\frac{c_{ab}}{m_a}\big(1+\frac{m_a}{m_b}\big)\int H_b \frac{\pr f_a}{\pr\bV} d^3v. \label{eq:Rab1}
\end{eqnarray}
Prescribing Maxwellian $f_a=(n_a/(\pi^{3/2}v_{\textrm{th} a}^3))\exp(-|\bV-\bu_a|^2/v_{\textrm{th} a}^2)$ with general velocity $\bu_a$
leads to the ``runaway'' effect addressed below in Section \ref{sec:runaway}. It is useful to first consider simplified situation
where the differences between drift velocities $\bu_a$ and $\bu_b$ are small. 
The Maxwellian $f_a$ is rewritten with
the variable $\by$ and variable $\bu=(\bu_b-\bu_a)/v_{\textrm{th} a}$, and expanded by assuming that $|\bu|\ll 1$, so that
\begin{equation}
f_a = \frac{n_a}{\pi^{3/2}v_{\textrm{th} a}^3} e^{-|\by\alpha+\bu|^2} \simeq \frac{n_a}{\pi^{3/2}v_{\textrm{th} a}^3} e^{-y^2\alpha^2}\big( 1-2\alpha \by\cdot\bu \big),
\end{equation}  
where $\alpha=v_{\textrm{th} b}/v_{\textrm{th} a}$. Then the derivative
\begin{equation}
\frac{\pr f_a}{\pr\bV} = -\frac{2 n_a}{\pi^{3/2}v_{\textrm{th} a}^4} e^{-y^2\alpha^2}\Big[ \bu+\alpha\by-2\alpha^2\by(\by\cdot\bu)\Big],
\end{equation}
and one needs to calculate
\begin{equation}
  \boldsymbol{R}_{ab} = \frac{4c_{ab}}{ m_a}\big(1+\frac{m_a}{m_b}\big)\frac{n_an_b}{\pi^{3/2} v_{\textrm{th} a}^4 v_{\textrm{th} b}}
  \underbrace{\int_{-\infty}^\infty \frac{\textrm{erf}(y)}{y} e^{-y^2\alpha^2}\Big[
      \bu+\alpha\by-2\alpha^2\by(\by\cdot\bu)\Big] d^3v}_{\bigcirc{ }\!\!\!\!\mbox{\small 1}+\bigcirc{ }\!\!\!\!\mbox{\small 2}+\bigcirc{ }\!\!\!\!\mbox{\small 3}},
  \label{eq:Rab}
\end{equation}
where we split the integral into three parts. The integration over $d^3v$ can be changed to $v_{\textrm{th} b}^3d^3y$.
We will use 
\begin{eqnarray}
&&  \int_0^\infty e^{-y^2\alpha^2} y \, \textrm{erf}(y) dy = \frac{1}{2\alpha^2\sqrt{1+\alpha^2}};\nn\\
&&  \int_0^\infty e^{-y^2\alpha^2} y^3 \, \textrm{erf}(y) dy = \frac{3\alpha^2+2}{4\alpha^4(1+\alpha^2)^{3/2}}.
\end{eqnarray}
The three integrals are then evaluated according to
\begin{eqnarray}
  \bigcirc{ }\!\!\!\!\mbox{\small 1} &=& \bu \int_{-\infty}^\infty \frac{\textrm{erf}(y)}{y} e^{-y^2\alpha^2} d^3v
  = \bu  v_{\textrm{th} b}^3 4\pi \int_0^\infty y \, \textrm{erf}(y) e^{-y^2\alpha^2} dy
  = \bu  v_{\textrm{th} b}^3 \frac{2\pi}{\alpha^2\sqrt{1+\alpha^2}};\nn\\
  \bigcirc{ }\!\!\!\!\mbox{\small 2} &=& \alpha \int_{-\infty}^\infty \by \frac{\textrm{erf}(y)}{y} e^{-y^2\alpha^2} d^3v =0;\nn\\
  \bigcirc{ }\!\!\!\!\mbox{\small 3} &=& -2\alpha^2 \int_{-\infty}^\infty \by(\by\cdot\bu) \frac{\textrm{erf}(y)}{y} e^{-y^2\alpha^2} d^3v
  = -\frac{2\alpha^2}{3}\bu \int_{-\infty}^\infty y\, \textrm{erf}(y) e^{-y^2\alpha^2} d^3v \nn\\
  &=& -\frac{8\pi}{3}\alpha^2 v_{\textrm{th} b}^3 \bu \int_0^\infty y^3 \, \textrm{erf}(y) e^{-y^2\alpha^2} dy
  = -\frac{8\pi}{3}\alpha^2 v_{\textrm{th} b}^3 \bu \frac{3\alpha^2+2}{4\alpha^4(1+\alpha^2)^{3/2}}, \label{eq:pretty2}
\end{eqnarray}
and so
\begin{equation}
  \bigcirc{ }\!\!\!\!\mbox{\small 1} + \bigcirc{ }\!\!\!\!\mbox{\small 3} = \bu v_{\textrm{th} b}^3 \frac{2\pi}{3\alpha^2 (1+\alpha^2)^{3/2}}
  = \bu \frac{2\pi}{3}\frac{v_{\textrm{th}a}^5 v_{\textrm{th}b}}{(v_{\textrm{th}a}^2+v_{\textrm{th}b}^2)^{3/2}}.
\end{equation}
The entire result (\ref{eq:Rab}) then can be written as (see for example equations (46)-(47) of \cite{Hinton1983})
\begin{equation}
\boldsymbol{R}_{ab} = \rho_a\nu_{ab}(\bu_b-\bu_a),\label{eq:Simp}
\end{equation}  
where the collisional frequency 
\begin{equation} \label{eq:timeGen}
\nu_{ab}=\tau^{-1}_{ab} = \frac{16}{3}\sqrt{\pi}\frac{n_b e^4 Z_a^2 Z_b^2 \ln\Lambda}{m_a^2(v_{\textrm{th}a}^2+v_{\textrm{th}b}^2)^{3/2}} \Big( 1+\frac{m_a}{m_b}\Big), 
\end{equation}
and the thermal speeds $v_{\textrm{th}a}^2=2T_a/m_a$. Note that $m_an_a\nu_{ab}=m_bn_b\nu_{ba}$ holds.
Collisional frequency (\ref{eq:timeGen}) is identical to equation (C2) of \cite{Schunk1977}; see equation (\ref{eq:timeSchunk}).

 It is useful to clarify the physical meaning of the collisional frequencies. Considering momentum equations for two species where
  all the spatial gradients are neglected, so that $\pr \bu_a/\pr t -(eZ_a/m_a)\bE = \boldsymbol{R}_{ab}/\rho_a$ and
  $\pr \bu_b/\pr t -(eZ_b/m_b)\bE = \boldsymbol{R}_{ba}/\rho_b$, subtracting them and defining difference $\delta\bu=\bu_b-\bu_a$ yields an
evolution equation
\begin{equation} \label{eq:Thierry313}
\frac{\pr \delta\bu}{\pr t} + \nu \delta\bu = e\bE \Big( \frac{Z_b}{m_b}-\frac{Z_a}{m_a}\Big); \qquad \nu=\nu_{ab}+\nu_{ba}. 
\end{equation}
 With no use of Maxwell's equations and instead assuming an applied (external) constant electric field and also constant collisional frequencies,
 an initial velocity difference $\delta\bu(0)$ evolves according to
\begin{equation}
\delta\bu(t) = \delta\bu (0) e^{-\nu t} + (1-e^{-\nu t}) \frac{e\bE}{\nu}\Big( \frac{Z_b}{m_b}-\frac{Z_a}{m_a}\Big).
\end{equation}
Approximately after time $\tau=1/\nu$ (which represents many small-angle collisions) the dependence on the initial
condition disappears and the difference between velocities  reaches a constant value
\begin{equation} \label{eq:Thierry315}
\bu_b -\bu_a =  \frac{e\bE}{\nu_{ab}+\nu_{ba}} \Big( \frac{Z_b}{m_b}-\frac{Z_a}{m_a}\Big) = \textrm{const.}
\end{equation}
 Provided that $Z_a/m_a\neq Z_b/m_b$, the collisional time $\tau=1/(\nu_{ab}+\nu_{ba})$ then can be interpreted as an average time that
  is required for particles ``a'' and ``b'' to experience (many small-angle) collisions, so that the difference between their average fluid velocities
  reaches a constant value proportional to the value of the applied (external) electric field $\bE$. For the particular case of
$Z_a/m_a=Z_b/m_b$, the velocities become equal regardless of the value of applied $\bE$. 

 For a particular case of a one ion-electron plasma $\bu_e-\bu_i = -e\bE/(\nu_{ei}m_e)$, which can be also directly obtained from the quasi-static electron
or ion momentum equations. Prescribing charge neutrality $n_e=Z_i n_i$ so that the current $\boldsymbol{j}=-e n_e (\bu_e-\bu_i)$ then yields 
relation $\boldsymbol{j}=\sigma \bE$ with the usual electrical conductivity $\sigma=1/\eta = e^2 n_e/(\nu_{ei} m_e)$, where $\sigma$ does not depend on the value
of current $j$ (because $j$ is assumed to be small).

\newpage  
\subsection{Energy exchange rates \texorpdfstring{$Q_{ab}$}{Qab}} \label{sec:Qab}
Similar calculations are used to obtain the energy exchange rates $Q_{ab}$, according to (\ref{eq:FP}).
It is beneficial to notice that $\textrm{Tr}\bD_{ab}=(4c_{ab}/m_a^2)H_b$ and so
\begin{equation} \label{eq:Qab}
  Q_{ab} = \frac{2c_{ab}}{m_a}\big(1+\frac{m_a}{m_b}\big) \int f_a \frac{\pr H_b}{\pr\bV}\cdot\bc_a d^3v
  +\frac{2c_{ab}}{m_a}\int f_a H_b d^3v;
\end{equation}  
\begin{equation}
\frac{\pr H_b}{\pr\bV} =\frac{n_b}{v_{\textrm{th} b}^2}\by \Big( \frac{1}{y^2}\frac{2}{\sqrt{\pi}}e^{-y^2} -\frac{1}{y^3}\textrm{erf}(y)\Big),
\end{equation}
and because $\bc_a=\by v_{\textrm{th} b} +\bu v_{\textrm{th} a}$ then
\begin{equation}
  \frac{\pr H_b}{\pr\bV}\cdot\bc_a =\frac{n_b}{v_{\textrm{th} b}^2}\Big(y^2v_{\textrm{th} b} +(\bu\cdot\by)v_{\textrm{th} a}\Big)
  \Big( \frac{1}{y^2}\frac{2}{\sqrt{\pi}}e^{-y^2} -\frac{1}{y^3}\textrm{erf}(y)\Big).
\end{equation}
Importantly, to correctly account for $|\bu|^2$ contributions, the $f_a$ has to be expanded further 
\begin{equation}
  f_a = \frac{n_a}{\pi^{3/2}v_{\textrm{th} a}^3} e^{-|\by\alpha+\bu|^2} \simeq \frac{n_a}{\pi^{3/2}v_{\textrm{th} a}^3} e^{-y^2\alpha^2}
  \Big( 1-(2\alpha \by\cdot\bu +|\bu|^2) +2\alpha^2(\by\cdot\bu)^2 \Big), \label{eq:f0cudo}
\end{equation}  
where $\alpha=v_{\textrm{th} b}/v_{\textrm{th} a}$. This distribution function yields
\begin{eqnarray}
  \int f_a H_b d^3v &=& \frac{2n_a n_b}{\sqrt{\pi} v_{\textrm{th} a}} \Big[ \frac{1}{\sqrt{1+\alpha^2}} -\frac{|\bu|^2}{3(1+\alpha^2)^{3/2}}\Big];\nn\\
  \int f_a \frac{\pr H_b}{\pr\bV}\cdot\bc_a d^3v &=& \frac{2n_a n_b}{\sqrt{\pi} v_{\textrm{th} a}} \Big[ -\frac{1}{(1+\alpha^2)^{3/2}}
    + \frac{|\bu^2|}{(1+\alpha^2)^{5/2}}\Big],
\end{eqnarray}
and the final result reads
\begin{equation} \label{eq:weird}
Q_{ab} = 3\rho_a\nu_{ab} \frac{T_b-T_a}{m_a+m_b}+\rho_a\nu_{ab} \frac{3}{2}\Big( \frac{m_b T_a}{m_b T_a+m_a T_b}-\frac{1}{3}\frac{m_b}{m_b+m_a}\Big)|\bu_b-\bu_a|^2,
\end{equation}
or equivalently
\begin{equation} \label{eq:weirdX}
Q_{ab} = 3\rho_a\nu_{ab} \frac{T_b-T_a}{m_a+m_b}+\rho_a\nu_{ab} \frac{m_b (3 T_a m_a +2 T_a m_b -T_b m_a)}{2 (T_b m_a+T_a m_b)(m_b+m_a)} |\bu_b-\bu_a|^2.
\end{equation}
\cite{Hinton1983} calculates only the first term, the thermal exchange rate (his equation (52);  see also \cite{Landau1936} for an ion-electron plasma).
Calculating $Q_{ab}+Q_{ba}=\rho_a\nu_{ab} |\bu_b-\bu_a|^2=(\bu_b-\bu_a)\cdot\boldsymbol{R}_{ab}$ yields energy conservation and the result
(\ref{eq:weird}) is well-defined. (Re-calculating $\boldsymbol{R}_{ab}$ with the further expanded $f_0$ (\ref{eq:f0cudo}) yields unchanged
result $\boldsymbol{R}_{ab} = \rho_a\nu_{ab}(\bu_b-\bu_a)$).
As a double check, expanding the more general expression for unrestricted drifts (\ref{eq:QabGen}) (by expansion $\Psi_{ab}=1-\epsilon^2$)
yields
\begin{equation} \label{eq:weird2}
  Q_{ab} = 3\rho_a\nu_{ab} \frac{T_b-T_a}{m_a+m_b} \Big(1-\frac{|\bu_b-\bu_a|^2}{\frac{2T_a}{m_a}+\frac{2T_b}{m_b}}\Big)
  +\rho_a\nu_{ab}\frac{m_b}{m_b+m_a} |\bu_b-\bu_a|^2.
\end{equation}
Results (\ref{eq:weird2}) and (\ref{eq:weird}) are equivalent, and valid for an unrestricted difference in temperature.      
After prescribing that the difference in temperatures is small simplifies the frictional part into
\begin{equation} 
  Q_{ab} = 3\rho_a\nu_{ab}\frac{T_b-T_a}{m_a+m_b}+\rho_a \nu_{ab} \frac{m_b}{m_a+m_b}|\bu_b-\bu_a|^2.
\end{equation}
This frictional part is derived elegantly in the Appendix of \cite{Braginskii1965}.

\newpage
\subsection{\texorpdfstring{$R_{ab}$}{Rab} and \texorpdfstring{$Q_{ab}$}{Qab} for unrestricted drifts \texorpdfstring{$\bu_b-\bu_a$}{ub-ua} (runaway effect)} \label{sec:runaway}
\begin{figure*}[!htpb]
  \includegraphics[width=0.42\linewidth]{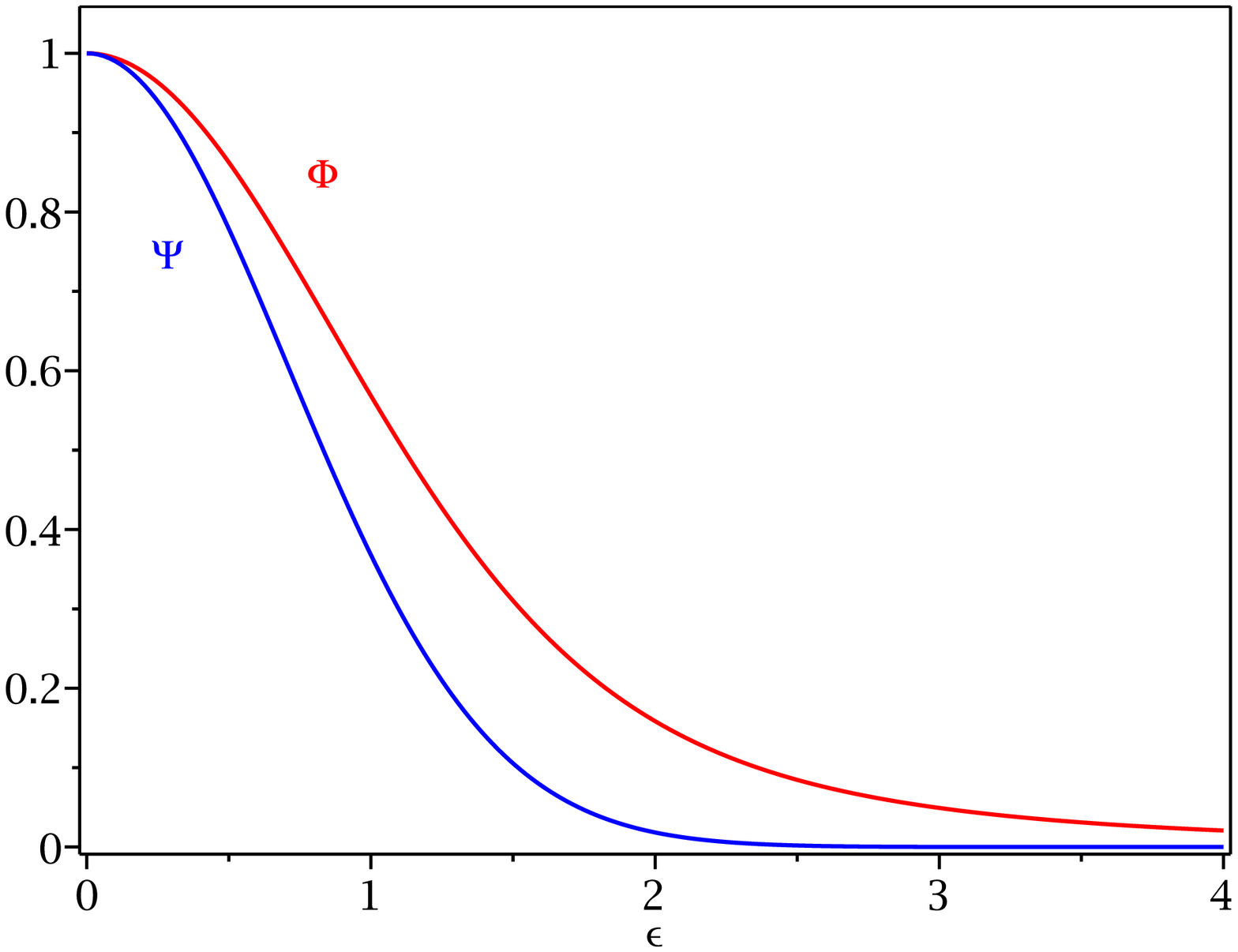}\hspace{0.1\textwidth}\includegraphics[width=0.42\linewidth]{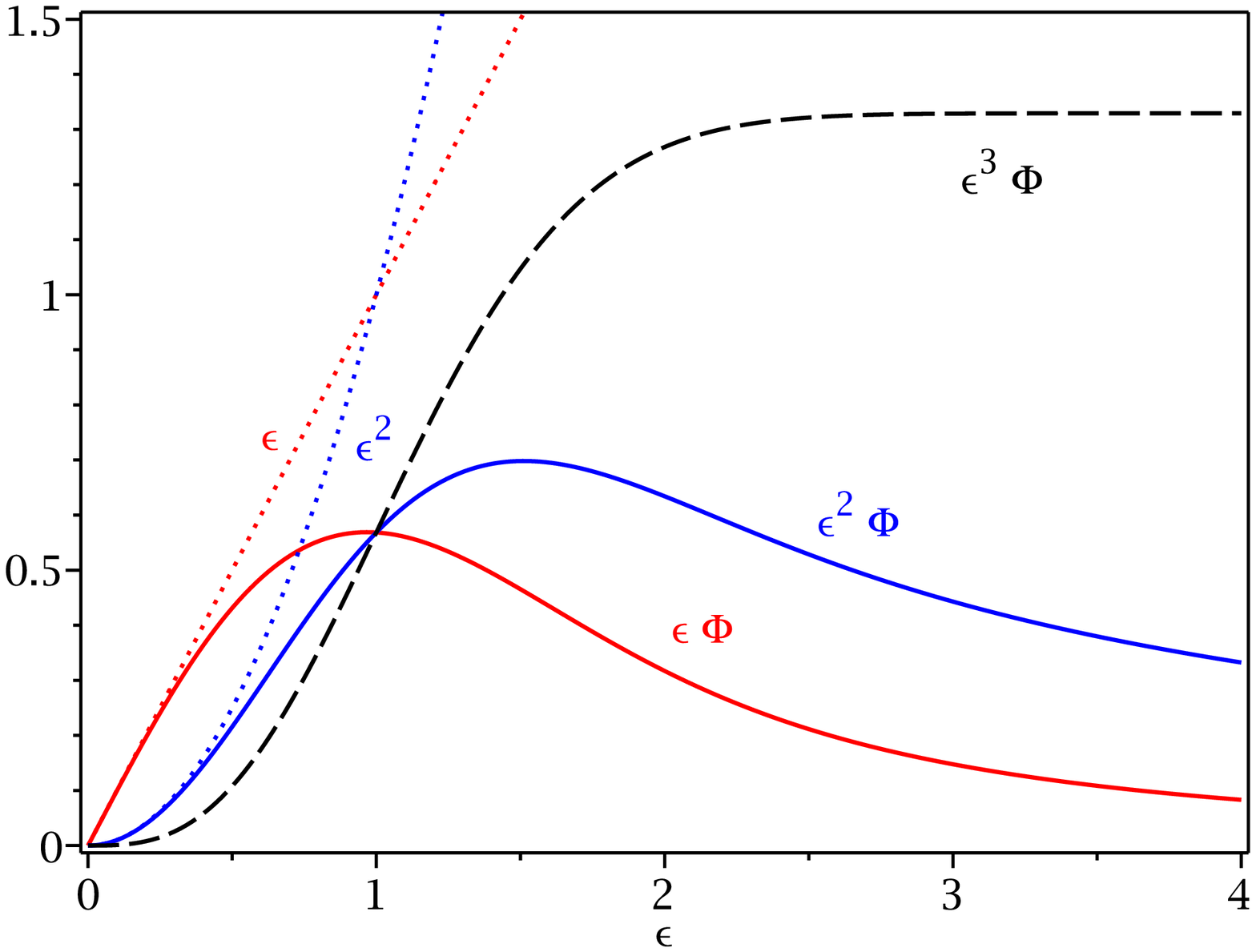}  
  \caption{Left panel: functions $\Phi_{ab}$ (red line) and $\Psi_{ab}$ (blue line), with respect to $\epsilon$ defined in equation (\ref{eq:QabGen}).
    Right panel: functions $\epsilon\Phi_{ab}\sim R_{ab}$ (red line) and $\epsilon^2\Phi_{ab}\sim Q_{ab}$ (blue line), where temperature is fixed.
    Corresponding approximations for small drifts with $\Phi_{ab}=1$ are also plotted (dotted lines).
    Function $\epsilon\Phi_{ab}$ reaches maximum 0.57 at $\epsilon=0.97$, and function $\epsilon^2\Phi_{ab}$ reaches maximum 0.70 at $\epsilon=1.51$.
    It is possible to conclude that the small drift approximation is reasonably accurate up to $\epsilon=0.5$, and
    that very small values $\epsilon\ll 1$ are actually not required.  Even though we did not calculate the runaway effect for higher order moments,
      out of curiosity we include a function $\epsilon^3\Phi_{ab}$ (black dashed line), which does not decrease to zero for large drifts but instead
      converges to a constant value 1.33.} \label{fig:Phi}
\end{figure*}

Here we want to calculate $\boldsymbol{R}_{ab}$ for a general Maxwellian distributions $f_a, f_b$, with no restriction
for the value of difference $\bu_b-\bu_a$. We follow \cite{Burgers1969} and \cite{Tanenbaum1967}. 
Instead of using the Rosenbluth potential $H_b$ and calculating (\ref{eq:Rab1}), it is easier to consider 
\begin{equation} \label{eq:intB}
\boldsymbol{R}_{ab} = 2\frac{c_{ab}}{m_a}\big(1+\frac{m_a}{m_b}\big) \int\int f_a(\bV)f_b(\bV') \frac{\bV'-\bV}{|\bV'-\bV|^3} d^3v d^3v'. 
\end{equation}
Additionally, instead of $\bV$ and $\bV'$, it feels more natural to use $\bV_a=\bV$ and $\bV_b=\bV'$.
It is useful to introduce vectors $\bx=\bV_b-\bV_a$ and $\bu=\bu_b-\bu_a$. The integral is then calculated by introducing ``center-of-mass'' velocity
\begin{eqnarray}
  \boldsymbol{C} = \frac{m_a \bV_a+m_b \bV_b}{m_a+m_b} -\frac{m_a\bu_a+m_b\bu_b}{m_a+m_b}
  +\frac{m_am_b}{(m_a+m_b)} \, \frac{T_b-T_a}{(m_b T_a +m_a T_b)}(\bu-\bx),
\end{eqnarray}
which transforms
\begin{equation}
f_a f_b = \frac{n_a n_b}{\pi^3 v_{\textrm{th} a}^3v_{\textrm{th} b}^3} \exp\Big( -\frac{|\bV_a-\bu_a|^2}{v_{\textrm{th} a}^2} -\frac{|\bV_b-\bu_b|^2}{v_{\textrm{th} b}^2}\Big), 
\end{equation}
into
\begin{eqnarray}
  f_a f_b &=& \frac{n_a n_b}{\pi^3 \tilde{\alpha}^3 \beta^3}\exp\Big( -\frac{|\boldsymbol{C}|^2}{\tilde{\alpha}^2} -\frac{|\bx-\bu|^2}{\beta^2}\Big),
\end{eqnarray}
with new thermal speeds
\begin{equation}
 \tilde{\alpha}^2 = \frac{2T_a T_b}{m_b T_a+m_a T_b}; \qquad \beta^2 = v_{\textrm{th} a}^2+v_{\textrm{th} b}^2. 
\end{equation}  
Importantly, $d^3v_a d^3v_b = d^3 C d^3 x$ (by calculating Jacobian).
For later calculations of more complicated integrals than (\ref{eq:intB}), useful transformations are
\begin{eqnarray}
  \bc_a &=& \boldsymbol{C} -\frac{m_b T_a}{m_bT_a+m_a T_b}(\bx-\bu);\nn\\
  \bc_b &=& \boldsymbol{C} +\frac{m_a T_b}{m_bT_a+m_a T_b}(\bx-\bu). \label{eq:Pic}
\end{eqnarray}  
The integral (\ref{eq:intB}) thus transforms into
\begin{equation} \label{eq:gpic2}
  \boldsymbol{R}_{ab} = 2\frac{c_{ab}}{m_a}\big(1+\frac{m_a}{m_b}\big)
  \frac{n_a n_b}{\pi^{3/2} \beta^3} \int  \frac{\bx}{x^3} e^{ -\frac{|\bx-\bu|^2}{\beta^2} } d^3x, 
\end{equation}
where we have already integrated over $d^3 C$.  One introduces reference frame in the $\bx$-space with
unit vectors $\hat{\boldsymbol{e}}_1, \hat{\boldsymbol{e}}_2, \hat{\boldsymbol{e}}_3$,
where the direction of vector
$\bu$ defines the axis $\hat{\boldsymbol{e}}_3=\bu/u$, so that
\begin{equation} \label{eq:gpic}
\bx = x \sin\theta \cos\phi \hat{\boldsymbol{e}}_1 + x\sin\theta\sin\phi \hat{\boldsymbol{e}}_2 + x\cos\theta \hat{\boldsymbol{e}}_3. 
\end{equation}
For example, integration of (\ref{eq:gpic}) over $\phi$ yields $\int_0^{2\pi} \bx d\phi = 2\pi x\cos\theta \hat{\boldsymbol{e}}_3$,
i.e. the result is in the direction of $\bu$. Furthermore, because $|\bx-\bu|^2=x^2+u^2-2xu\cos\theta$, the integration of (\ref{eq:gpic2}) 
over $\phi$ can be carried out, yielding
\begin{eqnarray}
  \int  \frac{\bx}{x^3} e^{ -\frac{|\bx-\bu|^2}{\beta^2}} d^3x = \frac{\bu}{u} 2\pi
  \int_0^\infty \int_0^\pi e^{ -\frac{|\bx-\bu|^2}{\beta^2}} \cos\theta \sin\theta dxd\theta.
\end{eqnarray}
To calculate that integral, it is useful to introduce (constant) $\epsilon=u/\beta$, and change the integration into variables
\begin{equation}
z=\frac{x}{\beta}-s; \qquad s=\epsilon \cos\theta,
\end{equation}
so that $|\bx-\bu|^2/\beta^2 = z^2-s^2+\epsilon^2$, yielding
\begin{eqnarray}
  \int  \frac{\bx}{x^3} e^{ -\frac{|\bx-\bu|^2}{\beta^2}} d^3x &=& \bu 2\pi
  \frac{e^{-\epsilon^2}}{\epsilon^3} \int_{-s}^\infty \int_{-\epsilon}^\epsilon s e^{-z^2+s^2} dz ds \nn\\
 &=&  \bu \pi^{3/2}
  \Big( \frac{\textrm{erf}(\epsilon)}{\epsilon^3}-\frac{2}{\sqrt{\pi}} \frac{e^{-\epsilon^2}}{\epsilon^2} \Big).
\end{eqnarray}
In the last integral, it is necessary to first integrate over $dz$ and then over $ds$, by using
\begin{eqnarray}
  \int_{-s}^\infty e^{-z^2} dz &=& \frac{\sqrt{\pi}}{2}\big( 1+\textrm{erf}(s) \big);\nn\\
  \int_{-\epsilon}^\epsilon s e^{s^2} \textrm{erf}(s)ds &=& e^{\epsilon^2}\textrm{erf}(\epsilon)-\frac{2}{\sqrt{\pi}}\epsilon.
\end{eqnarray}
The final result then reads
\begin{eqnarray}
  \boldsymbol{R}_{ab} &=& \rho_a\nu_{ab}(\bu_b-\bu_a) \Phi_{ab}; \label{eq:Rabx}\\
  \Phi_{ab} &=& \Big( \frac{3}{4}\sqrt{\pi}\frac{\textrm{erf}(\epsilon)}{\epsilon^3}-\frac{3}{2}\frac{e^{-\epsilon^2}}{\epsilon^2}\Big);
  \qquad \epsilon = \frac{|\bu_b-\bu_a|}{\sqrt{v_{\textrm{th} a}^2+v_{\textrm{th} b}^2}}, \label{eq:Thierry310}
\end{eqnarray}
recovering equation (26.4) of \cite{Burgers1969} and equation (25b) of \cite{Schunk1977}.
For small values $\epsilon\to 0$, the contribution $\Phi\to 1$ (more precisely $\Phi_{ab}=1-(3/5)\epsilon^2$), recovering the previous
result (\ref{eq:Simp}) with small drifts. However, for large values $\epsilon\gg 1$, the contribution $\Phi_{ab}$ decreases to zero
as $\Phi_{ab}=3\sqrt{\pi}/(4\epsilon^3)$ and thus for large differences in drifts $|\bu_b-\bu_a|$,
momentum exchange rates $\boldsymbol{R}_{ab}$ disappear for Coulomb collisions. The phenomenon is known as
the ``runaway effect'' \citep{Dreicer1959}. It is also possible to write
\begin{equation}
  \Phi_{ab} = \frac{3\sqrt{\pi}}{2\epsilon} \widetilde{G}_{ab}(\epsilon); \quad \textrm{where} \quad
  \widetilde{G}_{ab}(\epsilon)=\frac{\erf(\epsilon)}{2\epsilon^2}-\frac{e^{-\epsilon^2}}{\sqrt{\pi}\epsilon}
  =\frac{\textrm{erf}(\epsilon)-\epsilon \textrm{erf}\,'(\epsilon)}{2\epsilon^2}, \label{eq:Energy60}
\end{equation}  
where $\widetilde{G}_{ab}(\epsilon)$ is called the Chandrasekhar function (we use tilde to differentiate it from the
Rosenbluth potential $G_b$), and (\ref{eq:Rabx}) then becomes
\begin{equation}
  \boldsymbol{R}_{ab} = \frac{3}{2}\sqrt{\pi} \rho_a\nu_{ab}(v_{\textrm{th} a}^2+v_{\textrm{th} b}^2)^{\frac{1}{2}}
  \frac{\bu_b-\bu_a}{|\bu_b-\bu_a|} \widetilde{G}_{ab}(\epsilon).
\end{equation}  
In plasma books (e.g. \cite{HelanderSigmar2002}), the Chandrasekhar function is typically introduced in velocity space as
$\widetilde{G}(v/v_{\textrm{th} b})$, i.e. without drifts and before integration over $d^3v$.
The runaway effect is then explained on a population of electron species, which gets accelerated
by applied external electric field. Because for large
velocities $v$ frictional forces (collisions) decrease as $\widetilde{G}\sim v_{\textrm{th} b}^2/(2v^2)$,
the tail of the distribution function might depart and run away. In this sense, the runaway effect could be viewed as
a purely kinetic effect. Nevertheless, obviously analogous runaway effect exists in a fluid
description (i.e. after integration over $d^3v$), it is just represented through difference in drifts $\bu_b-\bu_a$
(which form a current $\boldsymbol{j}$).  For example, considering a one ion-electron plasma with an electric current $\boldsymbol{j}=-en_e (\bu_e-\bu_i)$,
  taking the electron momentum equation and neglecting for simplicity all the terms except of the external $\bE$ and $\boldsymbol{R}_{ei}$
  (including $\pr \bu_e/\pr t$ which neglects acceleration) yields a relation
\begin{equation}
\bE = \frac{\boldsymbol{R}_{ei}}{e n_e} = \eta \boldsymbol{j}; \qquad \eta = \frac{1}{\sigma} = \frac{\rho_e \nu_{ei}}{e^2 n_e^2} \Phi_{ei}, 
\end{equation}
 which agrees with equation (33.6) of \cite{Burgers1969}. 
The electrical resistivity $\eta$ now contains $\Phi_{ei}$ given by (\ref{eq:Thierry310}) with $\epsilon=j/(e n_e v_{\textrm{th} e})$.
For small values of current $j$, the $\eta$ is independent of $j$. 
The runaway effect means that with increasing current $j$ the electrical resistivity $\eta$ decreases, and for large current $j$ it becomes
$\eta=(3\sqrt{\pi}/4)e n_e \rho_e\nu_{ei} v_{\textrm{th} e}^3/j^3$. In reality the problem is much more complex when the acceleration
is considered, because subtracting two momentum equations, a general difference in velocities $\delta\bu = \bu_b-\bu_a$ now
evolves according to a non-linear differential equation
\begin{equation} \label{eq:Thierry314}
  \frac{\pr \delta\bu}{\pr t} + \nu \Phi_{ab}(\epsilon)\delta\bu = e\bE \Big( \frac{Z_b}{m_b}-\frac{Z_a}{m_a}\Big);\qquad
  \nu=\nu_{ab}+\nu_{ba}, 
\end{equation}
 which does not seem to be solvable analytically. Nevertheless (after studying the solutions for some time), it is possible to
  conclude that there exist two distinct classes of solutions, that are typically separated by the value of applied
  constant electric field $E$ with respect to a critical value $E_{\textrm{crit}}$, where the maximal frictional forces balance the electric forces.
  For $E<E_{\textrm{crit}}$ the solutions converge in time towards
  a situation where $\Phi_{ab}=1$, and one recovers
  evolution equation (\ref{eq:Thierry313}) with static solution (\ref{eq:Thierry315}). In contrast, for $E>E_{\textrm{crit}}$, the solutions evolve in time
  towards a situation with $\Phi_{ab}=0$, which can be shown for example by considering solutions where $\Phi_{ab}(\epsilon)$ is approximated with
  its asymptotic expansion. For very large values of $E$ one can
straightforwardly prescribe $\Phi_{ab}=0$, yielding a (collisionless) solution
\begin{equation}
  \bu_b-\bu_a = e\bE \Big( \frac{Z_b}{m_b}-\frac{Z_a}{m_a}\Big)\, t.
\end{equation}
 Thus, provided that $Z_a/m_a \neq Z_b/m_b$ is true, a stationary solution does not
  exist and the difference in velocities grows in time without bounds, before beam/stream plasma instabilities with associated development
  of turbulence (and in extreme cases eventually relativistic effects) restrict its further growth.
  For the particular case $Z_a/m_a=Z_b/m_b$ the runaway
  effect does not exist, and difference in velocities will converge to zero according to (\ref{eq:Thierry314}).
  The frictional forces $\epsilon \Phi_{ab}(\epsilon)$ are plotted as a red curve in the right panel of Figure \ref{fig:Phi}.
  They reach its maximum value $[\epsilon \Phi_{ab}(\epsilon)]_{\textrm{max}}=0.57$ at $\epsilon=0.97$ (often rounded as $\epsilon=1$).
  The critical electric field $\bE_{\textrm{crit}}$ is determined by making the maximum frictional forces equal 
  to the electric forces, so that (\ref{eq:Thierry314}) becomes $\pr \delta\bu/\pr t=0$, yielding
\begin{equation}
  \bE_{\textrm{crit}} = \underbrace{[\epsilon \Phi_{ab}(\epsilon)]_{\textrm{max}}}_{0.57} \sqrt{v_{\textrm{th} a}^2+v_{\textrm{th} b}^2}\,
  \frac{(\nu_{ab}+\nu_{ba})}{e} \frac{m_a m_b}{(Z_b m_a-Z_a m_b)} \frac{\bu_b-\bu_u}{|\bu_b-\bu_a|}.
\end{equation}
 Alternatively, one might use the Chandrasekhar function where $[\epsilon \Phi_{ab}]_{\textrm{max}} = (3/2)\sqrt{\pi} [\widetilde{G}_{ab}]_{\textrm{max}}$,
and $[\widetilde{G}_{ab}]_{\textrm{max}}=0.214$.
The runaway effect thus exist for 
\begin{eqnarray}
 E>E_{\textrm{crit}} &=& \underbrace{[\widetilde{G}_{ab}(\epsilon)]_{\textrm{max}}}_{0.214} \hat{E}_D;\\
\hat{E}_D &=& \frac{3\sqrt{\pi}}{2}\sqrt{v_{\textrm{th} a}^2+v_{\textrm{th} b}^2}\frac{(\nu_{ab}+\nu_{ba})}{e} \frac{m_a m_b}{|Z_b m_a-Z_a m_b|},  
\end{eqnarray}
 where $\hat{E}_D$ can be viewed as a generalized Dreicer electric field for two species with arbitrary masses, charges and temperatures.  
By further substituting for the collisional frequencies (we take $\ln\lambda$ to be constant)
\begin{equation} 
\hat{E}_D = 8\pi  \frac{(m_a+m_b)}{m_a m_b} \frac{(\rho_a +\rho_b)}{|Z_b m_a-Z_a m_b|}\frac{e^3 Z_a^2 Z_b^2 \ln\lambda}{(v_{\textrm{th} a}^2+v_{\textrm{th} b}^2)},
\end{equation}  
 which for an ion-electron plasma yields the usual Dreicer electric field
\begin{equation} \label{eq:Thierry316}
E_D = \frac{4\pi n_i e^3 Z_i^2 \ln\lambda}{T_e}.
\end{equation}
 In the paper of \cite{Dreicer1959} his reference field is defined as $E_c=E_D/2$, so in his notation the runaway effect exists for $E>0.43 E_c$ instead
  of $E>0.214 E_D$. In the most of recent literature definition (\ref{eq:Thierry316}) is used. It is sometimes incorrectly stated that the runaway effect
  exists for $E$ exceeding $E_D$, whereas the correct value as calculated by Dreicer is almost 5 times smaller.
  Note the dependence of (\ref{eq:Thierry316}) on $T_e$, meaning that for
  any given value of electric field the runaway effect will appear if the temperatures are sufficiently high.
  For $Z_a/m_a=Z_b/m_b$ the $\hat{E}_D$ becomes infinitely large and the runaway effect between these species is not present.
  For an ion-electron plasma the Dreicer electric field is also discussed for example by \cite{Tanenbaum1967} (p. 258) and \cite{Balescu1988} (p. 775).
  We found it useful to consider the situation for two arbitrary (charged) species.

 Similarly to $\boldsymbol{R}_{ab}$, the $Q_{ab}$ is obtained by calculating two integrals in (\ref{eq:Qab}), and the first integral yields
\begin{eqnarray}
  && \int f_a \frac{\pr H_b}{\pr\bV_a}\cdot\bc_a d^3 v_a = \int\int f_a f_b \frac{\bV_b-\bV_a}{|\bV_b-\bV_a|^3}\cdot \bc_a d^3 v_a d^3 v_b\nn\\
  && = -\, \frac{n_a n_b}{\pi^{3/2}\beta^3} \frac{m_b T_a}{m_b T_a+m_a T_b}\int \frac{\bx}{x^3}\cdot(\bx-\bu)
  e^{ -\frac{|\bx-\bu|^2}{\beta^2}} d^3x \nn\\
  && = -\, \frac{2n_a n_b}{\sqrt{\pi}\beta^3}v_{\textrm{th} a}^2 e^{-\epsilon^2},
\end{eqnarray}
where we have used
\begin{eqnarray}
  \int\frac{1}{x}  e^{ -\frac{|\bx-\bu|^2}{\beta^2}} d^3x &=& \pi^{3/2} \beta^2 \frac{\textrm{erf}(\epsilon)}{\epsilon};\\
  \int \frac{\bx}{x}\cdot(\bx-\bu)
  e^{ -\frac{|\bx-\bu|^2}{\beta^2}} d^3x &=& 2\pi \beta^2 e^{-\epsilon^2}.
\end{eqnarray}
The second integral in (\ref{eq:Qab}) yields 
\begin{eqnarray}
  \int f_a H_b d^3v_a = \int\int f_a f_b \frac{1}{|\bV_b-\bV_a|} d^3v_a d^3 v_b = \frac{n_a n_b}{\beta} \frac{\textrm{erf}(\epsilon)}{\epsilon}
  =\frac{n_a n_b}{\beta^3} \frac{\textrm{erf}(\epsilon)}{\epsilon^3}|\bu|^2.
\end{eqnarray}
The entire equation (\ref{eq:Qab}) then becomes
\begin{equation}
  Q_{ab} = \rho_a \nu_{ab} \Big[ -\frac{3T_a}{m_a} e^{-\epsilon^2}+\frac{3}{4}\sqrt{\pi}
    \frac{\textrm{erf}(\epsilon)}{\epsilon^3} \frac{m_b}{m_a+m_b}|\bu_b-\bu_a|^2\Big],
\end{equation}
and the difference in temperatures $T_b-T_a$ is not directly visible. Nevertheless, the solution can be rewritten into
\begin{eqnarray}
Q_{ab} &=& \rho_a\nu_{ab}\Big[ 3\frac{T_b-T_a}{m_a+m_b} \Psi_{ab} + \frac{m_b}{m_a+m_b}|\bu_b-\bu_a|^2 \Phi_{ab}\Big]; \label{eq:QabGen}\\
 \Psi_{ab}&=& e^{-\epsilon^2};\qquad \Phi_{ab} = \Big( \frac{3}{4}\sqrt{\pi}\frac{\textrm{erf}(\epsilon)}{\epsilon^3}-\frac{3}{2}\frac{e^{-\epsilon^2}}{\epsilon^2}\Big);
  \qquad \epsilon = \frac{|\bu_b-\bu_a|}{\sqrt{v_{\textrm{th} a}^2+v_{\textrm{th} b}^2}},\nn
\end{eqnarray}
recovering equation (26.8) of \cite{Burgers1969} and equation (25c) of \cite{Schunk1977}. Similarly to $R_{ab}$, for large differences in drifts, the $Q_{ab}$ disappears. 

It is of interest to explore the validity of results with small drifts, obtained in Sections \ref{sec:Rab} and \ref{sec:Qab}.
The functions $\Phi_{ab}$ and $\Psi_{ab}$ are plotted in the left panel of Figure \ref{fig:Phi}. Both functions are decreasing,
and thus in fluid models with the small drift approximation the effects of collisions are overestimated. 
We fix the temperature (so that $\nu_{ab}=\textrm{const.}$), and in the right panel of Figure \ref{fig:Phi} we
plot function $\epsilon\Phi_{ab}$ which corresponds to $R_{ab}$ (red line), and function $\epsilon^2\Phi_{ab}$ which corresponds to $Q_{ab}$ (blue line).
For large drifts $\epsilon\gg 1$ functions $\epsilon\Phi_{ab}\sim 3\sqrt{\pi}/(4\epsilon^2)$ and
$\epsilon^2\Phi_{ab}\sim 3\sqrt{\pi}/(4\epsilon)$.



\newpage
\subsection{Difficulties with Rosenbluth potentials} \label{sec:difficulties}
It is interesting to analyze, why it seems impossible to calculate the run-away effect for $\boldsymbol{R}_{ab}$
through the Rosenbluth potentials, and why one needs to use the ``center-of-mass'' transformation instead.
An attempt to calculate the run-away effect yields
\begin{eqnarray}
  \boldsymbol{R}_{ab} &=& m_a \int f_a \boldsymbol{A}_{ab} d^3v \nn\\
  &=& -4\frac{c_{ab}}{m_a}\big(1+\frac{m_a}{m_b}\big) \frac{n_a n_b v_{\textrm{th}b}}{\pi^{3/2} v_{\textrm{th}a}^3}
  \int e^{-|\alpha\by+\bu|^2} \frac{\by}{y}\underbrace{\Big(\frac{\erf(y)}{2y^2}- \frac{1}{\sqrt{\pi}}\frac{e^{-y^2}}{y}\Big)}_{\widetilde{G}_{ab}(y)} d^3y,
\end{eqnarray}
where $\alpha=v_{\textrm{th}b}/v_{\textrm{th}a}$ and $\bu=(\bu_b-\bu_a)/v_{\textrm{th}a}$, and we have
also identified the Chandrasekhar function.
First integrating over $d\phi$ where the direction of $\bu$ forms the axis $\hat{\boldsymbol{e}}_3=\bu/u$ yields
\begin{eqnarray}
&&  \int e^{-|\alpha\by+\bu|^2} \frac{\by}{y}  \Big(\frac{\erf(y)}{2y^2}- \frac{1}{\sqrt{\pi}}\frac{e^{-y^2}}{y}\Big) d^3y \nn\\
  &&  = 2\pi \frac{\bu}{u} e^{-u^2} \int_0^\infty \int_0^\pi e^{-\alpha^2 y^2} e^{-2\alpha y u\cos\theta} \cos\theta \sin\theta
   \Big(\frac{\erf(y)}{2}- \frac{y}{\sqrt{\pi}} e^{-y^2} \Big) dy d\theta. \label{eq:Cudo}
\end{eqnarray}
Then one can perform integration over $d\theta$, however, subsequent integration over $dy$ does not seem possible.
Or by attempting first integration over $dy$, by using substitutions $s=u\cos\theta$; $z=\alpha y+s$; so that $|\alpha\by+\bu|^2=z^2-s^2+u^2$ yields 
\begin{equation}
  (\ref{eq:Cudo}) =  2\pi \frac{\bu}{u^3} e^{-u^2} \int_s^\infty \int_{-u}^{u} s e^{+s^2} e^{-z^2}\Big[ \frac{1}{2}\erf\big(\frac{z-s}{\alpha}\big)
    - \frac{z-s}{\alpha\sqrt{\pi}}e^{-\frac{(z-s)^2}{\alpha^2}} \Big] dz ds, \label{eq:Cudo2}
\end{equation}  
and the 1D integrals over $dz$ again appear impossible to calculate. The problem is the ``drift'' ``s'', and also constants $\alpha$.
For example, the following indefinite integral is easily calculated by parts
\begin{equation}
\int e^{-(az+b)^2}\erf(az+b)dz = \frac{\sqrt{\pi}}{4a} \erf^2(az+b),
\end{equation}
but the result is not useful. Obviously, a different approach has to be used
to integrate over the Chandrasekhar function if $f_a^{(0)}$ is a Maxwellian with unrestricted drifts. 

Importantly, from Section \ref{sec:runaway} where the ``center-of-mass'' transformation is used,
we know that the correct answer has to be
\begin{equation}
  \int f_a^{(0)} \frac{\by}{y} \widetilde{G}_{ab}(y) d^3v
   \overset{!}{=} - n_a \frac{v_{\textrm{th}b}^2}{v_{\textrm{th}a}^2+v_{\textrm{th}b}^2} \frac{\bu_b-\bu_a}{|\bu_b-\bu_a|}\widetilde{G}_{ab}(\epsilon),
\end{equation}
where $\by=(\bV-\bu_b)/v_{\textrm{th}b}$; $\epsilon = |\bu_b-\bu_a|/\sqrt{v_{\textrm{th} a}^2+v_{\textrm{th} b}^2}$; $d^3v=v_{\textrm{th}b}^3d^3y$;  
or written in a full form
\begin{eqnarray}
&&  \frac{n_a}{\pi^{3/2}v_{\textrm{th}a}^3} \int e^{-\frac{|\bV-\bu_a|^2}{v_{\textrm{th}a}^2}} \frac{\by}{y}
  \Big( \frac{\erf(y)}{2y^2}-\frac{e^{-y^2}}{\sqrt{\pi}y} \Big) d^3v  \nn\\
&&  \overset{!}{=} - n_a \frac{v_{\textrm{th}b}^2}{v_{\textrm{th}a}^2+v_{\textrm{th}b}^2} \frac{\bu_b-\bu_a}{|\bu_b-\bu_a|}
  \Big(\frac{\erf(\epsilon)}{2\epsilon^2}-\frac{e^{-\epsilon^2}}{\sqrt{\pi}\epsilon}\Big).
\end{eqnarray}
Finally, written in perhaps the prettiest form when not referring to any physical quantities
(i.e. a form suitable for integral tables)
\begin{eqnarray}
&& \int_{-\infty}^{\infty} e^{-|\alpha\by+\bu|^2} \frac{\by}{y}
\Big( \frac{\erf(y)}{2y^2}-\frac{e^{-y^2}}{\sqrt{\pi}y} \Big) d^3y  \nn\\
&& \overset{!}{=} -\,\frac{\pi^{3/2}}{\alpha(1+\alpha^2)}\frac{\bu}{u}\Big( \frac{\erf(\epsilon)}{2\epsilon^2}-\frac{e^{-\epsilon^2}}{\sqrt{\pi}\epsilon} \Big);
\quad \textrm{where}\quad \epsilon=\frac{u}{\sqrt{1+\alpha^2}};\quad \alpha>0. \label{eq:beauty}
\end{eqnarray}
It is remarkable that the integral has such a striking symmetry, even though the integral seems impossible 
to calculate directly,
i.e. the integral ``transfers'' a Chandrasekar function in y-variable to a
Chandrasekar function in $\epsilon$-variable. The result seems well-defined
even for $\alpha<0$, so the restriction is $\alpha\neq 0$ and real (the integral is divergent for $\alpha=0$). Limit $u\to0$ yields zero.
The ``proof'' of (\ref{eq:beauty}) can be viewed as analogous when evaluating the
1D Gaussian integral $\int_{-\infty}^{\infty} e^{-x^2}dx$ through $\int\int e^{-(x^2+y^2)}dxdy$ in polar co-ordinates,
where here instead of integrating over $d^3v$, a trick is used to integrate over $d^3v d^3v'$.

\newpage
\section{8-moment model (heat flux and thermal force)} \label{sec:8momentM}
\setcounter{equation}{0}
To obtain collisional contributions with the heat flux, one uses the following 8-moment distribution function of Grad
\begin{equation}
  f_b (\bV')= \frac{n_b}{\pi^{3/2} v_{\textrm{th}b}^3} e^{-\frac{|\bc_b|^2}{v_{\textrm{th}b}^2}}
  \Big[ 1-\frac{m_b}{T_b p_b}\Big( 1-\frac{m_b |\bc_b|^2}{5T_b}\Big) \vecq_b\cdot\bc_b\Big]. \label{eq:f8}
\end{equation}
Calculations done by \cite{Burgers1969,Schunk1977,Killie2004} were performed by using the ``center-of-mass'' transformation described in Section \ref{sec:runaway}.
Here, to do something slightly different,  we verify the calculations by using the Rosenbluth potentials.
The route through Rosenbluth potentials has a great dis-advantage, that error functions are encountered even if we are interested
only in expressions with small  drift velocities (with respect to thermal velocities). This is because the Rosenbluth
potentials have to be derived with exact (\ref{eq:f8}), and not expanded for small drifts from the beginning. 
Nevertheless, the route has an advantage that it is possible to do a double-check in the middle of calculations, because there are identities that
the Rosenbluth potentials must satisfy. 

\subsection{Rosenbluth potentials}
By using the same variables $\bx=(\bV'-\bV)/v_{\textrm{th} b}$ and $\by=(\bV-\bu_b)/v_{\textrm{th} b}$ as before, so that
$\bc_b=(\bx+\by)v_{\textrm{th}b}$, we need to obtain Rosenbluth potentials
\begin{eqnarray}
  H_b (\bV) &=& \int \frac{f_b(\bV')}{|\bV'-\bV|}d^3v'\nn\\
  &=& \frac{n_b}{\pi^{3/2} v_{\textrm{th}b}} \int \frac{e^{-|\bx+\by|^2}}{x} \Big[ 1-\frac{m_bv_{\textrm{th}b}}{T_b p_b}\vecq_b\cdot(\bx+\by)
    \Big( 1-\frac{2}{5}|\bx+\by|^2\Big)\Big] d^3x;
\end{eqnarray}
\begin{eqnarray}
  G_b (\bV) &=& \int |\bV'-\bV| f_b(\bV') d^3v'\nn\\
  &=& \frac{n_b  v_{\textrm{th}b}}{\pi^{3/2}} \int x e^{-|\bx+\by|^2} \Big[ 1-\frac{m_bv_{\textrm{th}b}}{T_b p_b}\vecq_b\cdot(\bx+\by)
    \Big( 1-\frac{2}{5}|\bx+\by|^2\Big)\Big] d^3x.
\end{eqnarray}
It is possible to calculate the following integrals (directly obtainable with Maple in spherical geometry,
after the vector integrals containing $\bx$ are first integrated by hand over $d\phi$)
\begin{eqnarray}
  \int \frac{1}{x}e^{-|\bx+\by|^2} d^3x &=& \pi^{3/2} \frac{\erf(y)}{y};\\
\int \frac{1}{x}e^{-|\bx+\by|^2} \big(1-\frac{2}{5}|\bx+\by|^2\big) d^3x &=& \frac{2}{5} \pi^{3/2}\frac{\erf(y)}{y} +\frac{2}{5}\pi e^{-y^2};\\ 
\int \frac{\bx}{x}e^{-|\bx+\by|^2} \big(1-\frac{2}{5}|\bx+\by|^2\big) d^3x &=& -\frac{2}{5} \pi^{3/2} \by \frac{\erf(y)}{y},
\end{eqnarray}
and similarly
\begin{eqnarray}
\int x e^{-|\bx+\by|^2} d^3x &=& \pi^{3/2}\Big( y+\frac{1}{2y}\Big)\erf(y) +\pi e^{-y^2};\\  
\int x e^{-|\bx+\by|^2} \big(1-\frac{2}{5}|\bx+\by|^2\big) d^3x &=& \frac{2}{5} \pi^{3/2} y \erf(y) +\frac{2}{5}\pi e^{-y^2};\\ 
\int x\bx e^{-|\bx+\by|^2} \big(1-\frac{2}{5}|\bx+\by|^2\big) d^3x &=& -\frac{2}{5}  \by \Big[ \pi^{3/2}\Big( y+\frac{1}{4y^3} \Big)\erf(y)
  +\pi \Big(1-\frac{1}{2y^2}\Big) e^{-y^2}\Big].
\end{eqnarray}
This yields the final Rosenbluth potentials for the 8-moment model, in the following form
\begin{eqnarray}
  H_b (\bV) &=& \frac{n_b}{v_{\textrm{th}b}} \Big[ \frac{\erf(y)}{y} -\frac{2}{5}\frac{m_b v_{\textrm{th}b}}{T_b p_b}(\vecq_b\cdot\by) \frac{1}{\sqrt{\pi}}e^{-y^2}\Big];\\
  G_b (\bV) &=& n_b  v_{\textrm{th}b}\Big[ \Big( y+\frac{1}{2y}\Big)\erf(y) + \frac{1}{\sqrt{\pi}}e^{-y^2} \Big] \nn\\
  && + \frac{2}{5}\frac{n_b}{p_b} (\vecq_b\cdot\by) \Big[ \frac{\erf(y)}{2y^3}-\frac{1}{\sqrt{\pi}} \frac{e^{-y^2}}{y^2}\Big].
\end{eqnarray}
We will need a vector
\begin{eqnarray}
  \frac{\pr H_b}{\pr\bV} &=& \frac{n_b}{v_{\textrm{th}b}^2} \Big[ \by\Big( \frac{2}{\sqrt{\pi}} \frac{e^{-y^2}}{y^2} -\frac{\erf(y)}{y^3}\Big)
   - \frac{2}{5}\frac{m_bv_{\textrm{th}b}}{T_b p_b}  \Big(\vecq_b -2\by (\vecq_b\cdot\by)\Big)\frac{1}{\sqrt{\pi}} e^{-y^2} \Big], \label{eq:derHb}
\end{eqnarray}
and a matrix
\begin{eqnarray}
  \frac{\pr^2 G_b}{\pr \bV \pr \bV} &=& \frac{n_b}{v_{\textrm{th}b}}\Big( \bI-\frac{\by\by}{y^2}\Big) \Big[ \frac{1}{\sqrt{\pi}}\frac{e^{-y^2}}{y^2}
    +\big( \frac{1}{y}-\frac{1}{2y^3}\big)\erf(y)\Big]
  + \frac{n_b}{v_{\textrm{th}b}} \frac{\by\by}{y^2}\Big[ \frac{\erf(y)}{y^3}-\frac{2}{\sqrt{\pi}}\frac{e^{-y^2}}{y^2}\Big]\nn\\
  && +\frac{n_b m_b}{5 T_b p_b} \Big\{ \Big[ \vecq_b\by +\by\vecq_b +(\vecq_b\cdot\by)\Big( \bI-\frac{\by\by}{y^2}\Big)\Big]
  \Big[ -\frac{3}{2}\frac{\erf(y)}{y^5}+\frac{1}{\sqrt{\pi}}\big(\frac{2}{y^2}+ \frac{3}{y^4}\big)e^{-y^2} \Big]\nn\\
  &&  + (\vecq_b\cdot\by)\frac{\by\by}{y^2}\Big[\frac{6}{y^5}\erf(y)-\frac{4}{\sqrt{\pi}}\big(1+\frac{2}{y^2}+\frac{3}{y^4}\big)e^{-y^2} \Big] \Big\}. \label{eq:Diffx}
\end{eqnarray}
As a double check, applying $(\pr/\pr\bV)\cdot$ on (\ref{eq:derHb}) recovers $-4\pi f_b$, and applying 
$(1/2)\textrm{Tr}$ on (\ref{eq:Diffx}) recovers $H_b$. 
The dynamical friction vector then reads 
\begin{eqnarray}
\boldsymbol{A}_{ab} = 2\frac{c_{ab}}{m_a^2}\big(1+\frac{m_a}{m_b}\big) \frac{n_b}{v_{\textrm{th}b}^2} \Big[ \by\Big( \frac{2}{\sqrt{\pi}} \frac{e^{-y^2}}{y^2} -\frac{\erf(y)}{y^3}\Big)
   - \frac{2}{5}\frac{m_bv_{\textrm{th}b}}{T_b p_b}  \Big(\vecq_b -2\by (\vecq_b\cdot\by)\Big)\frac{1}{\sqrt{\pi}} e^{-y^2}\Big],
\end{eqnarray}
and after slight re-arrangement the diffusion tensor becomes
\begin{eqnarray}
  \bD_{ab} &=& 2\frac{c_{ab}}{m_a^2}\Big\{ \frac{n_b}{v_{\textrm{th}b}}\bI \Big[ \frac{1}{\sqrt{\pi}}\frac{e^{-y^2}}{y^2}
    +\big( \frac{1}{y}-\frac{1}{2y^3}\big)\erf(y)\Big]
  + \frac{n_b}{v_{\textrm{th}b}} \frac{\by\by}{y^2}\Big[ \big( \frac{3}{2y^3}-\frac{1}{y}\big)\erf(y)-\frac{3}{\sqrt{\pi}}\frac{e^{-y^2}}{y^2}\Big]\nn\\
  && +\frac{n_b m_b}{5 T_b p_b} \Big[ \vecq_b\by +\by\vecq_b +(\vecq_b\cdot\by)\bI\Big]
  \Big[ -\frac{3}{2}\frac{\erf(y)}{y^5}+\frac{1}{\sqrt{\pi}}\big(\frac{2}{y^2}+ \frac{3}{y^4}\big)e^{-y^2} \Big]\nn\\
  &&  + \frac{n_b m_b}{5 T_b p_b} (\vecq_b\cdot\by)\frac{\by\by}{y^2}\Big[\frac{15}{2}\frac{\erf(y)}{y^5}
  -\frac{1}{\sqrt{\pi}}\big(4+\frac{10}{y^2}+\frac{15}{y^4}\big)e^{-y^2} \Big] \Big\}.
\end{eqnarray}

\subsection{Momentum exchange rates \texorpdfstring{$\boldsymbol{R}_{ab}$}{Rab}}
Then similarly to $f_b$ according to (\ref{eq:f8}), one prescribes for species 'a'
\begin{equation}
  f_a (\bV)= \frac{n_a}{\pi^{3/2} v_{\textrm{th}a}^3} e^{-\frac{|\bc_a|^2}{v_{\textrm{th}a}^2}}
  \Big[ 1-\frac{m_a}{T_a p_a}\Big( 1-\frac{m_a |\bc_a|^2}{5T_a}\Big) \vecq_a\cdot\bc_a\Big],
\end{equation}
and introduces variable $\bu=(\bu_b-\bu_a)/v_{\textrm{th}a}$, so that $\bc_a=\by v_{\textrm{th}b} +\bu v_{\textrm{th}a}$. However, the resulting
integrals would yield the runaway effect, and were never evaluated. It is necessary to get rid of the runaway effect,
and approximate the $f_a$ with small drifts $u\ll 1$, and in the first step
\begin{eqnarray}
  f_a (\bV)
  &\simeq& \frac{n_a}{\pi^{3/2} v_{\textrm{th}a}^3} e^{-\alpha^2 y^2}\Big[ 1-2\alpha (\by\cdot\bu)-u^2+2\alpha^2(\by\cdot\bu)^2 \Big]
  \Big[ 1-\frac{m_a}{T_a p_a} \vecq_a\cdot \big(  \by v_{\textrm{th}b} +\bu v_{\textrm{th}a}   \big)\nn\\
    && \quad + \frac{m_a^2}{5 T_a^2 p_a}\vecq_a\cdot \big(  \by v_{\textrm{th}b} +\bu v_{\textrm{th}a}   \big)
    \big( y^2 v_{\textrm{th}b}^2 + 2\by\cdot\bu v_{\textrm{th}a} v_{\textrm{th}b} +u^2 v_{\textrm{th}a}^2\big) \Big], \label{eq:fa}
\end{eqnarray}
where $\alpha=v_{\textrm{th}b}/v_{\textrm{th}a}$. Distribution function (\ref{eq:fa}) needs to be further reduced
to the ``semi-linear approximation'', where the difference in temperatures is not restricted,
but one keeps only precision $o(\bu)$ and also neglects all the cross-terms such as $\vecq_a\cdot\bu$, keeping only
\begin{eqnarray}
  f_a (\bV)&\simeq& \frac{n_a}{\pi^{3/2} v_{\textrm{th}a}^3} e^{-\alpha^2 y^2}\Big[ 1-2\alpha(\by\cdot\bu)
    -\frac{m_a}{T_a p_a}(\vecq_a\cdot\by)v_{\textrm{th}b}\Big( 1 -\frac{2}{5} \alpha^2 y^2\Big)\Big].
\end{eqnarray}
We want to obtain
\begin{eqnarray}
  \boldsymbol{R}_{ab} = 2\frac{c_{ab}}{m_a}\big(1+\frac{m_a}{m_b}\big)\int f_a \frac{\pr H_b}{\pr\bV} d^3v,
\end{eqnarray}
and we split the calculation to two integrals of (\ref{eq:derHb}). The first integral $\sim\by$
calculates
\begin{eqnarray}
  &&  \int \by\Big( \frac{2}{\sqrt{\pi}} \frac{e^{-y^2}}{y^2} -\frac{\erf(y)}{y^3}\Big) f_a d^3v \nn\\
  && = \frac{n_a v_{\textrm{th}b}^3}{\pi^{3/2} v_{\textrm{th}a}^3 } 4\pi \Big[  \frac{\alpha}{3\alpha^2(1+\alpha^2)^{3/2}}\bu
    + \frac{m_a}{T_a p_a} \vecq_a v_{\textrm{th}b} \frac{1}{10\alpha^2(1+\alpha^2)^{5/2}}\Big],
\end{eqnarray}
where we have used
\begin{eqnarray}
  \int_0^\infty e^{-\alpha^2 y^2} y^4 \Big( \frac{2}{\sqrt{\pi}} \frac{e^{-y^2}}{y^2} -\frac{\erf(y)}{y^3}\Big) dy = - \frac{1}{2\alpha^2(1+\alpha^2)^{3/2}};\\
  \int_0^\infty e^{-\alpha^2 y^2} y^4 \big(1+\frac{2}{5}\alpha^2 y^2\big)\Big( \frac{2}{\sqrt{\pi}} \frac{e^{-y^2}}{y^2} -\frac{\erf(y)}{y^3}\Big) dy
  = - \frac{3}{10\alpha^2(1+\alpha^2)^{5/2}}, 
\end{eqnarray}
and the second part of (\ref{eq:derHb}) calculates
\begin{eqnarray}
 &&  \frac{2}{\sqrt{\pi}} \int e^{-y^2} \Big(\vecq_b -2\by (\vecq_b\cdot\by)\Big) f_a d^3v 
  = \frac{n_a v_{\textrm{th}b}^3}{\pi^{3/2} v_{\textrm{th}a}^3 } \vecq_b 2\pi \frac{\alpha^2}{(1+\alpha^2)^{5/2}}.
\end{eqnarray}
For a quick conversion to collisional frequencies, it is useful to write
\begin{eqnarray}
  \nu_{ab} &=& 
  \frac{8}{3\sqrt{\pi}} \frac{n_b}{v_{\textrm{th}a}^3(1+\alpha^2)^{3/2}} \frac{c_{ab}}{m_a^2}\big( 1+\frac{m_a}{m_b}\big).
\end{eqnarray}  
Putting the results together yields the final result
\begin{equation}
  \boldsymbol{R}_{ab} = \rho_a\nu_{ab}(\bu_b-\bu_a) + \nu_{ab} \frac{3}{5} \frac{\mu_{ab}}{T_{ab}} \Big( \vecq_a-\frac{\rho_a}{\rho_b}\vecq_b\Big), \label{eq:RabHeat}
\end{equation}
recovering equation (41b) of \cite{Schunk1977} (before derived by \cite{Burgers1969}). Alternatively $\mu_{ab}/T_{ab}=2/(v_{\textrm{th}a}^2+v_{\textrm{th}b}^2)$.
As a double check, $\boldsymbol{R}_{ab}=-\boldsymbol{R}_{ba}$ and for self-collisions $\boldsymbol{R}_{aa}=0$, as it should be. The contribution coming from
the heat flux is known as the \emph{thermal force}.

\subsection{Heat flux exchange rates}
To calculate the heat flux contributions, one needs to calculate
\begin{equation}
   \vecQ_{ab}^{(3)}\,' =\frac{\delta \vecq_{ab}\, '}{\delta t} = \frac{1}{2}\textrm{Tr}\bQ_{ab}^{(3)} -\frac{5}{2}\frac{p_a}{\rho_a}\boldsymbol{R}_{ab}
  -\frac{1}{\rho_a}\boldsymbol{R}_{ab}\cdot\bPi_a^{(2)}, 
\end{equation}  
where $\bPi_a^{(2)}=0$ for the 8-moment model (cross-term $\boldsymbol{R}_{ab}\cdot\bPi_a^{(2)}$ would be neglected anyway) and where
\begin{eqnarray}
 \frac{1}{2}\textrm{Tr}\bQ_{ab}^{(3)} &=& m_a \int f_a \Big[ (\boldsymbol{A}_{ab}\cdot\bc_a)\bc_a +\frac{1}{2}\boldsymbol{A}_{ab}|\bc_a|^2\Big] d^3v\nn\\
  && + m_a\int f_a \Big[ \frac{1}{2}(\textrm{Tr}\bD_{ab})\bc_a +\bD_{ab}\cdot\bc_a\Big] d^3v. \label{eq:Sch1}
\end{eqnarray}
We have used $\textrm{Tr}[\boldsymbol{A}\bc\bc]^S=2(\boldsymbol{A}\cdot\bc)\bc+\boldsymbol{A}|\bc|^2$, and because the diffusion tensor is symmetric
$\bD^S=2\bD$, and $\textrm{Tr}[\bD^S\bc]^S=2(\textrm{Tr}\bD)\bc+4\bD\cdot\bc$. By assuming no restriction on the temperature difference,
we have verified (with a great help of Maple) that the ``semi-linear'' heat flux contributions (45)-(49) of \cite{Schunk1977}
(derived before by \cite{Burgers1969}) are indeed correct for Coulomb collisions (with $z_{st}=3/5$, $z_{st}'=13/10$, $z_{st}''=2$ and also
$z_{st}'''=4$).  For Coulomb collisions, the final result (after subtraction of $\frac{5}{2}\frac{p_a}{\rho_a}\boldsymbol{R}_{a}$)
  is written in a compact form in Section \ref{sec:Schunk13mom}; see equation (\ref{eq:Thierry10}). 

In the ``linear approximation''  where the temperature differences are small, the result simplifies into 
\begin{eqnarray}
  \frac{1}{2}\textrm{Tr}\bQ_{ab}^{(3)} = \frac{\delta \vecq_{ab}}{\delta t}
  = \nu_{ab}\Big[-\vecq_a D_{ab (1)} + \vecq_b \frac{\rho_a}{\rho_b} D_{ab (4)} +p_a (\bu_b-\bu_a)\frac{m_b+ \frac{5}{2}m_a}{m_a+m_b}\Big],  
\end{eqnarray}  
where the introduced constants are defined in (\ref{eq:Dab1}), (\ref{eq:Dab2}). Alternatively, by summing over all the 'b' species and
separating the self-collisions
\begin{eqnarray}
  && \frac{1}{2}\textrm{Tr}\bQ_{a}^{(3)}
  =\frac{\delta \vecq_a}{\delta t} = -\frac{4}{5}\nu_{aa}\vecq_a - \sum_{b\neq a} \nu_{ab} \Big[  D_{ab (1)}\vecq_a-  D_{ab (4)}\frac{\rho_a}{\rho_b}\vecq_b
    -p_a (\bu_b-\bu_a)\frac{m_b+ \frac{5}{2}m_a}{m_a+m_b}\Big];\\
  &&  D_{ab (1)} = \frac{1}{(m_a+m_b)^2}\Big( 3m_a^2+\frac{1}{10}m_a m_b -\frac{1}{5}m_b^2\Big); \label{eq:Dab1}\\
  &&  D_{ab (4)} = \frac{1}{(m_a+m_b)^2}\Big(\frac{6}{5}m_b^2 -\frac{3}{2}m_a m_b\Big),\label{eq:Dab2}
\end{eqnarray}
recovering equations (41e)-(43) of \cite{Schunk1977}; see also equations (34)-(36) of \cite{Killie2004}. 
The entire heat flux contributions thus are
\begin{eqnarray}
  \vecQ_{a}^{(3)}\,' &=& \frac{\delta \vecq_{a}\,'}{\delta t} = \frac{1}{2}\textrm{Tr}\bQ_{a}^{(3)} -\frac{5}{2}\frac{p_a}{\rho_a}\boldsymbol{R}_{a} \nn\\
  &=& -\vecq_a \Big[ \frac{4}{5}\nu_{aa} +\sum_{b\neq a}\nu_{ab}\Big( D_{ab (1)} + \frac{3}{2}\frac{p_a}{\rho_a}\frac{\mu_{ab}}{T_{ab}}\Big) \Big]
  +\sum_{b\neq a}\vecq_b \nu_{ab}\frac{\rho_a}{\rho_b} \Big( D_{ab (4)} +\frac{3}{2}\frac{p_a}{\rho_a}\frac{\mu_{ab}}{T_{ab}}\Big)\nn\\
  && -\frac{3}{2}p_a \sum_{b\neq a} \nu_{ab} \frac{m_b}{m_a+m_b} (\bu_b-\bu_a), \label{eq:SchunkGen1}
\end{eqnarray} 
and enter the r.h.s. of evolution equation for the heat flux vector, for example in its simplest form
\begin{equation}
  \frac{d_a\vecq_a}{d t} +\Omega_a\bhat\times\vecq_a + \frac{5}{2}\frac{p_a}{m_a}\nabla T_a
  = \vecQ_{a}^{(3)}\,'.
\end{equation}
Importantly, in comparison to the BGK operator, the r.h.s also contains all the heat fluxes $\vecq_b$. Formally, it is still possible
to obtain a result for $\vecq_a$ in a quasi-static approximation, as a solution of equation
\begin{equation}
  \bhat\times\vecq_a +\frac{\bnu_a}{\Omega_a}\vecq_a  =  - \frac{\vec{\boldsymbol{a}}_a}{\Omega_a},
\end{equation}
where we defined
\begin{eqnarray}
  \bnu_{a} &=& \frac{4}{5}\nu_{aa} +\sum_{b\neq a}\nu_{ab}\Big( D_{ab (1)} + \frac{3}{2}\frac{p_a}{\rho_a}\frac{\mu_{ab}}{T_{ab}}\Big);\\
  \vec{\boldsymbol{a}}_a &=& \frac{5}{2}\frac{p_a}{m_a}\nabla T_a
  -\sum_{b\neq a}\vecq_b \nu_{ab}\frac{\rho_a}{\rho_b} \Big( D_{ab (4)} +\frac{3}{2}\frac{p_a}{\rho_a}\frac{\mu_{ab}}{T_{ab}}\Big)\nn\\
  && \quad +\frac{3}{2}p_a \sum_{b\neq a} \nu_{ab} \frac{m_b}{m_a+m_b} (\bu_b-\bu_a),
\end{eqnarray}
which has the following exact solution
\begin{equation}
  \vecq_a = -\frac{1}{\bnu_a} (\vec{\boldsymbol{a}}_a\cdot\bhat) \bhat
  -\frac{\bnu_a}{\Omega_a^2+\bnu_a^2} \vec{\boldsymbol{a}}_{a \perp}
  +\frac{\Omega_a}{\Omega_a^2+\bnu_a^2}\bhat\times\vec{\boldsymbol{a}}_a. \label{eq:SchunkHF}
\end{equation}
Nevertheless, the heat fluxes of various species are coupled. 

\subsection{One ion-electron plasma} \label{sec:BS-IEplasma}
Considering a one ion-electron plasma (so $n_e=Z_i n_i$) with small differences in temperature, and neglecting ratios $m_e/m_i$
the ion and electron heat fluxes de-couple. 
For the electron species $D_{ei (1)}=-1/5$, $D_{ei(4)}=6/5$, $\mu_{ei}=m_e$ and by using abbreviation $\delta\bu=\bu_e-\bu_i$ then
\begin{eqnarray}
  \boldsymbol{R}_e &=& -\rho_e \nu_{ei}\delta\bu +\nu_{ei} \frac{\rho_e}{p_e}\frac{3}{5}\vecq_e;\label{eq:Resch}\\
  \frac{\delta \vecq_e}{\delta t} &=& - \vecq_e\Big(\frac{4}{5}\nu_{ee}-\frac{1}{5}\nu_{ei} \Big) -\nu_{ei}p_e \delta\bu.
\end{eqnarray}
The entire heat flux contributions are
\begin{eqnarray}
 \vecQ_{e}^{(3)}\,'  &=& -\bnu_e \vecq_e +\frac{3}{2}\nu_{ei}p_e \delta\bu;\\
  \bnu_e &=& \frac{4}{5}\nu_{ee}+\frac{13}{10}\nu_{ei};\label{eq:Weird}\\
  \vec{\boldsymbol{a}}_e &=& \frac{5}{2}\frac{p_e}{m_e}\nabla T_e -\frac{3}{2}\nu_{ei}p_e \delta\bu,
\end{eqnarray}
yielding solution for the electron heat flux (split into thermal part and frictional part)
\begin{eqnarray}
\vecq_e^T &=& -\kappa_\parallel^e \nabla_\parallel T_e - \kappa_\perp^e \nabla_\perp T_e + \kappa_\times^e \bhat\times\nabla T_e; \label{eq:SchHF4}\\
  \vecq_e^u &=& +\frac{3}{2}\frac{\nu_{ei}}{\bnu_e} p_e \delta\bu_\parallel
  +\frac{3}{2}\frac{\bnu_e \nu_{ei}}{\Omega_e^2+\bnu_e^2} p_e \delta\bu_\perp
  -\frac{3}{2} \frac{\Omega_e \nu_{ei}}{\Omega_e^2+\bnu_e^2} p_e \bhat\times \delta\bu, \label{eq:SchFr1}
\end{eqnarray}
with thermal conductivities
\begin{equation}
 \kappa_\parallel^e = \frac{5}{2}\frac{p_e}{\bnu_e m_e}; \qquad
  \kappa_\perp^e = \frac{5}{2}\frac{p_e}{m_e}\frac{\bnu_e}{(\Omega_e^2+\bnu_e^2)}; \qquad
  \kappa_\times^e = \frac{5}{2}\frac{p_e}{m_e}\frac{\Omega_e}{(\Omega_e^2+\bnu_e^2)}. \label{eq:SchHF5}
\end{equation}
The thermal conductivities have the same form as the BGK conductivities. The difference is only that while for the BGK operator
$\bnu_e=\nu_{ee}+\nu_{ei}$, now we have to use (\ref{eq:Weird}). By using $\nu_{ee}=\nu_{ei}/(Z_i\sqrt{2})$  from equation (\ref{eq:Posible1})
\begin{equation}
\bnu_e = \big( \frac{1}{Z_i\sqrt{2}}\frac{4}{5}+\frac{13}{10} \big) \nu_{ei}; \quad \textrm{for}\quad Z_i=1: \quad \bnu_e=1.866 \nu_{ei}. \label{eq:AddS}
\end{equation}

The momentum exchange rates are also split to friction part and thermal part
\begin{eqnarray}
  \boldsymbol{R}_e^u &=& -\rho_e \nu_{ei} \Big[ \Big( 1-\frac{9}{10}\frac{\nu_{ei}}{\bnu_e}\Big)\delta\bu_\parallel
  + \Big( 1-\frac{9}{10}\frac{\bnu_e\nu_{ei}}{\Omega_e^2+\bnu_e^2}\Big)\delta\bu_\perp
  + \frac{9}{10}\frac{\Omega_e\nu_{ei}}{\Omega_e^2+\bnu_e^2}\bhat\times\delta\bu \Big];\\
  \boldsymbol{R}_e^T &=& -\frac{3}{2} \frac{\nu_{ei}}{\bnu_e} n_e \nabla_\parallel T_e
  -\frac{3}{2} \frac{\bnu_e \nu_{ei}}{\Omega_e^2+\bnu_e^2} n_e \nabla_\perp T_e +\frac{3}{2} \frac{\Omega_e\nu_{ei}}{\Omega_e^2+\bnu_e^2} n_e \bhat\times \nabla T_e.
   \label{eq:SchFr2}
\end{eqnarray}
In comparison, the \cite{Braginskii1965} result for $Z_i=1$ reads
\begin{eqnarray}
  \boldsymbol{R}_{e}^u &=&
  -\rho_e \nu_{ei} \Big[0.51 \delta\bu_\parallel + \Big( 1-\frac{6.42x^2+1.84}{x^4+14.79x^2+3.77}\Big) \delta\bu_\perp
    + \frac{x(1.70x^2+0.78)}{x^4+14.79x^2+3.77}\bhat\times\delta\bu \Big]; \nn\\
  \boldsymbol{R}_{e}^T &=& - 0.71 n_e \nabla_\parallel T_e - \frac{5.10x^2+2.68}{x^4+14.79x^2+3.77} n_e \nabla_\perp T_e
    + \frac{x((3/2)x^2+3.05)}{x^4+14.79x^2+3.77} n_e \bhat\times\nabla T_e, \label{eq:BragRx}
\end{eqnarray}
where $x=\Omega_e/\nu_{ei}$. The heat flux and associated thermal force of Burgers and Schunk therefore finally explain
the entire mathematical structure of Braginskii equations, i.e. all the terms are finally present, only the numerical values are different.  

Examining the obtained numerical values, for example in the limit of strong magnetic field with $Z_i=1$
(where for simplicity we neglect all the ratios $\nu_{ei}/\Omega_e$) yields
\begin{eqnarray}
  \boldsymbol{R}_{e} &=& -\rho_e \nu_{ei} ( 0.518\delta\bu_\parallel +\delta\bu_\perp ) -0.80 n_e \nabla_\parallel T_e;\nn\\
   \vecq_e^u &=& +0.80 p_e \delta\bu_\parallel,\label{eq:HFpica1}
\end{eqnarray}
which is very close to Braginskii values
\begin{eqnarray}
  \boldsymbol{R}_{e} &=& -\rho_e \nu_{ei} ( 0.513\delta\bu_\parallel +\delta\bu_\perp ) -0.71 n_e \nabla_\parallel T_e;\nn\\
   \vecq_e^u &=& +0.71 p_e \delta\bu_\parallel. \label{eq:HFpica2}
\end{eqnarray}
Note that both results (\ref{eq:HFpica1}), (\ref{eq:HFpica2}) contain the same symmetrical constants 0.8 and 0.71 in the
frictional heat flux $\vecq_e^u$ and the thermal force $\boldsymbol{R}_{e}^T$. This is known as the Onsager symmetry, and it is also valid
for a general magnetic field strength and a general charge, as can be seen by comparing (\ref{eq:SchFr1}) and (\ref{eq:SchFr2}).

Continuing with the strong magnetic field and examining the perpendicular heat conductivities yields ($Z_i=1$ for $\kappa_\perp^e$)
\begin{equation}
\kappa_\perp^e = 4.66 \frac{p_e\nu_{ei}}{m_e\Omega_e^2}; \quad \kappa_\times^e = \frac{5}{2}\frac{p_e}{m_e\Omega_e},
\end{equation}
and both match Braginskii exactly. Nevertheless, the parallel heat conductivity (which is independent of magnetic field strength; $Z_i=1$)
\begin{equation}
 \kappa_\parallel^e = 1.34\frac{p_e}{\nu_{ei} m_e},
\end{equation}
which is quite low in comparison to the Braginskii value of 3.16.

\subsection*{Ion species}
For ion species $D_{ie (1)}=3$, $D_{ie(4)}=-3m_e/(2m_i)$ and identical proton and electron temperatures,
momentum exchange rates (\ref{eq:RabHeat}) yield
\begin{equation}
\boldsymbol{R}_i = \rho_i \nu_{ie}\delta\bu -\nu_{ie} \frac{\rho_i}{p_e}\frac{3}{5}\vecq_e = -\boldsymbol{R}_e,
\end{equation}
and $\boldsymbol{R}_e$ was already calculated. Furthermore, collisional heat flux contributions (\ref{eq:SchunkGen1})-(\ref{eq:SchunkHF}) simplify into
\begin{eqnarray}
 \vecQ_{i}^{(3)}\,'  &=& -\bnu_i \vecq_i;\\
  \bnu_i &=& \frac{4}{5}\nu_{ii}+3\nu_{ie};\label{eq:Weird2}\\
  \vec{\boldsymbol{a}}_i &=& \frac{5}{2}\frac{p_i}{m_i}\nabla T_i,
\end{eqnarray}
where notably the electron heat flux $\vecq_e$ cancels out exactly for equal temperatures.  
Ion frequencies should be thus added according to
\begin{eqnarray}
  \bnu_i &=& \Big( \frac{4}{5}+3\frac{\sqrt{2}}{Z_i}\sqrt{\frac{m_e}{m_i}} \Big) \nu_{ii}; \quad \textrm{for}\quad Z_i=1: \quad \bnu_i=0.899 \nu_{ii};
  \label{eq:SchunkIonTime}\\
  &=& \frac{4}{5}\Big( 1+\frac{15}{2Z_i} \sqrt{\frac{m_e}{2m_i}}\Big)\nu_{ii}.\nn
\end{eqnarray}
The model of Burgers-Schunk yields ion heat flux
\begin{eqnarray}
\vecq_i &=& -\kappa_\parallel^i \nabla_\parallel T_i - \kappa_\perp^i \nabla_\perp T_i + \kappa_\times^i \bhat\times\nabla T_i, \label{eq:SchunkIon1}
\end{eqnarray}
with ion thermal conductivities
\begin{equation}
 \kappa_\parallel^i = \frac{5}{2}\frac{p_i}{\bnu_i m_i}; \qquad
  \kappa_\perp^i = \frac{5}{2}\frac{p_i}{m_i}\frac{\bnu_i}{(\Omega_i^2+\bnu_i^2)}; \qquad
  \kappa_\times^i = \frac{5}{2}\frac{p_i}{m_i}\frac{\Omega_i}{(\Omega_i^2+\bnu_i^2)}, \label{eq:SchunkIon2}
\end{equation}
where frequencies are added according to (\ref{eq:SchunkIonTime}).
Importantly, the ion-electron contributions are not completely negligible, and without them $\bnu_i = (4/5)\nu_{ii}=0.8\nu_{ii}$. 

However, in the work of \cite{Braginskii1965} the ion-electron collisions are neglected for the
ion heat fluxes and viscosities, and only ion self-collisions are accounted for. This can be seen from his ion coefficients which do
not depend on $Z_i$. Neglecting the ion-electron collisions, the model of Burgers-Schunk yields
\begin{equation}
 \kappa_\parallel^i = \frac{25}{8}\frac{p_i}{\nu_{ii} m_i}; \qquad
  \kappa_\perp^i = 2 \frac{p_i}{m_i}\frac{\nu_{ii}}{\Omega_i^2+(4/5)^2\nu_{ii}^2}; \qquad
  \kappa_\times^i = \frac{5}{2}\frac{p_i}{m_i}\frac{\Omega_i}{\Omega_i^2+(4/5)^2 \nu_{ii}^2}. \label{eq:SchunkIon3}
\end{equation}
For the parallel conductivity $\kappa_\parallel^i\sim 25/8=3.125$, in comparison to Braginskii $3.906$. In the strong magnetic field limit
\begin{equation}
  \kappa_\perp^i = 2 \frac{p_i}{m_i}\frac{\nu_{ii}}{\Omega_i^2}; \qquad
  \kappa_\times^i = \frac{5}{2}\frac{p_i}{m_i\Omega_i}. 
\end{equation}
and both match Braginskii exactly (!).
If ion-electron collisions are taken into account, these Burger-Schunk coefficients change into (for $Z_i=1$)
$\kappa_\parallel^i\sim 2.78$, $\kappa_\perp^i \sim 2.24$ and $\kappa_\times^i\sim 5/2$, and the perpendicular $\kappa_\perp^i$ would suddenly
not match Braginskii. 
It would not make sense that electron $\kappa_\perp^e$ matches Braginskii exactly (for strong B-field) and ion $\kappa_\perp^i$ does not, which
is a definitive indication that ion-electron collisions are neglected in Braginskii. 

Including the ion-electron collisions, the $\kappa_\perp^i$ in the strong B-limit reads
\begin{equation} \label{eq:beauty1}
\kappa_\perp^i = \frac{p_i \nu_{ii}}{m_i\Omega_i^2} \Big( 2+\frac{15}{Z_i}\sqrt{\frac{m_e}{2m_i}}\Big).
\end{equation}
Neglecting ion-electron collisions with respect to ion-ion (self) collisions, is analogous to neglecting
$0.1$ with respect to $0.8$, the contribution is not tiny. 

\newpage
\section{Comparison of various models with Braginskii (electrons)} \label{sec:Comparison}
\setcounter{equation}{0}
Focusing at the parallel direction, the momentum exchange rates $\boldsymbol{R}_{e \parallel}$ and electron heat flux $\vecq_{e\parallel}$
can be written in a general form
\begin{eqnarray}
  \boldsymbol{R}_{e \parallel} &=& -\alpha_0\rho_e\nu_{ei}\delta\bu_\parallel -\beta_0 n_e \nabla_\parallel T_e;\nn\\
  \vecq_{e\parallel} &=& +\beta_0^* p_e \delta\bu_\parallel- \gamma_0 \frac{p_e}{m_e\nu_{ei}} \nabla_\parallel T_e.
\end{eqnarray}
\cite{Braginskii1965} values of $\alpha_0;\beta_0=\beta_0^*$ and $\gamma_0$ are given in his Table 2, page 25. 
The model of \cite{Burgers1969}-\cite{Schunk1977} is given by
\begin{equation}
\alpha_0=1-\frac{9}{10}\frac{\nu_{ei}}{\bnu_e};\quad \beta_0=\beta_0^*=\frac{3}{2}\frac{\nu_{ei}}{\bnu_e}; \quad
  \gamma_0=\frac{5}{2}\frac{\nu_{ei}}{\bnu_e}; \qquad \bnu_e = \Big( \frac{1}{Z_i\sqrt{2}}\frac{4}{5}+\frac{13}{10} \Big) \nu_{ei},
\end{equation}
or equivalently
\begin{equation}
\alpha_0 = \frac{\sqrt{2}+Z_i}{\sqrt{2}+(13/4)Z_i}; \quad \beta_0=\beta_0^* = \frac{15 Z_i}{4\sqrt{2}+13Z_i}; \quad \gamma_0 = \frac{25Z_i}{4\sqrt{2}+13Z_i}.
\end{equation}  
The model of \cite{Killie2004} discussed in Section \ref{sec:Killie} yields
\begin{equation}
\alpha_0=1-\frac{9}{35}\frac{\nu_{ei}}{\bnu_e}; \quad
\beta_0=\frac{3}{7}\frac{\nu_{ei}}{\bnu_e}; \quad \beta_0^*=\frac{3}{2}\frac{\nu_{ei}}{\bnu_e};\quad \gamma_0=\frac{5}{2}\frac{\nu_{ei}}{\bnu_e};
\quad \bnu_e = \Big(\frac{1}{Z_i\sqrt{2}}\frac{16}{35}+\frac{11}{35} \Big) \nu_{ei}.
\end{equation}
Other included models are described bellow.

In Table \ref{eq:TableFr}, we compare the parallel friction force, in Table \ref{eq:TableTr} the parallel thermal force,
in Table \ref{eq:TablePar1} the parallel thermal heat flux (thermal conductivity $\kappa_\parallel^e$), and
in Table \ref{eq:TablePar2} the parallel frictional heat flux.
Furthermore, in Table \ref{eq:TablePerp} we compare $\kappa_\perp^e$ in the strong magnetic field limit.

\begin{table}[ht!]
\centering
\begin{tabular}{| c | c | c | c | c | c | c |}
  \hline
   $\parallel$ friction force $\boldsymbol{R}_e^u$ & $Z_i=1$   & $Z_i=2$     &  $Z_i=3$    &  $Z_i=4$   & $Z_i=16$ & $Z_i=\infty$ \\
  \hline
  Burgers-Schunk $(N=1)$    & 0.518      & 0.431         & 0.395       & 0.376  & 0.326  & 0.308  \\
  Killie et al.             & 0.597      & 0.460         & 0.391       & 0.349  & 0.231  & 0.182 \\
  \hline
  Braginskii $(N=2)$              & 0.513      & 0.4\bf{31}    & 0.39\bf{5}  & 0.375  & 0.319  & 0.2949 \\
  Landshoff $(N=4)$               & 0.50\bf{8} & 0.430         & 0.395       &        &        & 0.29455 \\
  Spitzer-H\"arm $(N=\infty)$     & 0.506      & 0.431         &             & 0.375  & 0.319  & 0.2945 \\
  \hline
\end{tabular}
\caption{Parallel friction force $\boldsymbol{R}_e^u=-\alpha_0\rho_e\nu_{ei}\delta\bu_\parallel$, coefficient $\alpha_0$ is plotted,
  or parallel electrical resistivity $\eta_\parallel=1/\sigma_\parallel=\alpha_0 m_e\nu_{ei}/(e^2 n_e)$. 
  The model of Burgers-Schunk is more precise than Killie et al.. The model of Landshoff for $N=1$ matches Burgers-Schunk and for
  $N=2$ it matches Braginskii. For $Z_i=1$, the value of Landshoff ($N=4$) is slightly corrected ($0.509\to0.508$, emphasized with bold font) from
  the more precise work of \cite{Kaneko1960}, and values of Landshoff for other $Z_i$ might be slightly incorrect.
  Values of Braginskii for $Z_i=2,3$ in his Table II are slightly incorrect, and we used values from 
  analytic expression (\ref{eq:BragCorr}), which now also match Landshoff ($N=2$).
  Braginskii value for $Z_i=16$ is also from (\ref{eq:BragCorr}).
  From \cite{KanekoTaguchi1978,KanekoYamao1980} and \cite{JiHeld2013}, the ``final'' value for $Z_i=1$ is $\alpha_0=0.50612$, and the result of Spitzer-H\"arm is correct.
  Note that by keeping $n_e$ and $T_e$ constant
  in the definition of $\nu_{ei}$, the friction force $\sim \alpha_0\nu_{ei}$ actually increases with increasing $Z_i$
  (and the electrical conductivity decreases).}
\label{eq:TableFr}
\end{table}

\begin{table}[ht!]
\centering
\begin{tabular}{| c | c | c | c | c | c | c |}
  \hline
   $\parallel$ thermal force $\boldsymbol{R}_e^T$ & $Z_i=1$   & $Z_i=2$     &  $Z_i=3$    &  $Z_i=4$   & $Z_i=16$ &  $Z_i=\infty$ \\
  \hline
  Burgers-Schunk     & 0.804      & 0.948  & 1.008  & 1.041  & 1.123  & 1.154  \\
  Killie et al.      & 0.672      & 0.901  & 1.015  & 1.085  & 1.281  & 1.364 \\
  \hline
  Braginskii         & 0.711      & 0.905  & 1.016  & 1.090  & 1.362  & 1.521 \\
  Landshoff (N=4)    & 0.7\bf{09} & 0.904  & 1.016  &        &        & 1.5005 \\
  Spitzer-H\"arm     & 0.703      & 0.908  &        & 1.092  & 1.346  & 3/2  \\
  \hline
\end{tabular}
\caption{Parallel thermal force $\boldsymbol{R}_e^T=-\beta_0 n_e \nabla_\parallel T_e$, coefficient $\beta_0$ is plotted.
  The model of Killie et al. is more precise than Burgers-Schunk. The model of Landshoff for $N=1$ matches Burgers-Schunk, and for
  $N=2$ it matches Braginskii. For $Z_i=1$ the Landshoff ($N=4$) value was slightly corrected ($0.710\to 0.709$) from Kaneko.  
  The final value for $Z_i=1$ from Kaneko et al. and Ji \& Held reads $\beta_0=0.70287$, and the Spitzer-H\"arm result is correct. }
\label{eq:TableTr}
\end{table}

\begin{table}[ht!]
  \centering
\begin{tabular}{| c | c | c | c | c | c | c |}
  \hline
  $\parallel$ heat conductivity $\kappa_\parallel^e$ & $Z_i=1$   & $Z_i=2$     &  $Z_i=3$    &  $Z_i=4$  & $Z_i=16$ & $Z_i=\infty$ \\
  \hline
  Burgers-Schunk     & 1.34      & 1.58  & 1.68  & 1.73    & 1.87   & 1.92  \\
  Killie et al.      & 3.92      & 5.25  & 5.92  & 6.33    & 7.47   & 7.95  \\
  \hline
  Braginskii         & 3.1616    & 4.890  & 6.064  & 6.920 & 10.334  & 12.471 \\
  Landshoff (N=4)    & 3.17\bf{8} & 4.902  & 6.069  &       &         & 13.572 \\
  Spitzer-H\"arm     & 3.203     & 4.960  &       & 6.983  & 10.629  & 13.581 \\ 
  \hline
\end{tabular}
\caption{Parallel electron heat conductivity $\kappa_\parallel^e = \gamma_0 p_e/(m_e\nu_{ei})$ (thermal heat flux $\vecq_e^T=-\kappa_\parallel^e \nabla_\parallel T_e$),
  coefficient $\gamma_0$ is plotted. The model of Killie et al. is a significant improvement over Burgers-Schunk.
  The model of Landshoff for $N=1$ matches Burgers-Schunk and for $N=2$ it approximately matches Braginskii.
  For $Z_i=1$ the Landshoff ($N=4$) value was slightly corrected ($3.175\to 3.178$) from Kaneko.
  The final value for $Z_i=1$ from Kaneko et al. and Ji \& Held reads $\gamma_0=3.2031$, and the Spitzer-H\"arm result is correct.
  Note that by keeping $n_e$ and $T_e$ constant in the definition of $\nu_{ei}$,
  the heat conductivity $\gamma_0/\nu_{ei}$ actually decreases with increasing $Z_i$.}
\label{eq:TablePar1}
\end{table}

\begin{table}[ht!]
  \centering
\begin{tabular}{| c | c | c | c | c | c | c |}
  \hline
  $\parallel$ frictional heat flux $\vecq_e^u$ & $Z_i=1$   & $Z_i=2$     &  $Z_i=3$    &  $Z_i=4$ & $Z_i=16$  & $Z_i=\infty$ \\
  \hline
  Killie et al.      & 2.35      & 3.15    & 3.55  & 3.80  & 4.48   & 4.77  \\
  Spitzer-H\"arm     & 0.699      & 0.888  &       & 1.089 & 1.346  & 3/2 \\
  \hline
\end{tabular}
\caption{Parallel electron frictional heat flux $\vecq_e^u=\beta_0^* p_e \delta\bu_\parallel$. For models of Burgers-Schunk, Braginskii
  and Landshoff the Onsager symmetry
  $\beta_0^*=\beta_0$ holds exactly with values given in Table \ref{eq:TableTr}. For the model of Spitzer-Harm the Onsager symmetry is
  satisfied only approximately, with the largest discrepancy for $Z_i=2$ of around $2\%$.
  For the model of Killie et al. the Onsager symmetry is broken, and the frictional heat flux values are quite large.}
\label{eq:TablePar2}
\end{table}

\begin{table}[ht!]
  \centering
\begin{tabular}{| c | c | c | c | c | c |}
  \hline
  $\perp$ heat conductivity $\kappa_\perp^e$ & $Z_i=1$   & $Z_i=2$     &  $Z_i=3$    &  $Z_i=4$  & $Z_i=\infty$ \\
  \hline
  Burgers-Schunk     & 4.664   & 3.957   & 3.721  & 3.604   & 3.25   \\
  Killie et al.      & 1.59    & 1.19    & 1.06   & 0.99    & 0.79\\
  Braginskii         & 4.664   & 3.957   & 3.721  & 3.604   & 3.25   \\
  \hline
\end{tabular}
\caption{Perpendicular electron heat conductivity $\kappa_\perp^e = \gamma_1' p_e \nu_{ei}/(m_e\Omega_e^2)$,
  in the limit of strong magnetic field (so the conductivity is small), coefficient $\gamma_1'$ is plotted.
  Braginskii values are from his Table II. 
  Interestingly, the Burgers-Schunk model matches Braginskii values exactly. In fact, both models yield
  the same analytic expression $\gamma_1'=(\sqrt{2}/Z_i)+13/4$, see (\ref{eq:BraggCorr1}),
  so the numerical comparison between Burgers-Schunk and Braginskii is a bit meaningless
  (and the reason why $Z_i=16$ value was omitted in our table).
  The table shows that the model of Killie et al. is imprecise.}
\label{eq:TablePerp}
\end{table}

We include the numerical model of \cite{SpitzerHarm1953} (see also \cite{Spitzer1962}), with their notation
discussed in Section \ref{sec:Spitzer}, which reads
\begin{equation}
  \alpha_0=\frac{3\pi}{32\gamma_E};\quad \beta_0=\frac{3}{2}\frac{\gamma_T}{\gamma_E};\quad \beta_0^*=4\frac{\delta_E}{\gamma_E}-\frac{5}{2};\quad
  \gamma_0=\epsilon\delta_T \frac{320}{3\pi},
\end{equation}
with numerical values of $\gamma_E,\gamma_T,\delta_E,\delta_T,\epsilon$ given by Table III in \cite{SpitzerHarm1953}. For Lorentzian plasma
($Z_i=\infty$) the coefficients are $\gamma_E=\gamma_T=\delta_E=\delta_T=1$ and $\epsilon=2/5$.
We also include the model of \cite{Landshoff1949,Landshoff1951}, who calculated several transport coefficients (with the inclusion of magnetic field)
before Spitzer and Braginskii, and studied convergence with increasing Laguerre polynomials from $N=1$ to $N=4$ (in his work $i=N+1$). The model is interesting because
for $N=1$ it matches the values of Burgers-Schunk, and for $N=2$ it matches Braginskii. His model can be figured out to be
\begin{equation}
  \alpha_0 = \frac{1}{Z_i}\big(\triangle_{00}/\triangle \big)^{-1}; \qquad \beta_0=\beta_0^*= \frac{5}{2}\frac{\triangle_{01}/\triangle}{\triangle_{00}/\triangle};
  \qquad \gamma_0 = \frac{25}{4}Z_i\Big(\frac{\triangle_{11}}{\triangle} -\frac{(\triangle_{01}/\triangle)^2}{\triangle_{00}/\triangle}\Big), 
\end{equation}
with coefficients from Table I of \cite{Landshoff1951}. We plot his highest-order model for $N=4$. The models of Landshoff were calculated with higher numerical
precision in the work of \cite{Kaneko1960}, where the following conversion has to be used 
\begin{equation}
  \alpha_0= \frac{1}{e^{\textrm{I}(0)}}; \qquad \beta_0 = \beta_0^*= -\,\frac{5}{2}\frac{b^{\textrm{I}(0)}}{e^{\textrm{I}(0)}};
  \qquad \gamma_0 = \frac{25}{4}\Big[ b^{\textrm{I}(-1)} -\frac{(b^{\textrm{I}(0)})^2}{e^{\textrm{I}(0)}} \Big],
\end{equation}  
with values in his Tables I,II,III. In his work $M=N+1$, and values for models from $N=1$ to $N=5$ are given, even though only for $Z_i=1$.
The model is easily comparable with \cite{Landshoff1951} because the same coefficients are given.
In our comparison tables, we thus slightly correct these $Z_i=1$ values of Landshoff ($N=4$) with the higher precise ones of Kaneko.   
In a later work of \cite{KanekoTaguchi1978,KanekoYamao1980} calculations with up to $M=50$ were made, and
the notation is changed into $b^{\textrm{I}(0)}\to b_{1}^{\textrm{I}(0)}$,
$b^{\textrm{I}(-1)}\to b_{1}^{\textrm{I}(1)}$. From their work and the recent work of \cite{JiHeld2013} who used up to 160 Laguerre polynomials,
the correct values for charge $Z_i=1$ read $\alpha_0=0.50612$; $\beta_0=0.70287$; $\gamma_0=3.2031$.

For the work of \cite{Balescu1988}, who was the first to recover Braginskii with the moment approach of Grad, the following conversion has to be used
\begin{equation}
  \alpha_0 = \frac{1}{\widetilde{\sigma}_\parallel}; \qquad
  \beta_0=\beta_0^*=-\,\sqrt{\frac{5}{2}}\frac{\widetilde{\alpha}_\parallel}{\widetilde{\sigma}_\parallel}; \qquad
  \gamma_0 = \frac{5}{2}\big(\widetilde{\kappa}_\parallel^e -\frac{\widetilde{\alpha}_\parallel^2}{\widetilde{\sigma}_\parallel}\big),
\end{equation}
with numerical values for $Z_i=1$ given on his page 239, Table 4.1. 
For his 13-moment model ($N=1$) the results are equal to Burgers-Schunk, and for his
21-moment model ($N=2$) the results are equal to Braginskii. However, for his 29-moment model ($N=3$) the coefficients of Balescu were
shown to be imprecise by \cite{JiHeld2013}, see their Table I, who were able to exactly pin-point analytic
errors in the collisional matrices of Balescu. That the Balescu $N=3$ values are indeed incorrect can be
quickly double-checked by comparison with the $M=4$ model of \cite{Kaneko1960}, from where the Balescu parameters
should be $\widetilde{\sigma}_\parallel=e^{\textrm{I}(0)}=1.964$; $\widetilde{\alpha}_\parallel=\sqrt{5/2}b^{\textrm{I}(0)}=-0.887$;
$\widetilde{\kappa}_\parallel^e=(5/2)b^{\textrm{I}(-1)}=1.666$, agreeing with the modern calculations of \cite{JiHeld2013}.

\subsection{Notation of Spitzer-H\"arm 1953} \label{sec:Spitzer}
The exact values of parallel transport coefficients (with the exception of parallel viscosity) were first numerically obtained by \cite{SpitzerHarm1953}.
Essentially, the perturbation $\phi_e$ (or $f^{(1)}_e$) around a Maxwellian
$f_e=f_e^{(0)}(1-\phi_e)$ that satisfies the Fokker-Planck equation was found numerically, and the obtained result was used to calculate the transport coefficients.  
No magnetic field is present in their work, and the results can
be interpreted as applying to unmagnetized plasmas, or to magnetized plasmas in the direction parallel to magnetic field lines.
Similarly to Braginskii (Chapter 2 \& 4), the paper treats a one ion-electron plasma (with $n_e=Z_i n_i$).

The notation of \cite{SpitzerHarm1953} can be very confusing. The results are given in a form 
\begin{eqnarray}
  \bj &=& \sigma \bE +\alpha \nabla T_e;\label{eq:SpitzerU1}\\
  \vecq^{\textrm{Spitzer}}_e &=& -\beta \bE -K \nabla T_e, \label{eq:SpitzerU2}
\end{eqnarray}
with coefficients $\sigma,\alpha,\beta,K$ given by their equations (33)-(36). These coefficients contain a quantity
$C^2$. This quantity is only defined by a sentence after equation (16) of their previous paper by \cite{Cohen1950},
which reads ``$C^2$ is the mean square electron velocity'',
meaning $C=\sqrt{3T_e/m_e}$ with the important factor of 3 present (we use the same notation as Braginskii,
with the Boltzmann constant equal to one).
Rewriting their coefficients in (\ref{eq:SpitzerU1}), (\ref{eq:SpitzerU2}) to our notation yields 
\begin{eqnarray}
  \sigma &=& \frac{32}{3\pi}\frac{e^2 n_e}{m_e\nu_{ei}} \gamma_E;\qquad 
  \alpha = \frac{16}{\pi} \frac{e n_e}{m_e \nu_{ei}} \gamma_T;\nn\\
  \beta &=&  \frac{128}{3\pi} \frac{e p_e}{m_e \nu_{ei}} \delta_E; \qquad
  K     = \frac{320}{3\pi} \frac{p_e}{m_e \nu_{ei}} \delta_T, \label{eq:SpitzerC}
\end{eqnarray}
where numerical values of $\gamma_E, \gamma_T, \delta_E, \delta_T$ are given in Table III of \cite{SpitzerHarm1953}.
Coefficients (\ref{eq:SpitzerC}) are essentially normalized with respect to a Lorentzian plasma $Z_i=\infty$ (meaning when
electron-electron collisions are negligible), in which case $\gamma_E=\gamma_T=\delta_E=\delta_T=1$.

Unfortunately, \cite{SpitzerHarm1953} do not define their $\vecq^{\textrm{Spitzer}}_e$ and only describe it as a ``the rate of flow of heat''.
The heat flux is also not defined in the book of \cite{Spitzer1962}, however he notes (equation 5.45), that from the thermodynamics of irreversible processes the model
closely satisfies
\begin{equation} \label{eq:SpitzerX}
\beta = \alpha T_e + \frac{5}{2}\frac{T_e}{e}\sigma.
\end{equation}
Equation (\ref{eq:SpitzerX}) should be the Onsager symmetry. In historical literature, there are three other major possibilities how to
define the heat flux, the first two choices are
\begin{eqnarray}
  \vecq^{\,***}_a &=& \frac{m_a}{2} \int \bV |\bV|^2 f_a d^3v = \vecq_a + \frac{5}{2}p_a\bu_a +\bu_a\cdot\bPi^{(2)}_a + \frac{\rho_a}{2}|\bu_a|^2\bu_a;\\
  \vecq^{\,**}_a &=& \int \bV\Big( \frac{m_a v^2}{2}-\frac{5}{2}T_a\Big) f_a d^3v = \vecq_a +\bu_a\cdot\bPi^{(2)}_a + \frac{\rho_a}{2}|\bu_a|^2\bu_a.
\end{eqnarray}
The nonlinear terms can be neglected. Spitzer is not using the second choice, and the first choice is almost correct, except that for
the electron heat flux only $(5/2)p_e\bu_e$ would be created, and not the whole current $\bu_e-\bu_i$.
The third choice is the definition of \cite{ChapmanCowling1939}, where the heat flux is defined with respect to \emph{average} velocity
of all the species $\langle\bu\rangle\equiv (\sum_a \rho_a\bu_a)/\sum_a\rho_a$ according to 
\begin{eqnarray}  
  \vecq_a^{\,*} &=& \frac{m_a}{2}\int (\bV-\langle\bu\rangle)|\bV-\langle\bu\rangle|^2 f_a d^3v = \vecq_a + \frac{5}{2}p_a \boldsymbol{w}_a
  +\frac{\rho_a}{2}|\boldsymbol{w}_a|^2 \boldsymbol{w}_a+\boldsymbol{w}_a\cdot\bPi_a^{(2)},
\end{eqnarray}
where $\boldsymbol{w}_a=\bu_a-\langle\bu\rangle$. For an ion-electron plasma $\langle\bu\rangle=\bu_i$ and $\boldsymbol{w}_e=\bu_e-\bu_i$.
Thus, to satisfy (\ref{eq:SpitzerX}) the correct interpretation seems to be
\begin{equation}
\vecq^{\textrm{Spitzer}}_e =  \vecq_e^{\,*} =\vecq_e+\frac{5}{2}p_e\delta\bu = \vecq_e -\frac{5}{2}\frac{T_e}{e}\bj,
\end{equation}  
where $\bj=-en_e\delta\bu$ and $\delta\bu=\bu_e-\bu_i$.

Result (\ref{eq:SpitzerU1}) should be viewed as part of the evolution equation for $\pr\bu_e/\pr t$ (here written in a steady state with all other terms neglected),
and substituting the electric field into (\ref{eq:SpitzerU2}) then yields
\begin{eqnarray}
  en_e \bE = \boldsymbol{R}_e &=& \frac{e n_e}{\sigma}\bj -e n_e \frac{\alpha}{\sigma} \nabla T_e;\nn\\
  \vecq_e &=& - \Big(\frac{\beta}{\sigma} -\frac{5}{2}\frac{T_e}{e}\Big)\bj - \epsilon K \nabla T_e; \quad
  \textrm{where}\quad \epsilon=1-\frac{\alpha\beta}{\sigma K}=1-\frac{3}{5}\frac{\delta_E \gamma_T}{\delta_T \gamma_E}.
\end{eqnarray}
The numerical coefficient $\epsilon$ is given in Table III of \cite{SpitzerHarm1953} as well. Or equivalently, by using (\ref{eq:SpitzerC})
\begin{eqnarray}
  \boldsymbol{R}_e &=& - \frac{3\pi}{32\gamma_E}\rho_e\nu_{ei}\delta\bu -\frac{3}{2}\frac{\gamma_T}{\gamma_E} n_e \nabla T_e;\nn\\
  \vecq_e &=& +\Big(4 \frac{\delta_E}{\gamma_E}-\frac{5}{2}\Big) p_e \delta\bu - \epsilon \delta_T \frac{320}{3\pi} \frac{p_e}{m_e \nu_{ei}} \nabla T_e.
\end{eqnarray}
In this form the results can be directly compared to Braginskii, with relations 
\begin{equation}
  \alpha_0=\frac{3\pi}{32\gamma_E};\quad \beta_0=\frac{3}{2}\frac{\gamma_T}{\gamma_E};\quad \beta_0^*=4\frac{\delta_E}{\gamma_E}-\frac{5}{2};\quad
  \gamma_0=\epsilon\delta_T \frac{320}{3\pi}.\nn
\end{equation}
The Onsager symmetry then reads
\begin{equation} \label{eq:SpitzerCheck}
\frac{3}{2}\gamma_T = 4 \delta_E-\frac{5}{2}\gamma_E,
\end{equation}
which the model satisfies approximately, and for the Lorentz case exactly. The largest difference appears for $Z_i=2$,
where the l.h.s of (\ref{eq:SpitzerCheck}) is 0.621 and the r.h.s. is 0.607, so Spitzer's claim that the equation (\ref{eq:SpitzerX})
is satisfied to about 1 part in thousand seems a bit exaggerated, or we are interpreting his results incorrectly. 
The model of \cite{SpitzerHarm1953} and \cite{Spitzer1962} 
is criticized in the monograph of \cite{Balescu1988}, Part 1, p. 266. 
Nevertheless, the coefficients $\alpha_0$, $\beta_0$ and $\gamma_0$ in the model of \cite{SpitzerHarm1953} are the correct answer, and
in comparison with \cite{KanekoTaguchi1978,KanekoYamao1980} or \cite{JiHeld2013}, these coefficients are valid for 3 decimal digits.
For numerical simulations that employ the heat flux of \cite{SpitzerHarm1953}, it seems logical to simply ignore the imprecise $\beta_0^*$ values,
and enforce the Onsager symmetry $\beta_0^*=\beta_0$ in their model by hand.

\subsection{Model of Killie et al. 2004} \label{sec:Killie}
Instead of the 8-moment distribution function of Grad (\ref{eq:f8}) used in the model of Burgers-Schunk,
\cite{Killie2004} argued it is better to use
\begin{equation} \label{eq:Thierry260}
f_a = f_{a}^{(0)} \Big[ 1-\frac{m_a^2 |\bc_a|^2}{5 T_a^2 p_a}\Big( 1-\frac{m_a |\bc_a|^2}{7 T_a}\Big) \vecq_a\cdot\bc_a\Big],
\end{equation}  
yielding collisional contributions (which we did not verify) for small temperature differences
\begin{eqnarray}
  \boldsymbol{R}_a = \rho_a \sum_b \nu_{ab}(\bu_b-\bu_a)
  + \sum_b\nu_{ab} \frac{3}{5}\frac{\mu_{ab}}{T_{ab}}\Big[ \vecq_a\Big( 1-\frac{5}{7}\frac{m_b}{m_a+m_b}\Big)
    -\vecq_b \frac{\rho_a}{\rho_b}\Big( 1-\frac{5}{7}\frac{m_a}{m_a+m_b}\Big) \Big],
\end{eqnarray}
and   
\begin{eqnarray}
  && \frac{1}{2}\textrm{Tr}\bQ_a^{(3)} = \frac{\delta \vecq_a}{\delta t}
  = -\frac{16}{35}\nu_{aa}\vecq_a - \sum_{b\neq a} \nu_{ab} \Big[ D_{ab}^{(1)}\vecq_a-D_{ab}^{(4)}\frac{\rho_a}{\rho_b}\vecq_b
    -p_a (\bu_b-\bu_a)\frac{m_b+\frac{5}{2}m_a}{m_a+m_b}\Big];\\
  && D_{ab}^{(1)} = \frac{1}{(m_a+m_b)^3}\Big( 3m_a^3-\frac{1}{2}m_a^2 m_b -\frac{2}{5}m_am_b^2-\frac{4}{35}m_b^3\Big);\\
  && D_{ab}^{(4)} = \frac{1}{(m_a+m_b)^3}\Big(\frac{6}{5}m_b^3 -\frac{171}{70}m_b^2 m_a -\frac{3}{7}m_b m_a^2\Big).
\end{eqnarray}
Similarly to Burgers-Schunk, they also provide equations for unrestricted temperature differences. 
Considering an ion-electron plasma yields $D_{ei}^{(1)}=-4/35$, $D_{ei}^{(4)}=6/5$ and
\begin{eqnarray}
  \boldsymbol{R}_e &=& -\rho_e \nu_{ei}\delta\bu +\nu_{ei} \frac{\rho_e}{p_e}\frac{6}{35}\vecq_e;\label{eq:Resch2}\\
  \frac{1}{2}\textrm{Tr}\bQ_e^{(3)} &=& \frac{\delta \vecq_e}{\delta t} = - \vecq_e\Big(\frac{16}{35}\nu_{ee}-\frac{4}{35}\nu_{ei} \Big) -\nu_{ei}p_e \delta\bu,
\end{eqnarray}  
with total collisional contributions
\begin{eqnarray}
  \vecQ_e^{(3)}\,' &=& -\bnu_e \vecq_e +\frac{3}{2}\nu_{ei}p_e \delta\bu;\\
  \bnu_e &=& \frac{16}{35}\nu_{ee}+\frac{11}{35}\nu_{ei};\\
  \vec{\boldsymbol{a}}_e &=& \frac{5}{2}\frac{p_e}{m_e}\nabla T_e -\frac{3}{2}\nu_{ei}p_e \delta\bu.
\end{eqnarray}
This yields the heat flux solution equivalent to equation (\ref{eq:SchHF4})-(\ref{eq:SchHF5}), with the only difference 
that the frequencies are now added according to
\begin{equation}
\bnu_e = \Big(\frac{1}{Z_i\sqrt{2}}\frac{16}{35}+\frac{11}{35} \Big) \nu_{ei}; \quad \textrm{for}\quad Z_i=1: \quad \bnu_e=0.6375 \nu_{ei}. \label{eq:AddK}
\end{equation}
The momentum exchange rates then read
\begin{eqnarray}
  \boldsymbol{R}_e^u &=& -\rho_e \nu_{ei} \Big[ \Big( 1-\frac{9}{35}\frac{\nu_{ei}}{\bnu_e}\Big)\delta\bu_\parallel
  + \Big( 1-\frac{9}{35}\frac{\bnu_e\nu_{ei}}{\Omega_e^2+\bnu_e^2}\Big)\delta\bu_\perp
  + \frac{9}{35}\frac{\Omega_e\nu_{ei}}{\Omega_e^2+\bnu_e^2}\bhat\times\delta\bu \Big]; \label{eq:Killie3}\\
  \boldsymbol{R}_e^T &=& -\frac{3}{7} \frac{\nu_{ei}}{\bnu_e} n_e \nabla_\parallel T_e
  -\frac{3}{7} \frac{\bnu_e \nu_{ei}}{\Omega_e^2+\bnu_e^2} n_e \nabla_\perp T_e +\frac{3}{7} \frac{\Omega_e\nu_{ei}}{\Omega_e^2+\bnu_e^2} n_e \bhat\times \nabla T_e,
   \label{eq:Killie2}
\end{eqnarray}
and direct comparison with Braginskii is done according to
\begin{equation}
\alpha_0=1-\frac{9}{35}\frac{\nu_{ei}}{\bnu_e}; \quad
\beta_0=\frac{3}{7}\frac{\nu_{ei}}{\bnu_e}; \quad \beta_0^*=\frac{3}{2}\frac{\nu_{ei}}{\bnu_e};\quad \gamma_0=\frac{5}{2}\frac{\nu_{ei}}{\bnu_e};
\quad \bnu_e = \Big(\frac{1}{Z_i\sqrt{2}}\frac{16}{35}+\frac{11}{35} \Big) \nu_{ei}.
\end{equation}
Examining the numerical values for $Z_i=1$, for example the parallel heat conductivity reads
$ \kappa_\parallel^e = 3.92 p_e/(\nu_{ei} m_e)$. This is a big improvement in the model of \cite{Killie2004}, the conductivity is almost 3 times larger than
the 1.34 value of Burgers-Schunk, and much closer to the correct value 3.20. 
Other results are (strong B-field, $Z_i=1$)
\begin{eqnarray}
  \boldsymbol{R}_{e} &=& -\rho_e \nu_{ei} ( 0.60\delta\bu_\parallel +\delta\bu_\perp ) -0.67 n_e \nabla_\parallel T_e;\nn\\
   \vecq_e^u &=& 2.35 p_e \delta\bu_\parallel,
\end{eqnarray}
and the thermal force value 0.67 is now closer to the correct value 0.70 as well. 
However, the frictional heat flux $\vecq_e^u$ is quite large (over 3 times larger than it should be, 2.35 vs 0.70).
Importantly, the Onsager symmetry between $\vecq_e^u$ and $\boldsymbol{R}_e^T$ is  broken,
which can be also seen from general results (\ref{eq:Killie2}), (\ref{eq:SchFr2}). 
Nevertheless, the model indeed improves the  parallel thermal heat flux and the parallel thermal force of Burgers-Schunk.

\clearpage
\section{10-moment model (viscosity)} \label{sec:10momentM}
\setcounter{equation}{0}
To calculate the collisional contributions for the stress-tensor with the Landau operator,
one uses the following 10-moment distribution function of Grad
\begin{equation}
  f_b (\bV')= \frac{n_b}{\pi^{3/2} v_{\textrm{th}b}^3} e^{-\frac{|\bc_b|^2}{v_{\textrm{th}b}^2}}
  \Big[ 1+\frac{m_b}{2T_bp_b} \bPi^{(2)}_b:\bc_b\bc_b \Big]. \label{eq:Dis10}
\end{equation}
As a reminder $\bPi_b^{(2)}:\bI=0$. By using symmetries and Gaussian integration it is possible to show that
\begin{eqnarray}
  && \int \bc_b \bc_b e^{-\frac{|\bc_b|^2}{v_{\textrm{th}b}^2}} d^3v' = \frac{\pi^{3/2}}{2}v_{\textrm{th}b}^5 \bI;\qquad
     \bPi_b^{(2)}:\int \bc_b\bc_b \bV' e^{-\frac{|\bc_b|^2}{v_{\textrm{th}b}^2}}d^3v' = 0; \nn\\
  && \bPi_b^{(2)}:\int \bc_b \bc_b \bc_b \bc_b e^{-\frac{|\bc_b|^2}{v_{\textrm{th}b}^2}} d^3v'
  = \frac{\pi^{3/2}}{2}v_{\textrm{th}b}^7  \bPi_b^{(2)}.
\end{eqnarray}
The last integral is a special case of (\ref{eq:formula2}).
Thus, the distribution function (\ref{eq:Dis10}) correctly reproduces density, fluid velocity, and full
pressure tensor $m_b\int \bc_b \bc_b f_b d^3c_b = p_b\bI+\bPi_b^{(2)}$, so the distribution function
is well defined.

\subsection{Rosenbluth potentials}
By using variables $\bx=(\bV'-\bV)/v_{\textrm{th} b}$ and $\by=(\bV-\bu_b)/v_{\textrm{th} b}$ 
with $\bc_b=(\bx+\by)v_{\textrm{th}b}$, we need to calculate Rosenbluth potentials
\begin{eqnarray}
  H_b (\bV) &=& \int \frac{f_b(\bV')}{|\bV'-\bV|}d^3v'\nn\\
  &=& \frac{n_b}{\pi^{3/2} v_{\textrm{th}b}} \int \frac{e^{-|\bx+\by|^2}}{x} \Big[ 1+\frac{\bPi^{(2)}_b}{p_b}:(\bx+\by)(\bx+\by)\Big] d^3x;
\end{eqnarray}
\begin{eqnarray}
  G_b (\bV) &=& \int |\bV'-\bV| f_b(\bV') d^3v'\nn\\
  &=& \frac{n_b  v_{\textrm{th}b}}{\pi^{3/2}} \int x e^{-|\bx+\by|^2} \Big[ 1 +\frac{\bPi^{(2)}_b}{p_b}:(\bx+\by)(\bx+\by) \Big] d^3x.
\end{eqnarray}
By using integrals (\ref{eq:H10}) and (\ref{eq:G10}), final results for the Rosenbluth potentials are
\begin{eqnarray}
  H_b &=& \frac{n_b}{v_{\textrm{th}b}}\Big\{ \frac{\erf(y)}{y}
  + \frac{\bPi^{(2)}_b}{p_b} :\by\by \Big[ \erf(y)\frac{3}{4y^5} - \frac{ e^{-y^2}}{\sqrt{\pi}}\Big( \frac{1}{y^2}+\frac{3}{2y^4}\Big)\Big]\Big\};\nn\\
  G_b &=& n_b v_{\textrm{th}b} \Big\{ \frac{e^{-y^2}}{\sqrt{\pi}}+\big(y+\frac{1}{2y}\big)\erf(y)
  +\frac{\bPi_b^{(2)}}{p_b}:\by\by \Big[ -\frac{3}{4\sqrt{\pi}}\frac{e^{-y^2}}{y^4} +\big(-\frac{1}{4y^3}+\frac{3}{8y^5}\big)\erf(y)\Big]\Big\}.
\end{eqnarray}
We will need derivative
\begin{eqnarray}
  \frac{\pr H_b}{\pr\bV} &=& \frac{n_b}{v_{\textrm{th}b}^2} \by \Big( \frac{2}{\sqrt{\pi}}\frac{e^{-y^2}}{y^2}-\frac{\erf(y)}{y^3}\Big) \nn\\
  && +\frac{n_b}{v_{\textrm{th}b}^2} \frac{2}{p_b}(\bPi^{(2)}_b\cdot\by)\Big[ \erf(y)\frac{3}{4y^5}
    - \frac{ e^{-y^2}}{\sqrt{\pi}}\Big( \frac{1}{y^2}+\frac{3}{2y^4}\Big)\Big]\nn\\
  && + \frac{n_b}{v_{\textrm{th}b}^2 p_b} (\bPi_b^{(2)} :\by\by) \by \Big[ \frac{2}{\sqrt{\pi}}\Big( \frac{1}{y^2}+\frac{5}{2y^4}+\frac{15}{4y^6}\Big)e^{-y^2}
    -\frac{15}{4}\frac{\erf(y)}{y^7} \Big],
\end{eqnarray}
As a double check, applying $\pr/\pr\bV\cdot$ at the last expression recovers $-4\pi f_b(\bV)$, where for example
\begin{equation}
  \frac{\pr}{\pr\bV}\cdot (\bPi^{(2)}_b\cdot\by) = 0; \qquad \frac{\pr}{\pr\bV}\cdot \big[(\bPi^{(2)}_b:\by\by)\by\big]
  = \frac{5}{v_{\textrm{th}b}} \bPi_b^{(2)}:\by\by.
\end{equation}
The entire dynamical friction vector for the 10-moment model then becomes
\begin{eqnarray}
  \boldsymbol{A}_{ab} &=& 2\frac{c_{ab}}{m_a^2}\big(1+\frac{m_a}{m_b}\big) \frac{n_b}{v_{\textrm{th}b}^2}\Big\{
  \by \Big( \frac{2}{\sqrt{\pi}}\frac{e^{-y^2}}{y^2}-\frac{\erf(y)}{y^3}\Big) \nn\\
  && +\frac{2}{p_b}(\bPi^{(2)}_b\cdot\by)\Big[ \erf(y)\frac{3}{4y^5}
    - \frac{ e^{-y^2}}{\sqrt{\pi}}\Big( \frac{1}{y^2}+\frac{3}{2y^4}\Big)\Big]\nn\\
  && + \frac{1}{p_b} (\bPi_b^{(2)} :\by\by) \by \Big[ \frac{2}{\sqrt{\pi}}\Big( \frac{1}{y^2}+\frac{5}{2y^4}+\frac{15}{4y^6}\Big)e^{-y^2}
    -\frac{15}{4}\frac{\erf(y)}{y^7} \Big]\Big\}.
\end{eqnarray}

For the diffusion tensor, to perform the subsequent analytic calculations in a clear way, it is useful to write the second Rosenbluth potential
$G_b$ by introducing $A_1$, $A_2$ 
\begin{eqnarray}
G_b &=& n_b v_{\textrm{th}b} \Big[ A_1 + \frac{1}{p_b} \big(\bPi_b^{(2)} :\by\by\big) A_2 \Big],
\end{eqnarray}
where
\begin{eqnarray}
  A_1 &=& \frac{e^{-y^2}}{\sqrt{\pi}} +\big(y+\frac{1}{2y}\big)\erf(y);\nn\\
  A_2 &=& -\frac{3}{4\sqrt{\pi}}\frac{e^{-y^2}}{y^4} +\big(-\frac{1}{4y^3}+\frac{3}{8y^5}\big)\erf(y).
\end{eqnarray}
The required derivatives then are
\begin{eqnarray}
  \frac{\pr G_b}{\pr\bV} = n_b\Big[ \frac{\by}{y}A_1' +\frac{2}{p_b}(\bPi_b^{(2)}\cdot\by)A_2
    + \Big( \frac{ \bPi_b^{(2)}}{p_b}:\by\by\Big) \frac{\by}{y}A_2'\Big],
\end{eqnarray}  
and
\begin{eqnarray}
  \frac{\pr G_b}{\pr\bV \pr\bV} &=& \frac{n_b}{v_{\textrm{th}b}}\Big\{
  \big(\frac{\bI}{y}-\frac{\by\by}{y^3}\big) A_1' +\frac{\by\by}{y^2} A_1''\nn\\
  && + \frac{2}{p_b}\Big[ \frac{\by}{y}(\bPi_b^{(2)}\cdot\by) + (\bPi_b^{(2)}\cdot\by) \frac{\by}{y}\Big] A_2'
  +\frac{2}{p_b} \bPi_b^{(2)} A_2 \nn\\
  && + \Big( \frac{ \bPi_b^{(2)}}{p_b}:\by\by\Big) \Big[  \big(\frac{\bI}{y}-\frac{\by\by}{y^3}\big)A_2' +\frac{\by\by}{y^2} A_2''\Big]\Big\}.
\end{eqnarray}
As a double check, applying $(1/2)\textrm{Tr}$ at the last expression recovers $H_b$.

After a slight re-arangement suitable for calculations, the entire diffusion tensor then becomes
\begin{eqnarray}
  \boldsymbol{D}_{ab} &=& 2\frac{c_{ab}}{m_a^2}\frac{n_b}{v_{\textrm{th}b}}\Big\{
  \bI \frac{A_1'}{y} +\frac{\by\by}{y^2} \big( A_1''-\frac{A_1'}{y}\big)\nn\\
  && + \frac{1}{p_b}\Big[ 2\by(\bPi_b^{(2)}\cdot\by) + 2(\bPi_b^{(2)}\cdot\by)\by +(\bPi_b^{(2)}:\by\by)\bI\Big] \frac{A_2'}{y}\nn\\
  && +\frac{2}{p_b} \bPi_b^{(2)} A_2 
   + \frac{1}{p_b}\big(\bPi_b^{(2)}:\by\by\big) \frac{\by\by}{y^2} \big( A_2''-\frac{A_2'}{y}\big) \Big\},
\end{eqnarray}
with ``coefficients''
\begin{eqnarray}
  \frac{A_1'}{y} &=& \big(\frac{1}{y}-\frac{1}{2y^3}\big)\erf(y)+\frac{1}{\sqrt{\pi}}\frac{e^{-y^2}}{y^2};\nn\\
  A_1''-\frac{A_1'}{y} &=& \big(-\frac{1}{y}+\frac{3}{2y^3}\big)\erf(y)-\frac{3}{\sqrt{\pi}}\frac{e^{-y^2}}{y^2};\nn\\
  A_2' &=& \big(\frac{3}{4y^4}-\frac{15}{8y^6}\big) \erf(y)+\frac{e^{-y^2}}{\sqrt{\pi}}\big(\frac{1}{y^3}+\frac{15}{4y^5}\big);\nn\\
  \frac{A_2'}{y} &=& \big(\frac{3}{4y^5}-\frac{15}{8y^7}\big) \erf(y)+\frac{e^{-y^2}}{\sqrt{\pi}}\big(\frac{1}{y^4}+\frac{15}{4y^6}\big);\nn\\
  A_2'' - \frac{A_2'}{y} &=& \big( -\frac{15}{4y^5}+\frac{105}{8y^7}\big)\erf(y)-\frac{e^{-y^2}}{\sqrt{\pi}}\big(\frac{2}{y^2}+\frac{10}{y^4}+\frac{105}{4y^6}\big).
\end{eqnarray}
Or explicitly in its entire form
\begin{eqnarray}
  \boldsymbol{D}_{ab} &=& 2\frac{c_{ab}}{m_a^2}\frac{n_b}{v_{\textrm{th}b}}\Big\{
  \bI \Big[\big(\frac{1}{y}-\frac{1}{2y^3}\big)\erf(y)+\frac{1}{\sqrt{\pi}}\frac{e^{-y^2}}{y^2}\Big]
   +\frac{\by\by}{y^2} \Big[ \big(-\frac{1}{y}+\frac{3}{2y^3}\big)\erf(y)-\frac{3}{\sqrt{\pi}}\frac{e^{-y^2}}{y^2}\Big]\nn\\
  && + \frac{1}{p_b}\Big[ 2\by(\bPi_b^{(2)}\cdot\by) + 2(\bPi_b^{(2)}\cdot\by)\by + (\bPi_b^{(2)}:\by\by)\bI\Big]
  \Big[  \big(\frac{3}{4y^5}-\frac{15}{8y^7}\big) \erf(y)+\frac{e^{-y^2}}{\sqrt{\pi}}\big(\frac{1}{y^4}+\frac{15}{4y^6}\big)\Big]\nn\\
  && +\frac{2}{p_b} \bPi_b^{(2)} \Big[ -\frac{3}{4\sqrt{\pi}}\frac{e^{-y^2}}{y^4} +\big(-\frac{1}{4y^3}+\frac{3}{8y^5}\big)\erf(y)  \Big]\nn\\
  &&+ \Big( \frac{\bPi_b^{(2)}}{p_b}:\by\by\Big)\frac{\by\by}{y^2}
  \Big[ \big( -\frac{15}{4y^5}+\frac{105}{8y^7}\big)\erf(y)-\frac{e^{-y^2}}{\sqrt{\pi}}\big(\frac{2}{y^2}+\frac{10}{y^4}+\frac{105}{4y^6}\big)\Big]\Big\}.
\end{eqnarray}

\subsection{Viscosity calculation}
For species ``a'', the distribution function in semi-linear approximation reads
\begin{eqnarray}
  f_a (\bV) = \frac{n_a}{\pi^{3/2}v_{\textrm{th}a}^3} e^{-\alpha^2 y^2 }\Big[ 1-2\alpha(\by\cdot\bu)+\frac{\alpha^2}{p_a} \bPi_a^{(2)}:\by\by\Big].
\end{eqnarray}
It can be seen that at the semi-linear level, there is no new contribution to the momentum equation. 
For the pressure tensor equation, we need to calculate the following collisional contributions
\begin{equation} \label{eq:ViscoP}
  \bQ_{ab}^{(2)} = m_a \int f_a \big[ \boldsymbol{A}_{ab}\bc_a\big]^S d^3v +m_a \int f_a \bD_{ab} d^3v,
\end{equation}
where we have used that the diffusion tensor is symmetric.
Starting with the second term, and using the derived formulas (\ref{eq:formula1})-(\ref{eq:formula2}),
integration over the diffusion tensor then yields
\begin{eqnarray}
m_a \int f_a \bD_{ab} d^3v &=& 2\frac{c_{ab}}{m_a^2} \frac{n_b}{v_{\textrm{th}b}} \frac{\rho_a}{\pi^{3/2}}\alpha^3\Big\{ 
 +\bI \frac{4\pi}{3}\int_0^\infty \big( 2A_1' y+A_1'' y^2 \big) e^{-\alpha^2 y^2} dy \nn\\
&& + \frac{\bPi_b^{(2)}}{p_b}8\pi \int_0^\infty \Big[ \frac{3}{5}y^3 A_2' +y^2A_2+\frac{1}{15}y^4 A_2''\Big] e^{-\alpha^2 y^2} dy\nn\\
&& + \frac{\bPi_a^{(2)}}{p_a}\frac{8\pi}{15}\alpha^2 \int_0^\infty y^4\big(A_1''-\frac{A_1'}{y}\big) e^{-\alpha^2 y^2} dy\Big\},
\end{eqnarray}
and further 1D integration brings the following result
\begin{eqnarray}
  m_a \int f_a \bD_{ab} d^3v &=& 2\frac{c_{ab}}{m_a^2} \frac{n_b}{v_{\textrm{th}b}} \frac{\rho_a}{\pi^{3/2}}\alpha^3\Big\{
  +\bI \frac{4\pi}{3} \frac{1}{\alpha^2 \sqrt{1+\alpha^2}}\nn\\
  && \quad -\frac{\bPi_b^{(2)}}{p_b} \frac{4\pi}{15} \frac{1}{(1+\alpha^2)^{3/2}} 
  -\frac{\bPi_a^{(2)}}{p_a} \frac{4\pi}{15} \frac{1}{\alpha^2 (1+\alpha^2)^{3/2}}\Big\}\nn\\
  &=& \rho_a\nu_{ab} \frac{m_b}{m_a+m_b} \Big[ \bI (v_{\textrm{th}a}^2+ v_{\textrm{th}b}^2)
    - \frac{\bPi_b^{(2)}}{p_b} \frac{v_{\textrm{th}b}^2}{5} - \frac{\bPi_a^{(2)}}{p_a} \frac{v_{\textrm{th}a}^2}{5}\Big]. \label{eq:Vis1}
\end{eqnarray}
Similarly, the first term in (\ref{eq:ViscoP}) calculates
\begin{eqnarray}
  m_a \int f_a \big[ \boldsymbol{A}_{ab}\bc_a\big]^S d^3v = \rho_a \nu_{ab} \Big[ -\bI v_{\textrm{th}a}^2
    + \frac{\bPi_b^{(2)}}{p_b} \frac{3}{5} \frac{v_{\textrm{th}b}^2 v_{\textrm{th}a}^2}{(v_{\textrm{th}a}^2+v_{\textrm{th}b}^2)}
    - \frac{\bPi_a^{(2)}}{p_a} \frac{5v_{\textrm{th}b}^2 +2v_{\textrm{th}a}^2}{5(v_{\textrm{th}a}^2+v_{\textrm{th}b}^2)}v_{\textrm{th}a}^2 \Big]. \label{eq:Vis2}
\end{eqnarray}
Adding (\ref{eq:Vis1})+(\ref{eq:Vis2}) yields final collisional contributions for the r.h.s. of the pressure tensor equation,
which can be written in the following convenient form
\begin{eqnarray}
  \bQ_{ab}^{(2)} &=& 2\frac{\rho_a \nu_{ab}}{m_a+m_b}(T_b-T_a)\bI
  -2\frac{m_a\nu_{ab}}{m_a+m_b}\frac{T_b}{T_{ab}}\Big(\bPi_a^{(2)}-\frac{T_an_a}{T_bn_b}\bPi_b^{(2)}\Big)\nn\\
  && -\frac{\nu_{ab}}{m_a+m_b}\Big[ \frac{6}{5}m_b -\frac{4}{5}\mu_{ab}\frac{T_b-T_a}{T_{ab}}\Big]
  \Big(\bPi_a^{(2)}+\frac{\rho_a}{\rho_b}\bPi_b^{(2)}\Big), \label{eq:Pi_noT} 
\end{eqnarray}
with reduced mass and reduced temperature
\begin{equation}
\mu_{ab} = \frac{m_a m_b}{m_a+m_b}; \qquad T_{ab} = \frac{m_a T_b + m_b T_a}{m_a+m_b}.\nn
\end{equation}
Introducing $\sum_b$ over all species, result (\ref{eq:Pi_noT}) identifies with equation (44) of \cite{Schunk1977} (derived before by Burgers).
It is valid in the semi-linear approximation, for unrestricted temperature differences.
For Coulomb collisions, viscosity calculated through the Rosenbluth potentials (for the Landau collisional operator) thus
yields the same result as the Boltzmann collisional operator. 
By explicitly separating the self-collisions
\begin{eqnarray}
  \bQ_a^{(2)} &=& \sum_b \bQ_{ab}^{(2)} = \frac{\delta \bp_a}{\delta t} \nn\\
  &=& -\,\frac{6}{5}\nu_{aa}\bPi_a^{(2)} +\sum_{b\neq a} \Big[ 2\frac{\rho_a \nu_{ab}}{m_a+m_b}(T_b-T_a)\bI
  -2\frac{m_a\nu_{ab}}{m_a+m_b}\frac{T_b}{T_{ab}}\Big(\bPi_a^{(2)}-\frac{T_an_a}{T_bn_b}\bPi_b^{(2)}\Big)\Big]\nn\\
  && -\sum_{b\neq a} \Big[\frac{\nu_{ab}}{m_a+m_b}\Big( \frac{6}{5}m_b -\frac{4}{5}\mu_{ab}\frac{T_b-T_a}{T_{ab}}\Big)
  \Big(\bPi_a^{(2)}+\frac{\rho_a}{\rho_b}\bPi_b^{(2)}\Big)\Big], \label{eq:Pi_noT2} 
\end{eqnarray}
where the ``famous'' $6/5$ constant is present. As a double check, calculating the energy exchange rates yields
\begin{equation}
Q_{ab} = \frac{1}{2}\textrm{Tr}\bQ_{ab}^{(2)} =  3\frac{\rho_a \nu_{ab}}{m_a+m_b}(T_b-T_a),
\end{equation}
as it should be. 

Collisional contributions for the stress-tensor thus are
\begin{eqnarray}
  \bQ_a^{(2)}\,' &=& \frac{\delta \bPi^{(2)}_a}{\delta t}=\bQ_a^{(2)} -\frac{\bI}{3}\textrm{Tr}\bQ_{a}^{(2)} \nn\\
  &=& -\,\frac{6}{5}\nu_{aa}\bPi_a^{(2)} -\sum_{b\neq a} \Big[ 
  2\frac{m_a\nu_{ab}}{m_a+m_b}\frac{T_b}{T_{ab}}\Big(\bPi_a^{(2)}-\frac{T_an_a}{T_bn_b}\bPi_b^{(2)}\Big)\Big]\nn\\
  && -\sum_{b\neq a} \Big[\frac{\nu_{ab}}{m_a+m_b}\Big( \frac{6}{5}m_b -\frac{4}{5}\mu_{ab}\frac{T_b-T_a}{T_{ab}}\Big)
  \Big(\bPi_a^{(2)}+\frac{\rho_a}{\rho_b}\bPi_b^{(2)}\Big)\Big], \label{eq:Pi_noTX}
\end{eqnarray}
and enter the r.h.s of its evolution equation, for example written in its simplest form
\begin{eqnarray}
&&  \frac{d_a\bPi^{(2)}_a}{dt} + \Omega_a \big(\bhat\times \bPi^{(2)}_a \big)^S  
    +p_a \bW_a = \frac{\delta \bPi^{(2)}_a}{\delta t}. \label{eq:Pi_simple2} 
\end{eqnarray}
Importantly, in the collisionless regime the r.h.s of (\ref{eq:Pi_simple2}) simply goes to zero. 
It is possible to write a general solution in quasi-static approximation, but the stress-tensors
of various species are coupled.

\subsection{Small temperature differences}
For a particular case of small temperature differences between species
\begin{eqnarray}
  \bQ_a^{(2)}=  \frac{\delta \bp_a}{\delta t} &=& -\,\frac{6}{5}\nu_{aa}\bPi_a^{(2)}\nn\\
 && -2\sum_{b\neq a} \frac{m_a\nu_{ab}}{m_a+m_b}\Big[ \bp_a-\frac{n_a}{n_b}\bp_b
  +\frac{3}{5} \frac{m_b}{m_a} \Big(\bPi_a^{(2)}+\frac{\rho_a}{\rho_b}\bPi_b^{(2)}\Big)\Big], \label{eq:Pi_noT5x} 
\end{eqnarray}
where one uses $\bp=p\bI+\bPi^{(2)}$, recovering equation (41d) of \cite{Schunk1977}.
Finally, for the stress tensor 
\begin{eqnarray}
  \bQ_a^{(2)}\,' =\frac{\delta \bPi^{(2)}_a}{\delta t} &=& -\,\frac{6}{5}\nu_{aa}\bPi_a^{(2)}
  -2 \sum_{b\neq a} 
  \frac{m_a\nu_{ab}}{m_a+m_b}\Big[ \big(1+\frac{3}{5}\frac{m_b}{m_a}\big)\bPi_a^{(2)}- \frac{2}{5}\frac{n_a}{n_b}\bPi_b^{(2)}\Big].
\end{eqnarray}
  
\subsection{One ion-electron plasma}
For a plasma consisting of one ion species and electrons, in the first step
\begin{eqnarray}
   \bQ_i^{(2)}\,' = \frac{\delta \bPi^{(2)}_i}{\delta t} &=& -\Big( \frac{6}{5}\nu_{ii}+2\nu_{ie}\Big) \bPi^{(2)}_i
  +\frac{4}{5}\nu_{ie} \frac{n_i}{n_e} \bPi^{(2)}_e;\\
   \bQ_e^{(2)}\,' =\frac{\delta \bPi^{(2)}_e}{\delta t} &=& -\frac{6}{5}\big(\nu_{ee}+\nu_{ei}\big)\bPi^{(2)}_e
  + \frac{4}{5}\nu_{ei}\frac{\rho_e}{\rho_i}\bPi^{(2)}_i.  
\end{eqnarray}
Nevertheless, because for example for the parallel viscosity the ion $\bPi^{(2)}_i$ is larger than the electron $\bPi^{(2)}_e$
by a factor of $\sqrt{m_i/m_e}$, the coupling is only weak and the last terms
in the above expressions can be for simplicity neglected. Then,
\begin{eqnarray}
  \bQ_i^{(2)}\,' = \frac{\delta \bPi^{(2)}_i}{\delta t} &=& -\bnu_i \bPi^{(2)}_i; \qquad
  \bnu_i = \frac{6}{5}\nu_{ii}+2\nu_{ie} = \frac{6}{5}\Big( 1+\frac{5}{3}\frac{\sqrt{2}}{Z_i}\sqrt{\frac{m_e}{m_i}} \Big)\nu_{ii}; \label{eq:VisB1}\\
   \bQ_e^{(2)}\,' =\frac{\delta \bPi^{(2)}_e}{\delta t} &=& -\bnu_e \bPi^{(2)}_e; \qquad
  \bnu_e = \frac{6}{5} (\nu_{ee}+\nu_{ei}) = \frac{6}{5}\Big( 1+\frac{1}{Z_i \sqrt{2}}\Big)\nu_{ei}.  \label{eq:VisB2}
\end{eqnarray}
In a quasi-static approximation, one derives the following viscosity coefficients
\begin{eqnarray}
  \eta_0^a &=& \frac{p_a}{\bnu_a}; \quad \eta_1^a = \frac{p_a\bnu_a}{4\Omega_a^2+\bnu_a^2};
  \quad \eta_2^a = \frac{p_a\bnu_a}{\Omega_a^2+\bnu_a^2};
\quad \eta_3^a = \frac{2p_a\Omega_a}{4\Omega_a^2+\bnu_a^2}; \quad \eta_4^a = \frac{p_a\Omega_a}{\Omega_a^2+\bnu_a^2},
\end{eqnarray}
which have the same form as the BGK viscosities. The difference is that while for the BGK operator
$\bnu_i=\nu_{ii}+\nu_{ie}$ and $\bnu_e=\nu_{ee}+\nu_{ei}$, here the frequencies have to be added according to 
(\ref{eq:VisB1}), (\ref{eq:VisB2}).

Importantly, because for ion viscosities \cite{Braginskii1965} neglected the ion-electron collisions,
direct comparision with Braginskii has to done with $\bnu_{i}=(6/5)\nu_{ii}$. Using this approximation, the parallel viscosities
of the Burger-Schunk model are 
\begin{equation}
  \eta_0^i = \frac{5}{6}\frac{p_i}{\nu_{ii}}; \qquad \eta_0^e = \frac{5}{6} \frac{Z_i\sqrt{2}}{(1+Z_i\sqrt{2})}\frac{p_e}{\nu_{ei}},
\end{equation}
where $5/6=0.83$, contrasting with Braginskii ion value of $0.96$.
Considering specific case $Z_i=1$ for the electron viscosity $\eta_0^e = 0.49 p_e/\nu_{ei}$, contrasting with Braginskii
value $0.73$. 

\subsection{Strong magnetic field limit}
Examining the strong magnetic field limit, viscosities for ions become
\begin{equation}
  \eta_1^i = \frac{3}{10}\frac{p_i\nu_{ii}}{\Omega_i^2}; \qquad \eta_2^i = \frac{6}{5}\frac{p_i\nu_{ii}}{\Omega_i^2};
  \qquad \eta_3^i = \frac{1}{2}\frac{p_i}{\Omega_i}; \qquad \eta_4^i = \frac{p_i}{\Omega_i}, \label{eq:fasci}
\end{equation}
(with relations $\eta_2^a = 4\eta_1^a$; $\eta_4^a = 2\eta_3^a$ valid for both electrons and ions).
All four viscosities match Braginskii exactly !
Similarly, for electrons in the strong magnetic field limit the Burgers-Schunk model yields
\begin{equation}
  \eta_1^e = \frac{3}{10}\Big(1+\frac{1}{Z_i\sqrt{2}}\Big)\frac{p_e\nu_{ei}}{\Omega_e^2};
  \qquad  \eta_2^e = \frac{6}{5}\Big(1+\frac{1}{Z_i\sqrt{2}}\Big)\frac{p_e\nu_{ei}}{\Omega_e^2};
 \qquad \eta_3^e = \frac{1}{2}\frac{p_e}{\Omega_e}; \qquad  \eta_4^e = \frac{p_e}{\Omega_e}. \label{eq:BragMatch}
\end{equation}
Evaluation for $Z_i=1$ yields $\eta_1^e= 0.51 p_e\nu_{ei}/\Omega_e^2$ and again all match
Braginskii exactly. If Braginksii provided electron viscosities for different $Z_i$ values,
all four viscosity coefficients (except of parallel $\eta_0$) would match his results exactly.  


 If ion-electron collisions are considered, the gyroviscosities $\eta_3^i$, $\eta_4^i$ given by (\ref{eq:fasci}) remain unchanged,
and the perpendicular viscosities become
\begin{equation}
  \eta_1^i = \frac{p_i \nu_{ii}}{\Omega_i^2}\frac{3}{10}\Big(1+\frac{5}{3}\frac{\sqrt{2}}{Z_i}\sqrt{\frac{m_e}{m_i}}\Big); \qquad
  \eta_2^i=\frac{p_i \nu_{ii}}{\Omega_i^2}\frac{6}{5}\Big(1+\frac{5}{3}\frac{\sqrt{2}}{Z_i}\sqrt{\frac{m_e}{m_i}}\Big), \label{eq:fasci2}
\end{equation}
where again $\eta_2^i=4\eta_1^i$ holds. That the result (\ref{eq:fasci2}) is indeed correct,
can be checked against the 2-Laguerre equation (89b) of \cite{JiHeld2013} when written in strong B-limit.
(Use $\zeta=(1/Z_i)\sqrt{m_e/m_i}$; $r_i=\Omega_i\hat{\tau}_{ii}$; $\eta_2^i=\hat{\eta}_2^i p_i\hat{\tau}_{ii}$,  
with conversion $\hat{\tau}_{ii}=\tau_{ii}/\sqrt{2}$ because we use Braginskii definition of $\tau_{ii}$; see Section \ref{section:ColFreq}).
Interestingly, the result is not changed in their 3-Laguerre model (or higher order models).
The same is true for the perpendicular heat conductivities $\kappa_\perp^a$.

\clearpage
\subsection{Table of integrals}
To calculate the first Rosenbluth potential $H_b$, we used the following integrals
\begin{eqnarray}
  \by\by \int \frac{1}{x} e^{-|\bx+\by|^2} d^3x &=& \by\by \pi^{3/2}\frac{\erf(y)}{y};\\
   \int \frac{\bx}{x}e^{-|\bx+\by|^2} d^3x &=& -\by \pi\Big[ \frac{e^{-y^2}}{y^2}+\sqrt{\pi} \Big(\frac{1}{y}-\frac{1}{2y^3}\Big)\textrm{erf}(y)\Big];\\
   \int \frac{\bx\bx}{x} e^{-|\bx+\by|^2} d^3x &=& \bI \pi \Big[ \frac{e^{-y^2}}{2y^2}
     +\frac{\sqrt{\pi}}{2}\erf(y)\Big(\frac{1}{y}-\frac{1}{2y^3}\Big)\Big]\nn\\
  && + \by\by \pi \Big[ e^{-y^2}\Big(\frac{1}{y^2}-\frac{3}{2y^4}\Big)
    +\sqrt{\pi}\erf(y) \Big( \frac{1}{y}-\frac{1}{y^3}+\frac{3}{4y^5}\Big)\Big], \label{eq:Why1}
\end{eqnarray}
and so
\begin{eqnarray}
\int \frac{(\bx+\by)(\bx+\by)}{x} e^{-|\bx+\by|^2} d^3x &=& \bI \pi \Big[ \frac{e^{-y^2}}{2y^2}
  +\frac{\sqrt{\pi}}{2}\erf(y)\Big(\frac{1}{y}-\frac{1}{2y^3}\Big)\Big]\nn\\
&& + \by\by \pi \Big[ \sqrt{\pi}\erf(y)\frac{3}{4y^5} - e^{-y^2}\Big( \frac{1}{y^2}+\frac{3}{2y^4}\Big)\Big] \label{eq:H10}. 
\end{eqnarray}
To calculate the second Rosenbluth $G_b$ we used
\begin{eqnarray}
  \by\by \int x e^{-|\bx+\by|^2} d^3x &=& \by\by\pi \Big[e^{-y^2}+ \sqrt{\pi}\big( y+\frac{1}{2y}\big)\erf(y)\Big];\\
  \int x\bx e^{-|\bx+\by|^2} d^3x &=& -\by \pi \Big[ \big(1+\frac{1}{2y^2}\big)e^{-y^2} +\sqrt{\pi}\big(y+\frac{1}{y}-\frac{1}{4y^3}\big)\erf(y)\Big];\\
  \int x\bx\bx e^{-|\bx+\by|^2} d^3x &=& \bI \pi \Big[ \big(\frac{1}{2}+\frac{1}{4y^2}\big)e^{-y^2}
    +\sqrt{\pi}\big(\frac{y}{2}-\frac{1}{8y^3}+\frac{1}{2y}\big)\erf(y)\Big]\nn\\
  && + \by\by\pi\Big[ \big(1+\frac{1}{y^2}-\frac{3}{4y^4}\big)e^{-y^2}
    +\sqrt{\pi}\big( y+\frac{3}{2y}-\frac{3}{4y^3}+\frac{3}{8y^5}\big)\erf(y)\Big],
\end{eqnarray}
and so
\begin{eqnarray}
 \int x(\bx+\by)(\bx+\by) e^{-|\bx+\by|^2} d^3x &=&  \bI \pi \Big[ \big(\frac{1}{2}+\frac{1}{4y^2}\big)e^{-y^2}
   +\sqrt{\pi}\big(\frac{y}{2}-\frac{1}{8y^3}+\frac{1}{2y}\big)\erf(y)\Big]\nn\\
 && +\by\by \pi\Big[ -\frac{3}{4y^4}e^{-y^2}+\sqrt{\pi}\big( -\frac{1}{4y^3}+\frac{3}{8y^5}\big)\erf(y)\Big] \label{eq:G10}.
\end{eqnarray}
To calculate the viscosity, the Rosenbluth potentials were integrated by the following scheme
\begin{eqnarray}
  \int \by\by f(y) e^{-\alpha^2 y^2} d^3y &=& \frac{\bI}{3}\int y^2 f(y) e^{-\alpha^2 y^2} d^3y = \bI \frac{4\pi}{3} \int_0^\infty y^4 f(y) e^{-\alpha^2 y^2}dy;
  \label{eq:formula1}\\
  \int \by(\bPi_b^{(2)}\cdot\by) f(y) e^{-\alpha^2 y^2} d^3y &=& \bPi_b^{(2)} \frac{4\pi}{3} \int_0^\infty y^4 f(y) e^{-\alpha^2 y^2} dy;\\
 \bPi_b^{(2)}: \int \by\by f(y) e^{-\alpha^2 y^2} d^3y &=& 0, 
\end{eqnarray}
where in our case functions $f(y)$ are well-behaved, so these integrals hold. 
Additionally, for any symmetric ($3\times 3$) matrix $\bA$ 
\begin{eqnarray}
  \bA: \int \by\by\by\by f(y) e^{-\alpha^2 y^2} d^3y &=&
  \Big[ \bA + (\textrm{Tr} \bA)\frac{\bI}{2}\Big] \frac{8\pi}{15}\int_0^\infty y^6 f(y) e^{-\alpha^2 y^2} dy, \label{eq:formula2}
\end{eqnarray}
and for the stress-tensor $\textrm{Tr} \bPi_b^{(2)}=0$ (the integral can be calculated for example by splitting
$\bA:\by\by$ explicitly to components, and then by using symmetries).
\newpage
\subsubsection{Spherical integration}
For example to obtain integrals (\ref{eq:Why1}),
one introduces orthogonal reference frame in the x-space with unit vectors $\hat{\boldsymbol{e}}_1, \hat{\boldsymbol{e}}_2, \hat{\boldsymbol{e}}_3$,
where the direction of $\by$ forms axis $\hat{\boldsymbol{e}}_3=\by/y$, so that 
\begin{equation}
\bx = x \sin\theta \cos\phi \hat{\boldsymbol{e}}_1 + x\sin\theta\sin\phi \hat{\boldsymbol{e}}_2 + x\cos\theta \hat{\boldsymbol{e}}_3, 
\end{equation}
which then allows to first perform integral over $d\phi$
\begin{eqnarray}
\int_0^{2\pi}\bx d\phi &=& 2\pi x\cos\theta \hat{\boldsymbol{e}}_3;\\
\int_0^{2\pi} \bx \bx d\phi &=& \pi x^2\sin^2\theta \bI +\pi x^2 (3\cos^2\theta-1)\hat{\boldsymbol{e}}_3\hat{\boldsymbol{e}}_3,
\end{eqnarray}
and then over $d\theta dx$.

\newpage
\section{Braginskii heat flux (11-moment model)} \label{sec:HeatFluxB}
\setcounter{equation}{0}
We use the usual \emph{reducible} Hermite polynomials with perturbation of the distribution function
$f_b=f_b^{(0)}(1+\chi_b)$  (see details in Appendix \ref{sec:Hermite})
\begin{equation} \label{eq:Xi11mom}
\chi_b =  \frac{1}{10}\tilde{h}_i^{b(3)} \tilde{H}_i^{b(3)} + \frac{1}{280}\tilde{h}_i^{b(5)} \tilde{H}_i^{b(5)},
\end{equation}
where
\begin{eqnarray}
\tilde{H}^{b(3)}_i &=& \delta_{jk}\tilde{H}^{b(3)}_{ijk}= \tc_i^b ( \tc_b^2-5);  \nn\\
\tilde{H}^{b(5)}_i &=& \delta_{jk}\delta_{lm} \tilde{H}^{b(5)}_{ijklm} =\tc_i^b ( \tc_b^4-14\tc_b^2+35). \label{eq:red_HFx}
\end{eqnarray}
For clarity of calculations, we here only consider the heat flux part of $\chi_b$ (i.e. 11-moment model) but the full 21-moment model can be implicitly assumed for the
final collisional contributions at the semi-linear level. The orthogonality relations are (species indices are dropped)
\begin{eqnarray}
&& \int \tilde{H}^{(3)}_{i} \tilde{H}^{(3)}_{j} \phi^{(0)} d^3\tc = 10 \delta_{i j};\qquad 
 \int \tilde{H}^{(5)}_i \tilde{H}^{(5)}_j \phi^{(0)} d^3\tc = 280\delta_{ij},
\end{eqnarray}
yielding  (\ref{eq:Xi11mom}). By using this perturbation $\chi_b$ one can directly calculate the heat flux vector and the 5th-order moment
vector
\begin{eqnarray}
  \vec{q}^{\,b}_i &=& \frac{m_b}{2}\int f_b c_ic^2 d^3c = \frac{p_b}{2}\sqrt{\frac{T_b}{m_b}}\tilde{h}^{b(3)}_i;\nn\\
  {X}^{b(5)}_i &=& m_b\int f_b c_ic^4 d^3c = p_b\frac{T_b}{m_b}\sqrt{\frac{T_b}{m_b}}\big(\tilde{h}^{b(5)}_i +14\tilde{h}^{b(3)}\big), \label{eq:defHF}
\end{eqnarray}  
or one can directly calculate Hermite moments
\begin{eqnarray}
&& \tilde{h}^{b(3)}_i = \frac{2}{p_b}\sqrt{\frac{m_b}{T_b}}\vec{q}^{\,b}_i;\nn\\   
&& \tilde{h}^{b(5)}_i = \frac{1}{p_b}\sqrt{\frac{m_b}{T_b}}\Big( \frac{m_b}{T_b} {X}^{b(5)}_i-28\vec{q}_i^{\, b} \Big). 
\end{eqnarray}
Note that we have chosen to define all the vectors and tensors (including $X^{b(5)}_i, \tilde{H}^{(3)}_i, \tilde{H}^{(5)}_i$ etc.) without any
additional normalization factors, so they are directly obtained from higher-order tensors by just applying contractions.
The sole exception is the heat flux vector which contains a factor of $1/2$, to match its usual definition.
As also noted after equation (\ref{eq:X5full}), the reminder of this
exception in the index notation is the arrow on the heat flux vector components $\vec{q}_i$. 
We will again use the Rosenbluth potentials, and not the center-of-mass transformation. However, this time we will keep working
with the Hermite fluid moments, which has a nice advantage that the expressions can be kept in partially dimensionless form. 

\subsection{Rosenbluth potentials}
By introducing
\begin{equation}
\tbc_b = \sqrt{\frac{m_b}{T_b}} (\bV'-\bu_b);\qquad |\bV'-\bV|=\sqrt{\frac{T_b}{m_b}}|\tbc_b-\tby|; \qquad \tby = \sqrt{\frac{m_b}{T_b}}(\bV-\bu_b),
\end{equation}  
so that our previously used $\by=\tby/\sqrt{2}$, the Rosenbluth potentials read
\begin{eqnarray}
  H_b (\bV) &=& \int \frac{f_b(\bV')}{|\bV'-\bV|}d^3v' = n_b \sqrt{\frac{m_b}{T_b}} \int \frac{\phi^{(0)}_b}{|\tbc_b-\tby|}(1+\chi_b)d^3\tc_b;\nn\\
  G_b (\bV) &=& \int |\bV'-\bV| f_b(\bV') d^3v' = n_b \sqrt{\frac{T_b}{m_b}} \int  |\tbc_b-\tby|\phi^{(0)}_b  (1+\chi_b)d^3\tc_b,
\end{eqnarray}
and calculate
\begin{eqnarray}
  H_b (\bV) &=& n_b \sqrt{\frac{m_b}{T_b}} \Big[ \frac{1}{\ty}\erf\Big(\frac{\ty}{\sqrt{2}}\Big)
    -\sqrt{\frac{2}{\pi}} \frac{e^{-\ty^2/2}}{10} \Big( \tby\cdot\tbh^{b(3)}+(\ty^2-5)\frac{\tby\cdot\tbh^{b(5)}}{28}\Big) \Big];\nn\\
  G_b (\bV) &=& n_b \sqrt{\frac{T_b}{m_b}} \Big[ \sqrt{\frac{2}{\pi}}e^{-\ty^2/2} +\big(\ty+\frac{1}{\ty}\big)\erf\big(\frac{\ty}{\sqrt{2}}\big) \nn\\
    && +\Big(\frac{\erf(\ty/\sqrt{2})}{5\ty^3} - \sqrt{\frac{2}{\pi}} \frac{e^{-\ty^2/2}}{5\ty^2} \Big)\tby\cdot\tbh^{b(3)}
    - \sqrt{\frac{2}{\pi}} \frac{e^{-\ty^2/2}}{140} \tby\cdot\tbh^{b(5)} \Big].
\end{eqnarray}
The derivatives calculate by using $\pr/\pr v_i = \sqrt{m_b/T_b}\pr/\pr \ty_i$ and
\begin{eqnarray}
  \frac{\pr H_b}{\pr \bV} &=& \frac{n_b m_b}{T_b} \Big[ \tby \Big( \sqrt{\frac{2}{\pi}}\frac{e^{-\ty^2/2}}{\ty^2}-\frac{\erf(\ty/\sqrt{2})}{\ty^3}\Big) \nn\\
    && -\,\sqrt{\frac{2}{\pi}} \frac{e^{-\ty^2/2}}{10}\Big( \tbh^{b(3)}-\tby (\tby\cdot\tbh^{b(3)})
    +(\ty^2-5)\frac{\tbh^{b(5)}}{28}-(\ty^2-7)\frac{\tby(\tby\cdot\tbh^{b(5)})}{28} \Big)\Big],
\end{eqnarray}
and by further applying $(\pr/\pr\bV)\cdot$ recovers $-4\pi f_b$. It is useful to write the second Rosenbluth potential as
\begin{equation}
G_b (\bV) = n_b \sqrt{\frac{T_b}{m_b}} \Big[ \tilde{A}_1 + \tilde{A}_3 \tby\cdot\tbh^{b(3)} + \tilde{A}_5\tby\cdot\tbh^{b(5)}\Big],
\end{equation}
where
\begin{eqnarray}
&&  \tilde{A}_1 = \sqrt{\frac{2}{\pi}}e^{-\ty^2/2} +\big(\ty+\frac{1}{\ty}\big)\erf\big(\frac{\ty}{\sqrt{2}}\big);\nn\\
&& \tilde{A}_3 = \frac{\erf(\ty/\sqrt{2})}{5\ty^3} - \sqrt{\frac{2}{\pi}} \frac{e^{-\ty^2/2}}{5\ty^2};\nn\\
&& \tilde{A}_5 = -\, \sqrt{\frac{2}{\pi}} \frac{e^{-\ty^2/2}}{140}, 
\end{eqnarray}  
so that the second derivative calculates easily 
\begin{eqnarray}
  \frac{\pr}{\pr\bV}\frac{\pr G_b}{\pr \bV} &=& n_b \sqrt{\frac{m_b}{T_b}} \bigg\{ \bI \frac{\tilde{A}_1'}{\ty}
  +\frac{\tby\tby}{\ty^2}\Big( \tilde{A}_1'' -\frac{\tilde{A}_1'}{\ty}\Big) \nn\\
  && +\Big( \tby\tbh^{b(3)} + \tbh^{b(3)}\tby +\bI (\tby\cdot\tbh^{b(3)})\Big) \frac{\tilde{A}_3'}{\ty}
  +\frac{\tby\tby}{\ty^2} (\tby\cdot\tbh^{b(3)}) \Big( \tilde{A}_3'' -\frac{\tilde{A}_3'}{\ty}\Big)\nn\\
   && +\Big( \tby\tbh^{b(5)} + \tbh^{b(5)}\tby +\bI (\tby\cdot\tbh^{b(5)})\Big) \frac{\tilde{A}_5'}{\ty}
  +\frac{\tby\tby}{\ty^2} (\tby\cdot\tbh^{b(5)}) \Big( \tilde{A}_5'' -\frac{\tilde{A}_5'}{\ty}\Big) \bigg\},
\end{eqnarray}
and applying $(1/2)\trace$ recovers $H_b$. The coefficients are
\begin{eqnarray}
  \tilde{A}_1' &=& \sqrt{\frac{2}{\pi}}\frac{e^{-\ty^2/2}}{\ty}+\big(1-\frac{1}{\ty^2}\big)\erf\big(\frac{\ty}{\sqrt{2}}\big);\nn\\
  \tilde{A}_3' &=& \sqrt{\frac{2}{\pi}}\big(\frac{1}{\ty}+\frac{3}{\ty^3}\big) \frac{e^{-\ty^2/2}}{5}
  -\frac{3}{5\ty^4}\erf\big(\frac{\ty}{\sqrt{2}}\big);\nn\\
  \tilde{A}_5' &=& \sqrt{\frac{2}{\pi}} \frac{\ty e^{-\ty^2/2}}{140};\nn\\
  \tilde{A}_1'' &=& -\, \sqrt{\frac{2}{\pi}} \frac{2}{\ty^2} e^{-\ty^2/2} +\frac{2}{\ty^3}\erf\big(\frac{\ty}{\sqrt{2}}\big);\nn\\
  \tilde{A}_3'' &=& -\, \sqrt{\frac{2}{\pi}} \big(1+\frac{4}{\ty^2}+\frac{12}{\ty^4}\big)\frac{e^{-\ty^2/2}}{5}
  +\frac{12}{5\ty^5}\erf\big(\frac{\ty}{\sqrt{2}}\big);\nn\\
  \tilde{A}_5'' &=& -\, \sqrt{\frac{2}{\pi}} (\ty^2-1)\frac{e^{-\ty^2/2}}{140},
\end{eqnarray}
and so
\begin{eqnarray}
  \tilde{A}_1'' -\frac{\tilde{A}_1'}{\ty} &=& -\, \sqrt{\frac{2}{\pi}} \frac{3}{\ty^2} e^{-\ty^2/2}
  -\big(\frac{1}{\ty}-\frac{3}{\ty^3}\big)\erf\big(\frac{\ty}{\sqrt{2}}\big);\nn\\
  \tilde{A}_3'' -\frac{\tilde{A}_3'}{\ty} &=& -\, \sqrt{\frac{2}{\pi}} \big(1+\frac{5}{\ty^2}+\frac{15}{\ty^4}\big)\frac{e^{-\ty^2/2}}{5}
  +\frac{3}{\ty^5}\erf\big(\frac{\ty}{\sqrt{2}}\big);\nn\\
  \tilde{A}_5'' -\frac{\tilde{A}_5'}{\ty} &=& -\, \sqrt{\frac{2}{\pi}} \frac{\ty^2 e^{-\ty^2/2}}{140}.
\end{eqnarray}


\subsection{Dynamical friction vector and diffusion tensor}
The dynamical friction vector thus reads
\begin{eqnarray}
  \boldsymbol{A}^{ab} &=& 2\frac{c_{ab}}{m_a^2}\big(1+\frac{m_a}{m_b}\big)
  \frac{n_b m_b}{T_b} \Big[ \tby \Big( \sqrt{\frac{2}{\pi}}\frac{e^{-\ty^2/2}}{\ty^2}-\frac{\erf(\ty/\sqrt{2})}{\ty^3}\Big) \nn\\
    && -\,\sqrt{\frac{2}{\pi}} \frac{e^{-\ty^2/2}}{10}\Big( \tbh^{b(3)}-\tby (\tby\cdot\tbh^{b(3)})
    +(\ty^2-5)\frac{\tbh^{b(5)}}{28}-(\ty^2-7)\frac{\tby(\tby\cdot\tbh^{b(5)})}{28} \Big)\Big], \label{eq:Energy750}
\end{eqnarray}  
and the diffusion tensor
\begin{eqnarray}
\bD^{ab} &=& 2\frac{c_{ab}}{m_a^2} n_b \sqrt{\frac{m_b}{T_b}} \bigg\{ \bI \frac{\tilde{A}_1'}{\ty}
  +\frac{\tby\tby}{\ty^2}\Big( \tilde{A}_1'' -\frac{\tilde{A}_1'}{\ty}\Big) \nn\\
  && +\Big( \tby\tbh^{b(3)} + \tbh^{b(3)}\tby +\bI (\tby\cdot\tbh^{b(3)})\Big) \frac{\tilde{A}_3'}{\ty}
  +\frac{\tby\tby}{\ty^2} (\tby\cdot\tbh^{b(3)}) \Big( \tilde{A}_3'' -\frac{\tilde{A}_3'}{\ty}\Big)\nn\\
   && +\Big( \tby\tbh^{b(5)} + \tbh^{b(5)}\tby +\bI (\tby\cdot\tbh^{b(5)})\Big) \frac{\tilde{A}_5'}{\ty}
  +\frac{\tby\tby}{\ty^2} (\tby\cdot\tbh^{b(5)}) \Big( \tilde{A}_5'' -\frac{\tilde{A}_5'}{\ty}\Big) \bigg\}, \label{eq:Energy751}
\end{eqnarray}
or in its entire beauty
\begin{eqnarray}
  \bD^{ab} &=& 2\frac{c_{ab}}{m_a^2} n_b \sqrt{\frac{m_b}{T_b}} \bigg\{ \bI  \Big[
  \sqrt{\frac{2}{\pi}}\frac{e^{-\ty^2/2}}{\ty^2}+\big(\frac{1}{\ty}-\frac{1}{\ty^3}\big)\erf\big(\frac{\ty}{\sqrt{2}}\big)\Big] \nn\\
  && +\frac{\tby\tby}{\ty^2} \Big[
    -\, \sqrt{\frac{2}{\pi}} \frac{3}{\ty^2} e^{-\ty^2/2}
  -\big(\frac{1}{\ty}-\frac{3}{\ty^3}\big)\erf\big(\frac{\ty}{\sqrt{2}}\big)\Big] \nn\\
  && +\Big( \tby\tbh^{b(3)} + \tbh^{b(3)}\tby +\bI (\tby\cdot\tbh^{b(3)})\Big) \Big[
   \sqrt{\frac{2}{\pi}}\big(\frac{1}{\ty^2}+\frac{3}{\ty^4}\big) \frac{e^{-\ty^2/2}}{5}
  -\frac{3}{5\ty^5}\erf\big(\frac{\ty}{\sqrt{2}}\big) \Big]\nn\\
  && +\frac{\tby\tby}{\ty^2} (\tby\cdot\tbh^{b(3)}) \Big[
    -\, \sqrt{\frac{2}{\pi}} \big(1+\frac{5}{\ty^2}+\frac{15}{\ty^4}\big)\frac{e^{-\ty^2/2}}{5}
  +\frac{3}{\ty^5}\erf\big(\frac{\ty}{\sqrt{2}}\big) \Big]\nn\\
  && +\Big( \tby\tbh^{b(5)} + \tbh^{b(5)}\tby +\bI (\tby\cdot\tbh^{b(5)})\Big) \Big[
      \sqrt{\frac{2}{\pi}} \frac{e^{-\ty^2/2}}{140}\Big]\nn\\
  && +\frac{\tby\tby}{\ty^2} (\tby\cdot\tbh^{b(5)}) \Big[ -\, \sqrt{\frac{2}{\pi}} \frac{\ty^2 e^{-\ty^2/2}}{140}\Big] \bigg\}.
\end{eqnarray}
As a reminder
\begin{equation}
\tby = \sqrt{\frac{m_b}{T_b}}(\bV-\bu_b); \qquad c_{ab} = 2\pi e^4 Z_a^2 Z_b^2\ln\Lambda
\end{equation}
\subsection{Distribution function for species ``a''} \label{sec:FA}
The general distribution function for species ``a'' reads
\begin{equation}
  f_a = f^{(0)}_a(1+\chi_a) = n_a\Big(\frac{m_a}{T_a} \Big)^{3/2} \phi^{(0)}_a (1+\chi_a); \qquad \phi^{(0)}_a = \frac{e^{-\tc^2_a/2}}{(2\pi)^{3/2}};
  \qquad \tc_a = \sqrt{\frac{m_a}{T_a}}(\bV-\bu_a),
\end{equation}
where the perturbation
\begin{equation} \label{eq:Xi11mom2}
\chi_a =  \frac{1}{10}\tilde{h}_i^{a(3)} \tilde{H}_i^{a(3)}(\tbc_a) + \frac{1}{280}\tilde{h}_i^{a(5)} \tilde{H}_i^{a(5)} (\tbc_a).
\end{equation}
To avoid the complicated run-away effect, the weight has to be expanded with small drifts, for example
by defining
\begin{equation}
  \tbu = (\bu_b-\bu_a)\sqrt{\frac{m_a}{T_a}}; \qquad
  \alpha=\frac{\sqrt{T_b/m_b}}{\sqrt{T_a/m_a}}; \qquad \tbc_a=\alpha\tby+\tbu,
\end{equation}
so that the expansion for small drifts
\begin{equation}
e^{-\tc_a^2/2} = e^{-|\alpha \tby+\tbu|^2/2} \simeq e^{-\alpha^2 \ty^2/2}\big(1-\alpha \tby\cdot\tbu\big).
\end{equation}
In comparison to our previously used normalization $\tby=\by\sqrt{2}$ and $\tbu=\bu\sqrt{2}$ and $\tbc_a=\bc_a\sqrt{2}/v_{\textrm{th} a}$.
The perturbation $\chi_a$ contains Hermite polynomials, and these also have to be expanded in the semi-linear approximation.
Importantly, after contraction with Hermite (fluid) moments 
\begin{eqnarray}
&&  \tilde{h}^{a(3)}_i \tilde{H}^{a(3)}_i (\tbc_a) \simeq \tilde{h}^{a(3)}_i \alpha \ty_i (\alpha^2\ty^2-5) = \tilde{h}^{a(3)}_i \tilde{H}^{a(3)}_i (\alpha \tby);\nn\\
  &&  \tilde{h}^{a(5)}_i \tilde{H}^{a(5)}_i (\tbc_a) \simeq \tilde{h}^{a(5)}_i \alpha \ty_i (\alpha^4 \ty^4 -14\alpha^2\ty^2+35)
  = \tilde{h}^{a(5)}_i \tilde{H}^{a(5)}_i (\alpha \tby), 
\end{eqnarray}
where all the drift $\tbu$ contributions such as $\tilde{h}^{a(3)}_i \tilde{u}_i$ are neglected in the semi-linear approximation.  
The expanded distribution function thus reads
\begin{equation}
f_a =  n_a\Big(\frac{m_a}{T_a} \Big)^{3/2} \frac{e^{-\alpha^2\ty^2/2}}{(2\pi)^{3/2}} (1-\alpha\tby\cdot\tbu +\chi_a),
\end{equation}
with perturbation
\begin{equation} \label{eq:Xi11mom3}
\chi_a =  \frac{1}{10}\tilde{h}_i^{a(3)} \tilde{H}_i^{a(3)}(\alpha\tby) + \frac{1}{280}\tilde{h}_i^{a(5)} \tilde{H}_i^{a(5)} (\alpha\tby).
\end{equation}
Integrals are evaluated with $d^3v=(T_b/m_b)^{3/2} d^3\ty$, so a useful shortcut is
\begin{equation} \label{eq:Thierry11}
\int f_a d^3v = n_a \alpha^3 \int \frac{e^{-\alpha^2\ty^2/2}}{(2\pi)^{3/2}} (1-\alpha\tby\cdot\tbu +\chi_a) d^3 \ty.
\end{equation}
Also, it is useful to express $c_{ab}$ directly through the collisional frequencies $\nu_{ab}$, according to
\begin{equation} \label{eq:Thierry12}
2 \frac{c_{ab} n_b}{m_a^2} \big(1+\frac{m_a}{m_b}\big)= 3\nu_{ab} \sqrt{\frac{\pi}{2}} (1+\alpha^2)^{3/2} \Big(\frac{T_a}{m_a}\Big)^{3/2}. 
\end{equation}  
\subsection{Momentum exchange rates \texorpdfstring{$R_{ab}$}{Rab}}
The momentum exchange rates calculate
\begin{eqnarray}
  \boldsymbol{R}_{ab} &=& m_a \int f_a \boldsymbol{A}^{ab} d^3v \nn\\
  &=& \nu_{ab} \rho_a \sqrt{\frac{T_a}{m_a}}\tbu + \frac{3}{5}\nu_{ab} \frac{\mu_{ab}}{T_{ab}}
  \Big[\frac{p_a}{2} \sqrt{\frac{T_a}{m_a}}\tbh^{a(3)}- \frac{\rho_a}{\rho_b} \frac{p_b}{2} \sqrt{\frac{T_b}{m_b}}\tbh^{b(3)}\Big]\nn\\
  && - \frac{3}{56}\nu_{ab} \Big(\frac{\mu_{ab}}{T_{ab}}\Big)^2
  \Big[ p_a \Big(\frac{T_a}{m_a}\Big)^{3/2} \tbh^{a(5)} - \frac{\rho_a}{\rho_b} p_b \Big(\frac{T_b}{m_b}\Big)^{3/2} \tbh^{b(5)}\Big], 
\end{eqnarray}
or expressed through usual fluid variables
\begin{eqnarray}
  \boldsymbol{R}_{ab} 
  &=& \nu_{ab} \rho_a (\bu_b-\bu_a) + \frac{3}{5}\nu_{ab} \frac{\mu_{ab}}{T_{ab}}
  \Big[\vecq^a- \frac{\rho_a}{\rho_b} \vecq^b\Big]\nn\\
  && - \frac{3}{56}\nu_{ab} \Big(\frac{\mu_{ab}}{T_{ab}}\Big)^2
  \Big[ \Big(\vecX^{a(5)}-28\frac{T_a}{m_a}\vecq^a\Big) - \frac{\rho_a}{\rho_b} \Big( \vecX^{b(5)}-28\frac{T_b}{m_b}\vecq^b  \Big)\Big]. 
\end{eqnarray}
Note that $\boldsymbol{R}_{ab}=- \boldsymbol{R}_{ba}$. An alternative form reads
\begin{eqnarray}
  \boldsymbol{R}_{ab} 
  &=& \nu_{ab} \rho_a (\bu_b-\bu_a) + \nu_{ab} \frac{\mu_{ab}}{T_{ab}}\Big[ \vecq^a\Big(\frac{3}{5}+\frac{3}{2}\frac{\mu_{ab}}{m_a}\frac{T_a}{T_{ab}}\Big)
  - \frac{\rho_a}{\rho_b} \vecq^b \Big( \frac{3}{5}+\frac{3}{2}\frac{\mu_{ab}}{m_b}\frac{T_b}{T_{ab}}\Big)\Big]\nn\\
  && - \frac{3}{56}\nu_{ab} \Big(\frac{\mu_{ab}}{T_{ab}}\Big)^2
  \Big[ \vecX^{a(5)} - \frac{\rho_a}{\rho_b} \vecX^{b(5)} \Big], 
\end{eqnarray}
or yet another one 
\begin{eqnarray}
  \boldsymbol{R}_{ab} 
  &=& \nu_{ab} \rho_a (\bu_b-\bu_a) + \nu_{ab} \frac{\mu_{ab}}{T_{ab}}\Big[ \vecq^a \frac{\frac{21}{10}T_am_b+\frac{3}{5}T_b m_a}{T_a m_b+T_b m_a}
  - \frac{\rho_a}{\rho_b} \vecq^b \frac{\frac{3}{5}T_am_b+\frac{21}{10}T_b m_a}{T_a m_b+T_b m_a} \Big]\nn\\
  && - \frac{3}{56}\nu_{ab} \Big(\frac{\mu_{ab}}{T_{ab}}\Big)^2
  \Big[ \vecX^{a(5)} - \frac{\rho_a}{\rho_b} \vecX^{b(5)} \Big]. 
\end{eqnarray}

\newpage
\subsection{Heat flux exchange rates}
We need to calculate collisional contributions for the heat flux 
\begin{eqnarray}
  \frac{1}{2}\textrm{Tr}\bQ_{ab}^{(3)} &=& \frac{\delta \vecq_{ab}}{\delta t} =
  m_a \int f_a \Big[ (\boldsymbol{A}_{ab}\cdot\bc_a)\bc_a +\frac{1}{2}\boldsymbol{A}_{ab}|\bc_a|^2\Big] d^3v\nn\\
  && \qquad \quad + m_a\int f_a \Big[ \frac{1}{2}(\textrm{Tr}\bD_{ab})\bc_a +\bD_{ab}\cdot\bc_a\Big] d^3v, \label{eq:HFexchange}
\end{eqnarray}
where the velocity
\begin{equation}
\bc_a = \sqrt{\frac{T_a}{m_a}}\tbc_a = \sqrt{\frac{T_a}{m_a}} (\alpha\tby+\tbu).
\end{equation}
Before attempting integration of (\ref{eq:HFexchange}), it is useful to apply the semi-linear approximation, which yields step-by-step
\begin{eqnarray}
  \boldsymbol{A}^{ab}\cdot\bc_a &\simeq& 2\frac{c_{ab}}{m_a^2}\big(1+\frac{m_a}{m_b}\big)
  \frac{n_b m_b}{T_b}\sqrt{\frac{T_a}{m_a}} \bigg[ (\alpha\ty^2+\tby\cdot\tbu) \Big( \sqrt{\frac{2}{\pi}}\frac{e^{-\ty^2/2}}{\ty^2}-\frac{\erf(\ty/\sqrt{2})}{\ty^3}\Big) \nn\\
    && -\,\sqrt{\frac{2}{\pi}} \frac{e^{-\ty^2/2}}{10} \alpha \Big( (\tby\cdot\tbh^{b(3)})(1-\ty^2)
    +\frac{(\tby\cdot\tbh^{b(5)})}{28}(8\ty^2-\ty^4-5) \Big)\bigg]; 
\end{eqnarray}
\begin{eqnarray}
  (\boldsymbol{A}^{ab}\cdot\bc_a)\bc_a &\simeq& 2\frac{c_{ab}}{m_a^2}\big(1+\frac{m_a}{m_b}\big)
  \frac{n_b m_b}{T_b} \frac{T_a}{m_a} \bigg[ \nn\\
    && +\alpha \Big(\alpha\ty^2\tby+\tby(\tby\cdot\tbu)+\ty^2\tbu \Big) \Big( \sqrt{\frac{2}{\pi}}\frac{e^{-\ty^2/2}}{\ty^2}-\frac{\erf(\ty/\sqrt{2})}{\ty^3}\Big) \nn\\
    && -\,\sqrt{\frac{2}{\pi}} \frac{e^{-\ty^2/2}}{10} \alpha^2 \Big( \tby(\tby\cdot\tbh^{b(3)})(1-\ty^2)
    +\frac{\tby(\tby\cdot\tbh^{b(5)})}{28}(8\ty^2-\ty^4-5) \Big)\bigg]. 
\end{eqnarray}
Furthermore, in the semi-linear approximation
\begin{equation}
|\bc_a|^2 \simeq \frac{T_a}{m_a}\big( \alpha^2\ty^2+2\alpha\tby\cdot\tbu\big),
\end{equation}
and thus
\begin{eqnarray}
  \boldsymbol{A}^{ab}|\bc_a|^2 &\simeq & 2\frac{c_{ab}}{m_a^2}\big(1+\frac{m_a}{m_b}\big)
  \frac{n_b m_b}{T_b} \frac{T_a}{m_a}\bigg[ \nn\\
    && + \alpha\Big(\alpha\ty^2\tby+2\tby(\tby\cdot\tbu)\Big)
    \Big( \sqrt{\frac{2}{\pi}}\frac{e^{-\ty^2/2}}{\ty^2}-\frac{\erf(\ty/\sqrt{2})}{\ty^3}\Big) \nn\\
    && -\,\sqrt{\frac{2}{\pi}} \frac{e^{-\ty^2/2}}{10} \alpha^2 \ty^2 \Big( \tbh^{b(3)}-\tby (\tby\cdot\tbh^{b(3)})
    +(\ty^2-5)\frac{\tbh^{b(5)}}{28}-(\ty^2-7)\frac{\tby(\tby\cdot\tbh^{b(5)})}{28} \Big)\bigg]. 
\end{eqnarray}
For the diffusion tensor
\begin{eqnarray}
\trace \bD^{ab} &=& 2\frac{c_{ab}}{m_a^2} n_b \sqrt{\frac{m_b}{T_b}} \bigg\{ \Big(2 \frac{\tilde{A}_1'}{\ty}
  +\tilde{A}_1'' \Big) \nn\\
  && + (\tby\cdot\tbh^{b(3)}) \Big( 4\frac{\tilde{A}_3'}{\ty} + \tilde{A}_3''\Big)
  +(\tby\cdot\tbh^{b(5)}) \Big( 4\frac{\tilde{A}_5'}{\ty} +\tilde{A}_5''\Big) \bigg\},
\end{eqnarray}
and in the semi-linear approximation
\begin{eqnarray}
  \frac{1}{2}(\trace \bD^{ab})\bc_a &\simeq& 2\frac{c_{ab}}{m_a^2}  \frac{n_b}{\alpha} \bigg\{
  (\alpha\tby+\tbu)\Big( \frac{\tilde{A}_1'}{\ty}
  +\frac{\tilde{A}_1''}{2} \Big) \nn\\
  && + \alpha\tby(\tby\cdot\tbh^{b(3)}) \Big( 2\frac{\tilde{A}_3'}{\ty} + \frac{\tilde{A}_3''}{2}\Big)
  +\alpha\tby (\tby\cdot\tbh^{b(5)}) \Big( 2\frac{\tilde{A}_5'}{\ty} +\frac{\tilde{A}_5''}{2}\Big) \bigg\};
\end{eqnarray}
\begin{eqnarray}
  \bD^{ab}\cdot\bc_a &\simeq& 2\frac{c_{ab}}{m_a^2} \frac{n_b}{\alpha} \bigg\{ \alpha\tby \tilde{A}_1''+\tbu \frac{\tilde{A}_1'}{\ty}
   +\frac{\tby (\tby\cdot\tbu)}{\ty^2}\Big( \tilde{A}_1'' -\frac{\tilde{A}_1'}{\ty}\Big) \nn\\
  && +\alpha \tbh^{b(3)} \ty \tilde{A}_3' +\alpha \tby(\tby\cdot\tbh^{b(3)}) \Big( \frac{\tilde{A}_3'}{\ty}+\tilde{A}_3''\Big)\nn\\
   && +\alpha \tbh^{b(5)} \ty \tilde{A}_5' +\alpha \tby(\tby\cdot\tbh^{b(5)}) \Big( \frac{\tilde{A}_5'}{\ty}+\tilde{A}_5''\Big)
  \bigg\}.
\end{eqnarray}
Collecting all the results together the first part of (\ref{eq:HFexchange}) becomes
\begin{eqnarray}
  (\boldsymbol{A}^{ab}\cdot\bc_a)\bc_a +\frac{1}{2} \boldsymbol{A}^{ab}|\bc_a|^2 &\simeq& 2\frac{c_{ab}}{m_a^2}\big(1+\frac{m_a}{m_b}\big)
  \frac{n_b }{\alpha^2} \bigg\{ \nn\\
    && +\alpha \Big(\frac{3}{2}\alpha\ty^2\tby+2\tby(\tby\cdot\tbu)+\ty^2\tbu \Big)
    \Big( \sqrt{\frac{2}{\pi}}\frac{e^{-\ty^2/2}}{\ty^2}-\frac{\erf(\ty/\sqrt{2})}{\ty^3}\Big) \nn\\
    && -\,\sqrt{\frac{2}{\pi}} \frac{e^{-\ty^2/2}}{10} \alpha^2 \Big( \frac{1}{2}\ty^2\tbh^{b(3)}+ \tby(\tby\cdot\tbh^{b(3)})(1-\frac{3}{2}\ty^2)\nn\\
    && +\frac{1}{2}\ty^2(\ty^2-5)\frac{\tbh^{b(5)}}{28}+\frac{\tby(\tby\cdot\tbh^{b(5)})}{28}\big(\frac{23}{2}\ty^2-\frac{3}{2}\ty^4-5\big) \Big)\bigg\};\label{eq:Thierry13} 
\end{eqnarray}
and the second part of (\ref{eq:HFexchange}) becomes
\begin{eqnarray}
  \frac{1}{2}(\trace \bD^{ab})\bc_a + \bD^{ab}\cdot\bc_a &\simeq& 2\frac{c_{ab}}{m_a^2}  \frac{n_b}{\alpha} \bigg\{
  \alpha\tby\Big( \frac{\tilde{A}_1'}{\ty} +\frac{3 \tilde{A}_1''}{2} \Big)
  +\tbu \Big( 2 \frac{\tilde{A}_1'}{\ty} + \frac{\tilde{A}_1''}{2}\Big)
  +\frac{\tby (\tby\cdot\tbu)}{ \ty^2}\Big( \tilde{A}_1'' -\frac{\tilde{A}_1'}{\ty}\Big) \nn\\
  && + \alpha \tbh^{b(3)} \ty \tilde{A}_3' + \alpha\tby(\tby\cdot\tbh^{b(3)}) \Big( 3\frac{\tilde{A}_3'}{\ty}+\frac{3}{2}\tilde{A}_3''\Big)\nn\\
   && + \alpha\tbh^{b(5)} \ty \tilde{A}_5' + \alpha\tby(\tby\cdot\tbh^{b(5)}) \Big( 3 \frac{\tilde{A}_5'}{\ty}+\frac{3}{2}\tilde{A}_5''\Big)\bigg\}.\label{eq:Thierry14}
\end{eqnarray}
Now (\ref{eq:HFexchange}) can be directly integrated, by again applying semi-linear approximation during integration. 
By using (\ref{eq:Thierry11}) \& (\ref{eq:Thierry12}) the entire collisional integral (\ref{eq:HFexchange}) can be written in a symbolic form 
\begin{eqnarray}
  \frac{1}{2}\textrm{Tr}\bQ_{ab}^{(3)} &=& m_a n_a 3 \alpha^3 \nu_{ab} \sqrt{\frac{\pi}{2}} \big(1+\alpha^2\big)^{3/2} \Big(\frac{T_a}{m_a}\Big)^{3/2} \times\nn\\
  && \int \frac{e^{-\alpha^2\ty^2/2}}{(2\pi)^{3/2}}
  \big(1-\alpha \tby\cdot\tbu +\chi_a\big) \Big[ \frac{1}{\alpha^2} \{\ref{eq:Thierry13}\}
    + \frac{1}{(1+\frac{m_a}{m_b})\alpha} \{\ref{eq:Thierry14}\} \Big] d^3\ty,
\end{eqnarray}
where \{\ref{eq:Thierry13}\} \& \{\ref{eq:Thierry14}\} represent only parts of the corresponding equations that are inside of curly brackets. 
 The final result of integration reads
\begin{eqnarray}
   \frac{1}{2}\textrm{Tr}\bQ_{ab}^{(3)} &=& \frac{\delta \vecq_{ab}}{\delta t}   
  = - \nu_{ab} p_a (\bu_b-\bu_a) U_{ab (1)} \nn\\
  && - \nu_{ab}  D_{ab (1)} \frac{p_a}{2}\sqrt{\frac{T_a}{m_a}}\tbh^{a(3)}
  + \nu_{ab}  D_{ab (2)} \frac{\rho_a}{\rho_b} \frac{p_b}{2}\sqrt{\frac{T_b}{m_b}}\tbh^{b(3)} \nn\\
  && +\nu_{ab} E_{ab (1)} p_a \sqrt{\frac{T_a}{m_a}}\tbh^{a(5)}   +\nu_{ab} E_{ab (2)} \frac{\rho_a}{\rho_b} p_b\sqrt{\frac{T_b}{m_b}}\tbh^{b(5)}, \label{eq:BragTT}
\end{eqnarray}
 with mass-ratio coefficients
\begin{eqnarray}
  U_{ab (1)} &=&  \frac{(4 T_a-11 T_b) m_a m_b-2 T_a m_b^2-5 T_b m_a^2}{2 (T_a m_b+T_b m_a) (m_b+m_a) };\nn\\
  D_{ab (1)} &=& -\, \frac{6 T_a^2 m_a m_b^2+2 T_a^2 m_b^3+21 T_a T_b m_a^2 m_b-5 T_a T_b m_a m_b^2-30 T_b^2 m_a^3-52 T_b^2 m_a^2 m_b}{10 (T_a m_b+T_b m_a)^2 (m_b+m_a)};\nn\\
  D_{ab (2)} &=& \frac{3 m_b T_a [(10 T_a -11 T_b) m_a m_b+4 T_a m_b^2-5 T_b m_a^2]}{10 (T_a m_b+T_b m_a)^2 (m_b+m_a)};\nn\\
  E_{ab (1)} &=& -\, \frac{3  T_a m_b [6 T_a^2 m_a m_b^2+2 T_a^2 m_b^3+27 T_a T_b m_a^2 m_b-11 T_a T_b m_a m_b^2-84 T_b^2 m_a^3-118 T_b^2 m_a^2 m_b]}{560 (T_a m_b+T_b m_a)^3 (m_b+m_a)};\nn\\
  E_{ab (2)} &=& -\, \frac{3  m_a m_b T_a T_b [16 T_a m_a m_b+10 T_a m_b^2-5 T_b m_a^2-11 T_b m_a m_b]}{112 (T_a m_b+T_b m_a)^3 (m_b+m_a)}.\nn\\
\end{eqnarray}
As a double-check, we have verified that neglecting the 5th-order Hermite moments $\tbh^{(5)}$ in (\ref{eq:BragTT}) yields a model
that matches Burgers-Schunk; see equations (45)-(49) in \cite{Schunk1977} (after prescribing there Coulomb collisions).
 For small temperature differences the mass-ratio coefficients simplify into
\begin{eqnarray}
  &&  U_{ab (1)} = -\, \frac{(5/2)m_a+m_b}{m_a+m_b};\nn\\
  &&  D_{ab (1)} = \frac{3m_a^2+\frac{1}{10}m_a m_b -\frac{1}{5}m_b^2}{(m_a+m_b)^2}; \qquad
    D_{ab (2)} = \frac{\frac{6}{5}m_b^2 -\frac{3}{2}m_a m_b}{(m_a+m_b)^2}; \nn\\
  &&  E_{ab (1)} = \frac{3}{560} \frac{m_b (84 m_a^2+7m_am_b -2 m_b^2)}{(m_a+m_b)^3};\qquad 
    E_{ab (2)} = \frac{15}{112} \frac{m_a m_b (m_a-2m_b)}{(m_a+m_b)^3}.
\end{eqnarray}
 The model is easily changed from Hermite moments to fluid moments by 
\begin{eqnarray}
  && \vecq_a = \frac{p_a}{2} \sqrt{\frac{T_a}{m_a}} \tbh^{a(3)}; \qquad  p_a \sqrt{\frac{T_a}{m_a}} \tbh^{a(5)} = \frac{m_a}{T_a}\vecX^{(5)}_a -28\vecq_a;\nn\\
  && \vecq_b = \frac{p_b}{2} \sqrt{\frac{T_b}{m_b}} \tbh^{b(3)}; \qquad  p_b \sqrt{\frac{T_b}{m_b}} \tbh^{b(5)} = \frac{m_b}{T_b}\vecX^{(5)}_b -28\vecq_b.
\end{eqnarray}
 The heat flux exchange rates become
\begin{eqnarray}
  \vec{\boldsymbol{Q}}^{(3)}_{ab}\,' &=& \frac{\delta \vecq_{ab}\,'}{\delta t} = \frac{1}{2}\trace \bQ_{ab}^{(3)} - \frac{5}{2}\frac{p_a}{\rho_a}\boldsymbol{R}_{ab}\nn\\
  &=&  -\nu_{ab} p_a (\bu_b-\bu_a) \hat{U}_{ab (1)} 
   - \nu_{ab}  \hat{D}_{ab (1)} \vecq_a
  + \nu_{ab}  \hat{D}_{ab (2)} \frac{\rho_a}{\rho_b} \vecq_b \nn\\
  && +\nu_{ab} \hat{E}_{ab (1)} \frac{\rho_a}{p_a}\vecX^{(5)}_a   -\nu_{ab} \hat{E}_{ab (2)} \frac{\rho_a}{\rho_b} \frac{\rho_b}{p_b} \vecX^{(5)}_b, \label{eq:BragTTT}
\end{eqnarray}
 with mass-ratio coefficients (introducing hat)
\begin{eqnarray}
  \hat{U}_{ab (1)} &=& U_{ab (1)} +\frac{5}{2};\nn\\
  \hat{D}_{ab (1)} &=& D_{ab (1)} +28 E_{ab (1)} +\frac{3}{2}\frac{T_a}{m_a}\frac{\mu_{ab}}{T_{ab}}+\frac{15}{4}\frac{T_a^2}{m_a^2}\frac{\mu_{ab}^2}{T_{ab}^2};\nn\\
  \hat{D}_{ab (2)} &=& D_{ab (2)} -28 E_{ab (2)} +\frac{3}{2}\frac{T_a}{m_a}\frac{\mu_{ab}}{T_{ab}}+\frac{15}{4}\frac{T_a T_b}{m_a m_b}\frac{\mu_{ab}^2}{T_{ab}^2};\nn\\
  \hat{E}_{ab (1)} &=& E_{ab (1)}+\frac{15}{112} \frac{T_a^2}{m_a^2} \frac{\mu_{ab}^2}{T_{ab}^2};\nn\\
  \hat{E}_{ab (2)} &=& -\Big(E_{ab (2)}-\frac{15}{112}\frac{T_a T_b}{m_a m_b} \frac{\mu_{ab}^2}{T_{ab}^2}\Big). 
\end{eqnarray}
 Introducing summation over all ``b'' species and separating the self-collisions, the final results are given by 
 (\ref{eq:FinalQ3}), (\ref{eq:Final_Q3t}).

\newpage
\subsection{5th-order moment exchange rates}
We need to calculate collisional contributions for the right hand side of the evolution equation for vector $X^{a(5)}_i$,
which is obtained by calculating
\begin{eqnarray}
\vec{\boldsymbol{Q}}^{(5)}_{ab} &=& \trace\trace\bQ^{(5)}_{ab} = \frac{\delta \vec{\boldsymbol{X}}^{(5)}_{ab}}{\delta t}= m_a \int \bc_a |\bc_a|^4 C_{ab}(f_a)d^3v\nn\\
&=& m_a \int f_a \Big( \boldsymbol{A}^{ab}|\bc_a|^4 +4(\boldsymbol{A}^{ab}\cdot\bc_a)|\bc_a|^2 \bc_a \Big) d^3v \nn\\
&& +m_a\int f_a \Big( 4(\bD^{ab}\cdot\bc_a)|\bc_a|^2+4(\bD^{ab}:\bc_a\bc_a)\bc_a +2(\trace\bD^{ab})|\bc_a|^2\bc_a\Big) d^3v. \label{eq:5exchange}
\end{eqnarray}
Again, before integration of (\ref{eq:5exchange}) it is useful to apply the semi-linear approximation, which calculates step by step
\begin{eqnarray}
  \boldsymbol{A}^{ab}|\bc_a|^4 &\simeq & 2\frac{c_{ab}}{m_a^2}\big(1+\frac{m_a}{m_b}\big)
  \frac{n_b m_b}{T_b} \Big(\frac{T_a}{m_a}\Big)^2 \bigg[ \nn\\
    && + \alpha\Big(\alpha^3\ty^4\tby+4\alpha^2\ty^2 \tby(\tby\cdot\tbu)\Big)
    \Big( \sqrt{\frac{2}{\pi}}\frac{e^{-\ty^2/2}}{\ty^2}-\frac{\erf(\ty/\sqrt{2})}{\ty^3}\Big) \nn\\
    && -\,\sqrt{\frac{2}{\pi}} \frac{e^{-\ty^2/2}}{10} \alpha^4 \ty^4 \Big( \tbh^{b(3)}-\tby (\tby\cdot\tbh^{b(3)})
    +(\ty^2-5)\frac{\tbh^{b(5)}}{28}-(\ty^2-7)\frac{\tby(\tby\cdot\tbh^{b(5)})}{28} \Big)\bigg]; 
\end{eqnarray}
\begin{eqnarray}
  (\boldsymbol{A}^{ab}\cdot\bc_a)\bc_a|\bc_a|^2 &\simeq& 2\frac{c_{ab}}{m_a^2}\big(1+\frac{m_a}{m_b}\big)
  \frac{n_b m_b}{T_b} \Big(\frac{T_a}{m_a}\Big)^2 \bigg[ \nn\\
    && +\alpha \Big(\alpha^3\ty^4\tby+\alpha^2 \ty^4\tbu +3\alpha^2\ty^2 \tby(\tby\cdot\tbu)\Big)
    \Big( \sqrt{\frac{2}{\pi}}\frac{e^{-\ty^2/2}}{\ty^2}-\frac{\erf(\ty/\sqrt{2})}{\ty^3}\Big) \nn\\
    && -\,\sqrt{\frac{2}{\pi}} \frac{e^{-\ty^2/2}}{10} \alpha^4 \ty^2 \Big( \tby(\tby\cdot\tbh^{b(3)})(1-\ty^2)
    +\frac{\tby(\tby\cdot\tbh^{b(5)})}{28}(8\ty^2-\ty^4-5) \Big)\bigg]; 
\end{eqnarray}
\begin{eqnarray}
\bD^{ab}:\bc_a\bc_a &\simeq& 2\frac{c_{ab}}{m_a^2} n_b \sqrt{\frac{m_b}{T_b}} \frac{T_a}{m_a}\bigg\{ 
  \Big(\alpha^2\ty^2 +2\alpha(\tby\cdot\tbu) \Big) A_1'' \nn\\
  && + \alpha^2\ty^2 (\tby\cdot\tbh^{b(3)}) \Big( \tilde{A}_3'' +2\frac{\tilde{A}_3'}{\ty}\Big) 
   + \alpha^2\ty^2 (\tby\cdot\tbh^{b(5)}) \Big( \tilde{A}_5'' +2\frac{\tilde{A}_5'}{\ty}\Big) \bigg\};
\end{eqnarray}
\begin{eqnarray}
  \bD^{ab}\cdot\bc_a|\bc_a|^2 &\simeq& 2\frac{c_{ab}}{m_a^2} \frac{n_b}{\alpha} \frac{T_a}{m_a}\bigg\{ \alpha^3\ty^2 \tby \tilde{A}_1''
  +\alpha^2\tbu \ty \tilde{A}_1'
   +\alpha^2 \tby (\tby\cdot\tbu)\Big( 3 \tilde{A}_1'' -\frac{\tilde{A}_1'}{\ty}\Big) \nn\\
   && +\alpha^3\ty^2 \Big[ \tbh^{b(3)} \ty \tilde{A}_3' + \tby(\tby\cdot\tbh^{b(3)}) \Big( \frac{\tilde{A}_3'}{\ty}+\tilde{A}_3''\Big)\Big]\nn\\
   && +\alpha^3\ty^2 \Big[ \tbh^{b(5)} \ty \tilde{A}_5' + \tby(\tby\cdot\tbh^{b(5)}) \Big( \frac{\tilde{A}_5'}{\ty}+\tilde{A}_5''\Big)\Big]
  \bigg\};
\end{eqnarray}
\begin{eqnarray}
\big(\bD^{ab}:\bc_a\bc_a\big)\bc_a &\simeq& 2\frac{c_{ab}}{m_a^2} \frac{n_b}{\alpha} \frac{T_a}{m_a}\bigg\{ 
  \Big(\alpha^3\ty^2 \tby +\alpha^2\ty^2\tbu +2\alpha^2 \tby(\tby\cdot\tbu)  \Big) A_1'' \nn\\
  && + \alpha^3\ty^2 \tby (\tby\cdot\tbh^{b(3)}) \Big( 2\frac{\tilde{A}_3'}{\ty} +\tilde{A}_3'' \Big) 
   + \alpha^3\ty^2 \tby (\tby\cdot\tbh^{b(5)}) \Big( 2\frac{\tilde{A}_5'}{\ty} + \tilde{A}_5''\Big) \bigg\};
\end{eqnarray}
\begin{eqnarray}
  2(\trace \bD^{ab})\bc_a |\bc_a|^2 &\simeq& 2\frac{c_{ab}}{m_a^2}  \frac{n_b}{\alpha} \frac{T_a}{m_a}\bigg\{
  \Big(\alpha^3 \ty^2 \tby+\alpha^2\ty^2\tbu +2\alpha^2\tby(\tby\cdot\tbu) \Big)\Big( 4\frac{\tilde{A}_1'}{\ty}
  +2\tilde{A}_1'' \Big) \nn\\
  && + \alpha^3 \ty^2 \tby(\tby\cdot\tbh^{b(3)}) \Big( 8\frac{\tilde{A}_3'}{\ty} + 2\tilde{A}_3'' \Big)
  +\alpha^3 \ty^2 \tby (\tby\cdot\tbh^{b(5)}) \Big( 8\frac{\tilde{A}_5'}{\ty} +2 \tilde{A}_5'' \Big) \bigg\}.
\end{eqnarray}
Collecting results together, the first part of (\ref{eq:5exchange}) becomes
\begin{eqnarray}
&&  \boldsymbol{A}^{ab}|\bc_a|^4+  4(\boldsymbol{A}^{ab}\cdot\bc_a)\bc_a|\bc_a|^2 \simeq 2\frac{c_{ab}}{m_a^2}\big(1+\frac{m_a}{m_b}\big)
  \frac{n_b}{\alpha^2} \frac{T_a}{m_a} \bigg\{ \nn\\
    && +\alpha \Big( 5\alpha^3\ty^4\tby+4\alpha^2 \ty^4\tbu +16\alpha^2\ty^2 \tby(\tby\cdot\tbu)\Big)
    \Big( \sqrt{\frac{2}{\pi}}\frac{e^{-\ty^2/2}}{\ty^2}-\frac{\erf(\ty/\sqrt{2})}{\ty^3}\Big) \nn\\
    && -\,\sqrt{\frac{2}{\pi}} \frac{e^{-\ty^2/2}}{10} \alpha^4 \ty^2 \Big( \ty^2 \tbh^{b(3)} +\tby(\tby\cdot\tbh^{b(3)})(4-5\ty^2)
    +\ty^2 (\ty^2-5)\frac{\tbh^{b(5)}}{28} \nn\\
    && \qquad +\frac{\tby(\tby\cdot\tbh^{b(5)})}{28}(39\ty^2-5\ty^4-20) \Big)\bigg\}, \label{eq:Thierry15}
\end{eqnarray}
and the second part of (\ref{eq:5exchange}) reads
\begin{eqnarray}
&&  4(\bD^{ab}\cdot\bc_a)|\bc_a|^2+4(\bD^{ab}:\bc_a\bc_a)\bc_a +2(\trace\bD^{ab})|\bc_a|^2\bc_a \simeq
  2\frac{c_{ab}}{m_a^2} \frac{n_b}{\alpha} \frac{T_a}{m_a}\bigg\{\nn\\
  &&\qquad  \alpha^3\ty^2 \tby \Big( 4\frac{\tilde{A}_1'}{\ty}+10\tilde{A}_1''\Big)
  +\alpha^2\ty^2\tbu \Big( 8\frac{\tilde{A}_1'}{\ty} +6\tilde{A}_1''\Big)
  +\alpha^2\tby(\tby\cdot\tbu)\Big(4\frac{\tilde{A}_1'}{\ty}+  24 \tilde{A}_1''\Big) \nn\\
  &&\qquad +4\alpha^3\ty^3 \tbh^{b(3)} \tilde{A}_3' +\alpha^3\ty^2\tby (\tby\cdot\tbh^{b(3)})\Big( 20\frac{\tilde{A}_3'}{\ty}+10\tilde{A}_3''\Big) \nn\\
  &&\qquad +4\alpha^3\ty^3 \tbh^{b(5)} \tilde{A}_5' +\alpha^3\ty^2\tby (\tby\cdot\tbh^{b(5)})\Big( 20\frac{\tilde{A}_5'}{\ty}+10\tilde{A}_5''\Big)\bigg\}. \label{eq:Thierry16}
\end{eqnarray}
Now (\ref{eq:5exchange}) can be integrated, and the entire collisional integral can be written in a symbolic form
\begin{eqnarray}
 \vec{\boldsymbol{Q}}^{(5)}_{ab} &=& m_a n_a 3 \alpha^3 \nu_{ab} \sqrt{\frac{\pi}{2}} \big(1+\alpha^2\big)^{3/2} \Big(\frac{T_a}{m_a}\Big)^{5/2} \times\nn\\
  && \int \frac{e^{-\alpha^2\ty^2/2}}{(2\pi)^{3/2}}
  \big(1-\alpha \tby\cdot\tbu +\chi_a\big) \Big[ \frac{1}{\alpha^2} \{\ref{eq:Thierry15}\}
    + \frac{1}{(1+\frac{m_a}{m_b})\alpha} \{\ref{eq:Thierry16}\} \Big] d^3\ty,
\end{eqnarray}
where \{\ref{eq:Thierry15}\} \& \{\ref{eq:Thierry16}\} represent only parts of the corresponding equations that are inside of curly brackets. 
 The integration yields
\begin{eqnarray}
  \vec{\boldsymbol{Q}}^{(5)}_{ab} &=& \trace\trace \bQ_{ab}^{(5)} = \frac{\delta \vec{\boldsymbol{X}}^{(5)}_{ab}}{\delta t}
  = \nu_{ab} \frac{T_a}{m_a}\Big\{ + p_a (\bu_b-\bu_a) U_{ab (2)}\nn\\
  && - F_{ab (1)} \frac{p_a}{2} \sqrt{\frac{T_a}{m_a}} \tbh^{a(3)}
  -  F_{ab (2)}  \frac{\rho_a}{\rho_b} \frac{p_b}{2}\sqrt{\frac{T_b}{m_b}}\tbh^{b(3)}\nn\\
  && -  G_{ab (1)} p_a \sqrt{\frac{T_a}{m_a}} \tbh^{a(5)} 
  + G_{ab (2)} \frac{\rho_a}{\rho_b} p_b \sqrt{\frac{T_b}{m_b}}\tbh^{b(5)} \Big\},
\end{eqnarray}
 with mass-ratio coefficients
\begin{eqnarray}
  U_{ab (2)} &=& -\, \frac{16 T_a^2 m_a m_b^2-8 T_a^2 m_b^3+56 T_a T_b m_a^2 m_b-52 T_a T_b m_a m_b^2-35 T_b^2 m_a^3-119 T_b^2 m_a^2 m_b}{(T_a m_b+T_b m_a)^2 (m_b+m_a)};\nn\\
  F_{ab (1)} &=& \big\{40 T_a^4 m_a m_b^3+8 T_a^4 m_b^4+180 T_a^3 T_b m_a^2 m_b^2+68 T_a^3 T_b m_a m_b^3+315 T_a^2 T_b^2 m_a^3 m_b
    +207 T_a^2 T_b^2 m_a^2 m_b^2 \nn\\
    && +700 T_a T_b^3 m_a^4+392 T_a T_b^3 m_a^3 m_b-280 T_b^4 m_a^4\big\} \big[5 (T_a m_b+T_b m_a)^3 (m_b+m_a) T_a\big]^{-1};\nn\\
    F_{ab (2)} &=& -\, \frac{3  T_a m_b \big[16 T_a^2 m_b^3+140 T_a T_b m_a^2 m_b+72 T_a T_b m_a m_b^2-35 T_b^2 m_a^3-119 T_b^2 m_a^2 m_b\big] }{5 (T_a m_b+T_b m_a)^3 (m_b+m_a)};\nn\\
    G_{ab (1)} &=&  -\, \big\{40 T_a^4 m_a m_b^4+8 T_a^4 m_b^5+220 T_a^3 T_b m_a^2 m_b^3+140 T_a^3 T_b m_a m_b^4 +495 T_a^2 T_b^2 m_a^3 m_b^2 \nn\\
    && +627 T_a^2 T_b^2 m_a^2 m_b^3+3640 T_a T_b^3 m_a^4 m_b+1916 T_a T_b^3 m_a^3 m_b^2-1400 T_b^4 m_a^5 \nn\\
    && -3304 T_b^4 m_a^4 m_b \big\} \big[280 (T_a m_b+T_b m_a)^4 (m_a+m_b)\big]^{-1};\nn\\
 G_{ab (2)} &=&  \frac{3  T_a T_b m_a^2 m_b\big[8 T_a^2 m_b^2-32 T_a T_b m_a m_b-28 T_a T_b m_b^2+5 T_b^2 m_a^2+17 T_b^2 m_a m_b\big]}{8 (T_a m_b+T_b m_a)^4 (m_a+m_b)}.\nn\\   
\end{eqnarray}
 For small temperature differences the mass-ratio coefficients simplify into
\begin{eqnarray}
&& U_{ab (2)} = \frac{35 m_a^2+28 m_a m_b+8 m_b^2}{(m_a+m_b)^2};\nn\\  
&& F_{ab (1)} = \frac{420 m_a^3+287 m_a^2 m_b+100 m_a m_b^2+8m_b^3}{5(m_a+m_b)^3};\nn\\  
&& F_{ab (2)} = \frac{3}{5}\frac{m_b(35 m_a^2-56m_a m_b -16m_b^2)}{(m_a+m_b)^3};\nn\\
&& G_{ab (1)} = \frac{1400 m_a^4 -1736 m_a^3 m_b -675 m_a^2 m_b^2 -172 m_a m_b^3-8m_b^4}{280 (m_a+m_b)^4};\nn\\
&& G_{ab (2)} = \frac{15}{8} \frac{m_a^2 m_b(m_a-4m_b)}{(m_a+m_b)^4}.  
\end{eqnarray}
 Rewritten with fluid moments, the exchange rates for the 5th-order moment become
\begin{eqnarray}
  \vec{\boldsymbol{Q}}^{(5)}_{ab}\,' &=& \vec{\boldsymbol{Q}}^{(5)}_{ab} -35\frac{p_a^2}{\rho_a^2} \boldsymbol{R}_{ab}
  = \frac{\delta \vec{\boldsymbol{X}}^{(5)}_{ab}\,'}{\delta t}
  = \nu_{ab}\Big\{ - \frac{p_a^2}{\rho_a} (\bu_b-\bu_a)  \hat{U}_{ab (2)}      \nn\\
  && -  \hat{F}_{ab (1)} \frac{p_a}{\rho_a} \vecq_a +\hat{F}_{ab (2)}  \frac{p_a}{\rho_a} \frac{\rho_a}{\rho_b} \vecq_b
  -\hat{G}_{ab (1)} \vecX^{(5)}_a - \hat{G}_{ab (2)} \frac{p_a}{p_b}\vecX^{(5)}_b \Big\}, 
\end{eqnarray}
 with mass-ratio coefficients (introducing hat)
\begin{eqnarray}
  \hat{U}_{ab (2)} &=& -\big( U_{ab (2)} -35 \big);\nn\\
  \hat{F}_{ab (1)} &=& F_{ab (1)} -28 G_{ab (1)} +35 \frac{T_a}{m_a}\frac{\mu_{ab}}{T_{ab}}V_{ab (1)};\nn\\
  \hat{F}_{ab (2)} &=& -\Big(F_{ab (2)} +28 G_{ab (2)} -35 \frac{T_a}{m_a}\frac{\mu_{ab}}{T_{ab}}V_{ab (2)}\Big);\nn\\
  \hat{G}_{ab (1)} &=& G_{ab (1)}-\frac{15}{8}\frac{T_a^2}{m_a^2}\frac{\mu_{ab}^2}{T_{ab}^2};\nn\\
  \hat{G}_{ab (2)} &=& -\Big( G_{ab (2)} -\frac{15}{8} \frac{T_a T_b}{m_a m_b}\frac{\mu_{ab}^2}{T_{ab}^2}\Big). 
\end{eqnarray}
 The final results are given by (\ref{eq:FinalQ5}), (\ref{eq:Final_Q5t}).


\newpage
\section{Braginskii viscosity (15-moment model)} \label{sec:BragVisc}
\setcounter{equation}{0}
We use polynomials derived from the \emph{reducible} Hermite polymials  (see details in Appendix \ref{sec:Hermite}),
with perturbation of the distribution function $f_b (\bV')=f_b^{(0)}(1+\chi_b)$
\begin{equation} \label{eq:Energy70}
\chi_b = \frac{1}{2}\tilde{h}^{b(2)}_{ij}\tilde{H}^{b(2)}_{ij}  +\frac{1}{28}\hat{h}^{b(4)}_{ij}\hat{H}^{b(4)}_{ij}.
\end{equation}
For clarity of calculations, we here only consider the viscous part of $\chi_b$ (i.e. the 15-moment model) but the full 22-moment model
 can be implicitly assumed for the final
collisional contributions at the semi-linear level. The Hermite polynomials are (dropping species index ``b'' for polynomials and velocities $\tbc$) 
\begin{eqnarray}
 \tilde{H}^{(2)}_{ij} &=& \tc_i \tc_j -\delta_{ij};\nn\\
 \tilde{H}^{(4)}_{ij} &\equiv& \tilde{H}^{(4)}_{ijkk} = \big( \tc_i\tc_j-\frac{\delta_{ij}}{3}\tc^2\big) (\tc^2-7) +\frac{\delta_{ij}}{3} \tilde{H}^{(4)};\nn\\
 \tilde{H}^{(4)} &\equiv& \tilde{H}^{(4)}_{iikk} = \tc^4-10\tc^2+15;\nn\\
 \hat{H}^{(4)}_{ij} &\equiv&  \tilde{H}^{(4)}_{ij} - \frac{\delta_{ij}}{3} \tilde{H}^{(4)} = \big( \tc_i\tc_j-\frac{\delta_{ij}}{3}\tc^2\big) (\tc^2-7).
\end{eqnarray}
The irreducible polynomials yield the same perturbation $\chi_b$. 
By using the perturbation (\ref{eq:Energy70}) one can calculate fluid moments $\Pi_{ij}^{b(2)}$, $\Pi_{ij}^{b(4)}$, 
or one can directly calculate Hermite moments 
\begin{eqnarray}
&&  \tilde{h}_{ij}^{b(2)} = \frac{1}{p_b}\Pi_{ij}^{b(2)};\qquad  \hat{h}_{ij}^{b(4)} = \frac{\rho_b}{p_b^2} \Pi_{ij}^{b(4)} - \frac{7}{p_b}\Pi_{ij}^{b(2)},\label{eq:Herm4}
\end{eqnarray}
yielding the same relations. Both $\tilde{h}_{ij}^{(2)}$ and $\hat{h}_{ij}^{(4)}$ are traceless (and $\tilde{h}_{ij}^{(2)}=\hat{h}_{ij}^{(2)}$).

  
\subsection{Rosenbluth potentials}
The notation reads
\begin{equation}
\tbc_b = \sqrt{\frac{m_b}{T_b}} (\bV'-\bu_b);\qquad |\bV'-\bV|=\sqrt{\frac{T_b}{m_b}}|\tbc_b-\tby|; \qquad \tby = \sqrt{\frac{m_b}{T_b}}(\bV-\bu_b),\nn
\end{equation}  
the Rosenbluth potentials are
\begin{eqnarray}
  H_b (\bV) &=& \int \frac{f_b(\bV')}{|\bV'-\bV|}d^3v' = n_b \sqrt{\frac{m_b}{T_b}} \int \frac{\phi^{(0)}_b}{|\tbc_b-\tby|}(1+\chi_b)d^3\tc_b;\nn\\
  G_b (\bV) &=& \int |\bV'-\bV| f_b(\bV') d^3v' = n_b \sqrt{\frac{T_b}{m_b}} \int  |\tbc_b-\tby|\phi^{(0)}_b  (1+\chi_b)d^3\tc_b,\nn
\end{eqnarray}
and further calculate
\begin{eqnarray}
  H_b (\bV) &=& n_b \sqrt{\frac{m_b}{T_b}} \Big\{ \frac{1}{\ty}\erf\Big(\frac{\ty}{\sqrt{2}}\Big)
  +\frac{1}{2} (\tbbh^{b(2)}:\tby\tby) \Big[ \frac{3}{\ty^5}\erf\Big(\frac{\ty}{\sqrt{2}}\Big)
    -\sqrt{\frac{2}{\pi}}\big(\frac{1}{\ty^2}+\frac{3}{\ty^4}\big)e^{-\ty^2/2}\Big] \nn\\
  && \qquad -\frac{1}{28} (\hbbh^{b(4)}:\tby\tby)  \sqrt{\frac{2}{\pi}} e^{-\ty^2/2} \Big\};\label{eq:HbVisc}\\
  G_b (\bV) &=& n_b \sqrt{\frac{T_b}{m_b}} \Big\{ \sqrt{\frac{2}{\pi}}e^{-\ty^2/2} +\big(\ty+\frac{1}{\ty}\big)\erf\big(\frac{\ty}{\sqrt{2}}\big) \nn\\
  && \qquad -\frac{1}{2}(\tbbh^{b(2)}:\tby\tby) \Big[ \frac{3}{\ty^4} \sqrt{\frac{2}{\pi}} e^{-\ty^2/2}
    +\big(\frac{1}{\ty^3}-\frac{3}{\ty^5}\big)\erf\Big(\frac{\ty}{\sqrt{2}}\Big) \Big] \nn\\
  && \qquad -\frac{1}{14} (\hbbh^{b(4)}:\tby\tby)\Big[ \sqrt{\frac{2}{\pi}}\big( \frac{1}{\ty^2}+\frac{3}{\ty^4}\big) e^{-\ty^2/2}
    -\frac{3}{\ty^5}\erf\Big(\frac{\ty}{\sqrt{2}}\Big) \Big] \Big\}.
\end{eqnarray}
The derivative of the first Rosenbluth potential becomes
\begin{eqnarray}
  \frac{\pr H_b}{\pr \bV} &=& \frac{n_b m_b}{T_b} \Big\{ \tby \Big( \sqrt{\frac{2}{\pi}}\frac{e^{-\ty^2/2}}{\ty^2}-\frac{\erf(\ty/\sqrt{2})}{\ty^3}\Big) \nn\\
  && +(\tbbh^{b(2)}\cdot\tby)\Big[ \frac{3}{\ty^5}\erf\Big(\frac{\ty}{\sqrt{2}}\Big)
    -\sqrt{\frac{2}{\pi}}\big(\frac{1}{\ty^2}+\frac{3}{\ty^4}\big)e^{-\ty^2/2}\Big] \nn\\
  && + \frac{1}{2} (\tbbh^{b(2)}:\tby\tby) \tby \Big[ -\frac{15}{\ty^7}\erf\Big(\frac{\ty}{\sqrt{2}}\Big)
    +\sqrt{\frac{2}{\pi}}\big(\frac{1}{\ty^2}+\frac{5}{\ty^4}+\frac{15}{\ty^6}\big) e^{-\ty^2/2}\Big]\nn\\
  && -\frac{1}{14} (\hbbh^{b(4)}\cdot\tby)  \sqrt{\frac{2}{\pi}} e^{-\ty^2/2} +\frac{1}{28} (\hbbh^{b(4)}:\tby\tby)\tby \sqrt{\frac{2}{\pi}} e^{-\ty^2/2}
  \Big\}. \label{eq:Hbder}
\end{eqnarray}
For the second Rosenbluth potential it is useful to use a form
\begin{eqnarray}
G_b(\bV) = n_b \sqrt{\frac{T_b}{m_b}}\Big[ \tilde{A}_1 +\tilde{A}_2 (\tbbh^{b(2)}:\tby\tby) +\tilde{A}_4 (\hbbh^{b(4)}:\tby\tby) \Big],
\end{eqnarray}
where 
\begin{eqnarray}
\tilde{A}_1 &=& \sqrt{\frac{2}{\pi}}e^{-\ty^2/2} +\big(\ty+\frac{1}{\ty}\big)\erf\big(\frac{\ty}{\sqrt{2}}\big);\nn\\
\tilde{A}_2 &=&  -\frac{1}{2} \Big[ \frac{3}{\ty^4} \sqrt{\frac{2}{\pi}} e^{-\ty^2/2}
  +\big(\frac{1}{\ty^3}-\frac{3}{\ty^5}\big)\erf\Big(\frac{\ty}{\sqrt{2}}\Big) \Big];\nn\\
\tilde{A}_4 &=& -\frac{1}{14} \Big[ \sqrt{\frac{2}{\pi}}\big( \frac{1}{\ty^2}+\frac{3}{\ty^4}\big) e^{-\ty^2/2}
    -\frac{3}{\ty^5}\erf\Big(\frac{\ty}{\sqrt{2}}\Big) \Big].
\end{eqnarray}
Its second derivative then calculates
\begin{eqnarray}
  \frac{\pr}{\pr \bV}\frac{\pr G_b}{\pr\bV} &=& n_b \sqrt{\frac{m_b}{T_b}}\Big\{
  \bI \frac{\tilde{A}_1'}{\ty}+\frac{\tby\tby}{\ty^2}\Big( \tilde{A}_1''-\frac{\tilde{A}_1'}{\ty}\Big)\nn\\
  &&  +\Big[ 2\tby (\tbbh^{b(2)}\cdot\tby) +2 (\tbbh^{b(2)}\cdot\tby)\tby +\bI (\tbbh^{b(2)}:\tby\tby)\Big] \frac{\tilde{A}_2'}{\ty}\nn\\
  && + 2\tilde{A}_2 \tbbh^{b(2)} + \frac{\tby\tby}{\ty^2}(\tbbh^{b(2)}:\tby\tby) \Big( \tilde{A}_2''-\frac{\tilde{A}_2'}{\ty}\Big) \nn\\
  &&  +\Big[ 2\tby (\hbbh^{b(4)}\cdot\tby) +2 (\hbbh^{b(4)}\cdot\tby)\tby +\bI (\hbbh^{b(4)}:\tby\tby)\Big] \frac{\tilde{A}_4'}{\ty}\nn\\
  && + 2\tilde{A}_4 \hbbh^{b(4)} + \frac{\tby\tby}{\ty^2}(\hbbh^{b(4)}:\tby\tby) \Big( \tilde{A}_4''-\frac{\tilde{A}_4'}{\ty}\Big)\Big\}, \label{eq:2ndG}
\end{eqnarray}
with coefficients
\begin{eqnarray}
  \tilde{A}_1' &=& \sqrt{\frac{2}{\pi}}\frac{e^{-\ty^2/2}}{\ty}+\big(1-\frac{1}{\ty^2}\big)\erf\big(\frac{\ty}{\sqrt{2}}\big);\nn\\
  \tilde{A}_2' &=& \big( \frac{1}{\ty^3}+\frac{15}{2\ty^5}\big)\sqrt{\frac{2}{\pi}} e^{-\ty^2/2}
  +\frac{3}{2}\big( \frac{1}{\ty^4}-\frac{{ 5}}{\ty^6}\big)\erf\big(\frac{\ty}{\sqrt{2}}\big) ;\nn\\
  \tilde{A}_4' &=& \frac{1}{14}\big(\frac{1}{\ty}+\frac{5}{\ty^3}+\frac{15}{\ty^5}\big)\sqrt{\frac{2}{\pi}} e^{-\ty^2/2}
  -\frac{15}{14\ty^6} \erf\big(\frac{\ty}{\sqrt{2}}\big);\nn\\
  \tilde{A}_1'' &=& -\,  \frac{2}{\ty^2} \sqrt{\frac{2}{\pi}} e^{-\ty^2/2} +\frac{2}{\ty^3}\erf\big(\frac{\ty}{\sqrt{2}}\big);\nn\\
  \tilde{A}_2'' &=& -\big( \frac{1}{\ty^2}+\frac{9}{\ty^4}+\frac{45}{\ty^6}\big)\sqrt{\frac{2}{\pi}}e^{-\ty^2/2}
  +\big( -\frac{6}{\ty^5}+\frac{45}{\ty^7}\big) \erf\big(\frac{\ty}{\sqrt{2}}\big);\nn\\
  \tilde{A}_4'' &=& -\, \frac{1}{14}\big( 1+\frac{6}{\ty^2}+\frac{30}{\ty^4}+\frac{90}{\ty^6}\big)\sqrt{\frac{2}{\pi}}e^{-\ty^2/2}
  +\frac{45}{7\ty^7}\erf\big(\frac{\ty}{\sqrt{2}}\big),
\end{eqnarray}
and
\begin{eqnarray}
  \tilde{A}_1'' -\frac{\tilde{A}_1'}{\ty} &=& -\,  \frac{3}{\ty^2} \sqrt{\frac{2}{\pi}} e^{-\ty^2/2}
  -\big(\frac{1}{\ty}-\frac{3}{\ty^3}\big)\erf\big(\frac{\ty}{\sqrt{2}}\big);\nn\\
  \tilde{A}_2'' -\frac{\tilde{A}_2'}{\ty} &=& -\big( \frac{1}{\ty^2}+\frac{10}{\ty^4}+\frac{105}{2\ty^6}\big)\sqrt{\frac{2}{\pi}} e^{-\ty^2/2}
  -\frac{15}{2}\big(\frac{1}{\ty^5}-\frac{7}{\ty^7}\big)\erf\big(\frac{\ty}{\sqrt{2}}\big);\nn\\
  \tilde{A}_4'' -\frac{\tilde{A}_4'}{\ty} &=& -\,\frac{1}{14}\big(1+\frac{7}{\ty^2}+\frac{35}{\ty^4}+\frac{105}{\ty^6}\big)\sqrt{\frac{2}{\pi}} e^{-\ty^2/2}
  +\frac{15}{2\ty^7}\erf\big(\frac{\ty}{\sqrt{2}}\big).
\end{eqnarray}
As a double-check, applying $(1/2)\trace$ on (\ref{eq:2ndG}) yields
\begin{eqnarray}
  \frac{1}{2}\trace \frac{\pr}{\pr \bV}\frac{\pr G_b}{\pr\bV} &=& n_b \sqrt{\frac{m_b}{T_b}} \frac{1}{2}\Big\{ \tilde{A}_1'' +2\frac{\tilde{A}_1'}{\ty}
  +(\tbbh^{b(2)}:\tby\tby)\big( \tilde{A}_2''+6 \frac{\tilde{A}_2'}{\ty}\big)\nn\\
  && +(\hbbh^{b(4)}:\tby\tby)\big( \tilde{A}_4''+6 \frac{\tilde{A}_4'}{\ty}\big)\Big\} = H_b,
\end{eqnarray}
recovering the first Rosenbluth potential (\ref{eq:HbVisc}). Similarly, applying $(\pr/\pr\bV)\cdot$ on (\ref{eq:Hbder}) recovers
$-4\pi f_b (\bV)$. Both Rosenbluth potentials seem to be calculated correctly.

\subsection{Dynamical friction vector and diffusion tensor}
The dynamical friction vector becomes
\begin{eqnarray}
  \boldsymbol{A}^{ab} &=& 2\frac{c_{ab}}{m_a^2}\big(1+\frac{m_a}{m_b}\big)
  \frac{n_b m_b}{T_b} \Big\{ \tby \Big( \sqrt{\frac{2}{\pi}}\frac{e^{-\ty^2/2}}{\ty^2}-\frac{\erf(\ty/\sqrt{2})}{\ty^3}\Big) \nn\\
  && +(\tbbh^{b(2)}\cdot\tby)\Big[ \frac{3}{\ty^5}\erf\Big(\frac{\ty}{\sqrt{2}}\Big)
    -\sqrt{\frac{2}{\pi}}\big(\frac{1}{\ty^2}+\frac{3}{\ty^4}\big)e^{-\ty^2/2}\Big] \nn\\
  && + \frac{1}{2} (\tbbh^{b(2)}:\tby\tby) \tby \Big[ -\frac{15}{\ty^7}\erf\Big(\frac{\ty}{\sqrt{2}}\Big)
    +\sqrt{\frac{2}{\pi}}\big(\frac{1}{\ty^2}+\frac{5}{\ty^4}+\frac{15}{\ty^6}\big) e^{-\ty^2/2}\Big]\nn\\
  && -\frac{1}{14} (\hbbh^{b(4)}\cdot\tby)  \sqrt{\frac{2}{\pi}} e^{-\ty^2/2} +\frac{1}{28} (\hbbh^{b(4)}:\tby\tby)\tby \sqrt{\frac{2}{\pi}} e^{-\ty^2/2}
  \Big\}, \label{eq:DynVec}
\end{eqnarray}
and the diffusion tensor
\begin{eqnarray}
  \bD^{ab} &=& 2\frac{c_{ab}}{m_a^2} n_b \sqrt{\frac{m_b}{T_b}}\Big\{
  \bI \frac{\tilde{A}_1'}{\ty}+\frac{\tby\tby}{\ty^2}\Big( \tilde{A}_1''-\frac{\tilde{A}_1'}{\ty}\Big)\nn\\
  &&  +\Big[ 2\tby (\tbbh^{b(2)}\cdot\tby) +2 (\tbbh^{b(2)}\cdot\tby)\tby +\bI (\tbbh^{b(2)}:\tby\tby)\Big] \frac{\tilde{A}_2'}{\ty}\nn\\
  && + 2\tilde{A}_2 \tbbh^{b(2)} + \frac{\tby\tby}{\ty^2}(\tbbh^{b(2)}:\tby\tby) \Big( \tilde{A}_2''-\frac{\tilde{A}_2'}{\ty}\Big) \nn\\
  &&  +\Big[ 2\tby (\hbbh^{b(4)}\cdot\tby) +2 (\hbbh^{b(4)}\cdot\tby)\tby +\bI (\hbbh^{b(4)}:\tby\tby)\Big] \frac{\tilde{A}_4'}{\ty}\nn\\
  && + 2\tilde{A}_4 \hbbh^{b(4)} + \frac{\tby\tby}{\ty^2}(\hbbh^{b(4)}:\tby\tby) \Big( \tilde{A}_4''-\frac{\tilde{A}_4'}{\ty}\Big)\Big\}, \label{eq:DiffTensor}
\end{eqnarray}
where $c_{ab} = 2\pi e^4 Z_a^2 Z_b^2\ln\Lambda$. 

\newpage
\subsection{Distribution function for species ``a''}
To avoid the complicated runaway effect, the distribution function $f_a(\bV) = f_a^{(0)}(1+\chi_a)$ has to be expanded for small drifts,
in the semi-linear approximation. 
Following the derivation and notation introduced in Section \ref{sec:FA}, the expanded distribution function becomes
\begin{equation}
f_a =  n_a\Big(\frac{m_a}{T_a} \Big)^{3/2} \frac{e^{-\alpha^2\ty^2/2}}{(2\pi)^{3/2}} (1-\alpha\tby\cdot\tbu +\chi_a),
\end{equation}
now with perturbation
\begin{equation}
\chi_a = \frac{1}{2}\tilde{h}^{a(2)}_{ij}\tilde{H}^{a(2)}_{ij}(\alpha\tby)  +\frac{1}{28}\hat{h}^{a(4)}_{ij}\hat{H}^{a(4)}_{ij}(\alpha\tby),
\end{equation}
where
\begin{eqnarray}
&&  \tilde{h}^{a(2)}_{ij} \tilde{H}^{a(2)}_{ij} (\alpha\tby) = \tilde{h}^{a(2)}_{ij} \alpha^2 \ty_i \ty_j;\nn\\
&&  \hat{h}^{a(4)}_{ij} \hat{H}^{a(4)}_{ij} (\alpha\tby) = \hat{h}^{a(4)}_{ij} \alpha^2 \ty_i \ty_j (\alpha^2 \ty^2 -7),
\end{eqnarray}
so the perturbation reads
\begin{equation}
\chi_a = \frac{\alpha^2}{2}(\tbbh^{a(2)}:\tby\tby) +\frac{\alpha^2}{28}(\hbbh^{a(4)}:\tby\tby)\big(\alpha^2\ty^2-7\big).
\end{equation}
As a reminder
\begin{equation*}
  \tbu = (\bu_b-\bu_a)\sqrt{\frac{m_a}{T_a}}; \qquad
  \alpha=\frac{\sqrt{T_b/m_b}}{\sqrt{T_a/m_a}}.
\end{equation*}

\subsection{Pressure tensor exchange rates}
We need to calculate collisional contributions for the r.h.s. of the pressure tensor equation, and these contributions read
\begin{equation} \label{eq:ViscoP2}
  \bQ_{ab}^{(2)} = m_a \int f_a \big[ \boldsymbol{A}_{ab}\bc_a\big]^S d^3v +m_a \int f_a \bD_{ab} d^3v.
\end{equation}
By emplying
\begin{equation*}
\bc_a = \sqrt{\frac{T_a}{m_a}} (\alpha\tby+\tbu),
\end{equation*}
in the semi-linear approximation
\begin{eqnarray}
  \boldsymbol{A}^{ab}\bc_a &\simeq& 2\frac{c_{ab}}{m_a^2}\big(1+\frac{m_a}{m_b}\big)
  \frac{n_b m_b}{T_b} \sqrt{\frac{T_a}{m_a}}\Big\{ \big(\alpha\tby\tby+\tby\tbu\big)
  \Big( \sqrt{\frac{2}{\pi}}\frac{e^{-\ty^2/2}}{\ty^2}-\frac{\erf(\ty/\sqrt{2})}{\ty^3}\Big) \nn\\
  && +\alpha(\tbbh^{b(2)}\cdot\tby)\tby \Big[ \frac{3}{\ty^5}\erf\Big(\frac{\ty}{\sqrt{2}}\Big)
    -\sqrt{\frac{2}{\pi}}\big(\frac{1}{\ty^2}+\frac{3}{\ty^4}\big)e^{-\ty^2/2}\Big] \nn\\
  && + \frac{\alpha}{2} (\tbbh^{b(2)}:\tby\tby) \tby\tby \Big[ -\frac{15}{\ty^7}\erf\Big(\frac{\ty}{\sqrt{2}}\Big)
    +\sqrt{\frac{2}{\pi}}\big(\frac{1}{\ty^2}+\frac{5}{\ty^4}+\frac{15}{\ty^6}\big) e^{-\ty^2/2}\Big]\nn\\
  && -\frac{\alpha}{14} (\hbbh^{b(4)}\cdot\tby)\tby  \sqrt{\frac{2}{\pi}} e^{-\ty^2/2} +\frac{\alpha}{28} (\hbbh^{b(4)}:\tby\tby)\tby\tby \sqrt{\frac{2}{\pi}} e^{-\ty^2/2}
  \Big\},
\end{eqnarray}
and
\begin{eqnarray}
  \big[\boldsymbol{A}^{ab}\bc_a\big]^S &\simeq& 2\frac{c_{ab}}{m_a^2}\big(1+\frac{m_a}{m_b}\big)
  \frac{n_b m_b}{T_b} \sqrt{\frac{T_a}{m_a}}\Big\{ \big(2\alpha\tby\tby+\tby\tbu +\tbu\tby\big)
  \Big( \sqrt{\frac{2}{\pi}}\frac{e^{-\ty^2/2}}{\ty^2}-\frac{\erf(\ty/\sqrt{2})}{\ty^3}\Big) \nn\\
  && +\alpha\Big((\tbbh^{b(2)}\cdot\tby)\tby +\tby(\tbbh^{b(2)}\cdot\tby) \Big)\Big[ \frac{3}{\ty^5}\erf\Big(\frac{\ty}{\sqrt{2}}\Big)
    -\sqrt{\frac{2}{\pi}}\big(\frac{1}{\ty^2}+\frac{3}{\ty^4}\big)e^{-\ty^2/2}\Big] \nn\\
  && + \alpha (\tbbh^{b(2)}:\tby\tby) \tby\tby \Big[ -\frac{15}{\ty^7}\erf\Big(\frac{\ty}{\sqrt{2}}\Big)
    +\sqrt{\frac{2}{\pi}}\big(\frac{1}{\ty^2}+\frac{5}{\ty^4}+\frac{15}{\ty^6}\big) e^{-\ty^2/2}\Big]\nn\\
  && -\frac{\alpha}{14} \Big( (\hbbh^{b(4)}\cdot\tby)\tby + \tby(\hbbh^{b(4)}\cdot\tby)  \Big)\sqrt{\frac{2}{\pi}} e^{-\ty^2/2}
  +\frac{\alpha}{14} (\hbbh^{b(4)}:\tby\tby)\tby\tby \sqrt{\frac{2}{\pi}} e^{-\ty^2/2}
  \Big\}.
\end{eqnarray}
The first term of (\ref{eq:ViscoP2}) is rewritten as
\begin{eqnarray}
  m_a \int f_a \big[ \boldsymbol{A}_{ab}\bc_a\big]^S d^3v
  = m_a n_a \alpha^3 \int \frac{e^{-\alpha^2\ty^2/2}}{(2\pi)^{3/2}} (1-\alpha\tby\cdot\tbu +\chi_a) \big[ \boldsymbol{A}_{ab}\bc_a\big]^S d^3 \ty,
\end{eqnarray}
and by using the following integrals
\begin{eqnarray}
  \int \tby\tby f(\ty) e^{-\alpha^2 \ty^2/2} d^3y &=& \bI \frac{4\pi}{3} \int_0^\infty \ty^4 f(\ty) e^{-\alpha^2 \ty^2/2}d\ty;\nn\\
  \int \tby(\tbbh^{b(2)}\cdot\tby) f(\ty) e^{-\alpha^2 \ty^2/2} d^3\ty &=& \tbbh^{b(2)} \frac{4\pi}{3} \int_0^\infty \ty^4 f(\ty) e^{-\alpha^2 \ty^2/2} d\ty;\nn\\
 \tbbh^{b(2)}: \int \tby\tby f(\ty) e^{-\alpha^2 \ty^2/2} d^3\ty &=& 0;\nn\\
  \tbbh^{b(2)}: \int \tby\tby\tby\tby f(\ty) e^{-\alpha^2 \ty^2/2} d^3\ty &=&
  \tbbh^{b(2)}\frac{8\pi}{15}\int_0^\infty \ty^6 f(\ty) e^{-\alpha^2 \ty^2/2} d\ty, \label{eq:formulaP}
\end{eqnarray}
and by further applying the semi-linear approximation it integrates 
\begin{eqnarray}
  m_a \int f_a \big[ \boldsymbol{A}_{ab}\bc_a\big]^S d^3v
  &=& \rho_a \nu_{ab} \Big[ -2\frac{T_a}{m_a}\bI + \frac{6}{5} \tbbh^{b(2)} \frac{T_a T_b}{T_a m_b+T_b m_a}
    -\frac{3}{7} \hbbh^{b(4)} \frac{m_a T_a T_b^2}{(T_a m_b+T_b m_a)^2} \nn\\
    && - \frac{2}{5}\frac{T_a}{m_a} \tbbh^{a(2)} \frac{2T_a m_b+5 T_b m_a}{T_a m_b+T_b m_a}
     + \frac{3}{35} \frac{T_a}{m_a}\hbbh^{a(4)} \frac{T_a m_b (2 T_a m_b +7 T_b m_a)}{(T_a m_b +T_b m_a)^2}\Big].
\end{eqnarray}
Similarly, the second term of (\ref{eq:ViscoP2}) integrates
\begin{eqnarray}
  m_a \int f_a  \bD_{ab} d^3v &=& \frac{\rho_a \nu_{ab}}{m_a+m_b} \Big[ \bI \frac{2}{m_a} (T_a m_b+T_b m_a)
    -\frac{2}{5} T_b \tbbh^{b(2)} +\frac{3}{35} \frac{m_a T_b^2}{T_a m_b+T_b m_a} \hbbh^{b(4)}\nn\\
  &&  -\frac{2}{5}\frac{T_a m_b}{m_a}\tbbh^{a(2)} +\frac{3}{35} \frac{m_b^2 T_a^2}{m_a (T_a m_b+T_b m_a)}\hbbh^{a(4)} \Big].
\end{eqnarray}
Adding the last two equations together finally yields
\begin{eqnarray}
  \bQ_{ab}^{(2)} &=& \frac{\rho_a \nu_{ab}}{m_a+m_b} \Big[ +2(T_b-T_a)\bI -K_{ab (1)} T_a \tbbh^{a(2)} +K_{ab (2)} T_b \tbbh^{b(2)}  \nn\\
  && +L_{ab (1)} T_a \hbbh^{a(4)} -L_{ab (2)} T_b \hbbh^{b(4)}  \Big], \label{eq:Qab2Brag}
\end{eqnarray}
with mass-ratio coefficients
\begin{eqnarray}
K_{ab (1)} &=& \frac{ 2 (2 T_a m_a m_b +3 T_a m_b^2 +5 T_b m_a^2 +6 T_b m_a m_b)}{5 (T_a m_b +T_b m_a) m_a};\nn\\
K_{ab (2)} &=& \frac{2 (3 T_a m_a +2 T_a m_b -T_b m_a)}{5 (T_a m_b +T_b m_a)};\nn\\
L_{ab (1)} &=& \frac{3 T_a m_b (2 T_a m_a m_b +3 T_a m_b^2 +7 T_b m_a^2 +8 T_b m_a m_b)}{35 (T_a m_b +T_b m_a)^2 m_a};\nn\\
L_{ab (2)} &=& \frac{ 3 m_a T_b (5 T_a m_a +4 T_a m_b -T_b m_a)}{35 (T_a m_b +T_b m_a)^2}. \label{eq:ViscGc}
\end{eqnarray}
As a partial double-check of the entire formulation, by neglecting the 4th-order Hermite moments $\hbbh^{(4)}$ in (\ref{eq:Qab2Brag})
it can be verified that the model is then equivalent to Burgers-Schunk; see equation (44) in \cite{Schunk1977}, or
our previous equation (\ref{eq:Pi_noT}).
For a particular case of small temperature differences, the  mass-ratio coefficients simplify into
\begin{eqnarray}
  K_{ab (1)} &=& \frac{2 (5 m_a +3 m_b)}{5 m_a};\qquad K_{ab (2)} = \frac{4}{5};\nn\\
  L_{ab (1)} &=& \frac{3 (7 m_a +3 m_b) m_b}{35 m_a (m_b+m_a)}; \qquad   L_{ab (2)} = \frac{12 m_a}{35 (m_a+m_b)},
\end{eqnarray}
and for self-collisions $K_{aa (1)} = 16/5$; $K_{aa (2)} = 4/5$; $L_{aa (1)} = 3/7$; $L_{aa (2)} = 6/35$.

\newpage
\subsection{Viscosity-tensor exhange rates}
 Collisional contributions for the viscosity-tensor $\bPi^{(2)}_a$ become
\begin{eqnarray}
 \bQ_{ab}^{(2)}\,' &=& \frac{\delta \bPi^{(2)}_{ab}}{\delta t} =  \bQ_{ab}^{(2)} - \frac{\bI}{3}\trace \bQ_{ab}^{(2)}\nn\\ 
 &=& \frac{\rho_a \nu_{ab}}{m_a+m_b} \Big[ -K_{ab (1)} T_a \tbbh^{a(2)} +K_{ab (2)} T_b \tbbh^{b(2)}  \nn\\
  && +L_{ab (1)} T_a \hbbh^{a(4)} -L_{ab (2)} T_b \hbbh^{b(4)}  \Big],
\end{eqnarray}
 and introducing summation over all ``b'' species and rewritten with fluid moments
\begin{eqnarray}
 \bQ_{a}^{(2)}\,' &=&  -\, \frac{21}{10}\nu_{aa} \bPi_a^{(2)} +\frac{9}{70} \nu_{aa}  \frac{\rho_a}{p_a}\bPi_a^{(4)} \nn\\
 && + \sum_{b\neq a} \frac{\rho_a \nu_{ab}}{m_a+m_b} \Big[ -\big(K_{ab (1)}+7 L_{ab (1)}\big) \frac{1}{n_a} \bPi_a^{(2)}
   +\big(K_{ab (2)}+7 L_{ab (2)}\big) \frac{1}{n_b} \bPi_b^{(2)} \nn\\
   && +L_{ab (1)} \frac{\rho_a}{n_a p_a}\bPi_a^{(4)}  -L_{ab (2)} \frac{\rho_b}{n_b p_b}\bPi_b^{(4)}\Big].
\end{eqnarray}
 It is useful to define (introducing hat)
\begin{equation}
\hat{K}_{ab (1)} = K_{ab (1)}+7 L_{ab (1)}; \qquad \hat{K}_{ab (2)} = K_{ab (2)}+7 L_{ab (2)},
\end{equation}
 and the final mass-ratio coefficients are given by (\ref{eq:HatK}).

\newpage
\subsection{4th-order moment exchange rates}
We need to calculate collisional contributions
\begin{eqnarray}
  \trace \bQ_{ab}^{(4)} &=& \frac{\delta \trace\br_{ab}}{\delta t} = m_a \int \bc_a\bc_a |\bc_a|^2 C_{ab}(f_a) d^3v \nn\\
  &=& m_a \int f_a \Big[ (\boldsymbol{A}^{ab}\bc_a)^S |\bc_a|^2 +2(\boldsymbol{A}^{ab}\cdot\bc_a)\bc_a\bc_a \Big] d^3v\nn\\
  && +m_a \int f_a \Big[ (\trace \bD^{ab})\bc_a\bc_a + \bD^{ab}|\bc_a|^2 +2\big( (\bD^{ab}\cdot\bc_a)\bc_a\big)^S \Big] d^3v. \label{eq:LastT}
\end{eqnarray}
There will be no $\tbu$ contributions at the end, and it is simpler to supress these from the beginning ($\tbu=0$),
and just use $\bc_a = \sqrt{T_a/m_a}\alpha\tby$. Then one evaluates step by step
\begin{eqnarray}
  \boldsymbol{A}^{ab}\cdot\bc_a &=& 2\frac{c_{ab}}{m_a^2}\big(1+\frac{m_a}{m_b}\big)
  \frac{n_b m_b}{T_b}\sqrt{\frac{T_a}{m_a}}\alpha
  \Big\{  \Big( \sqrt{\frac{2}{\pi}} e^{-\ty^2/2}-\frac{\erf(\ty/\sqrt{2})}{\ty}\Big) \nn\\
  && +(\tbbh^{b(2)}:\tby\tby)\Big[ -\, \frac{9}{2\ty^5}\erf\Big(\frac{\ty}{\sqrt{2}}\Big)
    +\frac{1}{2}\big(1+\frac{3}{\ty^2}+\frac{9}{\ty^4}\big)  \sqrt{\frac{2}{\pi}} e^{-\ty^2/2}\Big] \nn\\
  && +\frac{1}{28} (\hbbh^{b(4)}:\tby\tby)\big(-2+\ty^2\big) \sqrt{\frac{2}{\pi}} e^{-\ty^2/2}
  \Big\};
\end{eqnarray}
\begin{eqnarray}
 2\big( \boldsymbol{A}^{ab}\cdot\bc_a\big)\bc_a\bc_a &=& 2\frac{c_{ab}}{m_a^2}\big(1+\frac{m_a}{m_b}\big)
  \frac{n_b m_b}{T_b}\sqrt{\frac{T_a}{m_a}} \frac{T_a}{m_a}\alpha^3
  \Big\{  2\tby\tby\Big( \sqrt{\frac{2}{\pi}} e^{-\ty^2/2}-\frac{\erf(\ty/\sqrt{2})}{\ty}\Big) \nn\\
  && +(\tbbh^{b(2)}:\tby\tby)\tby\tby\Big[ -\, \frac{9}{\ty^5}\erf\Big(\frac{\ty}{\sqrt{2}}\Big)
    +\big(1+\frac{3}{\ty^2}+\frac{9}{\ty^4}\big)  \sqrt{\frac{2}{\pi}} e^{-\ty^2/2}\Big] \nn\\
  && +\frac{1}{14} (\hbbh^{b(4)}:\tby\tby)\tby\tby\big(-2+\ty^2\big) \sqrt{\frac{2}{\pi}} e^{-\ty^2/2}
  \Big\};
\end{eqnarray}
\begin{eqnarray}
  \big(\boldsymbol{A}^{ab}\bc_a\big)^S |\bc_a|^2 &=& 2\frac{c_{ab}}{m_a^2}\big(1+\frac{m_a}{m_b}\big)
  \frac{n_b m_b}{T_b} \sqrt{\frac{T_a}{m_a}} \frac{T_a}{m_a}\alpha^3 
  \Big\{ 2 \tby\tby
  \Big( \sqrt{\frac{2}{\pi}} e^{-\ty^2/2}-\frac{\erf(\ty/\sqrt{2})}{\ty}\Big) \nn\\
  && +\Big((\tbbh^{b(2)}\cdot\tby)\tby +\tby(\tbbh^{b(2)}\cdot\tby) \Big)\Big[ \frac{3}{\ty^3}\erf\Big(\frac{\ty}{\sqrt{2}}\Big)
    -\big( 1+\frac{3}{\ty^2}\big) \sqrt{\frac{2}{\pi}} e^{-\ty^2/2}\Big] \nn\\
  && + (\tbbh^{b(2)}:\tby\tby) \tby\tby \Big[ -\frac{15}{\ty^5}\erf\Big(\frac{\ty}{\sqrt{2}}\Big)
    +\big(1+\frac{5}{\ty^2}+\frac{15}{\ty^4}\big) \sqrt{\frac{2}{\pi}} e^{-\ty^2/2}\Big]\nn\\
  && -\frac{1}{14} \Big( (\hbbh^{b(4)}\cdot\tby)\tby + \tby(\hbbh^{b(4)}\cdot\tby)  \Big) \ty^2 \sqrt{\frac{2}{\pi}} e^{-\ty^2/2}
  +\frac{1}{14} (\hbbh^{b(4)}:\tby\tby)\tby\tby \ty^2\sqrt{\frac{2}{\pi}} e^{-\ty^2/2}
  \Big\},
\end{eqnarray}
and adding the last two results together
\begin{eqnarray}
  &&\big(\boldsymbol{A}^{ab}\bc_a\big)^S |\bc_a|^2 + 2\big( \boldsymbol{A}^{ab}\cdot\bc_a\big)\bc_a\bc_a \nn\\
  && = 2\frac{c_{ab}}{m_a^2}\big(1+\frac{m_a}{m_b}\big)
  \frac{n_b m_b}{T_b} \sqrt{\frac{T_a}{m_a}} \frac{T_a}{m_a}\alpha^3 
  \Big\{4\tby\tby\Big( \sqrt{\frac{2}{\pi}} e^{-\ty^2/2}-\frac{\erf(\ty/\sqrt{2})}{\ty}\Big) \nn\\
  && +\Big((\tbbh^{b(2)}\cdot\tby)\tby +\tby(\tbbh^{b(2)}\cdot\tby) \Big)\Big[ \frac{3}{\ty^3}\erf\Big(\frac{\ty}{\sqrt{2}}\Big)
    -\sqrt{\frac{2}{\pi}}\big( 1+\frac{3}{\ty^2}\big)e^{-\ty^2/2}\Big] \nn\\
  && +(\tbbh^{b(2)}:\tby\tby)\tby\tby\Big[ -\, \frac{24}{\ty^5}\erf\Big(\frac{\ty}{\sqrt{2}}\Big)
    +\big(2+\frac{8}{\ty^2}+\frac{24}{\ty^4}\big)  \sqrt{\frac{2}{\pi}} e^{-\ty^2/2}\Big] \nn\\
  && -\frac{1}{14} \Big( (\hbbh^{b(4)}\cdot\tby)\tby + \tby(\hbbh^{b(4)}\cdot\tby)  \Big) \ty^2 \sqrt{\frac{2}{\pi}} e^{-\ty^2/2}
  +\frac{1}{14} (\hbbh^{b(4)}:\tby\tby)\tby\tby\big(-2+2\ty^2\big) \sqrt{\frac{2}{\pi}} e^{-\ty^2/2}
  \Big\}. \label{eq:LastT1}
\end{eqnarray}  
Similarly for the diffusion tensor, calculating step by step
\begin{eqnarray}
  \bD^{ab}\cdot\bc_a &=& 2\frac{c_{ab}}{m_a^2} n_b \sqrt{\frac{m_b}{T_b}} \sqrt{\frac{T_a}{m_a}}\alpha \Big\{
  \tby \tilde{A}_1'' \nn\\
  &&  +(\tbbh^{b(2)}\cdot\tby)\Big(2\ty \tilde{A}_2' +2 \tilde{A}_2\Big)
   + \tby(\tbbh^{b(2)}:\tby\tby) \Big( \tilde{A}_2''+2\frac{\tilde{A}_2'}{\ty}\Big) \nn\\
  &&  +(\hbbh^{b(4)}\cdot\tby)\Big(2\ty \tilde{A}_4' +2 \tilde{A}_4\Big)
   + \tby(\hbbh^{b(4)}:\tby\tby) \Big( \tilde{A}_4''+2\frac{\tilde{A}_4'}{\ty}\Big)\Big\};
\end{eqnarray}
\begin{eqnarray}
  \trace \bD^{ab} &=& 2\frac{c_{ab}}{m_a^2} n_b \sqrt{\frac{m_b}{T_b}}\Big\{
   \tilde{A}_1''+2\frac{\tilde{A}_1'}{\ty}\nn\\
   &&  +(\tbbh^{b(2)}:\tby\tby) \Big( \tilde{A}_2''+6\frac{\tilde{A}_2'}{\ty}\Big)
   + (\hbbh^{b(4)}:\tby\tby) \Big( \tilde{A}_4''+6\frac{\tilde{A}_4'}{\ty}\Big)
   \Big\};
\end{eqnarray}
\begin{eqnarray}
  \trace \bD^{ab}\bc_a\bc_a &=& 2\frac{c_{ab}}{m_a^2} n_b \sqrt{\frac{m_b}{T_b}} \frac{T_a}{m_a} \alpha^2 \Big\{
   \tby\tby \Big( \tilde{A}_1''+2\frac{\tilde{A}_1'}{\ty} \Big) \nn\\
   &&  +(\tbbh^{b(2)}:\tby\tby)\tby\tby \Big( \tilde{A}_2''+6\frac{\tilde{A}_2'}{\ty}\Big)
   + (\hbbh^{b(4)}:\tby\tby)\tby\tby \Big( \tilde{A}_4''+6\frac{\tilde{A}_4'}{\ty}\Big)
   \Big\};
\end{eqnarray}
\begin{eqnarray}
  \bD^{ab}|\bc_a|^2 &=& 2\frac{c_{ab}}{m_a^2} n_b \sqrt{\frac{m_b}{T_b}}\frac{T_a}{m_a}\alpha^2 \Big\{
  \bI \ty\tilde{A}_1' +\tby\tby \Big( \tilde{A}_1''-\frac{\tilde{A}_1'}{\ty}\Big)\nn\\
  &&  +\Big[ 2\tby (\tbbh^{b(2)}\cdot\tby) +2 (\tbbh^{b(2)}\cdot\tby)\tby +\bI (\tbbh^{b(2)}:\tby\tby)\Big] \ty \tilde{A}_2' \nn\\
  && + 2 \ty^2 \tilde{A}_2 \tbbh^{b(2)} + \tby\tby(\tbbh^{b(2)}:\tby\tby) \Big( \tilde{A}_2''-\frac{\tilde{A}_2'}{\ty}\Big) \nn\\
  &&  +\Big[ 2\tby (\hbbh^{b(4)}\cdot\tby) +2 (\hbbh^{b(4)}\cdot\tby)\tby +\bI (\hbbh^{b(4)}:\tby\tby)\Big] \ty \tilde{A}_4' \nn\\
  && + 2 \ty^2 \tilde{A}_4 \hbbh^{b(4)} + \tby\tby(\hbbh^{b(4)}:\tby\tby) \Big( \tilde{A}_4''-\frac{\tilde{A}_4'}{\ty}\Big);
\end{eqnarray}
\begin{eqnarray}
  2\big[(\bD^{ab}\cdot\bc_a)\bc_a\big]^S &=& 2\frac{c_{ab}}{m_a^2} n_b \sqrt{\frac{m_b}{T_b}} \frac{T_a}{m_a}\alpha^2 \Big\{
  4\tby\tby \tilde{A}_1'' \nn\\
  &&  +2\big[(\tbbh^{b(2)}\cdot\tby)\tby \big]^S\Big(2\ty \tilde{A}_2' +2 \tilde{A}_2\Big)
   + 4\tby\tby(\tbbh^{b(2)}:\tby\tby) \Big( \tilde{A}_2''+2\frac{\tilde{A}_2'}{\ty}\Big) \nn\\
  &&  +2\big[(\hbbh^{b(4)}\cdot\tby)\tby \big]^S \Big(2\ty \tilde{A}_4' +2 \tilde{A}_4\Big)
   + 4\tby\tby(\hbbh^{b(4)}:\tby\tby) \Big( \tilde{A}_4''+2\frac{\tilde{A}_4'}{\ty}\Big)\Big\},
\end{eqnarray}
and adding the last three results together
\begin{eqnarray}
 && (\trace \bD^{ab})\bc_a\bc_a +\bD^{ab}|\bc_a|^2 + 2\big[ (\bD^{ab}\cdot\bc_a)\bc_a\big]^S \nn\\
  && = 2\frac{c_{ab}}{m_a^2} n_b \sqrt{\frac{m_b}{T_b}}\frac{T_a}{m_a}\alpha^2 \Big\{
  \bI \ty\tilde{A}_1' +\tby\tby \Big( 6\tilde{A}_1''+\frac{\tilde{A}_1'}{\ty}\Big)\nn\\
  &&  +2\big[(\tbbh^{b(2)}\cdot\tby)\tby \big]^S\Big(3\ty \tilde{A}_2' +2 \tilde{A}_2\Big) +\bI (\tbbh^{b(2)}:\tby\tby)\ty \tilde{A}_2' \nn\\
  && + 2 \ty^2 \tilde{A}_2 \tbbh^{b(2)} + \tby\tby(\tbbh^{b(2)}:\tby\tby) \Big( 6 \tilde{A}_2''+13\frac{\tilde{A}_2'}{\ty}\Big) \nn\\
  &&  +2\big[(\hbbh^{b(4)}\cdot\tby)\tby \big]^S\Big(3\ty \tilde{A}_4' +2 \tilde{A}_4\Big) +\bI (\hbbh^{b(4)}:\tby\tby)\ty \tilde{A}_4' \nn\\
  && + 2 \ty^2 \tilde{A}_4 \hbbh^{b(4)} + \tby\tby(\hbbh^{b(4)}:\tby\tby) \Big( 6 \tilde{A}_4''+13\frac{\tilde{A}_4'}{\ty}\Big)\Big\}.\label{eq:LastT2}
\end{eqnarray}
Now by using (\ref{eq:LastT1}), (\ref{eq:LastT2}), we are ready to calculate collisional integrals (\ref{eq:LastT}).
The first integral in (\ref{eq:LastT}) calculates
\begin{eqnarray}
&& m_a \int f_a \Big[ (\boldsymbol{A}^{ab}\bc_a)^S |\bc_a|^2 +2(\boldsymbol{A}^{ab}\cdot\bc_a)\bc_a\bc_a \Big] d^3v\nn\\
  && = \rho_a \nu_{ab} \frac{p_a^2}{\rho_a^2} \Big\{ -\bI  \frac{4 (2 T_a m_b +5 T_b m_a)}{(T_a m_b +T_b m_a)}
    - \tbbh^{b(2)} \frac{6 (3 T_a m_b-7 T_b m_a) T_b m_a}{5 (T_a m_b +T_b m_a)^2}\nn\\
    &&   + \hbbh^{b(4)} \frac{3 (T_a m_b -T_b m_a) T_b^2 m_a^2}{(T_a m_b +T_b m_a)^3} \nn\\
    && - \tbbh^{a(2)} \frac{4 (8 T_a^2 m_b^2 +28 T_a T_b m_a m_b +35 T_b^2 m_a^2)}{5 (T_a m_b +T_b m_a)^2}\nn\\
    && + \hbbh^{a(4)} \frac{2 (8 T_a^3 m_b^3 +36 T_a^2 T_b m_a m_b^2 +63 T_a T_b^2 m_a^2 m_b -70 T_b^3 m_a^3)}{35 (T_a m_b +T_b m_a)^3} \Big\}. \label{eq:Posled1}
\end{eqnarray}
The second integral in (\ref{eq:LastT}) calculates
\begin{eqnarray}
&&  m_a \int f_a \Big[ (\trace \bD^{ab})\bc_a\bc_a + \bD^{ab}|\bc_a|^2 +2\big( (\bD^{ab}\cdot\bc_a)\bc_a\big)^S \Big] d^3v\nn\\
  &&  =\rho_a \nu_{ab} \frac{p_a^2}{\rho_a^2} \Big\{ \bI \frac{4 (2 T_a m_b +5 T_b m_a)}{T_a (m_b+m_a)}
  + \tbbh^{b(2)} \frac{2 (11 T_a m_b -7 T_b m_a) T_b m_a}{5 T_a (T_a m_b +T_b m_a) (m_b +m_a)} \nn\\
  && - \hbbh^{b(4)} \frac{3 (23 T_a m_b -7 T_b m_a) T_b^2 m_a^2 }{35 T_a (T_a m_b +T_b m_a)^2 (m_b+m_a)}\nn\\
  && + \tbbh^{a(2)} \frac{2 (4 T_a^2 m_b^2 +21 T_a T_b m_a m_b +35 T_b^2 m_a^2)}{5 T_a (T_a m_b +T_b m_a) (m_b+m_a)} \nn\\
  && - \hbbh^{a(4)} \frac{m_b (T_a m_b +7 T_b m_a) (4 T_a m_b +19 T_b m_a)}{35 (T_a m_b +T_b m_a)^2 (m_b+m_a)} \Big\}. \label{eq:Posled2}
\end{eqnarray}
Adding (\ref{eq:Posled1}) and (\ref{eq:Posled2}) together then yields collisional contributions
\begin{eqnarray}
  \trace \bQ_{ab}^{(4)} &=& \rho_a \nu_{ab} \frac{p_a^2}{\rho_a^2} \Big\{
  + \bI \frac{4 (2 T_a m_b +5 T_b m_a) m_a}{(T_a m_b +T_b m_a) (m_b+m_a)} \frac{(T_b-T_a)}{T_a} \nn\\
  && - M_{ab (1)}\tbbh^{a(2)}  +M_{ab (2)} \tbbh^{b(2)} -N_{ab (1)}\hbbh^{a(4)} - N_{ab (2)} \hbbh^{b(4)}\Big\}, \label{eq:Posled3}
\end{eqnarray}
with  mass-ratio coefficients
\begin{eqnarray}
  M_{ab (1)} &=& \Big\{ 2 \big(16 T_a^3 m_a m_b^2 +12 T_a^3 m_b^3 +56 T_a^2 T_b m_a^2 m_b +31 T_a^2 T_b m_a m_b^2 +70 T_a T_b^2 m_a^3\nn\\
   && \quad +14 T_a T_b^2 m_a^2 m_b -35 T_b^3 m_a^3 \big)\Big\} \Big[5 T_a (T_a m_b +T_b m_a)^2 (m_b+m_a)\Big]^{-1};\nn\\
  M_{ab (2)} &=& -\, \frac{2 T_b m_a (9 T_a^2 m_a m_b -2 T_a^2 m_b^2 -21 T_a T_b m_a^2 -25 T_a T_b m_a m_b +7 T_b^2 m_a^2)}{5 (T_a m_b +T_b m_a)^2 T_a (m_b +m_a)};\nn\\
  N_{ab (1)} &=& -\, \Big\{16 T_a^3 m_a m_b^3 +12 T_a^3 m_b^4 +72 T_a^2 T_b m_a^2 m_b^2 +21 T_a^2 T_b m_a m_b^3 +126 T_a T_b^2 m_a^3 m_b\nn\\
 && \quad -54 T_a T_b^2 m_a^2 m_b^2 -140 T_b^3 m_a^4 -273 T_b^3 m_a^3 m_b\Big\} \Big[35 (T_a m_b +T_b m_a)^3 (m_b +m_a)\Big]^{-1};\nn\\
  N_{ab (2)} &=& -\, \frac{3 T_b^2 m_a^2 (35 T_a^2 m_a m_b +12 T_a^2 m_b^2 -35 T_a T_b m_a^2 -51 T_a T_b m_a m_b +7 T_b^2 m_a^2)}{35 (T_a m_b +T_b m_a)^3 T_a (m_b +m_a)}. \label{eq:Posled5}
\end{eqnarray}
For a particular case of small temperature differences between species the  mass-ratio coefficients simplify into
\begin{eqnarray}
  M_{ab (1)} &=& \frac{2 (35 m_a^2 +35 m_a m_b +12 m_b^2)}{5 (m_b +m_a)^2};\qquad
  M_{ab (2)} = \frac{4  m_a (7 m_a +m_b)}{5 (m_b +m_a)^2};\nn\\
  N_{ab (1)} &=& \frac{140 m_a^3 +7 m_a^2 m_b -25 m_a m_b^2 -12 m_b^3}{35 (m_b +m_a)^3};\qquad
  N_{ab (2)} = \frac{12  m_a^2 (7 m_a -3 m_b)}{35 (m_b +m_a)^3},
\end{eqnarray}
and for self-collissions $M_{aa (1)} = 41/5$; $M_{aa (2)} = 8/5$; $N_{aa (1)} = 11/28$; $N_{aa (2)} = 6/35$.

\subsection{Exchange rates \texorpdfstring{$\bQ_{a}^{(4)}\,'$}{Q4'}}
Applying trace at (\ref{eq:Posled3}) yields scalar
\begin{equation}
  \trace \trace \bQ_{ab}^{(4)} = \rho_a \nu_{ab} \frac{p_a^2}{\rho_a^2} \Big\{
  + 3 \frac{4 (2 T_a m_b +5 T_b m_a) m_a}{(T_a m_b +T_b m_a) (m_b+m_a)} \frac{(T_b-T_a)}{T_a}\Big\}, 
\end{equation}
and thus
\begin{eqnarray}
  \bQ^{(4)}_{ab}\,' &\equiv& \trace \bQ^{(4)}_{ab} -\frac{\bI}{3}\trace\trace \bQ^{(4)}_{ab}\nn\\
  &=& \rho_a \nu_{ab} \frac{p_a^2}{\rho_a^2} \Big[
   - M_{ab (1)}\tbbh^{a(2)}  +M_{ab (2)} \tbbh^{b(2)} -N_{ab (1)}\hbbh^{a(4)} - N_{ab (2)} \hbbh^{b(4)}\Big]. \label{eq:Posled4}
\end{eqnarray}
 Finally, introducing summation over all ``b'' species and rewritten with fluid moments
\begin{eqnarray}
  \bQ^{(4)}_{a}\,' &=& -\, \frac{53}{20} \nu_{aa} \frac{p_a}{\rho_a} \bPi^{(2)}_a - \frac{79}{140} \nu_{aa} \bPi^{(4)}_a
   + \sum_{b\neq a} \nu_{ab} \Big[ - \big(M_{ab (1)}-7N_{ab (1)}\big) \frac{p_a}{\rho_a} \bPi^{(2)}_a \nn\\
  &&  +\big(M_{ab (2)}+7N_{ab (2)}\big) \frac{p_a^2}{\rho_a p_b} \bPi^{(2)}_b -N_{ab (1)}\bPi^{(4)}_a
    - N_{ab (2)}\frac{p_a^2 \rho_b}{p_b^2\rho_a} \bPi^{(4)}_b \Big].
\end{eqnarray}
 It is useful to define (introducing tilde)
\begin{equation}
\hat{M}_{ab (1)} = M_{ab (1)}-7N_{ab (1)}; \qquad \hat{M}_{ab (2)} = M_{ab (2)}+7N_{ab (2)},
\end{equation}
 and the final mass-ratio coefficients are given by (\ref{eq:Posled20}).

\newpage
\section{Collisional contributions for scalar \texorpdfstring{$\widetilde{X}^{(4)}$}{\textasciitilde X4}} \label{sec:X4coll}
\setcounter{equation}{0}
Here we consider perturbation
\begin{equation}
\chi_b = \frac{1}{120} \tilde{h}^{b(4)} \tilde{H}^{b(4)},
\end{equation}
with Hermite polynomial $\tilde{H}^{(4)} = \tc^4-10\tc^2+15$  and Hermite moment $\tilde{h}^{b(4)} = \frac{\rho_b}{p_b^2} \widetilde{X}^{b(4)}$.
The Rosenbluth potentials become
\begin{eqnarray}
  H_b (\bV) &=& n_b \sqrt{\frac{m_b}{T_b}} \Big\{ \frac{1}{\ty}\erf\Big(\frac{\ty}{\sqrt{2}}\Big)
  +\frac{1}{120} \tilde{h}^{b(4)}(3-\ty^2)\sqrt{\frac{2}{\pi}} e^{-\ty^2/2} \Big\};\\
  G_b (\bV) &=& n_b \sqrt{\frac{T_b}{m_b}} \Big\{ \sqrt{\frac{2}{\pi}}e^{-\ty^2/2} +\big(\ty+\frac{1}{\ty}\big)\erf\big(\frac{\ty}{\sqrt{2}}\big) 
   -\frac{1}{60} \tilde{h}^{b(4)} \sqrt{\frac{2}{\pi}} e^{-\ty^2/2} \Big\},
\end{eqnarray}
and the dynamical friction vector and the diffusion tensor
\begin{eqnarray}
  \boldsymbol{A}^{ab} &=& 2\frac{c_{ab}}{m_a^2}\big(1+\frac{m_a}{m_b}\big)
  \frac{n_b m_b}{T_b} \Big\{ \tby \Big( \sqrt{\frac{2}{\pi}}\frac{e^{-\ty^2/2}}{\ty^2}-\frac{\erf(\ty/\sqrt{2})}{\ty^3}\Big) \nn\\
  && -\tby \frac{\tilde{h}^{b(4)}}{120} \sqrt{\frac{2}{\pi}}\big(5-y^2\big)e^{-\ty^2/2}\Big\}, \label{eq:Thierry120}
\end{eqnarray}
\begin{eqnarray}
  \bD^{ab} &=& 2\frac{c_{ab}}{m_a^2} n_b \sqrt{\frac{m_b}{T_b}}\Big\{
  \bI \frac{\tilde{A}_1'}{\ty}+\frac{\tby\tby}{\ty^2}\Big( \tilde{A}_1''-\frac{\tilde{A}_1'}{\ty}\Big)\nn\\
  &&+\big(\bI-\tby\tby\big) \frac{\tilde{h}^{b(4)}}{60} \sqrt{\frac{2}{\pi}} e^{-\ty^2/2}
  \Big\}. \label{eq:Thierry121}
\end{eqnarray}
The perturbation $\chi_a=(\tilde{h}^{a(4)}/120)(\alpha^4\ty^4-10\alpha^2\ty^2+15)$.
\subsection{Pressure tensor exchange rates}
It is sufficient to consider $\bc_a=\sqrt{T_a/m_a} \alpha\tby$, and so
\begin{eqnarray}
  \big[\boldsymbol{A}^{ab}\bc_a \big]^S &=& 2\frac{c_{ab}}{m_a^2}\big(1+\frac{m_a}{m_b}\big)
  \frac{n_b m_b}{T_b}\sqrt{\frac{T_a}{m_a}} \Big\{ 2\alpha\tby\tby \Big( \sqrt{\frac{2}{\pi}}\frac{e^{-\ty^2/2}}{\ty^2}-\frac{\erf(\ty/\sqrt{2})}{\ty^3}\Big) \nn\\
  && -2\alpha\tby\tby \frac{\tilde{h}^{b(4)}}{120} \sqrt{\frac{2}{\pi}}\big(5-y^2\big)e^{-\ty^2/2}\Big\},
\end{eqnarray}
which further integrates
\begin{eqnarray}
  m_a \int f_a \big[ \boldsymbol{A}^{ab}\bc_a\big]^S d^3v
  &=& \rho_a \nu_{ab} \bI \Big[ -2\frac{T_a}{m_a} -\tilde{h}^{b(4)}\frac{T_a T_b^2 m_a}{4(T_a m_b +T_b m_a)^2} \nn\\
    && - \tilde{h}^{a(4)} \frac{m_b T_a^2 (T_a m_b -4 T_b m_a)}{20 m_a (T_a m_b +T_b m_a)^2 }  \Big],
\end{eqnarray}
together with
\begin{eqnarray}
  m_a \int f_a  \bD_{ab} d^3v &=& \frac{\rho_a \nu_{ab}}{m_a+m_b} \bI\Big[ \frac{2}{m_a} (T_a m_b+T_b m_a)
    +\tilde{h}^{b (4)} \frac{T_b^2 m_a}{20 (T_a m_b +T_b m_a)} \nn\\
    && + \tilde{h}^{a(4)} \frac{m_b^2 T_a^2}{20 m_a (T_a m_b +T_b m_a)}   \Big].
\end{eqnarray}
Adding the last two results together yields collisional contributions
\begin{eqnarray}
  \bQ_{ab}^{(2)} &=& \frac{\rho_a \nu_{ab}}{m_a+m_b} \bI \Big[ +2(T_b-T_a)
    - T_b \tilde{h}^{b (4)} \frac{T_b m_a (5T_a m_a +4 T_a m_b -T_b m_a)}{20(T_a m_b +T_b m_a)^2} \nn\\
  &&  + T_a \tilde{h}^{a (4)} \frac{T_a m_b (5 T_b m_b +4 T_b m_a -T_a m_b  )}{20 (T_a m_b +T_b m_a)^2}\Big],
\end{eqnarray}
which can be written as
\begin{eqnarray}
  \bQ_{ab}^{(2)} &=& \frac{\rho_a \nu_{ab}}{m_a+m_b} \bI \Big[ +2(T_b-T_a)+ P_{ab (1)} T_a \tilde{h}^{a (4)}
    -P_{ab (2)} T_b \tilde{h}^{b (4)}     \Big], \label{eq:Thierry36}
\end{eqnarray}
with  mass-ratio coefficients
\begin{eqnarray}
  P_{ab (1)} = \frac{T_a m_b (5 T_b m_b +4 T_b m_a -T_a m_b  )}{20 (T_a m_b +T_b m_a)^2}; \qquad
  P_{ab (2)} = \frac{T_b m_a (5 T_a m_a +4 T_a m_b -T_b m_a  )}{20 (T_a m_b +T_b m_a)^2},
\end{eqnarray}
or for the particular case of small temperature differences
\begin{eqnarray}
  P_{ab (1)} = \frac{m_b}{5 (m_b+m_a)}; \qquad P_{ab (2)} = \frac{m_a}{5(m_b+m_a)}.
\end{eqnarray}
The pressure tensor exchange rates (\ref{eq:Thierry36}) are rewritten to fluid variables according to
\begin{eqnarray}
  \bQ_{ab}^{(2)} &=& \frac{\rho_a \nu_{ab}}{m_a+m_b} \bI \Big[ +2(T_b-T_a)+ P_{ab (1)}\frac{\rho_a}{n_a p_a} \widetilde{X}^{(4)}_a
    - P_{ab (2)} \frac{\rho_b}{n_b p_b} \widetilde{X}^{(4)}_b     \Big].
\end{eqnarray}
The energy exchange rates then become
\begin{eqnarray}
  Q_{ab} = \frac{1}{2}\trace \bQ_{ab}^{(2)} &=& \frac{\rho_a \nu_{ab}}{(m_a+m_b)}  \Big[ +3(T_b-T_a)+
    \frac{3}{2}P_{ab (1)}\frac{\rho_a}{n_a p_a} \widetilde{X}^{(4)}_a
    - \frac{3}{2} P_{ab (2)} \frac{\rho_b}{n_b p_b} \widetilde{X}^{(4)}_b  \Big],
\end{eqnarray}
and collisional contributions for the stress-tensor are
\begin{equation}
\bQ_{ab}^{(2)}\,' =  \bQ_{ab}^{(2)} - \frac{\bI}{3}\trace \bQ_{ab}^{(2)} =0. 
\end{equation}
The scalar perturbations $\widetilde{X}^{(4)}_a$ \& $\widetilde{X}^{(4)}_b$ thus do not modify the $\bQ_{ab}^{(2)}\,'$,
however they enter the conservation of energy.  The final model uses $\hat{P}_{ab (1)}=(3/2) P_{ab (1)}$ and
$\hat{P}_{ab (2)}=(3/2) P_{ab (2)}$,  and the result is written in Section \ref{sec:SummaryQ4}, equation (\ref{eq:Thierry38}).
The result is also shown in the Discussion, equation (\ref{eq:Thierry35}).

\subsection{4th-order moment exchange rates}
It is straightforward to calculate
\begin{eqnarray}
  &&\big(\boldsymbol{A}^{ab}\bc_a\big)^S |\bc_a|^2 + 2\big( \boldsymbol{A}^{ab}\cdot\bc_a\big)\bc_a\bc_a \nn\\
  && = 2\frac{c_{ab}}{m_a^2}\big(1+\frac{m_a}{m_b}\big)
  \frac{n_b m_b}{T_b} \Big(\frac{T_a}{m_a}\Big)^{3/2}\alpha^3 
  \Big\{4\tby\tby\Big( \sqrt{\frac{2}{\pi}} e^{-\ty^2/2}-\frac{\erf(\ty/\sqrt{2})}{\ty}\Big) \nn\\
 && \quad - 4\tby\tby \ty^2 (5-\ty^2) \frac{\tilde{h}^{b(4)}}{120} \sqrt{\frac{2}{\pi}} e^{-\ty^2/2}
  \Big\},
\end{eqnarray}
together with
\begin{eqnarray}
 && (\trace \bD^{ab})\bc_a\bc_a +\bD^{ab}|\bc_a|^2 + 2\big[ (\bD^{ab}\cdot\bc_a)\bc_a\big]^S \nn\\
  && = 2\frac{c_{ab}}{m_a^2} n_b \sqrt{\frac{m_b}{T_b}}\frac{T_a}{m_a}\alpha^2 \Big\{
  \bI \ty\tilde{A}_1' +\tby\tby \Big( 6\tilde{A}_1''+\frac{\tilde{A}_1'}{\ty}\Big)\nn\\
&& \quad +\Big[\bI \ty^2 +\tby\tby (7-6\ty^2)\Big]\frac{\tilde{h}^{b(4)}}{60} \sqrt{\frac{2}{\pi}} e^{-\ty^2/2} \Big\},
\end{eqnarray}
and integrate
\begin{eqnarray}
&& m_a \int f_a \Big[ (\boldsymbol{A}^{ab}\bc_a)^S |\bc_a|^2 +2(\boldsymbol{A}^{ab}\cdot\bc_a)\bc_a\bc_a \Big] d^3v\nn\\
  && = \rho_a \nu_{ab} \frac{p_a^2}{\rho_a^2} \bI \Big\{ -  \frac{4 (2 T_a m_b +5 T_b m_a)}{(T_a m_b +T_b m_a)}
  + \tilde{h}^{b(4)}\frac{ m_a^2 T_b^2 (2 T_a m_b -5 T_b m_a)}{2(T_a m_b +T_b m_a)^3} \nn\\
&& \quad + \tilde{h}^{a(4)} \frac{2 T_a^3 m_b^3 +9 T_a^2 T_b m_a m_b^2 +72 T_a T_b^2 m_a^2 m_b -40 T_b^3 m_a^3}{30 (T_a m_b +T_b m_a)^3 } \Big\},
\end{eqnarray}
together with
\begin{eqnarray}
&&  m_a \int f_a \Big[ (\trace \bD^{ab})\bc_a\bc_a + \bD^{ab}|\bc_a|^2 +2\big( (\bD^{ab}\cdot\bc_a)\bc_a\big)^S \Big] d^3v\nn\\
  &&  =\rho_a \nu_{ab} \frac{p_a^2}{\rho_a^2} \bI \Big\{  \frac{4 (2 T_a m_b +5 T_b m_a)}{T_a (m_b+m_a)}
  - \tilde{h}^{b (4)}\frac{T_b^2 m_a^2(2 T_a m_b -T_b m_a) }{2 T_a (T_a m_b +T_b m_a)^2 (m_b+m_a)} \nn\\
&&\quad  - \tilde{h}^{a (4)} \frac{ m_b (2 T_a^2 m_b^2 +T_a T_b m_a m_b +44 T_b^2 m_a^2)}{30 (T_a m_b +T_b m_a)^2 (m_b+m_a)} \Big\}.
\end{eqnarray}
Adding last two results together then yields collisional contributions
\begin{eqnarray}
  \trace \bQ_{ab}^{(4)} &=& \rho_a \nu_{ab} \frac{p_a^2}{\rho_a^2} \bI \Big\{
  +  S_{ab (0)}\frac{(T_b-T_a)}{T_a} - S_{ab (1)}\tilde{h}^{a(4)}  - S_{ab (2)} \tilde{h}^{b(4)}\Big\}, \label{eq:Thierry40}
\end{eqnarray}
with  mass-ratio coefficients
\begin{eqnarray}
  S_{ab (0)} &=& \frac{4  m_a (2 T_a m_b +5 T_b m_a)}{(T_a m_b +T_b m_a) (m_b+m_a)};\nn\\
  S_{ab (1)} &=& -\, \Big\{ m_a \big(2 T_a^3 m_b^3 +9 T_a^2 T_b m_a m_b^2 +6 T_a^2 T_b m_b^3 +72 T_a T_b^2 m_a^2 m_b
  +27 T_a T_b^2 m_a m_b^2 \nn\\
  && \quad -40 T_b^3 m_a^3 -84 T_b^3 m_a^2 m_b \big)\Big\} \Big[30 (T_a m_b +T_b m_a)^3 (m_b+m_a) \Big]^{-1};\nn\\
  S_{ab (2)} &=& -\, \frac{T_b^2 m_a^3 (2 T_a^2 m_b -5 T_a T_b m_a -6 T_a T_b m_b +T_b^2 m_a)}{2T_a (T_a m_b +T_b m_a)^3  (m_b+m_a)}.
\end{eqnarray}
For the particular case of small temperature differences
\begin{eqnarray}
  S_{ab (0)} &=& \frac{4 m_a (5 m_a +2 m_b)}{(m_b+m_a)^2};\nn\\
  S_{ab (1)} &=&  \frac{2 m_a(10 m_a^2 -7 m_a m_b -2 m_b^2)}{15 (m_b +m_a)^3};\qquad
  S_{ab (2)} =  \frac{2 m_a^3}{(m_b+m_a)^3},
\end{eqnarray}
and for self-collisions $S_{aa (1)} =  1/60$ and $S_{aa (2)} =  1/4$. 
Applying trace at (\ref{eq:Thierry40}) and changing to fluid moments yields 
\begin{equation}
  \trace \trace \bQ_{ab}^{(4)} = 3 \nu_{ab} \frac{p_a^2}{\rho_a} \Big\{
  +  S_{ab (0)}\frac{(T_b-T_a)}{T_a} -  S_{ab (1)} \frac{\rho_a}{p_a^2} \widetilde{X}^{(4)}_a
  -  S_{ab (2)} \frac{\rho_b}{p_b^2}\widetilde{X}^{(4)}_b\Big\}, 
\end{equation}
and collisional contributions for the stress-tensor $\bPi^{(4)}_a$ are
\begin{eqnarray}
  \bQ^{(4)}_{ab}\,' &\equiv& \trace \bQ^{(4)}_{ab} -\frac{\bI}{3}\trace\trace \bQ^{(4)}_{ab}= 0.
\end{eqnarray}

\subsection{Collisional contributions \texorpdfstring{$\widetilde{Q}^{(4)}_a\,'$}{\textasciitilde Q4'}}
Collisional contributions for the evolution equation $\widetilde{X}^{(4)}_a$, equation (\ref{eq:PPosled12}), then become
\begin{eqnarray}
\widetilde{Q}^{(4)}_{ab}\,' &=& \trace\trace \bQ^{(4)}_{ab} -20 \frac{p_a}{\rho_a}Q_{ab} \nn\\
&=& \nu_{ab} \Big\{ +\frac{p_a^2}{\rho_a} \frac{(T_b-T_a)}{T_a}\Big( 3 S_{ab (0)} -\frac{60m_a}{m_a+m_b}\Big)
-\widetilde{X}^{(4)}_a \Big( 3 S_{ab (1)} +\frac{30m_a}{m_a+m_b} P_{ab (1)}\Big)\nn\\
&& \qquad -\frac{p_a^2 \rho_b}{p_b^2 \rho_a}  \widetilde{X}^{(4)}_b \Big( 3S_{ab (2)}-\frac{30 m_a}{(m_a+m_b)} \frac{T_b}{T_a} P_{ab (2)}\Big) \Big\}.
\end{eqnarray}
It is useful to define
\begin{eqnarray}
  \hat{S}_{ab (0)} &=& -\Big(3 S_{ab (0)} -\frac{60m_a}{m_a+m_b}\Big);\nn\\
  \hat{S}_{ab (1)} &=& 3 S_{ab (1)} +\frac{30m_a}{m_a+m_b} P_{ab (1)};\nn\\
  \hat{S}_{ab (2)} &=& -\Big(3 S_{ab (2)}-\frac{30 m_a}{(m_a+m_b)} \frac{T_b}{T_a} P_{ab (2)}\Big),
\end{eqnarray}
and the final model then reads
\begin{eqnarray}
\widetilde{Q}^{(4)}_{ab}\,' &=& \trace\trace \bQ^{(4)}_{ab} -20 \frac{p_a}{\rho_a}Q_{ab} \nn\\
&=& \nu_{ab} \Big\{ -\frac{p_a^2}{\rho_a} \frac{(T_b-T_a)}{T_a} \hat{S}_{ab (0)}
-\widetilde{X}^{(4)}_a \hat{S}_{ab (1)} 
 +\frac{p_a^2 \rho_b}{p_b^2 \rho_a}  \widetilde{X}^{(4)}_b \hat{S}_{ab (2)}\Big\},
\end{eqnarray}
with  mass-ratio coefficients given by (\ref{eq:Thierry37}).

\newpage
\section{Coupling of two species} \label{sec:2species}
\setcounter{equation}{0}
Here we would like to emphasize the usefullness of the mutli-fluid formulation, which makes calculation of transport coefficients
straightforward. We consider two species with indices ``a'' and ``b''. Evolution equations for heat fluxes ``a'' become
\begin{eqnarray}
&&  \frac{d_a}{d t}\vecq_a + \Omega_a \bhat\times\vecq_a + \frac{5}{2}p_a \nabla \Big(\frac{p_a}{\rho_a}\Big)=  
    -   \Big[ 2\nu_{aa} +\nu_{ab} \hat{D}_{ab (1)} \Big]\vecq_a
  + \nu_{ab} \hat{D}_{ab (2)} \frac{\rho_a}{\rho_b}\vecq_b \nn\\
  &&  \qquad +    \Big[ \frac{3}{70}\nu_{aa}+\nu_{ab}\hat{E}_{ab (1)} \Big]\frac{\rho_a}{p_a}\vecX^{(5)}_a
   - \nu_{ab}   \hat{E}_{ab (2)} \frac{\rho_a}{p_b}\vecX^{(5)}_b -p_a \nu_{ab} (\bu_b-\bu_a) \hat{U}_{ab (1)};\label{eq:Num91}\\
&&  \frac{d_a}{d t}\vecX^{(5)}_a +\Omega_a\bhat\times\vecX^{(5)}_a+70\frac{p_a^2}{\rho_a}\nabla\Big(\frac{p_a}{\rho_a}\Big) 
=  -\Big[\frac{76}{5} \nu_{aa} + \nu_{ab} \hat{F}_{ab (1)} \Big] \frac{p_a}{\rho_a} \vecq_a
+ \nu_{ab} \hat{F}_{ab (2)}  \frac{p_a}{\rho_b}  \vecq_b \nn\\
&& \qquad -\Big[\frac{3}{35} \nu_{aa} + \nu_{ab} \hat{G}_{ab (1)}\Big]\vecX^{(5)}_a
-  \nu_{ab} \hat{G}_{ab (2)} \frac{p_a}{p_b}\vecX^{(5)}_b  - \frac{p_a^2}{\rho_a} \nu_{ab} (\bu_b-\bu_a)  \hat{U}_{ab (2)}, 
\end{eqnarray}
together with evolution equations for heat fluxes ``b''
\begin{eqnarray}
&&  \frac{d_b}{d t}\vecq_b + \Omega_b \bhat\times\vecq_b + \frac{5}{2}p_b \nabla \Big(\frac{p_b}{\rho_b}\Big)=  
    -   \Big[ 2\nu_{bb} +\nu_{ba} \hat{D}_{ba (1)} \Big]\vecq_b
  + \nu_{ba} \hat{D}_{ba (2)} \frac{\rho_b}{\rho_a}\vecq_a \nn\\
  &&  \qquad +    \Big[ \frac{3}{70}\nu_{bb}+\nu_{ba}\hat{E}_{ba (1)} \Big]\frac{\rho_b}{p_b}\vecX^{(5)}_b
   - \nu_{ba}   \hat{E}_{ba (2)} \frac{\rho_b}{p_a}\vecX^{(5)}_a +p_b \nu_{ba} (\bu_b-\bu_a) \hat{U}_{ba (1)};\\
&&  \frac{d_b}{d t}\vecX^{(5)}_b +\Omega_b\bhat\times\vecX^{(5)}_b+70\frac{p_b^2}{\rho_b}\nabla\Big(\frac{p_b}{\rho_b}\Big) 
=  -\Big[\frac{76}{5} \nu_{bb} + \nu_{ba} \hat{F}_{ba (1)} \Big] \frac{p_b}{\rho_b} \vecq_b
+ \nu_{ba} \hat{F}_{ba (2)}  \frac{p_b}{\rho_a}  \vecq_a \nn\\
&& \qquad -\Big[\frac{3}{35} \nu_{bb} + \nu_{ba} \hat{G}_{ba (1)}\Big]\vecX^{(5)}_b
-  \nu_{ba} \hat{G}_{ba (2)} \frac{p_b}{p_a}\vecX^{(5)}_a  + \frac{p_b^2}{\rho_b} \nu_{ba} (\bu_b-\bu_a)  \hat{U}_{ba (2)}, \label{eq:Num92}
\end{eqnarray}
where for  similar temperatures the  mass-ratio coefficients are given by (\ref{eq:FinalQ3c}), (\ref{eq:FinalQ5c})
and for arbitrary temperatures by (\ref{eq:Final_Q3t}), (\ref{eq:Final_Q5t}). 
The system is fully specified and after precribing
quasi-static approximation it can be solved. Unfortunatelly, the general analytic solution is too long to write down, even for the unmagnetized case. 
It is beneficial to consider a specific example. Nevertheless, the above system is a very powerfull tool, which allows one to obtain
transport coefficients between two different species, being it a two ion plasma, or a precise solutions for a specific ion-electron plasma without
neglecting $m_e/m_i$.

Similarly, the viscosity between two species is described by evolution equations for viscosity-tensors of species ``a''
\begin{eqnarray}
  && \frac{d_a}{dt} \bPi^{(2)}_a  +\Omega_a \big(\bhat\times \bPi^{(2)}_a \big)^S + p_a \bW_a
  =  -\, \frac{21}{10}\nu_{aa} \bPi_a^{(2)} +\frac{9}{70} \nu_{aa}  \frac{\rho_a}{p_a}\bPi_a^{(4)} \nn\\
 && \qquad + \frac{\rho_a \nu_{ab}}{m_a+m_b} \Big[ - \hat{K}_{ab (1)} \frac{1}{n_a} \bPi_a^{(2)}
   +\hat{K}_{ab (2)} \frac{1}{n_b} \bPi_b^{(2)} 
    +L_{ab (1)} \frac{\rho_a}{n_a p_a}\bPi_a^{(4)}  -L_{ab (2)} \frac{\rho_b}{n_b p_b}\bPi_b^{(4)}\Big]; \label{eq:Posled21x}\\
  && \frac{d_a}{dt} \bPi^{(4)}_a  +\Omega_a \big(\bhat\times \bPi^{(4)}_a \big)^S + 7 \frac{p_a^2}{\rho_a} \bW_a 
  =  -\, \frac{53}{20} \nu_{aa} \frac{p_a}{\rho_a} \bPi^{(2)}_a - \frac{79}{140} \nu_{aa} \bPi^{(4)}_a\nn\\
  &&\qquad + \nu_{ab} \Big[ - \hat{M}_{ab (1)} \frac{p_a}{\rho_a} \bPi^{(2)}_a 
    +\hat{M}_{ab (2)} \frac{p_a^2}{\rho_a p_b} \bPi^{(2)}_b -N_{ab (1)}\bPi^{(4)}_a
    - N_{ab (2)}\frac{p_a^2 \rho_b}{p_b^2\rho_a} \bPi^{(4)}_b \Big], \label{eq:Energy22xxx}
\end{eqnarray}
together with evolution equations for viscosity-tensors of species ``b''
\begin{eqnarray}
  && \frac{d_b}{dt} \bPi^{(2)}_b  +\Omega_b \big(\bhat\times \bPi^{(2)}_b \big)^S + p_b \bW_b
  =  -\, \frac{21}{10}\nu_{bb} \bPi_b^{(2)} +\frac{9}{70} \nu_{bb}  \frac{\rho_b}{p_b}\bPi_b^{(4)} \nn\\
 && \qquad + \frac{\rho_b \nu_{ba}}{m_a+m_b} \Big[ - \hat{K}_{ba (1)} \frac{1}{n_b} \bPi_b^{(2)}
   +\hat{K}_{ba (2)} \frac{1}{n_a} \bPi_a^{(2)} 
    +L_{ba (1)} \frac{\rho_b}{n_b p_b}\bPi_b^{(4)}  -L_{ba (2)} \frac{\rho_a}{n_a p_a}\bPi_a^{(4)}\Big]; \label{eq:Posled21xxx}\\
  && \frac{d_b}{dt} \bPi^{(4)}_b  +\Omega_b \big(\bhat\times \bPi^{(4)}_b \big)^S + 7 \frac{p_b^2}{\rho_b} \bW_b 
  =  -\, \frac{53}{20} \nu_{bb} \frac{p_b}{\rho_b} \bPi^{(2)}_b - \frac{79}{140} \nu_{bb} \bPi^{(4)}_b\nn\\
  &&\qquad + \nu_{ba} \Big[ - \hat{M}_{ba (1)} \frac{p_b}{\rho_b} \bPi^{(2)}_b 
    +\hat{M}_{ba (2)} \frac{p_b^2}{\rho_b p_a} \bPi^{(2)}_a -N_{ba (1)}\bPi^{(4)}_b
    - N_{ba (2)}\frac{p_b^2 \rho_a}{p_a^2\rho_b} \bPi^{(4)}_a \Big]. \label{eq:Energy22xxxx}
\end{eqnarray}
Here the heat fluxes (\ref{eq:Num91})-(\ref{eq:Num92}) and viscosities (\ref{eq:Posled21x})-(\ref{eq:Energy22xxxx}) are
de-coupled, but one can consider more precise solutions with coupling between heat fluxes and viscosities, similarly
to Section \ref{sec:Couplingg}.

\newpage
\subsection{Protons and alpha particles (unmagnetized)} \label{sec:2speciesAlpha}
As an example, we consider collisions between protons and alpha particles (fully ionized Helium with proton mass 4).
Protons will be ``a'' species and alpha particles  will be ``b'' species. For the ion coefficients, collisions with electrons are neglected in
an analogous fashion to \cite{Braginskii1965}.
By prescribing mass $m_b = 4 m_a$, the  mass-ratio coefficients with equal temperatures $T_a=T_b$ become
\begin{eqnarray}
  &&  \hat{D}_{ab (1)} = \frac{499}{125}; \qquad \hat{D}_{ab (2)} = \frac{396}{125}; \qquad \hat{E}_{ab (1)} = \frac{87}{875};
  \qquad  \hat{E}_{ab (2)} = \frac{9}{175};  \qquad  \hat{U}_{ab (1)} = \frac{6}{5};\nn\\
  &&  \hat{F}_{ab (1)} = \frac{7624}{125}; \qquad \hat{F}_{ab (2)} = \frac{4848}{125}; \qquad \hat{G}_{ab (1)} = -\,\frac{171}{125};
  \qquad \hat{G}_{ab (2)} = \frac{12}{25}; \qquad \hat{U}_{ab (2)} = 24,\nn\\
  &&  \hat{D}_{ba (1)} = \frac{2011}{500}; \qquad \hat{D}_{ba (2)} = \frac{117}{250}; \qquad \hat{E}_{ba (1)} = \frac{897}{14000};
  \qquad  \hat{E}_{ba (2)} = \frac{9}{700};  \qquad  \hat{U}_{ba (1)} = \frac{3}{10};\nn\\
  &&  \hat{F}_{ba (1)} = \frac{979}{50}; \qquad \hat{F}_{ba (2)} = \frac{1383}{125}; \qquad \hat{G}_{ba (1)} = \frac{8907}{7000};
  \qquad \hat{G}_{ba (2)} = \frac{3}{10}; \qquad \hat{U}_{ba (2)} = \frac{39}{5}.
\end{eqnarray}
By specifying charges $Z_a=1$; $Z_b=2$, the four different collisional frequencies are related by
\begin{equation}
\nu_{ba}=\frac{\rho_a}{\rho_b}\nu_{ab};\qquad \nu_{ab} = 8\frac{n_b}{n_a} \sqrt{\frac{2}{5}} \nu_{aa};  \qquad \nu_{bb} = 8\frac{n_b}{n_a}\nu_{aa},
\end{equation}
and we chose $\nu_{aa}$ as the reference frequency. Furthermore, applying the charge neutrality $n_a+2n_b=n_e$ we choose as
a reference normalized density $N_a \equiv n_a/n_e$ and express $n_b/n_e=(1-N_a)/2$. We also prescribe $\nabla T_a = \nabla T_b$.

\begin{figure*}[!htpb]
  \includegraphics[width=0.32\linewidth]{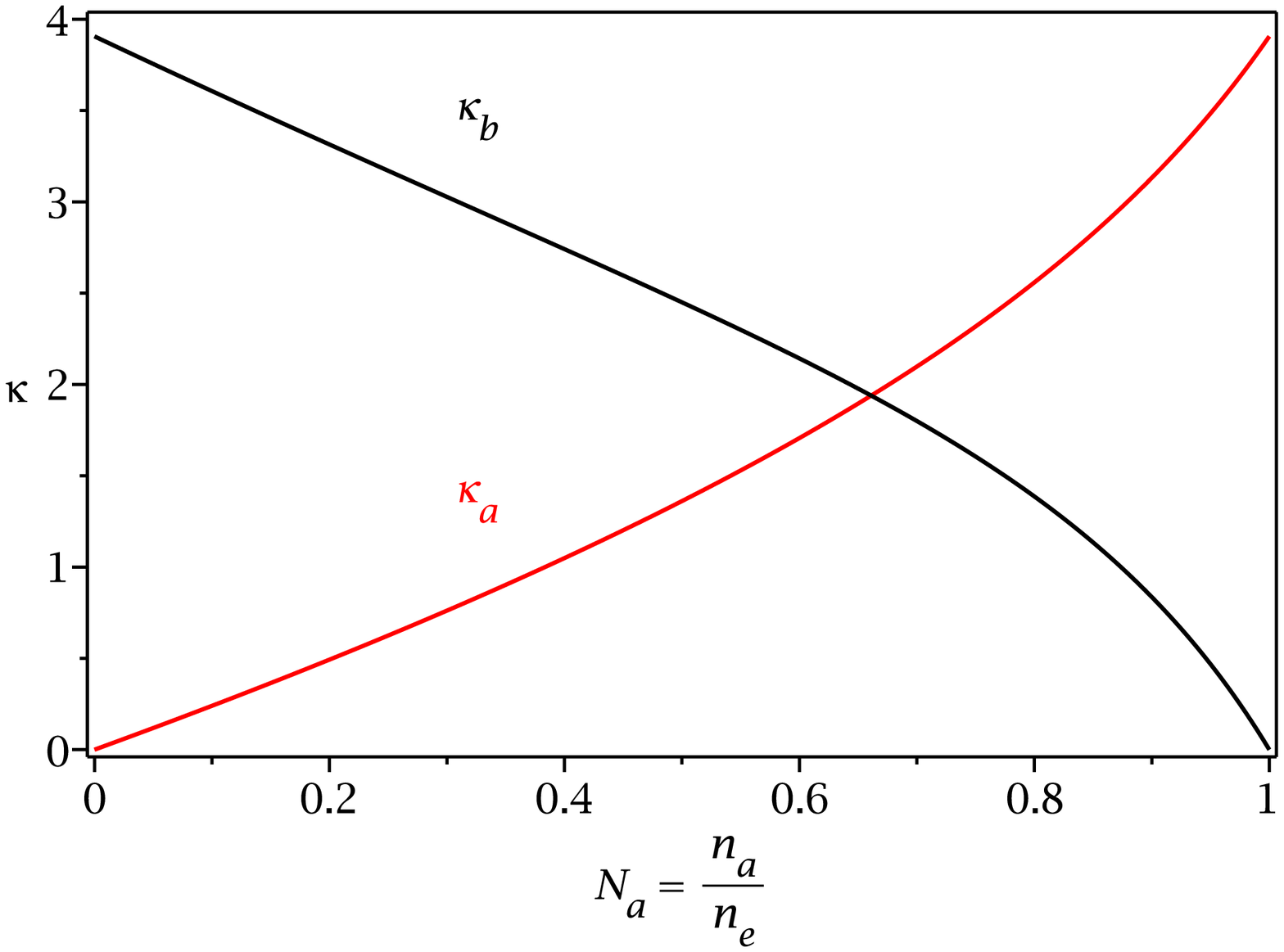}\hspace{0.02\textwidth}\includegraphics[width=0.32\linewidth]{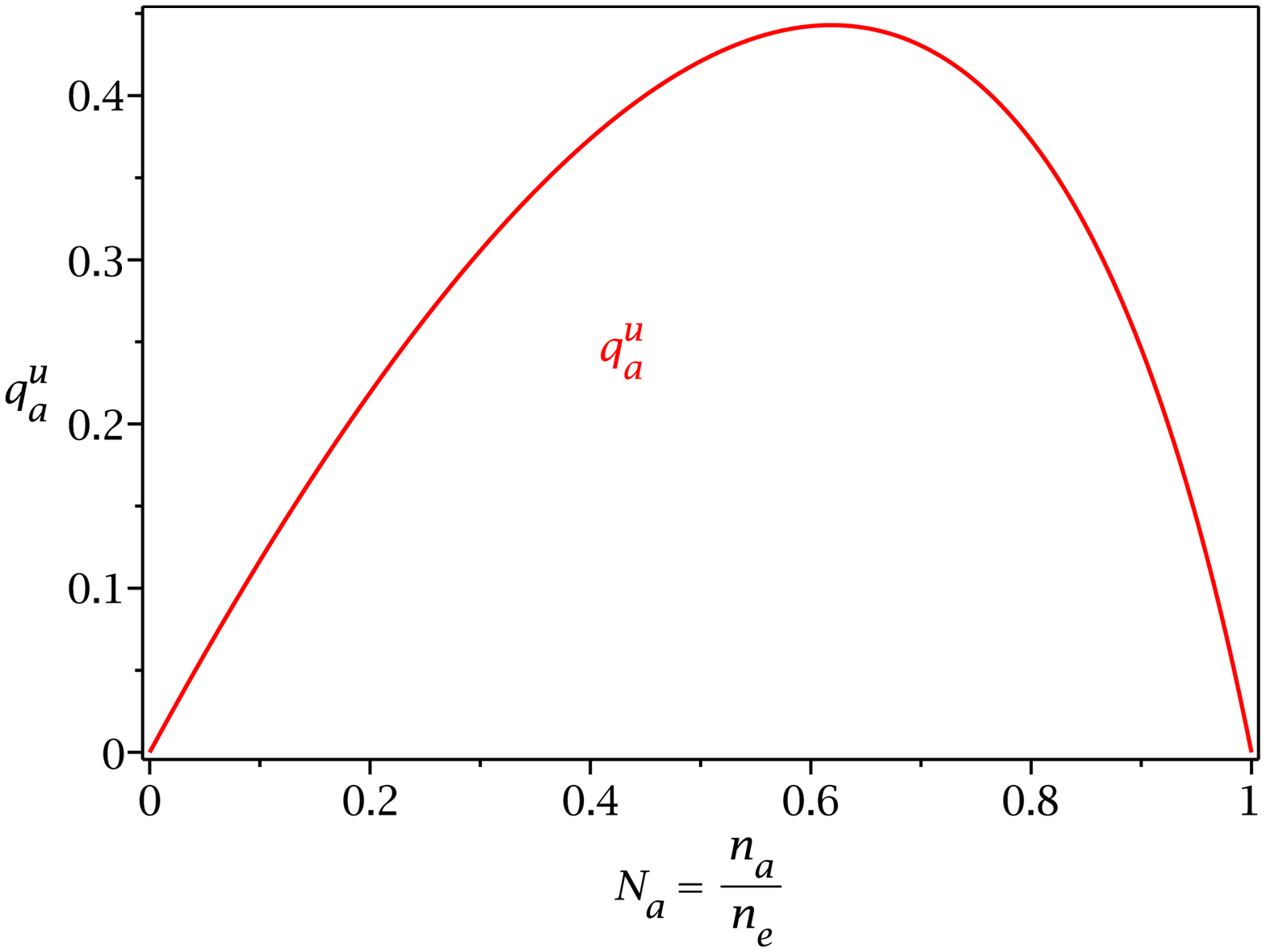}\hspace{0.02\textwidth}\includegraphics[width=0.32\linewidth]{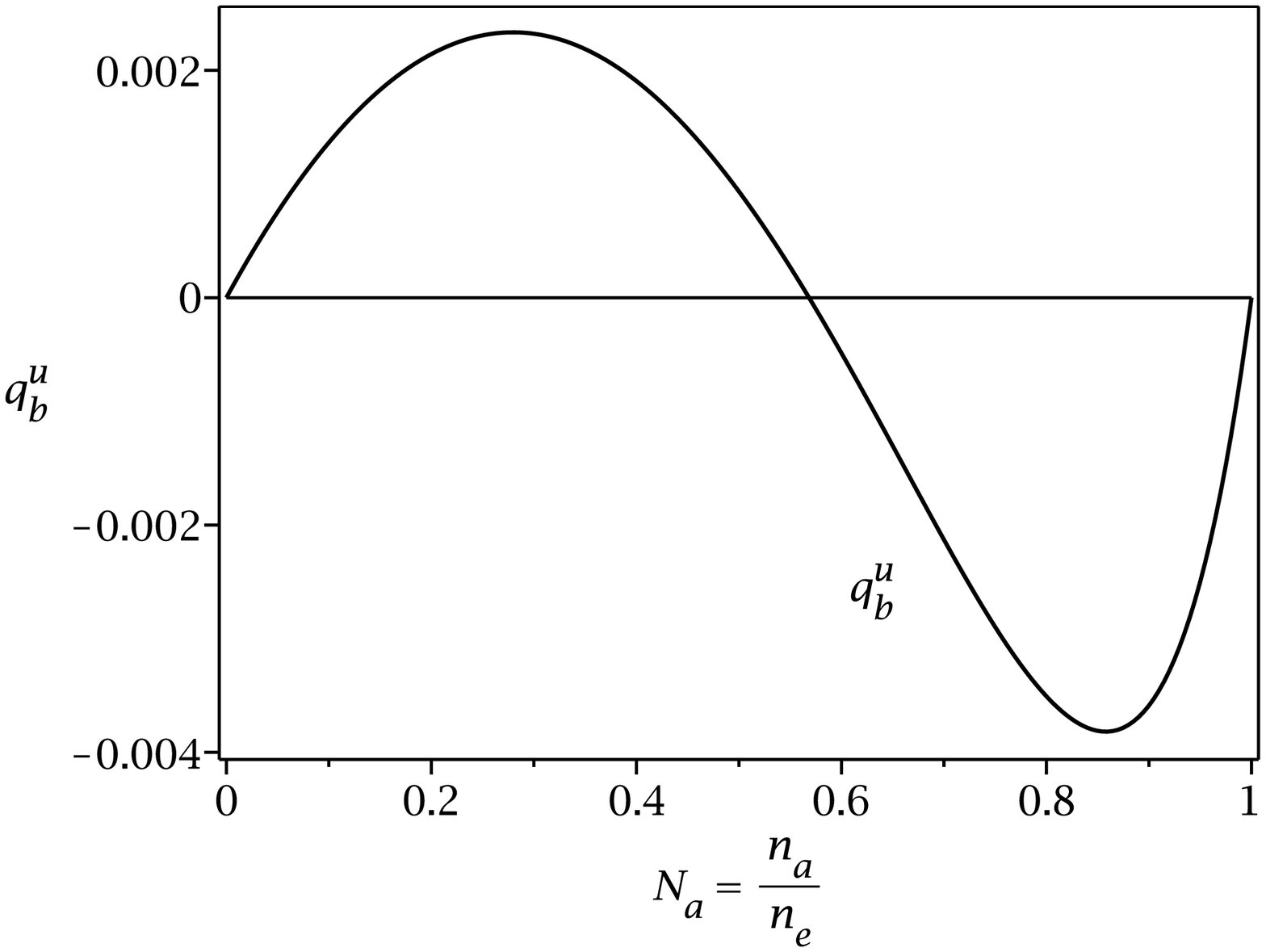}  
  \caption{Left panel: proton thermal conductivity $\hat{\kappa}_a$ (red) given by (\ref{eq:Num112}),
    and alpha-particles thermal conductivity $\hat{\kappa}_b$ (black) given by (\ref{eq:Num113}).
    Middle panel: proton frictional heat flux given by $\beta_{0a}$ (\ref{eq:Num110}).
    Right panel: alpha-particle frictional heat flux given by $\beta_{0b}$(\ref{eq:Num111}). Note the surprising change of sign
    of $\beta_{0b}$ for $N_a>0.57$. We have verified that the same effect is present in the simplified 13-moment model of
  \cite{Burgers1969}-\cite{Schunk1977}. } \label{fig:AP1}
\end{figure*}

Then solving the system yields (parallel) thermal heat fluxes $\vecq_a^T=-\kappa_a \nabla T_a$; $\vecq_b^T=-\kappa_b \nabla T_a$ with
thermal conductivities
\begin{equation} \label{eq:Num111xx}
\kappa_a = \frac{T_a n_a}{m_a \nu_{aa}} \hat{\kappa}_a; \qquad \kappa_b = \frac{T_a n_b}{m_b \nu_{bb}} \hat{\kappa}_b,
\end{equation}
and with normalized fully analytic values
\begin{eqnarray}
 \hat{\kappa}_a &=& N_a \bigg\{ \Big(- \frac{17989001}{10557600} \sqrt{10}+\frac{292708195}{54054912} \Big) N_a^3
  +\Big( \frac{2129490299}{675686400} \sqrt{10} -\frac{1032644005}{108109824} \Big) N_a^2 \nn\\
 && \qquad + \Big(- \frac{98252949}{45045760}\sqrt{10} +\frac{8035835}{1689216} \Big) N_a
  + \frac{51625}{70384} \sqrt{10} +\frac{3425}{140768} \bigg\}/\triangle_1;\\
  \hat{\kappa}_b &=& 32 (1-N_a)\bigg\{ \frac{125}{1024}+
  \Big(\frac{128513167}{2162196480}\sqrt{10} -\frac{166007075}{864878592} \Big) N_a^3 \nn\\
&&  +\Big(- \frac{67953383}{540549120} \sqrt{10} + \frac{386788475}{864878592} \Big) N_a^2
  +\Big( \frac{15671599}{216219648}\sqrt{10} -\frac{1540025}{4504576}  \Big) N_a   \bigg\}/\triangle_1;\\
  \triangle_1 &=& \Big[1+
    \Big(- \frac{722521001}{563072000} \sqrt{10} +\frac{14274588957}{3519200000} \Big) N_a^4
    +\Big(\frac{1043512703}{337843200} \sqrt{10} - \frac{8606493541}{879800000} \Big) N_a^3 \nn\\
&&  \qquad  +\Big(- \frac{23828129}{8798000} \sqrt{10} +\frac{15644893541}{1759600000} \Big) N_a^2
    +\Big(\frac{23828129}{26394000} \sqrt{10} -4 \Big) N_a \Big], \label{eq:Num300}
\end{eqnarray}
or with numerical values
\begin{eqnarray}
 \hat{\kappa}_a &=&  N_a
 \big[ 2.3438 +0.02684 N_a^3 +0.4144 N_a^2 - 2.1404 N_a  \big]/\triangle_1; \label{eq:Num112}\\
 \hat{\kappa}_b &=& 32 (1-N_a)
 \big[0.1221 -0.003988 N_a^3 +0.04968 N_a^2 -0.1127 N_a\big]/\triangle_1; \label{eq:Num113}\\
 \triangle_1 &=& 1-0.001559 N_a^4 -0.01485 N_a^3 +0.3266 N_a^2 -1.1451 N_a.
\end{eqnarray}
Note that $n_a/\nu_{aa}$ is independent of $n_a$, and that is why definitions (\ref{eq:Num111xx}) were chosen.
For the ``b'' species (alpha particles), the results are written in a form
so that it is easy to use $32 n_b/(\nu_{bb} m_b)=n_a/(\nu_{aa}m_a)$. 
As a double check, prescribing
\begin{eqnarray}
&&  N_a=1; \quad => \qquad  \kappa_a = \frac{T_a n_a}{\nu_{aa} m_a}\frac{125}{32};\qquad \kappa_b = 0;\nn\\
  &&  N_a=0; \quad => \qquad  \kappa_a = 0;\qquad
  \kappa_b = \frac{T_a n_b}{\nu_{bb} m_b} \frac{125}{32},
\end{eqnarray}  
as it should be. In general, thermal conductivities of a single ion plasmas compare
as $\kappa_a/\kappa_b=\sqrt{m_b/m_a}(Z_b/Z_a)^4$. In our case, thermal conductivity of pure alpha particles is
32 times smaller than of pure protons. The thermal conductivities $\hat{\kappa}_a,\hat{\kappa}_b$
are plotted in the left panel of Figure \ref{fig:AP1}.

The frictional heat fluxes read
\begin{eqnarray}
 \vecq_a^u &=& - T_a n_e (\bu_b-\bu_a) \beta_{0a}; \qquad  \vecq_b^u = - T_a n_e (\bu_b-\bu_a)\beta_{0b};\label{eq:Num110xxx}\\
 \beta_{0a} &=& N_a (1-N_a) \bigg\{
    \Big(-\frac{150058601}{43990000} +\frac{1522393}{1407680} \sqrt{10} \Big) N_a^3 \nn\\
  &&  +\Big(+ \frac{258658601}{43990000} -\frac{199422}{109975} \sqrt{10} \Big) N_a^2
    +\Big( -\frac{16290}{4399} +\frac{99711}{109975} \sqrt{10}\Big) N_a  + \frac{5430}{4399} \bigg\} / \triangle_1;\\
  \beta_{0b} &=& N_a (1-N_a) \bigg\{ \frac{7351}{1407680}\sqrt{10}+ 
    \Big(- \frac{54551}{22522880}\sqrt{10} +\frac{264247}{35192000}\Big) N_a^3\nn\\
&&    +\Big(\frac{289783}{22522880}\sqrt{10} -\frac{2663863}{70384000}\Big) N_a^2
    +\Big(-\frac{22053}{1407680}\sqrt{10} +\frac{2663863}{140768000}\Big) N_a  \bigg\} / \triangle_1,
\end{eqnarray}
where the denominator $\triangle_1$ is identical to (\ref{eq:Num300}), and with numerical values
\begin{eqnarray}
  \beta_{0a} &=& N_a (1-N_a) \big[1.2344 +0.008776 N_a^3 +0.1457 N_a^2 -0.8360 N_a\big] /\triangle_1;\label{eq:Num110}\\
  \beta_{0b} &=& N_a (1-N_a)  \big[0.01651 -0.0001504 N_a^3 +0.002839 N_a^2 -0.03062 N_a \big]/\triangle_1. \label{eq:Num111}
\end{eqnarray}
In both limits $N_a=0,1$ the frictional heat fluxes disappear. The frictional heat fluxes are plotted in the middle and right
panels of Figure \ref{fig:AP1}.

\subsection*{Viscosities}
\begin{figure*}[!htpb]
  \centering
  \includegraphics[width=0.32\linewidth]{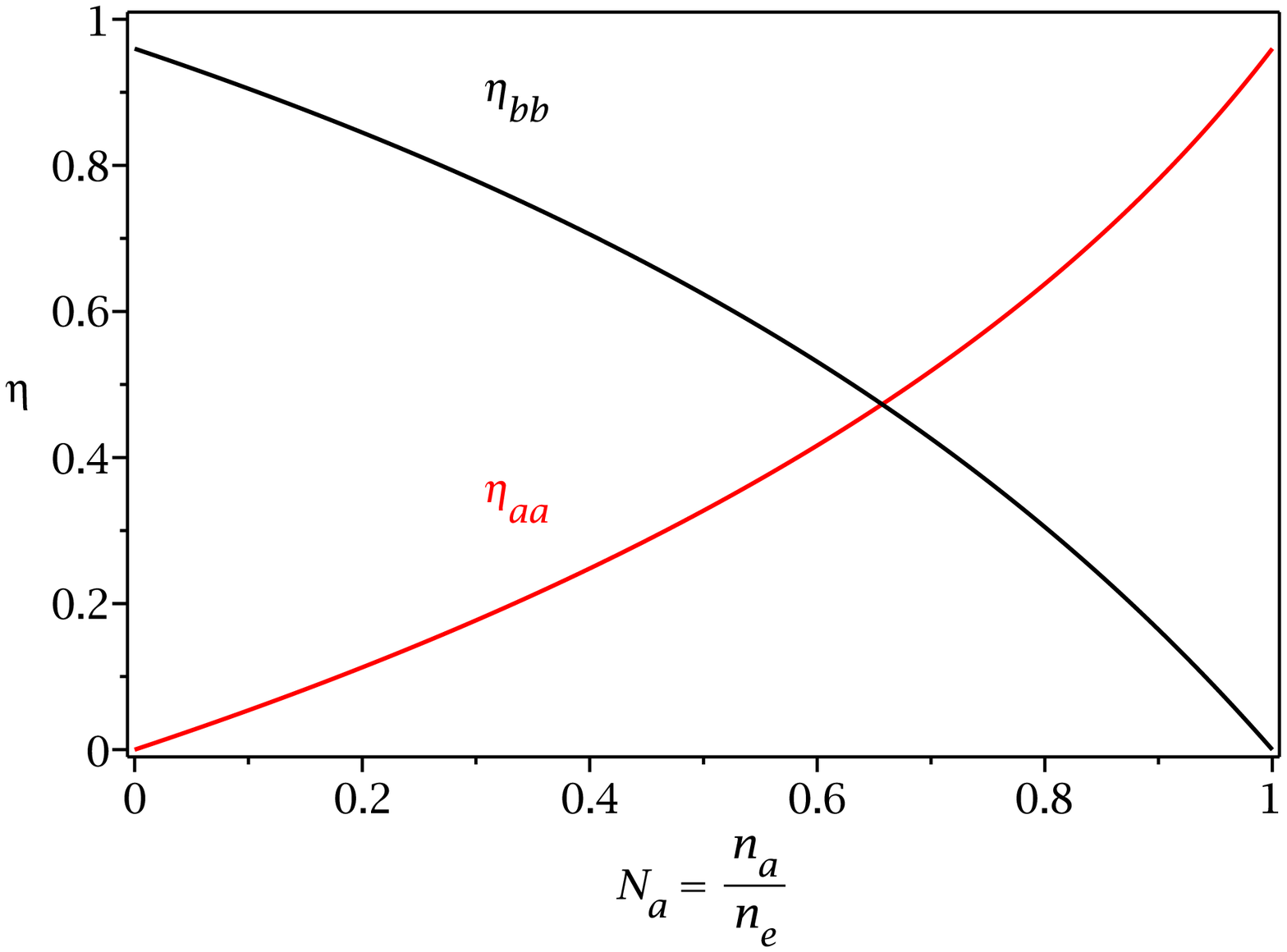}\hspace{0.02\textwidth}\includegraphics[width=0.32\linewidth]{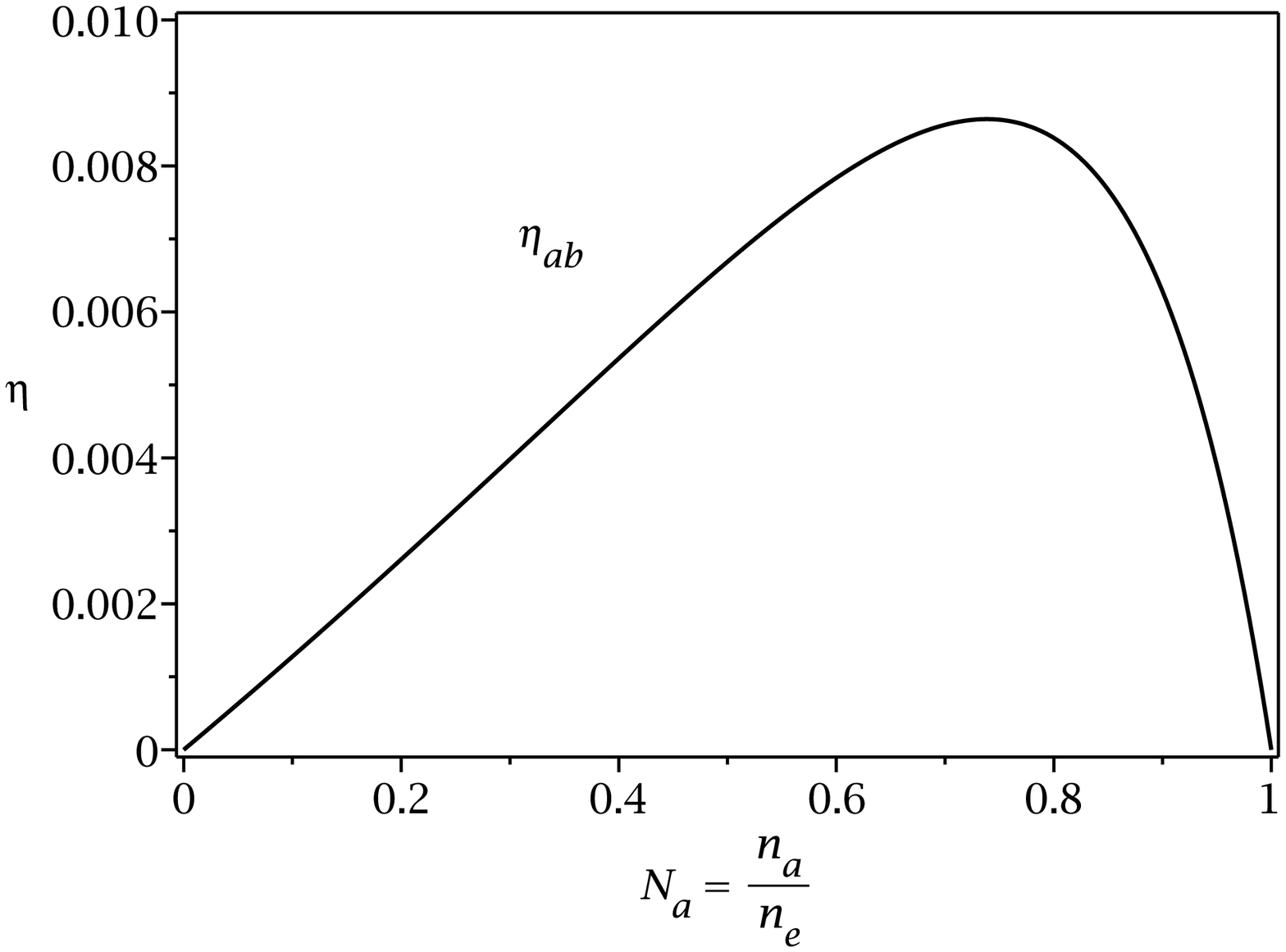}  
  \caption{Normalized viscosities of proton \& alpha-particle plasma, according to (\ref{eq:EnergyX}). Collisions with electrons are neglected,
    in an analogous fasion to Braginskii. Left panel: proton viscosity $\hat{\eta}_{aa}$ (red)
    and alpha-particle viscosity $\hat{\eta}_{bb}$ (black). Right panel: ``cross-viscosity'' $\hat{\eta}_{ab}$.} \label{fig:AP2}
\end{figure*}

One first calculates the required viscosity  mass-ratio coefficients, which for protons (``a'') and alpha particles (``b'') become
\begin{eqnarray}
  &&  \hat{K}_{ab (1)} = \frac{398}{25}; \qquad \hat{K}_{ab (2)} = \frac{32}{25}; \qquad
  {L}_{ab (1)} = \frac{228}{175};   \qquad  {L}_{ab (2)} = \frac{12}{175};  \nn\\
  &&  \hat{M}_{ab (1)} = \frac{934}{125}; \qquad \hat{M}_{ab (2)} = \frac{32}{125}; \qquad
  {N}_{ab (1)} = -\frac{8}{35};  \qquad {N}_{ab (2)} = -\frac{12}{875}; \nn\\
  &&  \hat{K}_{ba (1)} = \frac{323}{100}; \qquad \hat{K}_{ba (2)} = \frac{68}{25}; \qquad
  {L}_{ba (1)} = \frac{93}{700};   \qquad  {L}_{ba (2)} = \frac{48}{175};  \nn\\
  &&  \hat{M}_{ba (1)} = -\frac{368}{125}; \qquad \hat{M}_{ba (2)} = \frac{1424}{125}; \qquad
  {N}_{ba (1)} = \frac{256}{125};  \qquad {N}_{ba (2)} = \frac{192}{175}, 
\end{eqnarray}
and which enter evolution equations (\ref{eq:Posled21x})-(\ref{eq:Energy22xxxx}). For an unmagnetized plasma,
quasi-static solution of these equations then yields viscosity-tensors
\begin{eqnarray}
  \bPi^{(2)}_a &=& -\frac{p_a}{\nu_{aa}}\big[ \hat{\eta}_{aa} \bW_a + \hat{\eta}_{ab} \bW_b \big];\nn\\
  \bPi^{(2)}_b &=& -\frac{p_b}{\nu_{bb}}\big[ 8\hat{\eta}_{ab} \bW_a + \hat{\eta}_{bb} \bW_b \big], \label{eq:Num876}
\end{eqnarray}
with numerical values
\begin{eqnarray}
  \hat{\eta}_{aa} &=& N_a (-0.05464 N_a^3+0.3704 N_a^2 -0.7717 N_a +0.5173 ) /\triangle;\nn\\
  \hat{\eta}_{ab} &=& N_a (1-N_a) (0.001874 N_a^2 -0.008142 N_a +0.01248 ) /\triangle; \nn\\
  \hat{\eta}_{bb} &=& 8(1-N_a) (-0.01150 N_a^3+0.07862 N_a^2 -0.1729 N_a +0.11997) /\triangle;\nn\\
  \triangle &=& 1+0.03923 N_a^4 -0.3759 N_a^3 +1.2959 N_a^2 -1.8953 N_a. \label{eq:EnergyX}
\end{eqnarray}
Note that $p_a/\nu_{aa}=8 p_b/\nu_{bb}$ and the chosen form (\ref{eq:Num876}) emphasizes that the ``cross-viscosities'' $\hat{\eta}_{ab}$ are
  directly related. In general, viscosities of a pure single ion species compare as $\eta_{a}/\eta_{b}=\sqrt{m_a/m_b}(Z_b/Z_a)^4$,
so in our case the viscosity of pure alpha particles is 8 times smaller than of pure protons.
We provide only numerical values for solutions (\ref{eq:EnergyX}), nevertheless it can be shown that for $N_a=1$ the proton viscosity
$\hat{\eta}_{aa}=1025/1068$ and the same result is obtained for the alpha particle viscosity $\hat{\eta}_{bb}$ if $N_a=0$.
The ``cross-viscosity'' $\hat{\eta}_{ab}$ becomes zero for both $N_a=1$ and $N_a=0$. Results are plotted in Figure \ref{fig:AP2}. 

\newpage
\subsection{Deuterium \& tritium plasma (unmagnetized)} \label{sec:DTplasma}
Here we calculate another example of deuterium-tritium plasma, also considered by \cite{Simakov2016b}.
Plasma consisting of deuterium-tritium is probably the most efficient way to achieve
plasma fusion. It is for example being used in the JET machine (see e.g. \cite{Joffrin2019})
  and it will be used in ITER (www.iter.org/sci/FusionFuels). 
Of course, we do not consider peculiar complications
associated with the neoclassical toroidal geometry, our calculation is classical. 
Deuterium core consists of one proton and one neutron. Tritium core consists of one proton and two neutrons.  
Deuterium will be ``a'' species and tritium will be ``b'' species. Collisions with electrons are neglected. 
By prescribing $m_b = (3/2) m_a$, the  mass-ratio coefficients with equal temperatures $T_b=T_a$ become
\begin{eqnarray}
  &&  \hat{D}_{ab (1)} = \frac{1989}{500}; \qquad \hat{D}_{ab (2)} = \frac{324}{125}; \qquad \hat{E}_{ab (1)} = \frac{189}{2000};
  \qquad  \hat{E}_{ab (2)} = \frac{81}{1400};  \qquad  \hat{U}_{ab (1)} = \frac{9}{10};\nn\\
  &&  \hat{F}_{ab (1)} = \frac{13543}{250}; \qquad \hat{F}_{ab (2)} = \frac{5022}{125}; \qquad \hat{G}_{ab (1)} = -\, \frac{1373}{1400};
  \qquad \hat{G}_{ab (2)} = \frac{81}{100}; \qquad \hat{U}_{ab (2)} = \frac{99}{5},\nn\\
  &&  \hat{D}_{ba (1)} = \frac{521}{125}; \qquad \hat{D}_{ba (2)} = \frac{189}{125}; \qquad \hat{E}_{ba (1)} = \frac{78}{875};
  \qquad  \hat{E}_{ba (2)} = \frac{27}{700};  \qquad  \hat{U}_{ba (1)} = \frac{3}{5};\nn\\
  &&  \hat{F}_{ba (1)} = \frac{5832}{125}; \qquad \hat{F}_{ba (2)} = \frac{3672}{125}; \qquad \hat{G}_{ba (1)} = -\,\frac{307}{875};
  \qquad \hat{G}_{ba (2)} = \frac{18}{25}; \qquad \hat{U}_{ba (2)} = \frac{72}{5}.
\end{eqnarray}
Further specifying $Z_a=Z_b=1$, the collisional frequencies are related by
\begin{equation}
\nu_{ab} = \frac{n_b}{n_a}\sqrt{\frac{6}{5}}\nu_{aa}; \qquad \nu_{bb} = \frac{n_b}{n_a}\sqrt{\frac{2}{3}}\nu_{aa}, 
\end{equation}  
and the charge neutrality $n_a+n_b=n_e$ implies $n_b/n_e=1-N_a$, where $N_a=n_a/n_e$.
These  mass-ratio coefficients  and collisional frequencies are used in the system (\ref{eq:Num91})-(\ref{eq:Num92}).
We present quasi-static solutions only for the unmagnetized case, and we assume $\nabla T_a=\nabla T_b$.
Thermal heat fluxes $\vecq_a^T=-\kappa_a \nabla T_a$; $\vecq_b^T=-\kappa_b \nabla T_a$ are given by
\begin{equation}
\kappa_a = \frac{T_a n_a}{m_a \nu_{aa}} \hat{\kappa}_a; \qquad \kappa_b = \frac{T_a n_b}{m_b \nu_{bb}} \hat{\kappa}_b,
\end{equation}
and with numerical values
\begin{eqnarray}
  \hat{\kappa}_a &=& N_a \big(4.2135 -0.009780 N_a^3+0.06292 N_a^2 +1.4992 N_a\big)/\triangle;\nn\\
  \hat{\kappa}_b &=& \sqrt{3/2}(1-N_a) \big(3.1894 -0.001385 N_a^3+0.04936 N_a^2 +0.9845 N_a  \big)/\triangle;\nn\\
  \triangle &=& 1-0.0021475 N_a^4-0.01543 N_a^3 +0.01753 N_a^2 +0.4761 N_a, \label{eq:Num112x}
\end{eqnarray}
where one can also use $\sqrt{3/2} n_b/(m_b\nu_{bb})=n_a/(\nu_{aa}m_a)$.
The frictional heat fluxes are given by
\begin{eqnarray}
  \vecq_a^u &=& - T_a n_e (\bu_b-\bu_a) \beta_{0a}; \qquad  \vecq_b^u = + T_a n_e (\bu_b-\bu_a)\beta_{0b};\label{eq:Num110xx}\\
  \beta_{0a} &=& N_a (1-N_a) \big[ 0.81156 +0.010099 N_a^3 +0.098815 N_a^2 +0.50235 N_a\big] /\triangle;\label{eq:Num110x}\\
  \beta_{0b} &=& N_a (1-N_a)  \big[0.26178 +0.0088461 N_a^3+0.069351 N_a^2 +0.24742 N_a \big]/\triangle. \label{eq:Num111x}
\end{eqnarray}

\begin{figure*}[!htpb]
  \includegraphics[width=0.32\linewidth]{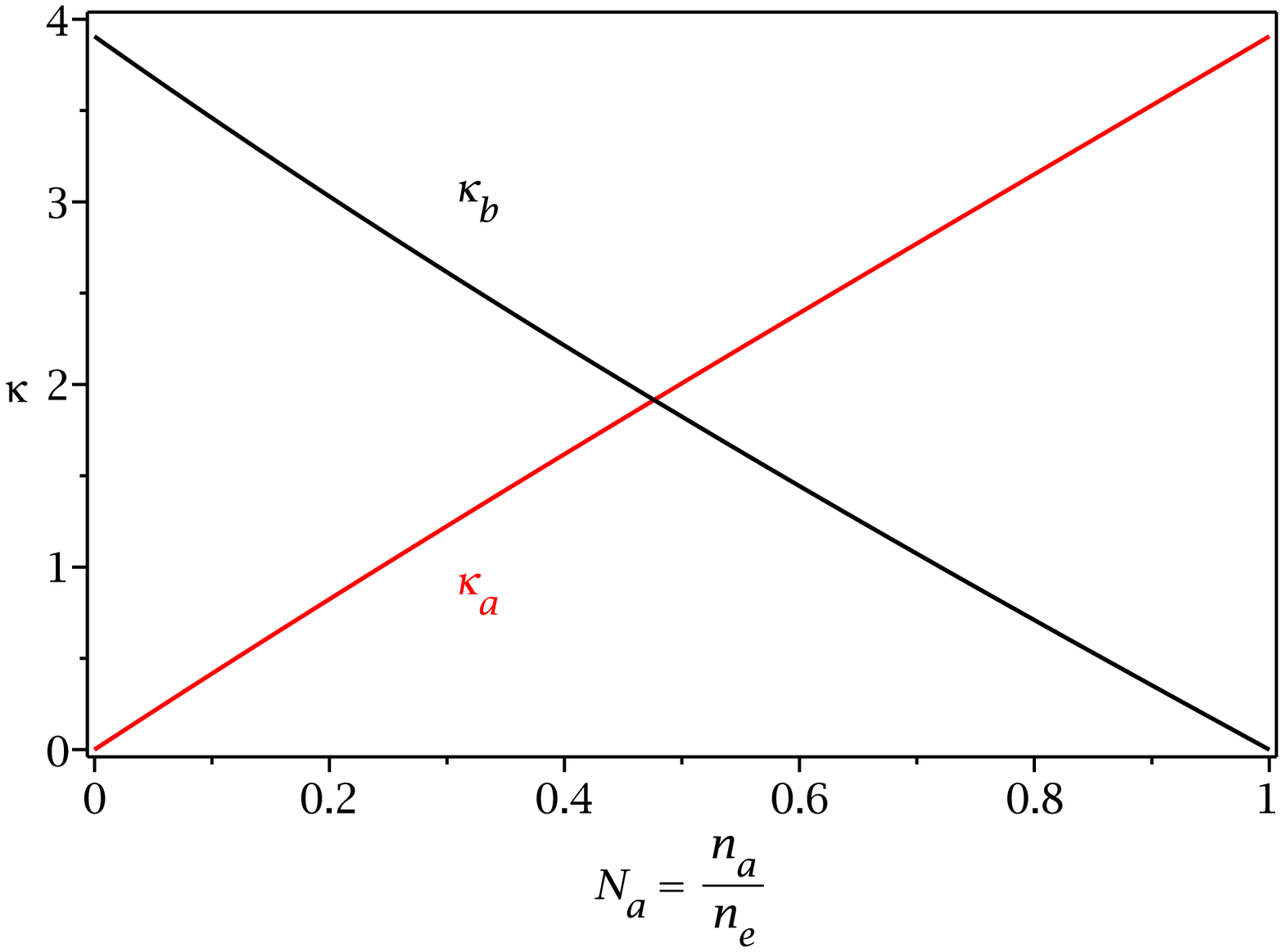}\hspace{0.02\textwidth}\includegraphics[width=0.32\linewidth]{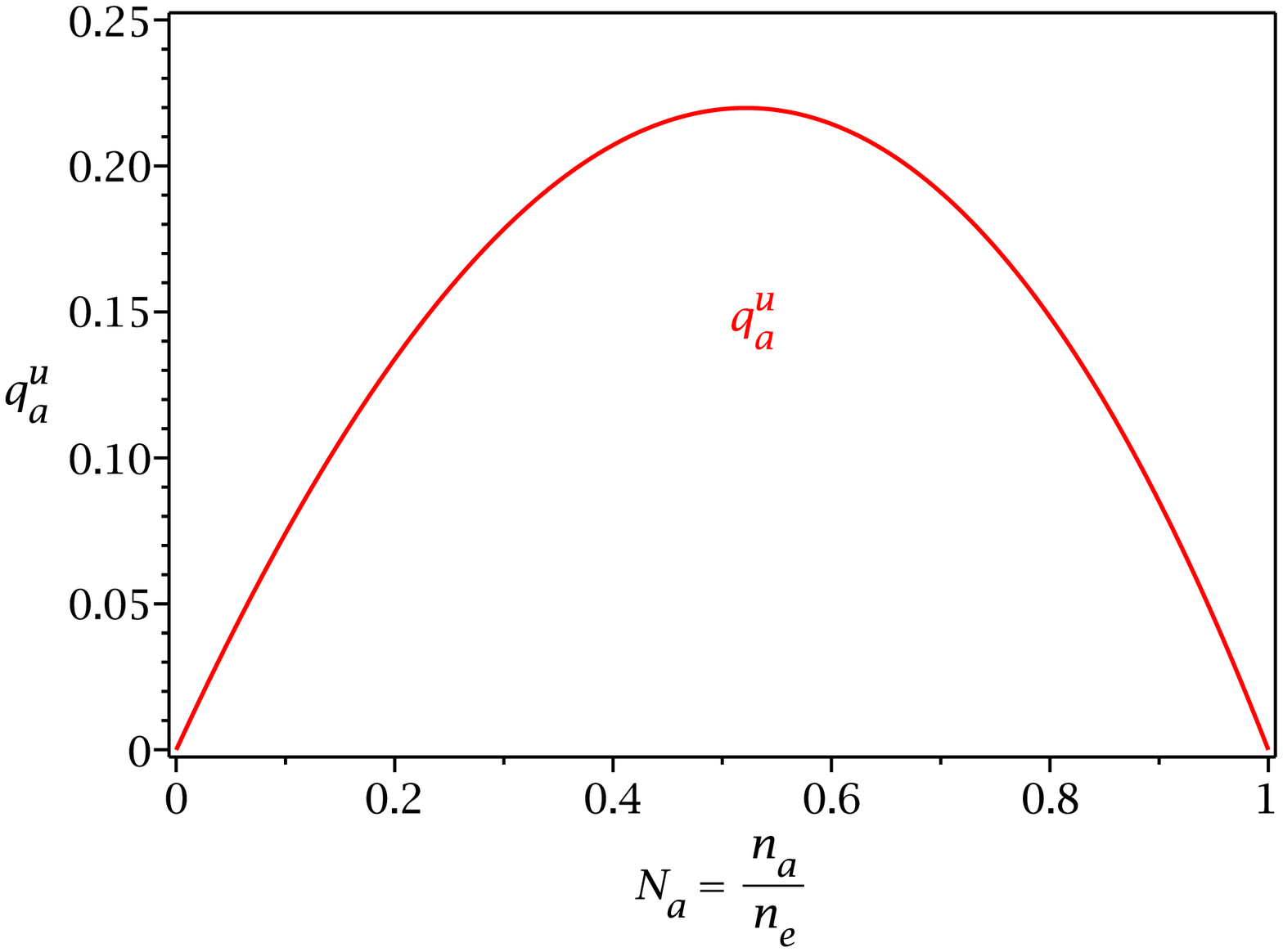}\hspace{0.02\textwidth}\includegraphics[width=0.32\linewidth]{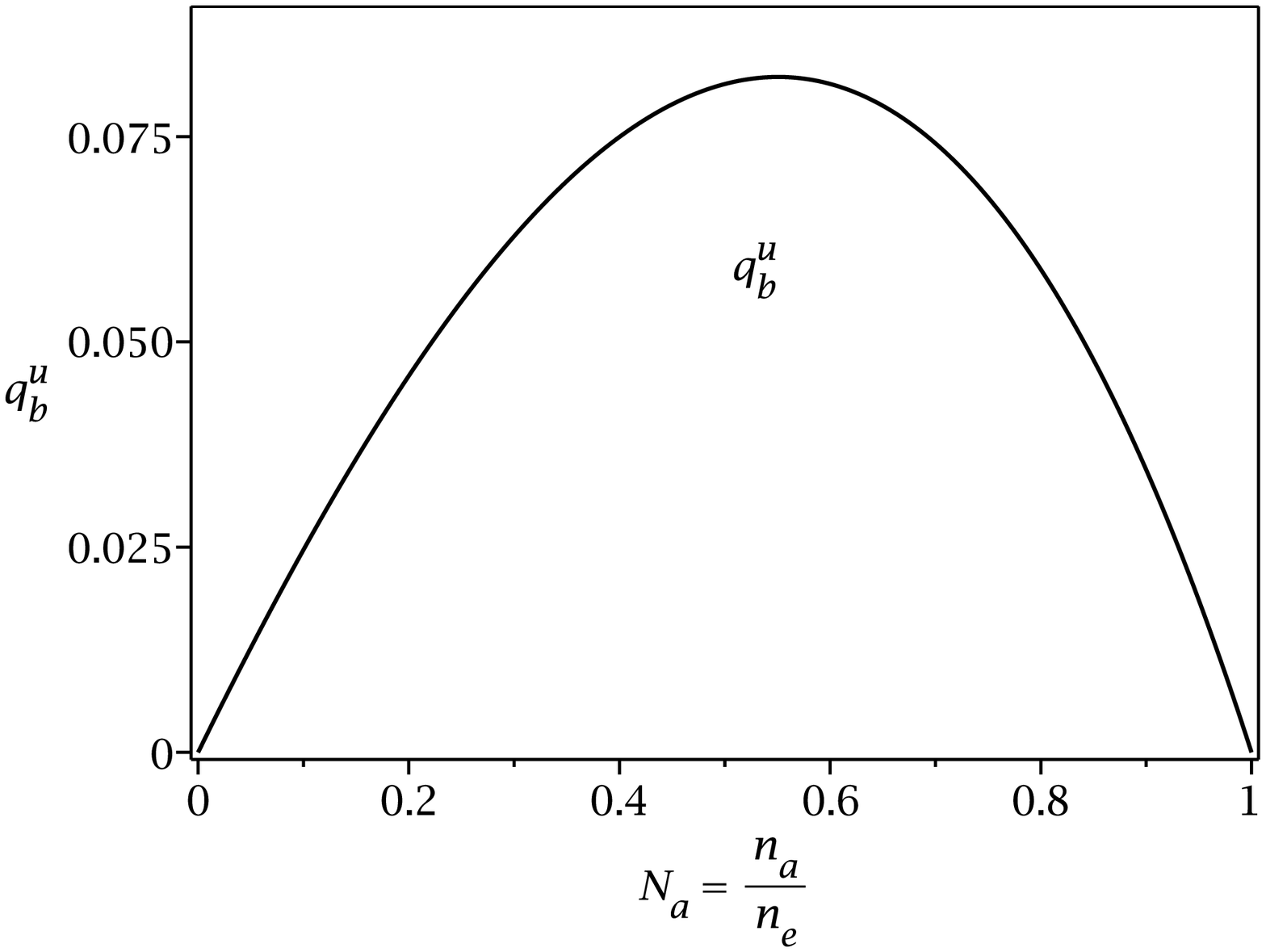}  
  \caption{Left panel: deuterium thermal conductivity $\hat{\kappa}_a$ (red)
    and tritium thermal conductivity $\hat{\kappa}_b$ (black), given by (\ref{eq:Num112x}).
    Middle panel: deuterium frictional heat flux given by $\beta_{0a}$ (\ref{eq:Num110x}).
    Right panel: tritium frictional heat flux given by $\beta_{0b}$(\ref{eq:Num111x}). Note that the frictional heat fluxes $\vecq_b^u$
    are defined with opposite signs in (\ref{eq:Num110xx}) and (\ref{eq:Num110xxx}).} \label{fig:DT1}
\end{figure*}


\subsection*{Viscosities} 
\begin{figure*}[!htpb]
  \centering
  \includegraphics[width=0.32\linewidth]{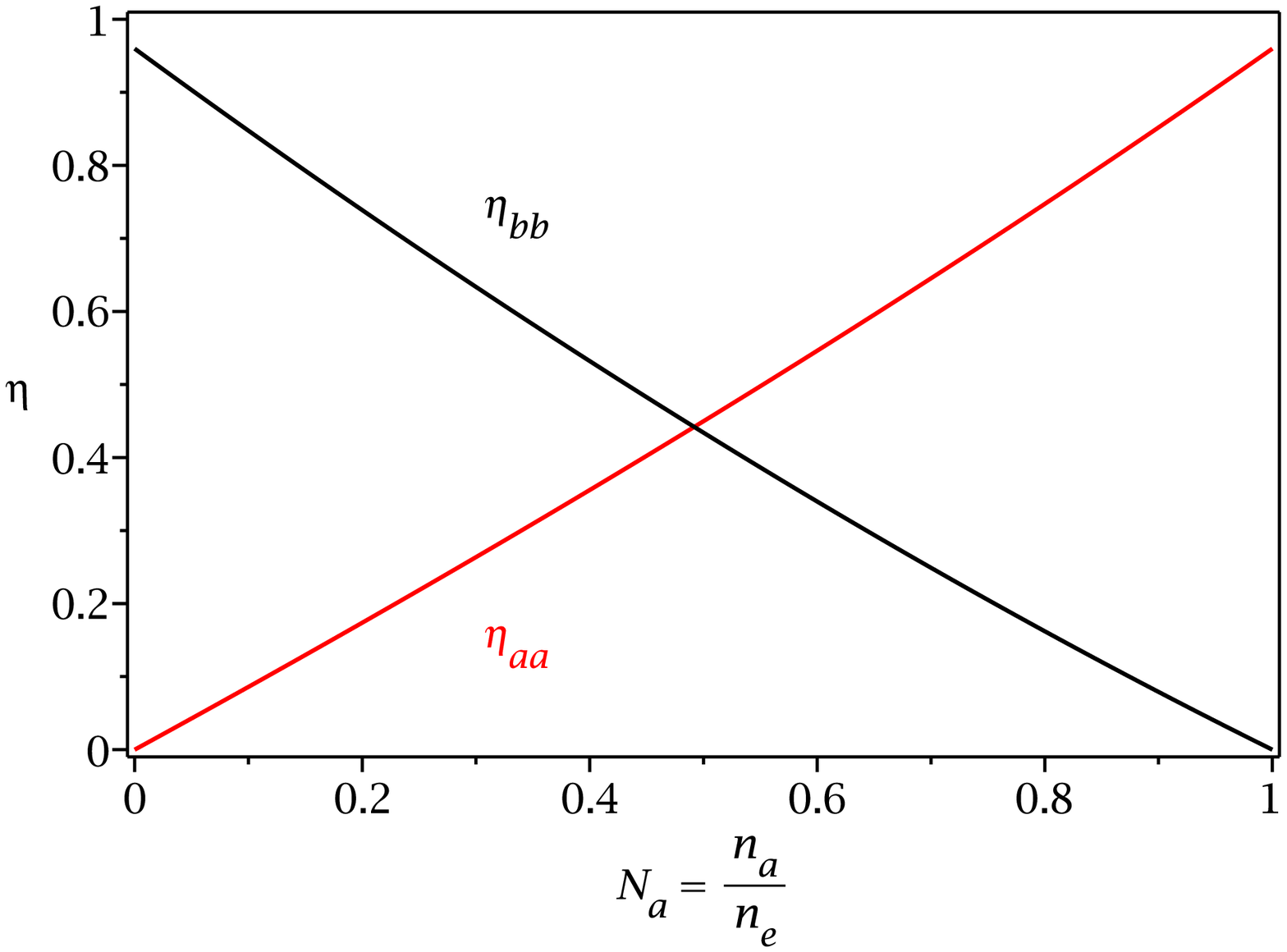}\hspace{0.02\textwidth}\includegraphics[width=0.32\linewidth]{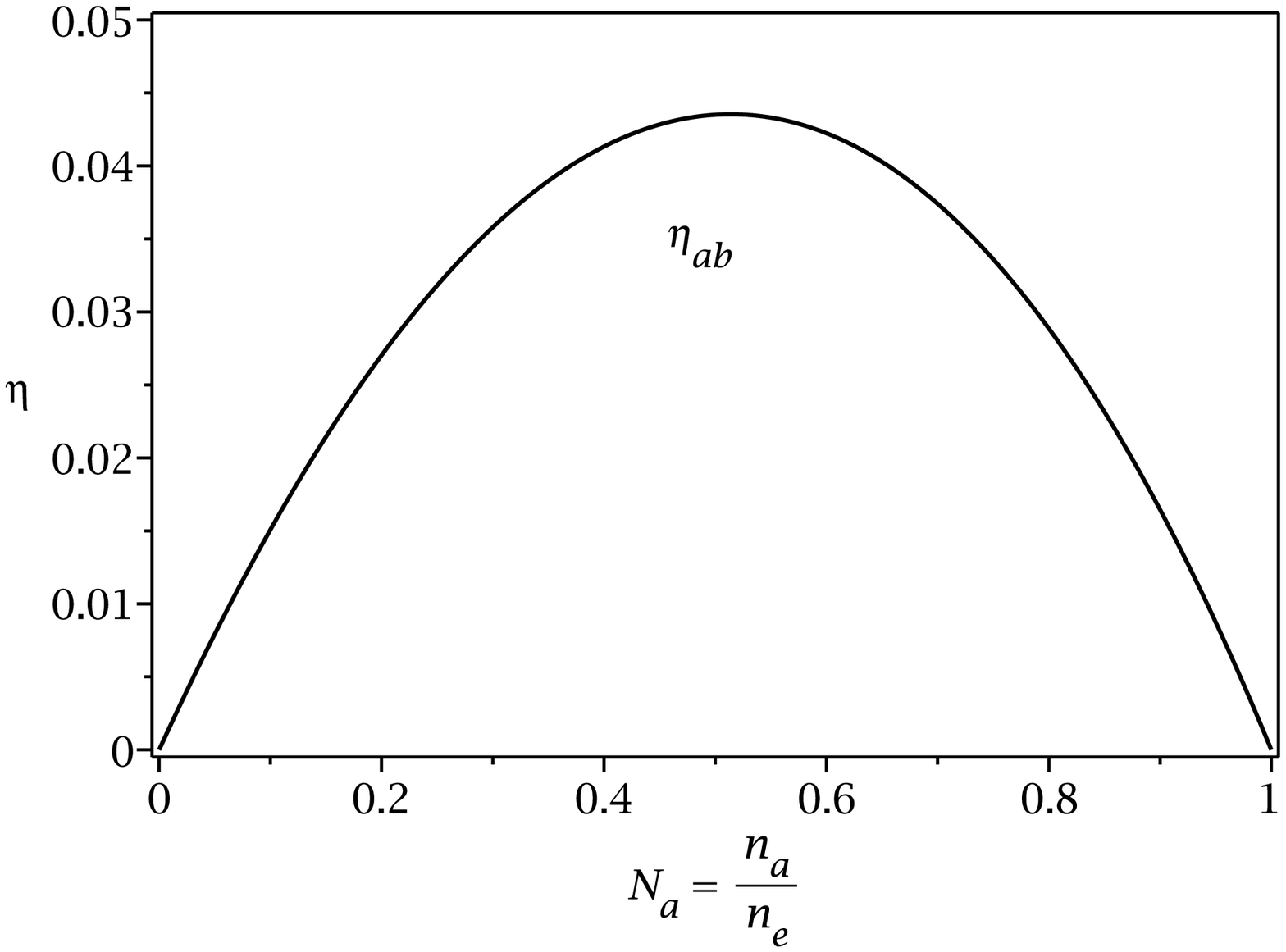}  
  \caption{ Viscosities of deuterium \& tritium plasma, according to (\ref{eq:EnergyXx}).
    Left panel: deuterium viscosity $\hat{\eta}_{aa}$ (red)
    and tritium viscosity $\hat{\eta}_{bb}$ (black). Right panel: ``cross-viscosity'' $\hat{\eta}_{ab}$.} \label{fig:DT2}
\end{figure*}

The required viscosity  mass-ratio coefficients for deuterium (``a'') and tritium (``b'') become
\begin{eqnarray}
  &&  \hat{K}_{ab (1)} = \frac{397}{50}; \qquad \hat{K}_{ab (2)} = \frac{44}{25}; \qquad
  {L}_{ab (1)} = \frac{207}{350};   \qquad  {L}_{ab (2)} = \frac{24}{175};  \nn\\
  &&  \hat{M}_{ab (1)} = \frac{166}{25}; \qquad \hat{M}_{ab (2)} = \frac{184}{125}; \qquad
  {N}_{ab (1)} = \frac{86}{875};  \qquad {N}_{ab (2)} = \frac{48}{875}; \nn\\
  &&  \hat{K}_{ba (1)} = \frac{124}{25}; \qquad \hat{K}_{ba (2)} = \frac{56}{25}; \qquad
  {L}_{ba (1)} = \frac{54}{175};   \qquad  {L}_{ba (2)} = \frac{36}{175};  \nn\\
  &&  \hat{M}_{ba (1)} = \frac{444}{125}; \qquad \hat{M}_{ba (2)} = \frac{24}{5}; \qquad
  {N}_{ba (1)} = \frac{702}{875};  \qquad {N}_{ba (2)} = \frac{324}{875}, 
\end{eqnarray}
and enter evolution equations (\ref{eq:Posled21x})-(\ref{eq:Energy22xxxx}). For an unmagnetized plasma the solutions 
read
\begin{eqnarray}
  \bPi^{(2)}_a &=& -\frac{p_a}{\nu_{aa}}\big[ \hat{\eta}_{aa} \bW_a + \hat{\eta}_{ab} \bW_b \big];\nn\\
  \bPi^{(2)}_b &=& -\frac{p_b}{\nu_{bb}}\big[ \sqrt{\frac{2}{3}}\hat{\eta}_{ab} \bW_a + \hat{\eta}_{bb} \bW_b \big], \label{eq:Num876x}
\end{eqnarray}
with numerical values
\begin{eqnarray}
  \hat{\eta}_{aa} &=& N_a (0.0046589 N_a^3+0.0064481 N_a^2 +0.17316 N_a +0.85048) /\triangle;\nn\\
  \hat{\eta}_{ab} &=& N_a (1-N_a) (0.0049729 N_a^2 +0.028578 N_a + 0.16621) /\triangle; \nn\\
  \hat{\eta}_{bb} &=& \sqrt{2/3} (1-N_a) (-0.0057061 N_a^3 -0.047294 N_a^2 -0.10519 N_a +1.17543) /\triangle;\nn\\
  \triangle &=& 1 +0.00017711 N_a^4  -0.00044516 N_a^3 -0.020987 N_a^2 +0.099409 N_a.  \label{eq:EnergyXx}
\end{eqnarray}
The solutions are written in a form so that one can directly use $\sqrt{2/3}p_b/\nu_{bb}=p_a/\nu_{aa}$, and are plotted in Figure (\ref{fig:DT2}).
To obtain more precise solutions one should include collisions with electrons (i.e. consider coupling between 3 species). Nevertheless, the
self-collisional values $1025/1068=0.96$ will only change to roughly 0.89, see for example equation (\ref{eq:Num777x}),
and the plotted viscosity profiles will not change much.

\newpage


\bibliographystyle{jpp}
\bibliography{hunana_mhd}

\end{document}